\documentclass[12pt,letter,openany,oneside,longbibliography]{book}
\usepackage{hyperref}
\usepackage{feynmf}
\usepackage[activeacute,english]{babel}
\usepackage{epsfig}
\usepackage{notoccite}
\usepackage{pst-all}
\usepackage{amsmath,amssymb}
\usepackage{amsfonts}
\usepackage{titlesec}
\usepackage{esint}
\usepackage{color}
\usepackage[nottoc,notlot,notlof]{tocbibind}
\usepackage{caption}
\usepackage{capt-of}
\usepackage[left=2.65cm, right=2.65cm, top=2.5cm, bottom=2.5cm]{geometry}
\usepackage{color}
\usepackage{calc}
\usepackage{titletoc}
\usepackage{pifont}
\usepackage{graphicx}
\usepackage{url}
\usepackage{relsize}
\usepackage{float}
\usepackage{mathtools,slashed}
\usepackage[section]{placeins}
\usepackage{caption}
\usepackage{subcaption}
\usepackage{fontenc}
\usepackage[final]{pdfpages}
\usepackage{epigraph}
\usepackage{quotchap}
\usepackage{braket}
\usepackage{lineno}
\usepackage{libertine}
\usepackage[backend=biber,hyperref=true,minbibnames=3,maxbibnames=3,maxcitenames=3,mincitenames=3,url=false,biblabel=brackets,style=phys]{biblatex}
\DeclareFieldFormat[article]{journaltitle}{\mkbibemph{#1}}
\DeclareFieldFormat[thesis]{title}{\mkbibemph{#1}}

\bibliography{PHDT.bib}
\DeclareFieldFormat[article]{journaltitle}{\mkbibemph{#1}}
\DeclareFieldFormat[thesis]{title}{\mkbibemph{#1}}
\hypersetup{
     colorlinks   = true,
	linkcolor={blue!50!black},
	citecolor={green!50!black},
	urlcolor={blue!80!black},
}
\usepackage{wrapfig}
\usepackage{mathtools,slashed}

\usepackage{epigraph}
\usepackage{quotchap}
\usepackage{braket}
\usepackage{lineno}
\usepackage{libertine}

\DeclareUnicodeCharacter{0301}{\'i}

\newcommand{\T}{\tilde}
\newcommand{\n}{\mathbf}
\newcommand{\be}{\begin{equation}}
\newcommand{\ee}{\end{equation}}
\newcommand{\bea}{\begin{eqnarray}}
\newcommand{\eea}{\end{eqnarray}}
\newcommand{\beas}{\begin{eqnarray*}}
\newcommand{\eeas}{\end{eqnarray*}}
\newcommand{\nn}{\nonumber}
\newcommand{\pp}{p_\parallel}
\newcommand{\pt}{p_\perp}

\newcommand{\ga}{\gamma_{\alpha}}
\newcommand{\gm}{\gamma_{\mu}}
\newcommand{\gn}{\gamma_{\nu}}
\newcommand{\psh}{\slashed{p}}
\newcommand{\ps}{\slashed{p}}
\newcommand{\rs}{\slashed{r}}
\newcommand{\ks}{\slashed{k}}
\newcommand{\ts}{\slashed{t}}
\newcommand{\ssl}{\slashed{s}}
\newcommand{\dpi}{(2\pi)}
\newcommand{\Op}{\mathcal{O}}
\newcommand{\gmn}{g^{\mu\nu}}
\newcommand{\gma}{g^{\mu\alpha}}
\newcommand{\gan}{g^{\alpha\nu}}

\newcommand{\eB}{\left |  q_fB\right |}
\newcommand{\etal}{\textit{et al.}}
\newcommand{\wq}{\omega_q}
\newcommand{\wwp}{\omega_p}
\newcommand{\wk}{\omega_k}
\newcommand{\slsh}{\slashed}

\newcommand{\np}{|\n{p}|}
\newcommand{\nk}{|\n{k}|}
\newcommand{\nq}{|\n{q}|}
\newcommand{\kt}{k_\perp}
\newcommand{\kp}{k_\parallel}

\newcommand{\bse}{\begin{subequations}}
\newcommand{\ese}{\end{subequations}}

\newcommand{\Pp}{\mathcal{P}^{\mu\nu}_\parallel}
\newcommand{\Pt}{\mathcal{P}^{\mu\nu}_\perp}
\newcommand{\Pcero}{\mathcal{P}^{\mu\nu}_0}
\newcommand{\factorglobal}{-\frac{i}{8\pi^2}g^2\int d^2x\,}
\newcommand{\pmu}{p^{\mu}}
\newcommand{\pnu}{p^{\nu}}
\newcommand{\kn}{k^\nu}
\newcommand{\km}{k^\mu}

\newcommand{\qt}{q_\perp}

\newcommand{\rhop}{\rho_\parallel}
\newcommand{\rhot}{\rho_\perp}
\newcommand{\B}{\mathcal{B}}
\newcommand{\p}{\parallel}
\newcommand{\Li}{\text{Li}}

\titleformat{\chapter}[display]
{\bfseries\huge}{
\vspace{1mm} \filleft \Large\chaptertitlename \resizebox{!}{1.2cm}{\bf\textup{{\thechapter}}}}
{15mm} \filcenter [\vspace{0.5mm} \vskip 3pt ]
\widenhead*{0pt}{\marginparsep}
\renewpagestyle{plain}{}
\newpagestyle{latex}[\sl\scriptsize]{
\headrule
\sethead[\thepage][\sectiontitle][]{\sectiontitle}{}{\small\thepage}}
\begin{document}



\begin{titlepage}

  \begin{center}
    \includegraphics[height=2.6cm]{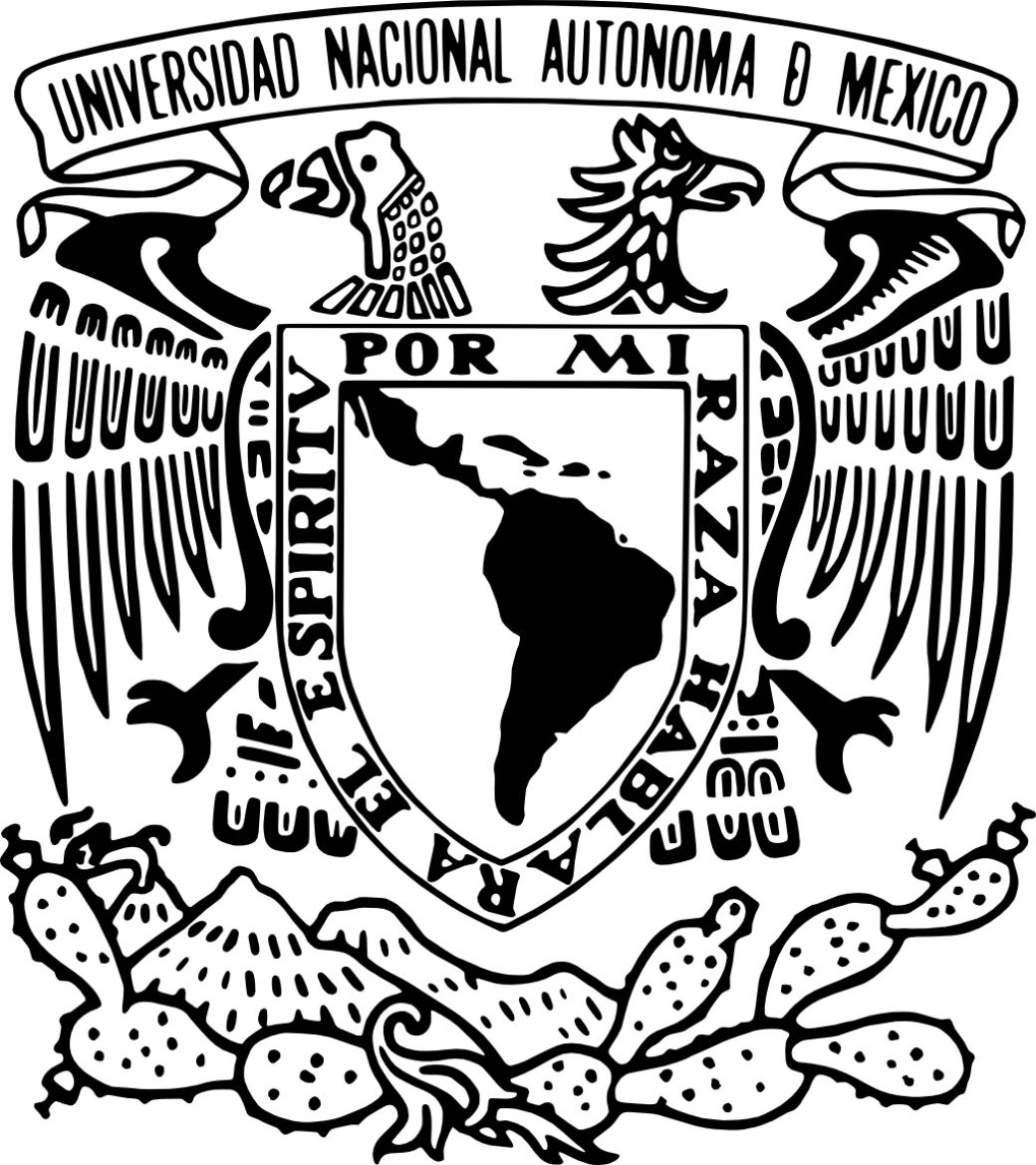}\\[10pt]
  \end{center}


  \begin{center}
    {\scshape \bfseries \large Universidad Nacional Aut\'onoma De M\'exico}\\
    {\scshape Posgrado en Ciencias F\'isicas}\\
    {\scshape Instituto de Ciencias Nucleares}

    \vspace{1.5cm}

    {\scshape \bfseries \large  Effects of Intense Magnetic Fields, High Temperature and Density on QCD-Related Phenomena}

    \vspace{2cm}

    \makebox[8cm][c]{\scshape \bfseries \Huge Tesis}\\[8pt]
    {\scshape Que para optar por el grado de:}\\[3pt]
    {\scshape Doctor en Ciencias (F\'isica)}\\[3pt]
    
     \vspace{1.5cm}
   {\scshape Presenta:}\\
    {\scshape \bfseries \large  Jorge David Casta\~no Yepes, M. Sc.}

    \vspace{1.5cm}

    {\scshape \bfseries Tutor Principal}\\ 
    {\scshape Dr. Jos\'e Alejandro Ayala Mercado (ICN-UNAM)}\vspace{0.5CM}\\
    {\scshape \bfseries Miembros del Comit\'e Tutor}\\ 
    {\scshape Dr. Sarira Sahu (ICN-UNAM)}\\  
    {\scshape Dr. Genaro Toledo S\'anchez (IF-UNAM)}

    \vspace{0.8cm}

   {\scshape Ciudad de M\'exico, Enero de 2021}
    
  \end{center}

\end{titlepage}


\includepdf[scale=0.85]{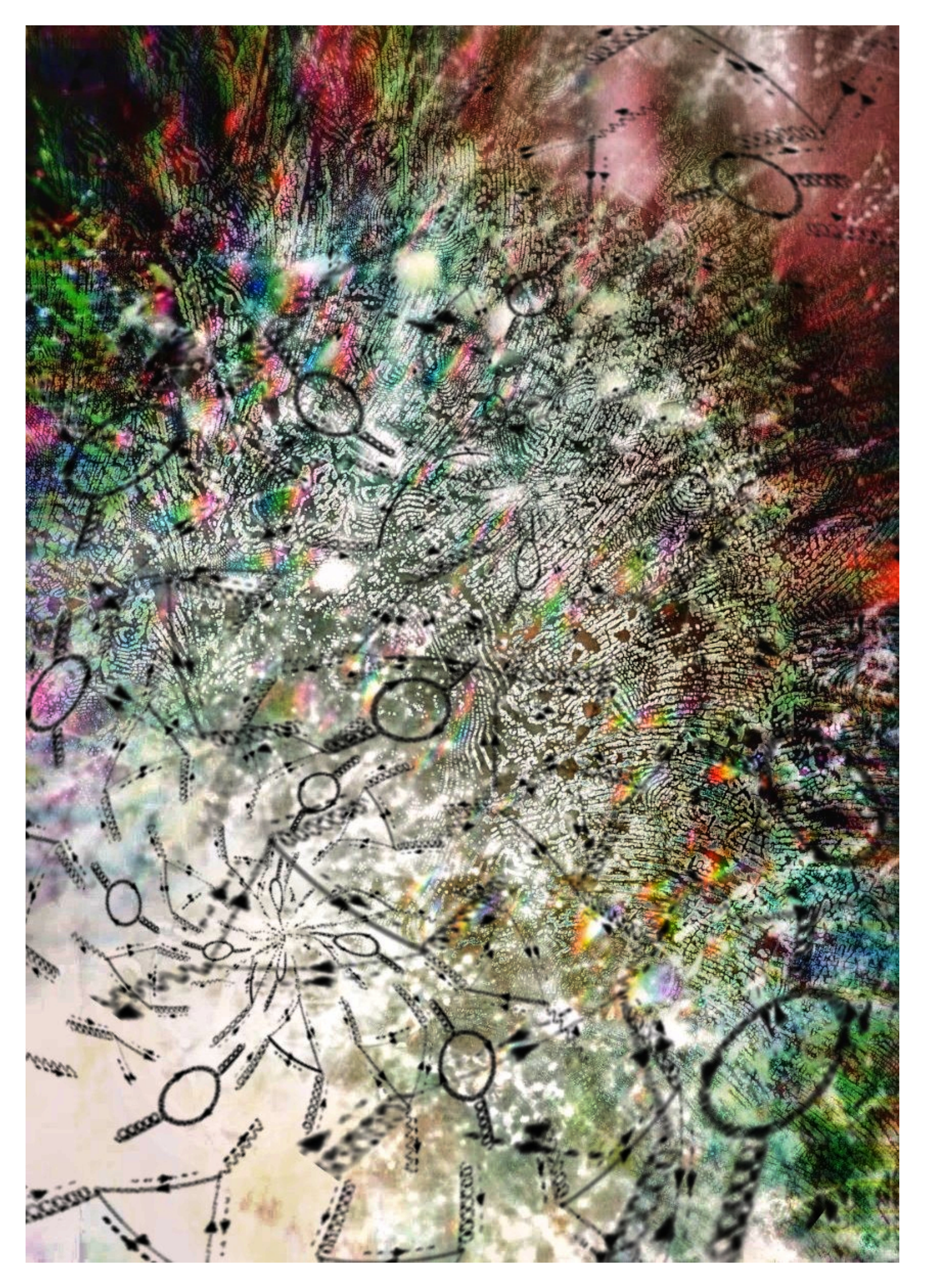}
 
\thispagestyle{empty}
\newpage
\chapter*{}
\begin{flushright}
\normalsize
\textit{To my parents and my sister.}
\vspace{8cm}
\\
\textit{``Study hard what interests you the most\\ in the most undisciplined, irreverent\\ and original manner possible.''}\\

---Richard Feynman.

\end{flushright}

\thispagestyle{empty}
\newpage
%
%
%
%


\frontmatter


\tableofcontents 
\listoffigures
\chapter*{Abstract}
\addcontentsline{toc}{chapter}{Abstract}

In this thesis, the effects of high-temperature, dense systems, and strong magnetic fields on Quantum Chromodynamics related phenomena are studied in different perspectives.  The primary motivation is provided by experimental setups in heavy-ion collisions at the RICH and LHC laboratories, were the strongly interacting matter reaches thermal, dense, and magnetized phases. 

The first problem to handle out is the QCD chiral symmetry restoration, which gives information about the phase transition between the hadronic matter to the Quark-Gluon plasma state. In order to make quantitative predictions, the Linear Sigma Model, coupled to quarks, is used as an effective model of low-energy QCD. In the framework of the thermal field theory, particularly, the imaginary-time formalism, the fermion, and boson effective potentials, up to the ring diagrams' contribution, were computed as a function of the temperature quark-chemical potential and the chiral condensate. Constrictions on the coupling constants and physical masses are imposed, leading to a sketch of the QCD-phase diagram and a plausible location of the so-called Critical End Point. The pressure is also computed, and the results show that the model reproduces, on average, the number of degrees of freedom. 

The second analysis is made at the intense magnetic field and high gluon density regimes. To improve the hydrodynamical calculations which fail to describe the photon invariant momentum distribution and the elliptic flow coefficient at low photon's momentum, two new channels for photon production are proposed: the gluon fusion and the gluon splitting. Those processes are demonstrated to be allowed by the presence of an intense magnetic field. The electromagnetic radiation produced by the named mechanism may be found at the pre-equilibrium stage called the Color Glass Condensate, given its coexistence with the intense magnetic fields in a heavy-ion collision. To consider a realistic space-temporal evolution of the magnetic background, the centrality, and the reaction volume, the proposal is complemented with simulations with the UrQMD formalism, showing a good agreement with the experimental data for photon's momentum below 1.5 GeV.

Finally, the screening properties of a thermomagnetic gluon medium are computed from the one-loop gluon polarization tensor. Starting from the fermion propagator written as a sum over Landau levels, the polarization tensor is computed for the magnetic field's arbitrary strength. It is shown that there are several thresholds associated with quark/antiquark pair-productions when the gluon's momentum resonates with a particular Landau level. Moreover, the birefringence phenomenon is found. On the other hand, the limits of Hard Thermal Loops and weak magnetic field show that the generated Debye mass changes according to the energy scales' hierarchy and the order of expansion for the magnetized fermion propagator. 

\chapter*{Resumen}
\addcontentsline{toc}{chapter}{Resumen}

En esta tesis, desde diversas perspectivas, se estudian los efectos de las temperaturas y densidades altas, as\'i como los campos magn\'eticos intensos en fen\'omenos relacionados con la Cromodin\'amica Cu\'antica (QCD). La motivaci\'on principal se fundamenta en las condiciones experimentales en colisiones de iones pesados de los laboratorios RICH y LHC, en donde la materia que interact\'ua fuertemente alcanza fases t\'ermicas, densas y magnetizadas.

El primer problema a tratar es la restauraci\'on dela simetr\'ia quiral de la QCD, la cual da informaci\'on acerca de la transici\'on  de fase de la materia hadr\'onica al estado de Plasma de Quarks y Gluones. Para hacer predicciones cuantitativas, el Modelo Sigma Lineal coplado con quarks es usado como modelo efectivo para QCD a baja energ\'ia. Mediante el formalismo de la teor\'ia t\'ermica de campos, en particular, el desarrollo en el tiempo imaginario, los potenciales efectivos para fermiones y bosones son calculados con correcciones de diagramas de anillo. Dchos potenciales resultan en fuciones de la temperatura, el potencial qu\'imico de quarks y el condensado quiral. Se imponen restricciones a las constantes de acoplamiento, que en conjunto con los valores de las masas f\'isicas, permiten hacer un esbozo del diagrama de fases para la QCD y dar una posible ubicaci\'on del Punto Cr\'itico. La presi\'on tambi\'en es calculada, mostrando que el modelo utilizado s\'olo reproduce en promedio los grados de libertad del sistema. 

El segundo an\'alisis se centra en el r\'egimen de campo magn\'etico intenso y alta densidad de gluones. Con el fin de mejorar los c\'alculos hidrodin\'amicos, los cuales fallan a la hora de describir la distribuci\'on invariante de fotones y el coeficiente de flujo el\'iptico, se proponen dos nuevos canales para la producci\'on de fotones: la fusi\'on y la divisi\'on de gluones. El formalismo muestra que dichos procesos est\'an permitidos siempre y cuando exista un campo magn\'etico de fondo, lo cual implica que este tipo de radiaci\'on proviene de etapa de pre-equilibrio denominada {\it Color Glass Condensate}, que coexiste con campos magn\'eticos intensos generados por la colisi\'on. Para considerar una evoluci\'on espacio-temporal realista del campo magn\'etico, de la centralidad y del volumen de interacci\'on, la propuesta es complementada con simulaciones en el marco del formalismo UrQMD, mostrando que los resultados tienen un buen acuerdo con los datos experimentales cuando el momento del fot\'on est\'a por debajo de 1.5 GeV.

Finalmente, las propiedades de apantallamiento para el glu\'on en un medio termomagn\'etico son calculadas a partir del tensor de polarizaci\'on a un bucle. Empezando por el propagador de fermin\'on escrito como una suma de niveles de Landau, el tensor de polarizaci\'on se obtiene para una intensidad arbitraria del campo magn\'etico. Se muestra que existen diferentes umbrales, los cuales se asocian a producci\'on de pares quark/antiquark cuando el momento del glu\'on resuena con alguno de los niveles de Landau. Adem\'as, se encuentra que el fen\'omeno de birrefringencia existe para este sistema. Por otro lado, en los l\'mites de alta temperatura ({\it Hard Thermal Loop}) y campo magn\'etico d\'ebil, se muestra que la masa de Debye cambia de acuerdo a la gerarqu\'ia entre las escalas de energ\'ia y el orden de la expansi\'on del propagador de fermi\'on magnetizado.  
\begin{savequote}[50mm]
``If each of my words were a drop of water, you would see through them and glimpse what I feel: gratitude, acknowledgment.''
\qauthor{--- Octavio Paz.}
\end{savequote}

\chapter{Acknowledgments}

I am indebted to Mexico: all my present (and future) career is thanks to this beautiful country, its people, and the support that they gave me. Words are not enough to express my gratitude.

To the Universidad Nacional Aut\'onoma de M\'exico for opening its doors and allowing me to be part of the highest house of studies in Latin America: it is an honor and a dream come true to call it my {\it alma mater}.

Of course, this process would not have been possible without the guidance of my supervisor Alejandro Ayala. Since my master's studies, he has become a unique mentor in academia and life. His patience, motivation, and friendship are inspirational, and he has become an example to follow. 

Many thanks to the Matter Under Extreme Conditions group members. In particular, to Malena, Alfredo Raya, Sa\'ul, Luis, Jordi, and Isabel. Their close collaboration over all these years was fundamental to the results presented in this thesis. 

I also want to thank professors Marcelo Loewe and Enrique Mu\~noz for allowing me to do two research stays at the Pontificia Universidad Cat\'olica de Chile. Their academic support and hospitality made these visits one of the most enriching experiences I ever had. 

To all my officemates, Turri, Omar, Edgar Rosas, Louis, Edgar Guzm\'an, Juan Carlos, Juan Jos\'e, Tonatiuh, Hugo, and Eduardo: physics, laughs, coffee and friends always will be an excellent combination. 

Thanks to Esa\'u Ortiz for the back cover art. {\it Now, the boutonniere may say ``Casta\~no'', but the hat says ``Ev\'u''}.

Thanks to Cristian, Erika, Elkinn, Medina, Ang\'elica, Juli\'an, Giovanna, and Ariadna: their presence makes all the things better. 

And of course, thanks to Alexis and his family for the support during the last two years: I'm in debt with them. Thank you for letting me in.

Finally, I cannot imagine a world without the company of Paps, Edith, Amelia, Nela, Lowen, Astra, Od\'in and Ori\'on. {\it Until one has loved an animal a part of one's soul remains unawakened.}~\footnote{Quote by Anatole France.}

\chapter*{Statement of Originality}
\addcontentsline{toc}{chapter}{Statement of Originality}

The calculations and results presented in this thesis are original and collaborative work. 

Chapter~\ref{Chap:QCD_phase_diagram} and Appendix~\ref{Ap:Veff_bosons_HT} cover my contributions to calculating the linear sigma model's thermodynamic potentials, as well as the discussion, guidance, and corrections made by the authors of the corresponding article in Ref.~\cite{ayala2016chiral}. 

Chapter~\ref{chap:Photons} and Appendices~\ref{ApPhaseFactorCalc},~\ref{ThreeQuarksinLLL},~\ref{ApM1Calculation}, and~\ref{Ap_M_squared} have my calculations for the matrix elements, photon's momentum invariant distribution, and elliptic flow corresponding to the gluon fusion mechanism. All the procedures were discussed, revised, and corrected by the authors of Ref.~\cite{PhysRevD.96.014023}. The computational details of Sec.~\ref{sec:gluon_splitting} were addressed by Jordi Salinas, Dr. Maria Elena Tejeda-Yeomans, and Dr. Isabel Dominguez Jimenez. My contribution consisted of the calculation of the gluon splitting channel and discussion of the results. All the authors of Ref.~\cite{Ayala:2019jey} agreed to consing such information in the present thesis. 

Chapter~\ref{Chap:Gluon_Pol_Tensor} and Appendices~\ref{ApAGluonPolTensorPRD} and~\ref{ApBGluonPolPRD} present my work at the academic stay in the Pontificia Universidad Cat\'olica de Chile. Those calculations were revised and discussed by all the authors of Ref.~\cite{PhysRevD.101.036016}. Section~\ref{sec:GluonPolTenThermoMag} is included because of discussions coming from the findings of Sec.~\ref{sec:Magnetized_Gluon_Polarization_Tensor_from_the_Sum_of_All_Landau_Levels}, which leads to expand and correct the results of previous work and is published in Ref.~\cite{RMFGluon}. Section~\ref{sec:GluonPolTenWeakField} is the result of collaborative work and is currently undergoing peer-review~\cite{ayala2020gluon}.

The other sections constitute the state-of-the-art around the topics presented. The figures were reproduced with their respective citations.

\chapter*{List of Publications}
\addcontentsline{toc}{chapter}{List of Publications}

This thesis is based on the following published articles:
\begin{enumerate}
\item {\it Chiral symmetry transition in the linear sigma model with quarks: Counting effective QCD degrees of freedom from low to high temperature.}\\
Alejandro Ayala, \underline{Jorge David Casta\~no-Yepes}, J. J. Cobos-Martinez, Saul Hernandez-Ortiz, Ana Julia Mizher, Alfredo Raya.\\
    Int. J. Mod. Phys. A {\bf 31}, 1650199 (2016).
    
\item {\it Prompt photon yield and elliptic flow from gluon fusion induced by magnetic fields in relativistic heavy-ion collisions.}\\
Alejandro Ayala, \underline{Jorge David Casta\~no-Yepes}, C. A. Domingez, L. A. Hernandez, Sa\'ul Hern\'andez-Ort\'iz, Mar\'ia Elena Tejeda-Yeomans.\\
Phys. Rev. D \textbf{96}, 014023 (2017).

\item{\it Centrality dependence of photon yield and elliptic flow from gluon fusion and splitting induced by magnetic fields in relativistic heavy-ion collisions.}\\
Alejandro Ayala, \underline{Jorge David Casta\~no-Yepes}, Isabel Dom\'inguez, Jordi Salinas San Mart\'in, Mar\'ia Elena Tejeda-Yeomans.\\
Eur .Phys. J. A {\bf 56}, 53 (2020).

\item {\it Gluon polarization tensor in a magnetized medium: Analytic approach starting from the sum over Landau levels.}\\
Alejandro Ayala, \underline{Jorge David Casta\~no-Yepes}, Enrique Mu\~noz, Marcelo Loewe.\\
Phys. Rev. D {\bf 101}, 036016 (2020).

\item {\it Thermal corrections to the gluon magnetic Debye mass}\\
Alejandro Ayala, \underline{Jorge David Casta\~no-Yepes},  C. A. Dominguez, S. Hernandez-Ortiz, L. A. Hernandez, M. Loewe, D. Manreza Paret, R. Zamora.\\
Rev. Mex. Fis. {\bf 66} (4) 446-461, (2020).
\end{enumerate}
and the manuscript under peer-review:
\begin{enumerate}
    \item {\it Gluon polarization tensor and dispersion relation in a weakly magnetized medium.}\\
    Alejandro Ayala, \underline{Jorge David Casta\~no-Yepes}, L. A. Hern\'andez, Jordi Salinas, R. Zamora.\\
    preprint arXiv:2009.00830.
\end{enumerate}

Also, participation in several meetings produced the following proceedings:
\begin{enumerate}
\item {\it Thermal photons from gluon fusion with magnetic fields.}\\
Alejandro Ayala, \underline{Jorge David Casta\~no-Yepes}, C. A. Dominguez, L. A. Hernandez.\\
EPJ Web of Conferences {\bf 141}, 02007 (2017).

\item {\it Using the Linear Sigma Model with quarks to describe the QCD phase diagram and to locate the critical end point.}\\
Alejandro Ayala, \underline{Jorge David Casta\~no-Yepes}, Jos\'e Antonio Flores, Sa\'ul Hern\'andez-Ort\'iz, L. A. Hern\'andez.\\
EPJ Web of Conferences \textbf{172}, 08002 (2018).

\item {\it Prompt photon yield and $v_2$ coefficient from gluon fusion induced by magnetic field in heavy-ion collision.}\\
Alejandro Ayala, \underline{Jorge David Casta\~no-Yepes}, C. A. Domingez, L. A. Hern\'andez, Sa\'ul Hern\'andez-Ort\'iz, Mar\'ia Elena Tejeda-Yeomans.\\
EPJ Web of Conferences \textbf{172}, 08004 (2018).

\item {\it New channels of prompt-photon production by magnetic fields in heavy-ion collisions}.\\
Alejandro Ayala, \underline{Jorge David Casta\~no-Yepes}, Isabel Dom\'inguez, Jordi Salinas San Mart\'in, Mar\'ia Elena Tejeda-Yeomans.\\
J. Phys. Conf. Ser. \textbf{1602}, 012014 (2020).
\end{enumerate}

And as product of collaborations in different topics, the following articles were published:
\begin{enumerate}
\item {\it In situ photoacoustic characterization for porous silicon growing: detection principles.}\\
C. F. Ramirez-Gutierrez, \underline{Jorge David Casta\~no-Yepes}, M. E. Rodriguez-Garcia.\\
J. Appl. Phys. {\bf 119}, 185103 (2016).

\item{\it Modeling the photoacoustic signal during the porous silicon formation.}\\
C. F. Ramirez-Gutierrez, \underline{Jorge David Casta\~no-Yepes}, M. E. Rodriguez-Garcia.\\
J. Appl. Phys. {\bf 121}, 025103 (2017).

\item{\it A comparative study on heat capacity, magnetization and magnetic susceptibility for a GaAs quantum dot with asymmetric confinement.}\\
\underline{J. D. Casta\~no-Yepes}, C. F. Ramirez-Gutierrez, H. Correa-Gallego, Edgar A. G\'omez.\\
Phys. E. \textbf{103}, 464-470 (2018).

\item{\it Perturbation theory for open quantum systems at the steady state.}\\ Edgar A. G\'omez, \underline{Jorge David Casta\~no-Yepes}, Saravana Prakash Thirumuruganandham\\
Results in Physics \textbf{10}, 353-355, (2018). 

\item{\it Porosity and roughness determination of porous silicon thin films by genetic algorithms.}\\ 
C. F. Ramirez-Gutierrez, \underline{Jorge David Casta\~no-Yepes}, M. E. Rodriguez-Garcia.\\
Optik \textbf{173}, 271-278 (2018).

\item{\it Optical interferometry and photoacoustics as   in-situ techniques to characterize the porous silicon formation: a review.}\\ C. F. Ramirez-Gutierrez, \underline{Jorge David Casta\~no-Yepes}, M. E. Rodriguez-Garcia.\\
Open Mater. Sci. \textbf{4}, 23–32 (2018).

\item{\it Effects of the interface roughness in the optical response of one-dimensional photonic crystals of porous silicon.}\\
I .A. Lujan-Cabrera, C. F. Ramirez-Gutierrez, \underline{Jorge David Casta\~no-Yepes}, M. E. Rodriguez-Garcia.\\
Phys. B. \textbf{560}, 133-139, (2019).

\item{\it Impact of a topological defect and Rashba spin-orbit interaction on the thermo-magnetic and optical properties of a 2D semiconductor quantum dot with Gaussian confinement.}\\
\underline{Jorge David Casta\~no-Yepes}, D. A. Amor-Quiroz, C. F. Ramirez-Gutierrez, and Edgar A. G\'omez.\\
Phys. E. \textbf{109}, 59-66, (2019).

\item{\it Super-statistical description of thermo-magnetic properties of a system of 2DGaAs quantum dots with gaussian confinement and Rashba spin–orbit interaction.}\\
\underline{Jorge David Casta\~no-Yepes}  and D. A. Amor-Quiroz.\\
Phys. A {\bf 548}, 123871 (2020).

\item{\it Optical intersubband properties of a core-shell semiconductor-topological insulator quantum dot described by $\theta$-Electrodynamics.}\\
\underline{Jorge David Casta\~no-Yepes}, O. J. Franca, C. F. Ramirez-Gutierrez, J. C. del Valle.\\
Phys. E. {\bf 123}, 114202 (2020).

\item{\it Comments on ``Superstatistical properties of the one-dimensional Dirac oscillator'' by Abdelmalek Boumali et al.}\\
\underline{Jorge David Casta\~no-Yepes}, I. A. Lujan-Cabrera, C. F. Ramirez-Gutierrez.\\
Phys. A., 125206 (2020)-- In Press.
\end{enumerate}

\cleardoublepage
\mainmatter
\chapter*{Introduction}
\addcontentsline{toc}{chapter}{Introduction}

In the study of nuclear and subnuclear matter, Quantum Chromodynamics (QCD) is the fundamental theory that describes the degrees of freedom related to the strong interaction, i.e., the quarks and gluons~\cite{GellMann1,Zweigbookhadrons}. From the beginning of the QCD formulation, two remarkable characteristics of that theory make it difficult to study from the theoretical and experimental perspectives: the {\it Asymptotic Freedom}, which difficults the calculations from a perturbative scheme for arbitrary energy scales~\cite{AsymptoticFreedom1,AsymptoticFreedom2}, and the {\it Color Confinement}, that forbids quarks and gluons to be isolated and independent entities~\cite{PhysRevD.10.2445,greensite2011introduction}. Even so, QCD has become an extremely successful theory, and all its predictions have been tested in particle accelerators over more than 50 years.

The QCD's theoretical vacuum properties are extensively studied, from the processes involving only quarks and gluons to the hadron's spectrum description. Experimental findings complement these studies, giving a global understanding of the strong interaction in normal conditions~\cite{greiner2007quantum}. However, relativistic heavy-ion experiments carried out at the BNL Relativistic Heavy-Ion Collider (RHIC) and CERN Large Hadron Collider (LHC) had found relevant information which proves the existence of a state of matter in where quarks and gluons are not confined to individual nucleons in the so-called \textit{Quark-Gluon Plasma} (QGP)~\cite{experiments1,experiments2,experiments3,experiments4,experiments5,experiments6}. In such experiments, there are several and remarkable differences between the results obtained when one considers proton-proton ($p+p$) and nucleus-nucleus ($A+A$) collisions~\cite{ALICEexperiment,azimutalcorrelationsALICE,AzimutalcorrelationsCMS,WeiLi,MultiplicityLHC}. In principle, the differences between both kinds of reactions can be attributed to the amount of matter which is involved in the collision: at first approximation, the proton can be considered as a point-like particle, and therefore, a $p+p$-collision does not create a strongly interacting medium which screens or interacts with the reaction products~\footnote{There are signatures of the QGP formation in \textit{pp}-collisions. See Refs.~\cite{QGPinpp1,QGPinpp2,QGPinpp3,QGPinpp4}}. Thus, the in-medium QCD's characterization is expected to be realized by analyzing the data obtained from \textit{AA}-experiments whose participants have a large number of nucleons (Au or Pb) which can interact total or partially in a region of a few Fermis long.

The energy deposited in the region created during heavy-ion collision is finally released into particle production so that one can expect collective dynamics and the possibility to study the theory in extreme conditions as high-temperature, large densities, and intense external magnetic fields. That situation becomes of great interest because, in principle, ultra-relativistic limits that give high temperatures and densities can reach the proper energy scales of QCD breaking the confinement of quarks and gluons, which brings insights into the evolution of the early universe. Nevertheless, because its features and the uncontrolled situation after the nuclear collision, the evolution of the strongly interacting matter passes through several phases, each characterized by energy scales, bulk parameters, and strength of its interactions. Three main phases are of particular interest: the pre-equilibrium state denominated color glass condensate or {\it Glasma}, the thermalized quark-gluon plasma, and the final stage of hadronization. The QCD asymptotic freedom and the color confinement play a crucial role in the theoretical description of such stages, particularly the very early one admits an perturbative approach. On the other hand, for example, high-density phases characterized by collective, non
linear phenomena need computational tools beyond the ordinary perturbation theory.

\begin{figure}[h]
    \centering
    \includegraphics[scale=0.75]{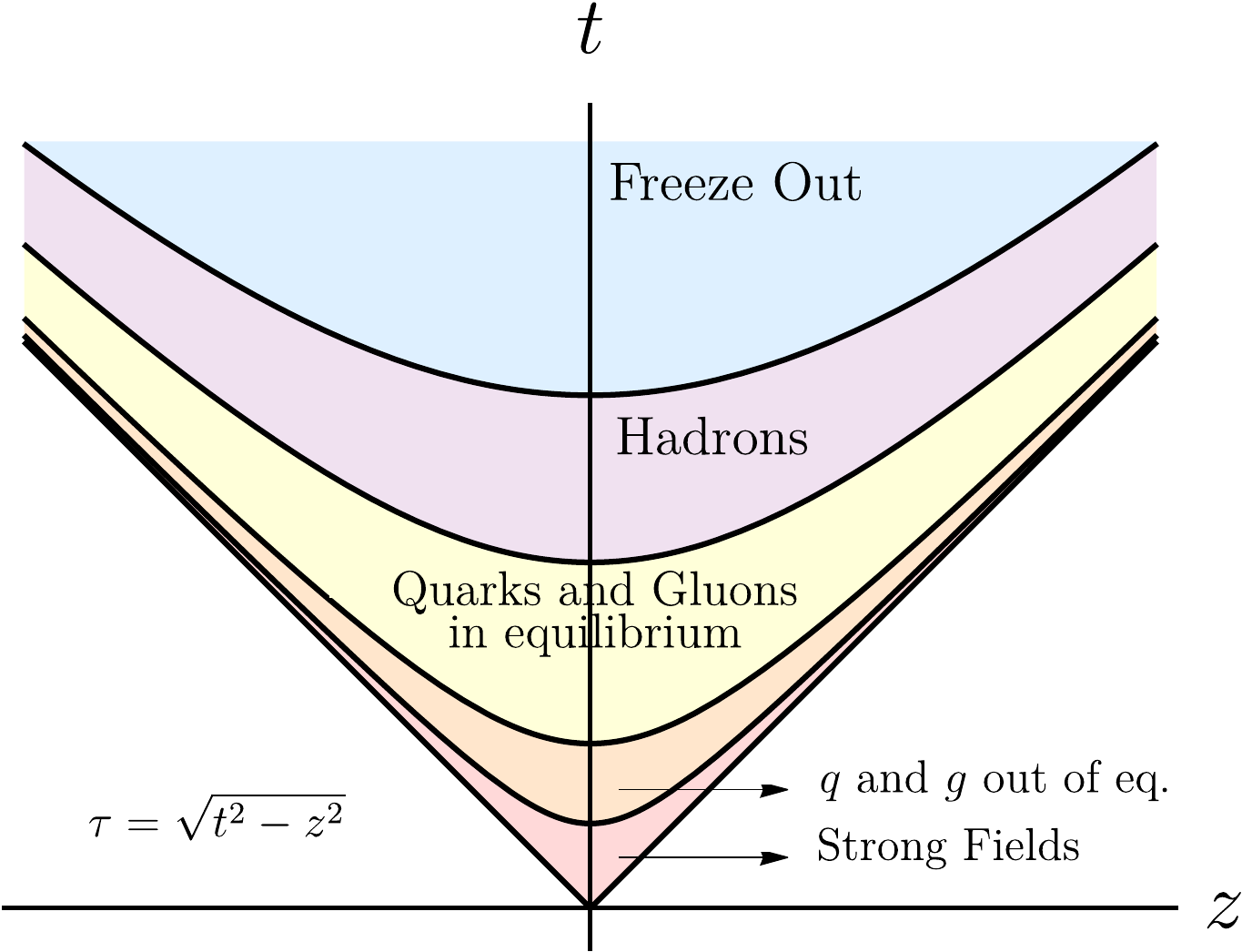}
    \caption{Lorentz diagram which schematizes the stages in a relativistic heavy-ion collision. The $z$-coordinate represents the collision axis, and the proper time $\tau$ characterizes the duration of each phase.}
    \label{Fig:Lorentz_diagram_HIC}
\end{figure}

The space-time evolution of a heavy-ion collision is schematized in Fig.~\ref{Fig:Lorentz_diagram_HIC}, giving a general description of the main QCD phases as follows~\cite{iancu2012qcd}:

Before the collision ($t<0$), in the center-of-mass frame, which corresponds to experiments carried out at LHC and RHIC, both nuclei are contracted longitudinally by a Lorentz factor $\gamma\sim100$, so that the colliding objects resemble two thin disks. In the described situation, disks have an abundant number of gluons with a small fraction $x\ll1$ of the longitudinal momentum, and therefore, by the Heisenberg's dispersion relations, the gluon density is increased by $1/x$ implying relatively sizeable transverse momentum. For example, in a typical Au-Au or Pb-Pb collisions $\kt\sim2$ GeV for $x=10^{-4}$; thus, given the QCD asymptotic freedom, the coupling between such gluons is weak.  This stage is known as the {\it color glass condensate} (CGC).

At the moment of the collision ($\tau=0$) the processes involving considerable momentum transference ($Q\gtrsim 
10$ GeV) evolve fastly ($\tau\sim 1/Q$), producing {\it hard particles} which have high energy and transverse momenta, such as hadronic jets, direct photons, dilepton pairs, heavy quarks, and vector bosons, which are easily identified at the final stage of the reaction. After the collision, at $\tau\simeq 0.2$ fm/c corresponding to a momentum scales $Q\sim 1$ GeV, the gluon content of each nucleus breaks into partonic constituents. The dynamics at this point is fundamental for final observables because most of the particles detected come from the hadronization of such partons. A gluonic high-density medium precedes the partonic production, which has an energy density estimated around $\epsilon\gtrsim15$ GeV/fm$^3$ for Pb-Pb collision at the LHC, corresponding to 3 times greater than in RHIC Au-Au experiments and three times larger than the nuclear matter. Such a stage is not in equilibrium; i.e., the temperature cannot be defined yet; nevertheless, it forms a boson condensate with particular features, which is the so-called {\it glasma}. 

The next step concerns the QGP identification directly: in a proton-proton collision, the produced particles have a weak interaction between them, leading to a rapid separation and independent dynamics until the hadronization stage. That means that in such collision, there is not an interacting medium for the liberated partons. Nevertheless, for heavy-ion reactions, the situation becomes more complicated, given that partons liberated by the collision do interact with a stronger coupling and exhibit collective phenomena. The experimental data shows that the named partonic matter reaches thermal equilibrium in a short thermalization time $\tau\sim 1$ fm/c, which is a characteristic of strong interactions between constituents making the perturbative methods inapplicable. The rapid thermalization regime is followed by the QGP phase, where quarks and gluons are practically free. From the fact that the overall system keeps expanding and cooling down, it is necessary to assume only local thermal equilibrium in a way that the temperature is spatially inhomogeneus. 

The transit from quarks and gluons to hadrons, i.e., the hadronization stage, occurs when the local temperature reaches the confinement temperature $T_c=150-180$ MeV (see Chapter~\ref{Chap:QCD_phase_diagram}) after a time of $\tau\sim 10$ fm/c. There exist a time window where the hadrons become a system relatively dense so that it is still possible to define a notion of temperature. Its duration is about $10\lesssim \tau\lesssim20$ fm/c, and it is identified as a {\it hadron gas} that continues expanding and cooling down. Finally, when $\tau\sim20$ fm/c, the hadron gas density decreases to a point where its constituents do not interact anymore so that the interaction rate is lower than the expansion rate. Such a scenario is known as the {\it freeze-out} regime. It is assumed that detected hadrons come from this stage, and the momentum distribution has a thermal signature of the last fluid-like phase just before the freeze-out. The data analysis shows that indeed the particle spectra are well described by Maxwell-Boltzmann thermal distributions, obtaining a freeze-out temperature $T_f\simeq 170$ MeV.

From the fact that the coupling strength of the several stages in the heavy-ion-collision varies considerably, it is logical to work with QCD effective models in each regime. Also, note that not only a coupling constant based description has to be made, but the active degrees of freedom have also be taken into account. The linear sigma model presented in Chapter~\ref{Chap:QCD_phase_diagram} is an excellent example of such effective degrees of freedom. In general terms, the theoretical framework currently adopted in the calculations is outlined in the following structure: 

\begin{itemize}
    \item At the very early stages, the dynamic is governed by gluons in a high density and not thermalized states: the CGC and the glasma state. Both are characterized by a semi-hard energy scale with $\alpha_s\sim0.3$, and although the perturbation theory is formally valid, an expansion in powers of $\alpha_s$ is not straightforwardly obtained given by it is necessary to sum an infinite class of Feynman graphs enhanced by the high energy and partonic occupation.  The latter gives rise to the so-called {\it CGC effective theory}, which takes into account fundamental aspects as the nuclei wave functions before the collision and the relevant phenomenology at the early stages of the reaction. Such formalism offers quasi-classical color fields as the relevant degrees of freedom.  A field point of view instead of a particle description obeys to the dense medium since the phase-space occupation numbers are large so that the particle overlap resembles coherent states which are well treated by classical fields.
    
    \item At the next stage, the system expands and becomes less dense; hence, the particle treatment is relevant.  This point is understood as a breaking of the classical fields into quarks and gluons. Moreover, given that the coupling evolution is continuous, it is expected that such degrees of freedom are weakly coupled and admit a kinetic theory model. That formalism suggests that the free path between two successive collisions is much longer than any other microscopic scale (because the small $\alpha_s$). Still, it cannot explain the rapid thermalization before the QGP formation.
    
    \item The locally thermalized QGP phase lies in a regime between the strong and weak coupling: the temperature estimation for such stage is around $T\sim500-600$ MeV so that $\alpha_s\sim 0.3-0.4$. Nonetheless, several effective models can be performed at this point, as well as lattice QCD simulations, which serve as a comparative guide. Of particular interest, thermodynamic quantities as the QGP pressure or energy density are well known from several calculation methods. In particular, if the coupling is weak, techniques such as the {\it Hard Thermal Loop (HTL)} effective theory have a good agreement with the lattice simulations. The latter is a version of the kinetic theory which assumes long-range interactions of thermalized quasi-particles in a picture where their masses and couplings are temperature-dependent. In that sense, the QGP is identified as a soft-interacting gas of thermal quarks and gluons. 
    
    \item Once the collision has reached the QGP phase, the system is assumed to be in local thermal equilibrium, given its continuous expansion. At this stage, the {\it Relativistic Hydrodynamics} is the adequate effective theory to describe the space-time evolution.
\end{itemize}

This thesis is devoted to studying QCD phenomena related to heavy-ion collisions' early and thermal stages, where the temperature, density, and magnetic fields are extreme. 

In the high temperature-baryonic density regimes, it is established that a phase transition achieves the transit from the hadronic matter to the QGP. Moreover, besides the confinement, the chiral symmetry breaking is present in the QCD formalism, leading to light Goldstone particles' description, identified as pions and a non-zero quark condensate. The phase transition from the QGP to hadronic degrees of freedom is then associated with the spontaneous symmetry breaking of the chiral symmetry allowing chiral condensates' formation. On the other hand, since the fundamental QCD's scale is $\Lambda_\text{QCD}\sim200$ MeV, it is expected that quarks and gluons would be liberated for temperatures around $T\sim\Lambda_\text{QCD}\sim\Op(10^{12})$ K and/or baryonic densities $n_B\sim\Lambda_\text{QCD}^3\sim 1$ fm$^{-3}$. The named critical values may be present in the deep interior of compact stellar objects like neutron stars, where the temperature is low, but the density is increased. Several efforts to observe and know the equation of state for those objects have been conducted to test the deconfinement predictions~\cite{Heiselberg:1999mq}. Clearly, the heavy-ion collisions provide the desired dense and hot mediums.

The nature and structure of the phase transition can be described from theoretical grounds. In the $T$-$n_B$ plane, the critical temperature $T_c$ is associated with the Hagedorn's limiting temperature $T_H$ using the Statistical Bootstrap Model (SBM), and the phase transition is identified as second-order~\cite{Cabibbo:1975ig,Hagedorn:1965st}. In a primitive approximation, the hadron resonance gas at finite temperature is a good starting point so that the density of states $\rho(m)$ (predominantly mesons) is proportional to $\exp\left(m/T_H\right)$, where $m$ is the resonance mass and $T_H\simeq0.19$ GeV is the Regge slope parameter~\cite{Fukushima:2008pe}. At the partition function level, the exponential growth is balanced by the Boltzmann factor; thus, the integration over $m$ is singular for $T>T_H$, yielding $T_H$'s identification as the Hagedorn temperature. A similar argument is applied to know the critical baryonic chemical potential $n_B^c$: the density of states and the Boltzmann factor are $\exp\left(m_B/T_H\right)$ and $\exp\left[-(m_B-\mu_B)/T\right]$, respectively. Hence, the limiting temperature is given by $T=\left(1-\mu_B/m_B\right)T_H$, which gives an estimate for the critical density at the low-temperature region: $T=0\to n_B^c\gtrsim1$ GeV. Now, given that the hadron picture breaks down above the limiting temperature is reasonable to argue a hadronic overlapping and percolating at $T_H$, resembling the physics behind the quark deconfinement~\cite{baym1979confinement,Satz:1998kg}. 

The phase transition associated with chiral symmetry restoration is identified as an evolution from a system with heavy constituent quarks to a state with light current quarks.~\cite{Hatsuda:1985eb}.  The order parameter is identified with the chiral condensate $\langle\bar{\psi}\psi\rangle$, which in the vacuum is $\langle\bar{\psi}\psi\rangle_0\simeq\left(0.24\text{GeV}\right)^3$. That value imposes a natural scale to conjecture the critical temperature and densities. It can be demonstrated that $\langle\bar{\psi}\psi\rangle$ decreases with the temperature and with the baryonic density, indicating a hadron melting in extreme environments~\cite{Gerber:1988tt,drukarev1991structure,Cohen:1991nk,Hatsuda:1991ez}. 
\begin{figure}
    \centering
    \includegraphics[scale=1]{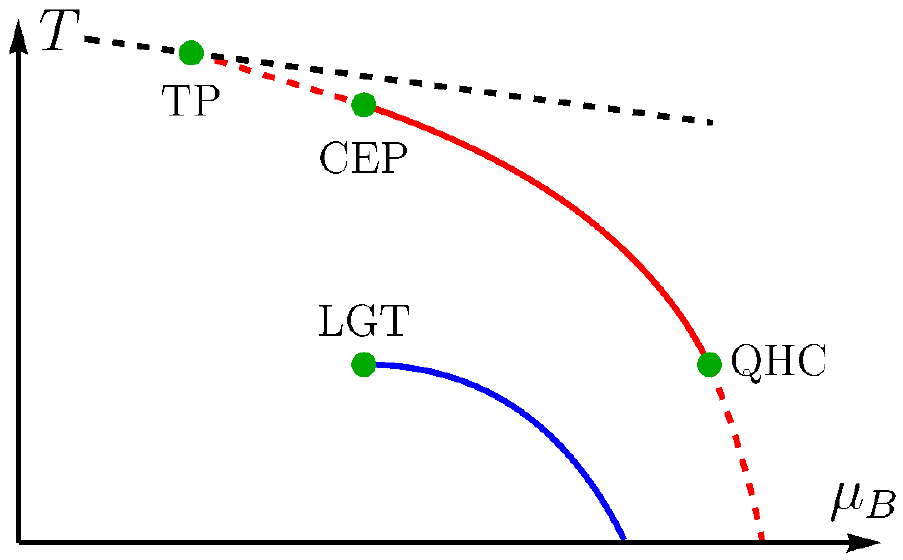}
    \caption{Critical points conjectured for the QCD phase diagram beyond the $\mu_B\sim T$ region: The Critical End Point (CEP), the critical point induced by the Quark-Hadron Continuity (QHC), the Liquid-Gas Transition of nuclear matter (LGT) and an approximate Triple Point (TP)~\cite{Fukushima:2010bq}.}
    \label{Fig:CriticalPoints}
\end{figure}

The phase diagram is not well established, and its main characteristics are conjectured from the theoretical and phenomenological guides. Nevertheless, there are arguments to make statements around some limit cases, mostly in the region of high temperature and small baryonic chemical potential $\mu_B\ll T$ and for asymptotically high density $\mu_B\gg\Lambda_\text{QCD}$. The several phases of the strongly interacting matter allow sketching the phase diagram passing from the {\it hadronic matter} to the QGP, and possible configurations that lead to {\it quarkyonic matter} and {\it color superconductivity} scenarios. In section~\ref{sec:QCD_phas_diag_general}, more details about such phases are discussed.

If there are several phases for the QCD matter, are there different critical points associated with each one? Figure~\ref{Fig:CriticalPoints} shows the proposed critical points in regions inaccessible for the current experiments and computational tools. Those points are characterized by $\mu_B\sim T$ and postulated from effective models. Chiral models predict the so-called QCD-{\it Critical End Point} (CEP) defined as the point where the phase transition associated with chiral symmetry restoration passes to be second-order ($\mu_B<\mu_\text{CEP}$) to be first-order ($\mu_B>\mu_\text{CEP}$)~\cite{AsakawaPhaseTran,BarducciPhasTran1,BarducciPhasTran2,BarducciPhasTran3,Wilczek:1992sf}. Another first-order phase line, which ends in a critical point for $T\ll\mu_B$, corresponds to cold systems where their degenerate flavors do not have a defined separation between the superfluid nuclear matter and the superconducting quark matter. That point is known as the {\it Quark--Hadron Continuity} critical point (QHC)~\cite{barrois1977superconducting,Bailin:1983bm,Iwasaki:1994ij,Alford:1997zt,Rapp:1997zu}.

The {\it Liquid-Gas Transition} (LGT) critical point is located when the liquid-gas first-order phase transition for low $T$ turns into second-order. The phases are described in terms of the nucleon properties: the nucleon mass is $m_N\simeq939$ MeV, and the isospin-symmetric systems have binding energy close to 16 MeV; thus, the baryon chemical potential of nuclear matter arises from $\mu_B=\mu_{NM}\simeq924$ MeV at $T=0$. In this value, the density $n_B$ varies from $n_B=0$ to the nuclear density $n_B=n_0=0.17$ fm$^{-3}$. The fragmentation into droplets with densities $n_0$ is allowed for $n_B<n_0$. Experimentally, heavy-ion collisions show that $\mu_\text{LGT}\sim\mu_{NM}$ and $T_\text{LGT}=15\sim20$ MeV~\cite{chomaz2002nuclear}. Finally, particle detections in collider experiments happens after the freeze-out, a thermally equilibrated gas of non-interacting mesons, baryons, and resonances, with associated $\mu_B$ running as a function of the collision energy. Fitting the particle ratios allows to extract the $(\mu_B,T)$ data for the freeze-out. Although such data do not correspond to a phase transition boundary, there are arguments to localize the sudden fix of chemical species near a phase transition~\cite{BraunMunzinger:2003zz}. Hence, the mesons dominate the freeze-out line for $\mu_B\ll m_N$, whereas for increasing $\mu_B$, baryonic states are excited, indicating a transition point where the latter surpasses the former. Analysis from the QCD-large $N_c$ limit suggests that finite baryon density accompanied by a growing pressure is found when $\mu_B$ is greater than the lowest baryon mass, giving rise to a cold, dense state called {\it quarkyonic matter}~\cite{McLerran:2007qj}. Figure~\ref{Fig:CriticalPoints} indicates that the phase diagram is separated into three regions: confined, deconfined and quarkyonic phases, where each line corresponds to a first-order boundary. The junction point of such lines is known as the {\it Triple Point} (TP)~\cite{Andronic:2009gj}. To find the transition lines, and the critical points are of great interest to the QCD-community. For that reason, the first part of this thesis is devoted to elucidating one of the named phase transitions: Chapter~\ref{Chap:QCD_phase_diagram} presents the QCD phase diagram obtained from the chiral symmetry restoration using the {\it Linear Sigma Model} coupled to quarks in the framework of the {\it Thermal Field Theory}. In particular, the analysis is focused on the CEP location as a function of the effective model coupling constants. 

The second part of the thesis focuses on the extreme magnetic fields and high-gluon density reached in the early stages of a heavy-ion collision and their consequences in the direct photon detection at the later stages. As stated above, all the information on the collision evolution comes from the analysis of the final hadronic states, so that the theoretical, as well as the experimental studies. To do this, it is necessary to take into account the multiple interactions (decays, early hadronizations, radiation, thermalization, among others), and final configurations. Hence, the system to study is very complex, making it necessary to find a clean probe containing factual information about the several configurations.

Photons are excellent candidates for testing the QCD properties: given that they don't have color charge, they cannot be scattered by the process in which the strong interaction dominates. Of course, photons would react with electromagnetic charges (quarks), but it is well established that at the first stages of the collision, the pre-equilibrium state is dominated by the presence of gluons (see Sec.~\ref{Sec:The_Color_Glass_Condensate}). On the other hand, at the equilibrium phases (thermalized QGP) and the hadronization stages, photons interact with all charged particles, and from Compton scattering or decays, more photons are produced. Fortunately, the current hydrodynamical and perturbative calculations, as well as the experimental efforts,  can take into account these mechanisms and differentiate them from each other~\cite{Ghiglieri, TurbideGale,Wambach,Rapp,ALICEPbPb,MassonALICE}. Therefore, in order to understand the main properties of strongly interacting matter under extreme conditions at a heavy-ion collision, the description of the photon production mechanisms is a major area of study. Currently, direct photon measurements have been carried out at RHIC and LHC~\cite{RHIC1,RHIC2,RHIC3,RHIC4,ALICEPbPb,ALICE1,ALICE2}, from where it is possible to identify several photon sources: prompt photons (from partonic Compton interactions), thermal (produced at the QGP and thermalized phases), and Hadronic decays (when the freeze-out is established)~\cite{ThermalPhotons1,ThermalPhotons2,TurbideGale,HadronicThermal1}.  Moreover, hydrodynamical calculations describe the photon invariant momentum distribution accurately for ranges of $\pt^\gamma\gtrsim 2$ GeV~\cite{hydro-photons1}. Nevertheless, when the photon production is decomposed in its angular distribution through the {\it flow coefficients}, the theoretical description fails: such incompatibility summarizes the so-called {\it direct photon puzzle}.  

The deviations of the predicted number of photons from the experimental data may be due to novel photon production mechanisms that were not yet considered. In search of those channels, the prompt radiation can be increased by an extra ingredient: the existence of large strength magnetic fields in peripheral heavy-ion collisions, with a magnitude of order of the hadronic scale ($|B|\sim m_\pi^2\sim 10^{18}$ Gauss)~\cite{intensity1,intensity2,intensity3,intensity4,intensity5}. Since a magnetic field induces the breaking of rotational invariance, this field is not only a source of an excess photon yield but also of an increase of the second harmonic coefficient of the Fourier expansion in the azimuthal photon distribution. These magnetic fields have a short duration ($\sim 0.5$ fm), and they are well localized in space, mostly in the participant's region (colliding hadrons) and in the transverse direction of the collision. On the other hand, the magnetic field peak intensity is found before the thermal phase (QGP) is created. During these early stages, its presence coincides with that CGC phase: a dense medium dominated by gluons with small longitudinal momentum fraction~\cite{mueller2002parton,iancu2004color,jalilian2006saturation,weigert2005evolution,Triantafyllopoulos:2005cn,Gelis:2010nm,Lappi:2010ek}. With this in mind, the present manuscript studies the photon production in stages where the magnetic field makes it possible the creation of photons from processes otherwise not allowed. In particular, attention is centered on channels which include gluons as initial states where the magnetic effects are encoded in the electromagnetic interaction of virtual quarks. Chapter~\ref{chap:Photons} describes the {\it photon puzzle} together with the {\it gluon fusion} and {\it gluon splitting} magnetic-induced photon production.  Both channels are computed at the lowest order in $\alpha_s$ by assuming a static and constant magnetic field. The invariant momentum distribution and the elliptic flow coefficient are computed analytically, and their contribution is understood as a yield over the hydrodynamical calculations. Moreover, to account for the centrality dependence of the parameters, the UrQMD formalism is used. The results are compared with experimental data from RHIC and ALICE, indicating an improvement at low photon transverse momentum.

As was stated, CGC phase coincides with the intense magnetic field generation, which is mostly present in the collision region; therefore, the dense gluonic medium is strongly magnetized. Besides the photon production channels opened by the field, the question about the magnetic effects on the CGC excitation spectrum naturally appears. That information is encoded in the {\it Polarization Tensor} from which the dynamic masses, dispersion relation, and medium's screening are analyzed. In fact, the one-loop gluon polarization tensor has been computed at zero temperature in standard perturbation theory, relating the vacuum QCD screening properties~\cite{peskin1995quantum}. Furthermore, that calculation is also made for finite temperature and magnetic field in the {\it Hard Thermal Loop} (the temperature is the hard scale) and {\it Lowest Landau Level} (the magnetic field is the dominant energy scale) approximations~\cite{hattori2018gluon,PhysRevD.45.4632,PhysRevLett.63.1129}. Breaking of both Lorentz and rotational symmetries shows that the gluon self-interaction becomes important for collective excitations with momenta $p\sim gT$, and for very intense magnetic fields, the Debye screening disappears, whereas the dynamical screening remains. Moreover, the magnetic field opens a threshold for pair production of masses $m_f$ when the gluon momentum component parallel to the magnetic field reaches the value $\pp^2=4m_f^2$~\cite{PhysRevD.83.111501}. In Chapter~\ref{Chap:Gluon_Pol_Tensor} the gluon polarization tensor is computed by taking into account all the possible energy scales combinations: zero temperature and arbitrary magnetic field, the mass or gluon momentum the lowest energy scale, and the weak field approximation. Such calculations show the role of temperature and magnetic field into the polarization orientation and the Debye's induced mass.

The above discussion leads to the present thesis motivation, which can be summarized in the following research questions:
\begin{enumerate}
    \item Is it possible to approximate the QCD-phase diagram and give a plausible location of the Critical End Point from the chiral symmetry restoration described by the {\it Linear Sigma Model}?
    
    \item Can the magnetic field in a dense gluon system open new channels for prompt photon production in order to improve the experimental data description?
    
    \item What is the impact of the temperature, the magnetic field, and the energy scales hierarchy on the gluon polarization tensor and the CGC screening properties?
\end{enumerate}

To answer these questions, the manuscript is organized as follows:

Chapter~\ref{Chap:QCD_phase_diagram}, in Sec.~\ref{sec:QCD_general}, introduces the QCD main aspects with a particular interest in its flavor and chiral symmetries, and asymptotic freedom. Section~\ref{sec:QCD_phas_diag_general} gives a general panorama of the state-of-art of the QCD-phase diagram, both from the theoretical and experimental perspectives. Section~\ref{sec:The_QCD_Phase_Diagram_from_the_Linear_Sigma Model} is about the Linear Sigma Model coupled to quarks and its symmetries, together with the spontaneous symmetry breaking and restoration from thermal and density effects. The results show that the transition lines and the CEP location are sensitive to the value of the effective coupling constants,  and the pressure is only approximated as an average description of the degrees of freedom. 

Chapter~\ref{chap:Photons} is about the {\it direct photon excess} or {\it photon puzzle}. In Sec.~\ref{SEC:Relativistic_Hydrodynamics_and_Elliptic_Flow}, an overall description of the hydrodynamical calculations and its accurate description of the elliptic flow for several experimental data is presented. Section~\ref{SEC:The_photon_puzzle} explains the {\it photon puzzle}'s origin. Sections~\ref{Sec:The_Color_Glass_Condensate} and~\ref{sec:Magnetic_field_HIC} are centered on the physics of the early stages in a heavy-ion collision, giving a review for the CGC phase and the magnetic field creation, respectively. Section~\ref{sec:Prompt Photon Production by Gluon Fusion in a Magnetized Medium} presents the photon production proposal from gluon fusion and the comparison with RHIC and ALICE data. Finally, in Sec.~\ref{sec:gluon_splitting}, the gluon splitting channel is computed and added to the gluon fusion one. In this section, the invariant momentum distribution and elliptic flow for the produced photons are improved using simulations in the UrQMD formalism, which consider the space-time evolution of the magnetic field, reaction volume, and the collision centrality. 

Chapter~\ref{Chap:Gluon_Pol_Tensor} is dedicated to studying the magnetic field and the temperature effects on the gluon polarization tensor. The energy scale hierarchy also plays a fundamental role in such a description. In Sec.~\ref{sec:Magnetized_Gluon_Polarization_Tensor_from_the_Sum_of_All_Landau_Levels}, the gluon polarization tensor is computed for an arbitrary magnetic field at zero temperature by resuming all the Landau Levels contribution to the fermion propagator. Section~\ref{sec:GluonPolTenThermoMag} computes the same object but in the lowest Landau Level approximation at finite temperature. Here, the comparison between the gluon mass and momentum with the temperature has significant consequences in the polarization tensor coefficients. Section~\ref{sec:GluonPolTenWeakField} studies the Debye mass and gluons dispersion relation in a weakly magnetized medium without temperature. 

Each chapter has its conclusions and the detailed calculations are presented at the Appendices.

\begin{savequote}[50mm]
``In every branch of knowledge, the progress is proportional to the amount of facts on which to build, and therefore to the facility of obtaining data.''
\qauthor{--- James Clerk Maxwell}
\end{savequote}

\chapter{The QCD Phase Diagram and the Critical End Point from the Linear Sigma Model}\label{Chap:QCD_phase_diagram}

This chapter is devoted to the phenomenological search of the QCD phase diagram and the so-called Critical End Point, which constitutes the point where the deconfinement and chiral phase transition pass from being second-order (at high temperatures and low densities) to be first order (at low temperatures and high densities). The analysis is based on the Linear Sigma Model  (L$\sigma$M) coupled with quarks in the thermal field theory framework that effectively models the QCD degrees of freedom and can be promoted to take into account the chiral symmetry restoration. 

The chapter starts with a general description of the QCD formalism, particularly the fundamental aspects of its field theory formalism, which covers Section~\ref{sec:QCD_general}. The symmetries of the QCD Lagrangian are presented in subsections~\ref{sec:flavor_symm} and~\ref{sec:chiral_symm}, corresponding to the flavor and chiral symmetry. The latter is of great importance in this work, given that the QCD phase diagram will be built in terms of the chiral symmetry restoration. Moreover, subsection~\ref{sec:Asymp_freedom} briefly exposes the QCD Asymptotic Freedom.

Section~\ref{sec:QCD_phas_diag_general} is about the generalities of the QCD phase diagram. Here, the description of the several phases that the strongly interacting matter pass through is accomplished from a theoretical and experimental perspective.

In Section~\ref{sec:The_QCD_Phase_Diagram_from_the_Linear_Sigma Model}, the Linear Sigma Model is presented as an attempt to describe the low-energy Hadronic physics. Subsection 1.2 demonstrates that at symmetry level, the L$\sigma$M Lagrangian has the same invariances as the QCD one; thus, the physical description of the QCD phase diagram can be achieved by such effective treatment. In subsection 1.3, the concept of hidden symmetry and Nambu-Goldstone theorem is discussed, leading to the implementation of the so-called Spontaneous Symmetry Breaking for the L$\sigma$M. Such treatment makes possible the identification of an order parameter as the chiral quark condensate. 

From subsections~\ref{sec:Effective_Potential_for_the_LsM} to~\ref{Effective_Potential_in_the_Low_Temperature_Approximation}, the effective thermodynamic potential is computed by the implementation of the thermal field theory formalism. In such treatment, the potential energy for bosons and fermions is calculated at tree-level and one-loop corrections. Moreover, ring contributions are also taken into account. The thermodynamic description is made by separating the high-temperature approximation, where the expressions become more manageable and analytically solvable, and the low-temperature regime, where the results are found numerically. 

Finally, in subsection~\ref{Sec:Locating_the_CEP_and_Computing the_Pressure}, the phase diagram is computed, and the CEP location is achieved by imposing constrictions to the coupling parameters $\lambda$ and $g$. The critical line at high-temperature and low quark-chemical potential follows the predictions computed in the Lattice QCD framework, whereas the CEP lies in a region compatible with other theoretical studies. The CEP identification depends on the coupling choice, which encodes the information about the degrees of freedom in the model. The results of the pressure support that asseveration: although the L$\sigma$M does not give the exact functional behavior of that quantity, the high and low-temperature approximations together with the value of the coupling constants reproduces on average the findings reported by Lattice QCD.

The results showed in this chapter are based in the following published manuscripts:
\begin{itemize}
    \item Article\\
    {\it Chiral symmetry transition in the linear sigma model with quarks: Counting effective QCD degrees of freedom from low to high temperature.}\\
    Alejandro Ayala, \underline{Jorge David Casta\~no-Yepes}, J. J. Cobos-Martinez, Saul Hernandez-Ortiz, Ana Julia Mizher, Alfredo Raya.\\
    Int. J. Mod. Phys. A {\bf 31}, 1650199 (2016).
    
    \item Proceeding\\
    {\it Using the Linear Sigma Model with quarks to describe the QCD phase diagram and to locate the critical end point}\\
    Alejandro Ayala, \underline{Jorge David Casta\~no-Yepes}, Jose Antonio Flores, Saul Hernandez-Ortiz, L. A. Hernandez.\\
    EPJ Web of Conferences \textbf{172}, 08002 (2018).
\end{itemize}

\section{Quantum Chromodynamics}\label{sec:QCD_general}

In order to describe the physics at the nuclear scale and related phenomena, the \textit{Quantum Chromodynamics} (QCD) has been developed as the fundamental theory which describes the strongly interacting matter~\cite{GellMann1,Zweigbookhadrons}. It is well established that QCD fundamental degrees of freedom are the \textit{quarks} and \textit{gluons}, which are the basic blocks of the so-called hadronic matter. The very nature of QCD does not allow for quarks and gluons to exist as isolated entities: the well known \textit{QCD Asymptotic Freedom}~\cite{AsymptoticFreedom1,AsymptoticFreedom2} demands that all the states in nature need to be colorless, i.e., strongly interacting matter can only exist as \textit{baryons} or \textit{mesons} or bound states of quarks, which are named \textit{hadrons}~\cite{GellMann2,Nambu}. Thus, a full understanding of the QCD dynamics can be constructed from two scenarios: one, concerning the (theoretical) study of free quarks and gluons and the other pointing at the hadronic (theoretical and experimental) physics. 

Formally, QCD is the $SU(3)$ non-Abelian gauge theory of the color charge, both quarks (fermions) and gluons (bosons) carry color charge. The fundamental representation of $SU(3)$ describes the former with a quark field $\psi_j^{(\alpha)}(x)$ with mass $m^{(\alpha)}$ where $\alpha=u,s,d,\cdots$ labels the flavor (the type of quark) and $j=1,2,3,\cdots,N_c$ is the color index. The gauge bosons are defined by the field $A_\mu^a$, wherein the adjoint representation of the color (gauge group) degrees of freedom $a=1,\cdots,N_c^2-1$, resulting in 8 color indices for gluons. Classical QCD is defined by the Lagrangian:
\bea
\mathcal{L}=\sum_\alpha\bar{\psi}_j^{\alpha}\left(i\slashed{D}_{jk}-m^{\alpha}\delta_{jk}\right)\psi_k^{\alpha}-\frac{1}{4}G_a^{\mu\nu}G^a_{\mu\nu},
\label{QCDLagrangian}
\eea
where the repeated color indices are summed over, and the gauge field tensor $G^a_{\mu\nu}$ is given by
\bea
G^a_{\mu\nu}=\partial_\mu A_\nu^a-\partial_\nu A_\mu^a-g_3f^{abc}A_\mu^b A_\nu^c,
\eea
with $g_3$ the $SU(3)$ coupling parameter. The coavariant derivative is given in terms of the group generators $\boldsymbol{\lambda}$, the Gell-Mann matrices, as follows:
\bea
\mathbf{D}_\mu\psi=\left(\partial_\mu+ig_3 A_\mu^a\frac{\boldsymbol{\lambda}_a}{2}\right)\psi,
\eea
and the structure constants $f^{abc}$ can be found from the commutation relations obeyed by the generators and the trace in group space:
\begin{subequations}
\bea
\left[\boldsymbol{\lambda}_a,\boldsymbol{\lambda}_b\right]=2if_{abc}\boldsymbol{\lambda}_c
\eea
and
\bea
\text{Tr}\left\{\boldsymbol{\lambda}_a\boldsymbol{\lambda}_b\right\}=2\delta_{ab}.
\eea
\end{subequations}

\subsection{QCD Flavor Symmetry}\label{sec:flavor_symm}

The symmetry of a Lagrangian is a fundamental piece for the analysis of a field theory, given that the information about the conserved quantities, as well as its breaking, have significant consequences in the physical predictions~\cite{coleman1988aspects,itzykson2012quantum,kaku1993quantum}. Symmetry arguments can introduce restrictions over the system constituents interactions or fix the form of the Lagrangian structure. Moreover, if the symmetry is related to the local dynamics (space-time dependent), the gauge theories are a natural result. On the other hand, if the symmetry is space-time independent (global symmetry), it can classify particles according to quantum numbers or predict the existence of massless particles.

The flavor sector of quarks in Eq.~(\ref{QCDLagrangian}) have a {\it flavor symmetry} which can be analyzed as follows~\cite{Sazdjian:2016hrz}: suppose that the number of all $N_f$ flavors have the same mass $m$, so that the quark Lagrangian becomes:
\bea
\mathcal{L}_q=\sum_\alpha\bar{\psi}^{\alpha}\left(i\slashed{D}-m\right)\psi_{\alpha}
\label{Lqflavorsymm},
\eea
where the color indexes are omitted. Equation~(\ref{Lqflavorsymm}) is invariant under infinitesimal global transformations $\eta^a$ given by:
\begin{subequations}
\bea
\delta\psi^{\alpha}=-\frac{i}{2}\delta\eta^a\left(\boldsymbol{T}^a\right)^\alpha_\beta\psi^\beta,
\eea
\bea
\delta\bar{\psi}^{\alpha}=\frac{i}{2}\delta\eta^a\bar{\psi}^\beta\left(\boldsymbol{T}^a\right)^\alpha_\beta,
\eea
and
\bea
\delta A_\mu=0.
\eea
\label{QCDflavorsymmetryeqs}
\end{subequations}
where $\boldsymbol{T}^a$ are $N_f\times N_f$ hermitian traceless matrices, which are the representatives of the group and follow the $SU(N_f)$ algebra:
\bea
[\boldsymbol{T}^a,\boldsymbol{T}^b]=if_{abc}\boldsymbol{T}^c,\text{ for } a,b,c=1,\ldots,N_f^2-1. 
\eea

From the Noether's theorem, the $N_f^2-1$ conserved currents are
\bea
j_\mu^a(x)=-\frac{i}{2}\frac{\partial\mathcal{L}_q}{\partial\left(\partial^\mu\psi^\alpha\right)}\left(\boldsymbol{T}^a\right)^\alpha_\beta\psi^\beta=\frac{1}{2}\bar{\psi}_\alpha\gamma_\mu\left(\boldsymbol{T}^a\right)^\alpha_\beta\psi^\beta,
\eea
so that
\bea
\partial^\mu j_\mu^a(x)=0.
\eea

The conserved charges will be
\bea
Q^a=\int d^3xj_0^a(x),
\eea
and they satisfy
\bea
\left[H,Q^a\right]=0,
\eea
is such a way that the corresponding operators satisfy the group algebra:
\bea
\left[Q^a,Q^b\right]=if_{abc}\,Q^c.
\eea

Thus, QCD is characterized by the existence of $N_f$ quark fields with equal masses, which are preserved after renormalization, given that the coupling constant is invariant under the group transformations. 

\subsubsection{Hadronic Approximate Flavor Symmetry}

A real description of the flavor symmetry comes about by considering not equal masses for the quarks. The Lagrangian of Eq.~(\ref{Lqflavorsymm}) in that situation reads:
\bea
\mathcal{L}_q=\sum_\alpha\bar{\psi}^{\alpha}\left(i\slashed{D}-m_{\alpha}\right)\psi_{\alpha},
\eea
whre the mass term is not invariant under de transformations of Eq.~(\ref{QCDflavorsymmetryeqs}), i.e., the Noether's currents are not conserved:
\bea
\partial^\mu j_\mu^a=-\frac{i}{2}\sum_{\alpha,\beta}\left(m_\alpha-m_\beta\right)\bar{\psi}_\alpha\left(\boldsymbol{T}^a\right)^\alpha_\beta\psi^\beta\neq0. 
\eea

Now, if the mass difference is smaller than the QCD energy scale $\Lambda_{\text{QCD}}\sim 300$ MeV, the term corresponding to the mass difference can be regarded as a perturbation of the Lagrangian with equal masses $\mathcal{L}_0$:
\bea
\mathcal{L}=\mathcal{L}_0+\Delta\mathcal{L}.
\eea

In the current state of the art, QCD has $N_f=6$, so that the mass difference between the light quarks $(u,d,s)$ and the heavy quarks $(c,b,t)$ is large ($> 1$ GeV); therefore, the approximate flavor symmetry can be expected only within the space of the three
light quarks. Hence the flavor symmetry is connected with the $SU(2)$ symmetry for $u$ and $d$, and with $SU(3)$ symmetry for $u,d,s$. 

Concerning the isospin symmetry, nucleons and kaons could be organized as doublets in $SU(2)$, and the pions in the triplet (adjoint) of $SU(3)$. Moreover, the mass difference in each multiplet is small which implies that $|m_u-m_d|\sim\text{few MeV}$.

On the other hand, for the $SU(3)$ symmetry, the baryons, and mesons can be grouped in several multiplets, so that in each one, the mass difference reads:
\bea
m_s-\frac{m_u+m_d}{2}\sim100~\text{MeV}\gg|m_u-m_d|\sim\text{few MeV}.
\eea

Finally, the heavy quark sector admits an approximate flavor symmetry when baryons are made of $q_hq_lq_l$ and mesons are made of $\bar{q}_h q_l$, where $q_h=c, b, t$ and $q_l=u,d,s$. 

\subsection{QCD Chiral Symmetry}\label{sec:chiral_symm}

The fermion sector of QCD admits other kinds of unitary transformations. In particular, the transformation mediated by the matrix $\gamma_5$ is of interest because it gives information about the parity of the field theory. To explore such symmetry in the quark sector, the axial flavor transformations in its infinitesimal form are:
\begin{subequations}
\bea
\delta\psi^{\alpha}=-\frac{i}{2}\delta\eta^a\left(\boldsymbol{T}^a\right)^\alpha_\beta\gamma_5\psi^\beta,
\eea
\bea
\delta\bar{\psi}^{\alpha}=-\frac{i}{2}\delta\eta^a\bar{\psi}^\beta\gamma_5\left(\boldsymbol{T}^a\right)^\alpha_\beta,
\eea
and
\bea
\delta A_\mu=0.
\eea
\label{QCDchiralsymmetryeqs}
\end{subequations}

By assuming equal mass for the quarks, the Lagrangian of Eq.~(\ref{QCDLagrangian}) is not invarian under the above transformations:
\bea
\delta\mathcal{L}_q=im\,\delta\eta^a\left(\boldsymbol{T}^a\right)^\alpha_\beta\gamma_5\psi^\beta,
\eea
thus, QCD is invariant under axial transformations only if the quarks are massless. For the unequal quark masses, the above equation is:
\bea
\delta\mathcal{L}_q=\frac{i}{2}\,\delta\eta^a\left(\left\{\mathbb{M},\boldsymbol{T}^a\right\}\right)^\alpha_\beta\gamma_5\psi^\beta,
\eea
where $\mathbb{M}=\text{diag}(m_1,m_2,\cdots,m_{N_f})$ and $\left\{\cdot,\cdot\right\}$ is the anticommutator.

To understand the meaning of the chiral symmetry, consider the massless case where the Lagrangian is invariant under the flavor and axial transformations, with conserved currents and charges given by:
\begin{subequations}
\bea
j_\mu^a(x)=\frac{1}{2}\bar{\psi}_\alpha\gamma_\mu\left(\boldsymbol{T}^a\right)^\alpha_\beta\psi^\beta\to Q^a=\int d^3xj_0^a(x),
\eea
and
\bea
j_{5\mu}^a(x)=\frac{1}{2}\bar{\psi}_\alpha\gamma_\mu\gamma_5\left(\boldsymbol{T}^a\right)^\alpha_\beta\psi^\beta\to Q^a_5=\int d^3xj_{5,0}^a(x),
\eea
\end{subequations}
which satisfies the following algebra:
\bea
\left[Q^a,Q^b\right]&=&if_{abc}\,Q^c\nn\\
\left[Q^a_5,Q^b_5\right]&=&if_{abc}\,Q^c\nn\\
\left[Q^a,Q^b_5\right]&=&if_{abc}\,Q^c_5. 
\eea

From the above, it is clear that the axial charges do not form a closed algebra, but if the right and left charges are defined as:
\bea
Q_R^a&\equiv&\frac{1}{2}\left(Q^a+Q_5^a\right)\nn\\
Q_L^a&\equiv&\frac{1}{2}\left(Q^a-Q_5^a\right),
\eea
the algebra is closed in both charges:
\bea
\left[Q^a_R,Q^b_R\right]&=&if_{abc}\,Q^c_R\nn\\
\left[Q^a_L,Q^b_L\right]&=&if_{abc}\,Q^c_L\nn\\
\left[Q^a_R,Q^b_L\right]&=&0.
\eea

Therefore, the right and left-handed charges define a subgroup of transformations, namely,
\bea
\text{Chiral Group}=SU(N_f)_R\otimes SU(N_f)_L.
\eea

At Lagrangian level, the chiral symmetry is manifested by separation of the fermions with definite chirality (i.e., handedness):
\bea
\mathcal{L}_q=\mathcal{L}_R+\mathcal{L}_L,
\eea
where
\bea
\mathcal{L}_{R,L}=i\bar{\psi}_{R,L}\slashed{D}\psi_{R,L}.
\eea

The Lagrangians $\mathcal{L}_{R,L}$ are invariant under the global chiral phase transformations:
\bea
\psi_{R,L}(x)\to\exp\left(-i\alpha_{R,L}\right)\psi_{R,L}(x),
\eea
with $\alpha_{R,L}$ real and constant phases. The total vector and axial charges
\bea
Q&=&Q_R+Q_L\nn\\
Q_5&=&Q_L-Q_R,
\eea
accounts for the number of left-handed and right-handed particles. 

Clearly, given the quark masses, the Lagrangian of Eq.~(\ref{QCDLagrangian}) is not chiral symmetric. That Lagrangian can be written by separating explicitly the term which breaks the symmetry, and for $N_f=2$ reads:
\bea
\mathcal{L}&=&\mathcal{L}_{\text{symmetric}}+\mathcal{L}_{\text{break}}\nn\\
&=&\mathcal{L}_{\text{symmetric}}+m_u\bar{\psi}_u\psi_u+m_d\bar{\psi}_d\psi_d\nn\\
&=&\mathcal{L}_{\text{symmetric}}+m_q\bar{\psi}\psi+\delta m_q\frac{\bar{\psi}_u\psi_u-\bar{\psi}_d\psi_d}{2},
\eea
where
\begin{subequations}
\bea
m_q=m_u+m_d,
\eea
\bea
\delta m_q=m_u-m_d
\eea
and
\bea
\bar{\psi}\psi=\frac{\bar{\psi}_u\psi_u+\bar{\psi}_d\psi_d}{2}.
\eea
\end{subequations}

The operator $\bar{\psi}\psi$ and its vacuum expectation value $\langle\bar{\psi}\psi\rangle$ (the so-called {\it chiral condensate}) characterize the magnitude of the symmetry breaking: if $\langle\bar{\psi}\psi\rangle=0$ the chiral symmetry is restored. In the following section, the generalities of the QCD phase diagram are described in terms of such condensate.

\subsection{Asymptotic Freedom}\label{sec:Asymp_freedom}
Compared with the other fundamental forces described by the Standard Model, Asymptotic Freedom is a feature belonging only to QCD, which establishes that the coupling constant of color $g$ decreases as the scale of renormalization $\mu_R$ is increased. Such behavior implies that the perturbative methods are not applicable in all the QCD scales. 

The running of the QCD coupling is understood from the analysis of the beta-function obtained from 1-loop diagrams~\cite{AsymptoticFreedom1,AsymptoticFreedom2,greiner2007quantum}:
\bea
\beta(g)=\mu_R\frac{dg(\mu_R)}{d\mu_R}=-\left(11-\frac{3}{2}N_f\right)\frac{g^3}{(4\pi)^2}-\left(102-\frac{38}{3}N_f\right)\frac{g^5}{(4\pi)^4}+\mathcal{O}(g^7),
\label{QCDbetafunction}
\eea
so that, by defining $\alpha_s(t)\equiv\bar{g}^2(t)/4\pi$, where $\bar{g}(t)$ is the running coupling constant with the constriction $\bar{g}(0)=g$, the asymptotic freedom is encoded in the following relation:
\bea
\alpha_s(\mu_R^2)=\frac{12\pi}{(33-2N_f)\ln\left(\mu_R^2/\Lambda^2\right)}\left[1-\frac{6(153-19N_f)}{(33-2N_f)^2}\frac{\ln\left(\ln\left(\mu_R^2/\Lambda^2\right)\right)}{\ln\left(\mu_R^2/\Lambda^2\right)}\right]+\cdots,
\label{alpha_s_runing_}
\eea
where $N_f$ is the number of flavors with mass less than $\sqrt{\mu_R^2}$, and $\Lambda$ is the scale at which $\bar{g}$ diverges around $\Lambda\equiv\Lambda_{QCD}\simeq200$ MeV.

The above results give essential properties of the vacuum and scales of QCD~\cite{donoghue2014dynamics}. First, the beta function comes from the divergent term of the one-loop corrections to the quark-gluon vertex. The sign on the leading-order in Eq.~(\ref{QCDbetafunction}) is negative for $N_f=6$ and positive if $N_f>16$ (it has severe implications in the quark family numbers). Thus, the screening resulting from virtual quark-antiquark pairs' spontaneous creation is overwhelmed by contributions from virtual gluons, which implies that the QCD vacuum is a paramagnetic medium and antiscreens the color charge. Second, although it is not a proven result, there is a considerable number of works that indicates that the QCD coupling continues growing when the energy is reduced, leading to the quark confinement phenomenon. Moreover, in comparison with Quantum Electrodynamics (QED), where the perturbative approaches are given in terms of the free parameter $\alpha_{em}(\mu_R^2=0)\simeq1/137$, in QCD the result of Eq.~(\ref{alpha_s_runing_}) implies that $\alpha_s$ is not free at all. Indeed, the running coupling is closely related to the mass scale $ \Lambda $; thus, the strong interaction description requires not only a Lagrangian but also a value of $ \Lambda $. In particular, QCD perturbation approaches can be applied for large mass scales $M^2$ with $\Lambda^2/M^2\ll1$, leaving other energy regimes for purely numerical treatment.

\section{Generalities of the QCD Phase Diagram}\label{sec:QCD_phas_diag_general}

Understanding the matter properties requires a full description of its components in terms of its thermodynamic quantities, such as temperature, density, and the number of particles. It is well known that modifications of the named parameters give rise to the so-called phase transitions: an abrupt and discontinuous change in the system's properties. Such phase transitions can be accompanied by a change of the aggregation state of the matter, i.e., from solid to liquid, gas, or plasma. A way to visualize the different phases of a system is by constructing a Phase Diagram, a graphical representation of the physical properties when the thermodynamic variables are modified, and the whole system reaches the equilibrium. In general, crossing a line in the diagram means a phase transition, and each line represents the coexistence of two or more phases.

It is well established that the strongly interacting matter can pass through different states (plasma, gas) and phases (chiral, hadronic, superconducting). Its description can give valuable information about the stages after the Big Bang and the general formation of the universe. Therefore, an experimental and theoretical construction of the QCD phase diagram (QCD-PD) is of great interest to the scientific community. Because of the asymptotic freedom of QCD, it is difficult to collect data from cosmological observations, since quarks and gluons are confined within hadrons. As was commented in previous sections, the relativistic heavy-ion collisions can break the confinement of such degrees of freedom, and therefore, those experiments are fundamental to build up the QCD-PD.
\begin{figure}[h]
    \centering
    \includegraphics[scale=0.45]{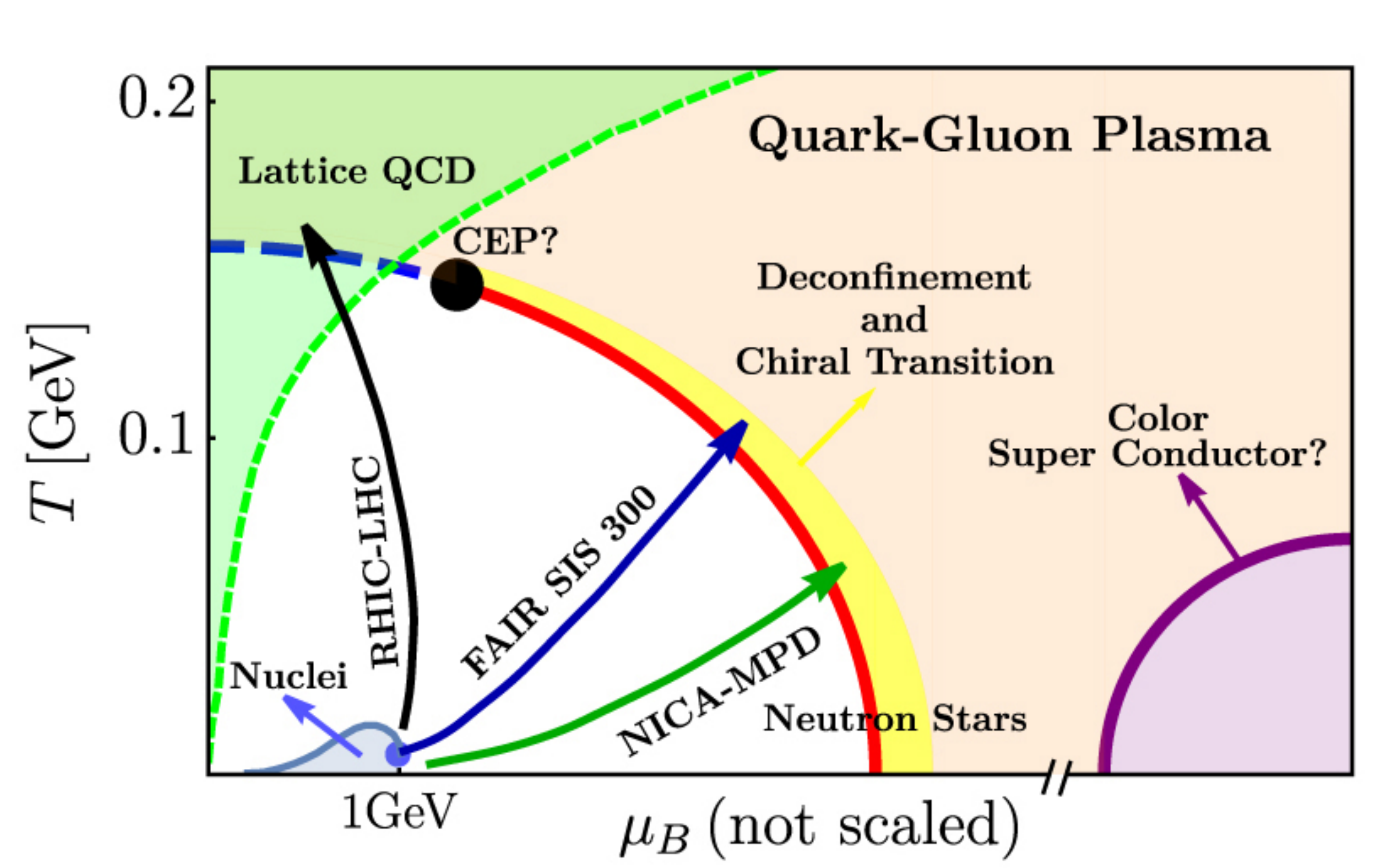}
    \caption{Schematic representation of the state-of-the-art for the QCD phase diagram on the $T$ and $\mu_B$ plane.}
    \label{Fig:QCD_Phase_Diagram_Sketch}
\end{figure}

Figure~\ref{Fig:QCD_Phase_Diagram_Sketch} shows a sketch which summarizes the current experimental, theoretical, and phenomenological knowledge about the main phases of the QCD sector when the temperature ($T$) and the net baryonic chemical potential ($\mu_B$) are considered. For low temperatures and low $\mu_B$, the quarks and gluons cannot exist as free; thus, the region on such a diagram represents the hadronic phase. Given that the light-quark chemical potential $\mu$ is related to the net baryonic chemical potential by $\mu_B=3\mu$, nuclear matter corresponds to $\mu_B\sim 3\mu$. In general terms, that low energetic region has been well described. There is a phase transition for $\mu_B=0$ and finite temperature, which has been extensively studied in numerical lattice simulations and depends on the number of colors and flavors~\cite{Svetitsky:1982gs, PhysRevD.29.338}. A first-order phase transition is found for $N_c=3$ and $N_f=0$ with a critical temperature $T_c\simeq270$ MeV~\cite{Fukugita:1989yw}. For $N_f>0$ light flavors, there is more than one phase transition.

For the high temperature and low baryonic chemical potential region, the information is provided by the experiments carried out at the BNL Relativistic Heavy Ion Collider (RHIC)~\cite{experiments1,experiments2,experiments3,experiments4,experiments5} and the
CERN Large Hadron Collider (LHC)~\cite{experiments6, BiroLHC}. For high temperatures, the QGP is formed, which corresponds to a phase where the quarks and gluons can be considered as free. For almost vanishing $\mu_B$, Lattice QCD calculations have shown that this change of phase is an analytical crossover and therefore, it is not a real phase transition~\cite{AokiLattice,YukalovPhaseTrans1, YukalovPhaseTrans2}. In that sense, a pseudo-critical temperature $T_c$ for $\mu_B=0$ is defined. Several values of $T_c$ have been reported by lattice calculations with 2+1 quark flavors: the MILC collaboration reported $T_c=169(12)(4)$ MeV~\cite{Tc1},  BNLRBC-Bielefeld collaboration obtained $T_c=193(7)(4)$ MeV~\cite{Tc2}, the Wuppertal-Budapest collaboration found the smaller number with $T_c=147(2)(3)$ MeV~\cite{Tc3,Tc4, Tc5}, the HotQCD collaboration calculated $T_c=154(9)$ MeV~\cite{Tc6,Tc7} and $T_c=155(1)(8)$ MeV~\cite{Tc8}.

The named lattice QCD calculations cannot be extrapolated to $\mu\neq 0$ because of the {\it sign problem}~\cite{singproblem}. This problem comes from the properties of the fermion determinant, which appears in the averages on QFT. If $M=(\slashed{D}+m+\mu\gamma_0)$ is the fermion matrix, it can be shown that
\bea
\det (\slashed{D}+m+\mu\gamma_0)=\left[\det (\slashed{D}+m-\mu^*\gamma_0)\right]^*,
\eea
which implies that the determinant is not real unless $\mu=0$ or purely imaginary, and therefore, the Monte Carlo sampling fails, given the oscillating nature of the averages. The green region in Fig.~\ref{Fig:QCD_Phase_Diagram_Sketch} schematizes the window of parameters in which lattice QCD can be applied in the present context.

The other extreme of the QCD phase diagram is the case $T\sim 0$ and $\mu_B>0$. In this case, different model approaches indicate that the transition along the horizontal axis is strongly first order~\cite{AsakawaPhaseTran,BarducciPhasTran1,BarducciPhasTran2,BarducciPhasTran3,BergerPhasTran,HalaszPhasTran,ScaveniusPhasTran,AntoniouPhasTran,HattaPhasTran}. Such approaches are made of effective models, assumption of the quarks masses, random matrix models or general thermodynamic considerations. Although the calculation methods are different, the general conclusion is that a deconfined phase is not only reached, but a chiral restoration transition is performed. The chiral phase transition is commonly described by the quark condensate $\langle\bar{\psi}\psi\rangle$ which can be written as~\cite{AyalaCEPrewiev}:
\bea
\langle\bar{\psi}\psi\rangle=\frac{\partial\Omega}{\partial m},
\eea
where $\Omega$ is free energy, and $m$ is the quark mass. The last object identifies the order parameter, because in the Nambu-Goldstone phase $\langle\bar{\psi}\psi\rangle\neq 0$, whereas for the Wigner-Weyl phase $\langle\bar{\psi}\psi\rangle=0$, with the identification $m=0$ as the chiral limit. To distinguish the chiral and deconfinement transitions is a matter of intense study. It has been suggested that magnetic fields can break the similarities between both phases~\cite{AnaChiral}, but recent lattice QCD calculations showed that magnetic fields up to $3.25$ GeV$^2$ do not give significant differences~\cite{chiralLATTICEQCD}. 

From the above, a line of first-order phase transitions at low temperatures, and high $\mu_B$ has to finish somewhere to give rise to a crossover line.  The starting point of the pseudo phase transition begins called the Critical End Point (CEP). To find it, from the theoretical point of view, lattice calculations have been extended to finite chemical potential by Taylor expansions or Fourier expansion of the grand canonical partition function, locating the CEP in the region $\left(\mu^{\text{CEP}}/T_c,T^{\text{CEP}}/T_c\right)\sim(1-1.4,0.9-0.95)$~\cite{PhysRevD.71.114014,deForcrand:2006ec,Sharma:2013hsa,Ding:2015ona}. The experimental search is led by the RHIC beam energy scan program. In the PHENIX and STAR experiments, the  Bose-Einstein correlations (BEC) as a signature for the possible formation of QGP has been reported for Au+Au collisions for $\sqrt{S_{NN}}=7.7, 11.5, 19.6, 27, 39$, and $62.4$ GeV~\cite{Mitchell:2013mza,wang2013star}. The low energy ensures that the stopping power is enough to increase $\mu_B$, and therefore, the exploration of a region far from the vertical axis in the phase diagram is achieved. Although the motivation of conserve non-vanishing baryon chemical potential is well established, the signals of QGP formation become weak~\cite{PhysRevLett.110.142301,PhysRevLett.109.152301}. Other observables such as event-by-event fluctuations of conserved charges, kurtosis or cumulants at the CERN, and future experiments as the Facility for Antiprotons and Ion Research (FAIR) in GSI and the Nuclotron-based Heavy-Ion Collider Facility NICA at JINR can give information about other unexplored regions in the QCD phase diagram. Nonetheless, the current knowledge is not conclusive about the existence and location of the CEP~\cite{OtherExperimentsCEP1,OtherExperimentsCEP2,OtherExperimentsCEP3,OtherExperimentsCEP4,OtherExperimentsCEP5}.

Finally, for asymptotically large $\mu_B\gg\Lambda_\text{QCD}$, the ground state is analyzed in terms of a weak-coupled system. Here, an analogy with condensed matter physics can be made: the quark matter behaves as electrons in a metal at low temperatures, so that the QCD-ground state should form Cooper pairs in the denominated Color Superconductivity (CSC) phase. Given that there is no single color and flavor, many Cooper pairing patterns may appear and give rise to several CSC kinds~\cite{Rajagopal:2000wf,Alford:2007xm,barrois1977superconducting,Bailin:1983bm}.

\section{The QCD Phase Diagram from the Linear Sigma Model Point of View}\label{sec:The_QCD_Phase_Diagram_from_the_Linear_Sigma Model}

As it was commented on, the description of the strongly interacting matter when the temperature and density (chemical potential) varies, is of great interest for the QCD phase diagram construction. There are two QCD phase transitions whose understanding represents a challenge: the restoration of chiral symmetry (chiral phase transition), and the deconfinement of quarks and gluons to form the so-called quark-gluon plasma. Currently, there is no agreement on what kind of relationship such phase transitions have and the order of each one. The experiments and theoretical calculations cannot answer if both phase transitions occur at the same temperature or if they manifest at independent $T_c$. Also, the order of transition varies if with the number of quarks and its masses: if the quarks are considered as massless, lattice calculations predict a second-order phase transition~\cite{PhysRevD.35.3972,PhysRevLett.59.1513,Gottlieb:1989mh}, whereas the effective models suggest the transition becomes first-order~\cite{Smilga:1996cm,Rajagopal:1995bc}. 

To understand the dynamics of the chiral restoration, the quark condensate plays the role of the order parameter: it passes from a certain value at the hadronic phase to approximately zero at the QGP phase. Such chiral symmetry breaking is essential to describe the hadronic sector, given that baryons have mass, and there is not parity doubled partners. In that sense, the breaking of that global symmetry implies the existence of massless Goldstone bosons which can be identified as pions (at a first approximation) so that in some region of parameters the symmetry is restored, i.e., the quark condensate disappears~\cite{Li:1999yx,Birse:1994cz,Hatsuda:2000by}. A mechanism to describe the latter is the so-called Spontaneous Symmetry Breaking (SSB), originally proposed by Nambu and Goldstone to describe the superconductivity in metals~\cite{Goldstone:1961eq,PhysRev.122.345,PhysRev.124.246}.  In general terms, in the SSB, although the Lagrangian remains with global symmetry, the vacuum-solutions are not symmetric, and therefore, if the system transits to one of these low-energy states, the named symmetry is broken around them. That formalism, together with the $SU(3)\times SU(3)$ algebra, can explain the SSB in the low energy sector of QCD. 

A Lagrangian which exhibits SSB is the Linear Sigma Model (L$\sigma$M). This model was constructed to study the chiral symmetry in the pion--nucleon system, and given that it has the same symmetries of QCD, it can be regarded as a low energy version of such theory~\cite{GellMann:1960np}. The Lagrangian of the L$\sigma$M coupled to quarks is given by:
\bea
\mathcal{L}&=&\frac{1}{2}\left(\partial_\mu\sigma\right)^2+\frac{1}{2}\left(\partial_\mu\boldsymbol{\pi}\right)^2+\frac{a^2}{2}\left(\sigma^2+\boldsymbol{\pi}^2\right)-\frac{\lambda}{4}\left(\sigma^2+\boldsymbol{\pi}^2\right)^2\nn\\
&+&i\bar{\psi}\gamma^\mu\partial_\mu\psi-g\bar{\psi}\left(\sigma+i\gamma_5\boldsymbol{\tau}\cdot\boldsymbol{\pi}\right)\psi,
\label{LagrangianLsM}
\eea
where $\psi$ is an SU(2) isospin doublet, $\boldsymbol{\pi}$ is an isospin triplet, and $\sigma$ is an isospin singlet. The neutral pion $\pi^0$ is taken as the third component of $\boldsymbol{\pi}$, the charged pions are defined as $\pi^{\pm}=(\pi_1\mp i\pi_2)/2$, and $\boldsymbol{\tau}=(\tau_1,\tau_2,\tau_3)$ where $\tau_i$ are the Pauli matrices. The mass squared $a^2$, as well as the coupling constants $\lambda$ and $g$, are chosen as positive.  

\subsection{Symmetries of the Linear Sigma Model}
As studied in Sec.~\ref{sec:QCD_general}, the symmetries play a crucial role in the physical description of the QCD phenomena. In that sense, if one desires to get close to the latter, the effective model necessarily has to inherit such symmetries. For that purpose, it is useful to rewrite the Lagrangian of Eq.~(\ref{LagrangianLsM}) in terms of the matrix field
\bea
\Sigma\equiv\sigma+i\boldsymbol{\tau}\cdot\boldsymbol{\pi},
\eea
so that
\bea
\sigma^2+\boldsymbol{\pi}^2=\frac{1}{2}\text{Tr}\left(\Sigma^{\dagger}\Sigma\right).
\eea

Therefore, in terms of the chiral fields $\psi_{R,L}$ the Lagrangian becomes
\bea
\mathcal{L}&=&\frac{1}{4}\text{Tr}\left(\partial_\mu\Sigma\,\partial^\mu\Sigma^{\dagger}\right)+\frac{a^2}{4}\text{Tr}\left(\Sigma^{\dagger}\Sigma\right)-\frac{\lambda}{16}\text{Tr}^2\left(\Sigma^{\dagger}\Sigma\right)+i\bar{\psi}_L\gamma^\mu\partial_\mu\psi_L+i\bar{\psi}_R\gamma^\mu\partial_\mu\psi_R\nn\\
&-&g\left(\bar{\psi}_L\Sigma\psi_R+\bar{\psi}_R\Sigma^\dagger\psi_L\right).
\eea

The above expression implies that the meson sector is invariant under rotations among the $\sigma$ and $\boldsymbol{\pi}$ fields. On the other hand, the right and left-handed fermions couples together only through interaction with the $\Sigma$ field.The total symmetry of the Lagrangian is then separated into a left and right invariances, namely:
\begin{subequations}
\bea
\psi_{R,L}\to\psi'_{R,L}=U_{R,L}\psi_{R,L},
\eea
and
\bea
\Sigma\to\Sigma'=U_L\Sigma\,U^\dagger_R,
\eea
\end{subequations}
which is regarded to an $SU(2)_L\times SU(2)_R$ symmetry with $U_{R,L}$ arbitrary $SU(2)$ matrices:
\bea
U_{R,L}=\exp\left(-i\boldsymbol{\alpha}_{R,L}\cdot\boldsymbol{\tau}/2\right).
\eea

Besides a $SU(2)$ isospin rotations on the left and right-handed fermions, the meson sector of the Lagrangian involves isospin rotations among the $\boldsymbol{\pi}$ fields plus transformations between the $\sigma$ and $\boldsymbol{\pi}$ fields given by
\begin{subequations}
\bea
\sigma\to\sigma'=\frac{1}{2}\text{Tr}\left(U_LU_R^\dagger\right)\sigma+\frac{i}{2}\text{Tr}\left(U_L\tau^kU_R^\dagger\right)\pi^k,
\eea
and
\bea
\pi^k\to\pi'^{k}=-\frac{i}{2}\text{Tr}\left(\tau^kU_LU_R^\dagger\right)\sigma+\frac{1}{2}\text{Tr}\left(\tau^kU_L\tau^l U_R^\dagger\right)\pi^l,
\eea
\end{subequations}
which in infinitesimal form reads:
\begin{subequations}
\bea
\sigma\to\sigma'\simeq\sigma+\frac{1}{2}\left(\boldsymbol{\alpha}_L-\boldsymbol{\alpha}_R\right)\cdot\boldsymbol{\pi},
\eea
and
\bea
\pi^k\to\pi'^{k}\simeq\pi^k-\frac{1}{2}\left(\alpha_L^k-\alpha_R^k\right)\sigma-\frac{1}{2}\epsilon^{klm}\pi^l\left(\alpha_L^m+\alpha_R^m\right).
\eea
\end{subequations}

The conserved currents asociated at the named invariances are:
\begin{subequations}
\bea
J_{L\mu}^k=\bar{\psi}_L\gamma_\mu\frac{\tau^k}{2}\psi_L-\frac{1}{2}\left(\sigma\partial_\mu\pi^k-\pi^k\partial_\mu\sigma\right)+\frac{1}{2}\epsilon^{klm}\pi^l\partial_\mu\pi^m,
\eea
and
\bea
J_{R\mu}^k=\bar{\psi}_R\gamma_\mu\frac{\tau^k}{2}\psi_R+\frac{1}{2}\left(\sigma\partial_\mu\pi^k-\pi^k\partial_\mu\sigma\right)+\frac{1}{2}\epsilon^{klm}\pi^l\partial_\mu\pi^m,
\eea
\end{subequations}
such that by combining them they give the conserved vector current:
\bea
V_\mu^k=J_{L\mu}^k+J_{R\mu}^k=\bar{\psi}\gamma_\mu\frac{\tau^k}{2}\psi+\epsilon^{klm}\pi^l\partial_\mu\pi^m,
\eea
and the conserved axial-vector current:
\bea
A_\mu^k=J_{L\mu}^k-J_{R\mu}^k=\bar{\psi}\gamma_\mu\gamma_5\frac{\tau^k}{2}\psi+\pi^k\partial_\mu\sigma-\sigma\partial_\mu\pi^k.
\label{axialcurrenLsM}
\eea

The presented results confirm that the L$\sigma$M has the same symmetries of the QCD Lagrangian; therefore, the former is regarded as an effective model of low-energy hadron dynamics. The next sections will show that chiral spontaneous symmetry breaking present in Eq.~(\ref{LagrangianLsM}) is a good approximation to describe the QCD phase diagram in terms of the restoration of the chiral symmetry.

\subsection{Spontaneous Symmetry Breaking of the Linear Sigma Model}
It is well established that certain Lagrangians contain {\it hidden} symmetries that occur when the ground state of the theory does not fulfill the invariance requirements.  Such realization appears when the Noether's conserved charge Q acts as a global symmetry transformation into the vacuum state:
\bea
\ket{0}\to e^{i\alpha\,Q}\ket{0},
\eea
where $\alpha$ is a continous parameter. The vacuum invariance implies that for all $\alpha$
\bea
e^{i\alpha\,Q}\ket{0}=\ket{0},
\eea
which is provided by
\bea
Q\ket{0}=0.
\eea

The above results show that the vacuum is unique so that if new states $\ket{\alpha}\neq\ket{0}$ appears from the action of Q, they are degenerate. The latter is easily proved from the Noether's theorem for the (time) conserved charge:
\bea
\dot{Q}=i[H,Q]=0,
\eea
but given that $H\ket{0}=E_0\ket{0}$, then
\bea
H\ket{\alpha}=H e^{i\alpha\,Q}\ket{0}=e^{i\alpha\,Q} H\ket{0}=E_0\ket{\alpha}.
\eea

Also, note that such degeneration is continuous because $\alpha$ is a continuous parameter. 

A family of continuous states which share the vacuum energy is interpreted as particles, given that they are excitations around the ground state. Moreover, the lowest energy in the spectrum is identified as the particle's mass; hence, the zero-energy excitations obtained from symmetry transformations must correspond to massless particles which their quantum numbers are associated with the conserved charge $Q$. That is known as the {\it Nambu-Goldstone theorem}~\cite{peskin1995quantum}. 

For the particular case of the L$\sigma$M, note that the potential energy of Eq.~(\ref{LagrangianLsM}) given by
\bea
V(\sigma,\pi)=-\frac{a^2}{2}\left(\sigma^2+\boldsymbol{\pi}^2\right)+\frac{\lambda}{4}\left(\sigma^2+\boldsymbol{\pi}^2\right)^2,
\eea
has a minimum located at the unique configuration $\sigma=\boldsymbol{\pi}=0$ for $a^2>0$, whereas if $a^2<0$ implies that the minimum is located at
\bea
v^2\equiv\sigma^2+\boldsymbol{\pi}^2=\frac{a^2}{\lambda},
\label{vevdef}
\eea
which defines a 3-sphere in the 4-dimensional space of the fields. Thus, the fields can acquire a configuration around one of those classical minima, independent of which direction it points, implying that the ground state is degenerated, yielding into a hidden symmetry. 

To simplify the calculations, the following ground state is selected:
\bea
\langle\sigma\rangle=v,\hspace{1cm}\langle\boldsymbol{\pi}\rangle=0,
\label{vev}
\eea
which implies that the $\boldsymbol{\pi}$ direction do not require any energy, so that the pions become massless or the so-called {\it Goldstone bosons}. The particular ground state configuration of Eq.~(\ref{vev}) has the same physics of the other vacuum states in such a way that one can perform a $\sigma$-field redefinition, i.e.,
\bea
\sigma\rightarrow \sigma+v,
\eea
and replacing into Eq.~(\ref{LagrangianLsM}), the Lagrangian density becomes:
\bea
\mathcal{L}&=&\frac{1}{2}\left(\partial_\mu\sigma\right)^2+\frac{1}{2}\left(\partial_\mu\boldsymbol{\pi}\right)^2+i\bar{\psi}\gamma^\mu\partial_\mu\psi-\frac{\lambda}{4}v^4+\frac{a^2}{2}v^2\nn\\
&-&\frac{1}{2}\left(3\lambda v^2-a^2\right)\sigma^2-\frac{1}{2}\left(\lambda v^2-a^2\right)\boldsymbol{\pi}^2-gv\bar{\psi}\psi\nn\\
&-&\frac{\lambda}{4}\left(\sigma^2+\boldsymbol{\pi}^2\right)^2-g\bar{\psi}\left(\sigma+i\gamma_5\boldsymbol{\tau}\cdot\boldsymbol{\pi}\right)\psi.
\label{LagLsMSSB}
\eea

The Lagrangians of Eqs.~(\ref{LagrangianLsM}) and~(\ref{LagLsMSSB}) have the same physical information because no additional field sources or interaction were added. However, the above Lagrangian possesses new features:
\begin{itemize}
    \item[(i)] The boson fields $\sigma$ and $\boldsymbol{\pi}$, and the fermion isospin doublet generate dynamic masses in terms of $v$, namely
    \begin{subequations}
    \bea
    m_\sigma^2=3\lambda v^2-a^2,
    \label{msigma}
    \eea
    \bea
    m_\pi^2=\lambda v^2-a^2,
    \label{mpi}
    \eea
    and
    \bea
    m_f^2=gv.
    \label{fermionmass}
    \eea
    \label{masses}
    \end{subequations}

   If the $v$ value is given by Eq.~(\ref{vevdef}), the pions are massless, the mass degeneracy $m_\sigma=m_\pi$ is not longer present and the normal isospin invariance asociated to axial current of Eq.~(\ref{axialcurrenLsM}) remains.

    \item[(ii)] The original SU(2) chiral symmetry is {\it spontaneously} broken by the term $m_f^2\bar{\psi}\psi$.
\end{itemize}

\begin{figure}
    \centering
    \includegraphics[scale=0.7]{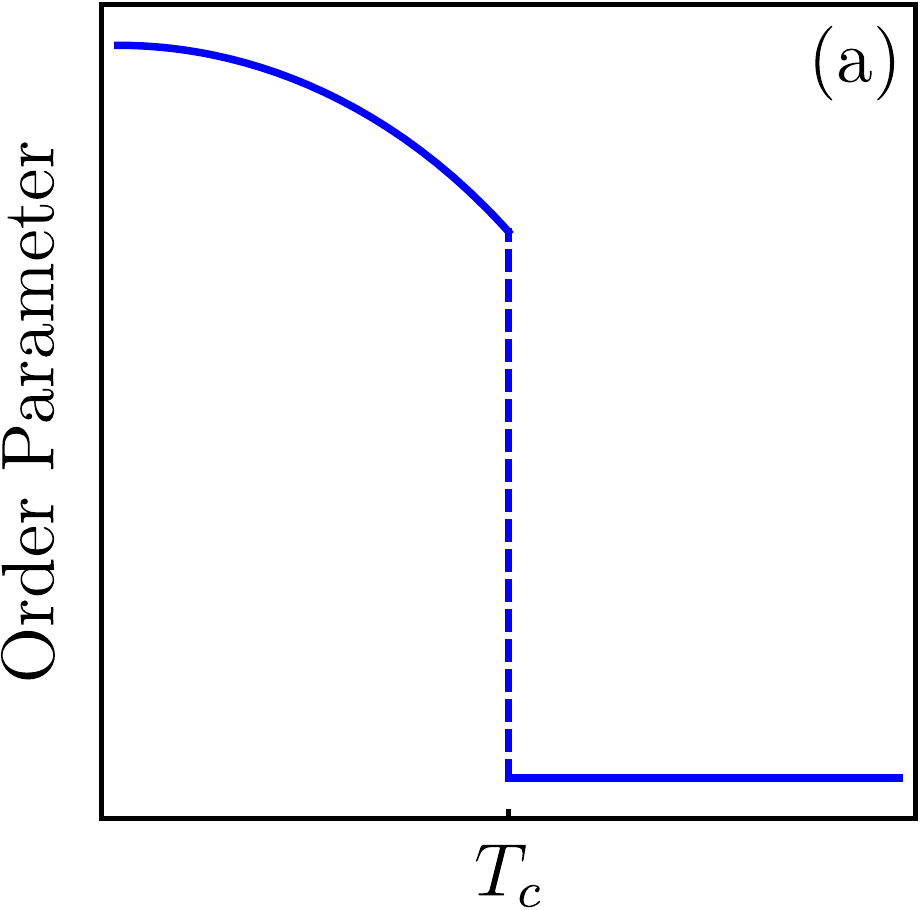}\hspace{2cm}\includegraphics[scale=0.7]{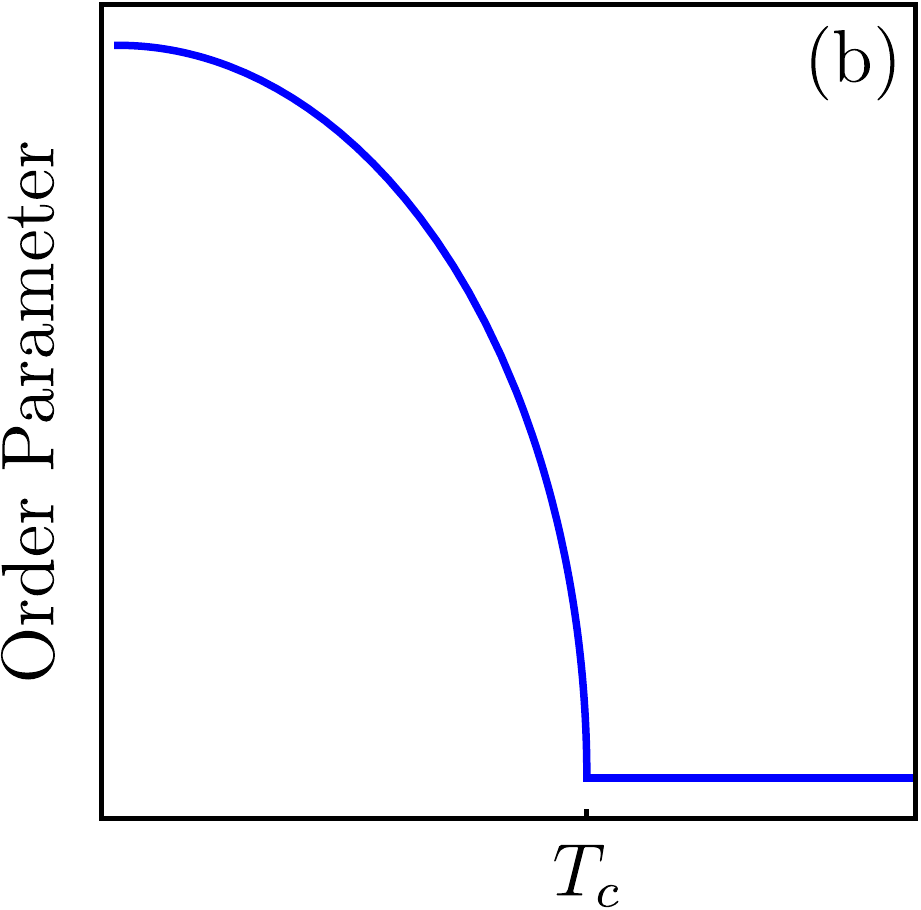}
    \caption{Order parameter for (a) a first-order phase transition and (b) a second-order phase transition. The transition occurs at the critical temperature $T_c$.}
    \label{Fig:order-parameter-phases-tran}
\end{figure}

Clearly, the theory returns to be chiral symmetric if $m_f=0$ which implies $v=0$. That scenario can be accomplished if $v$ is promoted as an order parameter, i.e., for some thermal configuration, there exists a transition from massive fermions ($v\neq0$) to a chiral phase of massless fermions ($v=0$). The order of the named phase transition is given by how $v$ goes to zero: if in the critical temperature $T_c$ the order parameter is discontinuous, the transition is first-order, but if $v$ tends continuously to zero, the transition is second-order or a crossover. Fig.~\ref{Fig:order-parameter-phases-tran} shows the order parameter's behavior in a first-order and second-order phase transition. 

\subsection{Effective Potential for the \texorpdfstring{L$\sigma$M}{}}\label{sec:Effective_Potential_for_the_LsM}

The chiral phase transition can be modeled by the Lagrangian of Eq.~(\ref{LagLsMSSB}) from the imaginary-time formalism of the thermal field theory. To implement a phase diagram in the $T-\mu$ plane, the quark chemical potential is included by assuming the conservation of the baryon number $\mathcal{Q}$, and therefore, the equilibrium system is described by the grand canonical partition function
\bea
\mathcal{Z}=\text{Tr}\left[ e^{-\beta\left(\mathcal{H}-\mu\mathcal{Q}\right)}\right],\text{ with }\beta=1/T, 
\eea
from which the thermodynamic potential is calculated.

It can be shown that such potential corresponds to the interacting part of the Lagrangian in Eq.~(\ref{LagLsMSSB}):
\bea
V_{\text{int}}&=&\frac{\lambda}{4}v^4-\frac{a^2}{2}v^2+\frac{\lambda}{4}\left(\sigma^2+\boldsymbol{\pi}^2\right)^2+g\bar{\psi}\left(\sigma+i\gamma_5\boldsymbol{\tau}\cdot\boldsymbol{\pi}\right)\psi,
\eea
so that the classical or {\it tree} level contribution is
\bea
V_{\text{cl}}=\frac{\lambda}{4}v^4-\frac{a^2}{2}v^2
\label{Vcl1},
\eea
and the quantum corrections are determined by
\bea
\mathcal{L}_{\text{int}}=\frac{\lambda}{4}\left(\sigma^2+\boldsymbol{\pi}^2\right)^2+g\bar{\psi}\left(\sigma+i\gamma_5\boldsymbol{\tau}\cdot\boldsymbol{\pi}\right)\psi.
\label{Lint}
\eea

In general terms, the thermal potential is a power series in $\lambda$. For this work, only the tree-level, the one-loop, and the ring contributions (plasma screening effects) will be taken into account for bosons and fermions. In the following, the expressions for the effective potentials, self-energy, and ring contributions are well-known results found in Ref.~\cite{LeBellac}.

First, the one-loop boson potential can be written as
\bea
V_b^{(1)}=T\sum_n\int\frac{d^3k}{\dpi^3}\ln D\left(\omega_n,\mathbf{k}\right)^{-1/2},
\label{Vb1def}
\eea
where
\bea
D\left(\omega_n,\mathbf{k}\right)=\frac{1}{\omega_n^2+\mathbf{k}^2+m_b^2},
\label{bosonpropag}
\eea
is the free propagator, $m_b$ is the boson's mass and $\omega_n=2n\pi T$ are the Matsubara's frequencies.

On the other hand, the one-loop correction for fermions at finite temperature and chemical potential $\mu$ is
\bea
V_f^{(1)}=-T\sum_n\int\frac{d^3k}{\dpi^3}\text{Tr}\left[\ln S\left(\widetilde{\omega}_n-i\mu,\mathbf{k}\right)^{-1}\right],
\label{Vf1def}
\eea
where
\bea
S\left(\widetilde{\omega}_n,\mathbf{k}\right)=\frac{1}{\gamma_0\widetilde{\omega}_n+\slashed{k}+m_f},
\label{fermionpropag}
\eea
is the propagator of a fermion with mass $m_f$ and Matsubara frequencies $\widetilde{\omega}_n=(2n+1)\pi T$.

\begin{figure}[h]
    \centering
    \includegraphics[scale=1.1]{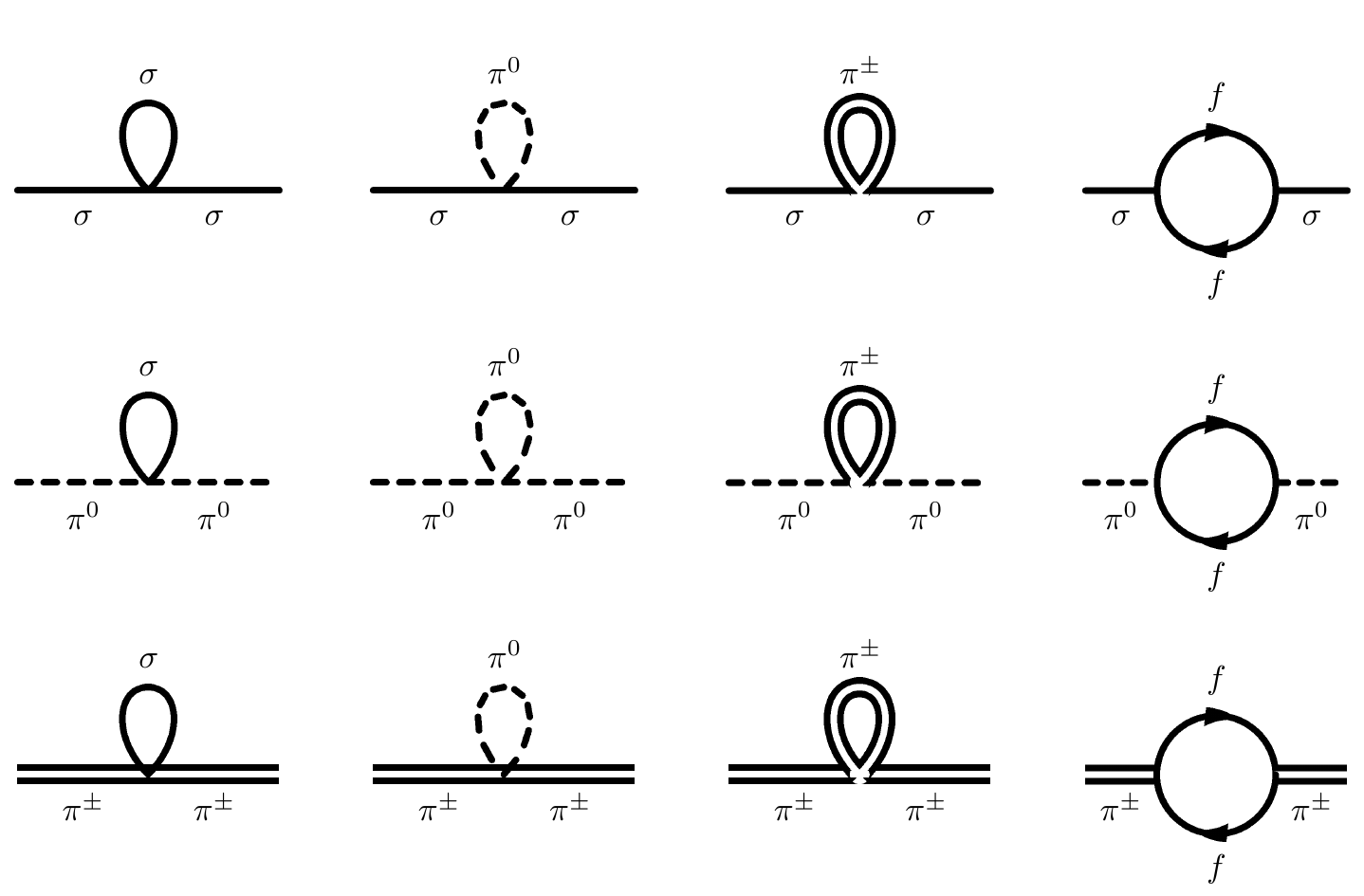}
    \caption{Self-energy Feynman diagrams for each boson species. From top to bottom, each line of diagrams contributes to $\Pi_\sigma$, $\Pi_{\pi^0}$, and $\Pi_{\pi^\pm}$, respectively. }
    \label{FIG:self_energy_bosons}
\end{figure}

Finally, the ring contribution is given by
\bea
V^{(\text{ring})}=\frac{T}{2}\sum_n\int\frac{d^3k}{\dpi^3}\ln\left[1+\Pi\left(m_b,T,\mu\right)D\left(\omega_n,\mathbf{k}\right)\right],
\label{Vringdef}
\eea
where $\Pi\left(m_b,T,\mu\right)$ is the boson's self-energy.

Figure.~\ref{FIG:self_energy_bosons} shows the diagrammatic representation of the therms contributing to $\Pi\left(m_b,T,\mu\right)$, such that each boson has a self-energy corresponding to a boson-boson and boson-fermion loops, and the total self-energy is given by
\bea
\Pi_l(T,\mu)=\Pi^b_l(T)+\sum_{j=u,d}\Pi_j^f(T,\mu),
\eea
where $l$ runs for the boson fields $(\sigma,\pi^0,\pi^{\pm})$ and $j$ for the two light quark flavors $(u,d)$. The explicit form of the pure bosonic part of $\Pi_l^b(T)$ is
\begin{subequations}
\bea
\Pi^b_\sigma=\lambda\left[3I\left(\sqrt{m_\sigma^2+\Pi^b_\sigma}\right)+2I\left(\sqrt{m_{\pi^\pm}^2+\Pi^b_{\pi^\pm}}\right)+I\left(\sqrt{m_{\pi^0}^2+\Pi^b_{\pi^0}}\right)\right],
\eea
\bea
\Pi^b_{\pi^0}=\lambda\left[I\left(\sqrt{m_\sigma^2+\Pi^b_\sigma}\right)+2I\left(\sqrt{m_{\pi^\pm}^2+\Pi^b_{\pi^\pm}}\right)+3I\left(\sqrt{m_{\pi^0}^2+\Pi^b_{\pi^0}}\right)\right],
\eea
\bea
\Pi^b_{\pi^\pm}=\lambda\left[I\left(\sqrt{m_\sigma^2+\Pi^b_\sigma}\right)+4I\left(\sqrt{m_{\pi^\pm}^2+\Pi^b_{\pi^\pm}}\right)+I\left(\sqrt{m_{\pi^0}^2+\Pi^b_{\pi^0}}\right)\right].
\eea
\label{bosonseflenergies}
\end{subequations}
where
\bea
I(x)=\frac{1}{2\pi^2}\int dk\frac{k^2}{\sqrt{k^2+x}}n\left(\sqrt{k^2+x}\right),
\eea
and $n(x)$ is the Bose-Einstein distribution. Moreover, the self-energy for the fermionic loop has the form
\bea
\Pi^f=-gT^2\sum_n\int\frac{d^3k}{\dpi^3}\text{Tr}\left[S\left(\widetilde{\omega}_n-i\mu,\mathbf{k}\right)S\left(\widetilde{\omega}_n-\omega-i\mu,\mathbf{k-p}\right)\right].
\eea

Equations~(\ref{Vcl1}), (\ref{Vb1def}), (\ref{Vf1def}), (\ref{Vringdef}) and~(\ref{bosonseflenergies}) define exactly at $\mathcal{O}(\lambda)$ the full thermodynamic information of the model. Still, as can be noticed, they have a non-trivial solution: for example, the expressions for the bosons self-energies are given by a self-consistent form, which implies a numeric treatment or the implementation of some approximations. In the following sections, high and low-temperature approaches will be treated.

\subsection{Effective Potential in the High Temperature Approximation}

From the phenomenology described at the beginning of this section, the current knowledge of the QCD phase diagram comes from experiments and lattice calculations in the high temperature-low chemical potential region. Therefore, by choosing an appropriate hierarchy of scales, the analysis can be simplified. To implement such a regime, for the first approximation, the temperature is taken as the dominant energy scale, i.e., $T\gg\mu\gg m_q$. The corresponding calculations to this scheme are shown in Appendix~\ref{Ap:Veff_bosons_HT}.

To simplify the expressions of the boson's self-energy, note that near the phase transition the quarks and boson masses can be neglected, and therefore, in Eq.~(\ref{bosonseflenergies}) this can be implemented by setting $m_i^2+\Pi_i=0$, which yields:
\bea
\Pi^b_\sigma=\Pi^b_{\pi^0}=\Pi_{\pi^{\pm}}^b=\frac{6\lambda}{2\pi^2}\int dk\frac{k}{e^{k/T}-1}=\frac{\lambda T^2}{2}.
\label{PibosonHT}
\eea

The fermionic self-energy can be computed without the scale hierarchy $T\gg\mu$ by assuming that near the critical temperature, the chiral symmetry is restored; namely, $m_f=0$, so that:
\bea
\Pi_j^f=-N_c\,g^2\frac{T^2}{\pi^2}\left[\text{Li}_2\left(-e^{\mu/T}\right)+\text{Li}_2\left(-e^{-\mu/T}\right)\right],
\label{PifermionHT}
\eea
where $N_c$ accounts for the number of colors for each quark, and $\text{Li}_2\left(x\right)$ is the second-order polylogarithm function. Thus, the total self-energy for $N_f$ quark-flavors becomes
\bea
\Pi_l(T,\mu)=\frac{\lambda T^2}{2}-N_fN_c\,g^2\frac{T^2}{\pi^2}\left[\text{Li}_2\left(-e^{\mu/T}\right)+\text{Li}_2\left(-e^{-\mu/T}\right)\right].
\label{PitotalHT}
\eea

For the bosonic effective potential of Eq.~(\ref{Vb1def}), the sum over Matsubara frequencies gives
\bea
V_b^{(1)}=\frac{1}{2\pi}\int dk\,k^2\left[\frac{\sqrt{k^2+m_b^2}}{2}+T\ln\left(1-e^{-\sqrt{k^2+m_b^2}/T}\right)\right],
\label{Vb1int}
\eea
which is composed of two pieces: one corresponding to the vacuum contribution and another one corresponding to the matter or $T$-dependent correction. The former can be regularized in the MS-scheme with renormalization scale $\widetilde{\mu}=a\,e^{-1/2}$, whereas the latter is computed from the approximation $m_b/T\ll1$. By taking only the dominant terms, the one-loop effective potential at high temperature is
\bea
V_{b,\text{HT}}^{(1)}=\sum_{i=\sigma,\boldsymbol{\pi}}\Bigg\{-\frac{m_i^4}{64\pi^2}\left[\ln\left(\frac{a^2}{4\pi T^2}\right)-\gamma_e+\frac{1}{2}\right]-\frac{\pi T^4}{90}+\frac{m_i^2 T^2}{24}-\frac{m_i^3T}{12\pi}\Bigg\},\quad
\label{VbHT}
\eea
where $\gamma_e$ is the Euler--Mascheroni constant.

The effective potential for fermions also have a piece of vacuum and a piece of matter, i.e., after summing the Matsubara frequencies the result is
\bea
\quad V_f^{(1)}&=&-\frac{1}{\pi^2}\int dk\,k^2\sqrt{k^2+m_f^2}\nn\\
&+&\frac{T}{\pi^2}\int dk\,k^2\Bigg\{\ln\left[1+e^{-\left(\sqrt{k^2+m_f^2}-\mu\right)/T}\right]+\ln\left[1+e^{-\left(\sqrt{k^2+m_f^2}+\mu\right)/T}\right]\Bigg\},\quad
\label{Vf1int}
\eea
so that by applying the MS-regularization scheme, and the approximations $m_f/T\ll 1$ and $\mu/T\ll 1$, the leading terms of this potential are
\bea
V_{f,\text{HT}}^{(1)}&=&\sum_{f=u,d}\Bigg\{\frac{m_f^4}{16\pi^2}\left[\ln\left(\frac{a^2}{4\pi T^2}\right)-\gamma_e+\frac{1}{2}-\psi^0\left(\frac{1}{2}+\frac{i\mu}{2\pi T}\right)-\psi^0\left(\frac{i\mu}{2\pi T}\right)\right]\nn\\
&-&8m_f^2T^2\left[\text{Li}_2\left(-e^{\mu/T}\right)+\text{Li}_2\left(-e^{-\mu/T}\right)\right]+32T^4\left[\text{Li}_4\left(-e^{\mu/T}\right)+\text{Li}_4\left(-e^{-\mu/T}\right)\right]\Bigg\}.\nn\\
\label{VfHT}
\eea

The plasma screening effects are computed from the lowest Matsubara frequency which is the dominant term~\cite{LeBellac}, so that Eq.~(\ref{Vringdef}) is simplified to
\bea
V^{(\text{ring})}_{\text{HT}}&=&\frac{T^2}{4\pi^2}\int dk\,k^2\Big\{\ln\left[k^2+m_b^2+\Pi\right]-\ln\left[k^2+m_b^2\right]\Big\}\nn\\
&=&\frac{T}{12\pi}\left[m_b^3-\left(m_i^2+\Pi\right)^{3/2}\right].
\label{VringHT}
\eea

By joining all the pieces, the first-order effective potential up to the ring diagrams for the L$\sigma$M with $N_c$ quark colours at high temperature is
\bea
V^{\text{eff}}_{\text{HT}}&=&-\frac{a^2}{2}v^2+\frac{\lambda}{4}v^4\nn\\
&-&\sum_{i=\sigma,\boldsymbol{\pi}}\Bigg\{\frac{m_i^4}{64\pi^2}\left[\ln\left(\frac{a^2}{4\pi T^2}\right)-\gamma_e+\frac{1}{2}\right]+\frac{\pi T^4}{90}-\frac{m_i^2 T^2}{24}+\frac{T}{12\pi}\left(m_i^2+\Pi\right)^{3/2}\Bigg\}\quad\nn\\
&+&N_c\sum_{f=u,d}\Bigg\{\frac{m_f^4}{16\pi^2}\left[\ln\left(\frac{a^2}{4\pi T^2}\right)-\gamma_e+\frac{1}{2}-\psi^0\left(\frac{1}{2}+\frac{i\mu}{2\pi T}\right)-\psi^0\left(\frac{i\mu}{2\pi T}\right)\right]\quad\nn\\
&-&8m_f^2T^2\left[\text{Li}_2\left(-e^{\mu/T}\right)+\text{Li}_2\left(-e^{-\mu/T}\right)\right]+32T^4\left[\text{Li}_4\left(-e^{\mu/T}\right)+\text{Li}_4\left(-e^{-\mu/T}\right)\right]\Bigg\},\nn\\
\label{VtotalHT}
\eea
where the boson and fermion masses are given by Eq.~(\ref{masses}), consequently, the effective potential is an explicit function of the order parameter $v$.

\subsection{Effective Potential in the Low Temperature Approximation}\label{Effective_Potential_in_the_Low_Temperature_Approximation}

The high-temperature approximation results is a useful simplification to perform the momentum integration in the effective potentials, resulting in analytical expressions. If one desires to get information about the chiral symmetry restoration when $T\sim\mu$ or $T<\mu$, numerical integration must be performed. Nevertheless, the integrals can be manipulated to obtain simpler integrals. 

The first convenient manipulation is for the ring contribution of Eq.~(\ref{Vringdef}), by taking into account the sum over the bosonic fields:
\bea
V^{(\text{ring})}=\frac{T}{2}\sum_{i=\sigma,\boldsymbol{\pi}}\sum_{n=-\infty}^{\infty}\int\frac{d^3k}{\dpi^3}\ln\left[1+\Pi_i\left(m_i,T,\mu\right)D\left(\omega_n,\mathbf{k}\right)\right].
\eea

Replacing the expression for $D(\omega_n,\boldsymbol{k})$ from Eq.~(\ref{bosonpropag}) it is easy to show that
\bea
V^{(\text{ring})}&=&\frac{T}{2}\sum_{i=\sigma,\boldsymbol{\pi}}\sum_{n=-\infty}^{\infty}\int\frac{d^3k}{\dpi^3}\ln\left[\frac{k^2+m_i^2+\omega_n^2+\Pi_i}{k^2+m_i^2+\omega_n^2}\right]\nn\\
&=&\frac{T}{2}\sum_{i=\sigma,\boldsymbol{\pi}}\sum_{n=-\infty}^{\infty}\int\frac{d^3k}{\dpi^3}\Big[\ln\left(k^2+m_i^2+\omega_n^2+\Pi_i\right)-\ln\left(k^2+m_i^2+\omega_n^2\right)\Big].
\eea

Now, the one-loop of Eq.~(\ref{Vb1def}) can be written as
\bea
V_b^{(1)}&=&T\sum_{i=\sigma,\boldsymbol{\pi}}\sum_{n=-\infty}^{\infty}\int\frac{d^3k}{\dpi^3}\ln D\left(\omega_n,\mathbf{k}\right)^{-1/2}\nn\\
&=&\frac{T}{2}\sum_{i=\sigma,\boldsymbol{\pi}}\sum_{n=-\infty}^{\infty}\int\frac{d^3k}{\dpi^3}\ln\left(k^2+m_i^2+\omega_n^2\right),
\eea
therefore, 
\bea
V_b^{(1)}+V^{(\text{ring})}&=&\frac{T}{2}\sum_{i=\sigma,\boldsymbol{\pi}}\sum_{n=-\infty}^{\infty}\int\frac{d^3k}{\dpi^3}\ln\left(k^2+m_i^2+\omega_n^2+\Pi_i\right)\nn\\
&=&T\sum_{i=\sigma,\boldsymbol{\pi}}\sum_{n=-\infty}^{\infty}\int\frac{d^3k}{\dpi^3}\ln\left[D^{(\text{ring})}\right]^{-1/2}.
\label{Vb+Vr}
\eea

The above equation implies that the calculation of $V_b^{(1)}+V^{(\text{ring})}$ can be performed as if the integral contains a simple boson propagator with the replacement $m_i^2\rightarrow m_i^2+\Pi_i$. 

Note that Eq.~(\ref{Vb+Vr}) can be manipulated in the form
\bea
V_b^{(1)}+V^{(\text{ring})}&=&T\sum_{i=\sigma,\boldsymbol{\pi}}\sum_{n=-\infty}^{\infty}\int\frac{d^3k}{\dpi^3}\int dm^2\frac{\partial}{\partial m^2}\ln\left[D^{(\text{ring})}\right]^{-1/2}\nn\\
&=&-\frac{T}{2}\sum_{i=\sigma,\boldsymbol{\pi}}\sum_{n=-\infty}^{\infty}\int\frac{d^3k}{\dpi^3}\int dm^2\left[D^{(\text{ring})}\right]^{-1}\frac{\partial}{\partial m^2}\left[D^{(\text{ring})}\right]\nn\\
&=&\frac{1}{2}\sum_{i=\sigma,\boldsymbol{\pi}}\int\frac{d^3k}{\dpi^3}\int dm^2\,T\sum_{n=-\infty}^{\infty}\frac{1}{\omega_n^2+k^2+m_i^2+\Pi_i},
\eea
and from the identity~\cite{LeBellac}
\bea
T\sum_{n=-\infty}^{\infty}\frac{1}{\omega_n^2+\omega^2}=\frac{1}{2\omega}\left(1+\frac{2}{e^{\omega/T}-1}\right),
\eea
it follows that
\bea
V_b^{(1)}+V^{(\text{ring})}&=&\frac{1}{2}\sum_{i=\sigma,\boldsymbol{\pi}}\int\frac{d^3k}{\dpi^3}\int dm^2\frac{1}{2\sqrt{k^2+m_i^2+\Pi_i}}\left[1+\frac{1}{\exp\left(\frac{\sqrt{k^2+m_i^2+\Pi_i}}{T}\right)-1}\right]\nn\\
&=&\sum_{i=\sigma,\boldsymbol{\pi}}\int\frac{d^3k}{\dpi^3}\left[\frac{\omega_i}{2}+T\ln\left(1-e^{-\omega_i/T}\right)\right],
\label{Vb+Vrfinal}
\eea
where $\omega_i^2=k^2+m_i^2+\Pi_i$.

For the self-energies, the fermionic contribution is taken to be the same as Eq.~(\ref{PifermionHT}). This choice is adequate for the temperatures of interest, namely, close to the phase transition, given that $T$ can still be considered large compared to the fermion mass. The same argument applies to the effective fermionic potential at low temperatures.

At the bosonic sector, the temperature cannot be taken small compared to the mass parameter $a$. Therefore, none expansion can be performed. Nevertheless, from the fact that the neutral and charged pions have the same mass as a consequence of isospin symmetry, the set of Eqs.~(\ref{bosonseflenergies}) is reduced to the more straightforward system:
\begin{subequations}
\bea
\Pi^b_\sigma=\lambda\left[3I\left(\sqrt{m_\sigma^2+\Pi^b_\sigma}\right)+3I\left(\sqrt{m_{\pi}^2+\Pi_{\pi}^b}\right)\right],
\eea
\bea
\Pi_\pi^b=\lambda\left[I\left(\sqrt{m_\sigma^2+\Pi^b_\sigma}\right)+5I\left(\sqrt{m_{\pi}^2+\Pi_{\pi}^b}\right)\right],
\eea
\label{selfenergybosonLT}
\end{subequations}
which leads to the low-temperature self-energy interaction
\bea
\Pi_l=\Pi^b_l-N_cN_f\,g^2\frac{T^2}{\pi^2}\left[\text{Li}_2\left(-e^{\mu/T}\right)+\text{Li}_2\left(-e^{-\mu/T}\right)\right].
\label{PiLT}
\eea

By collecting Eqs.~(\ref{VfHT}), (\ref{Vb+Vr}), and~(\ref{PiLT}), and after mass renormalization at the
scale $\widetilde{\mu}=a\,e^{-1/2}$, the effective potential in the low temperature approximation is
\bea
V^{\text{eff}}_{\text{LT}}&=&-\frac{a^2}{2}v^2+\frac{\lambda}{4}v^4-\sum_{i=\sigma,\boldsymbol{\pi}}\Bigg\{\frac{\left(m_i^2+\Pi_i\right)^2}{64\pi^2}\left[\ln\left(\frac{4\pi a^2}{m_i^2+\Pi_i}\right)-\gamma_e+\frac{1}{2}\right]\nn\\
&+&\frac{T}{2\pi^2}\int dk\,k^2\ln\left[1-\exp\left(-\frac{\sqrt{k^2+m_i^2+\Pi_i}}{T}\right)\right]\Bigg\}\quad\nn\\
&+&N_c\sum_{f=u,d}\Bigg\{\frac{m_f^4}{16\pi^2}\left[\ln\left(\frac{a^2}{4\pi T^2}\right)-\gamma_e+\frac{1}{2}-\psi^0\left(\frac{1}{2}+\frac{i\mu}{2\pi T}\right)-\psi^0\left(\frac{i\mu}{2\pi T}\right)\right]\quad\nn\\
&-&8m_f^2T^2\left[\text{Li}_2\left(-e^{\mu/T}\right)+\text{Li}_2\left(-e^{-\mu/T}\right)\right]+32T^4\left[\text{Li}_4\left(-e^{\mu/T}\right)+\text{Li}_4\left(-e^{-\mu/T}\right)\right]\Bigg\}.\nn\\
\label{VtotalLT}
\eea

Equations~(\ref{VtotalHT}) and~(\ref{VtotalLT}) help to describe a wide range of the $T-\mu$ plane. The parameters $a$, $\lambda$, and $g$ are chosen by comparison with other works and by lattice inspired calculations. That will be explained in the next subsection.

\subsection{Locating the CEP and Computing the Pressure}\label{Sec:Locating_the_CEP_and_Computing the_Pressure}

The discussed framework enables the study of the chiral phase transition when the temperature and the quark-chemical potential vary independently. It is worthy of mentioning that the present formalism only reproduces effective or average values of the thermodynamic observables along the transition curve, which will be appreciated at the pressure's behavior. Thus, the results given by the Linear Sigma Model have to be improved for a theory that takes into account the fundamental QCD degrees of freedom, i.e., quarks and gluons. In the following, the results presented need to be understood as a rough approximation to the real CEP location.

\begin{figure}
    \centering
    \includegraphics[scale=0.7]{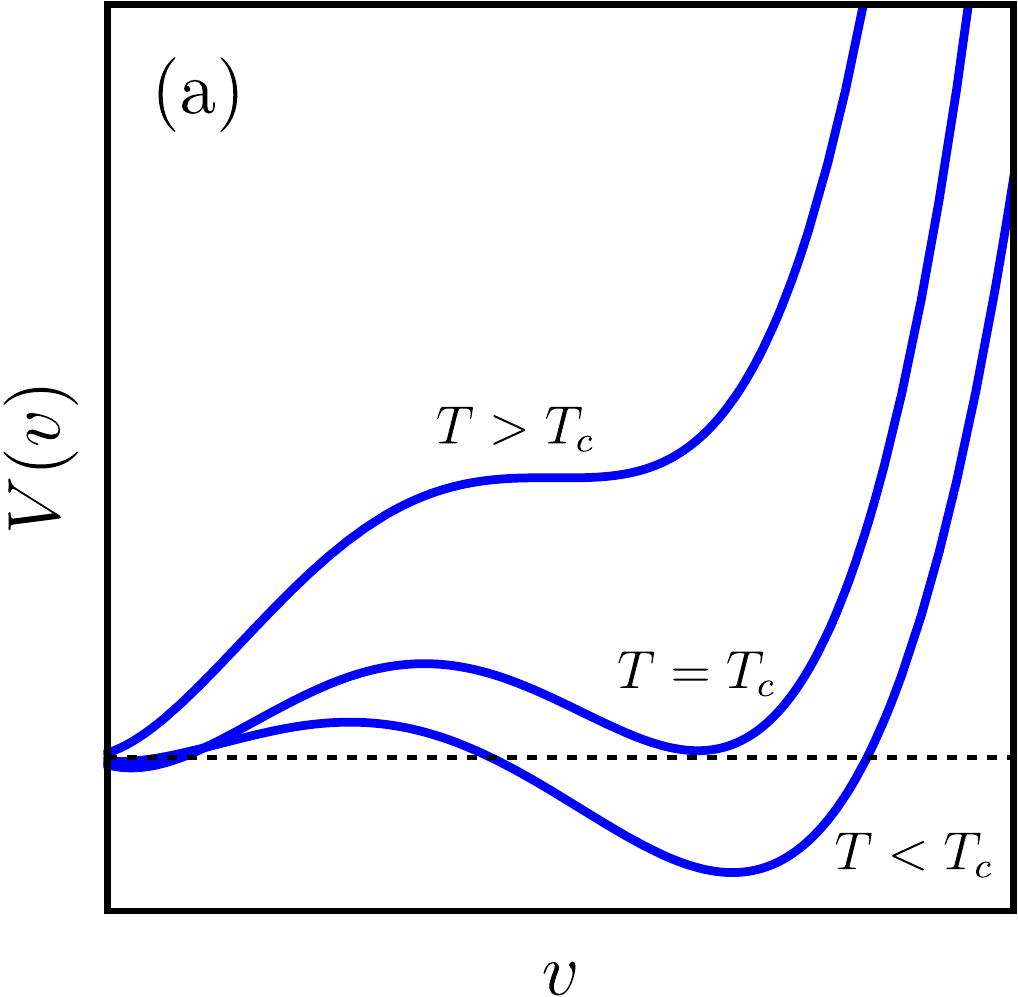}\hspace{1cm} \includegraphics[scale=0.7]{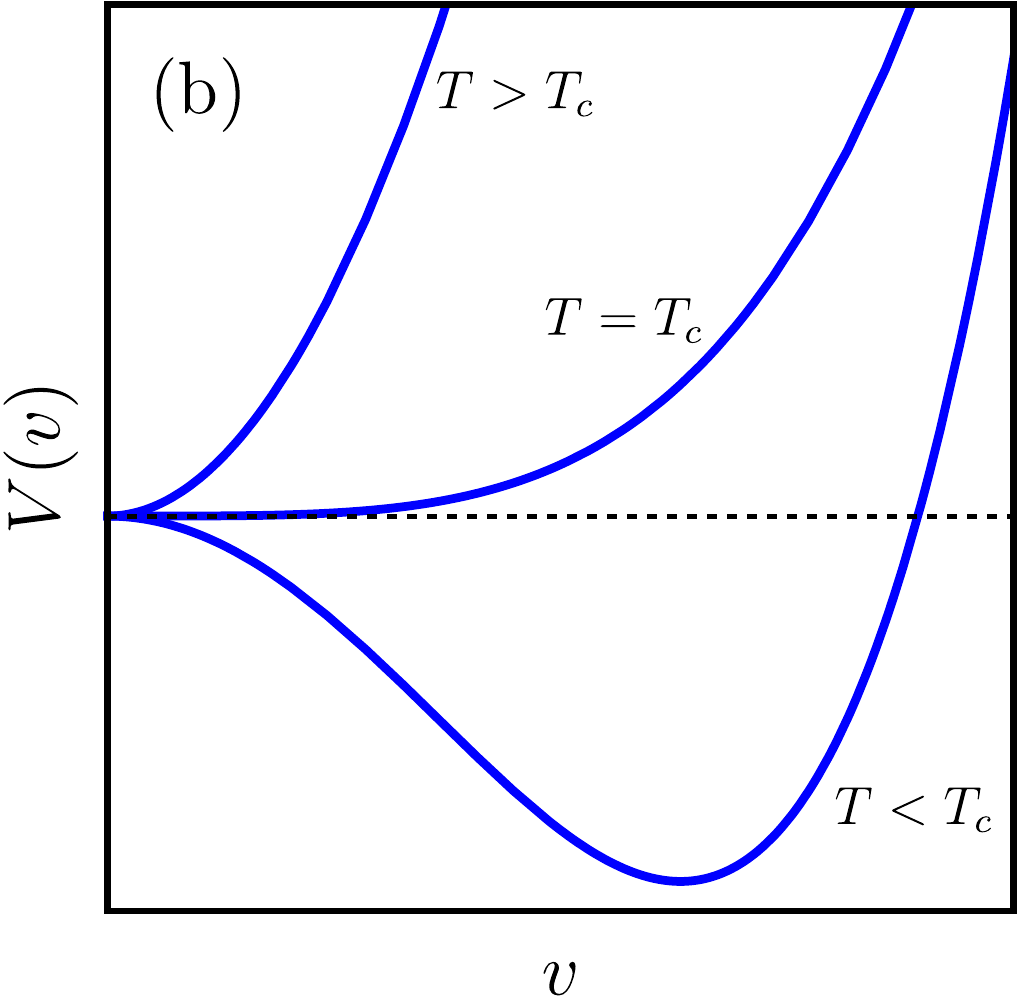}
    \caption{Thermodynamic potential $V(v)$ as a function of the order parameter $v$ for temperatures lower, equal, and greater than the critical temperature $T_c$. The phase transition is classified as (a)~first-order and (b)~second-order.}
    \label{Fig:V_order_phase_tran}
\end{figure}

Classification of the phase transition order is given in terms of the value of the order parameter, as discussed in Fig.~\ref{Fig:order-parameter-phases-tran}. This is achieved by following the behavior of the effective potential $V^{eff}$ when the critical parameters $T_c$ and $\mu_c$ are reached. From a pure thermodynamic description, the system prefers to be in the lowest energy state, which is translated as the minimum of the potential. In that sense, there are three possible cases as a function of the thermal parameters, as is shown in Fig.~\ref{Fig:V_order_phase_tran}:
\begin{itemize}
    \item[(i)] The system has a temperature $T<T_c$, and the symmetry is broken.
    \item[(ii)]The system has a temperature $T=T_c$, and the symmetry is restored.
    \item[(iii)]The system has a temperature $T>T_c$, and the symmetry remains.
\end{itemize}

Of particular interest is when the thermal parameters coincide with the critical ones, i.e., when $T=T_c$. It is well established that there are two configurations for the free energy (effective potential) at the critical line when it is analyzed as a function of the order parameter~\footnote{It can be argued the existence of phase transitions of a superior order in terms of the Ehrenfest classification, which takes into account the $n$-th derivative of a thermodynamic potential. As an example, the 2D Chern-insulators present a third-order topological phase transition. However, there is neither broken symmetry nor local order parameters involved in this transition. Therefore, its discussion is beyond the scope of this thesis.}. Figure~\ref{Fig:V_order_phase_tran}(a) shows two stable minima located at $v=0$ and $v\neq 0$, which implies that although the symmetry is restored, the order parameter can have two possible values. By following the thermal evolution of those minima, one can notice that the stable minimum evolves continuously from $v\neq0$ at $T<T_c$ to $v=0$ and $v\neq 0$ for $T=T_c$, and remains to be $v=0$ for $T>T_c$. That evolution corresponds to Fig.~\ref{Fig:order-parameter-phases-tran}(a), and therefore, it describes a first-order phase transition. On the other hand, the evolution of the stable minimum in Fig~\ref{Fig:V_order_phase_tran}(b) implies a single value of $v$ which goes continuously from $v\neq 0$ for $T<T_c$ to $v=0$ for $T=T_c$ and continues in $v=0$ for $T>T_c$, thus, by comparing with Fig.~\ref{Fig:order-parameter-phases-tran}(b) such behaviour correspond to a second-order phase transition. Hence, the critical parameters on the chiral phase transition and its order will be reported by following the shape of $V^{\text{eff}}$ given by Eqs.~(\ref{VtotalHT}) and~(\ref{VtotalLT}), in terms of the vacuum expectation value $v$.

The L$\sigma$M has three free parameters, namely, $a$, $\lambda$ and $g$, which need to be fixed to the current phenomenology to give an appropriate description of the phase transition. The parameter $a$ is fixed by the bosons vacuum masses, which is implemented by solving for $a$ the Eqs.~(\ref{msigma}) and~(\ref{mpi}), that is:
\bea
a=\sqrt{\frac{m_\sigma^2-3m_\pi^2}{2}}.
\label{arel1}
\eea

Furthermore, to connect with the lattice data at $\mu=0$, notice that the thermal boson masses are given by
\bea
m_b^2(T)=m_{b,\text{vac}}^2+\Pi_b(T,\mu=0),
\eea
where $m_{b,\text{vac}}^2$ corresponds to Eqs.~(\ref{msigma}) and~(\ref{mpi}) and $\Pi_b(T,\mu=0)$ is the high temperature approximation for the self-energy of Eq.~(\ref{PitotalHT}). Explicitly, they have the form
\begin{subequations}
\bea
m_\sigma^2(T)=3\lambda v^2-a^2+\frac{\lambda T}{2}+\frac{N_f N_cg^2T^2}{6},
\label{msigmaT}
\eea
\bea
m_\pi^2(T)=\lambda v^2-a^2+\frac{\lambda T}{2}+\frac{N_f N_cg^2T^2}{6}.
\label{mpiT}
\eea
\label{bosonthermalmasses}
\end{subequations}

At the phase transition ($T=T_c$), the curvature of the effective potential vanishes for $v=0$. Since the boson thermal masses are proportional to this curvature, they also vanish at $v=0$. Therefore, from Eqs.~(\ref{bosonthermalmasses}) a relation of the model parameters and the critical temperature $T_c$ is obtained
\bea
a=T_c\sqrt{\frac{\lambda}{2}+\frac{N_f N_cg^2}{6}}.
\label{arel2}
\eea

The present work takes into account only two quark flavors in the chiral limit, so it is adequate to compare with lattice simulations for $N_f=2+1$. Then, the value of the critical temperature at $\mu=0$ is taken to be $T_c\simeq 170$ MeV~\cite{Maezawa:2007fd}. The Eqs.~(\ref{arel1}) and~(\ref{arel2}) provide constrictions to the coupling $\lambda$ and $g$ through experimental/computational constants.
\begin{figure}[h]
    \centering
    \includegraphics[scale=0.65]{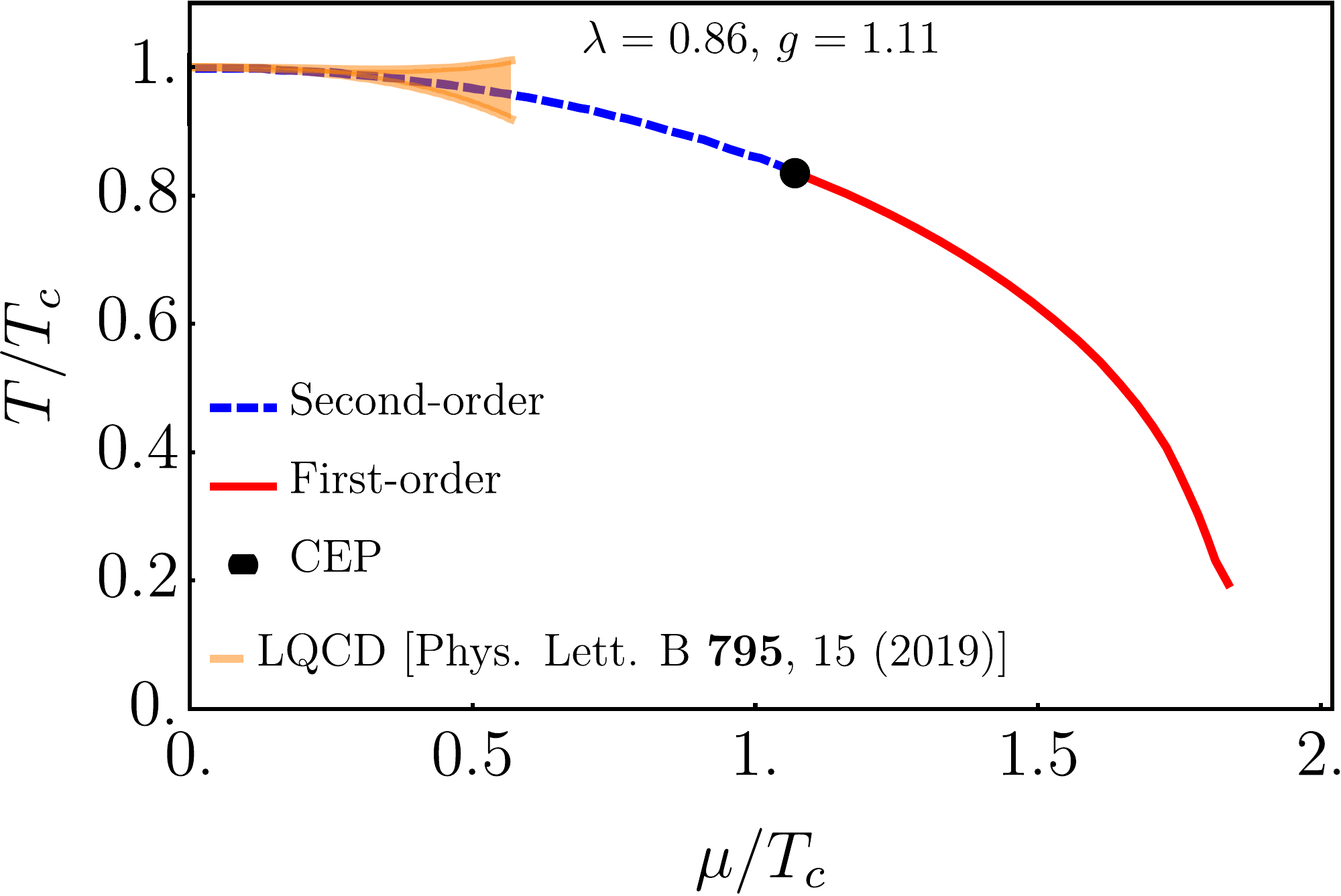}
    \caption{Phase diagram computed from the high-temperature approximation. The orange band is crossover transition region found by the LQCD Taylor expansion method~\cite{latticecrossover2019}.}
    \label{Fig:Phase_Diagram_HT}
\end{figure}

Figure~\ref{Fig:Phase_Diagram_HT} shows the phase diagram obtained from the high-temperature approximation. To obtain it, $\lambda$ and $g$ were chosen in such a way they satisfy Eqs.~(\ref{arel1}) and~(\ref{arel2}) and have a value $\sim\mathcal{O}(1)$. It is worth mentioning that in order to obtain a second-order phase transition at $\mu=0$, it is necessary that $g^2> \lambda $, which implies that the fermionic sector is more important at high densities. As can be noticed, the model reproduces a crossover at low chemical potentials in agreement with the results found by the LQCD Taylor expansion method~\cite{latticecrossover2019}. The second-order phase transition line continues to $\mu/T_c\sim1$ and gives rise to first-order phase transitions. The CEP is between those two regimes, and the effective potential predicts a continuous transit between them. Nonetheless, given that the analysis is made step by step in the values of $\mu$ and $T$, its exact location lies in a tiny band around the black dot showed in the phase diagram. Moreover, the knee-shape at low temperatures and high densities is a consequence of the non-applicability of the approximation $T\gg\mu$; however, the critical line follows the desired behavior.

The Critical End Point is found to be located at $\left(\mu_{\text{CEP}}/T_c,T_\text{CEP}/T_c\right)\sim\left(1.07016, 0.834889\right)$ or $\left(\mu_{\text{CEP}},T_\text{CEP}\right)\sim\left(181.927\text{ MeV}, 141.931\text{ MeV}\right)$. This value is close around the extensions of lattice QCD, which reports the CEP location in the range $\left(\mu^{\text{CEP}}/T_c,T^{\text{CEP}}/T_c\right)\sim(1.0-1.4,0.9-0.95)$~\cite{PhysRevD.71.114014,deForcrand:2006ec,Sharma:2013hsa,Ding:2015ona} and at the upper limit of calculations with effective models where $\left(\mu^{\text{CEP}}/T_c,T^{\text{CEP}}/T_c\right)\sim(1.0-2.0,0.4-0.8)$~\cite{PhysRevD.77.034024,PhysRevD.76.074023,PhysRevD.77.065016,PhysRevD.88.056004,PhysRevD.81.016007,PhysRevD.78.034034,PhysRevD.77.014006,PhysRevD.77.096001}.

\begin{figure}[h]
    \centering
    \includegraphics[scale=0.65]{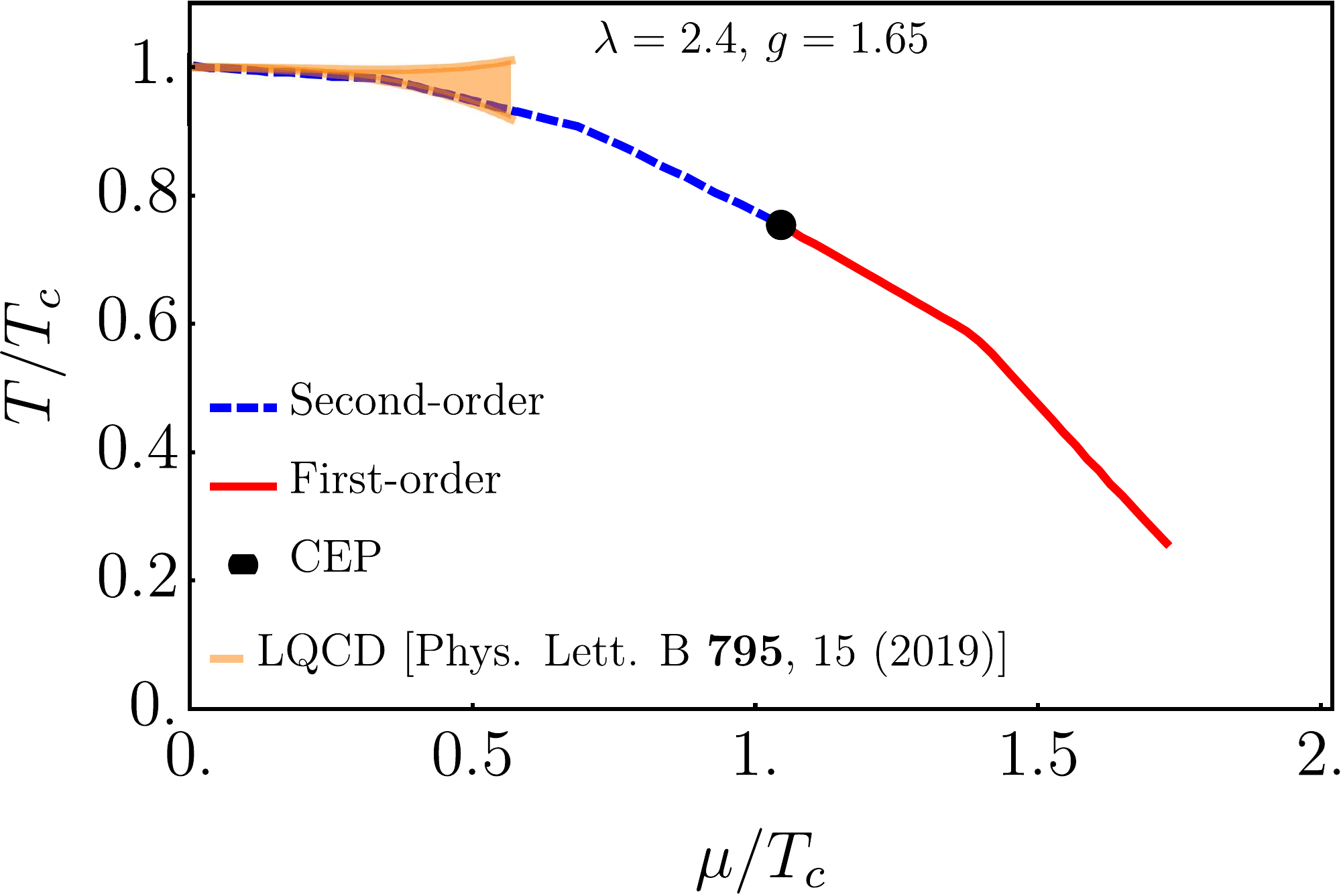}
    \caption{Phase diagram computed from the low-temperature approximation. The orange band is crossover transition region found by the LQCD Taylor expansion method~\cite{latticecrossover2019}.}
    \label{Fig:Phase_Diagram_LT}
\end{figure}

To fix the coupling constant values at the low-temperature approximation, the restriction $T_c/a=1/2$ was imposed, which means to set with $m_\sigma\simeq 540$ MeV. Also, the condition that the CEP is near the values obtained at the high-temperature approximation constrains the parameters $\lambda$ and $g$. That gives $\lambda=2.4$ and $g=1.65$. Its difference compared with the high-temperature sets of couplings is a measure of the change in the effective degrees of freedom in each phase. Figure~\ref{Fig:Phase_Diagram_LT} shows the phase diagram resulting from those considerations. As well as in the high-temperature regime, the crossover and the first-order lines were obtained. The former has a good correspondence with the lattice predictions and the latter starts inside the region where other effective models suggest the CEP position. The transition line smoothness is lost because of the numerical integration to solve the self -consistent system of  Eqs.~(\ref{selfenergybosonLT}) and~(\ref{VtotalLT}), but in essence, the phase diagram obtained from both temperature criteria are the same. 

\begin{figure}
    \centering
    \includegraphics[scale=0.78]{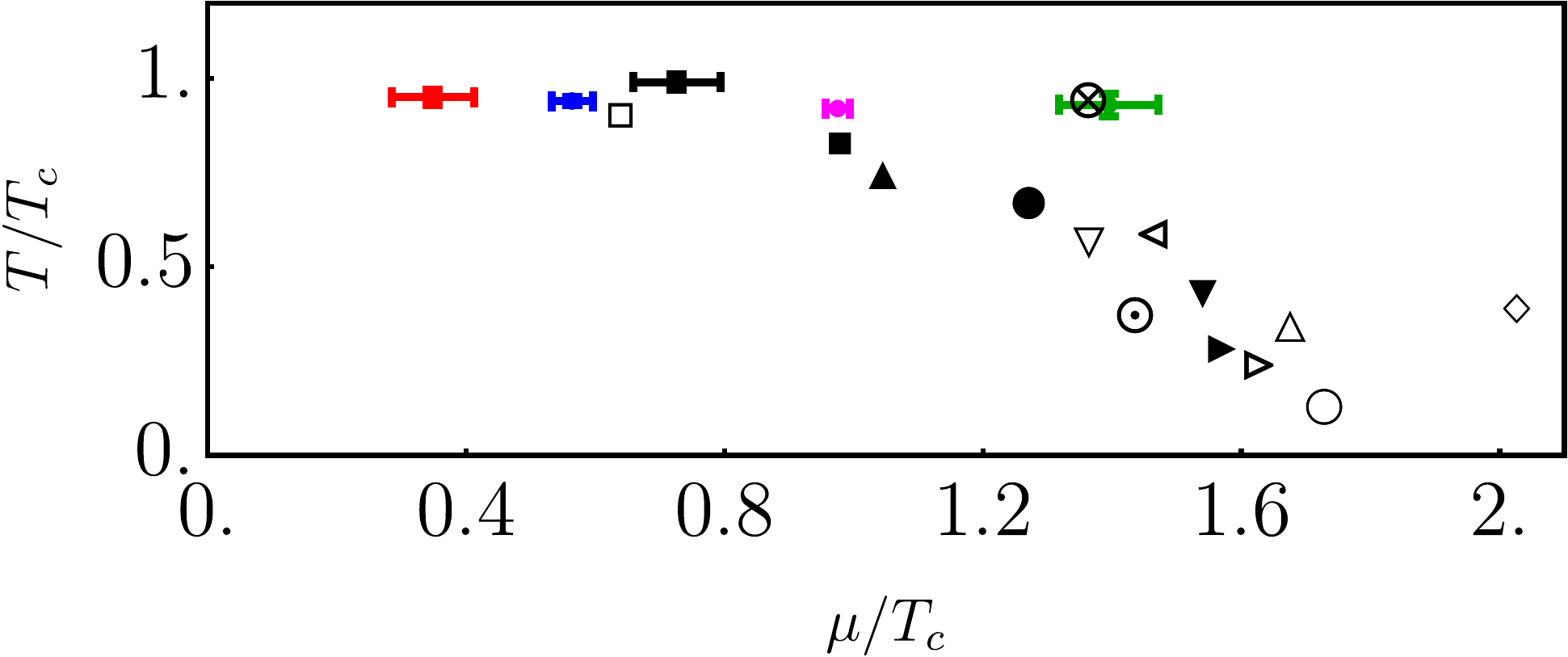}
    \caption{CEP location from different approaches. The points with error bars come from LQCD extensions and they are taken  from Refs.~\cite{PhysRevD.71.114014} (red),~\cite{PhysRevD.78.114503} (blue),~\cite{fodor2004critical} (black),~\cite{deForcrand:2006ec} (magenta), and~\cite{fodor2002lattice} (green). The black filled symbols correspond to the low-temperature approach of Eq.~(\ref{VtotalLT}) when the coupling constants $(\lambda,g)$ are: $(2.2,1.7)$ for $\blacksquare$, $(2.4,1.65)$ for $\blacktriangle$, $(2.1,1.725)$ for $\bullet$, $(2.5,1.7)$ for $\blacktriangledown$, and $(2.7,1.6)$ for $\blacktriangleright$. The empty symbols are the effective models predictions: $\square$ from Ref.~\cite{AyalaSuperStatistics1}, $\bigotimes$ from Ref.~\cite{PhysRevD.88.056004},  $\bigtriangledown$ from Ref.~\cite{PhysRevD.77.014006}, $\bigodot$ from Ref.~\cite{PhysRevD.81.016007}, $\vartriangleleft$ from Ref.~\cite{PhysRevD.79.014018}, $\vartriangleright$ from Ref.~\cite{PhysRevD.78.034034}, $\bigtriangleup$ from Ref.~\cite{PhysRevD.77.096001}, $\bigcirc$ from Ref.~\cite{Ayala:2019skg}, and $\lozenge$ from Ref.~\cite{PhysRevD.77.065016}.}
    \label{Fig:CEP_vs_lambda_and_g}
\end{figure}

The effective model admits a window of values for $\lambda$ and $g$ in such a way that they respect the Eqs.~(\ref{arel1}) and~(\ref{arel2}) and reproduces the CEP location close to the other works predictions. Figure~\ref{Fig:CEP_vs_lambda_and_g} shows the CEP locations computed from the low-temperature result of Eq.~(\ref{VtotalLT}) when the coupling constants are changed (black filled symbols), compared with LQCD extensions (points with error bars), and results taken from other models (empty white symbols). It can be noticed that the change of $\lambda$ and $g$ moves considerably the CEP in such a way that if the boson interaction is stronger than the fermionic one, the first-order phase transition begins at lower temperatures and large quark densities. Moreover, this coupling scenario lies in a region of critical points reported by calculations made from effective estimates based on Nambu–Jona-Lasinio type models with Polyakov loop for 2+1 quarks~\cite{PhysRevD.77.096001,PhysRevD.77.014006,PhysRevD.78.034034,PhysRevD.81.016007,PhysRevD.88.056004}, the linear sigma model with $N_f=3$~\cite{PhysRevD.79.014018}, and perturbation theory~\cite{PhysRevD.77.065016}. Remarkably, the CEP obtained from lower values of the couplings is closer to the LQCD extensions. Thus, the correct choice of degrees of freedom for bosons and fermions moves the phase line, which provides at the model with versatility in terms of the change of few parameters in order to reproduce current or future physical results.
\begin{figure}
    \centering
    \includegraphics[scale=0.4]{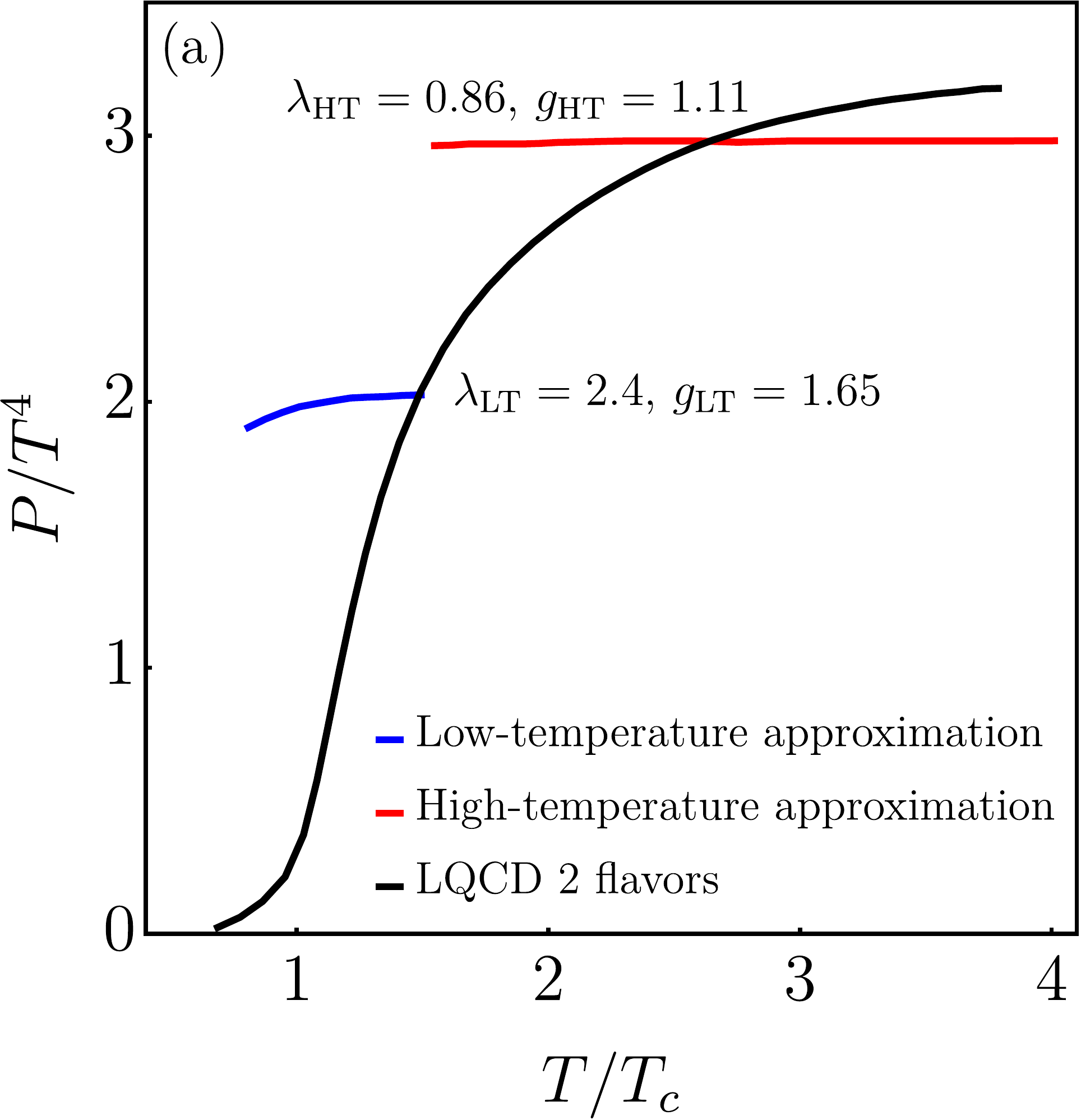}\hspace{0.5cm}\includegraphics[scale=0.4]{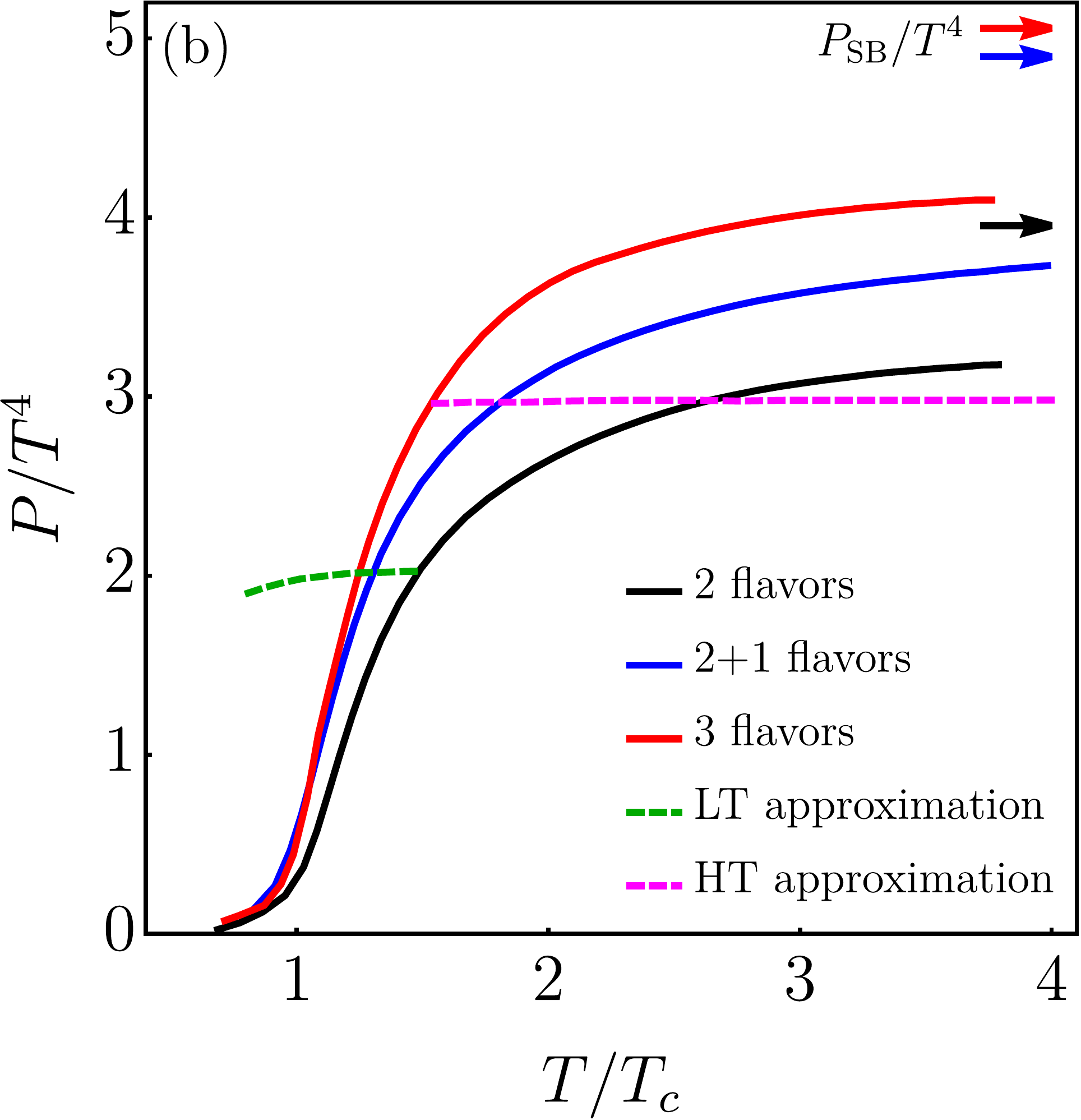}
    \caption{Pressure in units of $T^4$. (a) Comparison between the results obtained from the low and high temperature-approximations and the LQCD data for two light flavors. (b) Pressure from LQCD simulations for 2 and 3 light flavors and 2 light + 1 heavy quark, compared with the named approaches. The arrows indicate the Stefan-Boltzmann gas or the ideal gas limit. The data were taken from Ref.~\cite{karsch2000pressure}.}
    \label{Fig:PressureLsM}
\end{figure}

To test the accuracy of the model in contrast with well-defined observables, the pressure is computed for the low and high-temperature results. Such variable is related to the effective potential as:
\bea
P=-V^{\text{eff}}(v=0).
\eea

Figure~\ref{Fig:PressureLsM}(a) shows the pressure for $\mu=0$ in units of $T^4$ from the results of the present effective model, compared with LQCD data of 2 light flavors~\cite{karsch2000pressure}. The coupling constants were chosen for the low- temperature (LT) potential as $\lambda_{\text{LT}}=2.4$ and $g_{\text{LT}}=1.65$, and for the high-temperature (HT) approximation as $\lambda_{\text{HT}}=0.86$ and $g_{\text{HT}}=1.11$. As can be noticed, the prediction is a constant value for such a thermodynamic variable, with a sudden jump from the LT regime to the HT description. 

That behavior can be understood in terms of QCD degrees of freedom (DOF): it is expected that bulk thermodynamic observables like pressure or energy density change abruptly from the hadronic matter at LT to the deconfined phase at HT. Both limits are usually modeled by
\bea
\frac{P}{T^4}=\left\{\begin{matrix}
 \frac{3\pi^2}{90},&T\rightarrow0 \\ 
\left(16+\frac{21}{2}N_f\right)\frac{\pi^2}{90}, & T\rightarrow\infty,
\end{matrix}\right.
\eea
where $N_f$ is the number of flavors~\cite{lourencco2002quark,karsch2000pressure}. The HT limit is the pressure in the so-called Stephan-Boltzman ($P_\text{SB}$) or ideal gas limit.

Figure~\ref{Fig:PressureLsM}(b) shows the pressure normalized to $T^4$ when the number of degrees of freedom (DOF) corresponding to the number of quarks is changed, together with the effective pressure of the present formalism. The number of DOF has a considerable impact on $P$, indicating that the QCD phase transition at high temperature is deconfined. However,  the transition takes place at rather small values of the pressure, and for temperatures $T\geqslant2T_c$, it is close to the ideal gas limit where the light DOF can be identified. Thus, the quasi-constant nature of $P$ computed from the L$\sigma$M is interpreted as an average description of the sudden change in the number of DOF. Moreover, the values of the coupling constants have a similar interpretation: the discontinuity in $P$ around $T_c$ is due to the depth of the potential of Eq.~(\ref{VtotalHT}), i.e., for smaller couplings the effective potential becomes predominantly negative and therefore the pressure rises. Hence, the LsM couplings reflect how the model can effectively incorporate the change in the DOF from the hadronic to the partonic description.

\section{Conclusion}
The QCD chiral symmetry restoration was studied from the L$\sigma$M as an effective hadronic model. Using the thermal field theory formalism, the boson and fermion potential energies were computed, as well as the self-energy corrections and ring diagrams up to the one-loop approximation. As was discussed, given the hierarchy between the energy scales, namely, the temperature, chemical potential, and fermion mass, the formalism was separated into two regimes: one when the temperature is the dominant scale and other when the chemical potential grows. In the former, the expression has an analytical description, whereas, in the latter, a numeral treatment is implemented. 

From the fact that the Linear Sigma Model presents the Spontaneous Symmetry Breaking, the $\sigma$- expectation value $v$ is chosen as the order parameter of the theory. Thus, the phase transition and its order were identified whit condition $v=0$ is reached from some configuration of  $T$ and $\mu$. Moreover, the free coupling constant $g$ and $\lambda$ are set by phenomenological constraints at $\mu=0$, where Lattice QCD simulations define the critical temperature $T_c$ with two-light quarks flavors. 

The results showed in Figs.~\ref{Fig:Phase_Diagram_HT} and~\ref{Fig:Phase_Diagram_LT}, constructed from the high and low-temperature approximations, respectively, suggest that the formalism gives a QCD phase diagram with the desired properties, and the location of the Critical End Point depending on the coupling constants values.  Moreover, as Fig.~\ref{Fig:CEP_vs_lambda_and_g} indicates, the CEP's location yields in a $T$-$\mu$ region comparable with other models, and Lattice QCD extensions. Therefore, the implemented scenario gives a general description of the chiral symmetry restoration without an involved mathematical formalism.   

Finally, the low-energy and effective nature of the Linear Sigma Model cannot give accurate results obtained from a pure QCD treatment. Such a limitation is well depicted in Fig.~\ref{Fig:PressureLsM} when the pressure computed from LsM is compared with lattice QCD simulations, which have a high level of confidence given that they are computed when $\mu=0$. Nonetheless, the sudden change of degrees of freedom is described as an average of the values of the coupling constants $g$ and $\lambda$.

\begin{savequote}[50mm]
``It is more important to have beauty in one's equations than to have them fit experiment.''
\qauthor{--- Paul Dirac}
\end{savequote}

\chapter{The Direct Photon Excess}\label{chap:Photons}

In this chapter, the effects of intense magnetic field on the photon production of a heavy-ion collision are studied, in particular, the photon sources coming from gluon fusion and gluon splitting channels are proposed to improve the explanation of the experimental data for the invariant momentum distribution and the elliptic flow coefficient $v_2$ reported by ALICE and PHENIX experiments. To take into account the high gluon abundance and the peak of the magnetic field produced in peripheral collisions, the proposed processes are assumed to take place in the early stages of the reaction in the so-called {\it shattered glasma} state. With the purpose of giving a general description of the problem, the theoretical framework, and of explain the assumptions of the calculations, this chapter is organized as follows:

Section~\ref{SEC:Relativistic_Hydrodynamics_and_Elliptic_Flow} is devoted to giving a general discussion about the relativistic hydrodynamical calculations applied to the understanding of the particle production and elliptic flow after a heavy-ion collision experiment. In Sec.~\ref{SEC:The_photon_puzzle} the problem statement is presented: the deviations of the hydrodynamical calculations of the photon production and elliptic flow from the experimental data reported for ALICE and PHENIX in the region of photon's low momentum. Such a problem is known as the {\it photon puzzle}. The physics of the early stages on a heavy-ion collision is explained in Sec.~\ref{Sec:The_Color_Glass_Condensate} from which the named photon production channels are inspired. In particular, the high gluon occupation number and its implications on the parton distribution functions are discussed. Section~\ref{sec:Magnetic_field_HIC} presents the main aspects of the magnetic fields in heavy-ion collisions.

The photon production by gluon fusion is proposed and computed in Sec.~\ref{sec:Prompt Photon Production by Gluon Fusion in a Magnetized Medium}.  whereas the gluon splitting mechanism is discussed in Sec.~\ref{sec:gluon_splitting}. The comparison with the experimental data is performed with phenomenological parameters for the former but computed with the  ultrarelativistic quantum molecular dynamics (UrQMD) for the later.  In both cases, the analytic calculation is carried out by considering the magnetic field as the dominant energy scale and a gluon occupation modeled by a Bose-Einstein-like distribution function. 

The calculations and results presented in this chapter are based on the following manuscripts:
\begin{itemize}
    \item Alejandro Ayala, Jorge David Casta\~no-Yepes, C. A. Dominguez, L. A. Hern\'andez, Saul Hern\'andez-Ortiz, and Maria Elena Tejeda-Yeomans, {\it Prompt photon yield and elliptic flow from gluon fusion induced by magnetic fields in relativistic heavy-ion collisions}, Phys. Rev. D {\bf 96}, 014023 (2017); Erratum Phys. Rev. D {\bf 96}, 119901 (2017).
    
    \item Alejandro Ayala, Jorge David Casta\~no-Yepes, Isabel Dominguez Jimenez, Jordi Salinas San Martin, Maria Elena Tejeda-Yeomans, {\it Centrality dependence of photon yield and elliptic flow from gluon fusion and splitting induced by magnetic fields in relativistic heavy-ion collisions}, Eur. Phys. J. A {\bf 56}, 53 (2020).
    
    \item Alejandro Ayala, Jorge David Casta\~no-Yepes, C. A. Dominguez, L. A. Hernandez, {\it Thermal photons from gluon fusion with magnetic fields},
    EPJ Web of Conferences {\bf 141}, 02007 (2017).

    \item Alejandro Ayala, Jorge David Casta\~no-Yepes, C. A. Dominguez, L. A. Hern\'andez, Saul Hern\'andez-Ortiz, and Maria Elena Tejeda-Yeomans, {\it Prompt photon yield and $v_2$ coefficient from gluon fusion induced by magnetic field in heavy-ion collision}, EPJ Web of Conferences \textbf{172}, 08004 (2018).

    \item Alejandro Ayala, Jorge David Casta\~no-Yepes, Isabel Dominguez Jimenez, Jordi Salinas San Martin, Maria Elena Tejeda-Yeomans, {\it New channels of prompt-photon production by magnetic fields in heavy-ion collisions}, J. Phys. Conf. Ser. \textbf{1602}, 012014 (2020).
\end{itemize}
\section{Relativistic Hydrodynamics and Elliptic Flow}~\label{SEC:Relativistic_Hydrodynamics_and_Elliptic_Flow}

The characterization of the QGP is of great interest for the understanding of the matter which participates in the strong interaction. As it was mentioned, in the technologies of the current experiments based on heavy-ion collisions, there is not a direct control on the dynamics of the participants in each measurement. Hence, the collected signals may have information from stages before and after of the QGP formation. That is why the photons become an excellent (and necessary) probe to test such a state of matter: they do not interact strongly and that feature enables them to escape from the system freely; thus, the photon measurements contain relevant information about the space-temporal evolution of the colliding system.

For each kind of particle produced at the collision, the final momentum distribution is related to the average number of particles $N$ and its Fourier decomposition in the azimuthal angle $\phi$

\bea
E\frac{d^3N}{d^3p}=\frac{1}{2\pi}\frac{dN}{p_Tdp_Tdy}\left[1+2\sum_{n=1}^\infty v_n\cos n\left(\phi-\Psi_n\right)\right],
\label{vndef}
\eea
where $p_\perp$ is the transverse moment, $y$ is the rapidity with respect to the reaction plane and $\Psi_n$ is an angle that defines the reaction plane which is related to spatial orientation~\cite{hidroLuzum}. The flow coefficients $v_n$ can include information about the pseudorapidity $\eta$ as well as momentum dependence~\cite{GaleHadrons,flowPoskanzer}. The hydrodynamical models allow making interpretations of the flow coefficients obtained from the experimental measurements~\cite{GaleHadrons,Heinz}.

The relativistic hydrodynamical calculations analyze the flow of conserved quantities, such as energy, momentum, or other charges. Mathematically, the main equations of the formalism are
\bea
\partial_\mu T^{\mu\nu}&=&0\nn,\\
\partial_\mu J_i^{\mu}&=&0,
\label{HydroEqs}
\eea
where $T^{\mu\nu}$ is the stress-energy tensor and $J_i^{\mu}$ denotes other conserver charges (net baryon number, electric charge, etc). In a differential form, these equation reads
\bea
\frac{d}{dt}\int_V d^3x\;T^{0\nu}&=&-\int_{\partial V}dS_i\; T^{i\nu},\nn\\
\frac{d}{dt}\int_V d^3x \;J_B^0&=&-\int_{\partial V}dS_i\;J_B^i,
\label{hidrodifferentialform}
\eea
where the direct volume dependence is established. The form of Eqs.~(\ref{hidrodifferentialform}) accepts a direct interpretation in terms of the time variation of conserved quantities from incident currents in a finite volume. Therefore, the hydrodynamical framework is centered on the study of the dynamics of the currents~\cite{hydrointroduction}. 

The application of the Relativistic Hydrodynamics (RH) to the study of the QGP arises from the fact that RH is a bulk theory, where the system is assumed to be made of a large number of particles in equilibrium. Their characterization can be reached by a few variables, such as the temperature, collective velocity, and chemical potential. These quantities control energy density, momentum density, and charge density, respectively.  Although the heavy-ion reaction is not static or in equilibrium, it is possible to argue local equilibrium by assuming the expansion rate is much slower than the microscopic interaction rate. In those conditions, the local temperature $T(t,\bf{x})$, chemical potential $\mu(t,\bf{x})$, and collective velocity $u^\mu(t,\bf{x})$ are close to describing Lagrange multipliers to fix the average energy, momentum and net charge, giving rise to a theory of conserved laws. 

The conversion from fluid-viscous dynamics to a particle description of the collision requires \textit{particlization} approaches which are hybrids of ideal or dissipative hydrodynamics (for the early partonic stages) and transport/dilute gas models for the hadronization time scales~\cite{HuovinenHydro,BassHydro1,TeaneyHydro,HiranoHydro,ChihoHydro,BassHydro2,BleicherHydro,NaraHydro,SorgeHydro,AdamHydro}. The matching is made on a constant energy-density or constant temperature hypersurface and the invariant momentum distribution of Eq.~(\ref{vndef}) is usually found by the Cooper-Frye formula~\cite{CooperFryeHydro}
\bea
E\frac{d^3N_i}{d^3p}=p^\mu d\sigma_\mu(x)f_i(x,\bf{p}),
\label{CooperFrye}
\eea
where  $d\sigma_\mu(x)$ is the local surface element vector that is normal to the hypersurface, $x$ is the spacetime, $\bf{p}$ is the momentum and $f_i$ is the phase space density for particle species $i$, which is related to Eqs.~(\ref{HydroEqs}) through
\bea
T^{\mu\nu}(x)=\int\frac{d^3p}{E}p^\mu p^\nu f(x,\bf{p}).
\eea

The stress-energy tensor, as well as the phase-space distribution, varies according to the model requirements, and once the  such models for $T^{\mu\nu}$ and $f_i$ are chosen, the hydrodynamical calculations work as follows:
\begin{figure}[t]
    \centering
    \includegraphics[scale=0.47]{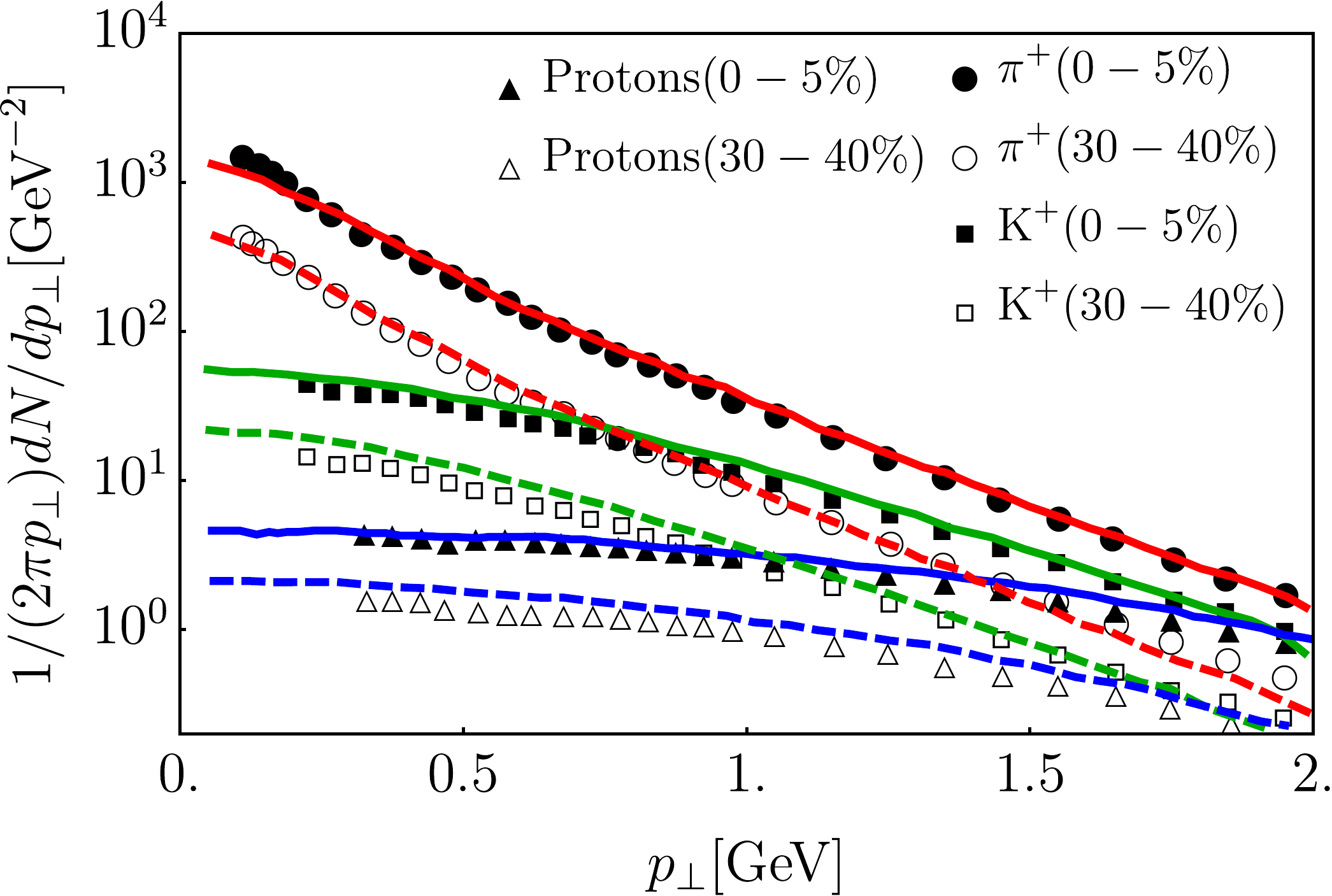}
    \caption{Transverse momentum spectra for pions, kaons and protons, as a function of the transverse momentun $p_\perp$ for two centralities: $0-5\%$ and $30-40\%$. The red (pions), green (kaons), and blue (protons) lines correspond to the predictions of the hydrodynamical calculations by Ryu \etal~\cite{PaquetPRL}. The continuous and dashed lines are referred to the maxima and minimal centralities, respectively. The experimental data are taken from Ref.~\cite{FlowALICE} (ALICE Collaboration) and Refs.~\cite{FlowCMS1,FlowCMS2} (CMS Collaboration).} 
    \label{Fig:RyuHadronMomentaSpectra}
\end{figure}

\begin{itemize}
    \item[1.] \textbf{Initial Condition:} The QGP needs to be presented in some initial configuration. Generally, the Glauber model as a Monte-Carlo approximation for the geometrical distribution of the energy density deposited in the transverse plane is used~\cite{GlauberModel}. Also, the Color Glass Condensate is useful. The last one will be discussed in Sec.~\ref{Sec:The_Color_Glass_Condensate}. The system size and centrality are defined in this steep.
    
    \item[2.]  \textbf{Pre-Equilibrium:} A hydrodynamical calculation cannot be accomplished unless the medium reaches some degree of local momentum isotropization. In its inputs, the relaxation time and the strength of the interaction after the collision are crucial~\cite{PreEquilibrium1,PreEquilibrium2,PreEquilibrium3}. 
    
    \item[3.] \textbf{Hydrodynamical Evolution} The fluid expansion starts with constrains as shear and bulk viscosities, and with the definition of local temperature. The system can interact with external fields (electric, magnetic). Finally, by solving the relations (\ref{HydroEqs}), the equation of state of QGP is achieved, which gives relevant information about the medium (phase transitions, momentum anisotropies).
    
    \item[4.] \textbf{\textit{Particlization} and final states:} The particlization in the Cooper-Frye formalism of Eq.~(\ref{CooperFrye}) allows describing the final states as hadrons, and therefore, the usual techniques of particle and field theory physics can be used. These final states carry on the relevant information of the QGP. 
\end{itemize}

\begin{figure}
    \centering
    \includegraphics[scale=0.47]{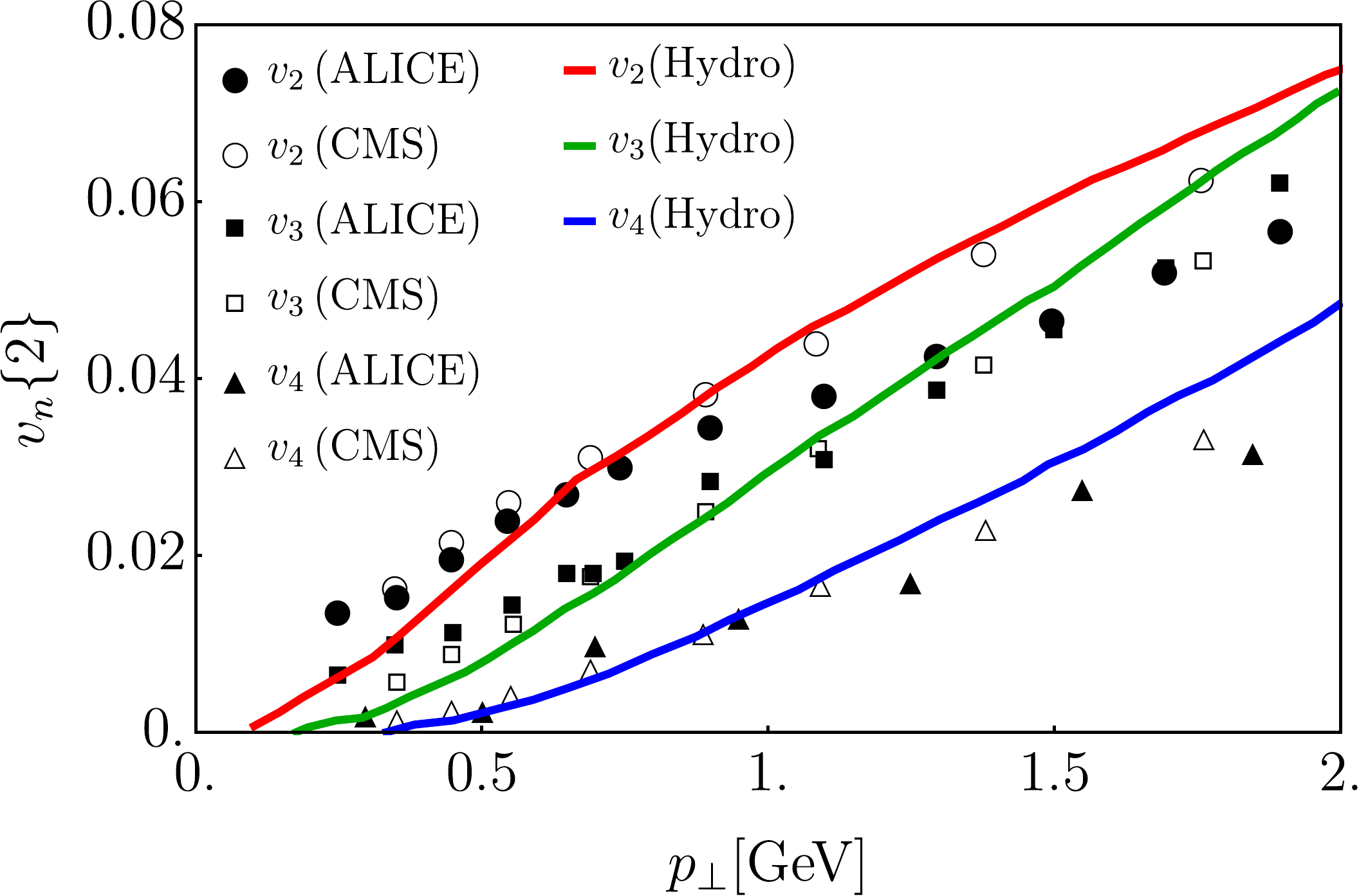}
    \caption{Charged particle flows $v_2$, $v_3$ and $v_4$ as a function of the transverse momentun $p_\perp$ for the centralities $0-5\%$. The red, green, and blue correspond to the predictions of the hydrodynamical calculations of Ryu \etal~\cite{PaquetPRL}. The experimental data are taken from Ref.~\cite{FlowALICE} (ALICE Collaboration) and Refs.~\cite{FlowCMS1,FlowCMS2} (CMS Collaboration).}
    \label{Fig:RyuFlowCoeff}
\end{figure}
As an example of HD-calculations for heavy-ion collisions, the work of Ryu \etal~is a good starting point~\cite{PaquetPRL}. Their initial conditions are determined by using the IP-Glasma model~\cite{IPGlasma1,IPGlasma2}, i.e., an impact-parameter-dependent saturation model for gluons (Glasma), combined with the classical Yang-Mills description of initial Glasma fields. The thermalization time is set to $\tau_0=0.4$ fm. The stress-energy tensor is chosen as a composition of a free-ideal part and a dissipative contribution:
\bea
T^{\mu\nu}=\underbrace{\epsilon u^\mu u^\nu-\Delta^{\mu\nu}P_0(\epsilon)}_{\text{ideal}}+\underbrace{\pi^{\mu\nu}-\Delta^{\mu\nu}\Pi}_{\text{dissipative}},
\label{PaquetEqs}
\eea
where $\epsilon$ is the local energy-density, $P_0(\epsilon)$ is the thermodynamics pressure, $u^{\mu}$ is the four-velocity, $\Pi$ is the bulk viscous pressure, $\pi^{\mu\nu}$ is the shear-stress tensor, and the projector operator is $\Delta^{\mu\nu}=g^{\mu\nu}-u^\mu u^\nu$. The named parameters were selected as follows: $P_0$ was fixed according to the equation of state taken from lattice QCD calculations in the chemical equilibrium of a hadronic gas at low temperatures~\cite{P0Lattice}, whereas $\Pi$ and $\pi^{\mu\nu}$ were found by solving relaxation-type equations of the kinetic theory which include non-linear terms such that couple them. Finally, the momentum distribution of hadrons is calculated via the Cooper-Frye formalism.

Figure~\ref{Fig:RyuHadronMomentaSpectra} shows the results obtained from the above calculations for the $p_T$ spectra of pions, kaons, and protons in two centrality classes. Moreover, Fig.~\ref{Fig:RyuFlowCoeff} shows the charged particle's flow coefficients extracted from the transverse momentum spectra. Both hydrodynamical predictions are compared with the ALICE and CMS experimental data~\cite{FlowALICE,FlowCMS1,FlowCMS2}.  As it can be noticed, the viscosity, the thermalization assumptions, and the pre-equilibrium saturated state are fundamental to have a good description of the observables. Therefore, it can be concluded that RH is a useful phenomenological framework to analyze the findings in a heavy-ion reaction. 

\section{The Direct Photon Puzzle}~\label{SEC:The_photon_puzzle}
As it was commented on the last section, the HD calculations give excellent results when its predictions are compared with the experimental data of the hadronic final states. In terms of the heavy-ion reaction characterization, the hadron spectra can carry on the information of several stages in the collision. Additionally, such information is encoded in a considerable kind of asymptotic states (protons, kaons, pions, etc.). Therefore, the measured signal is plagued with noise, and this makes data analysis difficult. Photons are a clean and abundant source of information within the above described scenario: given that they do not participate in the strong interaction, a photon produced in the QGP phase escapes from the medium without significant scattering processes. In such a way, photons constitute excellent probes to test the strongly interacting matter. 
\begin{figure}[h]
    \centering
    \includegraphics[scale=0.47]{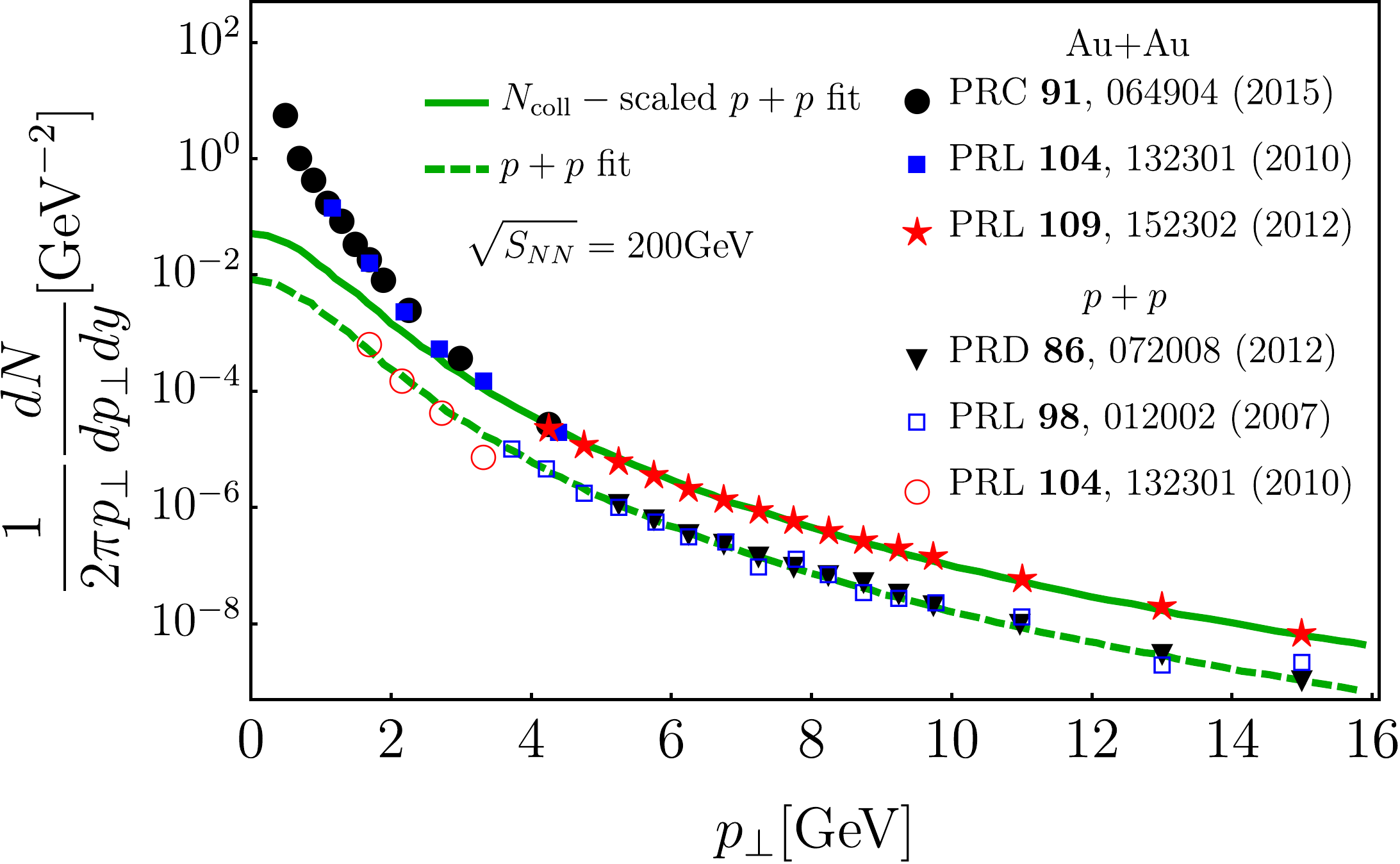}
    \caption{Experimental direct $p_\perp$ photon spectra reported from different PHENIX measurements and compared with the $p+p$ fit (dashed line) and with the $p+p$ scaled by the corresponding average number of binary collisions $N_{\text{coll}}$ (continuous line). Only the central value is shown. For details see Refs.~\cite{pQCD2,RHIC1,RHIC3,PHENIX2012,PHENIX2007}.}
    \label{Fig:Exp_direct_photon_PHENIX}
\end{figure}

\begin{figure}[h]
    \centering
    \includegraphics[scale=0.47]{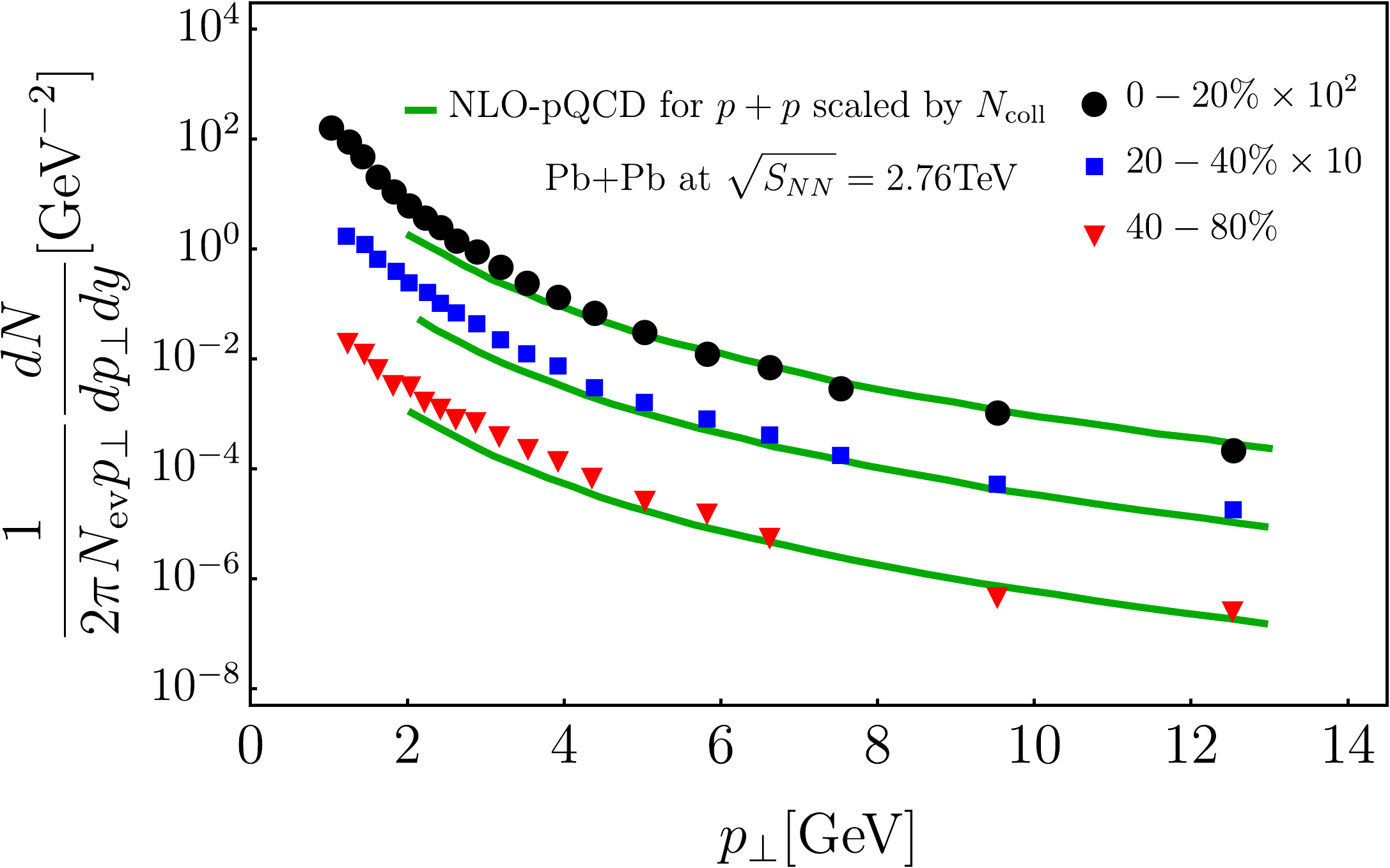}
    \caption{Experimental direct $p_\perp$ photon spectra reported by the ALICE collaboration in Pb+Pb collisions at different centralities. The data is compared to NLO pQCD predictions for the direct photon yield in $p+p$ collisions, which are scaled by the number of binary nucleon collisions $N_{\text{coll}}$ for each centrality class. Only the central value is shown. For details see Ref.~\cite{ALICEPbPb}. }
    \label{Fig:Exp_direct_photon_ALICE}
\end{figure}

The photon invariant distribution measurements have been reported by the PHENIX~\cite{RHIC1,RHIC2,RHIC3,RHIC4} and ALICE~\cite{ALICEPbPb,ALICE1,ALICE2} experiments, for low and intermediate transverse momentum in Au-Au, Cu-Cu and Pb-Pb collisions for different centralities. Figures~\ref{Fig:Exp_direct_photon_PHENIX} and~\ref{Fig:Exp_direct_photon_ALICE} shows the invariant $p_\perp$ photon spectra for the the so-called \textit{direct photons}, i.e., the measurements without photons that come from hadronic decays or final stages. Each collaboration compares the experimental results with theoretical calculations using perturbative-QCD (pQCD) by considering a scaling of the binary collisions for protons. As can be noticed, the PHENIX results in Fig.~\ref{Fig:Exp_direct_photon_PHENIX} show an enhancement above the direct photon production in comparison to the $p+p$ results. This enhancement is found at low transverse momenta ($p_\perp=3$GeV), where the named calculations do not reproduce the overall data behavior. For the ALICE results show in Fig.~\ref{Fig:Exp_direct_photon_ALICE}, the increase in direct photon production at low $p_\perp$ is also observed. Moreover, the yield over the pQCD calculations is present for different centralities, which makes it evident that other radiation sources have to be taken into account. In that way, the invariant momentum distribution has a slope which suggests a foot print of a thermal phase, and therefore, photons must be created in an equilibrated medium. 

To give a phenomenological explanation of the low pt deviations, Paquet \etal~have developed a hydrodynamical approach to study several photon sources in a thermalized medium, namely the QGP~\cite{hydro-photons1}. Their work is based on phenomenology which describes the stress-energy tensor of Eq.~(\ref{PaquetEqs}), namely, an initial state made of a color glass condensate, where the gluonic fields evolve according to the Yang-Mills equations. Here, the thermalization initial time is set to $\tau_0=0.4$ fm, and the viscous properties of the QGP are taken into account.


\begin{figure}[t]
     \centering
     \begin{subfigure}[b]{0.4\textwidth}
         \centering
         \includegraphics[scale=0.45]{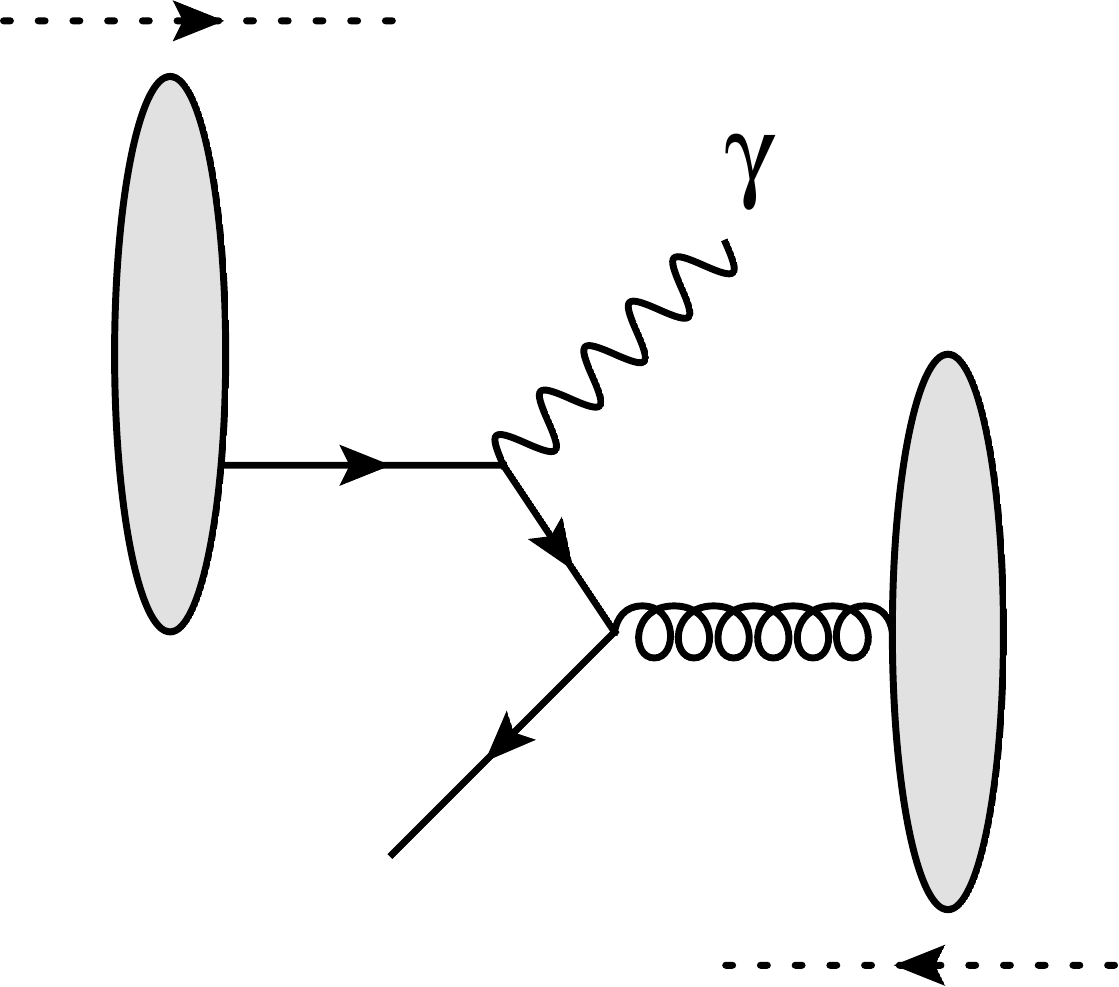}
         \caption{}
         \label{Fig:Promp_photons_sketch}
     \end{subfigure}
     \begin{subfigure}[b]{0.55\textwidth}
         \centering
         \includegraphics[scale=0.5]{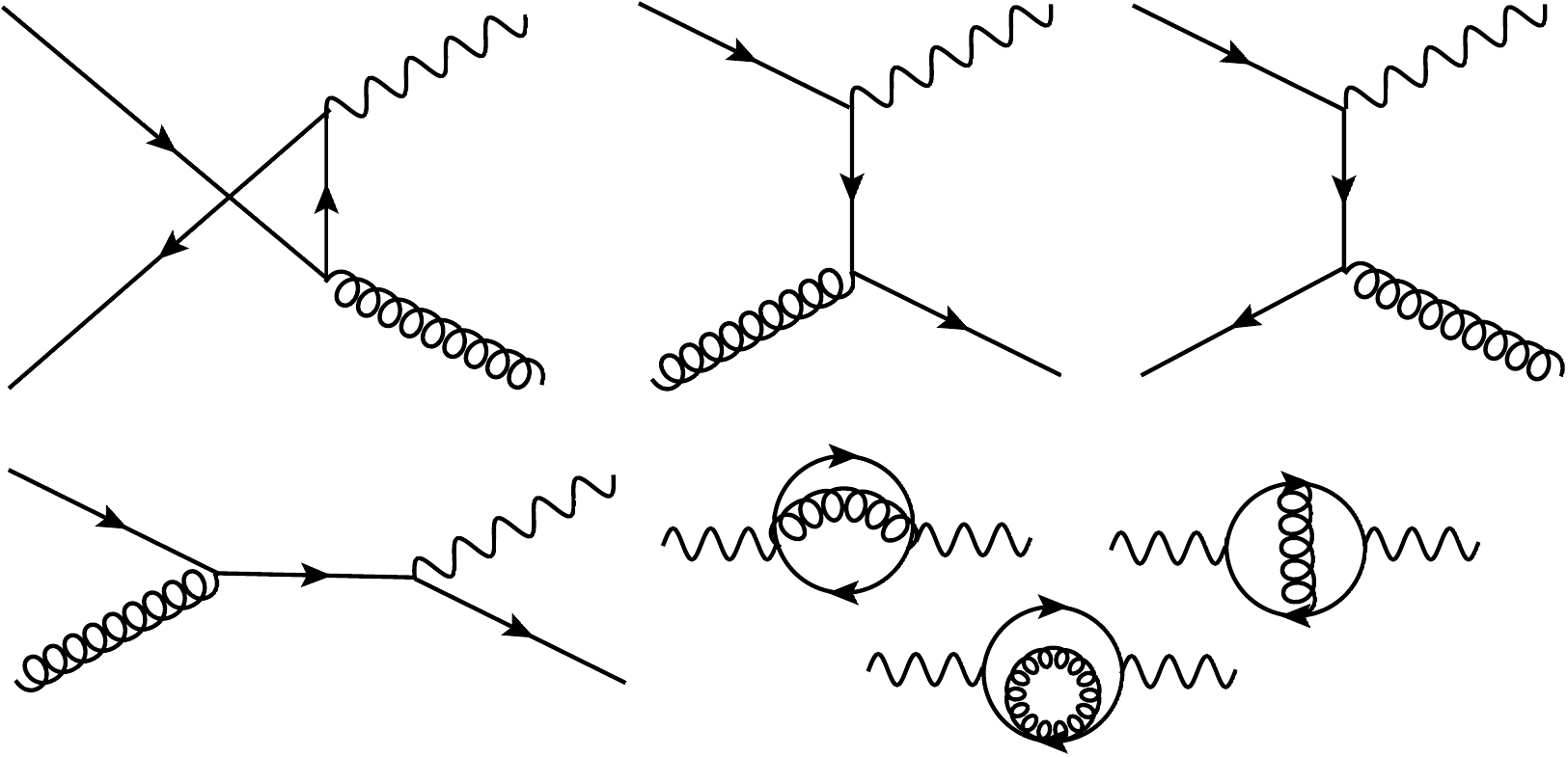}
         \caption{}
         \label{Fig:Thermal_partonic}
     \end{subfigure}
     \\
     \vspace{0.4cm}
     \begin{subfigure}[b]{0.55\textwidth}
         \centering
         \includegraphics[scale=0.5]{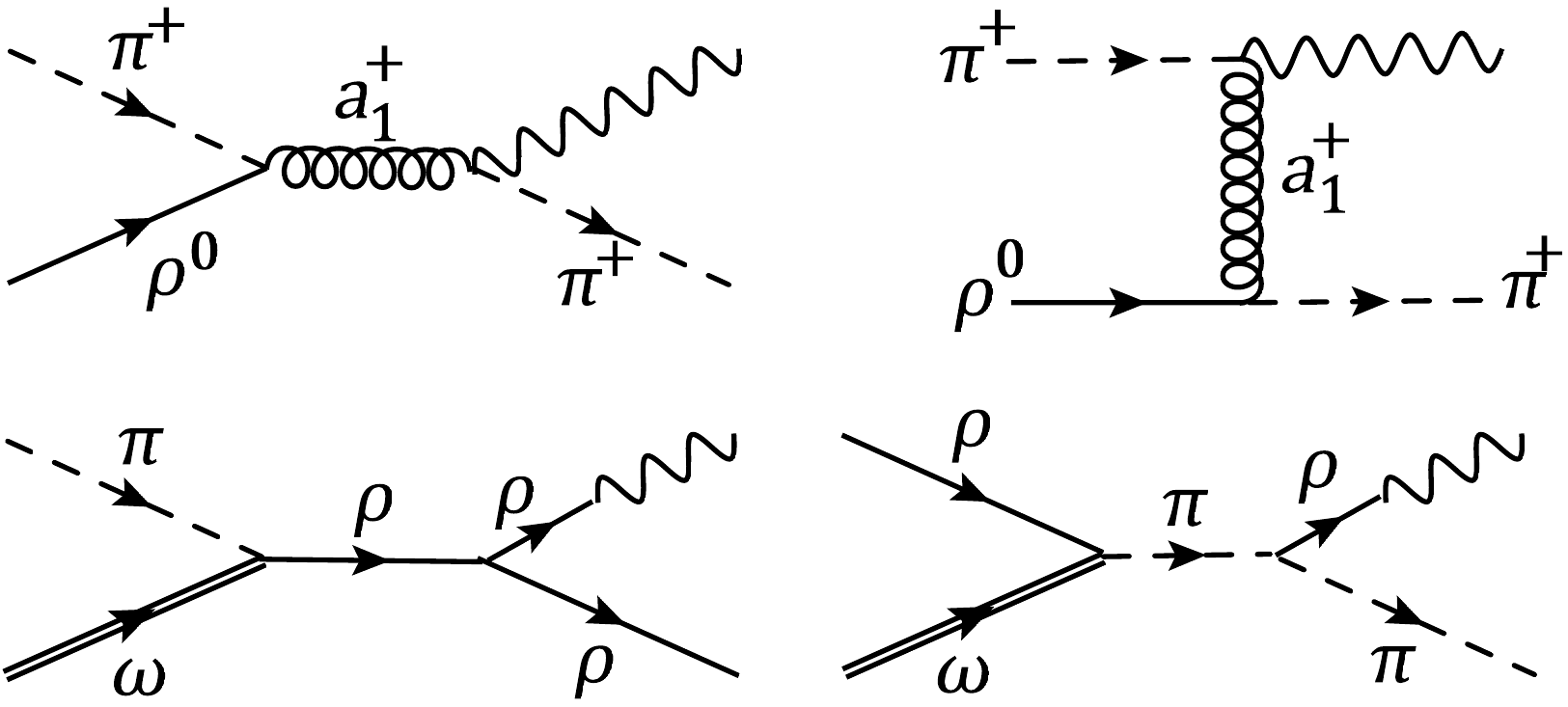}
         \caption{}
         \label{Fig:Thermal_meson}
     \end{subfigure}
      \begin{subfigure}[b]{0.4\textwidth}
         \centering
         \includegraphics[scale=0.45]{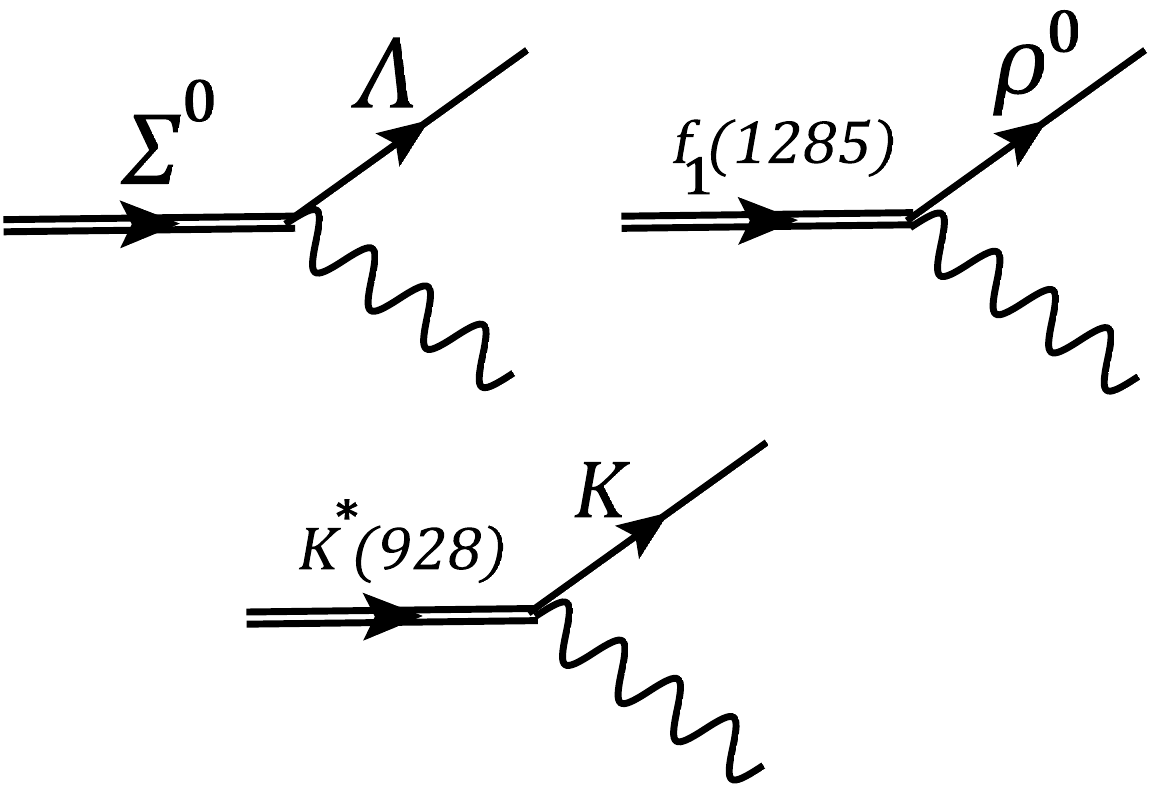}
         \caption{}
         \label{Fig:NC_decays}
     \end{subfigure}
        \caption{Scheme of some relevant processes for photon production in the Paquet's hydrodynamical model: (a) prompt photons, (b) thermal photons from QGP~\cite{ThermalPhotons1,ThermalPhotons2}, (c) thermal photons for mesonic reactions~\cite{TurbideGale,HadronicThermal1}, and (d) non-cocktail decays. }
        \label{Fig:Photon_Sources_Diagrams}
\end{figure}
The photon sources coming from the named calculations are classified in each stage of the heavy-ion collision:
\begin{itemize}
    \item {\bf Prompt Photons:} This refers to photons produced in the early stages of the ion collision through partonic Compton interactions, quark-antiquark annihilations, and bremsstrahlung radiation. Figure~\ref{Fig:Promp_photons_sketch} schematizes the origin of a prompt photon. All prompt photon's sources from hadronic interactions can be known by pQCD-techniques and basically the number of photons can be scaled by the binary proton-proton collisions through proton-parton distribution and parton-to-photon fragmentation functions at next-to-leading order (NLO) ~\cite{pQCD2,PromptPhotonsSources,AurenchePromptPhotons1,AversaPromptPhotons,AurenchePromptPhotons2,ChatrchyanPromptPhotons}. 
    
    \item {\bf Thermal Photons:} This refers to photons produced in the QGP phase which in the partonic sector are calculated by thermal field theory techniques at LO and NLO ~\cite{ThermalPhotons1,ThermalPhotons2}. Those calculations includes near-collinear bremsstrahlung and inelastic pair annihilation contributions. For hadrons with temperatures lower or comparable with the corssover temperature, the photons are calculated by mesonic reactions~\cite{TurbideGale,HadronicThermal1}. Also, the barionic contributions in the form $NN\rightarrow NN\gamma$ and $\pi N\rightarrow\pi N\gamma$, whith $N$ a nucleon, are taken into account. Figures~\ref{Fig:Thermal_partonic} and~\ref{Fig:Thermal_meson} schematizes some of the processes which contributes to the named calculations. 
    
    \item {\bf Noncocktail Hadronic Decay Photons:} These are photons coming from the early hadronization stages and which have not been subtracted from the photon production of the final hadronic interactions. Such a subtraction is known as the \textit{cocktail}, which is made of $(\pi^0,\eta,\rho,\omega,\eta',\phi)$ for LHC and $(\pi^0,\eta,\eta')$ for RHIC~\cite{cocktail1,cocktail2}. For both LCH and RHIC, the noncocktail sources are dominated by the channels $\Sigma^0\rightarrow\Lambda\gamma$, $f_1(1285)\rightarrow\rho^0\gamma$ and $K^*(928)\rightarrow K\gamma$. Also, the channel $\pi^0\rightarrow e^{+}e^{-}\gamma$ is considered as a noncocktail contribution at RHIC given that in comparison with ALICE it is not subtracted. Fig.~\ref{Fig:NC_decays} illustrates the channels which contributes to the NC radiation. 
\end{itemize}
%

\begin{figure}
    \centering
        \includegraphics[scale=0.377]{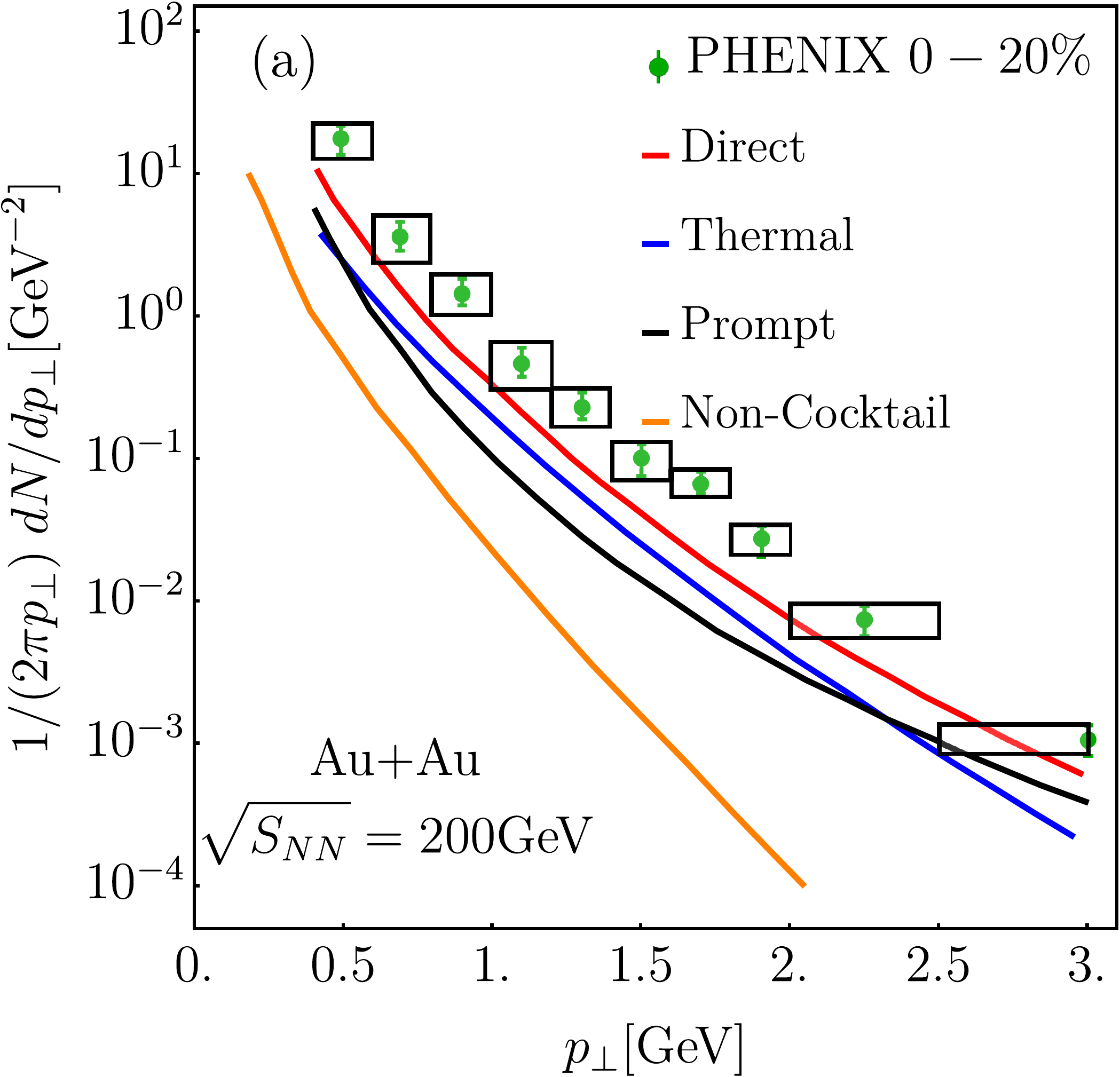}\hspace{0.37cm}\includegraphics[scale=0.37]{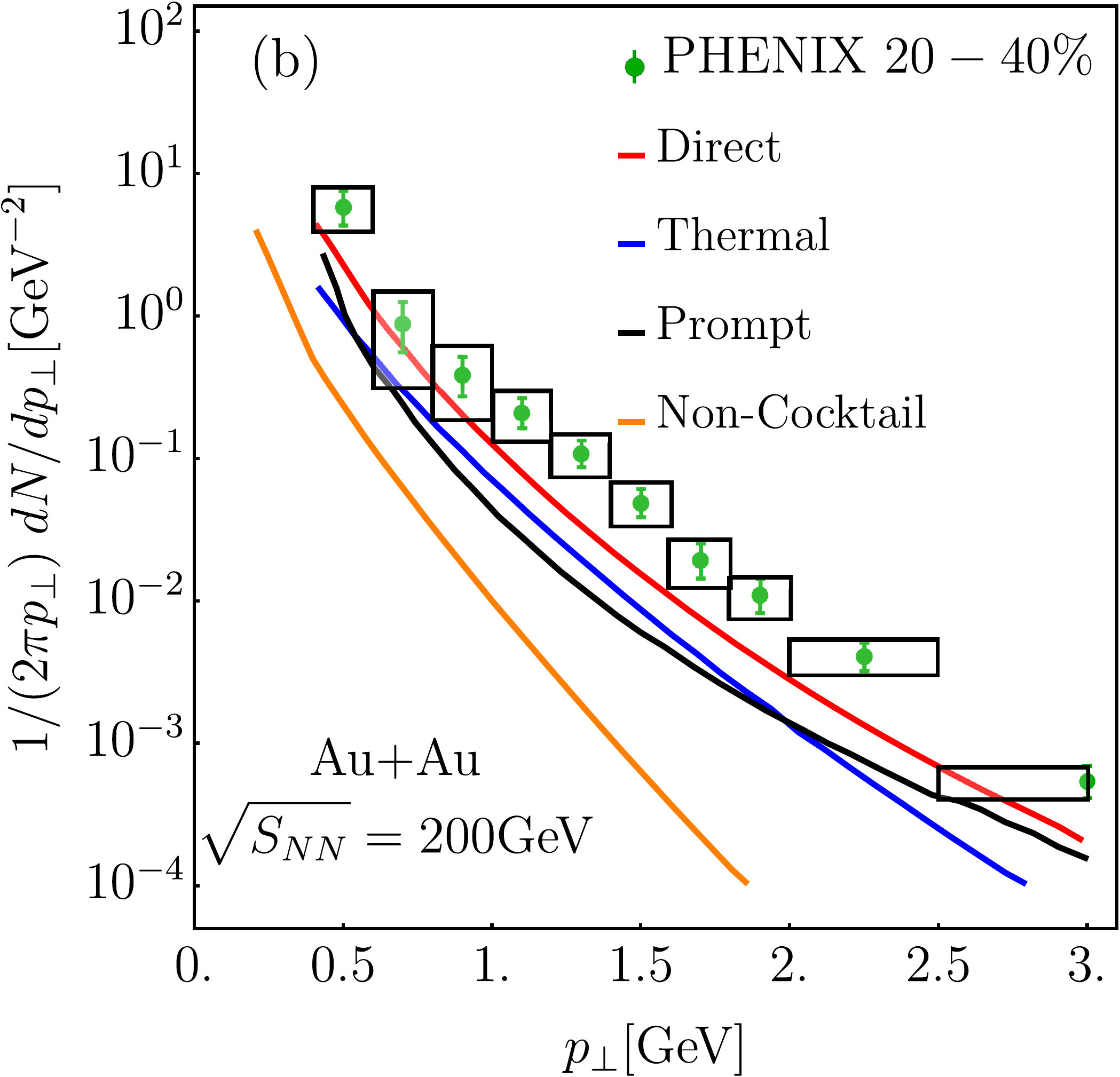}
    \caption{Hydrodinamical calculation of the direc photon spectra for Au+Au collisions by Paquet \etal~from Ref.~\cite{hydro-photons1} in two centrality ranges: (a) $0-20\%$ and (b) $20-40\%$. The calculations are compared with the experimental data from PHENIX~\cite{RHIC3}. }
    \label{Fig:Paquet_and_PHENIX}
\end{figure}

\begin{figure}[h]
    \centering
    \includegraphics[scale=0.5]{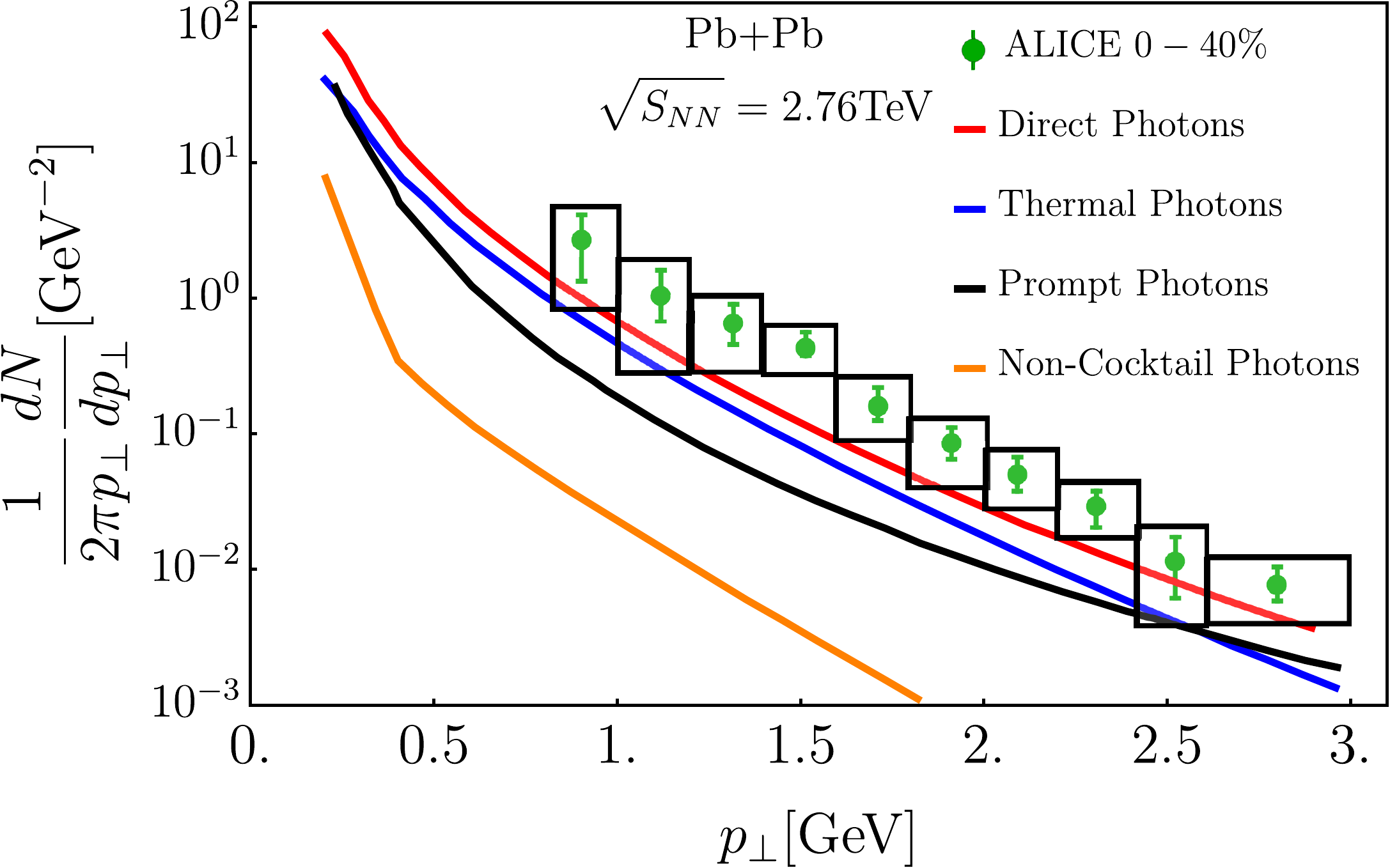}
    \caption{Hydrodinamical calculation of the direct photon spectra for Pb+Pb collisions by Paquet \etal~from Ref.~\cite{hydro-photons1} for the centrality $0-40\%$. The calculations are compared with the experimental data from ALICE~\cite{ALICE1}. }
    \label{Fig:Paquet_and_ALICE}
\end{figure}

Figure~\ref{Fig:Paquet_and_PHENIX} shows the photon invariant momentum distribution of the Paquet's hydrodynamical calculations compared with the RHIC's data for Au-Au collisions at $\sqrt{S_{NN}}=200$GeV. The direct photons are obtained by adding the contributions of the prompt, thermal, and \textit{non-cocktail} photons.  As can be noticed from the two centrality cases, the calculation for $p_\perp\sim 1.5$ GeV underestimates the PHENIX data central points roughly by a factor of 3. The same behavior is found when the model is compared against the ALICE data for Pb-Pb collisions at $\sqrt{S_{NN}}=2.76$TeV in Fig.~\ref{Fig:Paquet_and_ALICE}. It is clear that the theoretical predictions cannot reproduce the experimental data unless the error bars of statistical and systematic uncertainties are taken into account.
%

\begin{figure}[H]
    \centering
        \includegraphics[scale=0.38]{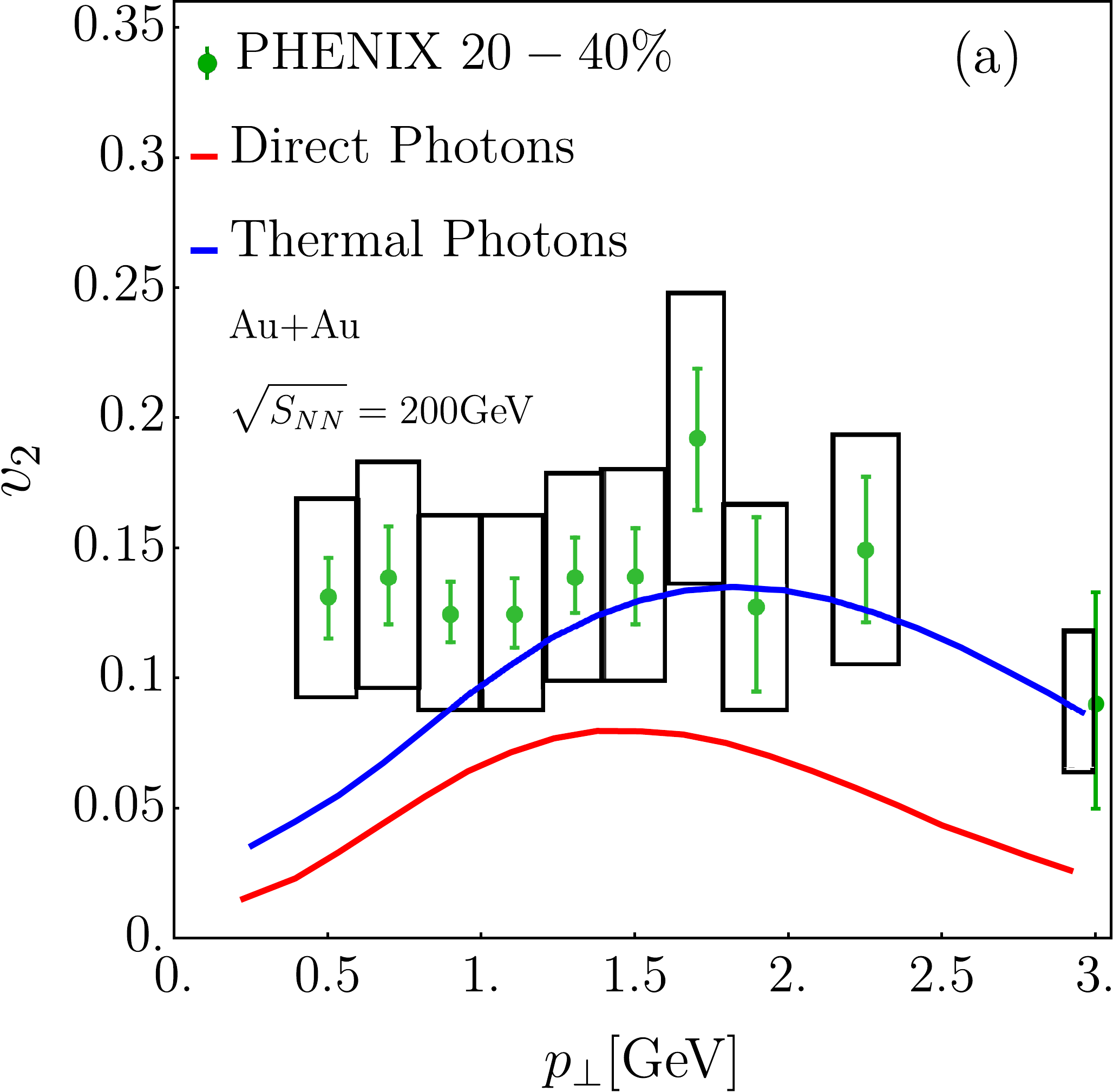}\hspace{0.4cm}\includegraphics[scale=0.38]{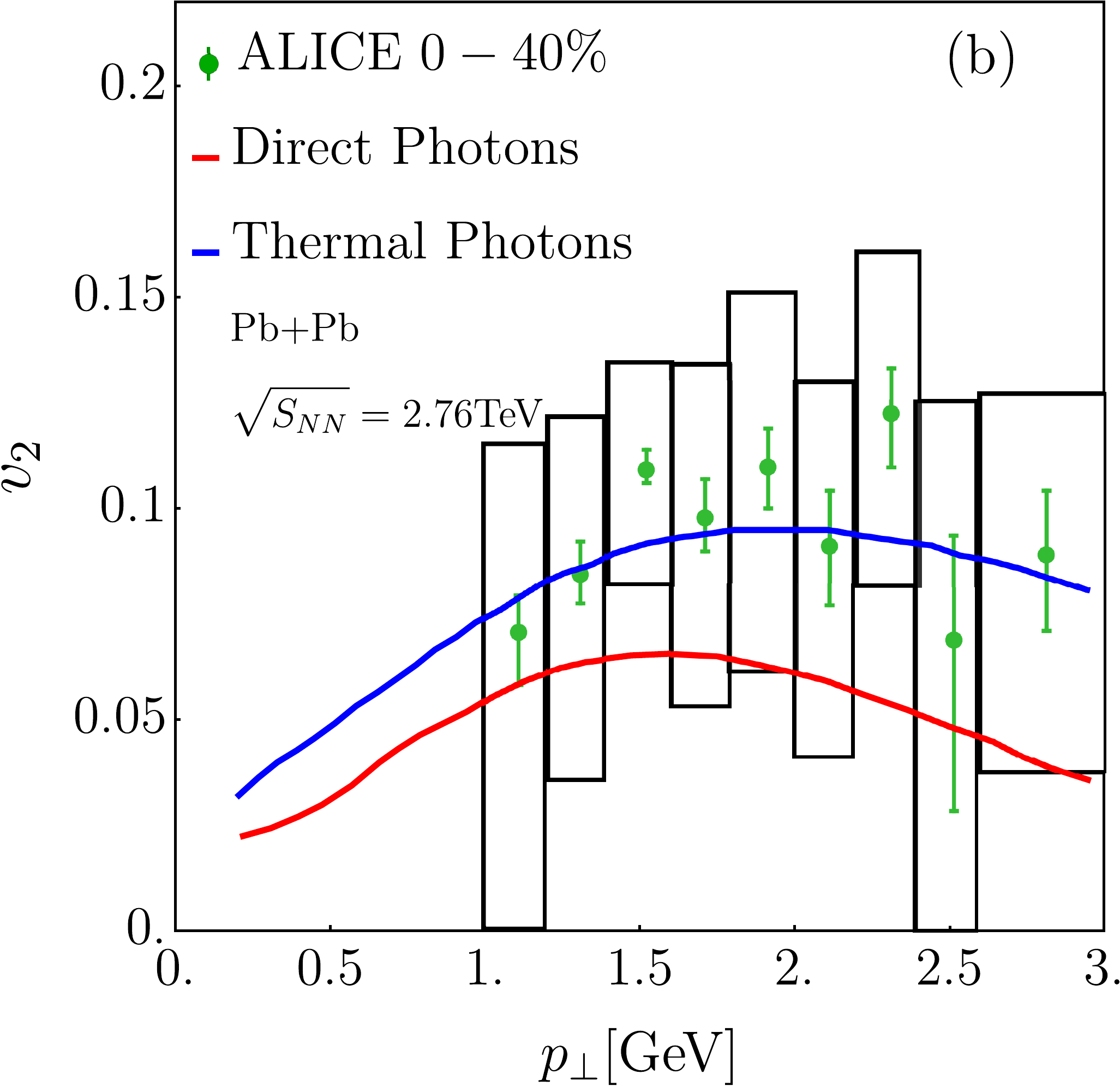}
    \caption{Hydrodinamical calculation of the flow coefficient $v_2$ for the direct and thermal photons by Paquet \etal~\cite{hydro-photons1}, compared with (a) PHENIX data with centrality $20-40\%$~\cite{cocktail2,bannier2014systematic} and (b) ALICE data with centrality $0-40\%$~\cite{ALICE1}. }
    \label{Fig:Paquet_v2_PHENIX_ALICE}
\end{figure}

On the other hand, when the flow coefficient $v_2$ is calculated from Eq.~(\ref{vndef}) and compared with the experimental data by PHENIX and ALICE for different centralities, the theoretical predictions cannot describe the overall behavior of this observable. As is shown in Fig.~\ref{Fig:Paquet_v2_PHENIX_ALICE} the predicted direct photons anisotropy underestimates the measured spectrum even by taking into account the large error bars. This problem is know as the {\it The Direct Photon Puzzle}: a challenge to describe the direct photon spectra from which the elliptic flow coefficient $v_2$ can be obtained in such a manner that both approximations fit the experimental measurements.

It is worth to mention that for both experiments, the hydrodynamical curve for the thermal photons gives a good description of the elliptic flow. Nevertheless, the harmonic coefficient $v_2$ need to be constructed from a weighted average, namely:
\bea
v_2(p_\perp)=\frac{\sum_i\frac{dN^{(i)}}{dp_\perp}(p_\perp)v_2^{(i)}(p_\perp)}{\sum_i\frac{dN^{(i)}}{dp_\perp}(p_\perp)},
\label{weightedaverage}
\eea
where the sum runs for the $i$-th mechanism of photon production. In that sense, although the thermal photons have an intrinsic flow close to the data, the corresponding weight reduces their contribution to the total flow. In this chapter, I discuss the prompt photon production at the early stages of a relativistic heavy-ion collision, therefore, it is necessary to give several relevant aspects of those initial times.
%


\section{The Color Glass Condensate (CGC)}\label{Sec:The_Color_Glass_Condensate}

\subsubsection{Deep Inelastic Scattering and Parton Distribution Functions}

As it was commented on the Introduction, at the early stages of a heavy-ion collision, the hadronic and partonic wave functions behave in a particular way, due the Lorentz contraction. Collision times are characterized by a high- density, state of QCD, with its principal characteristic being a large number of gluons. 

The hadronic structure depends on the scale used to be tested, encoded in the scattering process' kinematics, that is, although the cross-section is a boost-invariant object, its interpretation in terms of partons depends on the energy scale at which this is computed. The valence quarks bounded by gluons in a colorless state is a good description of a hadron in the rest frame, so that there is a probability to generate quark-antiquark pairs as virtual particles. Such fluctuations coming from the gluon exchange are non-perturbative because they occur at energy and momentum scales of order $\Lambda_{\text{QCD}}$ where the coupling constant is $\Op(1)$, and the binding is strong. On the other hand, the typical scale for vacuum fluctuations (namely quark-antiquarks pumped from the vacuum) is also $\Lambda_{\text{QCD}}$; therefore, in the rest frame is not possible to speak of a {\it hadron structure}, given that vacuum fluctuations are not distinguishable from hadronic (partonic) fluctuations.

If the hadron is viewed in a boosted reference frame along $p=p_z$ with Lorentz factor $\gamma\gg1$, the 4-momentum passes from being $p^\mu=(M,0,0,0)$ to $p^\mu=(M,0,0,p)$, so that in the so-called Infinite Momentum Frame (IMF), $E=\sqrt{M^2+p^2}\simeq p$, the life-time of the hadronic fluctuations is:
\bea
\Delta t_{\text{boost}}=\gamma\Delta t_{\text{rest}}\sim\frac{\gamma}{\Lambda_{\text{QCD}}},
\label{Delta_t_boost}
\eea
which implies that the boost-invariant vacuum fluctuations with life-time $\Delta t_{\text{vacuum}}\sim\Lambda_{\text{QCD}}^{-1}$ can be well differentiated from the hadronic fluctuations. The time given in Eq.~(\ref{Delta_t_boost}) is larger compared to a typical collision time, hence, the fluctuations are recognized as individual free particles called {\it partons}. That situation allows factorizing the cross-section for hadron-hadron collisions into the product of {\it parton-distribution functions} and {\it partonic cross-sections}. The former is the probability to find a parton with particular kinematical characteristics in the hadron wave-function, and the latter describes the collision between partons of projectile and target.

For a large range of energy scales, the partonic cross-section may be computed analytically. Still, by its definition, the parton distribution function cannot be calculated in a perturbative framework. Nonetheless, depending on the resolution of Deep Inelastic Scattering (DIS) experiments, some features of such processes are approximated by perturbation theory.

\begin{figure}
    \centering
    \includegraphics[scale=0.56]{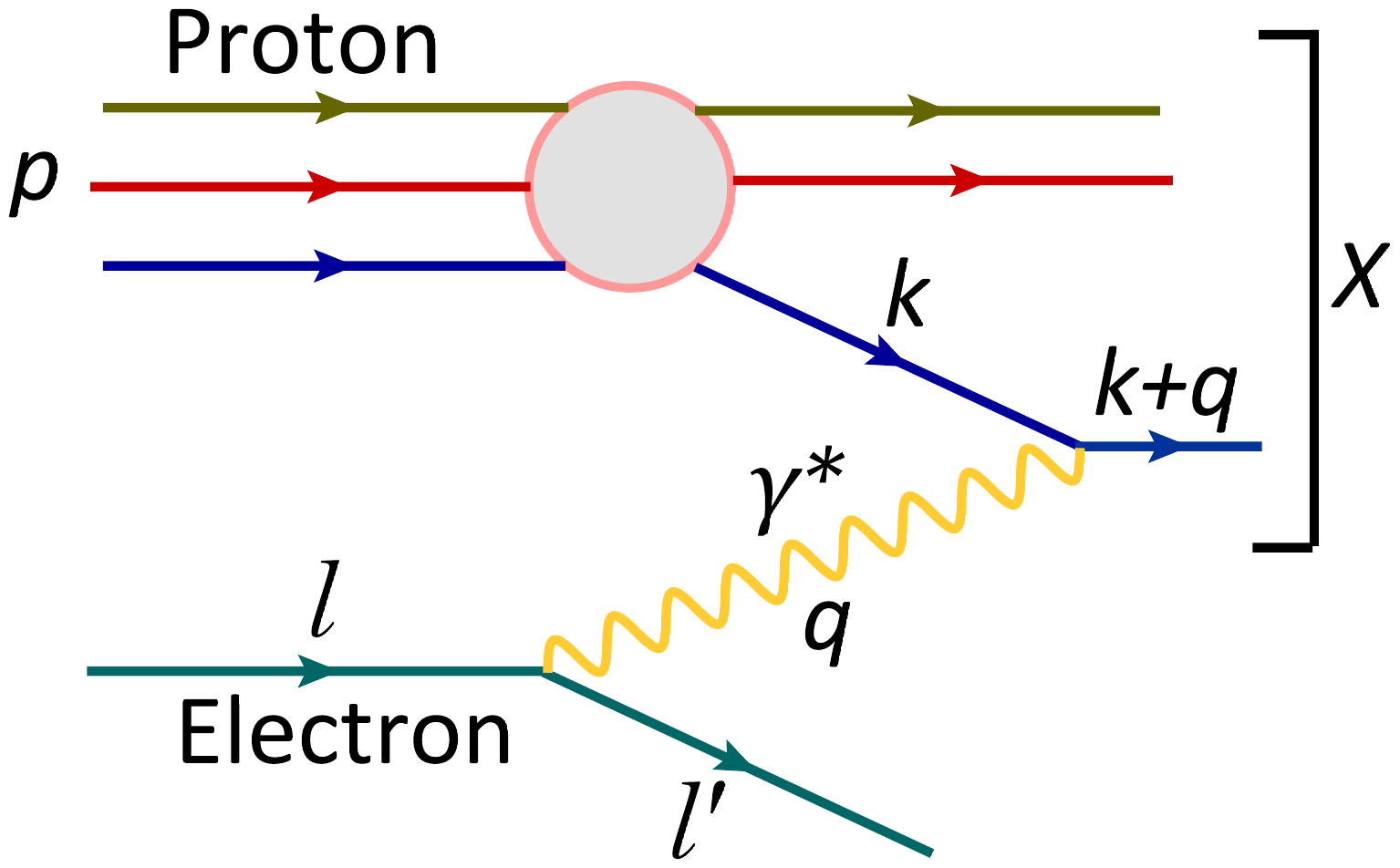}
    \caption{Deep Inelastic Scattering: an electron with 4-momentum $l_\mu$ is scattered by a proton with 4-momentum $p_\mu$ exchanging a virtual photon $\gamma^*$ with 4-momentum $q_\mu$. If the proton is broken the final hadronic states are represented by the set $X$.}
    \label{Fig:DIS}
\end{figure}

Figure~\ref{Fig:DIS} shows the DIS of an electron with 4-momentum $l_\mu$ is scattered by a proton with 4-momentum $p_\mu$ exchanging a virtual photon $\gamma^*$ with 4-momentum $q_\mu$. The final electron has a 4-momentum $l'_\mu=l_\mu-q_\mu$ and the photon is considered space-like so that
\bea
q^2=-2l\cdot l'=-2 E_l E_{l'}(1-\cos\theta_{l l'})\equiv -Q^2,
\eea
where the quantity $Q^2>0$ is called the {\it virtuality}. The space-like photon exchanged for a proton in the IMF measures partons localized in the transverse plane in the area $\Sigma~\sim 1/Q^2$. Such partons carry a longitudinal momentum $k_z=xp$ with $x$ the Bjorken variable given by~\cite{Bjorken:1976mk}:
\bea
x\equiv\frac{Q^2}{2(p\cdot q)}=\frac{Q^2}{s^2+Q^2-M^2},
\eea
where $s\equiv(p+q)^2$ is the invariant energy squared. The invariant variables $x$ and $Q$ totally determine the partons transverse length $1/Q$ which participates into the scattering. Note that that parton needs to be a quark or antiquark, given that the photon does no couples to gluons.

The energy difference at the emission photon vertex gives the {\it transverse resolution} of the virtual photon, i.e.,
\bea
\Delta E=|\mathbf{k}+\mathbf{q}|-|\mathbf{k}|=\sqrt{(xp)^2+\qt^2}-xp\simeq\frac{\qt^2}{2xp},
\eea
where the 4-momentum of the space-like photon is chosen as $q_\mu=(0,\mathbf{q}_\perp,0)$ and $Q^2=\qt^2$. Therefore, the collision time is
\bea
\Delta t_\text{coll}\simeq\frac{2xp}{\qt^2}.
\eea

To be tested by the virtual photon, the quark (antiquark) excitation must have a lifetime $\Delta t_\text{fluc}>\Delta t_\text{coll}$. The former is estimated in the IMF by its longitudinal momentum $k_z=xp$ for quarks near of the mass-shell condition ($k_\mu\simeq xp_\mu$, $k^2\simeq0)$ boosted by a Lorentz factor $\gamma=k_z/\kt$, so that
\bea
\Delta t_\text{fluc}=\gamma \Delta t_\text{rest}\simeq\frac{2k_z}{\kt^2}=\frac{2xp}{\kt^2},
\eea
hence, to satisfy $\Delta t_\text{fluc}>\Delta t_\text{coll}$ it follows that $\kt^2<Q^2$ which means the DIS resolution measures partons with transverse momentum smaller than virtuality $Q$. 

The above analysis admits an expression for the DIS cross-section given by:
\bea
\sigma_{\gamma^*p}=\frac{4\pi^2\alpha_\text{em}}{Q^2}F_2(x,Q^2),
\eea
which is built from the elementary cross-section in the photon absorption by a quark weighted by the structure-function:
\bea
F_2(x,Q^2)=\sum_f e_f^2x\left[q_f(x,Q^2)+\bar{q}_f(x,Q^2)\right],
\eea
with
\bea
q_f(x,Q^2)\equiv\frac{dN_f(Q^2)}{dx}=\int^Q d^2\mathbf{\kt}\frac{dN_f}{dxd^2\mathbf{\kt}},
\eea
so that $q_f(x,Q^2)dx$ is the number of quarks ($\bar{q}_f$ for antiquarks) with flavor $f$ with longitudinal momentum fraction between $x$ and $x+dx$. Note that even the condition $\kt^2<Q^2$ is always satisfied by partons with $\kt\sim\Lambda_\text{QCD}$ if $Q^2\gg\Lambda_\text{QCD}^2$, which implies the structure function $F_2(x,Q^2)$ must be independent of $Q^2$ satisfying the {\it Bjorken scaling}. Nevertheless, the experimental data shows that the Bjorken scaling is valid for $x\gtrsim 0.1$~\cite{Gribov:1972ri,altarelli1977asymptotic,dokshitzer1977calculation}. The radiative processes give the phenomenology behind that result: although the transverse momenta for valence quarks are around $\Lambda_\text{QCD}$, virtual excitations with large $\kt$ can be generated by bremsstrahlung. Moreover, even when they have short lifetimes, there is a window of values for $\kt$ that contribute to DIS~\cite{ellis2003qcd}. 

The current experiments can lead to collisions with small $x$ understood as a high-energy regimen where $s\gg Q^2$. Consider a parton-parton collision inside a hadronic scattering: in the Center of Mass (COM) frame, each parton has the 4-momentum $k_i^\mu=x_ip_i^\mu+k_{i\perp}^\mu$, where $p_1^\mu=(p,0,0,p)$ and $p_2^\mu=(p,0,0,-p)$ so that $p=\sqrt{s}/2$. If two particles are produced bt the partonic collision, with transverse momenta $\mathbf{p}_{a\perp}$ and $\mathbf{p}_{b\perp}$ the energy-momentum conservation implies:
\bea
\mathbf{p}_{a\perp}+\mathbf{p}_{b\perp}=\mathbf{k}_{1\perp}+\mathbf{k}_{2\perp},
\eea
and
\bea
x_{1}&=&\frac{p_{a \perp}}{\sqrt{s}} e^{y_{a}}+\frac{p_{b \perp}}{\sqrt{s}} e^{y_{b}},\nn\\
x_{2}&=&\frac{p_{a \perp}}{\sqrt{s}} e^{-y_{a}}+\frac{p_{b \perp}}{\sqrt{s}} e^{-y_{b}},
\eea
where $y_i$ is the rapidity.

For RHIC and LHC, the average hadronic momentum at the latter stages is aroun 1 GeV, so that the total number of produced hadrons (multiplicity) has $\pt\leq 2$ GeV; therefore, by considering $p_{a,b\perp}\sim1$ GeV and for central rapidities $y_{a,b}\sim 0$, the longitudinal momentum fraction is~\cite{Carminati:2004fp,Cortese:2005qfz,kharzeev2001manifestations}:
\bea
x_i&\simeq& 10^{-2}\quad\text{for}\quad\sqrt{s}=200~\text{GeV (RHIC)},\nn\\
x_i&\simeq& 4\times10^{-4}\quad\text{for}\quad\sqrt{s}=2.76~\text{TeV (LHC)}.
\eea

\subsubsection{Gluon Distribution Functions}
As was mentioned, gluons do not interact with photons, which implies that gluonic excitations must come from radiative processes. In perturbative QCD, gluons are produced via bremsstrahlung, so that they are collinear and have $x\ll1$ and relatively small transverse momentum~\cite{iancu2012qcd,Gelis:2012ri}. 

\begin{figure}[h]
    \centering
    \includegraphics[scale=0.65]{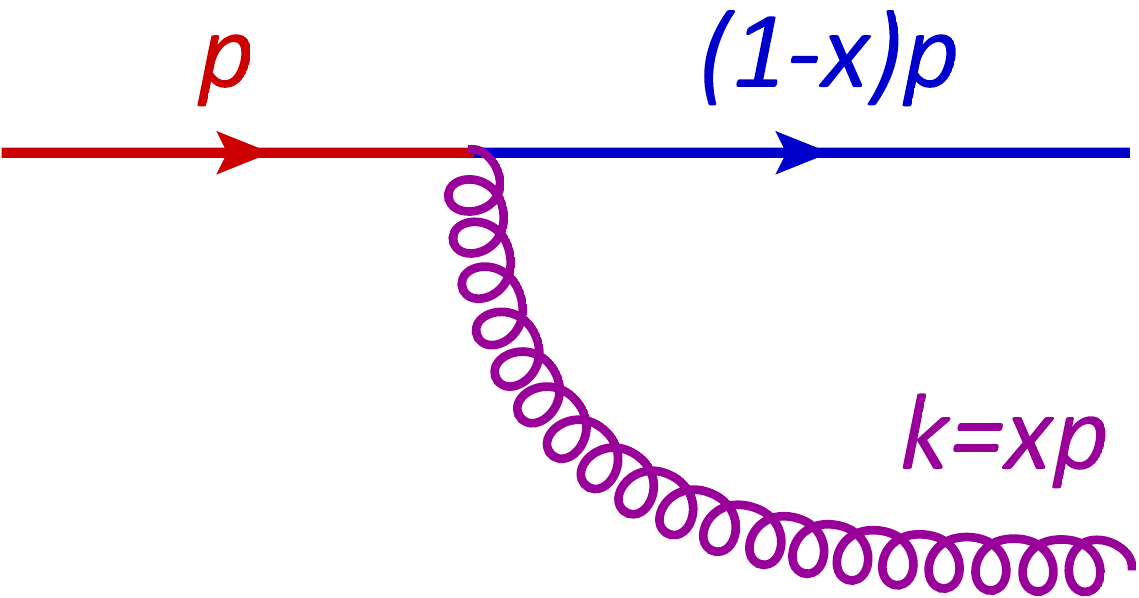}
    \caption{Gluon bremsstrahlung from a quark in perturbative QCD at lowest order.}
    \label{Fig:gluonbrems}
\end{figure}

Figure~\ref{Fig:gluonbrems} shows the elemental diagram for gluon bremsstrahlung from a quark in perturbative QCD at lowest order. The parent quark with momentum $p$ emits a gluon with longitudinal momentum $k=xp$ and transverse momentum $\kt$, so that an the end of the process, there is a quark with longitudinal momentum $(1-x)p$ and transverse momentum $-\kt$. By considering $x\ll1$ and a LO in $\alpha_s$ that gluon production channel has a differential probability given by~\cite{prosper2012techniques}:
\bea
\mathrm{d} P_{\mathrm{Brem}} \simeq C_{R} \frac{\alpha_{s}\left(k_{\perp}^{2}\right)}{\pi^{2}} \frac{\mathrm{d}^{2} k_{\perp}}{k_{\perp}^{2}} \frac{\mathrm{d} x}{x},
\label{Pbremss}
\eea
where $C_{R}$ is the Cassimir in the $SU(N_c)$ representation. The last equation presents two fundamental features: it is singular in the full collinear limit ($\kt\to0$) and the soft momentum regimen ($x\to0$). Moreover, if that differential probability is compared with the quark emission, the parton production enhancement is only present at the collinear and not in the small $x$ region. On the other hand, although a power of alpha suppresses Eq.~(\ref{Pbremss}), given that such coupling decreases for $\kt^2\gg\Lambda_\text{QCD}$, the large phase-space for the gluon radiation compensates such reduction: it is of order $\ln\left(Q^2/\Lambda_\text{QCD}^2\right)$ for the emission of any parton with $\kt\ll Q$, and of order $\ln(1/x)$ for the emission of a gluon with longitudinal momentum fraction $\xi$ in the interval $x\ll\xi\ll1$. Thus, for high-energy regimes where $Q^2\gg\Lambda_\text{QCD}^2$ or $x\ll1$, the radiative processes are not suppressed and needs to be resumed to all orders.

From the above, depending on the $Q^2$ and $x$ values, the quark and gluon emission are mixed or not. The parton creation is an effect depending on the value of $Q^2$. For increasing virtuality, more partons are visible, i.e., their density enhances. Moreover, a large $Q^2$ implies decreasing in $\alpha_s$ in such a way that partons interact weakly. Nevertheless, even if $\alpha_s$ is small, successive emissions need to be taken into account when $\ln Q^2$ is large so that $\alpha_s\ln Q^2\sim\Op(1)$. That condition implies the resummation of higher-order terms. The DGLAP equation usually describes it in order to obtain information about the parton distribution inside a hadron~\cite{Gribov:1972ri,altarelli1977asymptotic,dokshitzer1977calculation}. 

\begin{figure}
    \centering
    \includegraphics[scale=0.5]{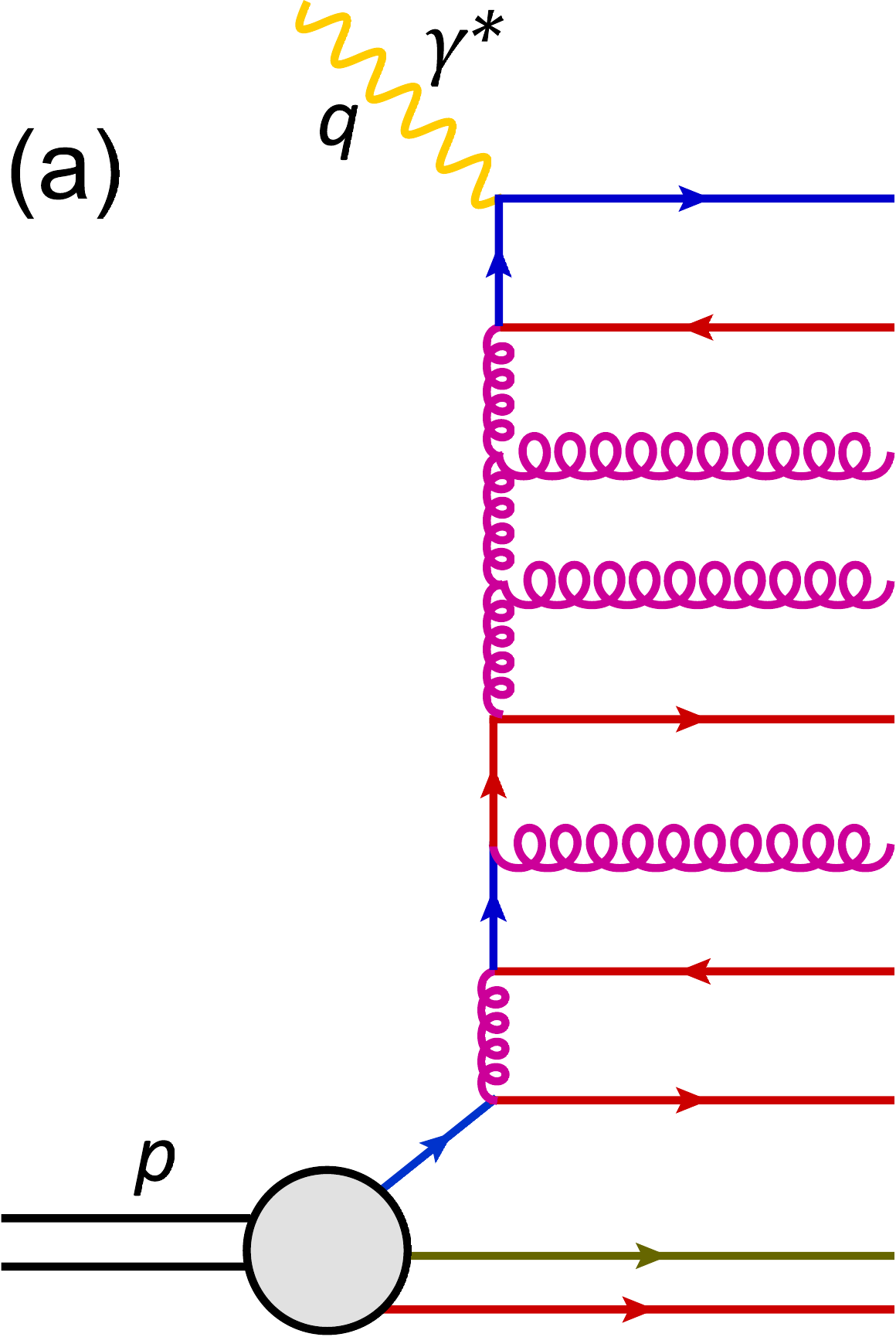}\hspace{2cm}\includegraphics[scale=0.5]{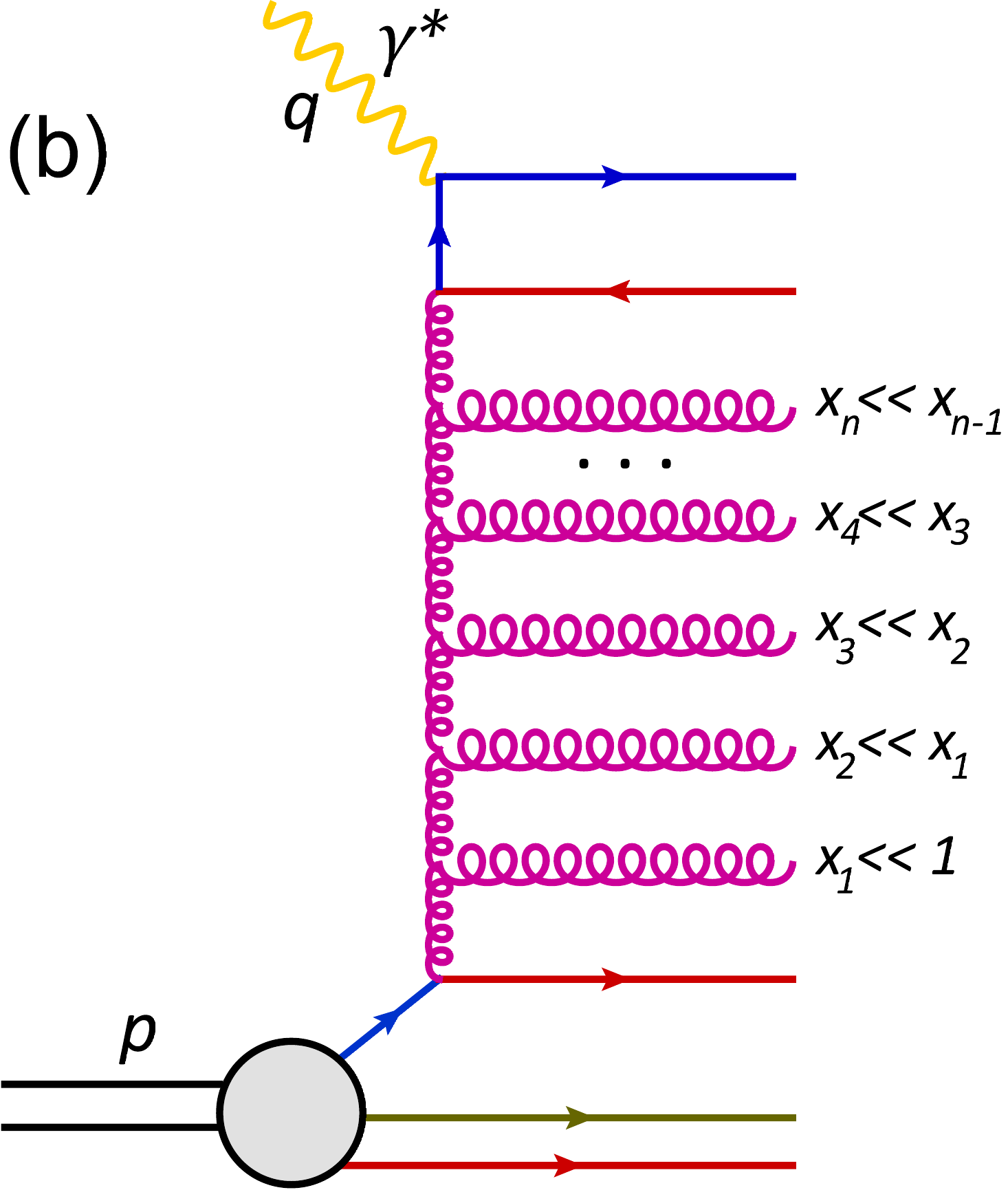}
    \caption{Parton production described by pQCD in a DIS configuration: (a) Evolution for high $Q^2$ at fixed $x$ ({\it diluted system}) (b) Evolution for high $Q^2$ and small $x$ ({\it dense system}).}
    \label{Fig:parton_evolution}
\end{figure}

The DGLAP formalism generates an increase of parton density when $Q^2$ grows, being typically proportional to $\ln Q^2$. However, the transverse area occupied by each parton is $\Sigma\sim1/Q^2$, so that, for fixed and not too small $x$, an increment of density is compensated by the tiny area of partons, and the system becomes dilute. Figure~\ref{Fig:parton_evolution}(a) schematizes that situation: there is a mixture of quarks and gluons from the virtual photon's exchange. On the other hand, increasing energy implies a decrease in $x$. It implies that the transverse area for partons $\Sigma\sim1/xp$ enhances, and corrections of the order $\alpha_s\ln(1/x)\sim1$ are important. Moreover, from Eq.~(\ref{Pbremss}), small $x$ means a high gluon production, and therefore, the system is a dense gluonic medium~\cite{iancu2002colour,mclerran2011cgc}. Figure~\ref{Fig:parton_evolution}(b) shows the dense system. 

The small $x$ behaviour can be computed from Eq.~(\ref{Pbremss}), given that the emission probability is the same as the number of gluons, so that at leading order in $\alpha_s$:
\bea
x \frac{\mathrm{d} N_{g}}{\mathrm{d} x}\left(Q^{2}\right)=\frac{\alpha_{s} C_{F}}{\pi} \int_{\Lambda_\text{QCD}^2}^{Q^2} \frac{\mathrm{d} k_{\perp}^{2}}{k_{\perp}^{2}}=\frac{\alpha_{s} C_{F}}{\pi} \ln \left(\frac{Q^{2}}{\Lambda_{\mathrm{QCD}}^{2}}\right),
\label{xdNg/xd}
\eea
where the infrared cutoff is introduced in order to account the confinement at low energies, where the parton has a minimal virtuality of order $\Lambda_{\mathrm{QCD}}^{2}$. Also, from the fact that Eqs.~(\ref{Pbremss}) and~(\ref{xdNg/xd}) are computed for $x\ll1$, it is valid to assume that another gluon with $x_1\ll x\ll1$ can be emitted, and consequent gluons with longitudinal momentum fractions $x_2\ll x_1\ll x$, $x_3\ll x_2\ll x_1,\cdots, x_n\ll x_{n-1}\ll x_{n-2}$ as is illustrated in Fig.~\ref{Fig:parton_evolution}(b). By adding $n$-th gluon emissions, Eq.~(\ref{Pbremss}) implies a contribution proportional to
\bea
\left(\frac{\alpha_s N_c}{\pi}\right)^n\int_{x}^{1} \frac{\mathrm{d} x_{n}}{x_{n}} \int_{x_{n}}^{1} \frac{\mathrm{d} x_{n-1}}{x_{n-1}} \cdots \int_{x_{2}}^{1} \frac{\mathrm{d} x_{1}}{x_{1}}=\frac{1}{n !}\left(\frac{\alpha_s N_c}{\pi} \ln \frac{1}{x}\right)^{n}.
\eea
\begin{figure}[h]
    \centering
    \includegraphics[scale=0.37]{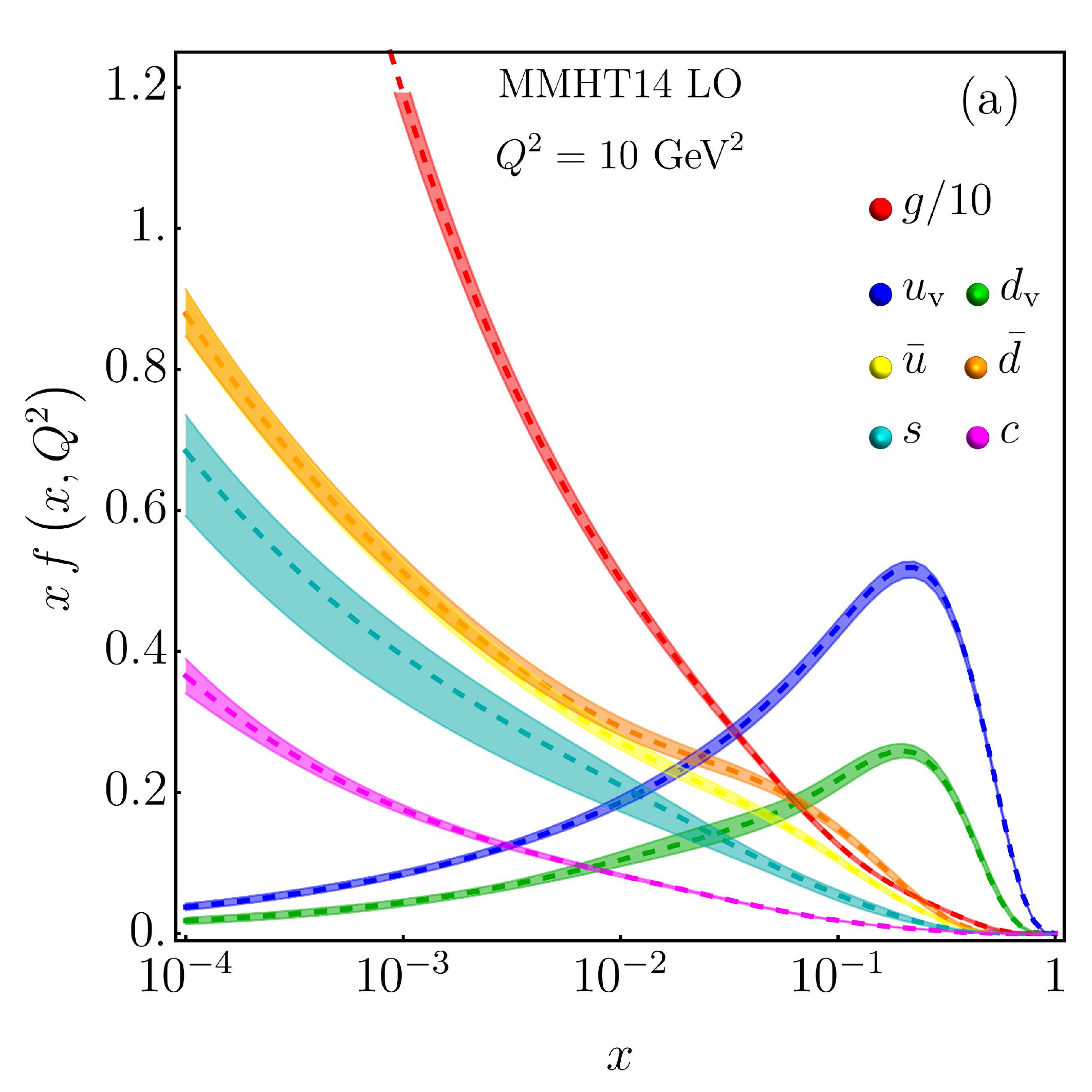}\hspace{0.3cm}\includegraphics[scale=0.37]{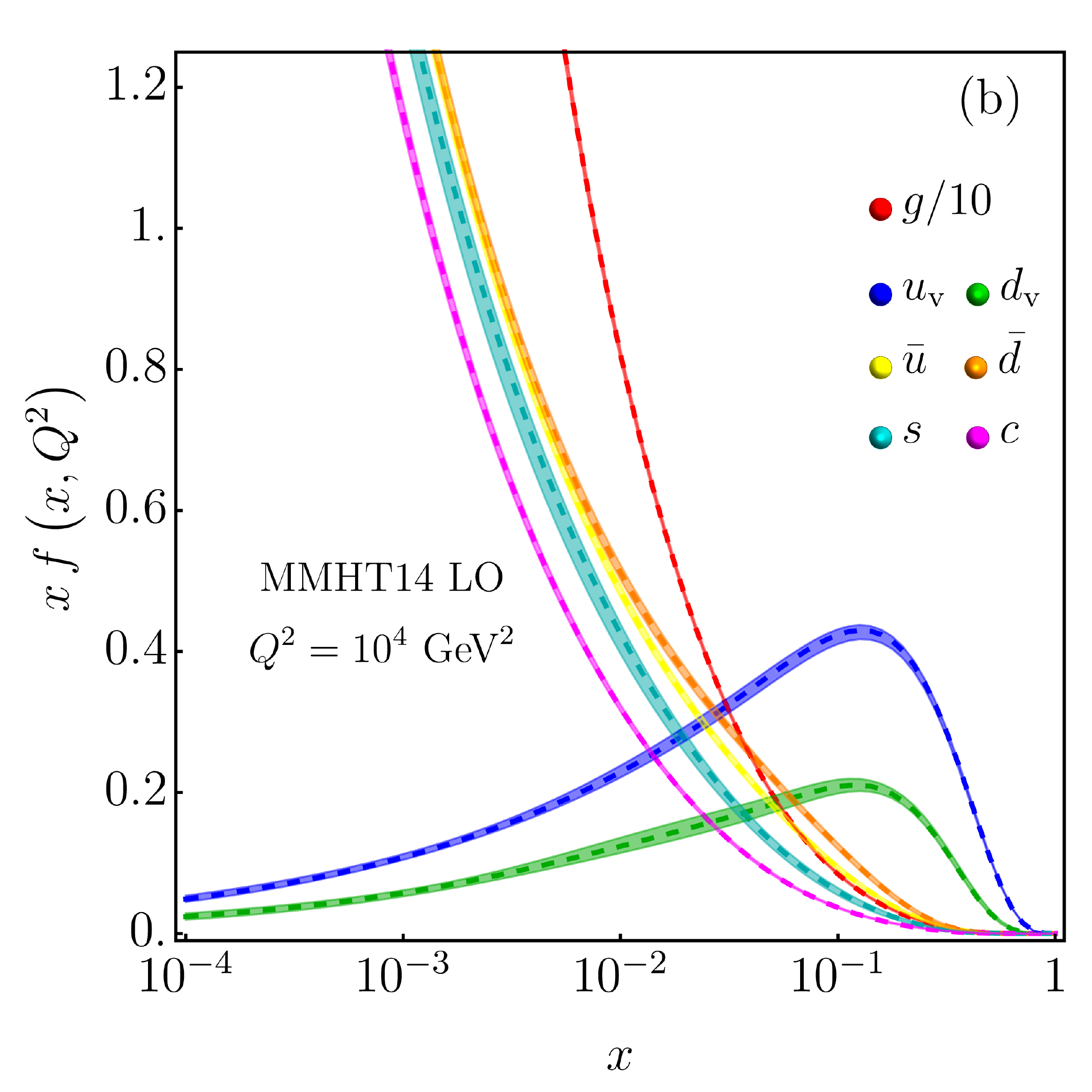}
    \caption{Parton distribution functions for the proton in the LHC for computed at LO as a function of the momentum fraction $x$ at the scales (a) $Q^2=10$ GeV$^2$ and (b) $Q^2=10^4$ GeV$^2$. In each plot, the PDFs are shown for gluons ($g$), valence quarks ($u_\text{v}, d_\text{v}$), sea-quarks ($\bar{u}, \bar{d}$), strange quarks ($s$) and charm quarks ($c$). The error bands correspond to a confidence--level uncertainty of 68$\%$. Note that the gluon PDF is scaled by a factor of 10.}
    \label{Fig:PDFs_LO}
\end{figure}

For a large amount of gluons, the latter equation gives an exponential function, so that the {\it unintegrated gluon distribution} is
\bea
x \frac{\mathrm{d} N_{g}}{\mathrm{d} x \mathrm{d} k_{\perp}^{2}} \sim \frac{\alpha_{s} C_{F}}{\pi} \frac{1}{k_{\perp}^{2}} \exp\left(\omega \frac{\alpha_s N_c}{\pi} Y\right),
\eea
where $\omega=4\log 2$ and $Y\equiv 1/x$ is the rapidity difference between the parent and the emitted gluon. A complete argumentation is found in the BFKL equation, where the kinematics of each gluon is considered, and arguments beyond a power counting are presented~\cite{lipatov1976reggeization,kuraev1977pomeranchuk,balitskii1978pomeranchuk}. The {\it coherence time} in the BFKL formalism is defined as the parton lifetime $\Delta t\simeq 2k_z/\kt^2$ results proportional to $x$, which means that the first gluons in a cascade can be visualized as static color sources. Although the average color charge of the system is zero, the overlapping of color sources leads to fluctuations of color charge density, which increases with $1/x$.
\begin{figure}
    \centering
    \includegraphics[scale=0.37]{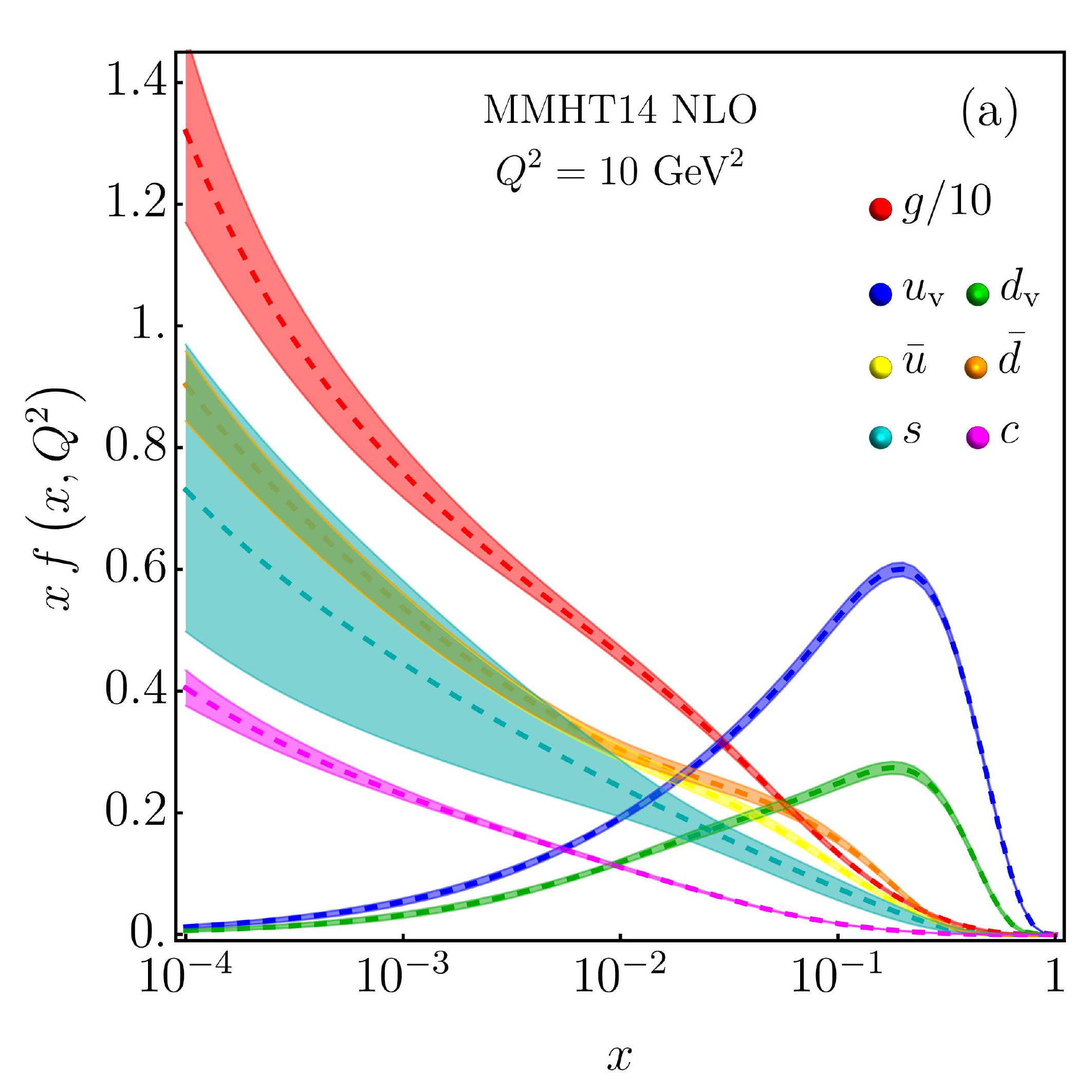}\hspace{0.3cm}\includegraphics[scale=0.37]{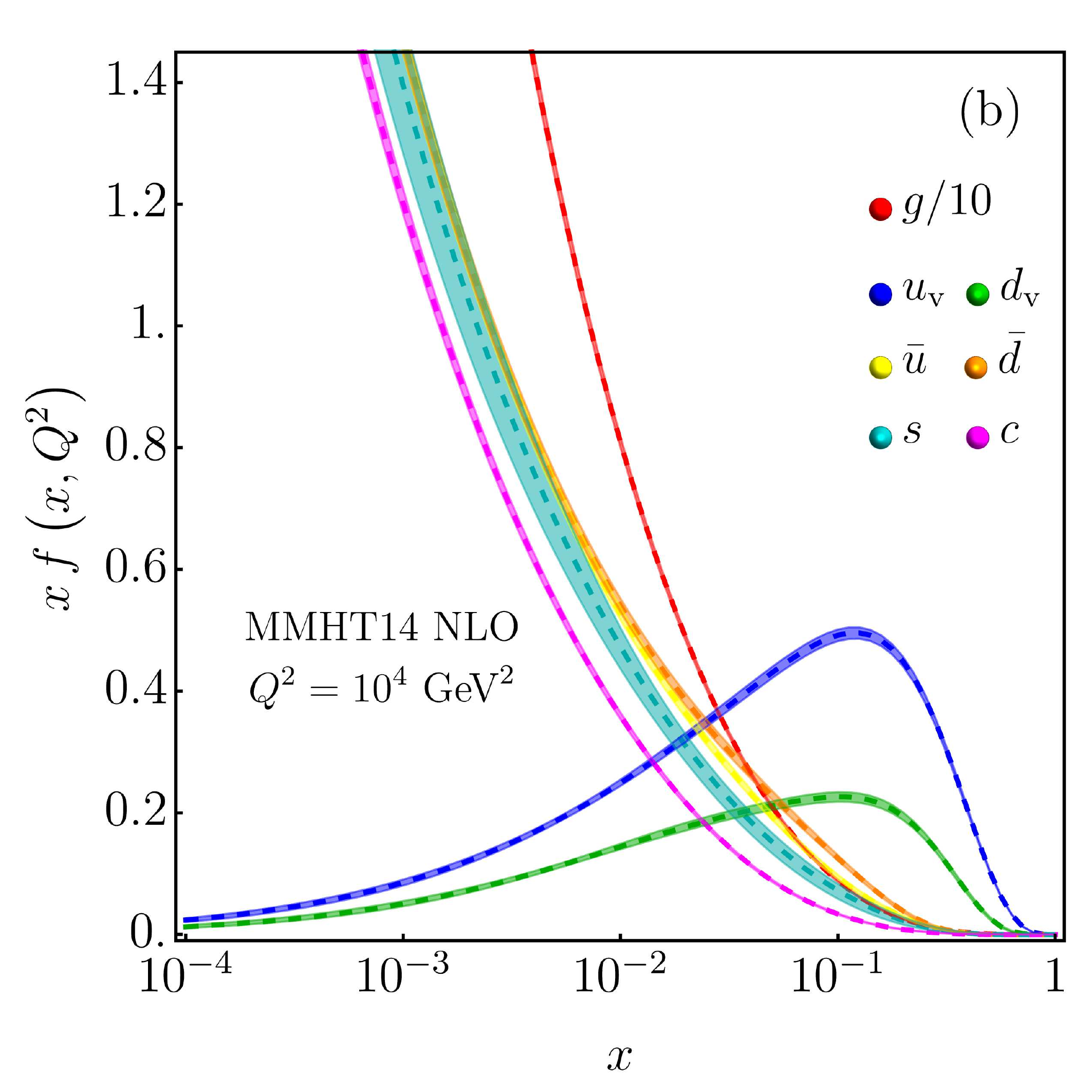}
    \caption{Parton distribution functions for the proton in the LHC for computed at NLO as a function of the momentum fraction $x$ at the scales (a) $Q^2=10$ GeV$^2$ and (b) $Q^2=10^4$ GeV$^2$. In each plot, the PDFs are shown for gluons ($g$), valence quarks ($u_\text{v}, d_\text{v}$), sea-quarks ($\bar{u}, \bar{d}$), strange quarks ($s$) and charm quarks ($c$). The error bands correspond to a confidence--level uncertainty of 68$\%$. Note that the gluon PDF is scaled by a factor of 10.}
    \label{Fig:PDFs_NLO}
\end{figure}

\begin{figure}
    \centering
    \includegraphics[scale=0.37]{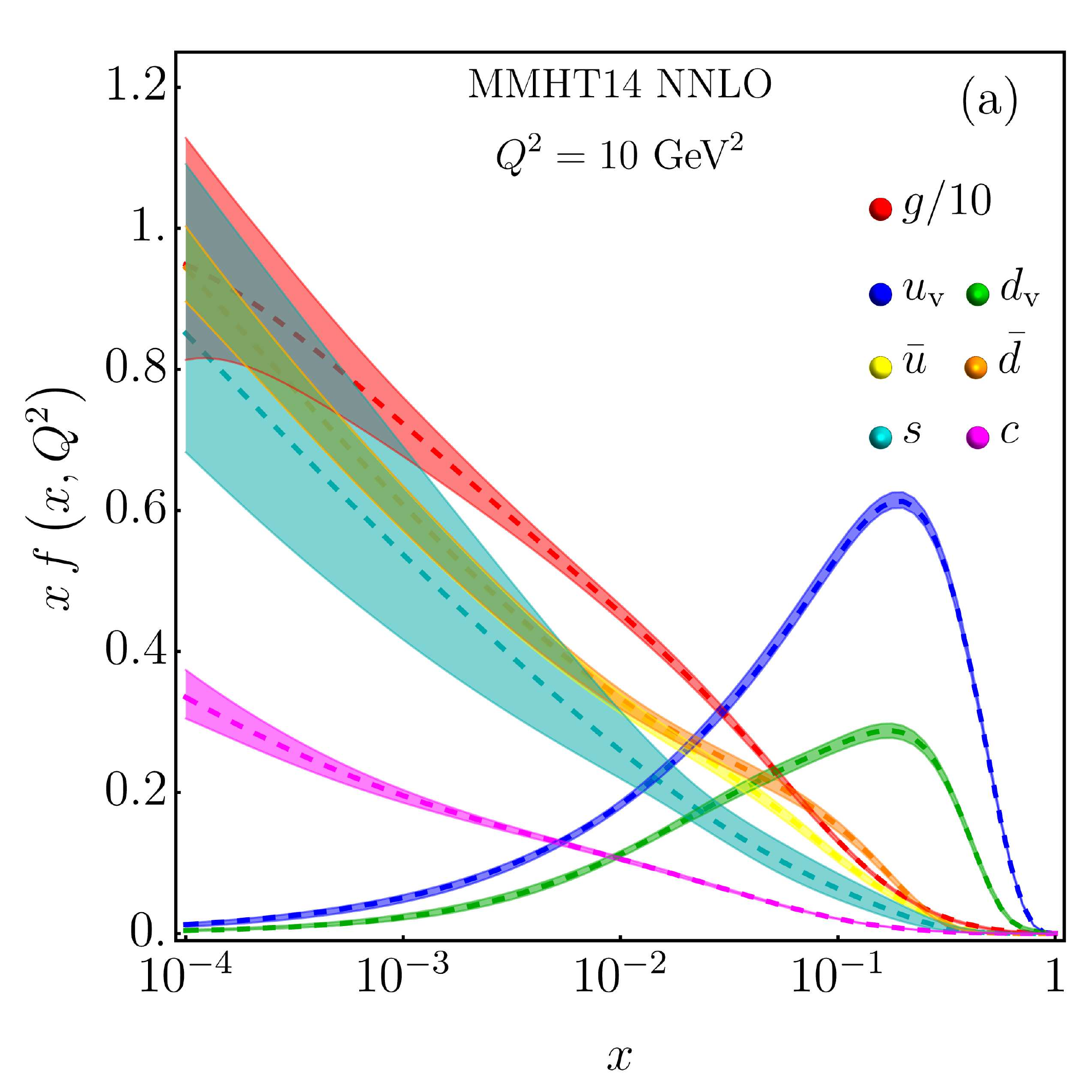}\hspace{0.3cm}\includegraphics[scale=0.37]{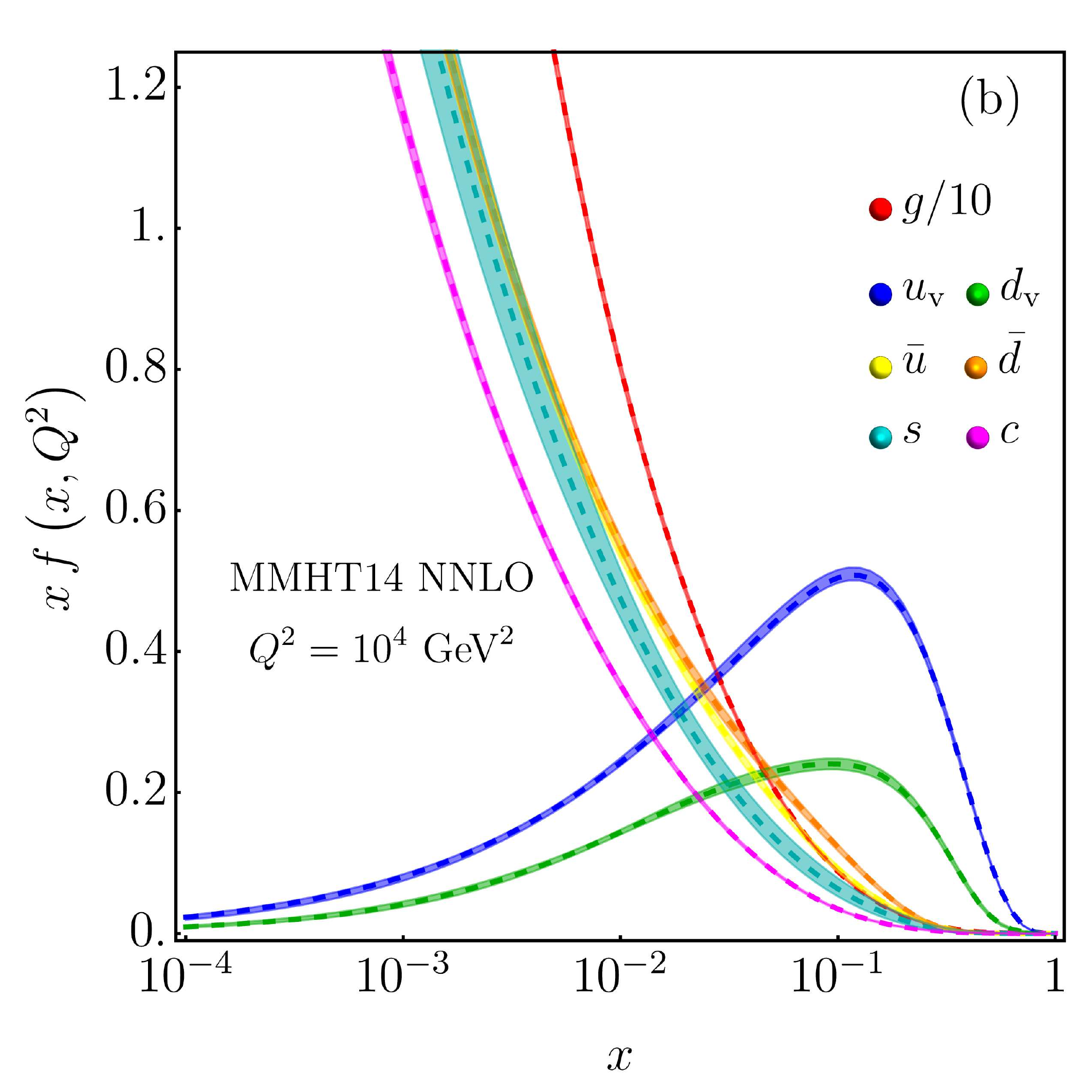}
    \caption{Parton distribution functions for the proton in the LHC for computed at NNLO as a function of the momentum fraction $x$ at the scales (a) $Q^2=10$ GeV$^2$ and (b) $Q^2=10^4$ GeV$^2$. In each plot, the PDFs are shown for gluons ($g$), valence quarks ($u_\text{v}, d_\text{v}$), sea-quarks ($\bar{u}, \bar{d}$), strange quarks ($s$) and charm quarks ($c$). The error bands correspond to a confidence--level uncertainty of 68$\%$. Note that the gluon PDF is scaled by a factor of 10.}
    \label{Fig:PDFs_NNLO}
\end{figure}

Figures~\ref{Fig:PDFs_LO},~\ref{Fig:PDFs_NLO} and~\ref{Fig:PDFs_NNLO} show the parton distribution functions for the proton as a function of the momentum fraction $x$ computed in the LHC at LO, NLO, and NNLO, respectively.  The analysis is made from hard scattering data at the scales $Q^2=10$ GeV$^2$ and $100$ GeV$^2$ witting a 68$\%$ of confidence--level uncertainty, and the running coupling constant reference is chosen as $\alpha_s(M_Z^2)=0.135$ (for LO), $\alpha_s(M_Z^2)=0.120$ (for NLO) and $\alpha_s(M_Z^2)=0.118$ (for NNLO), where $M_Z$ is the $Z$-boson mass.~\cite{Harland-Lang:2014zoa}. In those figures, the PDFs are shown for gluons ($g$), valence quarks ($u_\text{v}, d_\text{v}$), sea-quarks ($\bar{u}, \bar{d}$), strange quarks ($s$) and charm quarks ($c$). Note that gluon production enhances at small $x$, which is indicated by scaling the corresponding PDF by a factor of 10. Moreover, as it was commented on, given the dense regime when the energy is increased, higher orders in the perturbation theory must be added in order to compute an accurate gluon distribution function. In particular, by comparison between Figs.~\ref{Fig:PDFs_LO}(a),~\ref{Fig:PDFs_NLO}(a) and~\ref{Fig:PDFs_NNLO}(a), it is clear that the gluon production reaches a saturation point when $x$ decreases and more terms in the perturbative series are considered. Moreover, in Figs.~\ref{Fig:PDFs_LO}(b),~\ref{Fig:PDFs_NLO}(b) and~\ref{Fig:PDFs_NNLO}(b), a saturation point is not found, and the corrections provided by the NLO and NNLO do no change drastically the PDFs behavior.  

The data is conclusive: at small $x$ (or high-energy reactions), the proton is made mostly by gluons. Nevertheless, the results for large $Q^2$ suggest a non-saturation value, which has no physical grounds. In order to obtain a saturation regime, it is necessary to separate the gluon evolution. First, the $Q^2$-evolution gives a picture of non-interacting gluons yielding linear effects in the BFKL formalism. However, even a considerable energy exchange from the virtual photons to partons leads a negligible interaction between the hadron constituents at the dilute regime, the $Y$-evolution (or $1/x$) implies a high-gluon density medium, and the coupling becomes significant, giving rise to non-linear effects~\cite{gribov1983semihard,mueller1986gluon,albacete2014gluon}. 

To obtain an estimate of the saturation scale, one considers the dilute and dense regimes of gluon occupation numbers: in the former, $n\ll1$ which is turned into order $n~\sim\Op(1)$ when gluons become to overlap, but given that their production is suppressed by $\alpha_s\ll1$, their interactions are still weak. Increasing the density implies interactions of order $\alpha_s\sim 1$ and a large density of $n\sim\Op(1/\alpha_s)$, where the non-linear effects stop the growth in the gluon number. Thus, from the gluon occupation number $n(x,Q^2)$ defined as
\bea
n(x,Q^2)&\equiv&\frac{\#\text{ of gluons with given $x$}\times\text{ occupied are by one gluon}}{\text{transverse area of the hadron}}\nn\\
&\simeq&\frac{x g(x,Q^2)}{R^2 Q^2},
\eea
where it is assumed that the area per gluon is $\Sigma_g\sim 1/Q^2$ and the transverse (invariant) area of the hadron is $\Sigma_{\perp}^H\sim\pi R^2$, the gluon saturation scale $Q_s(x)$ is estimated to be~\cite{gribov1983semihard,PhysRevD.49.2233}:
\bea
n\left(x, Q^{2}=Q_{s}^{2}(x)\right) \sim \frac{1}{\alpha_{s}} \Longrightarrow Q_{s}^{2}(x) \simeq \alpha_{s} \frac{x g\left(x, Q_{s}^{2}(x)\right)}{R^{2}}.
\eea

The above equation implies that gluons with momentum $\kt<Q_s(x)$ correspond to large occupation numbers but in a saturation scale such as they do not increase the number of gluons when $x$ is decreased. On the other hand, gluons with $\kt\gg Q_s(x)$ remains in the dilute regime.

This analysis yields the basic concepts of the so-called Color Glass Condensate (CGC) as a state of the high-energy hadronic matter, which can be summarized as follows\footnote{A detailed description from the theoretical and experimental perspectives can be found at the cited references throughout the current section and complemented with Refs.~\cite{mueller2002parton,iancu2004color,jalilian2006saturation,weigert2005evolution,Triantafyllopoulos:2005cn,Gelis:2010nm,Lappi:2010ek}. Although the CGC theory is of particular interest, its full treatment is outside the scope of this work.}:
\begin{itemize}
    \item At high-energetic reactions, the hadron is most dominated by the presence of gluons, so that the medium is {\it colored} by the non-Abelian charges of $SU(3)$. 
    
    \item The system behaves like a glass: the coherence time discussion shows that gluons with larger $x$ are viewed as static sources ({\it fast gluons})  by the radiated gluons with small $x$ ({\it slow gluons}). However, fast gluons can vary in a long time window set by their longitudinal momentum. Hence, for short time scales, the system is like a ``solid'', but it is like a ``liquid'' for large times: that is the main characteristic of glass.
    
    \item The high saturated gluon occupation number adds its color charges coherently. The corresponding saturation scale is a multiparticle state of bosons in the Bose-Einstein's condensate phase. 
\end{itemize}

The CGC state is the starting point of the photon production mechanism proposed in this chapter: the gluons will be treated as on-shell particles. Their high-density makes them weakly coupled, and perturbation theory is then applicable for scales $Q_s^2(x)\gg\Lambda_\text{QCD}^2$.

\section{Magnetic Fields in a Relativistic Heavy-Ion Collision}\label{sec:Magnetic_field_HIC}
In Chapter~\ref{Chap:QCD_phase_diagram}, the effects of high temperatures and baryonic densities were studied as a key to understand QCD matter under extreme conditions, particularly the chiral symmetry restoration and the QCD-phase diagram were presented. On the other hand, Sec.~\ref{Sec:The_Color_Glass_Condensate} shows that ultrarelativistic systems have severe implications for the parton distribution functions, yielding a gluonic dense medium known as the CGC. In principle, this novel dynamics can be measured in heavy-ion collision experiments, which give rise to another phenomenon that probes the QCD in an additional limiting scenario: the presence of intense magnetic fields in a deconfined medium. In order to introduce the properties of such magnetic fields, this section is devoted to its theoretical study. The framework presented here summarizes Refs.~\cite{intensity2,intensity3,intensity4,intensity5} and an extension of these results is given in Sec.~\ref{subsec:Magnetic_field_from_UrQMD}. 

The existence of strongly interacting matter in the presence of intense magnetic fields has become of interest, given its implications in fundamental theoretical and experimental phenomena. The so-called magnetic catalysis, where the magnetic field acts as a catalyst for the dynamical fermion mass generation from flavor symmetry breaking, is an excellent example~\cite{Klimenko:1991he,Gusynin:1994re,miranskypropagador2,Preis:2010cq,Gusynin:1995gt}. Also, the QCD-topological invariants encoded in the winding number related to symmetries $\mathcal{P}$ and $\mathcal{CP}$ are affected by external intense magnetic fields~\cite{intensity1}. Moreover, it has been suggested that signals of the deconfinement and the chiral phase transition may be found by measuring the imbalance of charged particles emission along the direction of the magnetic field provided by a nonzero chirality~\cite{intensity1,PhysRevD.78.074033}. Similarly, intense magnetic fields have an impact in the QCD-phase diagram~\cite{Agasian:2008tb,Fraga:2008qn,PhysRevD.71.023004,Ayala:2015lta}, the spontaneous creation of axial currents~\cite{Son:2004tq,Metlitski:2005pr}, the formation of $\pi^0$ domain walls~\cite{Son:2007ny}, and the color superconducting phases~\cite{Alford:1999pb,Ferrer:2005vd}, among others~\cite{Okorokov:2019bsi,zhong2014systematic,Skokov:2013axa,Voskresensky:1980nk,Sadooghi:2009fi}.

The idea behind the intense magnetic fields created in heavy-ion collisions is simple: for non-central events, i.e., collisions with an impact parameter $b$, the non-colliding hadrons (protons) move as free charges with opposite directions. The electric current associated with these protons leads to the creation of a magnetic field perpendicular to the protons direction of motion. The magnetic field evolution for hadron collisions starts with the Boltzmann equation for the phase-space distribution function $f(x,p)$:
\bea
p^{\mu} \frac{\partial}{\partial x^{\mu}} f=C[f],
\eea
where $C[f]$ is the so-called {\it collision integral} which connects the dynamics of the colliding particles, where the particles are considered on-shell with 4-momentum $p^\mu$ and 4-position $x^\mu$. The effects of a brackground electromagnetic field can be achieved by introducing the field tensor $F_{\mu\nu}$:
\bea
p^{\mu}\left( \frac{\partial}{\partial x^{\mu}}-F_{\mu\nu} \frac{\partial}{\partial p^{\nu}}\right) f=C[f].
\eea

If the particles are constrained by an interaction potential $U$, the equation for a test particle with distribution $f(\mathbf{r},\mathbf{p},t)$ is
\bea
\left[\frac{\partial}{\partial t}+\left(\frac{\mathbf{p}}{p_{0}}+\nabla_{\mathbf{p}} U\right) \nabla_{\mathbf{r}}+\left(-\nabla_{\mathbf{r}} U+e \mathbf{E}+e \mathbf{v} \times \mathbf{B}\right) \nabla_{\mathbf{p}}\right]f=C_\text{coll}(f,f_1,\cdots,f_N).
\eea

Besides the Boltzmann equation, the charged particle dynamics is determined by the Maxwell (field) equations:
\bea
\nabla \times \mathbf{E}=-\frac{1}{c} \frac{\partial \mathbf{B}}{\partial t}, \quad \nabla \cdot \mathbf{B}=0,
\eea
so that for a point-like charge modelled by the sources:
\bea
\rho(\mathbf{r}, t)&=&e\,\delta[\mathbf{r}-\mathbf{r}(t)],\nn\\
\mathbf{j}(\mathbf{r}, t)&=&e\,\mathbf{v}(t) \delta[\mathbf{r}-\mathbf{r}(t)],
\eea
the electric and magnetic field background is found to be
\begin{subequations}
\bea
e \mathbf{E}=\frac{\operatorname{sign}(e) \alpha \mathbf{R}(t)\left(1-v^{2} / c^{2}\right)}{\left[(\mathbf{R}(t) \cdot \mathbf{v} / c)^{2}+R^{2}(t)\left(1-v^{2} / c^{2}\right)\right]^{3 / 2}}
\eea
and
\bea
e \mathbf{B}=\frac{\operatorname{sign}(e) \alpha[\mathbf{v} \times \mathbf{R}(t)]\left(1-v^{2} / c^{2}\right)}{c\left[(\mathbf{R}(t) \cdot \mathbf{v} / c)^{2}+R^{2}(t)\left(1-v^{2} / c^{2}\right)\right]^{3 / 2}},
\eea
\end{subequations}
where $\alpha=e^2/4\pi$. The above equations are the well-known Lienard-Wiechert fields for a moving particle. 

The presented formalism to account for the created electromagnetic fields is more on less standard, but several computational approaches address variations in the event generation. The Ultrarelativistic Quantum Molecular Dynamics (UrQMD) is one of them, and it is discussed in Sec.~\ref{subsec:Magnetic_field_from_UrQMD}. The Hadron String Dynamics (HSD) is also used, which corresponds to Kadanoff-Baym kinetic equations, where the nuclear collisions are modeled as quasiparticles with finite width~\cite{Ehehalt:1996uq,Cassing:1999es}.  The latter was used by Voronyuk {\it et al.} to describe the space-time evolution of the magnetic field in non-central collisions~\cite{intensity3}, but other approaches from general considerations on the rapidity $Y$ and charge distribution density $\varrho_\pm\left(\mathbf{x}_\perp'\right)$ are implemented. For example, the charge distribution density inside the contracted nucleus is usually taken as:
\bea
\varrho_{\pm}\left(\boldsymbol{x}_{\perp}^{\prime}\right)=\frac{2}{\frac{4}{3} \pi R^{3}} \sqrt{R^{2}-\left(\boldsymbol{x}_{\perp}^{\prime} \pm \boldsymbol{b} / 2\right)^{2}} \theta_{\pm}\left(\boldsymbol{x}_{\perp}^{\prime}\right),
\eea
with the normalization condition
\bea
\int \mathrm{d}^{2} \boldsymbol{x}_{\perp}^{\prime} \varrho_{\pm}\left(\boldsymbol{x}_{\perp}^{\prime}\right)=1,
\eea
where $\boldsymbol{x}_{\perp}^{\prime}$ is a vector transverse to the beam axis ($\boldsymbol{e}_{y}$) and the nuclei are considered as {\it pancake} shaped. The function $\theta_{\pm}\left(\boldsymbol{x}_{\perp}^{\prime}\right)$ are the nuclei's projections over the transverse plane separating spectators (no colliding hadrons) and participants (hadrons with at leats one collision) so that:
\bea
\theta_{\pm}\left(\boldsymbol{x}_{\perp}^{\prime}\right)=\theta\left[R^{2}-\left(\boldsymbol{x}_{\perp}^{\prime} \pm \boldsymbol{b} / 2\right)^{2}\right].
\eea

The latter formalism is used by Kharzeev {\it et al.} by writing the Lienard-Wiechert potentials as function of $Y$ as follows~\cite{intensity1}:
\bea
e \boldsymbol{B}(\boldsymbol{x})=Z \alpha \sinh (Y) \frac{\left(\boldsymbol{x}_{\perp}^{\prime}-\boldsymbol{x}_{\perp}\right) \times \boldsymbol{e}_{y}}{\left[\left(\boldsymbol{x}_{\perp}^{\prime}-\boldsymbol{x}_{\perp}\right)^{2}+(t \sinh Y-z \cosh Y)^{2}\right]^{3 / 2}},
\eea
where $Z$ is the nucleus charge. 
\begin{figure}
    \centering
    \includegraphics[scale=0.38]{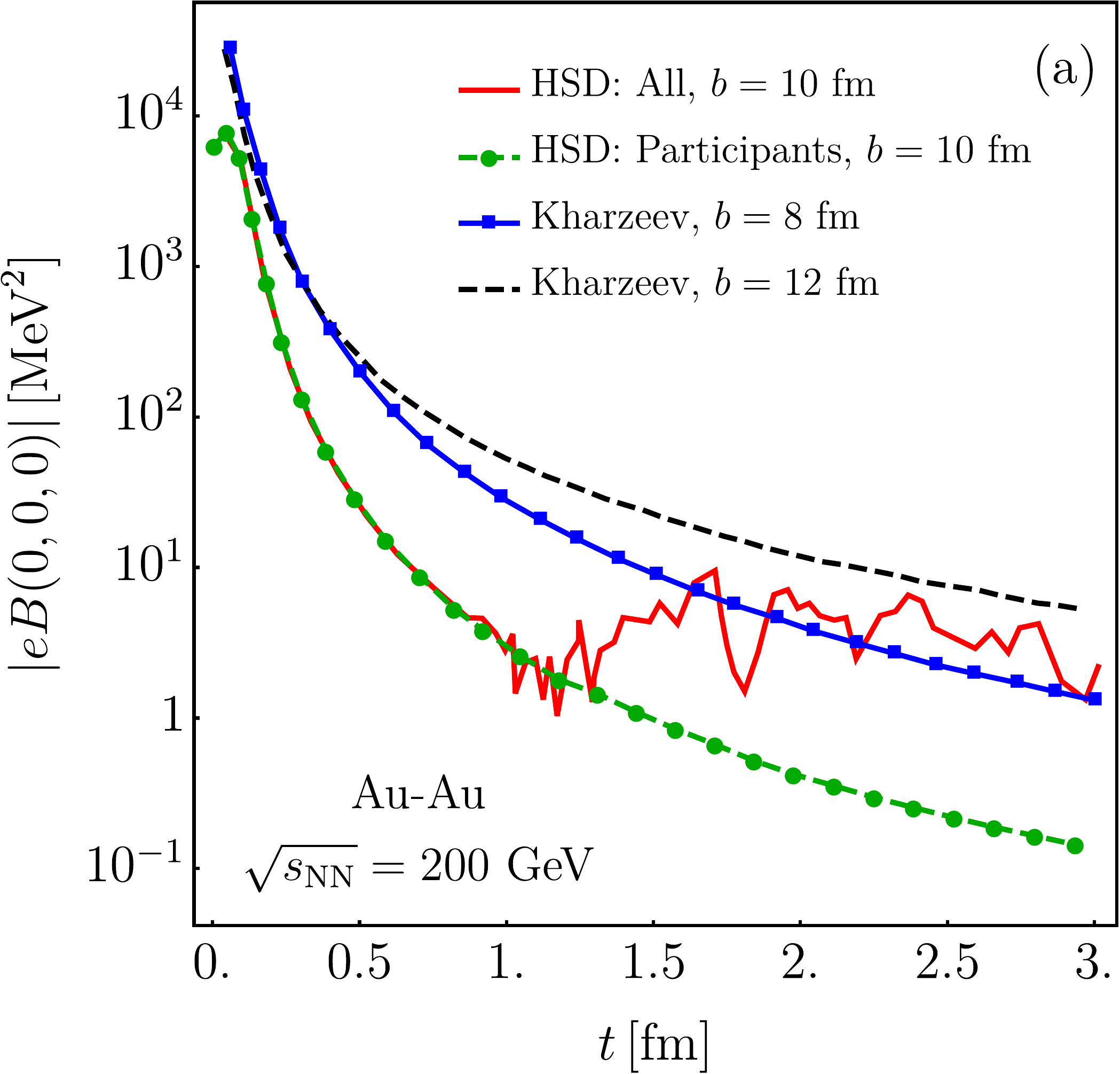}\hspace{0.5cm}\includegraphics[scale=0.38]{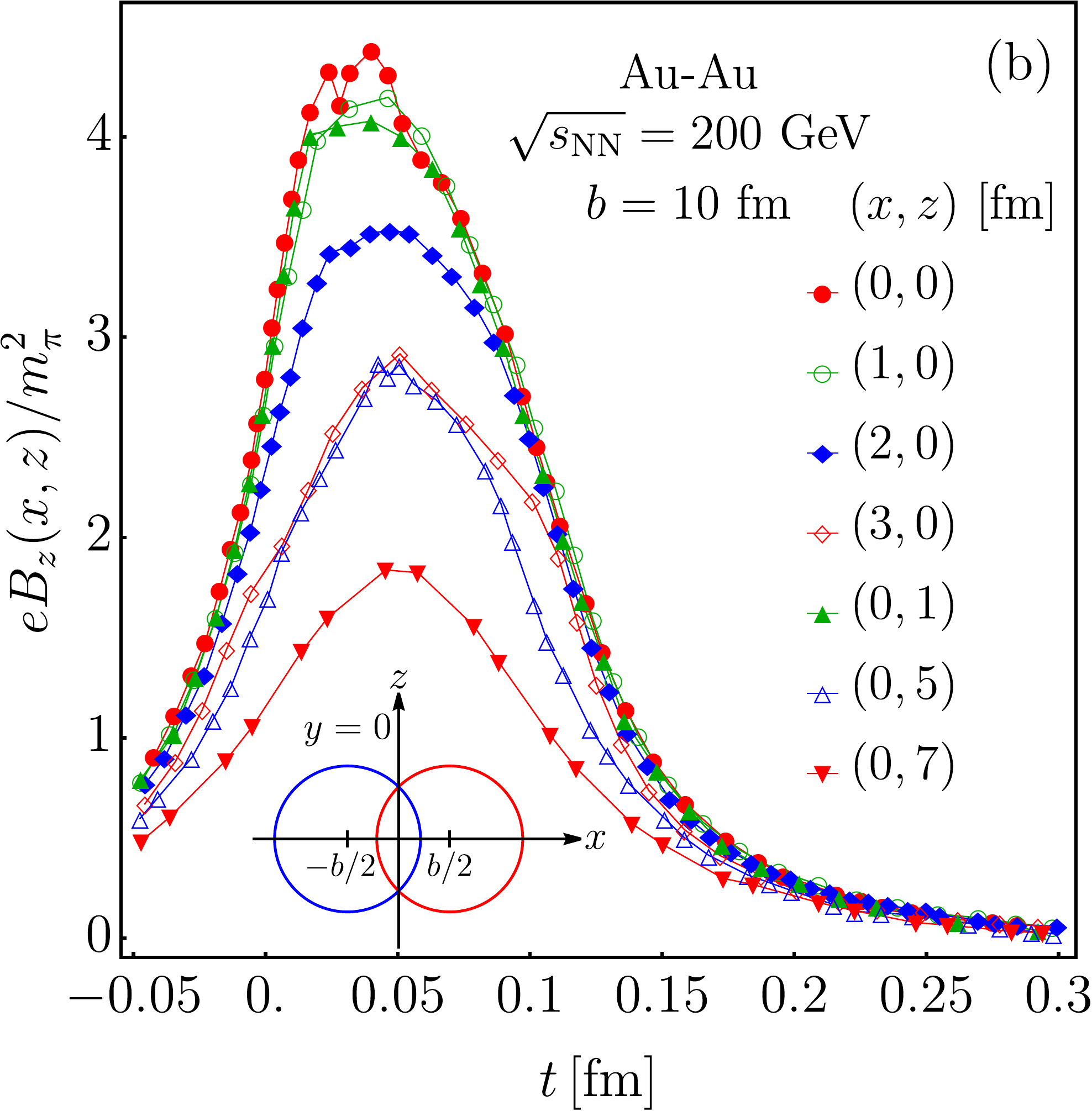}
    \caption{(a) Magnetic field space--time evolution at the center of the collision computed in Refs.~\cite{intensity1,intensity3} when the impact parameter $b$ is changed. (b) Magnetic field in units of the squared pion mass $m_\pi^2$ in $z$-direction for several points $(x,z)$ in the plane perpendicular to the charge's motion with $b=10$ fm~\cite{intensity3}. The inset represents the geometry in the transverse plane ($y=0$) of a noncentral heavy-ion collision.}
    \label{Fig:HSD_spacetime}
\end{figure}

In terms of the charge distribution density, the total magnetic field is given by:
\bea
\boldsymbol{B}=\boldsymbol{B}_{s}^{+}+\boldsymbol{B}_{s}^{-}+\boldsymbol{B}_{p}^{+}+\boldsymbol{B}_{p}^{-},
\eea
where $\boldsymbol{B}_{s}^{\pm}$ and $\boldsymbol{B}_{p}^{\pm}$ is the magnetic field created by spectators and participants moving in directions $\pm\boldsymbol{e}_{y}$, respectively. By defining the proper time $\tau=(t^2-y^2)^{1/2}$ and the space-time rapidity $\eta=\frac{1}{2}\log\left[(t+y)/(t-y)\right]$, the spectators' field is:
\bea
e \boldsymbol{B}_{s}^{\pm}\left(\tau, \eta, \boldsymbol{x}_{\perp}\right)&=&\pm Z \alpha \sinh \left(Y_{0} \mp \eta\right)\\
&\times&\int \mathrm{d}^{2} \boldsymbol{x}_{\perp}^{\prime} \varrho_{\pm}\left(\boldsymbol{x}_{\perp}^{\prime}\right)\left[1-\theta_{\mp}\left(\boldsymbol{x}_{\perp}^{\prime}\right)\right]\frac{\left(\boldsymbol{x}_{\perp}^{\prime}-\boldsymbol{x}_{\perp}\right) \times \boldsymbol{e}_{y}}{\left[\left(\boldsymbol{x}_{\perp}^{\prime}-\boldsymbol{x}_{\perp}\right)^{2}+\tau^{2} \sinh \left(Y_{0} \mp \eta\right)^{2}\right]^{3 / 2}},\nn
\eea
where is assumed that spectators do no scatter and continues traveling with the initial beam's rapidity $Y_0$.

The participant's contribution is taken only from the original colliding hadrons, i.e., the produced particles after the reaction are neglected. Such consideration is validated because produced particles expand spherically, and the magnetic field created by them is minimal. Moreover, the number of charged particles (positive and negative) remains approximately equal than before the collision. Therefore, by taking into account the normalized distribution of the participants along the collision axis:
\bea
f(Y)=\frac{a}{2 \sinh \left(a Y_{0}\right)} e^{a Y}, \quad-Y_{0} \leq Y \leq Y_{0},
\eea
where the parameter $a\approx1/2$ is obtained from experimental data~\cite{Adler:2004zn}, the contribution of the participants to the magnetic field is
\bea
e \boldsymbol{B}_{p}^{\pm}\left(\tau, \eta, \boldsymbol{x}_{\perp}\right)&=&\pm Z \alpha \int \mathrm{d}^{2} \boldsymbol{x}_{\perp}^{\prime} \int_{-Y_{0}}^{Y_{0}} \mathrm{d} Y f(Y) \sinh (Y \mp \eta) \varrho_{\pm}\left(\boldsymbol{x}_{\perp}^{\prime}\right) \theta_{\mp}\left(\boldsymbol{x}_{\perp}^{\prime}\right)\nn\\
&\times&\frac{\left(\boldsymbol{x}_{\perp}^{\prime}-\boldsymbol{x}_{\perp}\right) \times \boldsymbol{e}_{y}}{\left[\left(\boldsymbol{x}_{\perp}^{\prime}-\boldsymbol{x}_{\perp}\right)^{2}+\tau^{2} \sinh (Y \mp \eta)^{2}\right]^{3 / 2}}.
\eea

Figure~\ref{Fig:HSD_spacetime} shows the space-time evolution of the magnetic field in the approaches thus discussed for simulations of Au-Au collisions at $\sqrt{s_\text{NN}}=200$ GeV with several impact parameters. In Fig.~\ref{Fig:HSD_spacetime}(a), the magnetic field at the collision's center from the HSD and Kharzeev's formalism is compared. The former separates the magnetic field originated by spectators from the spectators and participants. Both situations imply a rapid decay in the magnetic field strength with a peak of high intensity at the beginning of the collision. Moreover, Fig.~\ref{Fig:HSD_spacetime}(b) shows the magnetic field's spatial profile in the transverse plane. The center's variation to other directions is less than 20$\%$, except for the boundaries of the nucleus overlap $x\approx b/2\sim5$ fm. Hence, the magnetic field is very homogeneous in the $z$-direction.  Note that in Fig.~\ref{Fig:HSD_spacetime}(b), the vertical axis is normalized to the squared pion mass $m_\pi^2$, giving an estimate of the magnetic field strength $m_\pi^2\sim10^{18}$ Gauss.

The above results give the main features of the magnetic field in a heavy-ion collision: it has a very high intensity but is short-lived. That lifetime is crucial to the Chiral Magnetic Wave (CMW) framework to explain the elliptic flow dependence asymmetry for charged particles~\cite{Burnier:2012ae}. The CMW formalism needs magnetic fields with a minimum of 4 fm duration, but the models and data present a lifetime around 0.1 fm~\cite{Skokov1}. Models to improve the magnetic field pulse time based on the effects of the QCD-medium conductivity have been proposed. Still, for realistic values of the parameters, there is not an appreciable difference~\cite{intensity4}. 

The presented discussion seeds the proposal for a source of photons in the early stages of a heavy-ion collision: a dense medium made of on-shell gluons with a strong magnetic background opens the photon production channel by gluon fusion and gluon splitting. The development of those channels and their impact on the photon invariant momentum distribution and the elliptic flow is presented in the following sections. 

\section{Prompt Photon Production by Gluon Fusion in a Magnetized Medium}\label{sec:Prompt Photon Production by Gluon Fusion in a Magnetized Medium}
As was commented on the previous sections, at the early stages of a heavy-ion collision two important scenarios have to be taken into account: the high gluon abundance and the existence of intense magnetic fields. In order to take into account such phenomenology, the photon production from gluon fusion is proposed as a source of prompt photons. The lowest order process in the strong, and electromagnetic couplings comes from an amplitude made out of a quark triangle diagram with two gluons and one photon attached to each one of the vertices of the triangle. Given that the electromagnetic interaction differentiates the particle charge, both quark and anti-quark triangles have to be considered. 

The magnetic field acts over the fermionic loop, and therefore, the quark propagator in coordinate representation is
\bea
S(x,x')=\Phi(x,x')\int\frac{d^4p}{\dpi^4}e^{-ip\cdot(x-x')}S(p),
\label{coodspaceprop}
\eea
where the {\it phase factor} is given by
\bea
\Phi(x,x')=\exp\left\{i|q_f|\int_{x'}^xd\xi^\mu\left[A_\mu+\frac{1}{2}F_{\mu\nu}(\xi-x')^\nu\right]\right\},
\label{PhaseFactorDef}
\eea
with $|q_f|=n |e|$, where $n$ and $e$ are the fractional quark's charge and electron charge, respectively~\cite{PhysRev.82.664}. The Fourier transform of the translationally invariant part of the propagator in the Schwinger's proper-time formalism is given by
\bea
i S(p)&=&\int _{0}^{\infty}\frac{d\tau}{\cos(|q_fB|\tau)}\exp\left\{i\tau\left[p_\parallel^2-p_\perp^2\frac{\tan(\eB\tau)}{\eB\tau}-m_f^2+i\epsilon\right]\right\}\nn\\
&\times&\left[\left(\cos(\eB\tau)+\gamma_1\gamma_2\sin(\eB\tau)\right)(m_f+\psh_\parallel)-\frac{\psh_\perp}{\cos(\eB\tau)}\right],
\label{iS(p)def}
\eea
where $m_f$ is the fermion mass. To simplify the calculations, a constant and homogeneous magnetic field along the $\hat{z}$-direction is chosen. Such a magnetic field, in the symmetric gauge, can be obtained from the vector potential
\bea
A^\mu=\frac{B}{2}(0,-y,x,0).
\label{Asymmetric}
\eea

Also, for a 4-momentum $p^{\mu}$ its parallel and transverse parts with respect to the magnetic field are defined as
\begin{subequations}
\bea
\pp^\mu=(p_0,0,0,p_3)\rightarrow \pp^2=p_0^2-p_3^2,
\eea
\bea
\pt^\mu=(0,p_1,p_2,0)\rightarrow\pt^2=p_1^2+p_2^2,
\eea
thus
\bea
p^2=\pp^2-\pt^2.
\eea
\end{subequations}

\begin{figure}
    \centering
    \includegraphics[scale=0.45]{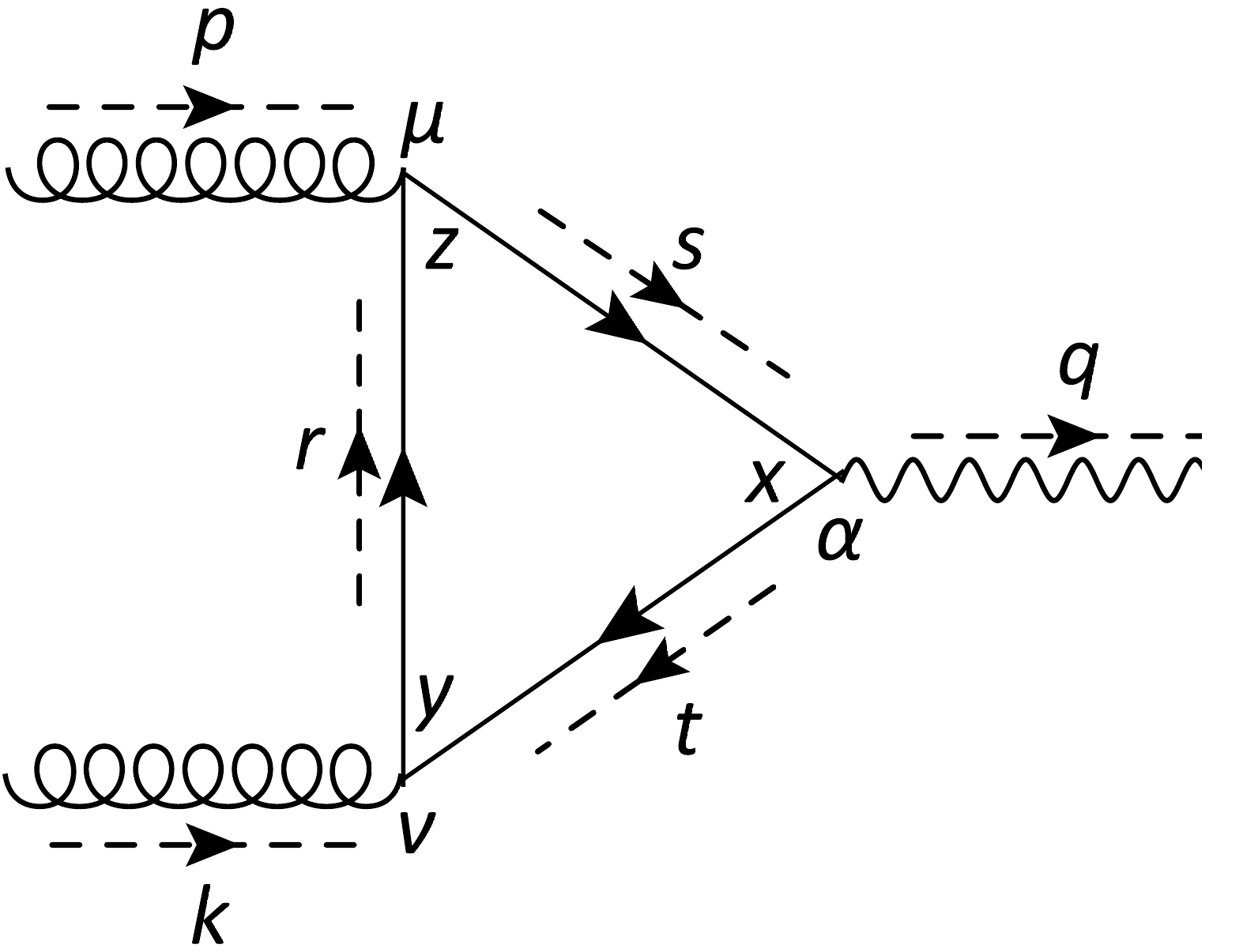}\includegraphics[scale=0.45]{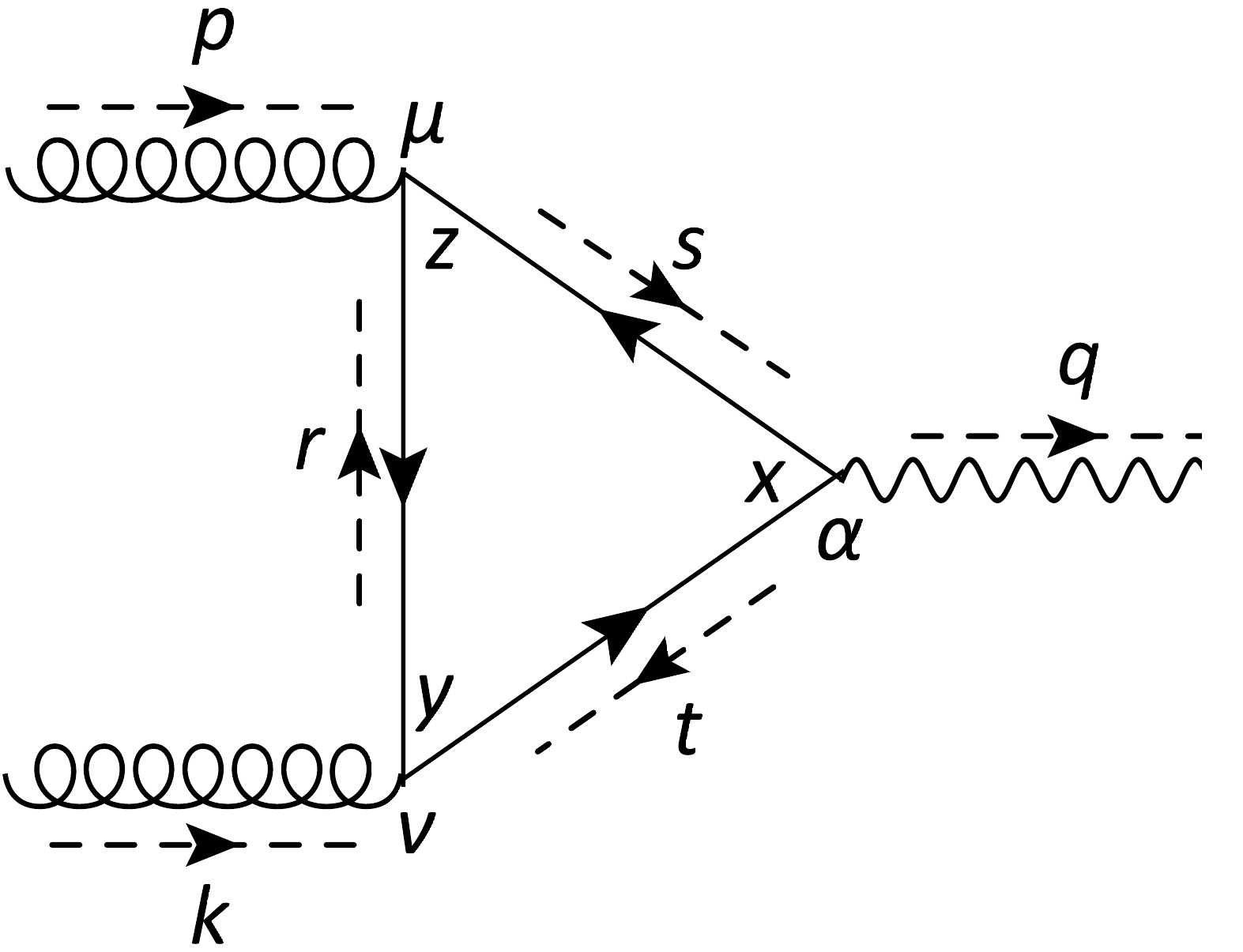}
    \caption{Feynman diagrams contributing to the production of a photon with four-momentum $q^\mu$ from the fusion of gluons with four-momentum $p^\mu$ and $k^\mu$. Note the charge flow in each diagram, which differentiates the quark from the antiquark. The fermion lines represent the magnetized propagator of Eq.~(\ref{iS(p)def}).}
    \label{Fig:QuarAntiQuarkTriangle}
\end{figure}

The Feynman diagrams which contribute to the process where two gluons with four-momentum $p^\mu$ and $k^\mu$ produced a photon with four-momentum $q^\mu$ are represented in Fig.~\ref{Fig:QuarAntiQuarkTriangle}. The quark lines into the triangle describe the magnetized fermion propagator so that one diagram schematics the quark's contribution whereas the other the antiquark's part. Hence, the total amplitude is given by
\bea
   \widetilde{{\mathcal{M}}}&=&-\int \!d^4x\,d^4y\,d^4z\int\!\frac{d^4r}{(2\pi)^4}
   \frac{d^4s}{(2\pi)^4}\frac{d^4t}{(2\pi)^4}e^{-it\cdot (y-x)}e^{-is\cdot (x-z)}e^{-ir\cdot (z-y)}e^{-ip\cdot z}e^{-ik\cdot y}e^{iq\cdot x}\nonumber\\
   &\times&
   \Big\{
   {\mbox{Tr}}\left[ iq_f\gamma_\alpha iS(s) ig\gamma_\mu t^c iS(r) ig\gamma_\nu t^d iS(t) \right]+
   {\mbox{Tr}}\left[ iq_f\gamma_\alpha iS(t) ig\gamma_\nu t^d iS(r) ig\gamma_\mu t^c iS(s) \right]
   \Big\}
   \nonumber\\
   &\times&\Phi(x,y)\Phi(y,z)\Phi(z,x)\epsilon^\mu(\lambda_p)\epsilon^\nu(\lambda_k)\epsilon^{\alpha}(\lambda_q),\quad
   \label{amplitude}
\eea
where $t^c,t^d$ are Gell-Mann matrices. The polarization vectors for the gluons and the photon are $\epsilon^\mu(\lambda_p)\epsilon^\nu(\lambda_k)$, and $\epsilon^{\alpha}(\lambda_q)$, respectively.

The product of phase factor has the form
\bea
\Phi(x,y)\Phi(y,z)\Phi(z,x)=\exp\left[-i\frac{\eB}{2}\epsilon_{mj}(z-x)_m(x-y)_j\right],
\label{phasefactor}
\eea
where $m,j$ are indices in the transverse space, and $\epsilon_{mj}$ is the 2D-Levi-Civita symbol~\cite{Bhattacharya:2005zu,kuznetsov2003electroweak,PhysRevD.91.016007}. The formal deduction of the Eq.~(\ref{phasefactor}) can be found in Appendix~\ref{ApPhaseFactorCalc}. 

The quark propagator is expressed in the well-know sum over Landau levels~\cite{miranskypropagador1,miranskypropagador2}:
\bea
iS(p)=i e^{-p_\perp^2/\eB}\sum_{n=0}^{+\infty}(-1)^n\frac{D_n(q_fB,p)}{p_\parallel^2-m_f^2-2n\left |  q_fB\right |},
\label{fermionpropdef}
\eea
where $L_n^\alpha(x)$ are the associated Laguerre polynomials, the numerator is
\bea
D_n(q_fB,p)=2(\psh_\parallel+m_f)\left[\mathcal{O}^{-}L_n^0\left(\frac{2p_\perp^2}{\left | q_fB \right|}\right)-\mathcal{O}^{+}L_{n-1}^0\left(\frac{2p_\perp^2}{\left | q_fB \right |}\right)\right]+4\psh_\perp L_{n-1}^1\left(\frac{2p_\perp^2}{\left | q_fB \right |}\right).\nn\\\quad
\label{DnDef}
\eea
and the operators $\mathcal{O}^{\pm}$ are given by
 \bea
 \mathcal{O}^{\pm}=\frac{1}{2}\left(1\pm i\gamma^1\gamma^2\right).
 \label{OpDef}
 \eea
 
 Usually, these operators are written by specifying the sign of the electric charge as
 \bea
  \mathcal{O}^{\pm}=\frac{1}{2}\left[1\pm i\text{sign}\left(q_fB\right)\gamma^1\gamma^2\right],
 \eea
so that $\mathcal{O}^{+}$ and $\mathcal{O}^{-}$ are exchanged in Eq.~(\ref{DnDef}) when the propagator passes from describing a fermion to describing an antifermion. Given that the definition of ``particle'' and ``antiparticle'' is a matter of choice it doesn't matter which sign of $\left(q_fB\right)$ is selected for the ``particle'', because the contribution of the two fermions is considered. 

To compute analytically Eq.~(\ref{amplitude}), the magnetic field is taken as the dominant energy scale. If the gluons have soft momentum and by the fact that the system has not reached thermal equilibrium, it is a suitable approximation. Therefore, it is possible to set $m_f=0$ and work with the contributions of the first Landau levels, i.e., with the Lowest Landau Level (LLL) and the first excited Landau level (1LL):
\begin{subequations}
\bea
iS^{\text{LLL}}(p)=i\frac{e^{-p_\perp^2/\eB}}{p_\parallel^2}\ps_\parallel\mathcal{O}^{-}
\label{propLLL1}
\eea

\bea
iS^{\text{1LL}}(p)=-2i\frac{e^{-p_\perp^2/\eB}}{p_\parallel^2-2\left |q_fB\right |}\left[\left(1-\frac{2p_\perp^2}{\left | q_fB \right |}\right)\ps_\parallel\mathcal{O}^{-}-\ps_\parallel\mathcal{O}^{+}+2\ps_\perp\right].
\label{propagadores}
\eea
\end{subequations}
\begin{figure}[t]
\begin{center}
\quad\includegraphics[scale=.35]{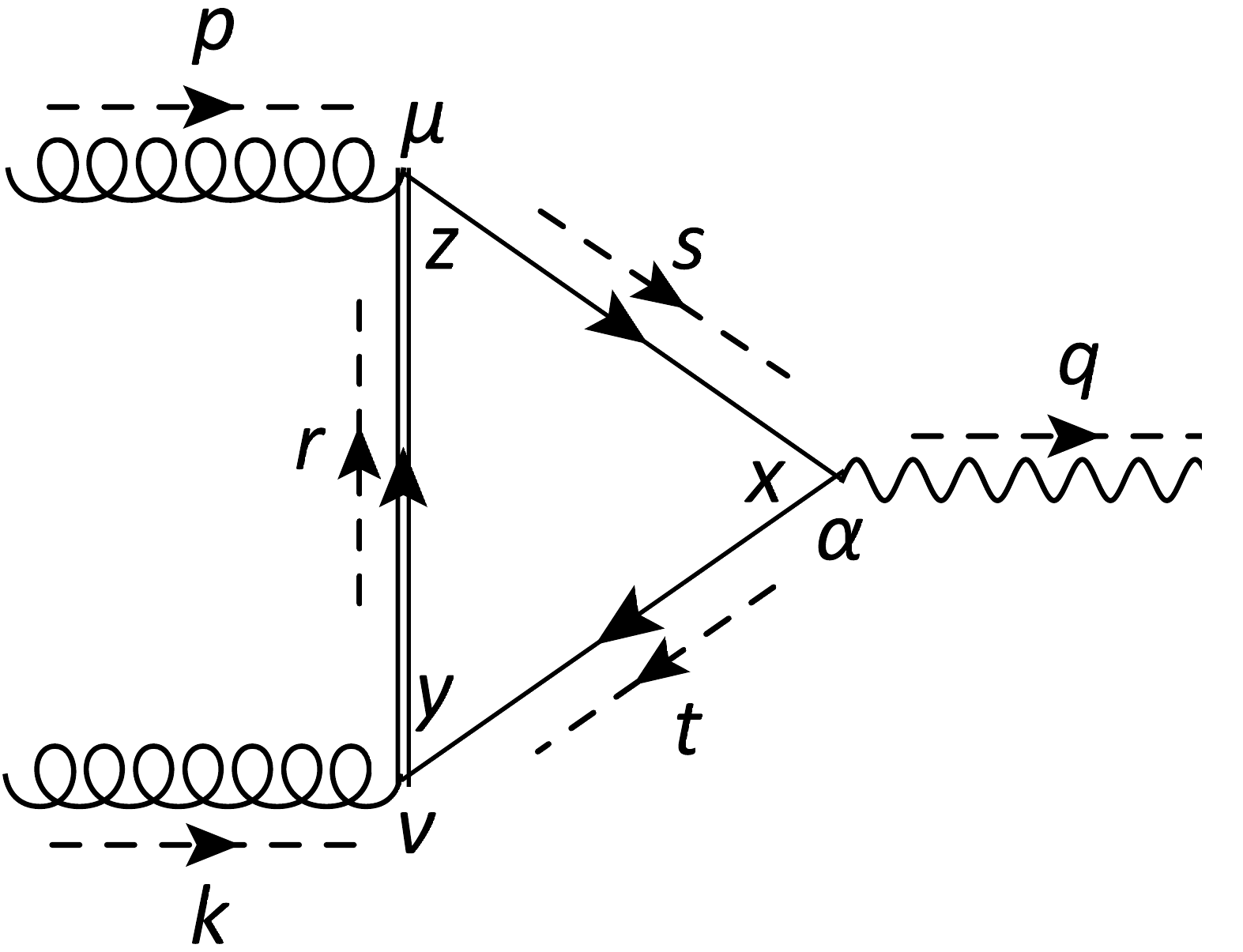}\includegraphics[scale=.35]{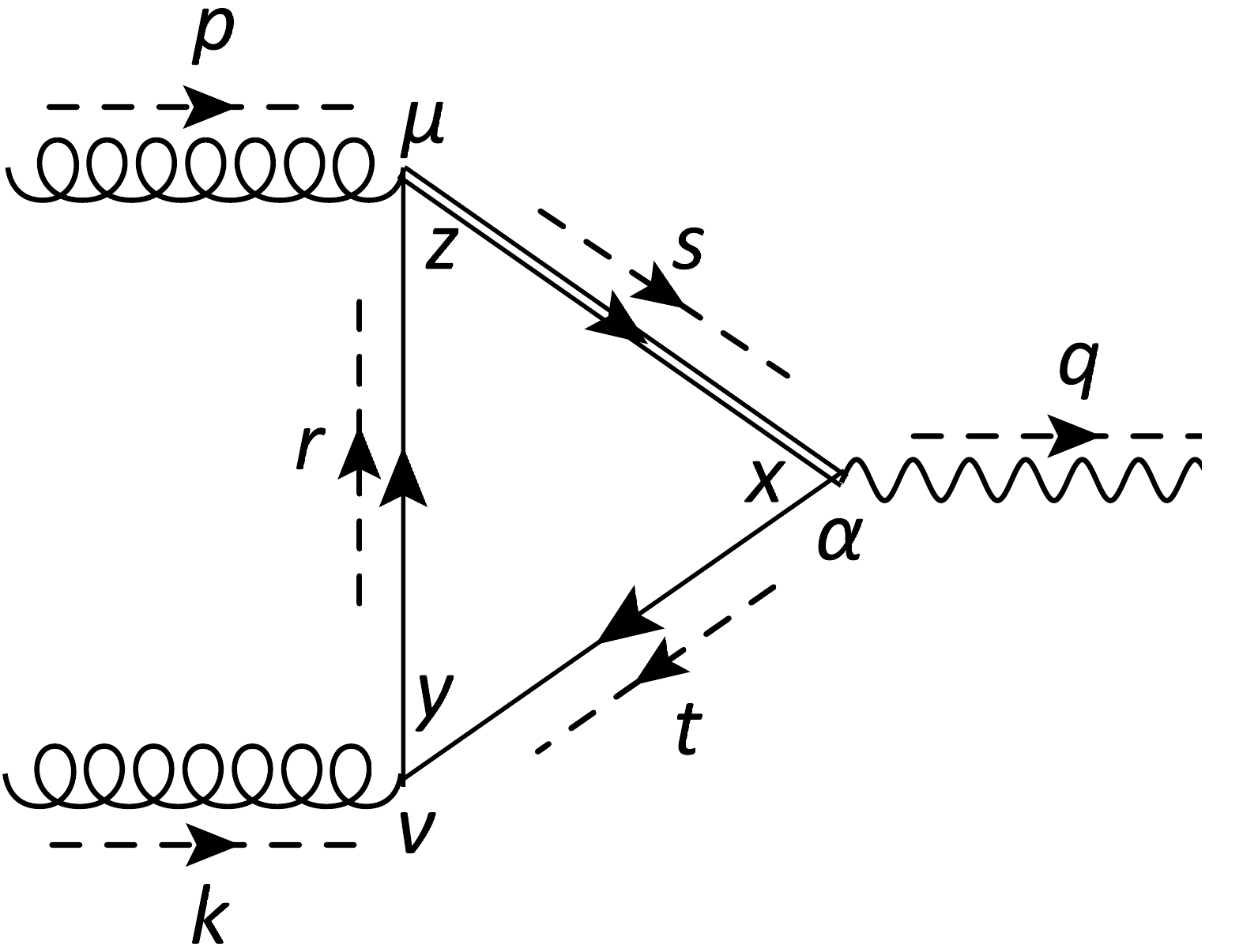}\includegraphics[scale=.35]{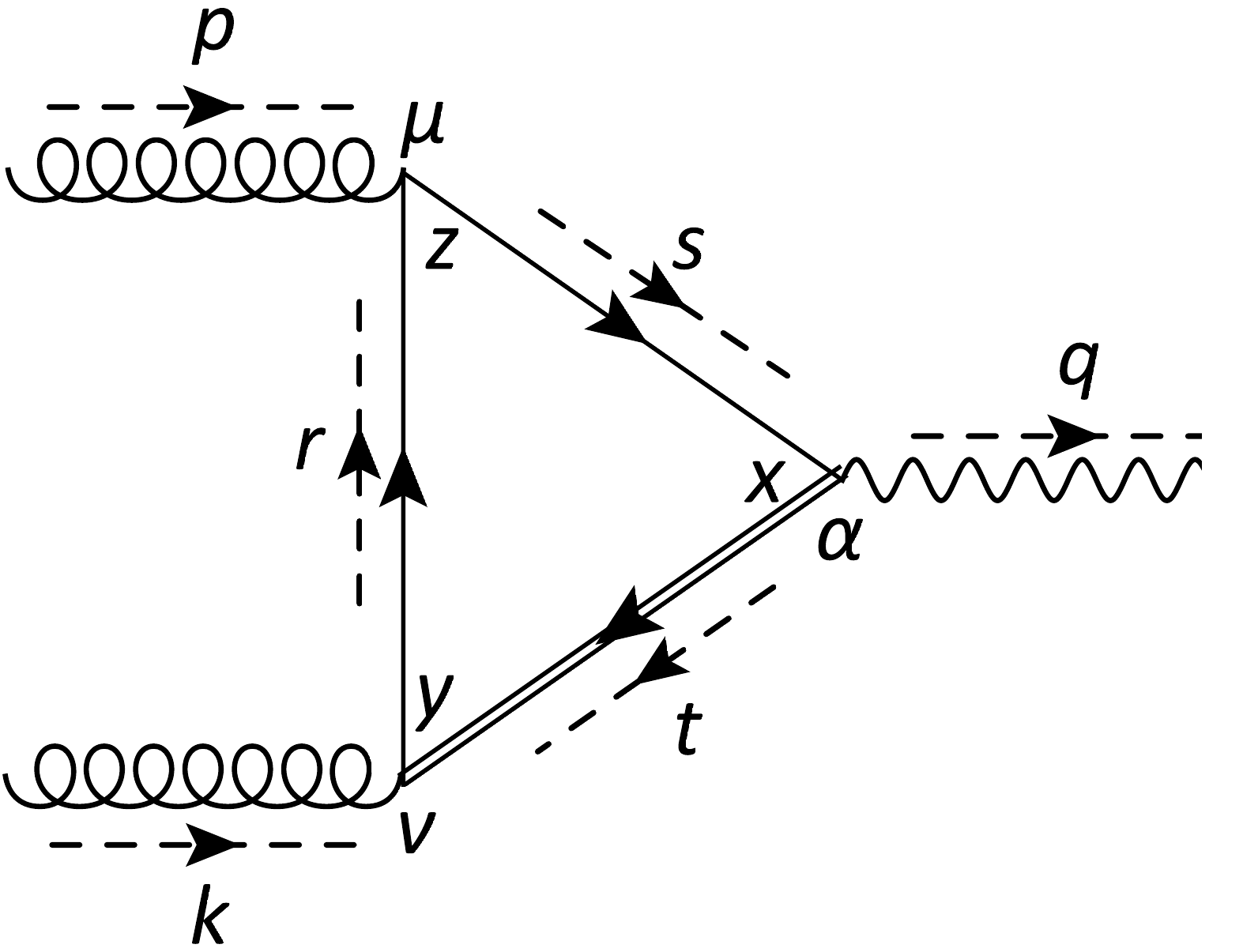}\\\vspace{0.3cm}
\includegraphics[scale=.35]{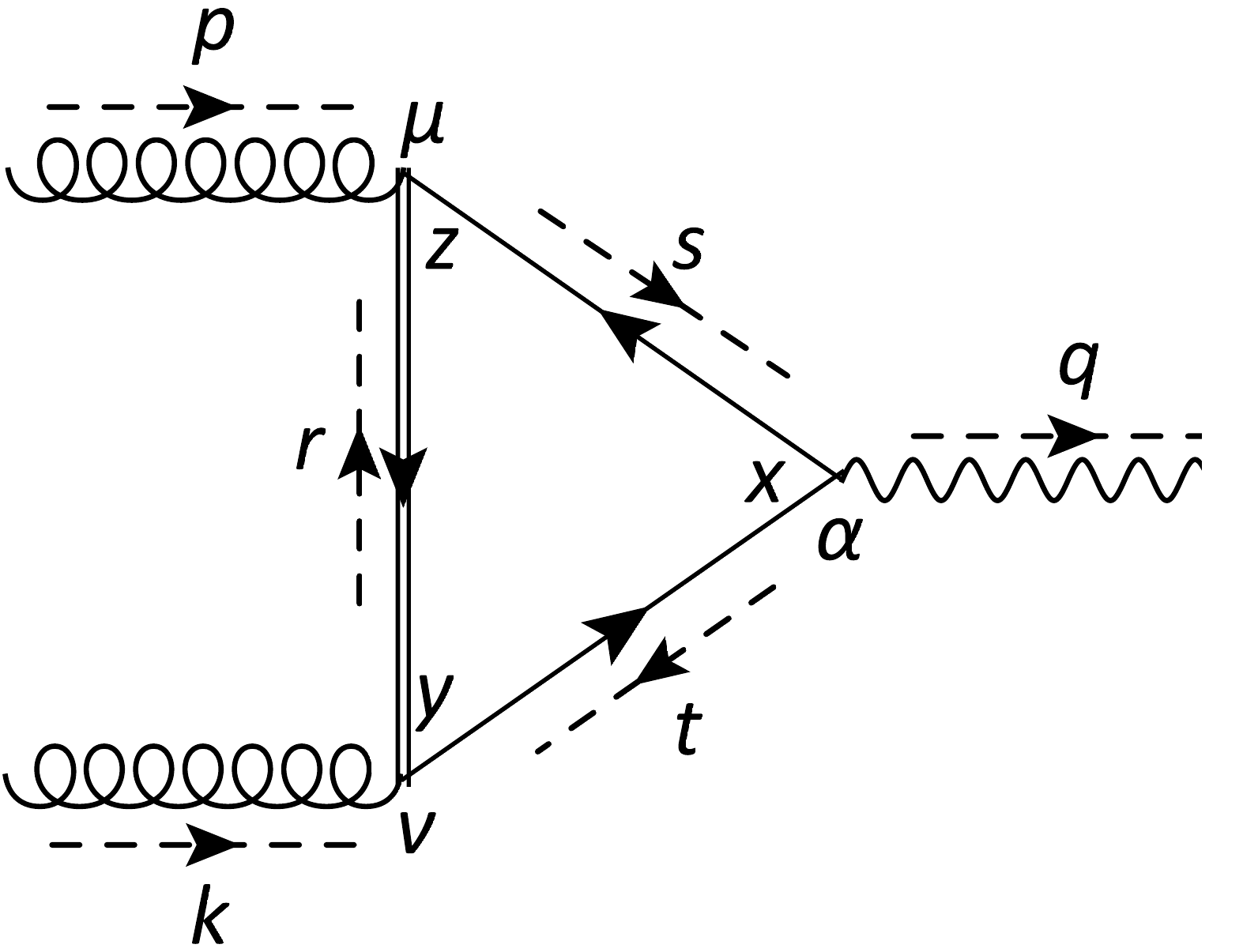}\includegraphics[scale=.35]{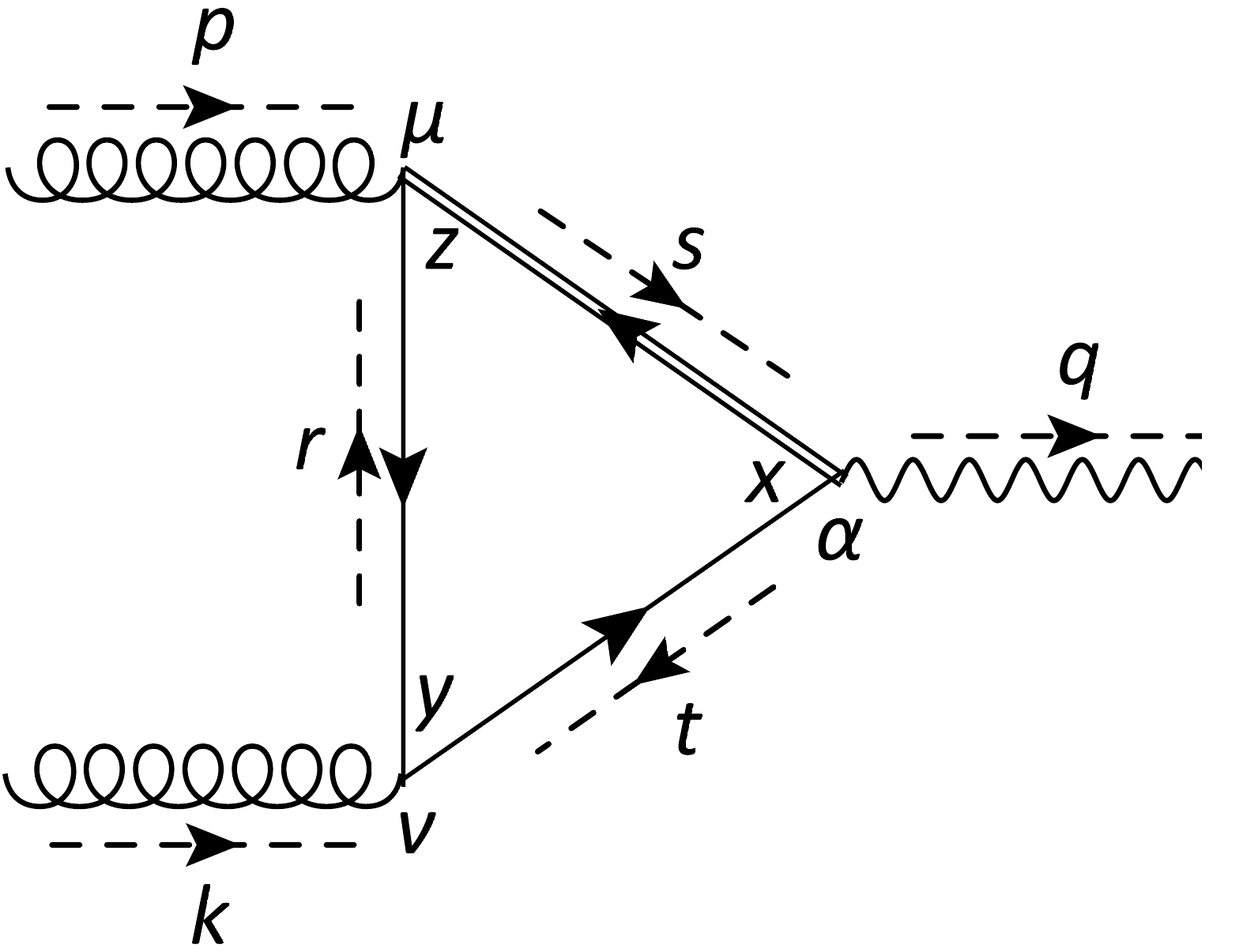}\includegraphics[scale=.35]{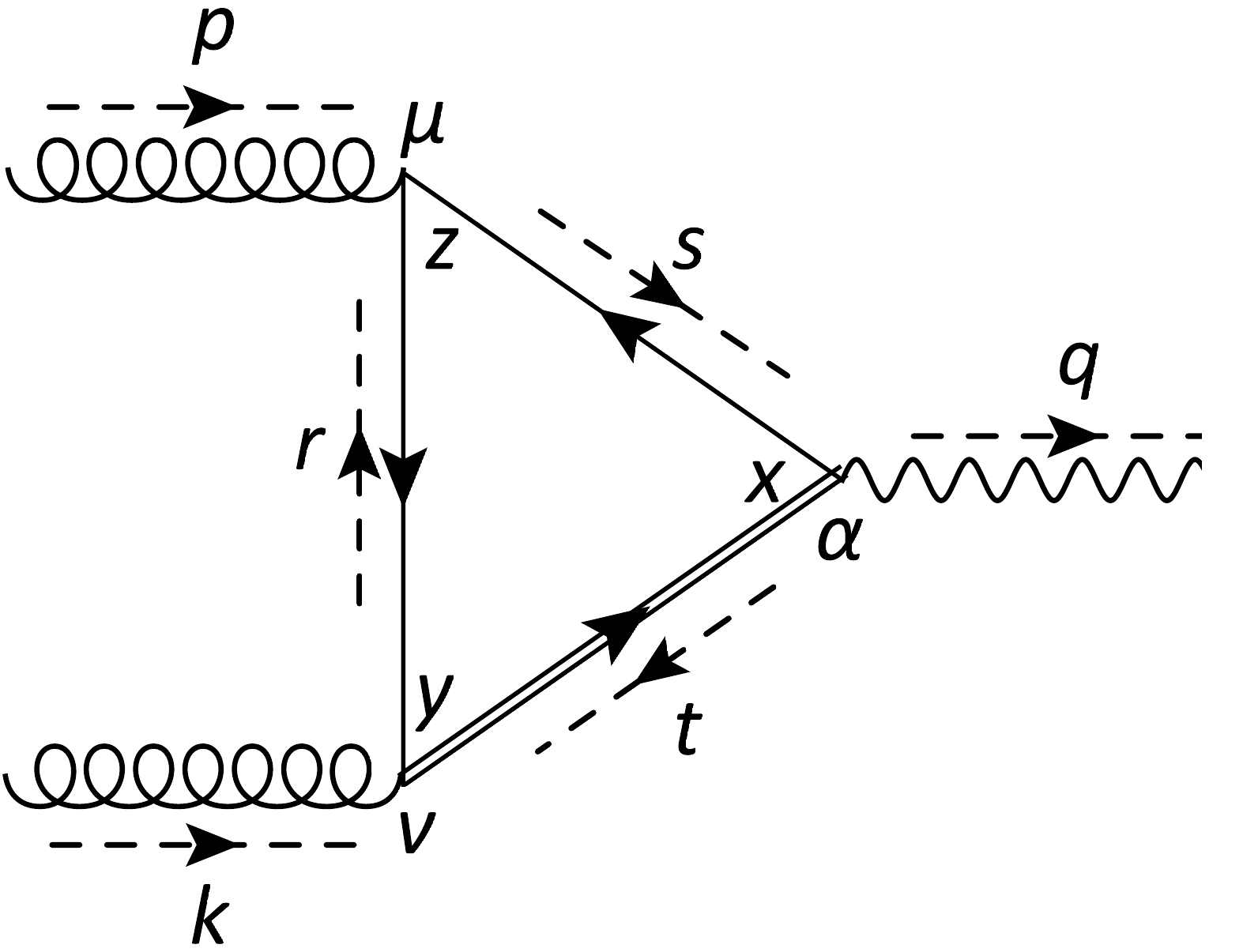}\quad
\caption{Feynman diagrams contributing to the process $gg\rightarrow\gamma$ in the presence of a magnetic field. The double lines represent that quark propagator in the 1LL, whereas the single lines represent the propagator in the LLL. The arrows in the propagators are the charge flow and the arrows on the sides represent the momentum direction.}
\label{Fig:Diag1}
\end{center}
\end{figure}

It is necessary to consider at least one quark in the 1LL since if all the fermions are in the LLL the fermionic traces vanishes, and the process is not allowed (see Appendix~\ref{ThreeQuarksinLLL}). The Feynman diagrams to be considered are shown in Fig.~\ref{Fig:Diag1}, where the quark/antiquark contribution is differentiated by the charge flow direction. Then, the leading order contribution of the amplitude reads
\bea
\widetilde{{\mathcal{M}}}&=&8i\dpi^4\delta^{(4)}\left(q-k-p\right)
\delta^{cd}|q_f| g^2\nonumber\\
&\times&\int\frac{d^4r}{\dpi^4}\frac{d^4s}{\dpi^4}\frac{d^4t }{\dpi^4}\;\epsilon^{\mu}(\lambda_p)\epsilon^\nu(\lambda_k)\epsilon^{\alpha}(\lambda_q)\nonumber\\
&\times&\int d^4w\;d^4l\, e^{-il(r-t-k)}e^{-iw(r-s+p)}\exp\left\{-i\frac{|q_fB|}{2}\epsilon_{mj}w_ml_j\right\}\exp\left\{-\frac{r_\perp^2+s_\perp^2+t_\perp^2}{|q_fB|}\right\}\nonumber\\
&\times&\text{Tr}\left\{\frac{\gamma_1\gamma_2\ga\slsh{t}_\perp\gn\slsh{r}_\parallel\gm^{\parallel}\slsh{s}_\parallel}{r_\parallel^2s_\parallel^2\left(t_\parallel^2-2\left| q_fB\right|\right)} +\frac{\gamma_1\gamma_2\gm\slsh{s}_\perp\ga\slsh{t}_\parallel\gn^{\parallel}\slsh{r}_\parallel}{t_\parallel^2r_\parallel^2\left(s_\parallel^2-2\left| q_fB\right|\right)}+\frac{\gamma_1\gamma_2\gn\slsh{r}_\perp\gm\slsh{s}_\parallel\ga^{\parallel}\slsh{t}_\parallel}{s_\parallel^2t_\parallel^2\left(r_\parallel^2-2\left| q_fB\right|\right)}\right\}.
\label{matrixelem}
\eea

A second important approximation is to consider that in the denominators of Eq.~(\ref{matrixelem}) the  term $|q_fB|$ is large with respect to the loop momenta, which implies that the calculations will be valid only for low photon energies. In this way, by considering that $2|q_fB|\gg t_\parallel^2,\ s_\parallel^2,\ r_\parallel^2$. After a straightforward calculation, the matrix element can be written as
\bea
\widetilde{\mathcal{M}}&=&-i\dpi^4\delta^{(4)}(q-k-p)\frac{q_fg^2\delta^{cd}e^{f(\pt,\kt)}}{32\pi\dpi^8}\epsilon_\mu(\lambda_p)\epsilon_\nu(\lambda_k)\epsilon_\alpha(\lambda_q)\nn\\
&\times&\left\{\left(\gma_\parallel-\frac{p^\mu_\parallel p^\alpha_\parallel}{p_\parallel^2}\right)h^\nu(a)-\left(\gmn_\parallel-\frac{p^\mu_\parallel p^\nu_\parallel}{p_\parallel^2}\right)h^\alpha(a)+\left(\gmn_\parallel-\frac{k^\mu_\parallel k^\nu_\parallel}{k_\parallel^2}\right)h^\alpha(b)\right.\nn\\
&-&\left.\left(\gan_\parallel-\frac{k^\alpha_\parallel k^\nu_\parallel}{k_\parallel^2}\right)h^\mu(b)+\left(\gan_\parallel-\frac{q^\alpha_\parallel q^\nu_\parallel}{q_\parallel^2}\right)h^\mu(c)-\left(\gma_\parallel-\frac{q^\mu_\parallel q^\alpha_\parallel}{q_\parallel^2}\right)h^\nu(c)\right\},
\label{matrint}
\eea
where 
\begin{subequations}
\bea
h^{\mu}(a)=-\frac{i}{\pi}\epsilon_{ij}a^ig^{j\mu}_\perp, 
\eea
\bea
a_i=p_i + 2k_i + i\epsilon_{im}p_m,
\eea
\bea
b_i=2p_i + k_i - i\epsilon_{im}k_m,
\eea
\bea
c_i=k_i - p_i + i\epsilon_{im}(p_m + k_m),
\eea
and
\bea
f\left(\pt,\kt\right)=\frac{1}{8|q_fB|}\left[p_m-k_m+i\epsilon_{mj}(p_j+k_j)\right]^2-\frac{1}{2|q_fB|}\left(p_m^2+k_m^2+2i\epsilon_{jm}p_mk_j\right).\nn\\
\eea
\label{habc_def}
\end{subequations}

For details see Appendix~\ref{ApM1Calculation}.

To obtain the invariant photon momentum distribution, the squared of Eq.~(\ref{matrint}) is required. Here, the three lightest flavors are considered. Moreover, in view of the tensor structure for the matrix elements, only a longitudinal polarization for the gluons and the photon needs to be evaluated. Thus, the process probability is given by
\bea
\sum_{\text{pol},f}|\mathcal{\widetilde{M}}|^2=\dpi^4\delta^{(4)}(q-k-p)\mathcal{V}\Delta\tau\sum_{\text{pol},f}|\mathcal{M}|^2,
\label{Msquareddiractdeltafac}
\eea
where
\bea
\sum_{\text{pol},f}|\mathcal{M}|^2=\frac{2\alpha_{\text{em}}\alpha_{\text{s}}^2q_\perp^2}{\pi\omega_q^2}\sum_{f}q_f^2\left(2\omega_p^2+\omega_k^2+\omega_p\omega_k\right)\exp\left\{-\frac{q_\perp^2}{\eB\omega_q^2}\left(\omega_p^2+\omega_k^2+\omega_p\omega_k\right)\right\}.\nn\\
\label{Msquared}
\eea

The factor $\mathcal{V}\Delta\tau$ represents the space-time volume where the reaction takes place, i.e., the overlap spatial region ${\mathcal{V}}(t)$ at time $t$ and the time interval $\Delta\tau$ where the magnetic field can be taken as having a constant intensity $B(t)$. The full calculation is presented in Appendix~\ref{Ap_M2_calculation_exp}.

In order to write Eq.~(\ref{Msquared}) which only depends on the photon momentum $q_\perp^2$, the energy-momentum conservation for massless gluons and photons was implemented without considering the dispersive properties of the magnetized medium. A simple calculation of such conserved quantity implies that the gluons and the photon have parallel spatial momenta. Explicitly, this condition is taken as
\bea
   p^\mu&=&\omega_p(1,\hat{p})
   =\left(\omega_p/\omega_q\right)q^\mu,\nonumber\\
   k^\mu&=&\omega_k(1,\hat{k})
   =\left(\omega_k/\omega_q\right)q^\mu,
\label{energymomentumconservation}
\eea
which can be relaxed if the magnetic mass acquired by the gluons is computed. This is the main motivation for the study of the Gluon Polarization Tensor presented in Chapter~\ref{Chap:Gluon_Pol_Tensor}. 

\subsection{Photon invariant momentum distribution and elliptic flow
}

The magnetic invariant photon momentum distribution is easily computed from the probability amplitude above discussed. Explicitly, it is given by the expression:
\bea
\omega_q\frac{dN^{\mbox{\tiny{mag}}}}{d^3q}=\frac{\chi{\mathcal{V}}\tau_s}{2(2\pi)^3}
\int\frac{d^3p}{\dpi^32\omega_p}\int\frac{d^3k}{\dpi^32\omega_k}
n(\omega_p)n(\omega_k)\dpi^4\delta^{(4)}\left(q-k-p\right)\sum_{\mbox{\small{pol}},f}|{\mathcal{M}}|^2,\nn\\
\label{invdist}
\eea
where $n(\omega)$ represent the distribution of gluons obtained from the shattered {\it glasma}, and $\chi$ is a factor to account for the fact that the overlap region in a semicentral collision is not the full nuclear volume. At first approximation, the gluon distribution can be modeled as a Bose-Einstein-like number occupation, so that temperature is replaced by the saturation energy scale $\Lambda_s$~\cite{mclerran2014glasma}:
\bea
n(\omega)=\frac{\kappa}{e^{\omega/\Lambda_s}-1}.
\eea

Therefore, Eq.~(\ref{invdist}) gives (see Appendix~\ref{Nmag_calc_exp_1}):
\bea
\frac{1}{2\pi\omega_q}\frac{dN^{\text{mag}}}{d\omega_q}&=&\chi\mathcal{V}\Delta\tau\frac{\alpha_{\text{em}}\alpha_\text{s}^2\pi}{2\dpi^6\omega_q}\sum_f
q_f^2 \int_0^{\omega_q} d\omega_p\left(2\omega_p^2+\omega_q^2-\omega_p\omega_q\right)e^{-g_f(\omega_p,\omega_q)}\nn\\
&\times&\Big{\{}I_0\left[g_f(\omega_p,\omega_q)\right]-I_1\left[g_f(\omega_p,\omega_q)\right]\Big{\}}n(\wwp)n(|\wq-\wwp|),
\label{yieldexpl}
\eea
where $I_0$, $I_1$ are the modified Bessel function of the first kind, and
\bea
g_f(\omega_p,\omega_q)=\frac{\omega_p^2+\omega_q^2-\omega_p\omega_q}{2\eB}.
\label{funcg}
\eea
\begin{figure}[H]
    \centering
    \includegraphics[scale=0.83]{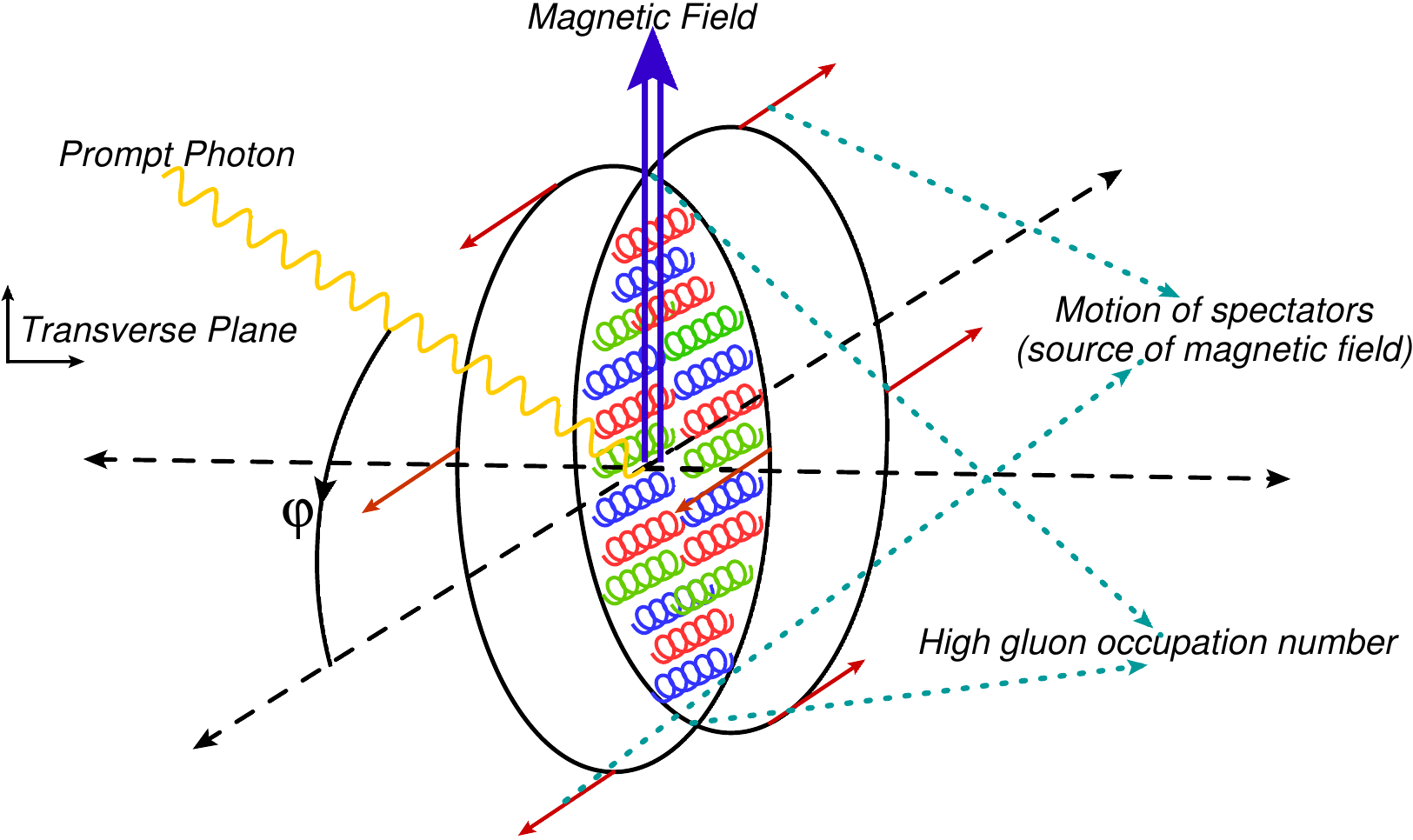}
    \caption{Geometry of the heavy-ion collision: The magnetic field lies on the transverse plane and the photon momentum $q_\perp$ lies in the reaction plane.}
    \label{Fig:Photon_production_geometry}
\end{figure}

The elliptic flow coefficient $v_2$ is obtained from the Fourier decomposition of Eq.~(\ref{yieldexpl}). It is necessary to point out that the magnitude of the photon momentum $q_\perp$, which is transverse to the magnetic field is defined with the geometry presented in Fig.~\ref{Fig:Photon_production_geometry}. Thus, azimuthal distribution with respect to the reaction plane is
\bea
   \frac{dN^{\mbox{\tiny{mag}}}}{d\varphi}=\frac{N^{\mbox{\tiny{mag}}}}{2\pi}
   \left[1+\sum_{n=1}^\infty 2v_n(\omega_q)\cos(n\varphi)\right],
\label{Fourier}
\eea
where the total number of photons, $N^{\mbox{\tiny{mag}}}$, is given by
\bea
N^{\mbox{\tiny{mag}}}=\int d^3q\frac{dN^{\mbox{\tiny{mag}}}}{d^3q}.
\eea

The calculation of $v_2$ for a central rapidity can be found in Appendix~\ref{v2_calc_exp_1} and its final expression is
\bea
v_2^{\text{mag}}(\omega_q)&=&\chi\frac{\alpha_{\text{em}}\alpha_\text{s}^2\pi\mathcal{V}\Delta\tau}{2\dpi^5 N^{\mbox{\tiny{mag}}} }\sum_f
q_f^2\int_0^{\omega_q} d\omega_q^\prime \int_0^{{\omega_q}^\prime} d\omega_p 
\left(2\omega_p^2+{\omega^\prime_q}^2-\omega_p\omega_q^\prime \right)e^{-g_f(\omega_p,\omega_q^\prime)}\nn\\
&\times&
\left\{I_0\left[g_f(\omega_p,\omega_q)\right]-\left[1+\frac{1}{g_f(\omega_p,\omega_q)}\right]I_1\left[g_f(\omega_p,\omega_q)\right]\right\}n(\wwp)n(|\wq^\prime-\wwp|).\nn\\
\label{v2expl}
\eea
\begin{figure}
    \centering
   \includegraphics[scale=0.377]{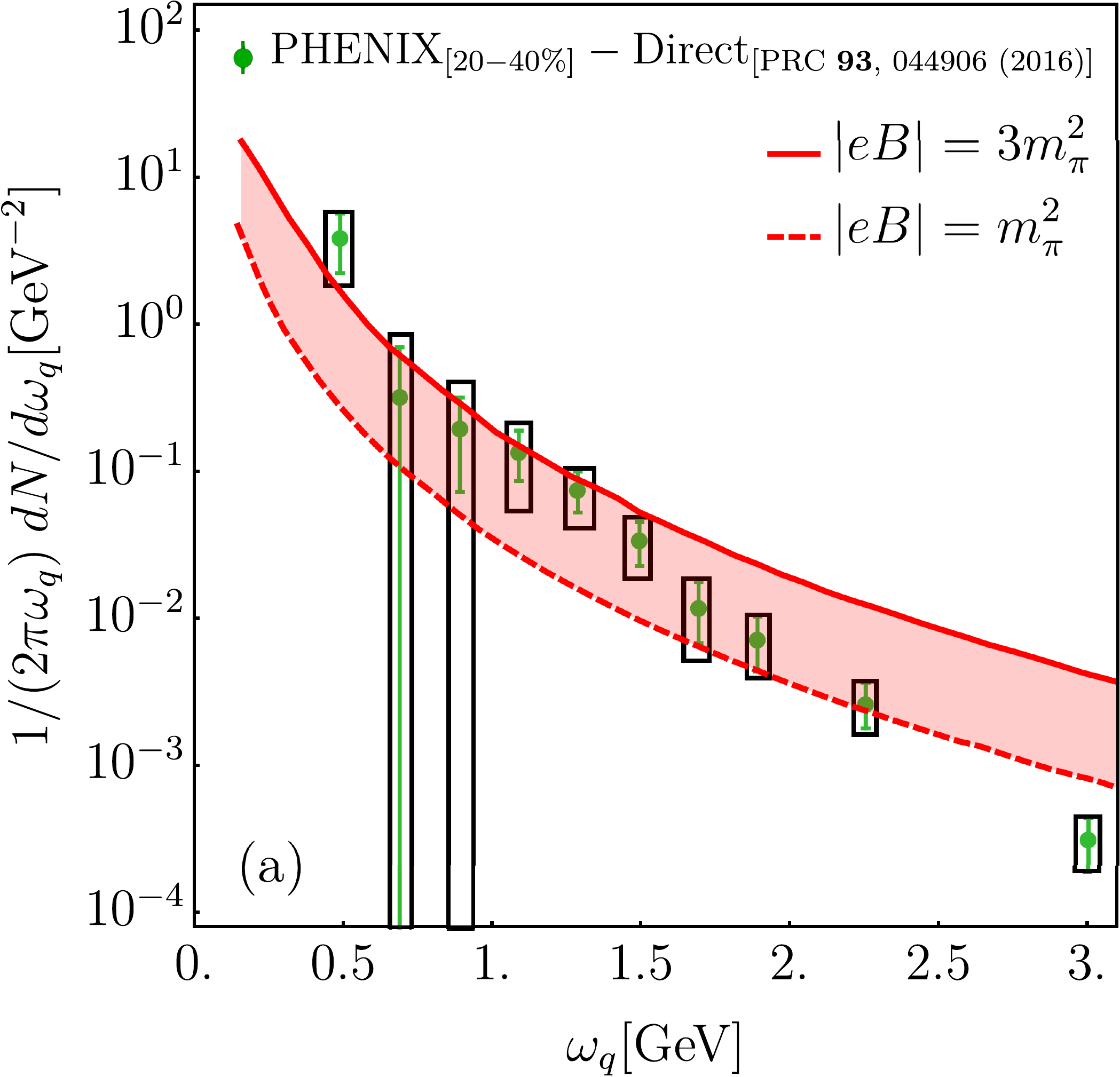}\hspace{0.2cm}\includegraphics[scale=0.37]{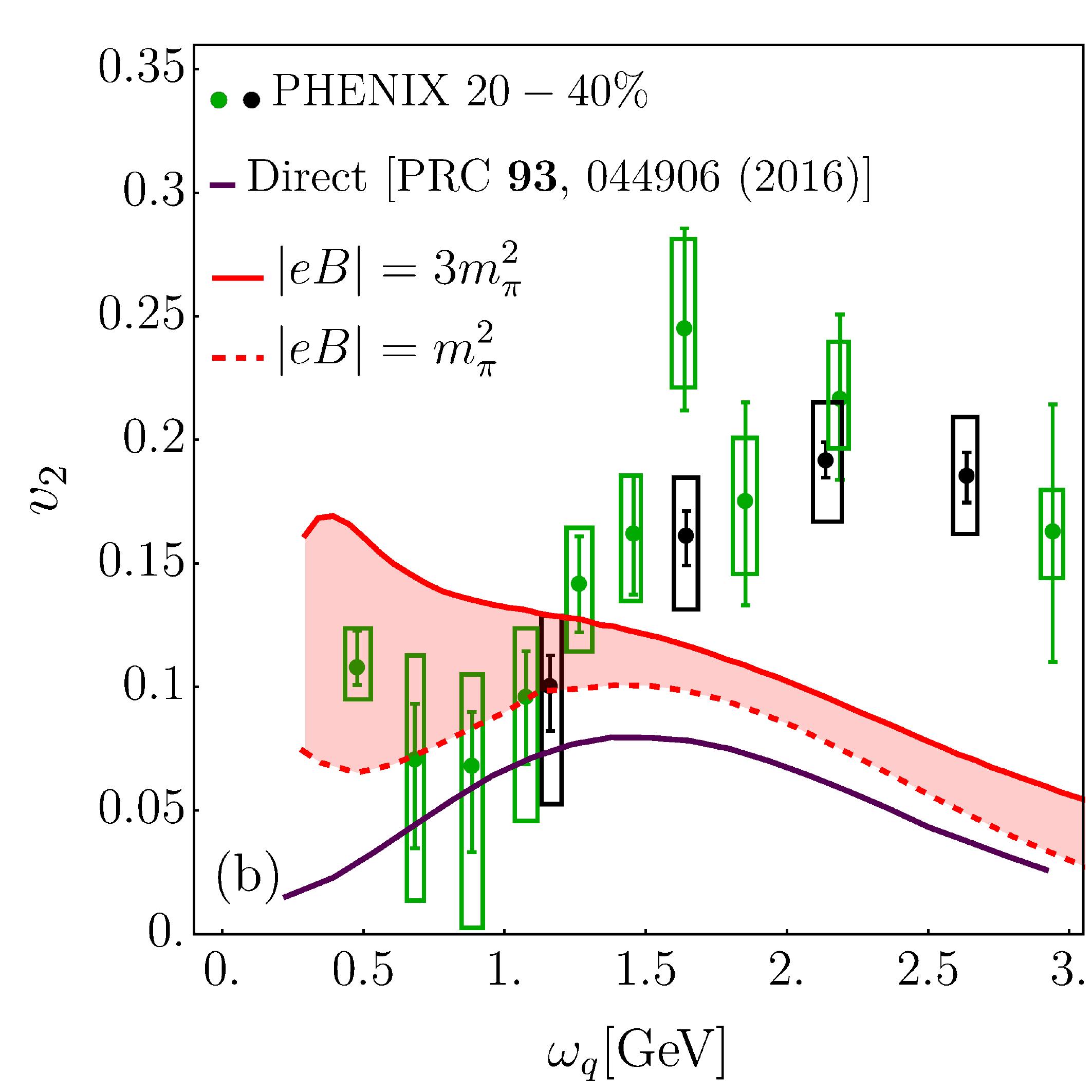}
    \caption{(a) Photon invariant momentum distribution obtained from Eq.~(\ref{yieldexpl}) compared with the difference between the PHENIX data~\cite{RHIC3} and direct photons from Ref.~\cite{hydro-photons1}. (b) Harmonic coefficient $v_2$ computed from Eq.~(\ref{v2pesado}) compared with PHENIX data~\cite{RHIC4} and the elliptic flow of the direct photons from Ref.~\cite{hydro-photons1}. The green (black) points and errors are experimental measurements with the conversion (calorimeter) method. In both figures, the band shows variations of the parameter $|eB|$ within the indicated ranges.}
    \label{Fig:yield_and_v2_beta_0}
\end{figure}

The results for the photon production and elliptic flow are shown in Fig.~\ref{Fig:yield_and_v2_beta_0}. The plots are built as a function of the photon transverse momentum $q_\perp=\omega_q$, for central
rapidity and the centrality range 20$\%–-40\%$. The value of parameters are: $\alpha_s=0.3$, $\Lambda_s=2$ GeV, $\kappa=3$, $\Delta\tau=1.5$ fm, $\chi=0.8$, $\mathcal{V}=4\pi R^2/3$, with $R=7$ fm (corresponding to the Au nuclear radius), and the external magnetic field intensity is in the range $m_\pi^2\leq |eB|\leq 3m_\pi^2$.

In Fig.~\ref{Fig:yield_and_v2_beta_0}(a) the photon invariant momentum distribution obtained from Eq.~(\ref{yieldexpl}) is compared with the difference between the PHENIX and the direct photons of Ref.~\cite{hydro-photons1}, therefore, the magnetic contribution has to be understood as an excess (missed photons) over calculations without magnetic field effects. As can be seen, the magnetic contribution helps to better describe the experimental data for intermediate and low energies, but for higher values of the photon transverse moment. The latter can be attributed to a combination of the gluon distribution which does not contain a power like tail which is known to better describe the numerical solutions for this kind of distribution~\cite{PhysRevLett.87.192302}, and to the low momentum approximation $2|q_fB|\gg t_\parallel^2,\ s_\parallel^2,\ r_\parallel^2$.

On the other hand, Fig.~\ref{Fig:yield_and_v2_beta_0} (b) displays the harmonic flow coefficient $v_2$ for the same parameter configuration of the photon yield. To compare with experimental data, having as a baseline a state-of-the-art calculation accounting for many of the well-described sources of photons, it was taken into account that the $v_2$ is a weighted average given by
\bea
v_2(\omega_q) &=& \frac{
\frac{dN^{\mbox{\tiny{mag}}}}{d\omega_q}(\omega_q)
v_2^{\text{mag}}(\omega_q)+\frac{dN^{\mbox{\tiny{direct}}}}{d\omega_q}(\omega_q)
v_2^{\text{direct}}(\omega_q)
}{\frac{dN^{\mbox{\tiny{mag}}}}{d\omega_q}(\omega_q)
+\frac{dN^{\mbox{\tiny{direct}}}}{d\omega_q}(\omega_q)},
\label{v2pesado}
\eea
where $N^{\mbox{\tiny{direct}}}(\omega_q)$ and $v_2^{\text{mag}}(\omega_q)$ are obtained from Ref.~\cite{hydro-photons1}.

For low photon momenta, the magnetic contribution improves the harmonic flow description, so that, in comparison with the pure direct photon formalism (purple line), the addition of magnetic photons lies in the error bars reported by PHENIX. Moreover, for a magnetic field $|eB|\sim m_\pi^2$, there is a desirable functional behavior that reproduces the overall data distribution. The good fit together with the functional form can be explained from the fact that the existence of a magnetic field breaks the translational symmetry, thus, a preferential direction to elliptic flux is created. It cannot be expected that for higher photon energies the predictions are in agreement with the experiment, for the same reason discussed in the yield description: the simplified denominators of Eq.~(\ref{matrixelem}) and the absence of a tail for the gluon distribution preclude the possibility to get a correct description in that regime. 

\begin{figure}[t]
    \centering
   \includegraphics[scale=0.377]{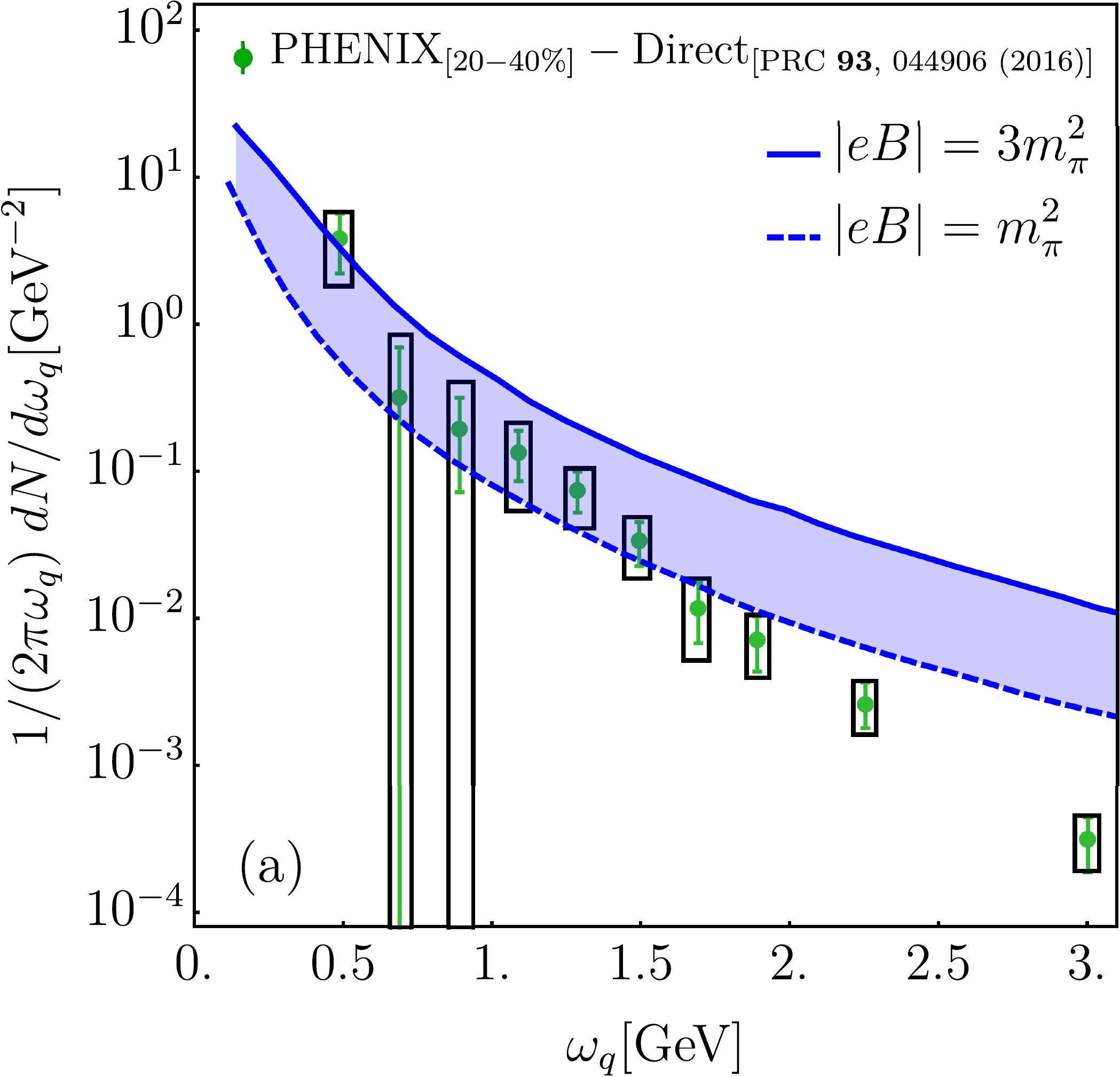}\hspace{0.2cm}\includegraphics[scale=0.37]{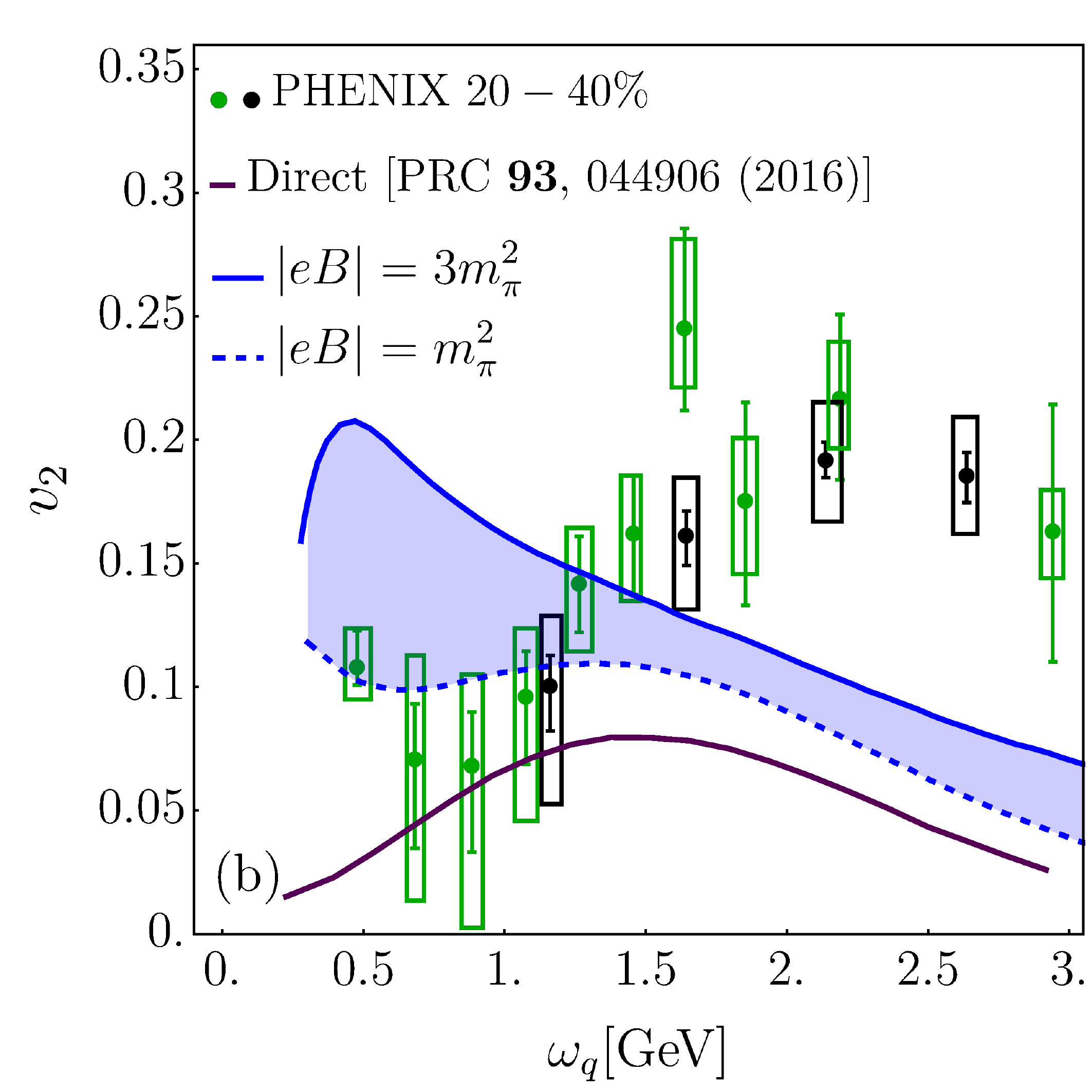}
    \caption{Impact of the flow velocity with $\beta=0.25$ for (a) Photon invariant momentum distribution obtained from Eq.~(\ref{yieldexpl}) compared with the difference between the PHENIX data~\cite{RHIC3} and direct photons from Ref.~\cite{hydro-photons1}. (b) Harmonic coefficient $v_2$ computed from Eq.~(\ref{v2pesado}) compared with PHENIX data~\cite{RHIC4} and the elliptic flow of the direct photons from Ref.~\cite{hydro-photons1}. The green (black) points and errors are measures with the conversion (calorimeter) method. In both figures, the band shows variations of the parameter $|eB|$ within the indicated ranges.}
    \label{Fig:yield_and_v2_beta_025}
\end{figure}

In a heavy-ion collision, the particle flow can be attributed to fluctuations in the internal fire-ball velocities~\cite{mclerran2014glasma,PhysRevC.97.034915,PhysRevC.86.024911,PhysRevLett.122.122302}. The sudden change of pressure in different directions between the interaction
region and vacuum when the glasma is shattered is an important factor to take into account, therefore, in order to study the impact of its expansive effects, a flow velocity factor is introduced in the expression through
\begin{subequations}
\bea
\omega_{p,k}\rightarrow (p,k)\cdot u,
\eea
with a constant four-velocity given by
\bea
u^\mu=\gamma(1,\beta),\;\gamma=\frac{1}{\sqrt{1-\beta^2}}.
\eea
\end{subequations}

Figure~\ref{Fig:yield_and_v2_beta_025} displays the photon invariant momentum distribution and the harmonic coefficient $v_2$ when the flow velocity of Eq.~(\ref{v2pesado}) is considered with a value of $\beta=0.25$. The inclusion of this velocity shifts the momentum to higher energies and increases the response in the magnetic field dependence for both observables. Hence, lower magnetic fields can generate the same behavior with the inclusion of the velocity flow.  As it was expected, the cases with and without flow cannot ensure the data description for high photon energies because of the simplifications named before. 

It is necessary to point out that the results presented in Figs.~\ref{Fig:yield_and_v2_beta_0} and~\ref{Fig:yield_and_v2_beta_025} are obtained by considering qualitatively three fundamental scenarios in the proposed phenomenology: the magnetic field is constant over all collision time, the space-temporal volume is assumed fixed and the centrality is not computed from first principles. Those assumptions allow having only a rough picture of the gluon fusion mechanism. In the next section, these issues are avoided by simulating computationally the heavy-ion collision, which gives the evolution in time and space of the magnetic field as well as the correct interaction volume and corresponding centrality. 

Nevertheless, the prompt photon production by gluon fusion mediated by a magnetized medium seems to be a reasonable source of photon radiation to explain the discrepancy between the current calculations and the experimental measurements of the invariant momentum distribution. Certainly, to have a better picture and impact of the proposed mechanism, in particular, the $v_2$ tail at higher momenta, the energy scale restricted to a high magnetic field has to be relaxed, which implies to include the contribution of several Landau Levels. This is a work in process.

\section{Centrality Dependence of the Photon Production by Gluon Fusion and Gluon Splitting in a Magnetized Medium}~\label{sec:gluon_splitting}

In this section, the centrality dependence of the photon production by the process $gg\rightarrow\gamma$ (gluon fusion) as well as the gluon splitting, i.e., $g\rightarrow g\gamma$ is presented. The latter comes from the fact that at order $\alpha_{\text{em}}\alpha_s^2$ this process has to be taken into account.

As was commented on the previous section, the results obtained from the gluon fusion in a magnetized medium can only give a rough description for the observables of interest, given that the centrality is modeled by a phenomenological parameter and the magnetic field evolution together with the space-temporal interaction volume were considered constant. Here, an estimate of the upper limit for the time evolution of the magnetic field intensity and the volume produced by the collision participants and spectators can be obtained,  and therefore, a more realistic scenario can be achieved. 

\begin{figure}[t]
\begin{center}
\quad\includegraphics[scale=.35]{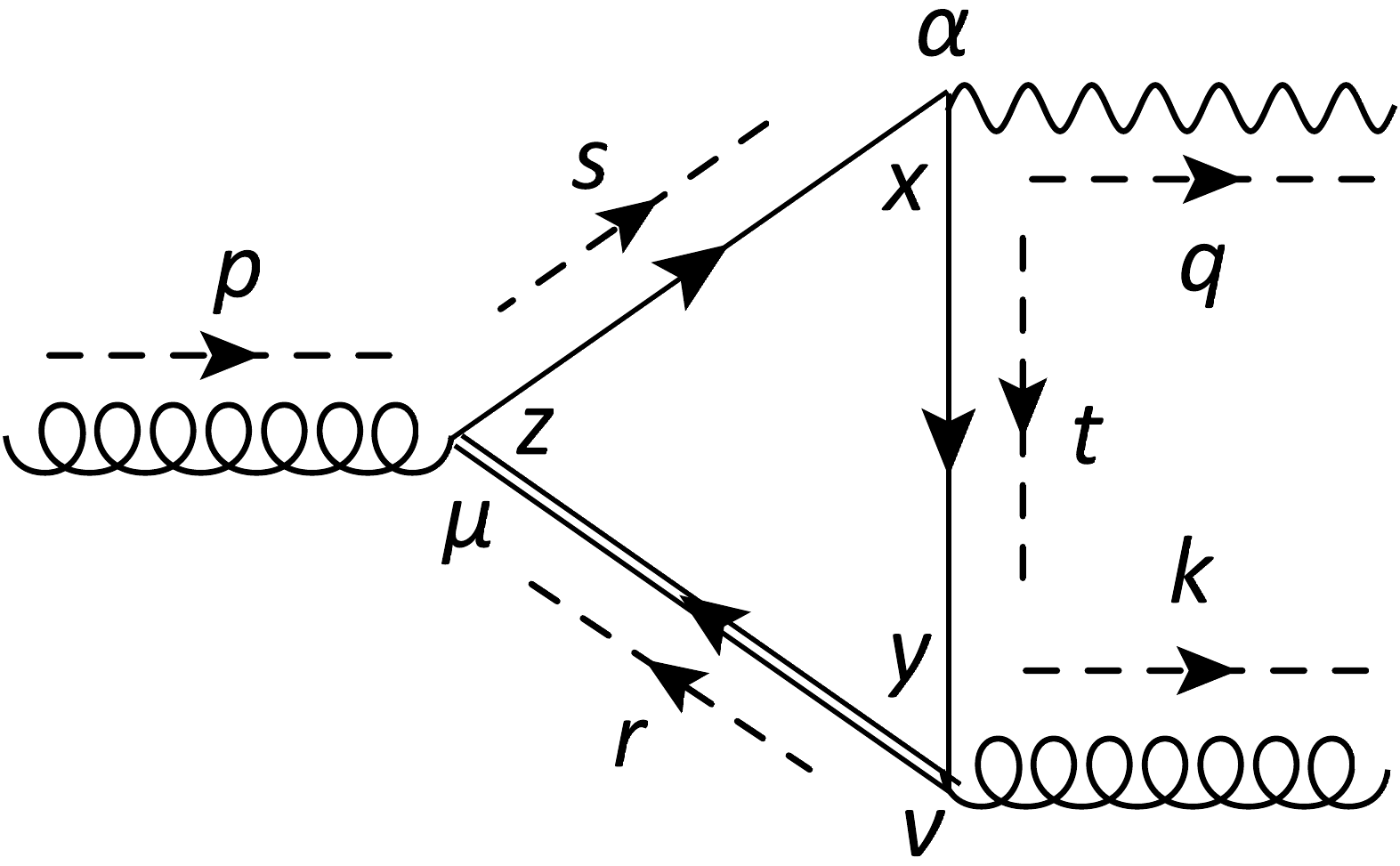}\includegraphics[scale=.35]{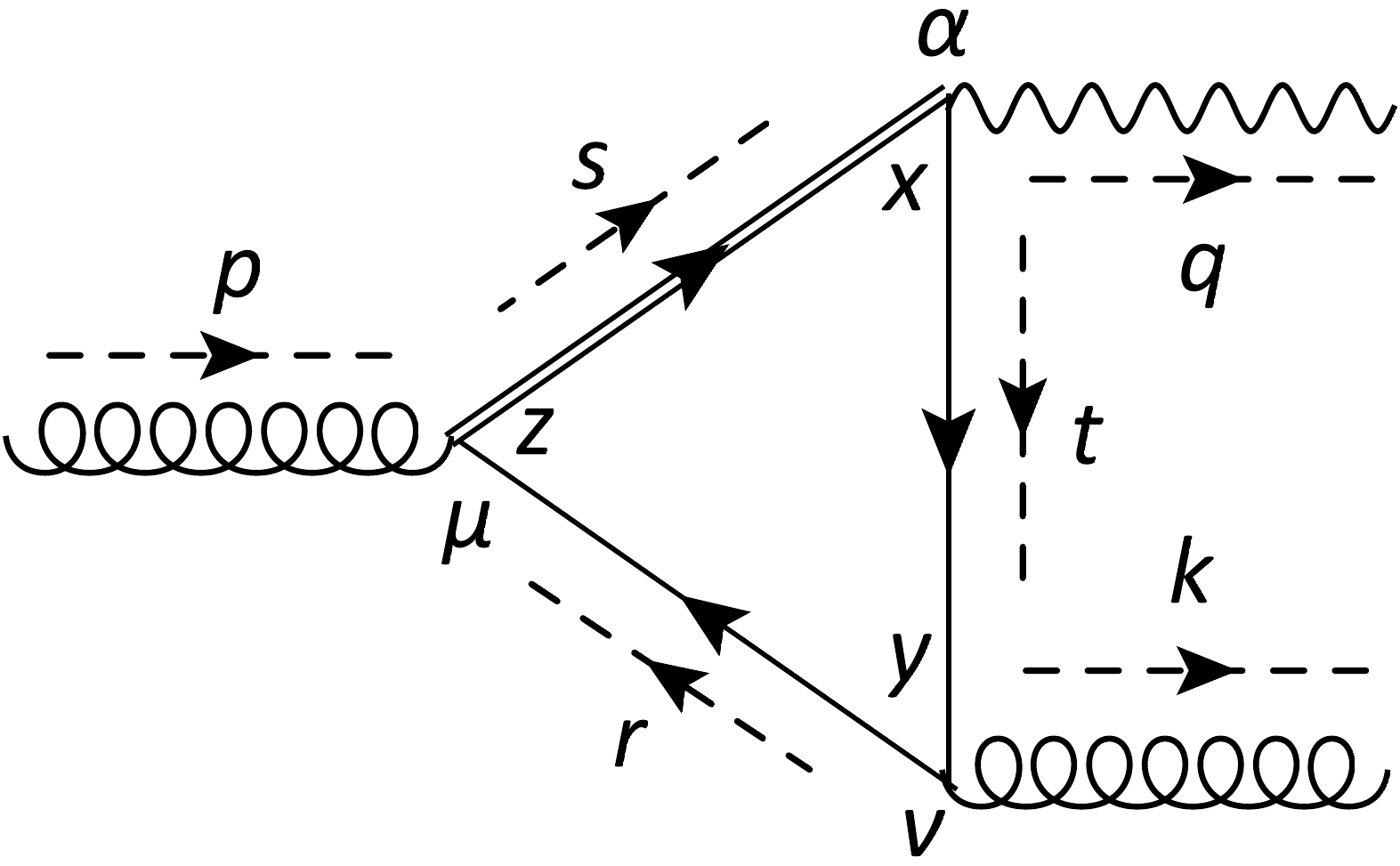}\includegraphics[scale=.35]{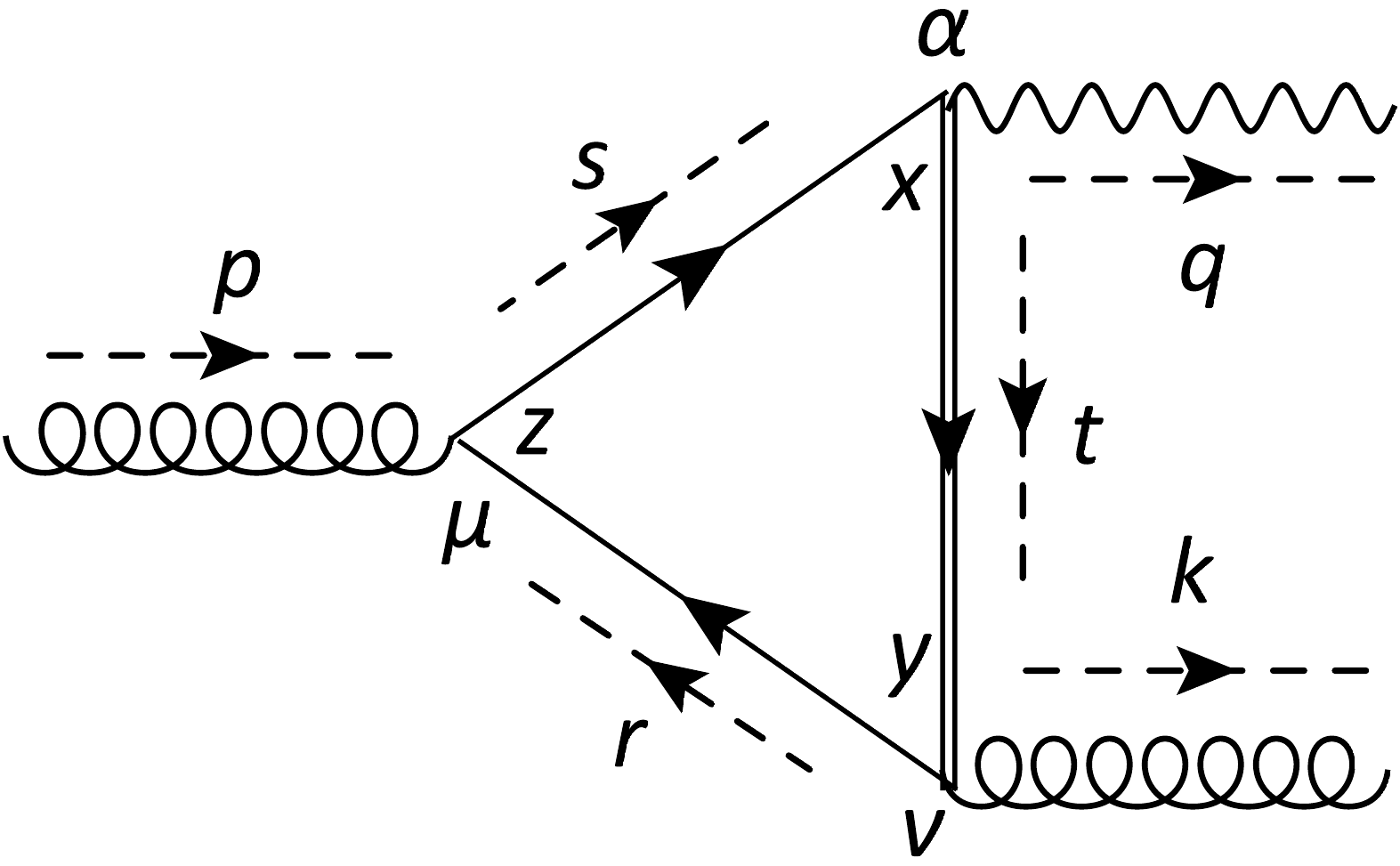}\\\vspace{0.3cm}
\includegraphics[scale=.35]{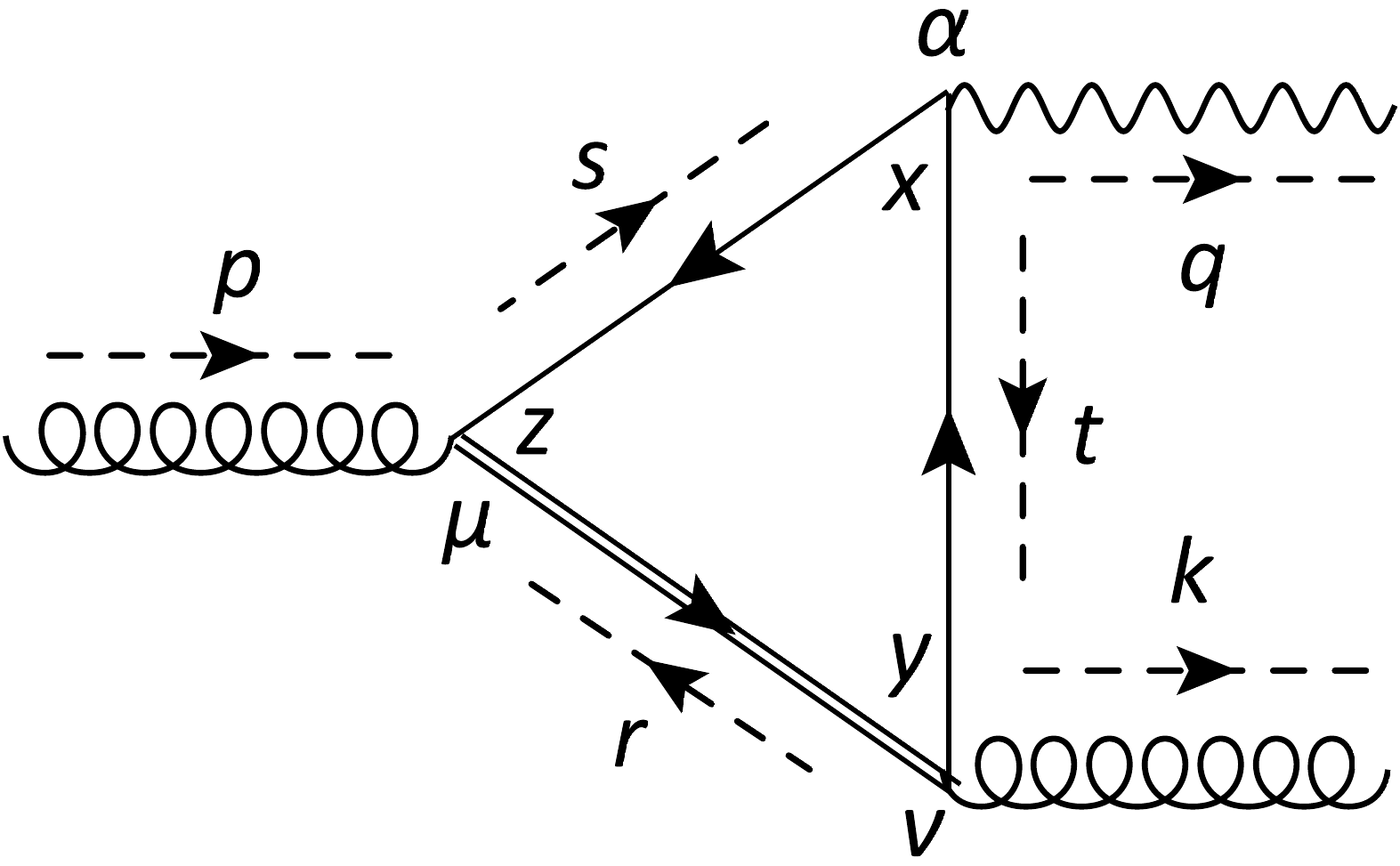}\includegraphics[scale=.35]{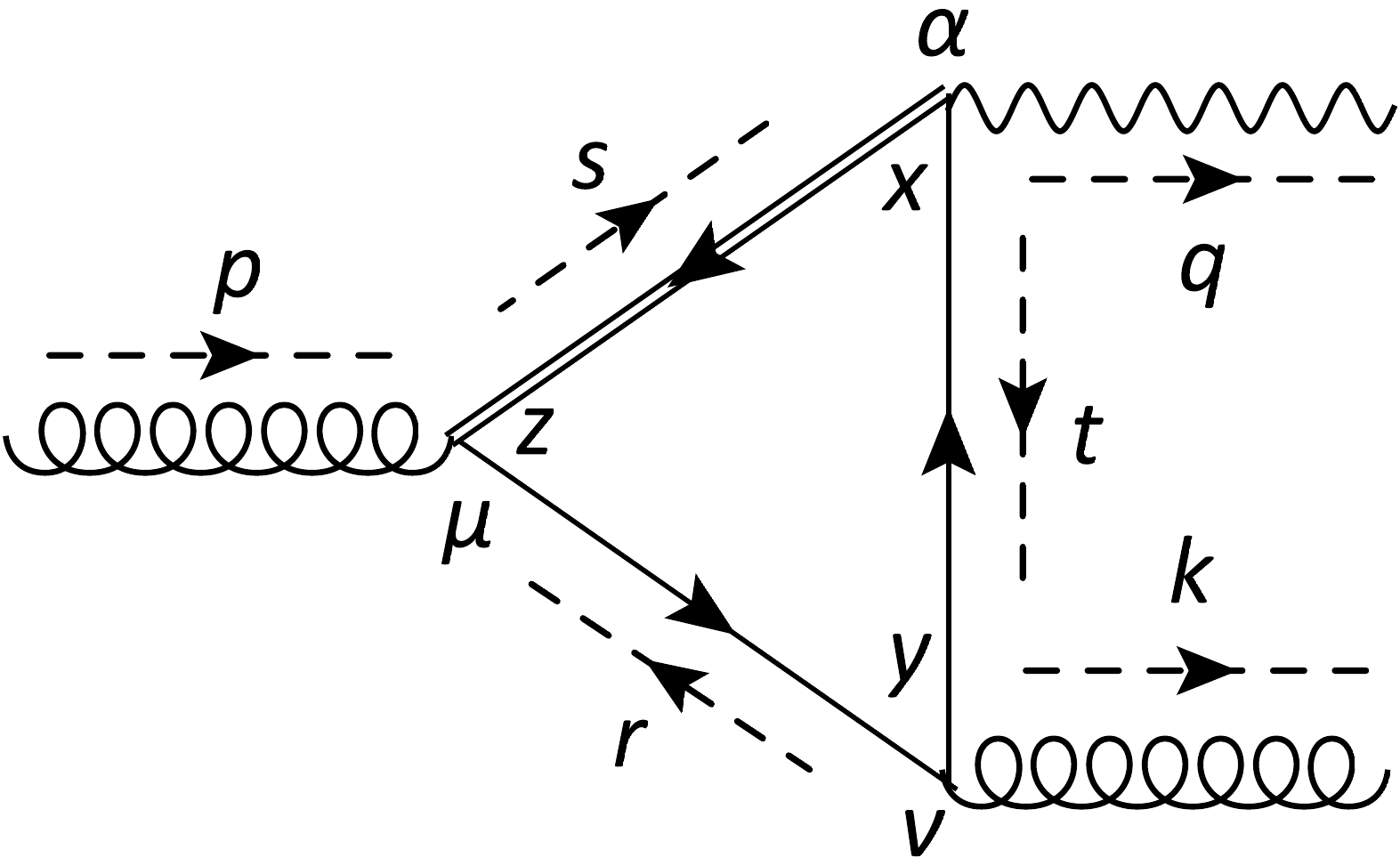}\includegraphics[scale=.35]{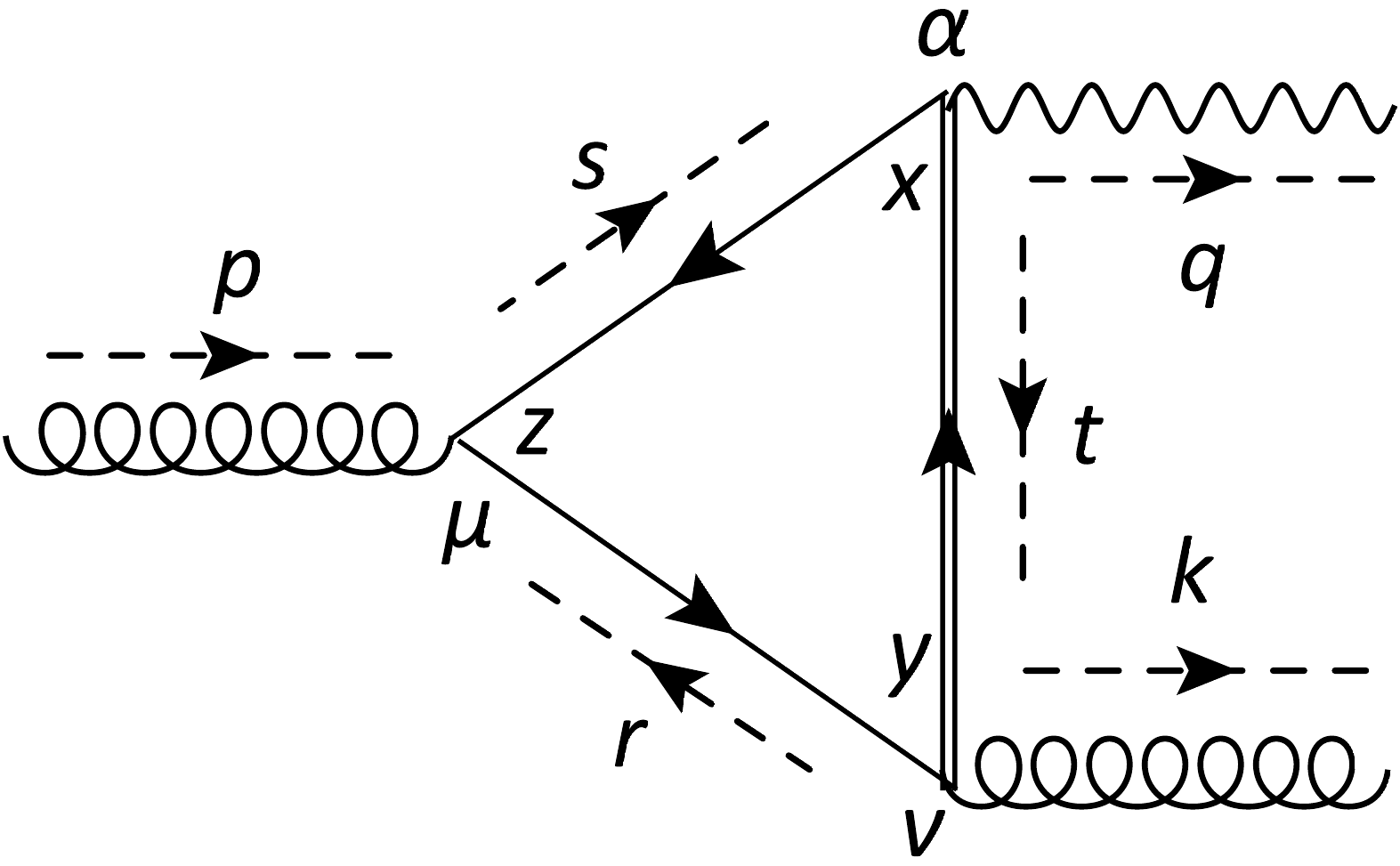}\quad
\caption{Feynman diagrams contributing to the process $g\rightarrow g\gamma$ in the presence of a magnetic field. The double lines represent that quark propagator in the 1LL, whereas the single lines represent the propagator in the LLL. The arrows in the propagators are the charge flow and the arrows on the sides represent the momentum direction.}
\label{Fig:Diag2}
\end{center}
\end{figure}

Figure~\ref{Fig:Diag2} shows the Feynman diagrams wich contributes to the process $g\rightarrow g\gamma$, at the same order in $\eB$ than the process $gg\rightarrow\gamma$, namely, in the fermionic triangle two quarks are in the LLL and one in the 1LL.

\subsection{Photon invariant momentum distribution and elliptic flow from gluon fusion and gluon splitting channels}
%

By a direct application of Feynman rules it is evident that the matrix element $\mathcal{M}_{gg\rightarrow\gamma}$ is related to $\mathcal{M}_{g\rightarrow g\gamma}$ by the crossing symmetry
\bea
\mathcal{M}_{g\rightarrow g\gamma}(p,k,q)=\mathcal{M}_{gg\rightarrow\gamma}(p,-k,q),
\label{Mgg_gamma=Mg_ggamma}
\eea
where $\mathcal{M}_{gg\rightarrow\gamma}$ is given by 
Eq.~(\ref{matrint}).

To find the total photon invariant momentum distribution, both processes have to be squared, and after summing and averaging over the initial state and summing over the final state particle polarization and color, they have to be added. To do this, note that there aren't any diagrams which indicate interference between the pair of channels, therefore, they can be summed incoherently, obtaining
\bea
{\overline{\sum_{\mathrm{\small{c, p,f}}}}}|\widetilde{\mathcal{M}}|^2&=&\mathcal{V}\Delta\tau\dpi^4{\overline{\sum_{\mathrm{\small{c,p,f}}}}}
\left[\delta^{(4)}\left(q-k-p\right)|\mathcal{M}_
{gg\rightarrow\gamma}|^2+ \delta^{(4)}\left(q+k-p\right)|{\mathcal{M}}_
{g\rightarrow g\gamma}|^2\right],\nn\\
\label{sumpol1}
\eea
%


The energy-momentum conservation gives the same condition for the particle momenta of Eq.~(\ref{energymomentumconservation}), thus, the obervables can be computed in terms of the energy components $(\omega_p,\omega_k,\omega_q)$ so that the three particles have parallel spatial-momentum. The invariant photon momentum distribution is given by
\bea
\omega_q\frac{dN^{\mbox{\tiny{mag}}}}{d^3q}&=&\frac{\dpi^4{\mathcal{V}}\Delta \tau}{2(2\pi)^3}
\int\frac{d^3p}{\dpi^32\omega_p}\int\frac{d^3k}{\dpi^32\omega_k}\Bigg{\{}\delta^{(4)}\left(q-k-p\right)n(\omega_k){\overline{\sum_{\mathrm{\small{c,p,f}}}}}|{\mathcal{M}}_
{gg\rightarrow\gamma}|^2 \nonumber \\
&& +\delta^{(4)}\left(q+k-p\right)\left[1+n(\omega_k)\right]{\overline{\sum_{\mathrm{\small{c,p,f}}}}}|{\mathcal{M}}_
{g\rightarrow g\gamma}|^2\Bigg{\}}\,n(\omega_p),\quad
\label{invdist2}
\eea
where the initial state gluon with energy $\omega_k$ comes weighed with an occupation factor $n(\omega_k)$, whereas the final state one comes weighed with an enhanced occupation factor $1+n(\omega_k)$. Explicitly, the last equation reads
\bea
&&\frac{1}{2\pi\omega_q}\frac{dN^{\text{mag}}}{d\omega_q}=\mathcal{V}\Delta\tau\frac{\alpha_{\text{em}}\alpha_\text{s}^2\pi}{2\dpi^6\omega_q}\sum_f
q_f^2
\int_0^{\omega_q} d\omega_p\left(2\omega_p^2+\omega_q^2-\omega_p\omega_q\right)e^{-g_f(\omega_p,\omega_q)}\nn\\
&\times&
\Big{\{}I_0\left[g_f(\omega_p,\omega_q)\right]-I_1\left[g_f(\omega_p,\omega_q)\right]\Big{\}}
\Big{\{}n(|\wq-\wwp|)+\left[1+n(|\wq-\wwp|)\right]\Big{\}}\,n(\wwp),\nn\\
\label{yieldexpl2}
\eea
where $g_f(\omega_p,\omega_q)$ is given in Eq.~(\ref{funcg}). The deduction of Eq.~(\ref{yieldexpl2}) is presented in Appendix~\ref{Mk=M-k}. Moreover, the expression for the total $v_2$ can be found and reads
\bea
&&v_2^{\text{mag}}(\omega_q)=\frac{\alpha_{\text{em}}\alpha_\text{s}^2\pi\mathcal{V}\Delta\tau}{2\dpi^5 N^{\mbox{\tiny{mag}}} }\sum_f
q_f^2\int_0^{\omega_q} d\omega_q^\prime \int_0^{{\omega_q}^\prime} d\omega_p 
\left(2\omega_p^2+{\omega^\prime_q}^2-\omega_p\omega_q^\prime \right)e^{-g_f(\boldsymbol{\omega})}\nn\\
&\times&
\left\{I_0\left[g_f(\boldsymbol{\omega})\right]-\left[1+\frac{1}{g_f(\boldsymbol{\omega})}\right]I_1\left[g_f(\boldsymbol{\omega})\right]\right\}
\left\{n(|\wq^\prime-\wwp|) + \left[1+n(|\wq^\prime -\wwp|)\right]\right\}n(\wwp), \nn \\
\label{v2expl2}
\eea
with $g_f(\boldsymbol{\omega})=g_f(\omega_p,\omega_q^\prime)$.

\subsection{Magnetic Field Strength and Space Time Volume from UrQMD}\label{subsec:Magnetic_field_from_UrQMD}

The UrQMD is a microscopic model that simulates ultra-relativistic heavy-ion collisions in the energy range from 100 MeV to 200 GeV~\cite{bass1998microscopic,bleicher1999relativistic}. The model has integrated a Monte Carlo simulation package for $p+p$, $p+A$ and $A+A$ interactions.

In general terms, the model solves a large set of coupled integro-differential equations for the time-evolution of the phase-space densities $f_i(x,p)$, where $x$ and $p$ represents the classical position and momentum of the $i$-ts particle specie ($N,\Lambda,\Delta$, etc). The density $f_i$ is assumed to satisfy a non-relativistic Boltzmann equation: 
\bea
\frac{df_i(x,p)}{dt}=\frac{\partial p}{\partial t}\frac{\partial f_i(x,p)}{\partial p}+\frac{\partial x}{\partial t}\frac{\partial f_i(x,p)}{\partial x}+\frac{\partial f_i(x,p)}{\partial t}=\textbf{St}f_i(x,p)
\eea
where $\textbf{St}f_i(x,p)$ represents the collision term or particle source which connects the species $i$ and $j$. Such terms contains information of covariant propagation hadrons on classical trajectories, stochastic binary scatterings, color string formation and resonance decays.

For $A+A$ collisions, the model considers binary and ternary interactions by taking into account the real part of the in-medium G-matrix, which includes a non-relativistic density-dependent Skyrmion-type potential given by
\bea
V=\frac{1}{2!}t_1\sum_{i\neq j}\delta\left(\mathbf{x}_i-\mathbf{x}_j\right)+\frac{1}{3!}t_2\sum_{i\neq j\neq k}\delta\left(\mathbf{x}_i-\mathbf{x}_j\right)\delta\left(\mathbf{x}_j-\mathbf{x}_k\right),
\eea
as well as Coulomb and Yukawa interactions for more than fifty baryon species and five meson nonetts (45 mesons). Moreover, the exchange of electric and baryonic charge, strangeness and four momentum is included for baryon-baryon collisions at low energies in the $t$-channel. The meson-baryon and the meson-meson interactions appears when $\sqrt{s}>3$ GeV through fromation and decay of resonances in the $s$-channel, whereas the $t$-channel for meson-baryon and meson-meson are explicitly given above $\sqrt{s}=6$ GeV. The anti-particles are implemented by the charge conjugation which ensures baryon-antibaryion symmetry. 

\begin{figure}[H]
    \centering
    \includegraphics[scale=0.41]{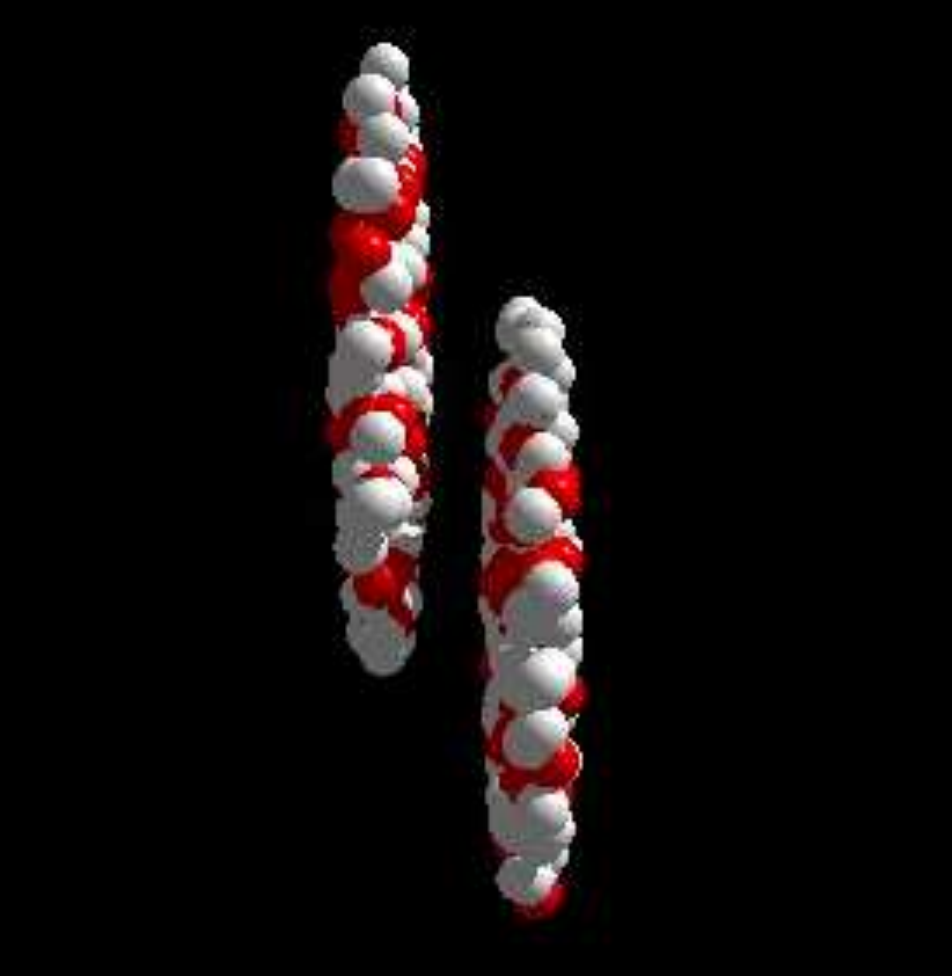}\,\includegraphics[scale=0.41]{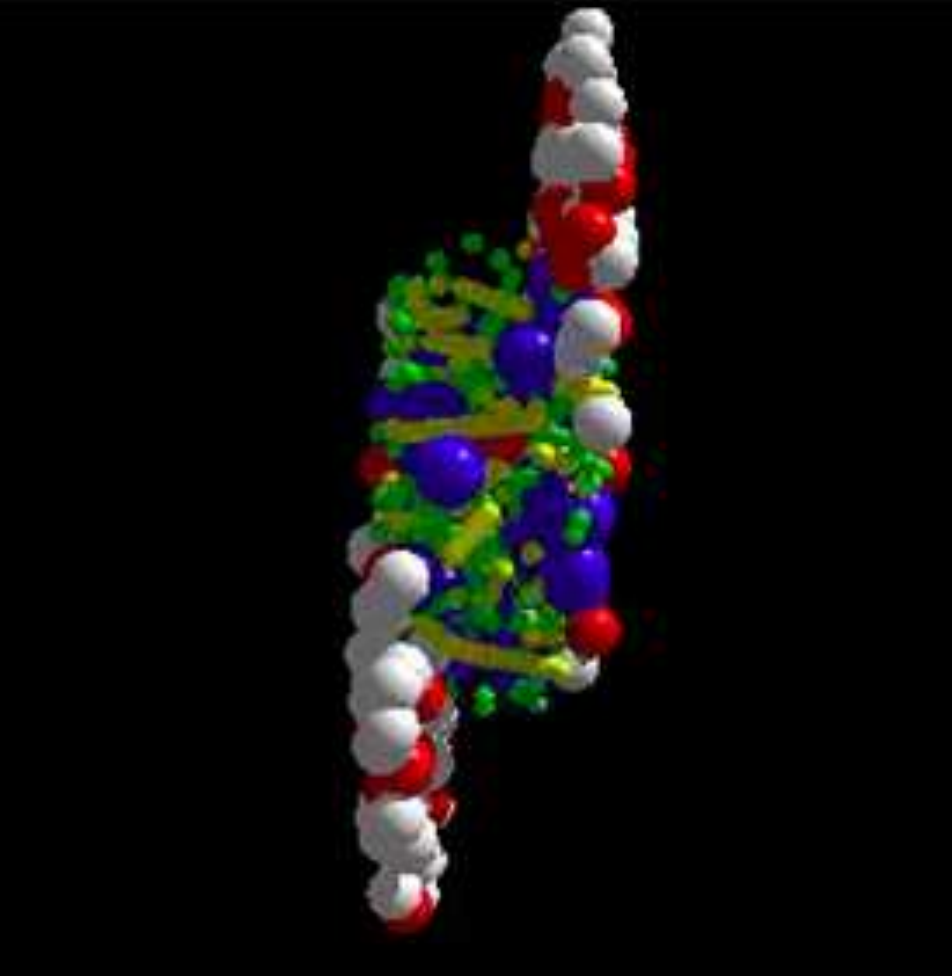}\,\includegraphics[scale=0.41]{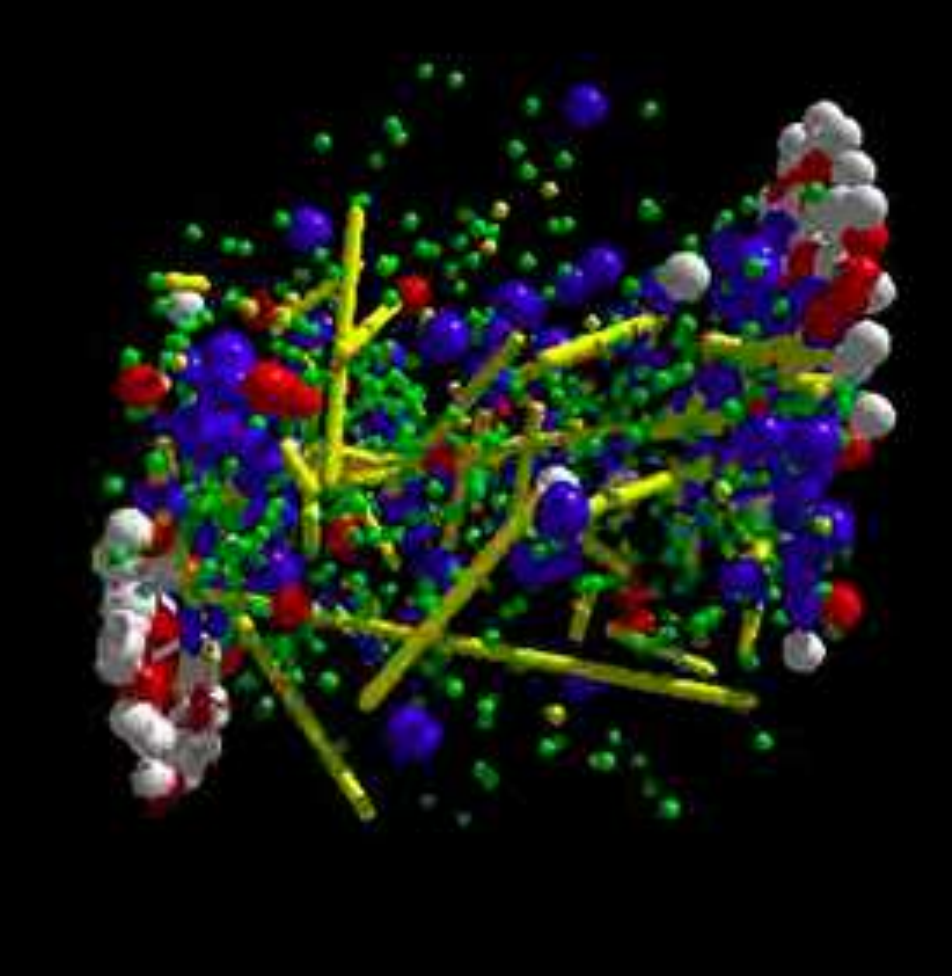}\,\includegraphics[scale=0.41]{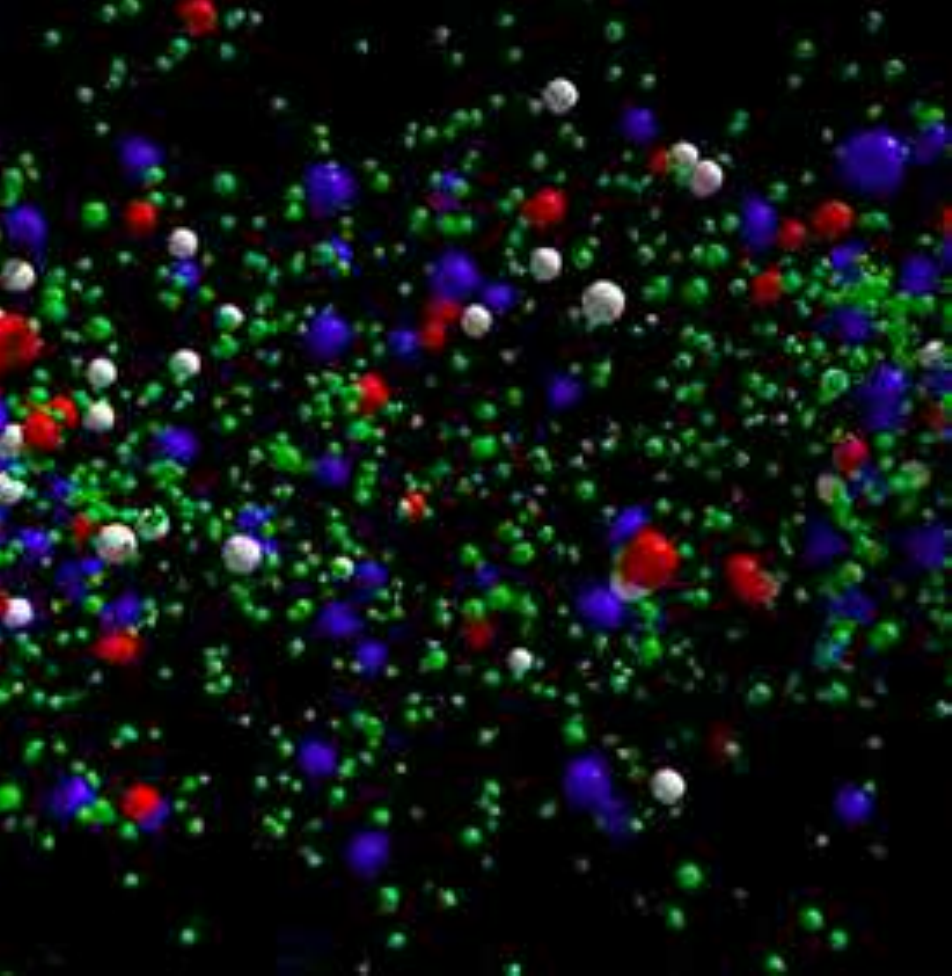}\quad
    \caption{Stages of a UrQMD simulation for a Pb+Pb collision at CERN/SPS energies.}
    \label{FIG:URQMD_Collision}
\end{figure}

The cross-sections enter as a pure geometrical input: if $b$ is the impact parameter of two hadrons, the collision will occurs if $b<\sqrt{\sigma_{\text{tot}}/\pi}$, where $\sigma_{\text{tot}}$ is the total cross-section for both hadrons, which depends on the isospins of colliding particles, their flavor and the center of mass energy. Such total and elastic cross-sections are modeled (above of the resonance energy $p_\text{lab}>2$ GeV/c ) by parameters of the CERN-HERA fit, given by
\bea
\sigma_{\text{tot,el}}(p)=A+Bp^n+C\ln^2 p+D\ln p,
\eea
where $p$ is the laboratory momentum~\cite{PhysRevD.54.1}. If there are no experimental data to give a reasonable fit for the cross-sections, the additive quark model and detailed balance arguments are used to extrapolate such functions. Figure~\ref{FIG:URQMD_Collision} shows the several stages of a heavy-ion collision simulated with UrQMD. As can it noticed, the Lorentz contraction of both nuclei is taken into account, as well as the color sting formation, centrality (nucleons which do not collide), and the hadronic final states. 

For the present work, the UrQMD was used to compute the space-time evolution of the interaction volume and the magnetic field strength by varying the centrality class and the colliding species. This procedure allows performing a systematic study of the photon excess coming from gluon fusion and splitting with a better estimate of the space-time volume $\mathcal{V}\Delta\tau$ and the time profile for the magnetic field which consist to an improvement to the results presented in the last section.

First, the magnetic field is computed event-by-event at the point $\mathbf{x}$ at time $t$ by using the Lienard-Wiechert potential generated by non-accelerated charges moving along the beam direction, namely
\bea
e\mathbf{B}(\mathbf{x},t) = \alpha_{em} \sum_{j}  
\frac{(1-v_j^{2})~\mathbf{v}_j\times \mathbf{R}_j}{ R_j^{3}\left[1-\frac{(\mathbf{v}_i\times\mathbf{R}_j)^{2}}{R_j^{2}}\right]^{3/2}},
\label{magneticfield}
\eea
where $\mathbf{R}_j = \mathbf{x} - \mathbf{x}_j(t)$, $\mathbf{x}_j(t)$ is the position of the $j$-th charge moving with velocity $\mathbf{v}_j$, $R_j$ is the magnitude of $\mathbf{R}_j$ and the sum runs over charged particles in each event~\cite{landau1971classical}. The charges which produce the magnetic field are classified into two categories: participants and spectators. The former is defined as the number of nucleons that undergo at least one collision, and therefore,  nucleons that do not participate in any collision and travel undeflected close to the beam direction, are defined as spectators. Furthermore, the centrality is defined as the fraction of the geometrical cross-section that overlaps for a given impact parameter $b$. This quantity is proportional to $b^2/R_A^2$, where $R_A$ is the radii of the nucleus that collide.

\begin{figure}
    \centering
    \includegraphics[scale=0.34]{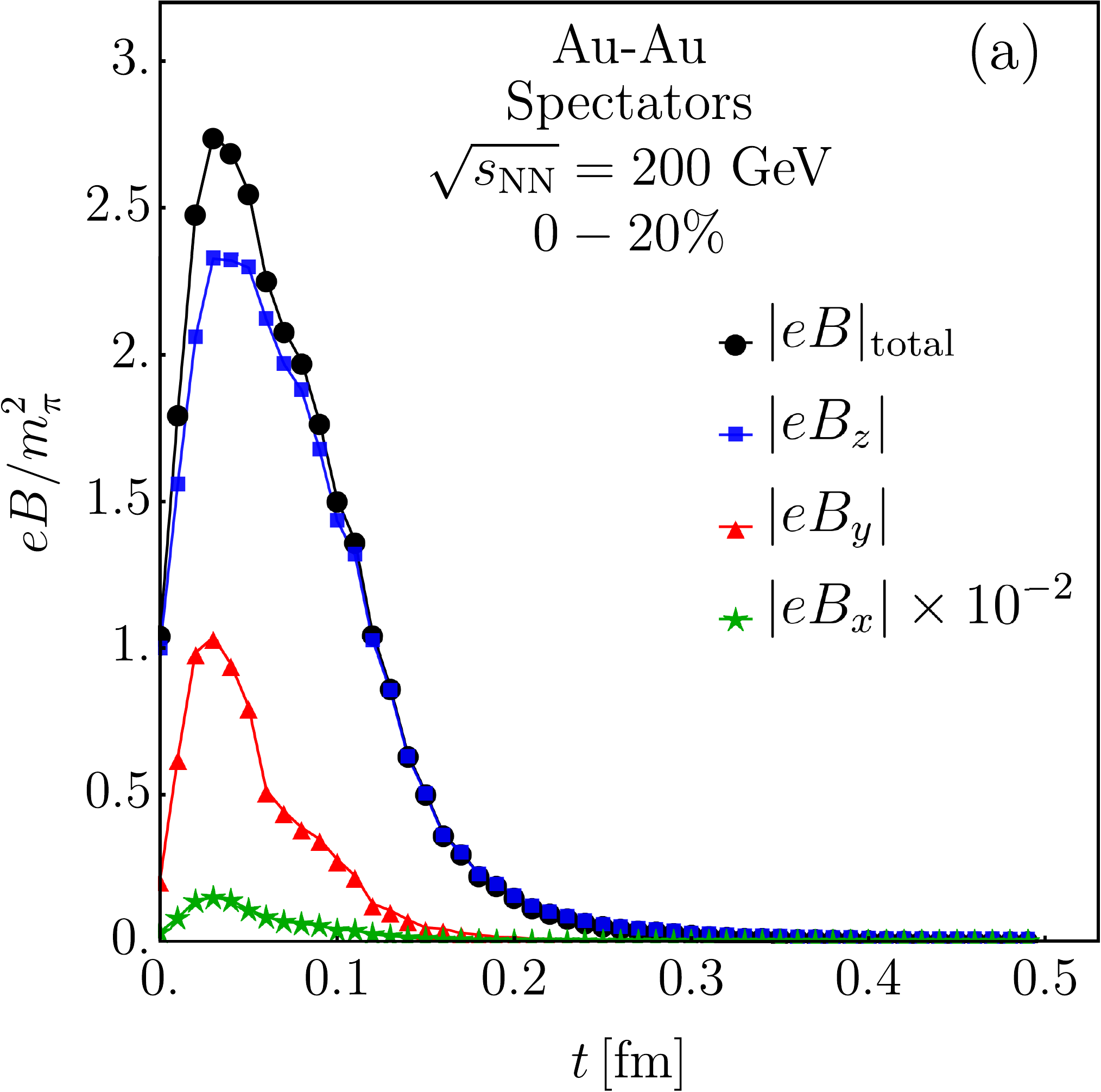}\hspace{2cm}\includegraphics[scale=0.34]{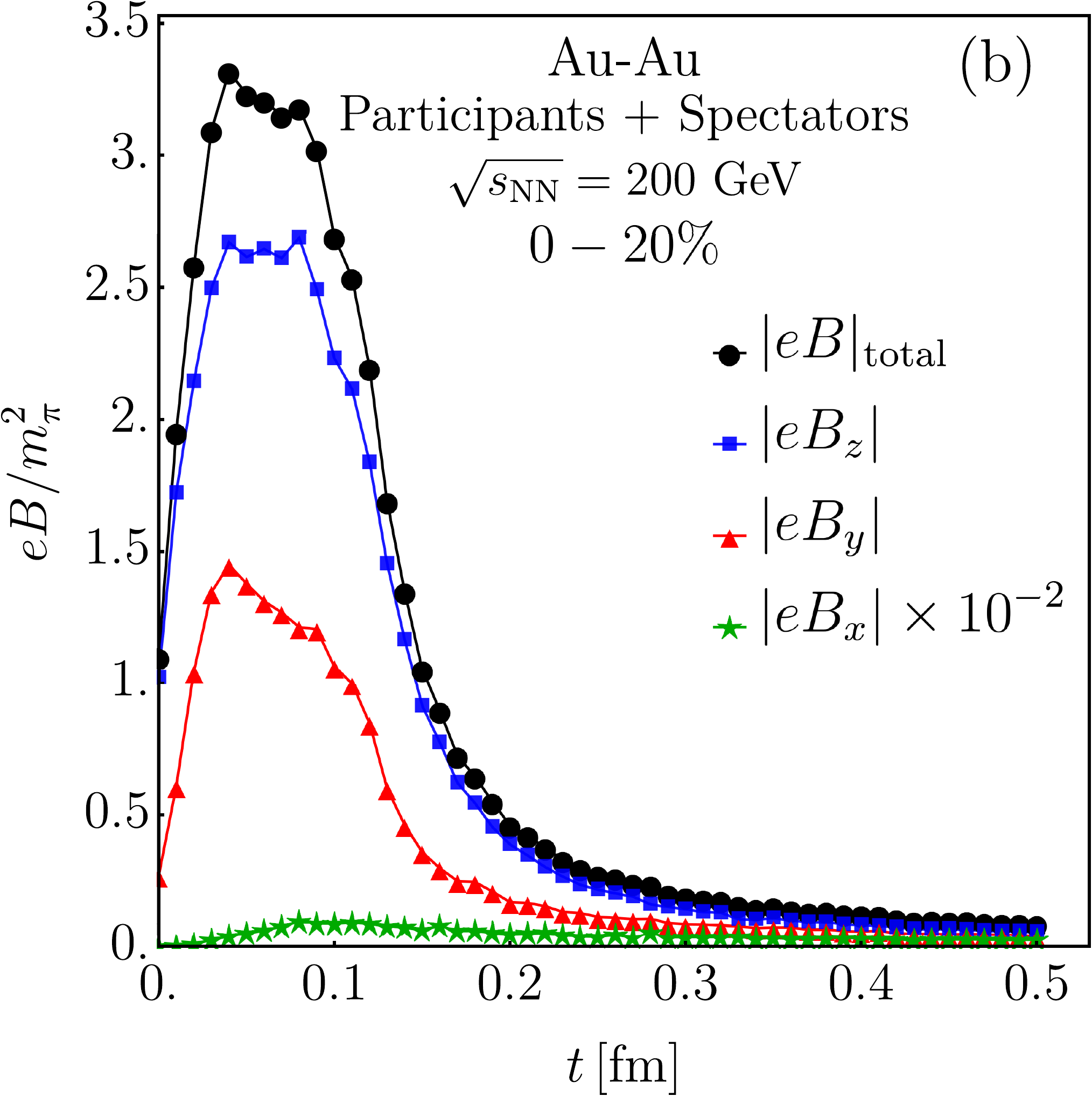}\\
    \vspace{0.4cm}
    \includegraphics[scale=0.34]{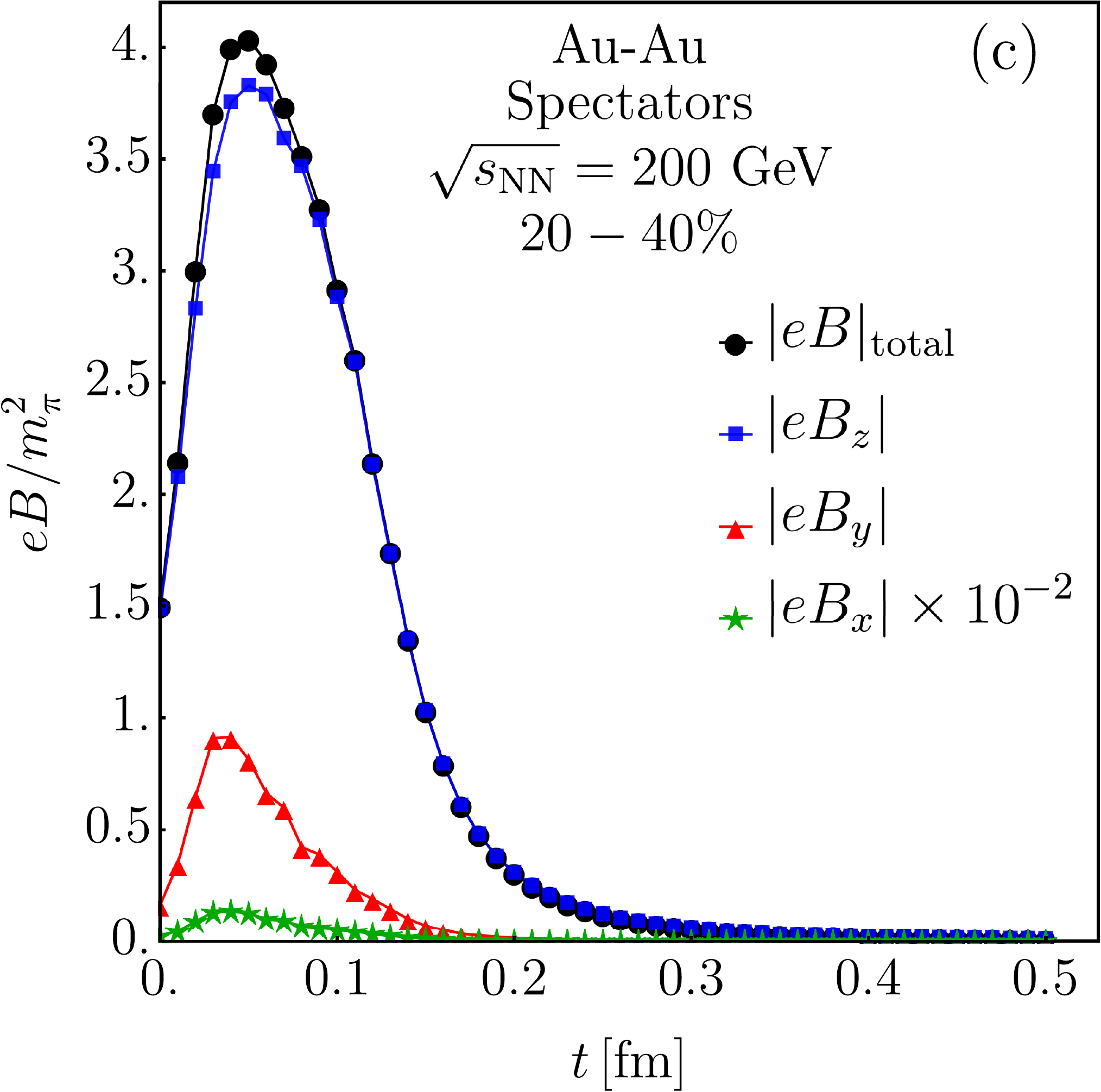}\hspace{2cm}\includegraphics[scale=0.34]{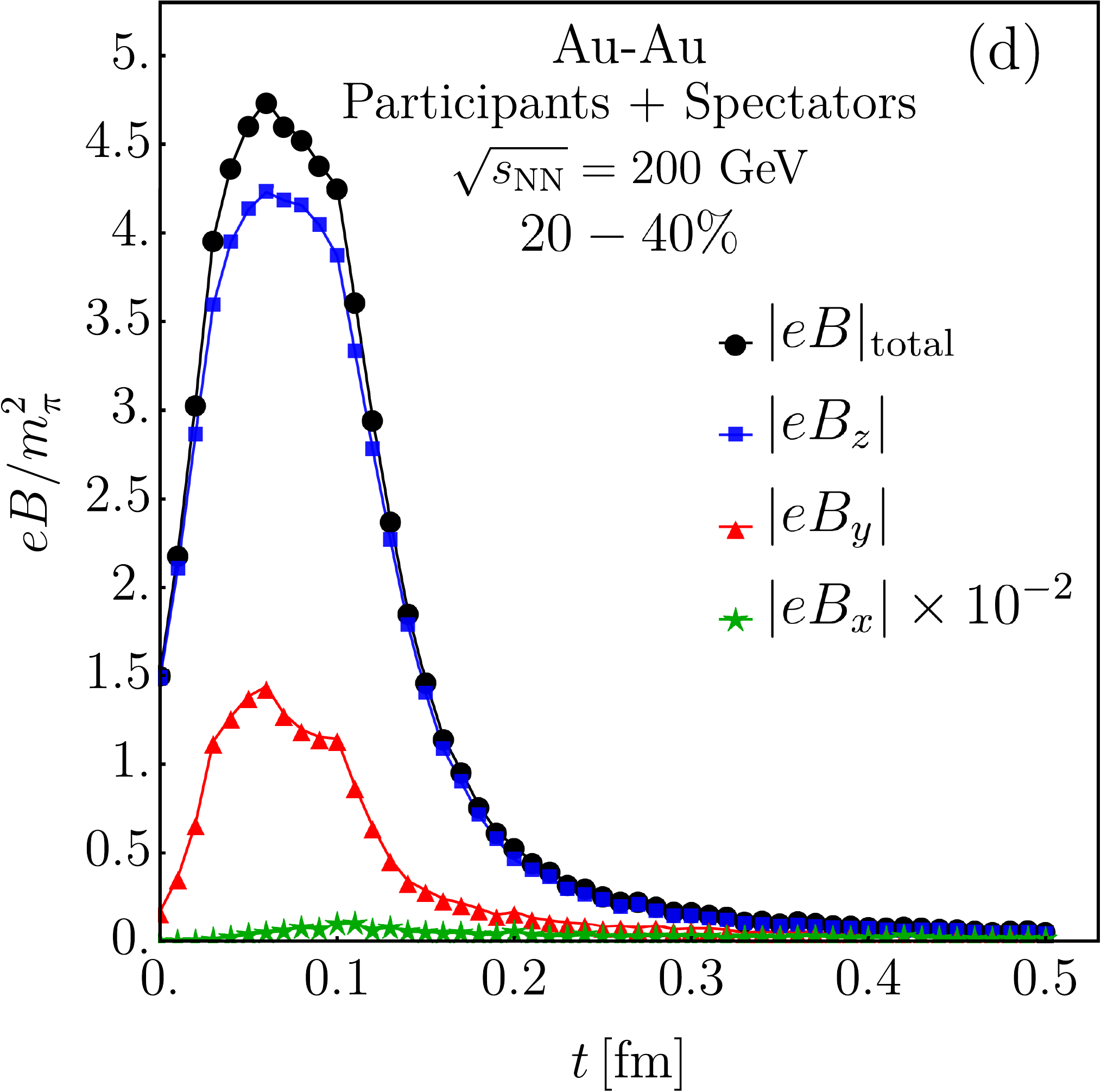}\\
    \vspace{0.4cm}
    \includegraphics[scale=0.34]{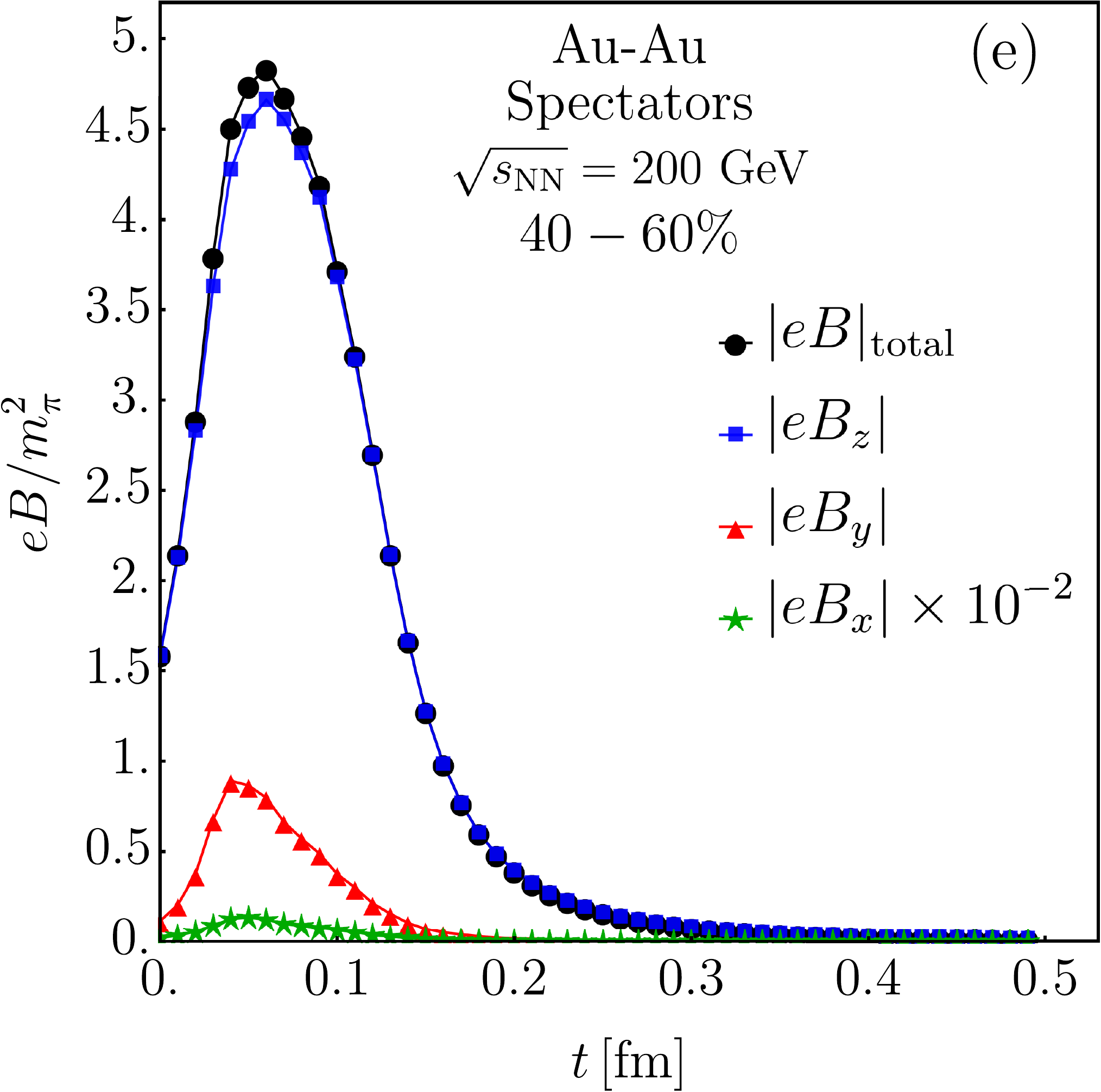}\hspace{2cm}\includegraphics[scale=0.34]{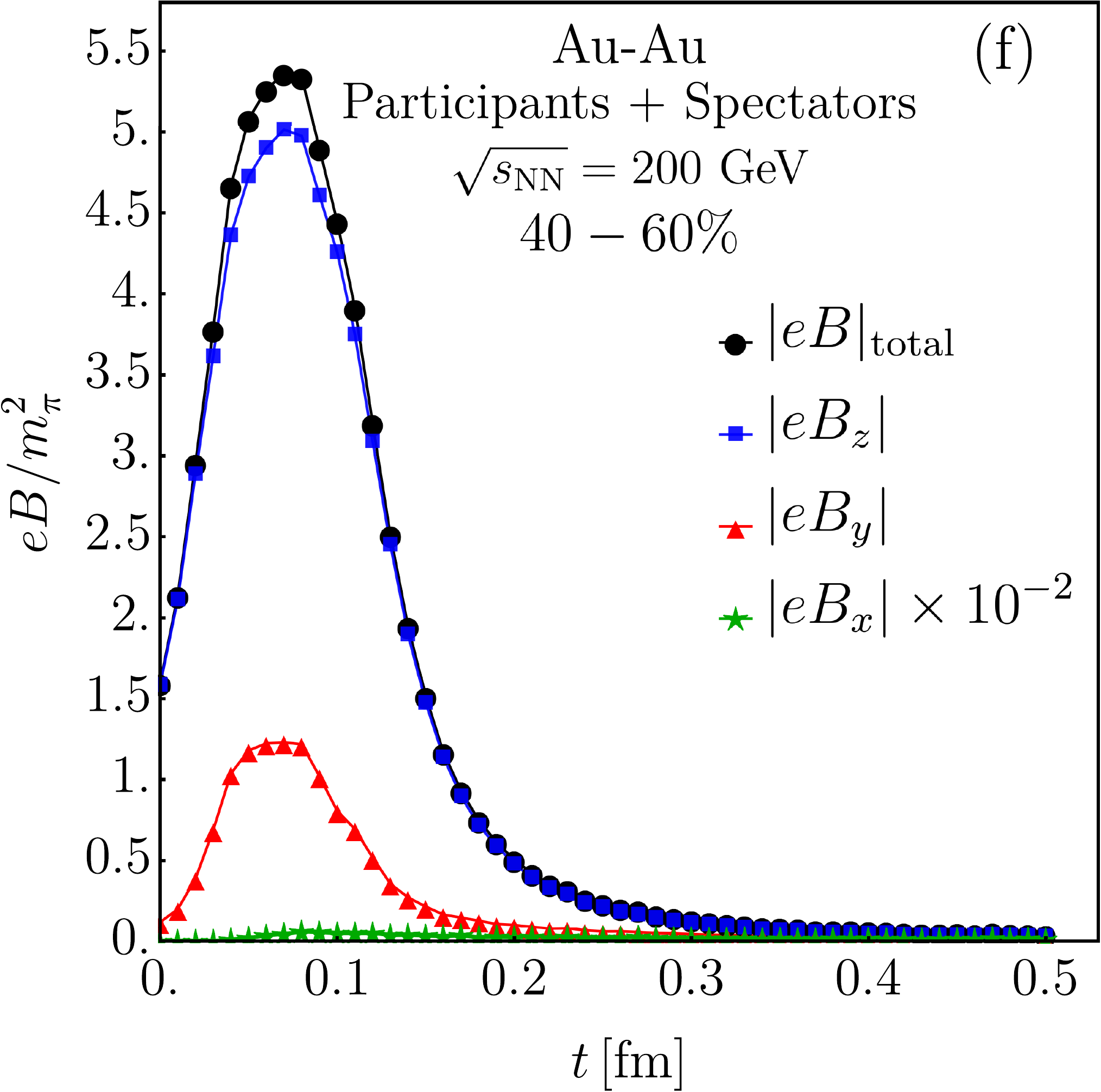}
    \caption{Magnetic field strength for directions parallel ($\hat{x}$) and perpendicular ($\hat{y}$,$\hat{z}$) to the collision axis for three centrality classes in an Au-Au reaction. The magnetic field is produced by spectators [(a), (c), (e)] and participants + spectators [(b), (d), (f)].}
    \label{FIG:Bxyz}
\end{figure}

Figure~\ref{FIG:Bxyz} shows the magnetic field strength at the center of the interaction region, i.e., $\mathbf{x}=0$, produced by the spectators [(a), (c), (e)] and participants + spectators [(b), (d), (f)] for different centrality classes in an Au-Au reaction for $\sqrt{s_{\text{NN}}}=200$ GeV. As it was expected, central collisions ($0-20\%$) have fewer spectators, therefore, in comparison with the peripheral classes, the magnetic field produced reach lower values. Moreover, the UrQMD allows calculating the spatial components for the magnetic field, being $|eB|_x$ and ($|eB|_y$, $|eB|_z$) the parallel and perpendicular directions with respect to the collision axis., respectively.

The results are in agreement with the fact that in the beam-direction the magnetic field has to be small whereas its magnitude should be appreciable for the other components. Figures~\ref{FIG:Bxyz} (a), (c), (e) implies that in peripheral collisions the magnetic field is predominantly oriented at the $z$-direction, but for central ones, it has a comparable component at the $y$-axis. This will impact on the observable description, given that the calculation assumes a magnetic field in one preferential direction.
\begin{figure}[t]
    \centering
    \includegraphics[scale=0.37]{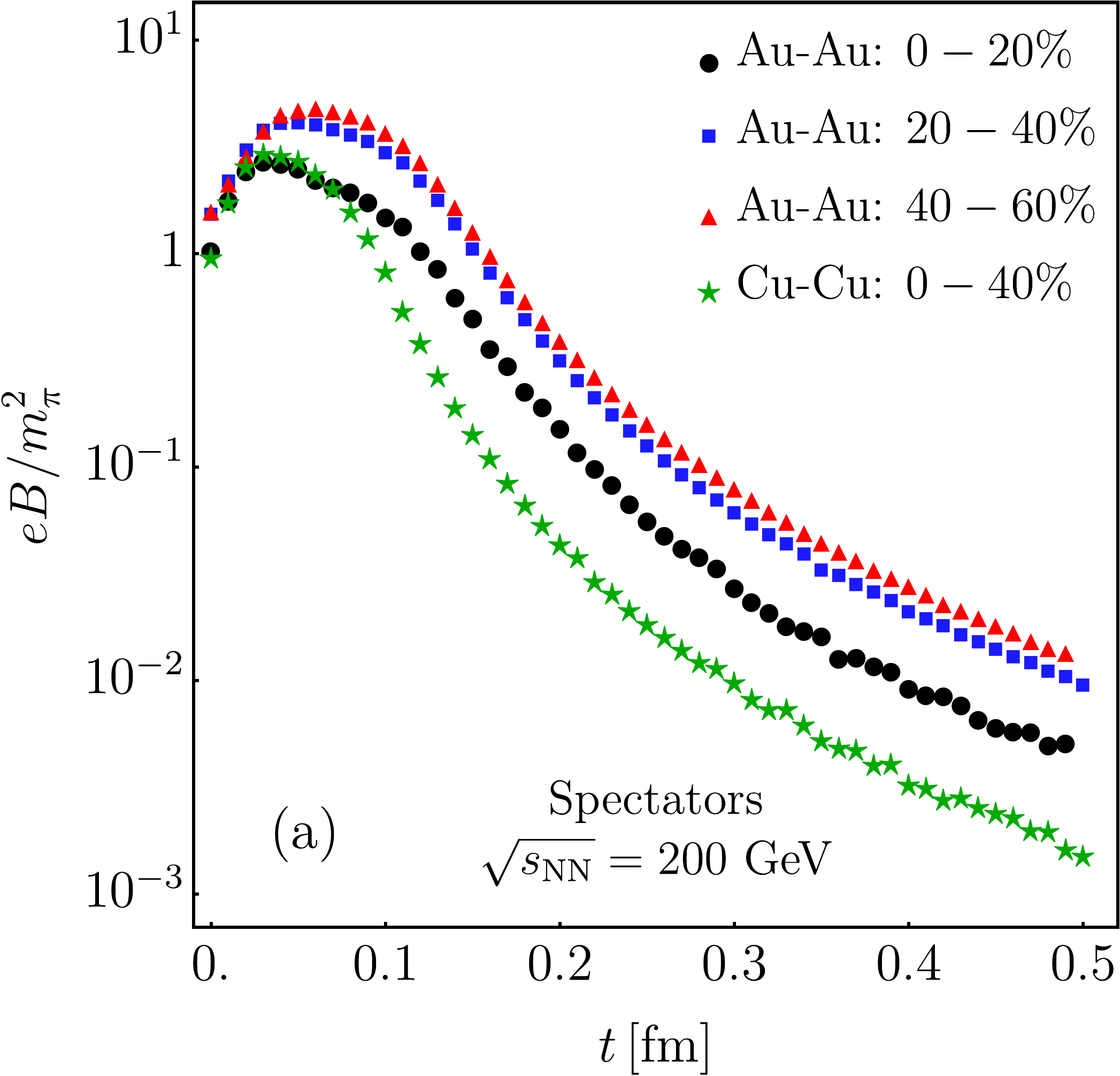}\hspace{0.42cm} \includegraphics[scale=0.37]{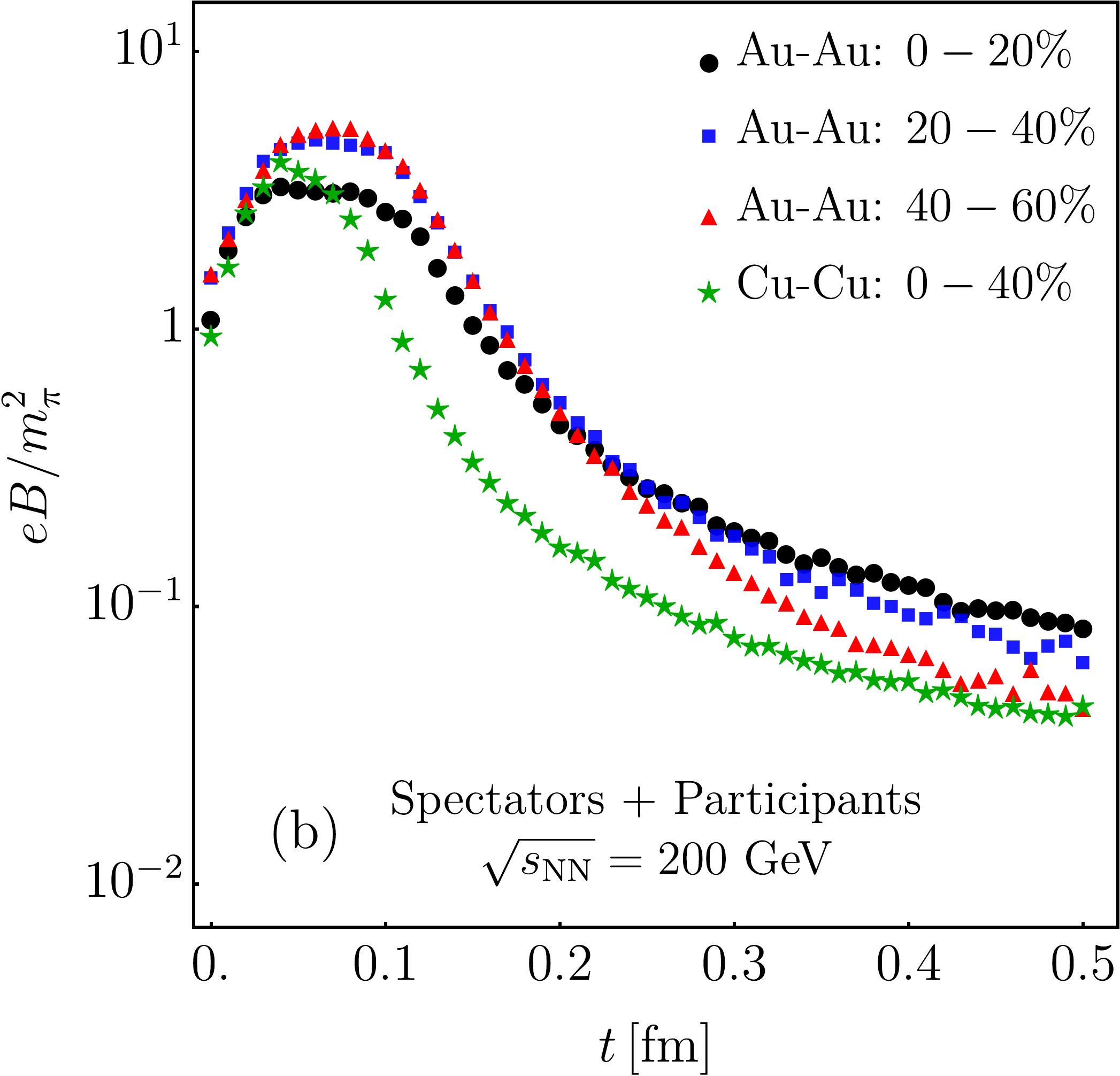}
    \caption{Magnetic field intensity in units of pion mass squared ($m_\pi^2$) at the center of the interaction region ($\mathbf{x}=0$) for several centrality classes and two colliding species with $\sqrt{s_{\text{NN}}}=200$ GeV.  The magnetic field is assumed to be produced by (a) spectators and (b) spectators + participants. }
    \label{Fig:eB_spectators_and_participants}
\end{figure}

In contrast to the field generated by the spectators, in Figs.~\ref{FIG:Bxyz} (b), (d), (f) it can be seen the magnetic field strength created by spectators and participants at $\mathbf{x}=0$. Although in peripheral collisions the main contribution is still is in the $z$-direction, the peak of intensity has an appreciable contribution in the $y$-axis. Thus, it can be expected deviations in the yield or the harmonic coefficient when the model takes into account both sources. 

To complete the panorama and contrast it with the field produced by other colliding species, Fig.~\ref{Fig:eB_spectators_and_participants} shows the total magnetic field produced by Au-Au collisions at different centralities, compared with Cu-Cu simulations with a centrality of $0-40\%$, at $\sqrt{s_{\text{NN}}}=200$ GeV. In agreement with the previous results, peripheral collisions imply a higher magnetic field, so much for participants in Fig.~\ref{Fig:eB_spectators_and_participants}(a) as well as for participants and spectators in Fig.~\ref{Fig:eB_spectators_and_participants}(b). Furthermore, the fact that in Cu-Cu collisions the magnetic field is understood in terms of the number of charges which can be spectators or participants: copper has 29 protons whereas gold has 79. Nevertheless, the considered centrality class produces a similar average field strength for the former compared to the latter for the most central collisions.

As it was commented on Sec.~\ref{sec:Magnetic_field_HIC}, a fundamental characteristic of the magnetic field in heavy-ion collisions is its short duration.  As can be inferred from Figs.~\ref{FIG:Bxyz} and~\ref{Fig:eB_spectators_and_participants}, the magnetic pulse is short-lived and it disappears around $t\sim0.3$ fm.  This means that possible thermal effects, implied by hydrodynamical simulations, start on average after the time interval when the magnetic pulse is important. Hence, this window of time allows calculating the interaction volume where the magnetic effects are relevant. Such volume is computed with the formula:
\begin{equation}
\mathcal{V}(t) = 2 t\pi r_\text{A}^2 \left(\frac{N_\text{part}}{2N}\right)^{2/3},
\label{volume}
\end{equation}
which corresponds to two times the volume of a cylinder with a circular basis of radius $r_\text{A}$ and height $t$ (the time of the magnetic pulse duration). The last term accounts for the overlap area in the collision region and can be justified by considering the cross-sections from Fig.~\ref{FIG:geometricalCrossSection}: (a) the geometric cross-section $\sigma_{\text{geom}}$, defined as the area of the circle with radius $r_1+r_2$ in a maximum peripheral collision, and (b) the cross-section $\sigma_{b}$ for a peripheral collision with impact parameter $b$ ~\cite{jerzy2008introduction}. Such a fraction reads:
\begin{figure}
    \centering
    \includegraphics[scale=0.4]{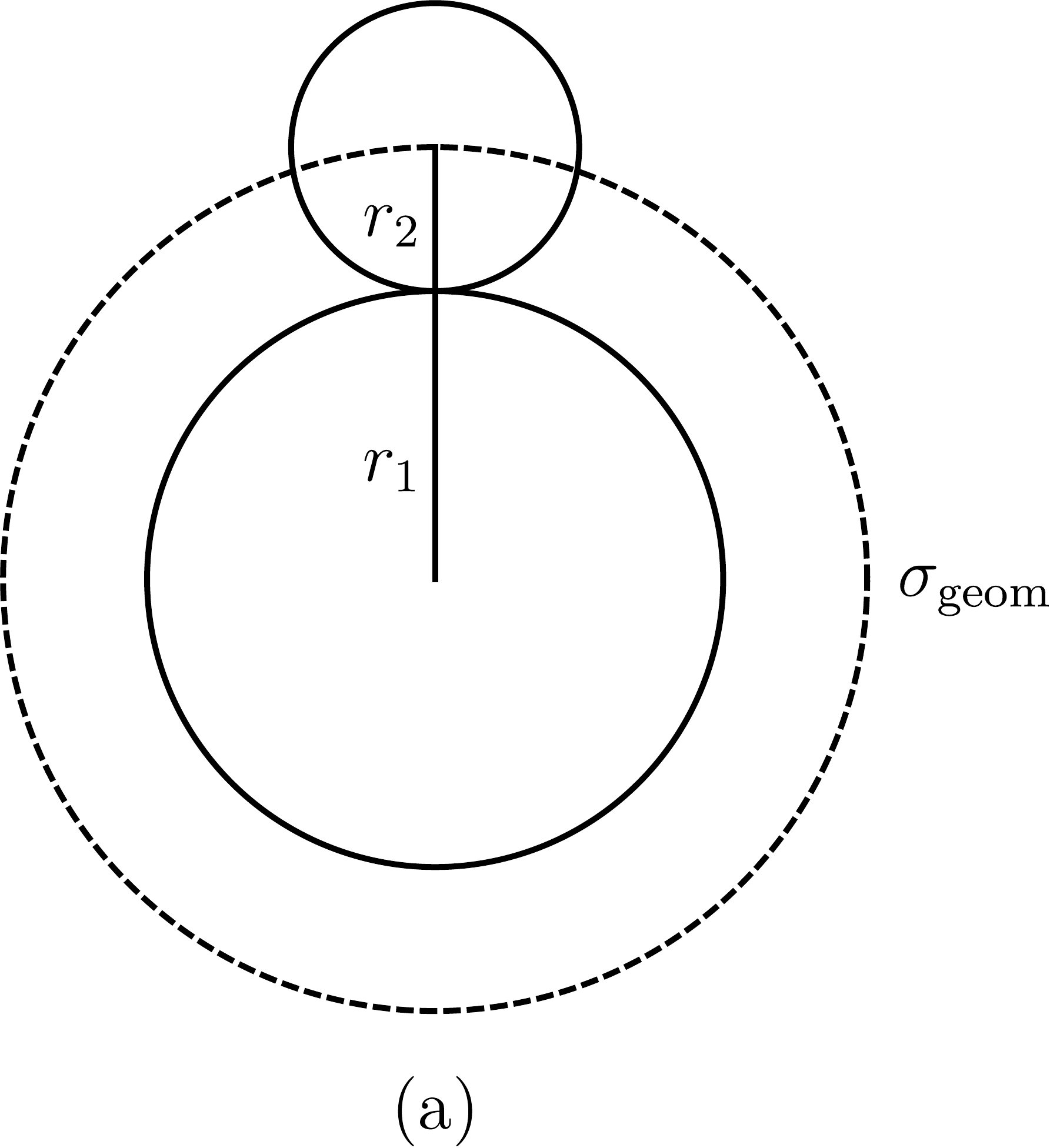}\hspace{3cm}\includegraphics[scale=0.4]{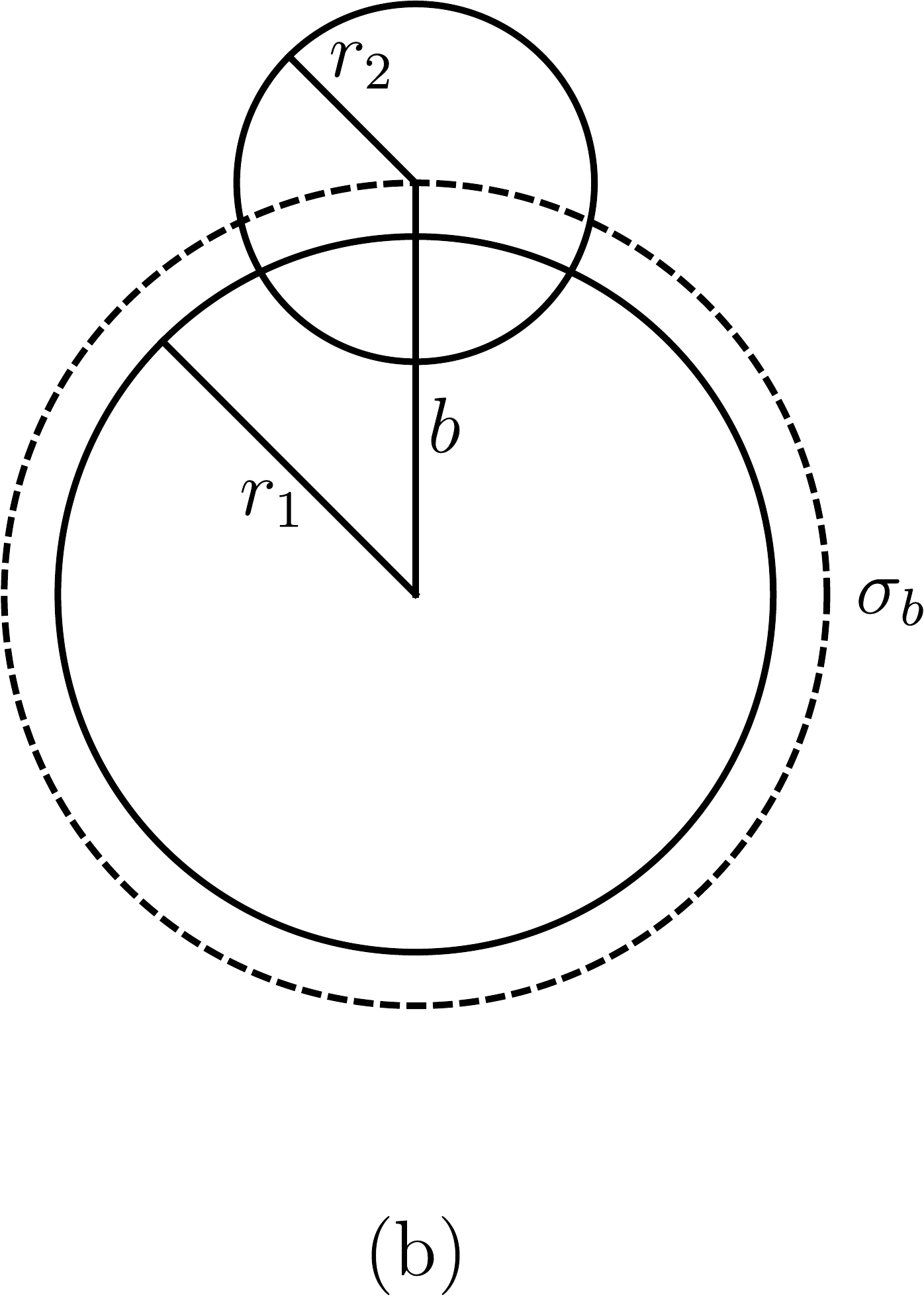}
    \caption{Cross-sections (dashed lines) for two colliding nuclei of radii $r_1$ and $r_2$ in (a) a maximal peripheral collision ({\it geometrical cross-section}), and (b) a peripheral collision with impact parameter $b$.}
    \label{FIG:geometricalCrossSection}
\end{figure}
\bea
f=\frac{\sigma_b}{\sigma_{\text{geom}}}=\frac{\pi b^2}{\pi\left(r_1+r_2\right)^2}.
\eea

If the colliding nuclei are of the same species, then $r_1=r_2=R$, and by assuming that $\sigma_b$ describes an effective nucleus of radius $b$, the relation of the nuclear radius $r=r_0 N^{1/3}$, whit $N$ is the number of nucleons per ion and $r_0\approx1.25$ fm, yields
\bea
f=\left(\frac{b}{2R}\right)^2=\left(\frac{N_{\text{part}}}{2N}\right)^{2/3},
\label{CrossSectionFraction}
\eea
where it was assumed that the effective nucleus has a number of particles that participate in the collision given by $N_{\text{part}}$. A more detailed analysis can be performed in order to get a better estimate of $f$, in terms of the nuclear overlap integral and generalizations of the geometric cross-section~\cite{vogt1999relation}.

\begin{figure}
    \centering
    \includegraphics[scale=0.37]{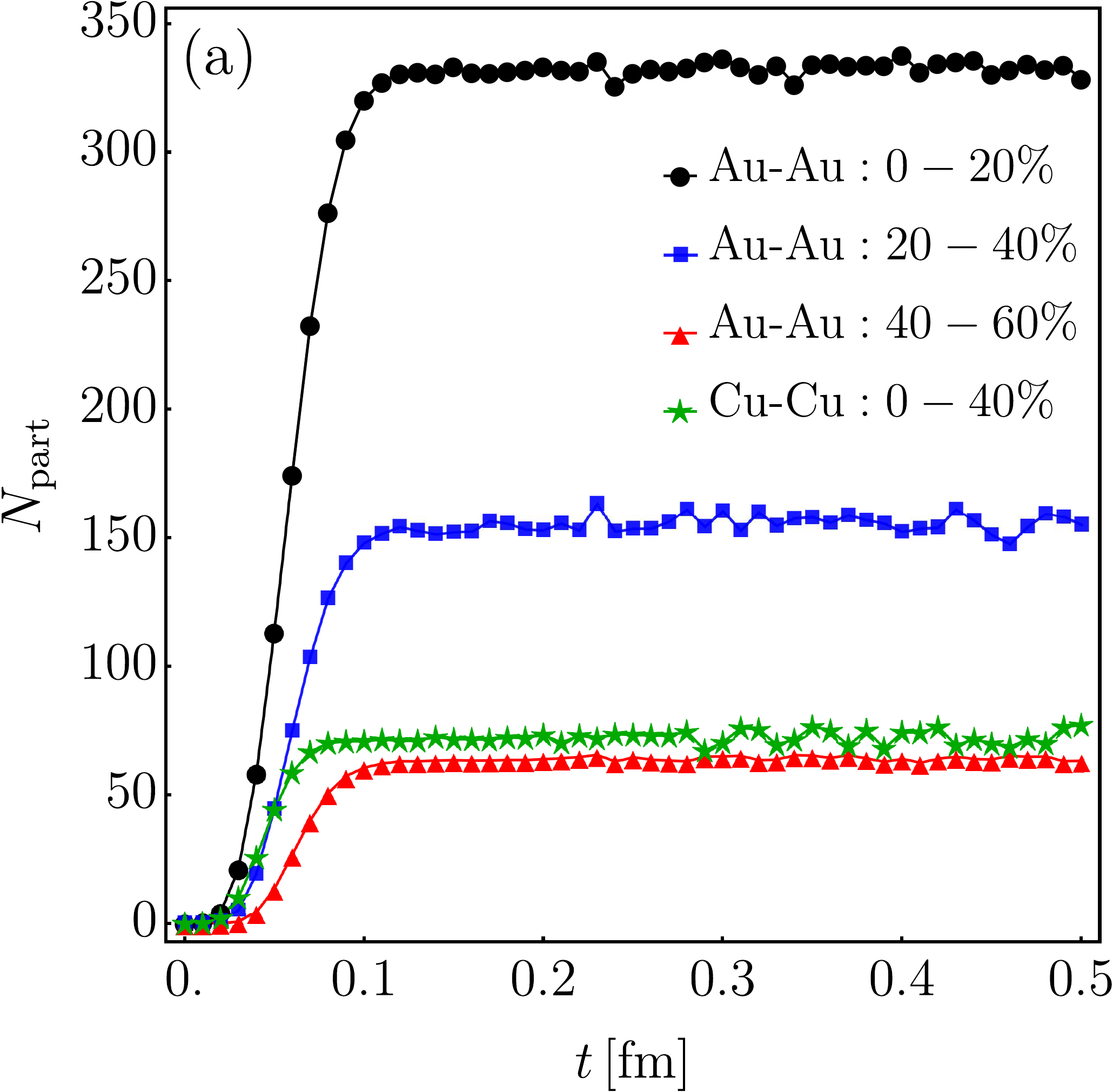}\hspace{1cm}\includegraphics[scale=0.37]{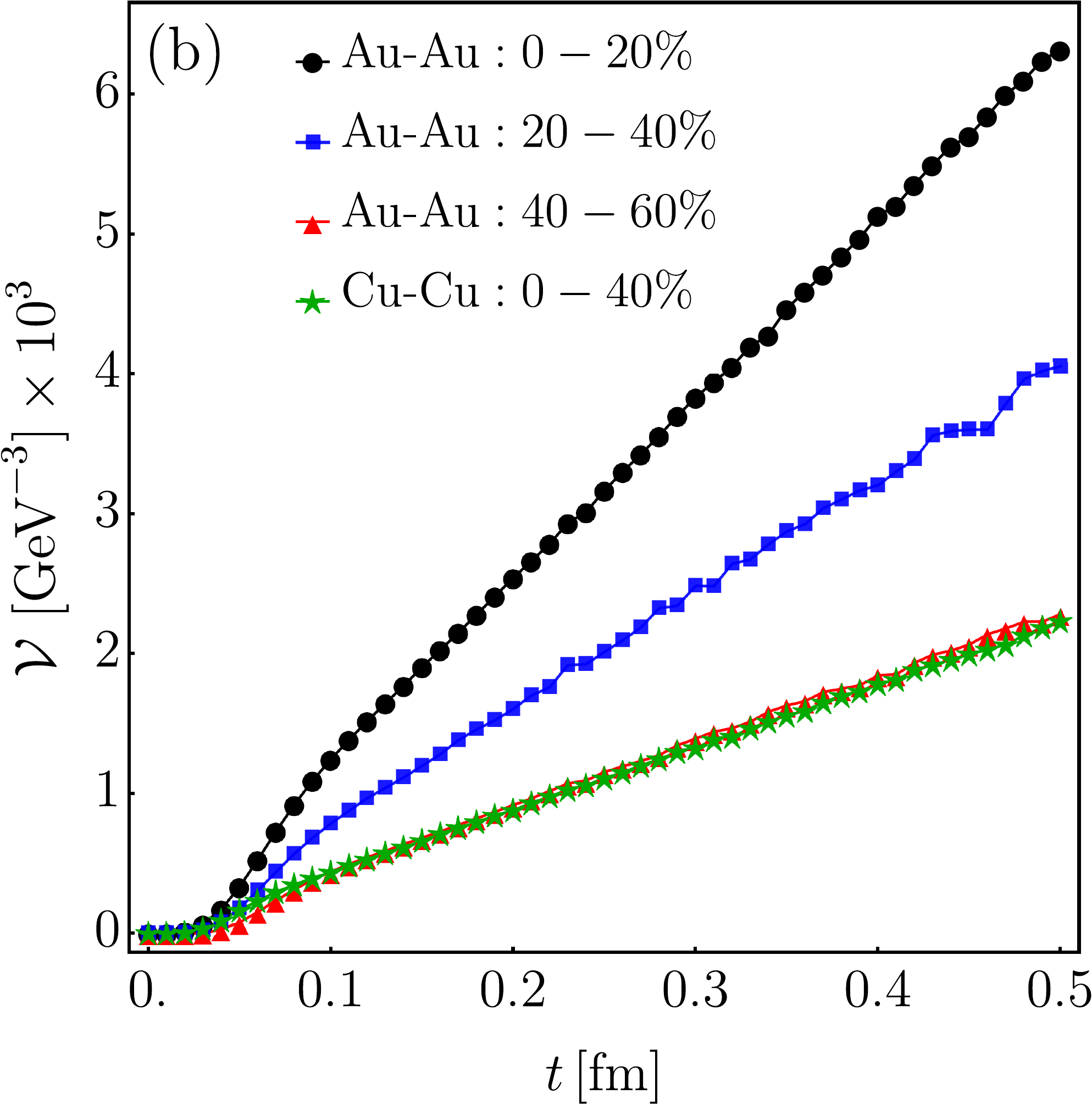}
    \caption{(a) Number of participants as function of the time for Au-Au and Cu-Cu collisions in different centrality classes. (b) Interaction volume eAu-Au and Cu-Cu collisions in different centrality classes. Both results are obtained by considering $\sqrt{s_\text{NN}}=200$ GeV.}
    \label{Fig:Npart_and_Volume}
\end{figure}

Figure~\ref{Fig:Npart_and_Volume}(a) shows the number of participants $N_{\text{part}}$ as a function of the collision time, obtained from UrQMD simulations for Au-Au collisions for the centrality classes 0-20\%, 20-40\% and 40-60\%, and for Cu-Cu collisions with centrality 0-40\% at $\sqrt{s_\text{NN}}=200$ GeV. As was expected, central collisions (0-20$\%$) have a higher number of participants than peripheral ones (40-60\%), and the number of particles that participates of at least one collision suddenly increases from zero (at the beginning of the collision) to a certain value in which such number saturates.

The time evolution of the interaction volume $\mathcal{V}$ computed from Eq.~(\ref{volume}) is shown in Fig.~\ref{Fig:Npart_and_Volume}(b) for the same centrality classes and energy of Fig.~\ref{Fig:Npart_and_Volume}(a). A linear increase of $\mathcal{V}$ as a function of $t$ is expected from Eq.~(\ref{volume}) so that the slope is related to the number of participants, which has a step-function form.  Interestingly, although Cu ions are smaller than Au nuclei, peripheral Cu-Cu reactions in the range of 0-40\% have basically the same interacting volume that Au-Au collisions in a region of smaller centrality. The impact of the magnetic field and volume on the photon invariant momentum distribution and the elliptic flow coefficient is presented in the next section. 

\subsection{Photon Invariant Distribution and Harmonic Coefficient with UrQMD simulations}
\begin{figure}[t]
    \centering
    \includegraphics[scale=0.37]{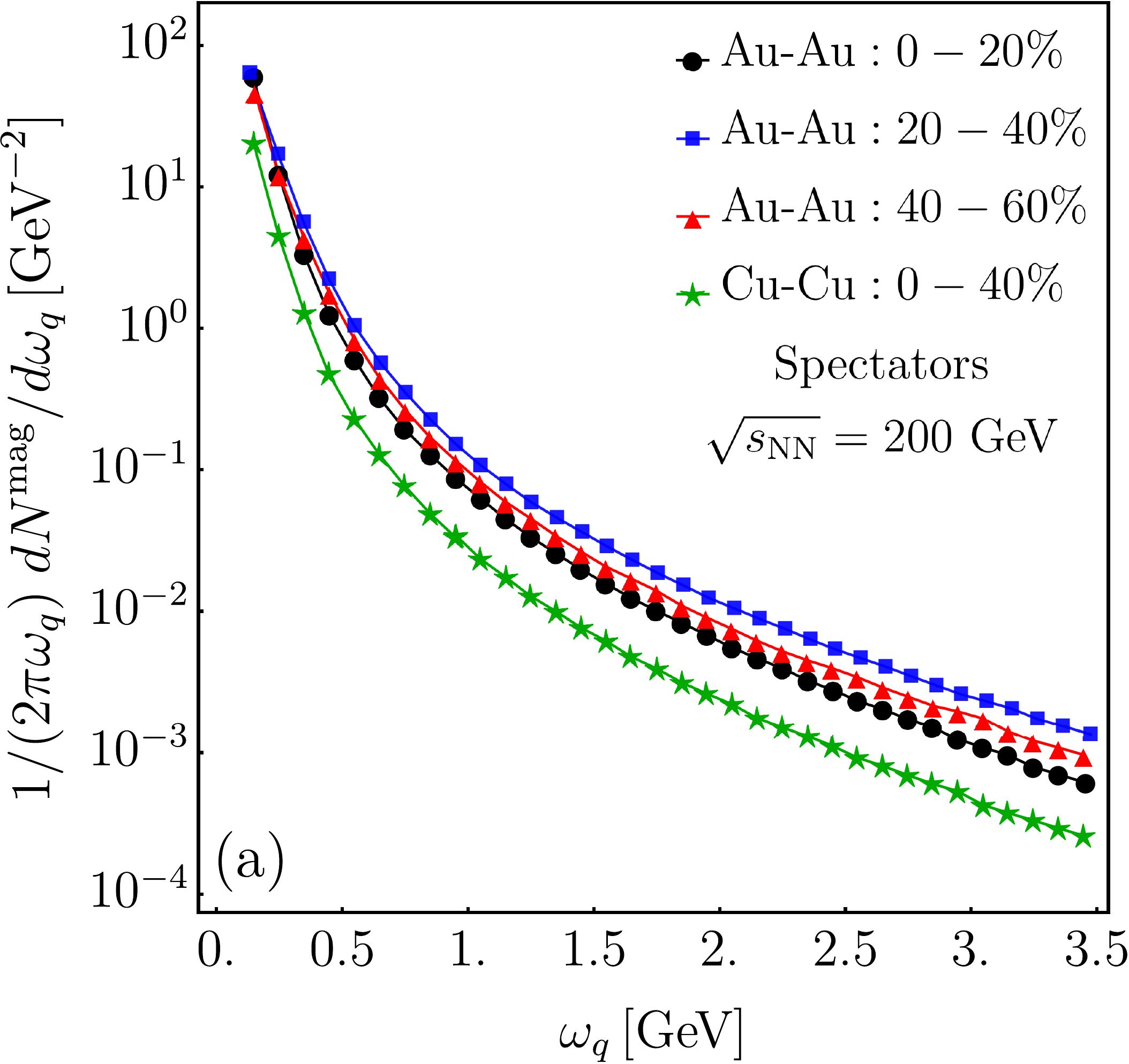}\hspace{0.3cm} \includegraphics[scale=0.37]{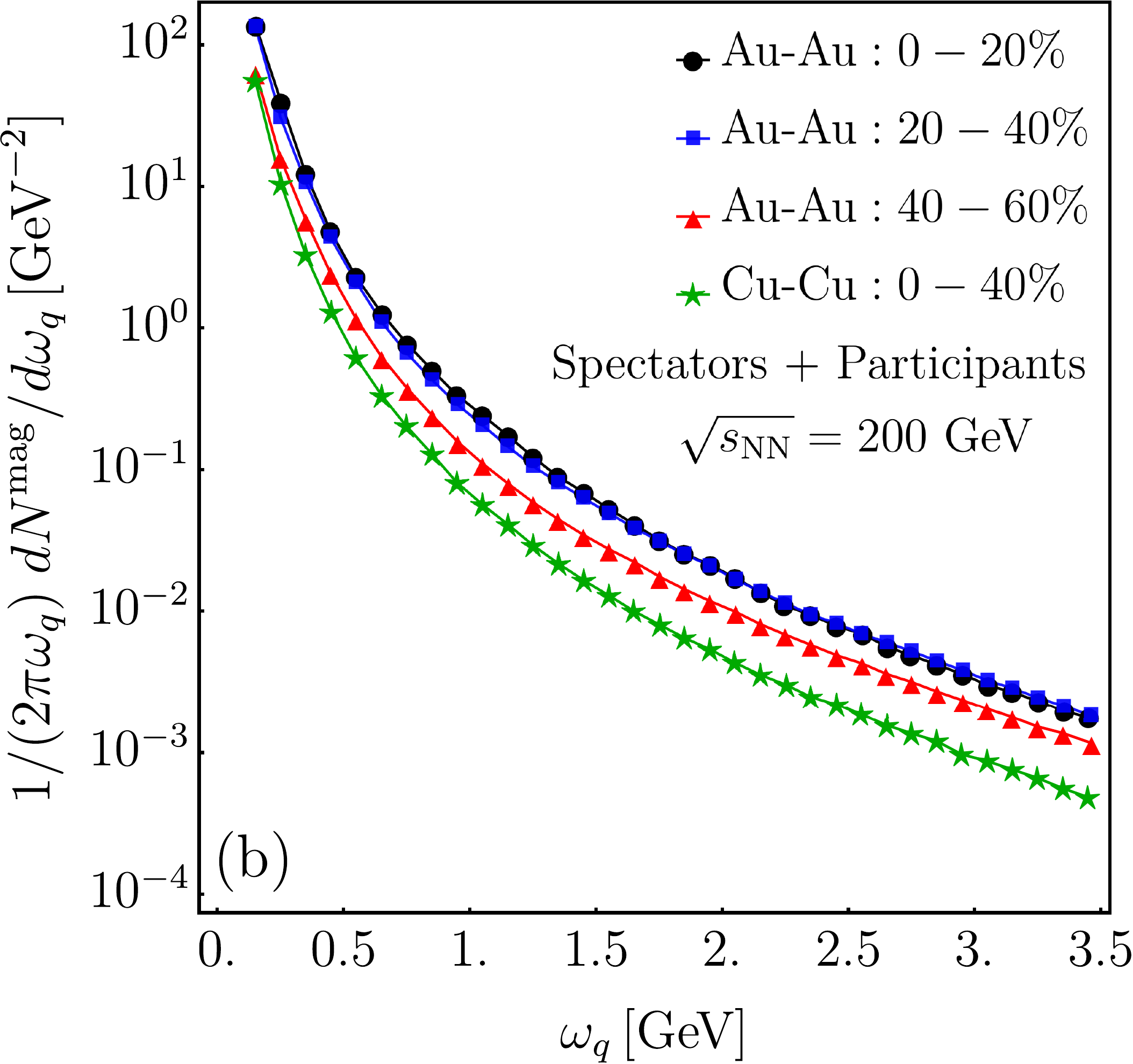}
    \caption{Invariant momentum distribution
    Au+Au collisions in the 0-20\%, 20-40\% and 40-60\% centrality classes and Cu+Cu collisions in the 0-40\% centrality class at $\sqrt{  s_{\text{NN}} }=200$ GeV considering the magnetic field generated by (a) participants and (b) participants and spectators.}
    \label{Fig:YieldFromUrQMD}
\end{figure}

%
In order to compute the invariant momentum distribution and the elliptic flow from the processes $gg\rightarrow\gamma$ and $g\rightarrow g\gamma$, UrQMD results for the magnetic field and interaction volume (centrality) were included into the phenomenology described by Eqs.~(\ref{yieldexpl2}) and~(\ref{v2expl2}). The rest of the parameters are chosen as follows: $\alpha_s=0.3 $, $\Lambda_s=2 $ GeV, and $\eta=3$. Figure~\ref{Fig:YieldFromUrQMD}   hows the invariant momentum distribution and for Au+Au collisions at $\sqrt{  s_{\text{NN}} }=200$ GeV in the 0-20\%, 20-40\% and 40-60\% centrality ranges and Cu+Cu collisions in the 0-40\% centrality range at the same energy. The magnetic field contribution is assumed to be generated by (a)spectators and (b) spectators + participants. 

The total photon yield is calculated by adding the yields for all time intervals within $\Delta\tau$, i.e., if $\Delta\tau$ is divided in small intervals of size $\Delta t_i$, the final output is given by
\bea
\frac{dN^{\text{mag}}}{d\omega_q} &=& \sum_{i=1} \left[\frac{dN^{\text{mag}}}{d\omega_q}\right]_i,
\eea
where $\left[\frac{dN^{\text{mag}}}{d\omega_q}\right]_i$ is the yield corresponding to the $ith$-time interval $\Delta t_i$ (0.01 fm), computed from Eq.~(\ref{yieldexpl}). The results confirm that a higher magnetic field intensity implies an enhancement of photon production. Nevertheless, despite the result of Eq.~(\ref{yieldexpl}) which indicates that the volume and centrality implies a hierarchy between the momentum distribution with different centrality classes, that is to say, less $\mathcal{V}$ and $\chi$ give small number of photons, Figs.~\ref{Fig:YieldFromUrQMD}(a) and~\ref{Fig:YieldFromUrQMD}(b) do not reflect such behavior. This can be understood from the fact that the integration volume has an opposite effect in centrality classes in with respect to the magnetic field. Therefore, the combination of both effects does not show a monotonically increasing (or decreasing) yield when the centrality is changed. Moreover, even when the Cu-Cu collisions has less interaction volume than Au-Au, the similarity between their magnetic fields (Fig.~\ref{Fig:eB_spectators_and_participants}), $N_{\text{part}}$ and $\mathcal{V}$ (Fig.~\ref{Fig:Npart_and_Volume}) make the photon production to be of the same order for the two species.
\begin{figure}
    \centering
   \includegraphics[scale=0.38]{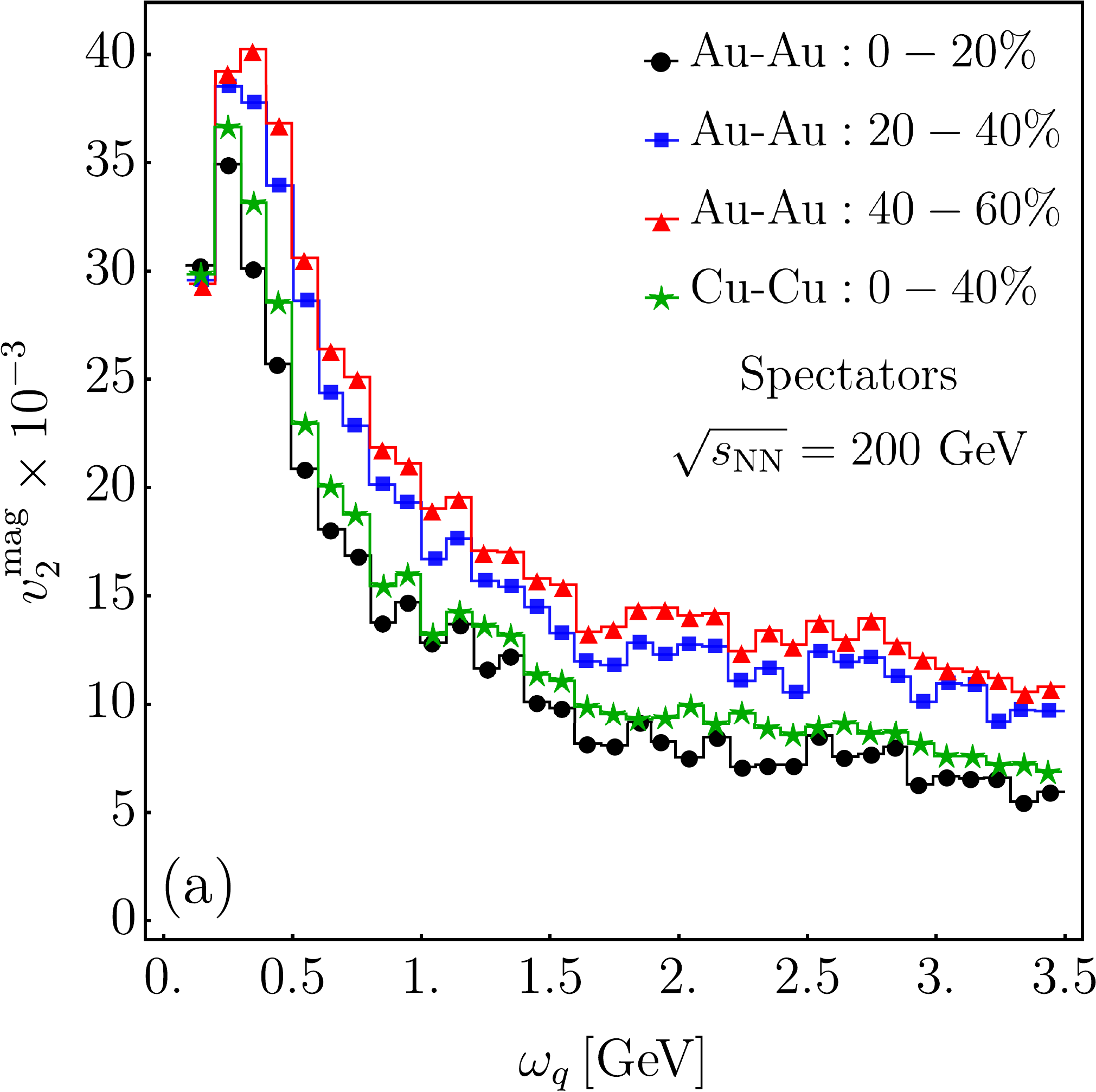}\hspace{0.6cm}\includegraphics[scale=0.38]{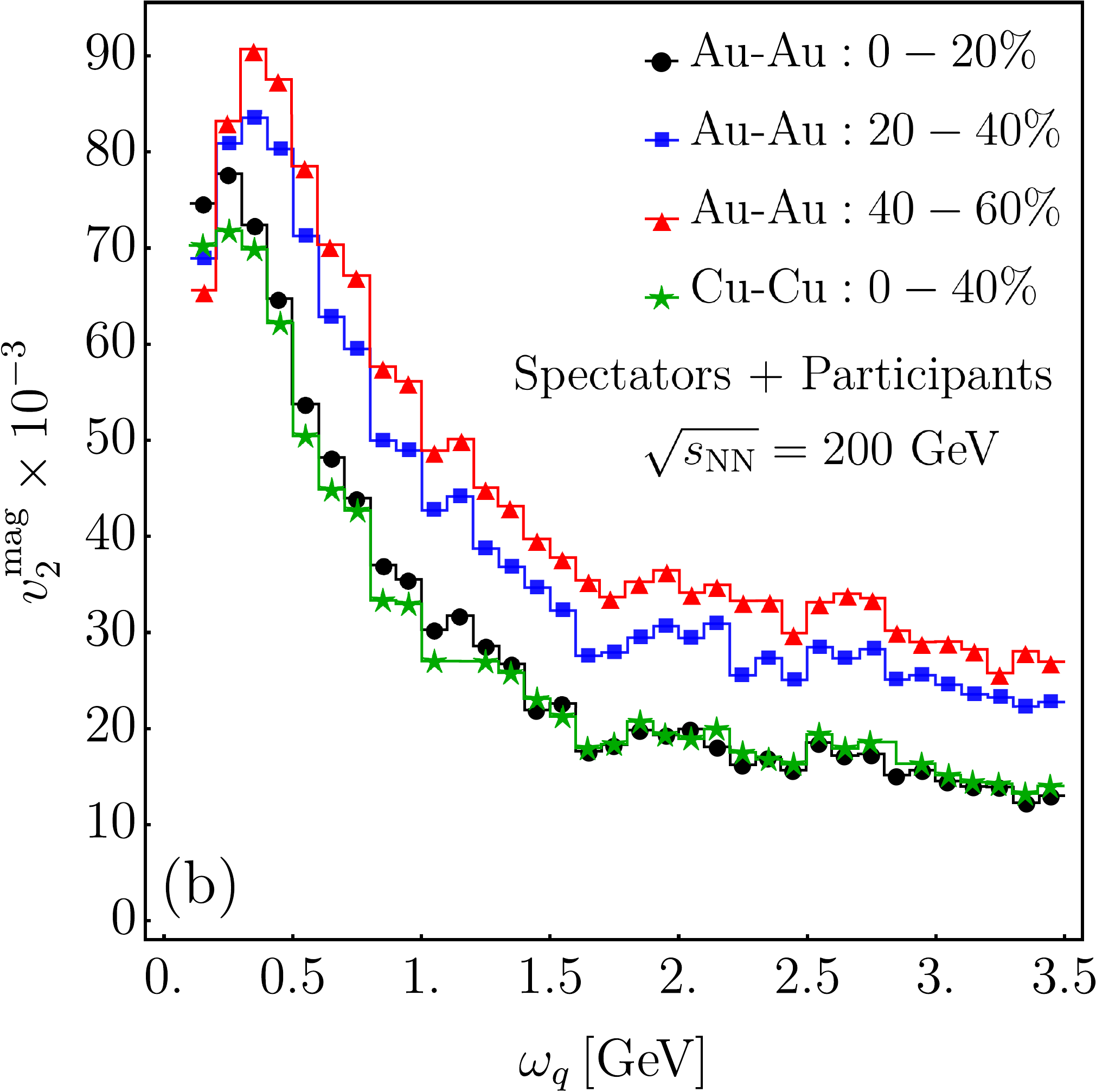}
    \caption{Harmonic coefficient $v_2^{\text{mag}}$ 
    for Au+Au collisions in the 0-20\%, 20-40\% and 40-60\% centrality classes and Cu+Cu collisions in the 0-40\% centrality class at $\sqrt{  s_{\text{NN}} }=200$ GeV considering the magnetic field generated by (a) participants and (b) participants and spectators.}
    \label{Fig:v2_UrQMD}
\end{figure}

Figure~\ref{Fig:v2_UrQMD} displays the pure magnetic contribution to the harmonic coefficient v2, for Au+Au collisions in the 0-20\%, 20-40\% and 40-60\% centrality classes and Cu+Cu collisions in the 0-40\% centrality class at $\sqrt{  s_{\text{NN}} }=200$ GeV. As it was previously mentioned, the elliptic flow needs to be considerer as a weighted average. In that way, by accounting the time-variation of the yield, the elliptic flow takes the form
\bea
v_2^{\text{mag}}(\omega_q) &=& \frac{
\sum_{i=1} \left[
\frac{dN^{\mbox{\tiny{mag}}}}{d\omega_q}(\omega_q)
\right]_i 
[v_2^{\text{mag}}(\omega_q)]_i
}{\sum_{i=1} \left[\frac{dN^{\mbox{\tiny{mag}}}}{d\omega_q}\right]_i}
\eea
where $[v_2^{\text{mag}}(\omega_q)]_i$ is the harmonic coefficient corresponding to the $ith$-time interval $\Delta\tau_i$, given by Eq.~(\ref{v2expl}). Notice that including the contribution to the magnetic field strength coming from the spectators and participants, produces an increase of $[v_2^{\text{mag}}(\omega_q)]$. Here, the hierarchy between centrality classes at least for Au-Au collisions is clear, but in comparison with Cu-Cu simulations, there is still a non-direct relation between centrality and the curve height. Again, it can be argued that this effect comes from internal competitions among the parameters in each time interval.
%
\subsection{Comparison with PHENIX data}
\begin{figure}
    \centering
    \includegraphics[scale=0.37]{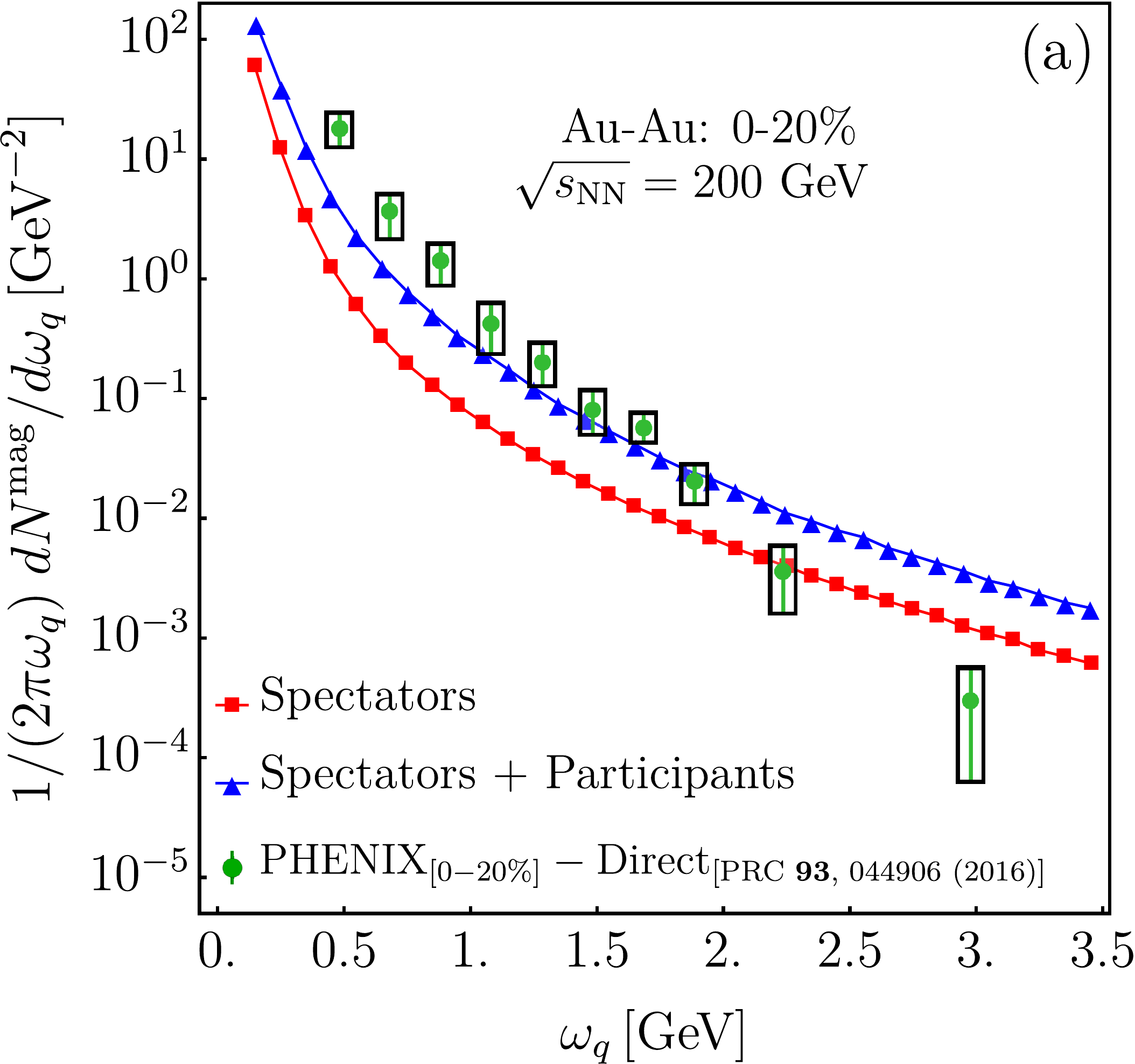}\hspace{0.3cm} \includegraphics[scale=0.365]{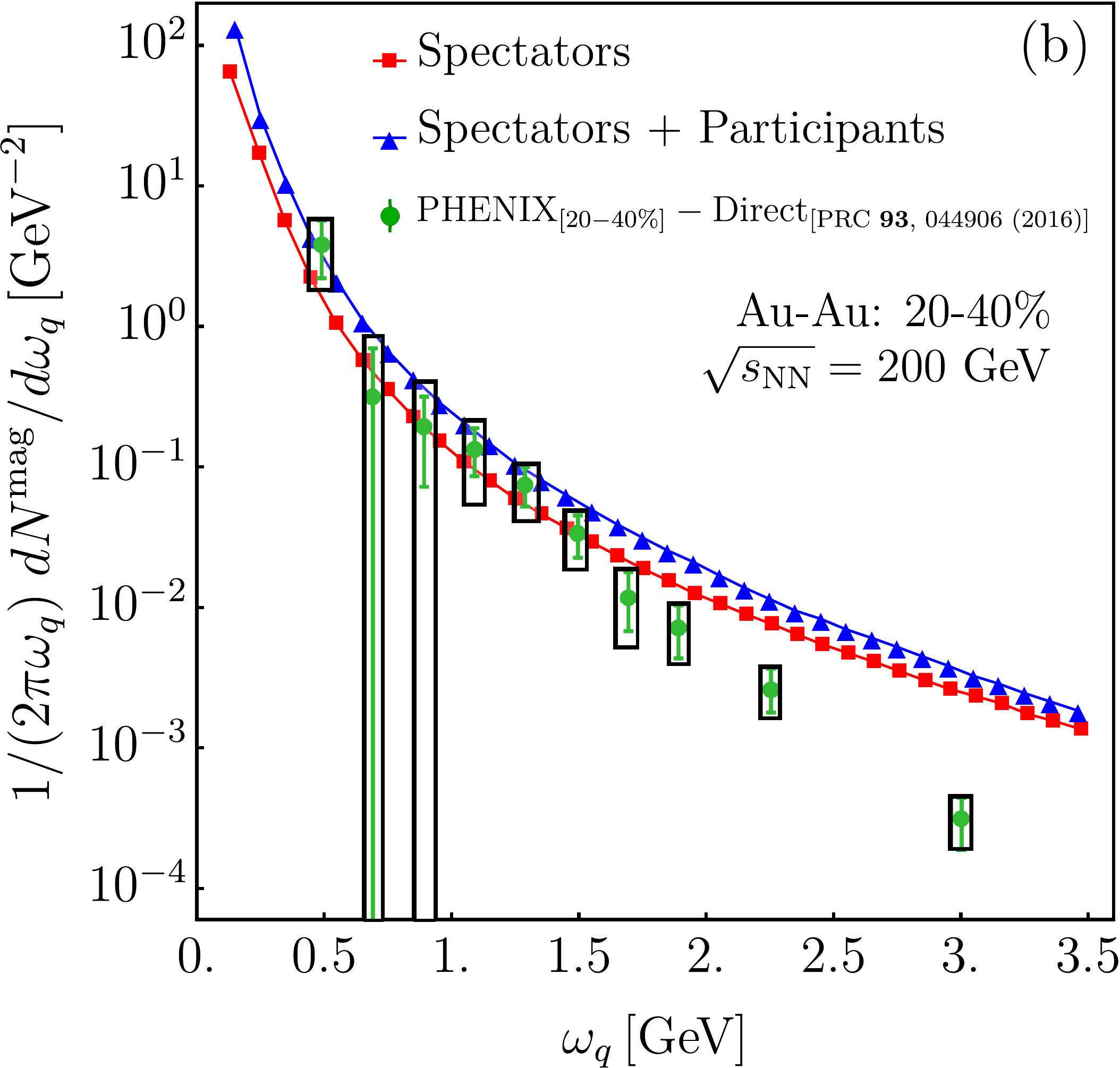}
    \caption{Difference between PHENIX data~\cite{experimentsyield,RHIC3} and the hydrodynamical calculation of Ref.~\cite{hydro-photons1} for Au-Au collisions at $\sqrt{s_{NN}}=200$ GeV compared to the calculation of the invariant yield considering contribution to the magnetic field strength produced by the spectators, and spectators + participants for (a) the 0-20\% and (b) 20-40\% centrality classes.}
    \label{Fig:YieldComparisonUrQMD}
\end{figure}

%
Figure~\ref{Fig:YieldComparisonUrQMD} shows the difference between PHENIX data~\cite{experimentsyield,RHIC3} and the hydrodynamical calculation of Ref.~\cite{hydro-photons1} compared to invariant yield produced by the processes $gg\rightarrow\gamma$ and $g\rightarrow g\gamma$ considering the contribution to the magnetic field strength produced by the spectators and spectators+participants in Au-Au collisions at $\sqrt{  s_{\text{NN}} }=200$ GeV for (a)the 0-20\% and (b) the 20-40\% centrality classes.

\begin{figure}[t]
    \centering
    \includegraphics[scale=0.45]{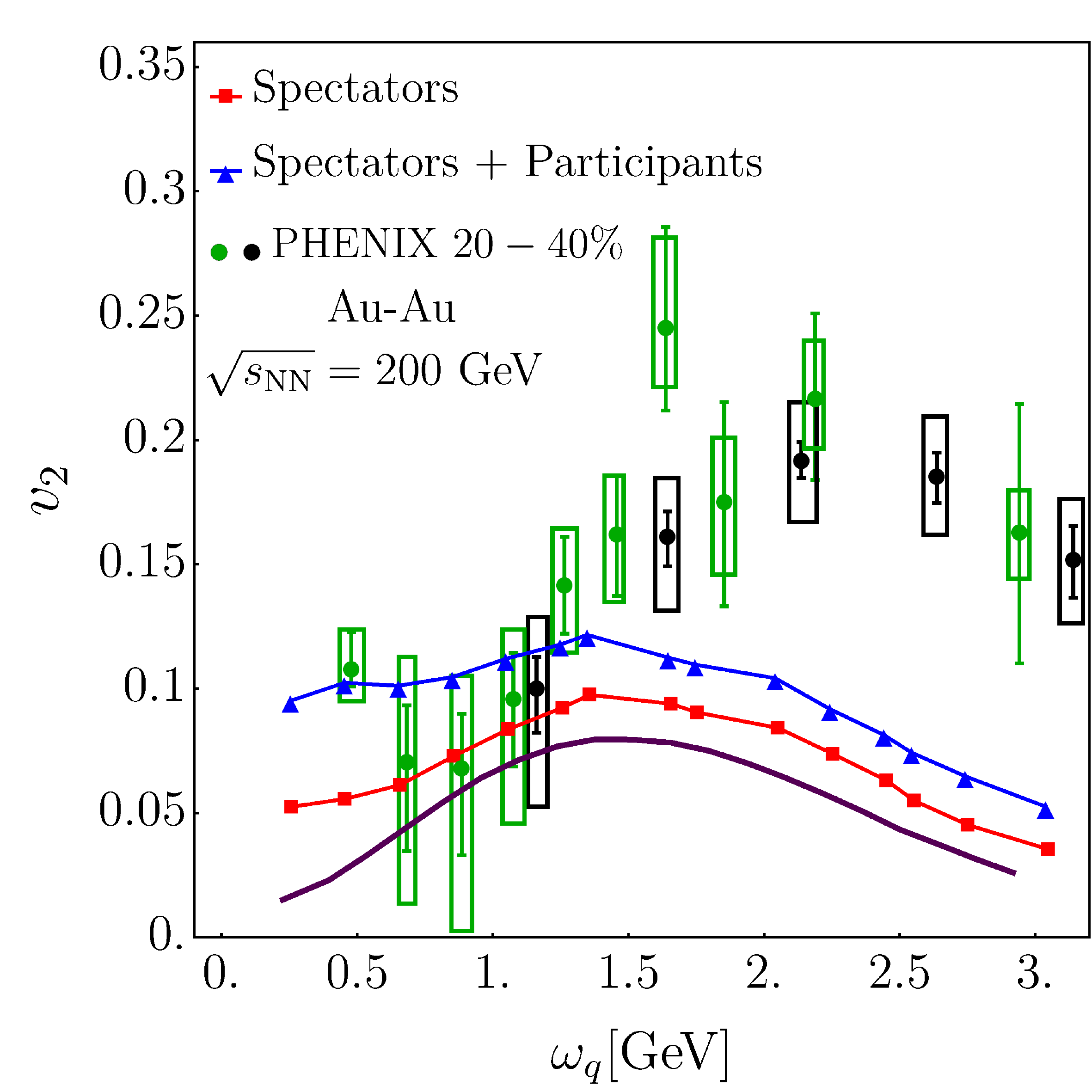}
    \caption{Harmonic coefficient $v_2$ computed from Eq.~(\ref{v2weightedUrQMD}) compared with PHENIX data~\cite{RHIC4} and the elliptic flow of the direct photons from Ref.~\cite{hydro-photons1}. The green (black) points and errors are experimental measurements with the conversion (calorimeter) method corresponding to Au+Au collisions at $\sqrt{s_{NN}}=200$ GeV in the 20-40\% centrality class. The red (blue) curve takes into account the magnetic elliptic flow when the field is generated by the spectators (spectators+participants).}
    \label{Fig:v2UrQMDFinal}
\end{figure}

It can be noticed that when considering the magnetic field produced only by spectators the theoretical yield compares better to peripheral than to central collisions. In fact, for peripheral collisions, both scenarios of magnetic field generation (spectators and spectators+participants) give a better data description. Such behavior comes from the orientation and intensity of the total magnetic field in that kind of reactions, which is predominantly aligned to the $z$-axis for non-central simulations [Figs.~\ref{FIG:Bxyz}(c) and (d)], but has an important component in the $y$-axis when the centrality is recovered  [Fig.~\ref{FIG:Bxyz}(a)]. Additionally, the magnitude of the magnetic field produced in central collisions at the transverse plane direction is lower than the produced in the peripheral reactions. Then, the theoretical assumption for the present calculation, $\mathbf{B}=B\mathbf{\hat{z}}$, and the hierarchy of the energy scales are not well justified for all the centrality classes.  Moreover, as can be seen from Fig.~\ref{FIG:Bxyz} if only the spectators magnetic field is considered, the named approximation is accurate.

Figure~\ref{Fig:v2UrQMDFinal} shows $v_2$ as a weighted average accounting for the magnetic and direct photons:
\bea
v_2(\omega_q)=\frac{
\sum_{i=1}^m \left[
\frac{dN}{d\omega_q}
\right]_i 
[v_2^{\text{mag}}(\omega_q)]_i
   +
   \frac{dN^{\mbox{\tiny{direct}}}}{d\omega_q}(\omega_q)\
   v_2^{\mbox{\tiny{direct}}}(\omega_q)} 
   {\sum_{i=1}^m \left[
\frac{dN}{d\omega_q}
\right]_i 
   + 
   \frac{dN^{\mbox{\tiny{direct}}}}{d\omega_q}(\omega_q)},
   \label{v2weightedUrQMD}
\eea
where $dN^{\mbox{\tiny{direct}}}/d\omega_q$ and $v_2^{\mbox{\tiny{direct}}}$ are the ($\omega_q$-dependent) spectrum and second harmonic coefficient of direct photons from Ref.~\cite{RHIC4}, respectively, compered to the experimental data for Au+Au collisions at $\sqrt{s_{NN}}=200$ GeV in the 20-40\% centrality class from Ref.~\cite{experimentsyield}.

For the sake of completeness, the magnetic field generated by both spectators and participants+spectators is also shown. Notice that the magnetic field contribution improves the agreement with experimental data for the low part of the spectrum helping to describe the rise of $v_2$ as the photon energy decreases. The results indicate that a magnetic field created by participants+spectators are far from the central value of the experimental data, but it gives a better description of most data points at low transverse momentum. As it was mentioned in Sec.~\ref{sec:Prompt Photon Production by Gluon Fusion in a Magnetized Medium}, the model cannot reproduce the data at higher photon energies, given that the formalism in the presented form assumes that the magnetic field is the dominant energy scale. Nonetheless, the addition of processes driven by magnetic fields allows enhancing the hydrodynamical predictions. The current calculations can be improved by relaxing the named approximation and by considering an arbitrary direction of the magnetic field as well as a better glasma occupation function. 

\section{Conclusion}

In this chapter, the photon production from gluon fusion and gluon splitting was proposed as a magnetic-field driven channel to improve the current state-of-art of the so-called {\it photon puzzle} presented in Sec.~\ref{SEC:The_photon_puzzle}. The collision's early stages provide the phenomenology, which allows proposing such a mechanism: as it was exposed in Sec.~\ref{Sec:The_Color_Glass_Condensate}, the ultra-relativistic limit reached by the colliding ions implies that gluons dominate the parton distribution function inside the constituent hadrons. Such gluons with small longitudinal momentum fractions (high-energy regime) are promoted to the mass-shell by processes involving larger virtualities. Therefore, a perturbative analysis of real particles is allowed. Moreover, as shown in Sec.~\ref{sec:Magnetic_field_HIC}, the production of intense magnetic fields at the beginning of the collision makes the Color Glass Condensate a strongly magnetized medium. The scales associated with the production of a large number of small momentum gluons coming from the shattering of the glasma. Additionally, a magnetic field's existence gives an initial asymmetry that evolves to an azimuthal anisotropy encoded in an appreciable elliptic flow at low photon momenta. 

By working to lowest order in perturbation theory, the channels presented in Figs.~\ref{Fig:Diag1} and~\ref{Fig:Diag2}, together with the assumption that in a strong magnetic field, the virtual quarks remain in the lowest and first-excited Landau levels, the probability amplitude of each process is computed analytically. It is worth to mention that the magnetic field is chosen as the dominant energy scale so that the approximation $2|q_fB|\gg t_\parallel^2,\ s_\parallel^2,\ r_\parallel^2$ is applied. The invariant momentum distribution and the elliptic flow were computed so that the former has to be understood as a yield over the hydrodynamical calculations, and the latter is an average-weighted term. 

In the first approximation, the magnetic field and the reaction volume were taken as constants, and an external parameter introduces the centrality. Figure~\ref{Fig:yield_and_v2_beta_0} shows that photons coming from the gluon fusion improve the hydrodynamic's framework's theoretical predictions considerably (within the error bar uncertainties). If a flow velocity factor is introduced in order to account for the medium expansion, the agreement with the experimental data becomes better, as shown in Fig.\ref{Fig:yield_and_v2_beta_025}. However, given the strong-field approximation, the theoretical curve falls to describe the data above $\pt>1$ GeV. That may be because the gluon distribution does not contain a power--like tail, which is known better to define the numerical solutions for this kind of distribution. 

In order to obtain predictions in a more realistic scenario, in Sec.~\ref{sec:gluon_splitting}, the space-time evolution of the magnetic field, the reaction volume, and their centrality dependence were simulated with the UrQMD model. Additionally, the gluon splitting mechanism is added to the gluon fusion, given that such diagrams belong to the same order in perturbation theory. As it was expected, the simulated magnetic field is found to be a magnitude around $1m_\pi^2\lesssim\eB\lesssim 4m_\pi^2$, and a lifetime around 0.2 fm. The centrality and the colliding species are fundamental to the magnetic field's strength and the number of participants, as Figs.~\ref{FIG:Bxyz} and~\ref{Fig:Npart_and_Volume} suggest. The results of Figs.~\ref{Fig:YieldComparisonUrQMD} and~\ref{Fig:v2UrQMDFinal} show again an agreement for the lower part of the spectra, and it is better for peripheral collisions. The comparison improves when the magnetic field strength includes the contribution of both spectators and participants. Hence, the magnetic field-driven photon emission for a magnetized and dense glasma appears to be an excellent candidate for better explaining the experimental data within a clear and well-accepted phenomenology. 

At this point, it is necessary to mention that the presented calculation still has to be improved. First, the gluon occupation number may depend on the space-temporal coordinates and have a functional form beyond a simple Bose-Einstein like statistics. Moreover, the approximation fails when considering a magnetic field with constant orientation: the UrQMD simulations show a non-negligible magnetic field contribution at other directions for some centralities. Also, the condition $2|q_fB|\gg t_\parallel^2,\ s_\parallel^2,\ r_\parallel^2$ have to be relaxed to calculate the matrix elements. Those improvements are being currently explored and will be reported elsewhere.

\begin{savequote}[75mm]
``Is the purpose of theoretical physics to be no more than a cataloging of all the things that can happen when particles interact with each other and seperate? Or is it to be an understanding at a deeper level in which there are things that are not directly observable (as the underlying quantized fields are) but in terms of which we shall have a more fundamental understanding?''
\qauthor{--- Julian Schwinger}
\end{savequote}

\chapter{Gluon Polarization Tensor in Magnetized and Thermomagnetic Media}\label{Chap:Gluon_Pol_Tensor}

In the last chapter, the photon production from gluon fusion and gluon splitting in a magnetized medium seems essential to improve the experimental data. The core of such a description is the virtual quarks interaction with the external magnetic field, which opens the particle production channel.  From that point of view, before a gluon fuses or splits, such particles can be strongly affected by the magnetic field since virtual quarks can drive processes with a single initial and final gluon. In particular, the self-energy or polarization tensor leads to modifications in the dispersion relation. The existence of a magnetic mass breaks the usual mass-shell condition, which is crucial for the analytic calculation of the probability amplitude. Moreover, the validity of Eq.~(\ref{energymomentumconservation}) which ensures that the three-momentum for gluons and photons has to be parallel is no longer valid; therefore, one can expect more involved integrals for the photon invariant momentum distribution and the elliptic flow. 

The polarization tensor encodes the interaction of the gluon (and photon) with the magnetized medium in a perturbative approximation. Attempts to compute the self-energy were performed in the strong field limit: the fermion propagator is written as a sum of Landau Levels so that only the lowest Landau level contributes to the tensor~\cite{PhysRevD.83.111501,PhysRevD.94.114034}. Such calculation is analytical and finite. On the other hand, when the magnetic field is not the dominant energy scale, all the Landau levels must be taken into account. These kinds of calculations are approximately performed with numerical methods by truncating the sum at a desired order in the magnetic field~\cite{hattori2013vacuum,Ishikawa:2013fxa}. However, to gain a more in-depth insight, an analytical approach for the infinite sum over Landau levels is desirable.

Although the gluon abundance in thermal stages of heavy-ion collisions is lower than in the glasma state, the thermal corrections to the gluon mass encoded in the polarization tensor are also interesting, which has been computed and extensively studied both at zero and finite temperature~\cite{PhysRevD.100.096006,PhysRevD.63.073010,dittrich1985effective,hattori2018gluon}. Nonetheless, in a thermalized or magnetic medium, the energy scale comparison can also be given in terms of the square of the gluon momentum components as well as on the fermion mass. The boost and Lorentz invariance breaks imply a tensor with dependence on the parallel or transverse momentum separately. Thus, the magnetic mass can be different if some direction's momentum is small compared with other projections. 

This chapter is divided into two parts: the first one is dedicated to computing the gluon polarization tensor in a magnetized medium by performing the analytic sum over all Landau levels. The tensor structure is reorganized, obtaining a self-energy written in terms of a transverse basis in the sense of the Ward-Takahashi identity.  Such manipulation leads to the appearance of spurious terms that are not transverse, and they are appropriately eliminated. The existence of spurious tensors implies that the usual projection method can be plagued with non-physical contributions. Finally, the gluon polarization tensor is computed analytically for the strong and weak field limits. 

The second part is devoted to computing the Debye mass or modifications on the gluon dispersion relation. In that calculation, the gluon is assumed to be in a thermalized and strongly magnetized medium. The analysis is performed by considering hierarchies between the fermion mass, the momentum squared, the temperature, and the magnetic field, namely $p^2\ll m_f^2\ll T^2$ and $m_f^2\ll p^2\ll T^2$. The results show that depending on which momentum component goes to zero first, the tensorial basis's coefficients are different. 

The information presented here is based on the findings of the following articles~\cite{PhysRevD.101.036016,RMFGluon,ayala2020gluon}:
\begin{itemize}
    \item Alejandro Ayala, Jorge David Casta\~no-Yepes, M. Loewe and Enrique Mu\~noz, {\it Gluon polarization tensor in a magnetized medium: Analytic approach starting from the sum over Landau levels}, Phys. Rev. D \textbf{101}, 036016 (2020).
    
    \item Alejandro Ayala, Jorge Casta\~no-Yepes, C. A. Dominguez, S. Hernandez-Ortiz, L. A. Hernandez, M. Loewe, D. Manreza Paret, and R. Zamora, {\it Thermal corrections to the gluon magnetic Debye mass}, Rev. Mex. Fis. {\bf 66} (4), 446–-461 (2020).
    
    \item Alejandro Ayala, Jorge David Casta\~no-Yepes, L. A. Hern\'andez, Jordi Salinas, R. Zamora.{\it Gluon polarization tensor and dispersion relation in a weakly magnetized medium.}. preprint arXiv:2009.00830.
\end{itemize}
\section{Magnetized Gluon Polarization Tensor from the Sum of All Landau Levels}\label{sec:Magnetized_Gluon_Polarization_Tensor_from_the_Sum_of_All_Landau_Levels}
%
\begin{figure}[H]
    \centering
    \includegraphics[scale=0.33]{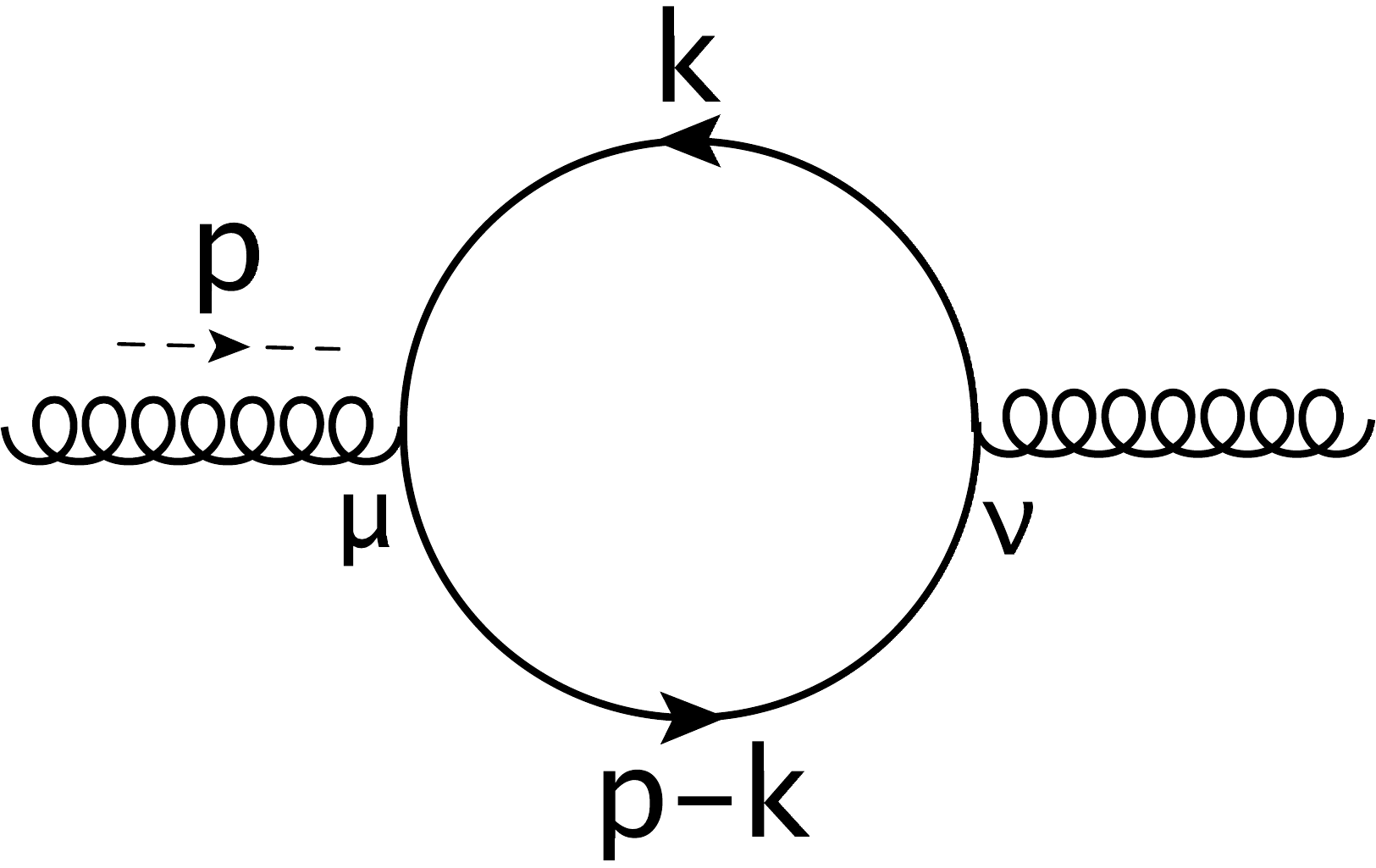}
    \caption{One-loop diagram representing the gluon polarization
tensor.}
    \label{diagrama}
\end{figure}

Starting from from the one-loop contribution to the gluon polarization tensor, which is depicted in Fig.~\ref{diagrama}, the object to be calculated is given explicitly by
\bea
i\Pi^{\mu\nu}_{ab}=-\frac{1}{2}\int\frac{d^4k}{\dpi^4}\text{Tr}\left\{igt_b\gamma^\nu iS^{(n)}(k)igt_a\gamma^\mu iS^{(m)}(q)\right\}+{\mbox{C.C.}},
\label{Pidef}
\eea
where  $g$ is the strong coupling, and C.C. refers to the {\it charge conjugate} contribution, that is, the contribution where the flow of charge within the loop is in the opposite direction. The factor $1/2$ accounts for the symmetry factor, which in the presence of the external magnetic field comes about given that the two contributing diagrams, with the opposite flow of charge, are not equivalent.

The quark propagator is denoted by $S(k)$, and $t_{a,b}$ are the generators of the color group in the fundamental representation. The fermion propagator in the presence of a magnetic field $\vec{B}=B\hat{z}$ can be written in terms of a sum over Landau levels in the same way that in Eq.~(\ref{fermionpropdef}), which has the form~\cite{miranskypropagador1,miranskypropagador2}:
\bea
iS(p)=i e^{-\pt^2/\eB}\sum_{n=0}^{+\infty}(-1)^n\frac{D_n(q_fB,p)}{p_\p^2-m_f^2-2n\left |  q_fB\right |},
\label{fermionpropdefPi}
\eea
where
$m_f$ and $q_f$ are the quark mass and electric charge, respectively, and
\bea
D_n(q_fB,p)&=&2(\psh_\p+m_f)\mathcal{O}^{-}L_n^0\left(\frac{2\pt^2}{\left | q_fB \right|}\right)-2(\psh_\p+m_f)\mathcal{O}^{+}L_{n-1}^0\left(\frac{2\pt^2}{\left | q_fB \right |}\right)\nn\\
&+&4\psh_\perp L_{n-1}^1\left(\frac{2\pt^2}{\left | q_fB \right |}\right).
\label{DnPi}
\eea
with $L_n^\alpha(x)$ the generalized Laguerre polynomials whose index $n$ label the $n$-th Landau level, and 
\bea
\mathcal{O}^{(\pm)}=\frac{1}{2}\left[1\pm i\gamma^1\gamma^2\text{sign}(q_f B)\right].
\label{Op}
\eea

The convention for the squared four-momentum $p^\mu$, expressed in terms of the square of its parallel and perpendicular (with respect to the magnetic field direction) components, is given by 
\bea
   p^2=p_\parallel^2-\pt^2=(p_0^2-p_3^2)-(p_1^2+p_2^2).
\eea

By replacing Eqs. (\ref{Pidef}) and (\ref{fermionpropdefPi}), after performing the sum over all Landau levels, the gluon polarization tensor can be written in terms of four tensor structures, given by
\bea
\!\!\!\!\!\!i\Pi^{\mu\nu}=
\factorglobal f_0\left(x_1,x_2\right)\sum_{i=1}^4f_i^{\mu\nu}(x_1,x_2),
\label{tensorfirst}
\eea
where on the right-hand side, the factor $\delta_{ab}$ coming from using the relation $\Tr (t^at^b)=\delta_{ab}/2$ was ommited, and correspondingly, for notation simplicity, removed the color indices on the left-hand side. Moreover, $(x_1,x_2)\in(0,\infty)$ are Schwinger parameters, with $d^2x=dx_1dx_2$ and 
\begin{subequations}
\bea
f_0\left(x_1,x_2\right)&=&\exp\left[\frac{x_1 x_2}{x_1+x_2}\pp^2-m_f^2(x_1+x_2)\right]\nn\\
&\times&\exp\left[-\frac{\tanh(\eB x_1)\tanh(\eB x_2)}{\tanh(\eB x_1)+\tanh(\eB x_2)}\frac{\pt^2}{\eB}\right],
\label{f0}
\eea

\bea
f_1^{\mu\nu}(x_1,x_2)&=&\eB\coth\left[\eB(x_1+x_2)\right]\nn\\
&\times&\left[\left(\frac{x_1x_2}{(x_1+x_2)^3}\pp^2+\frac{m_f^2}{x_1+x_2}\right)\gmn_\parallel-\frac{2x_1x_2}{(x_1+x_2)^3}\pmu_\parallel\pnu_\parallel\right],\nn\\
\label{f1}
\eea

\bea
f_2^{\mu\nu}(x_1,x_2)&=&\eB\frac{\cosh\left[\eB(x_2-x_1)\right]}{\sinh\left[\eB(x_1+x_2)\right]}\left[\frac{x_1x_2}{(x_1+x_2)^3}\pp^2+\frac{m_f^2}{x_1+x_2}+\frac{1}{(x_1+x_2)^2}\right]\gmn_\perp,\nn\\
\label{f2}
\eea

\bea
f_3^{\mu\nu}(x_1,x_2)&=&\frac{\eB}{2(x_1+x_2)^2\sinh^2\left[\eB (x_1+x_2)\right]}\nn\\
&\times&\Big[x_1\sinh(2\eB x_2)+x_2\sinh(2\eB x_1)\Big]\left(\pmu_\parallel\pnu_\perp+\pnu_\parallel\pmu_\perp\right),
\label{f3}
\eea

\bea
f_4^{\mu\nu}(x_1,x_2)&=&\frac{\eB^2}{(x_1+x_2)\sinh^2\left[\eB(x_1+x_2)\right]}\nn\\
&\times&\left[
\left(1-\frac{\tanh(\eB x_1)\tanh(\eB x_2)}{\eB\left[\tanh(\eB x_1)+\tanh(\eB x_2)\right]}\pt^2\right)\gmn\right.\nn\\
&-&\gmn_\perp-\left.\frac{2\tanh(\eB x_1)\tanh(\eB x_2)}{\eB\left[\tanh(\eB x_1)+\tanh(\eB x_2)\right]}\pmu_\perp\pnu_\perp\right].
\label{f4}
\eea
\label{fs}
\end{subequations}
For calculation details, see Appendix~\ref{ApAGluonPolTensorPRD}.

\subsection{Tensor Basis}
In order to give a direct physical interpretation of the obtained results, the correct choice of a tensorial basis is fundamental. That basis has to obey the following fundamental properties:
\begin{itemize}
    \item All the tensors should be symmetric under the exchange of its Lorentz indices.
    
    \item It has to be constructed from external products of independent four-vectors which are related to the physical observables. 
\end{itemize}

From the above, the gluon polarization tensor will be described in terms of the gluon four-momentum $p^\mu$, the magnetic field direction $b^\mu$, and the metric tensor $g^{\mu\nu}$. Given that there are no more dynamical variables in the present problem, the named vectors have to be enough to construct the tensorial basis. 

The propagator of Eq.~(\ref{fermionpropdefPi}) comes from an analysis in where the external magnetic field is assumed to be oriented in the $z$-axis so that without loss of generality one can choose along this direction. Moreover, due to the breaking of Lorentz invariance by the vector $b^\mu$ (magnetic field), it is convenient to split the metric itself into parallel and perpendicular (with respect to the magnetic field direction) components, that is
\bea
g^{\mu\nu}=g^{\mu\nu}_\parallel + g^{\mu\nu}_\perp,
\label{metric}
\eea
where 
\bea
g^{\mu\nu}_\parallel={\mbox{diag}}(1,0,0,-1),
\label{metricpara}
\eea
and
\bea
g^{\mu\nu}_\perp={\mbox{diag}}(0,-1,-1,0).
\label{metricperp}
\eea

Thus, the most general symmetric tensor can be constructed out of combinations of the independent tensors
\bea
p^\mu p^\nu,\  b^\mu b^\nu,\ p^\mu b^\nu + p^\nu b^\mu,\ g^{\mu\nu}.
\label{possible}
\eea

However, notice that in QCD, $\Pi^{\mu\nu}$ must satisfy the generalized Ward-Takahashi identity namely, the {\it transversality} condition~\cite{peskin1995quantum}
\bea
p_\mu p_\nu \Pi^{\mu\nu}=0.
\label{WTI}
\eea

Therefore, since Eq.~(\ref{WTI}), implies a relation between the coefficients of the tensors to express $\Pi^{\mu\nu}$, only three {\it transverse} tensors turn out to be independent. Explicitly, the polarization tensor can be written as
\bea
\Pi^{\mu\nu} = a\ A^{\mu\nu} + b\ B^{\mu\nu} + c\ C^{\mu\nu} + d\ D^{\mu\nu}.
\eea
where the constants in front of each tensorial structure are determined from the gauge invariance of Eq.~(\ref{WTI}), i.e.,
\bea
p_{\mu} p_{\nu} \Pi^{\mu\nu} &=& a \left(p_{\mu} p_{\nu} A^{\mu\nu}\right)+ b \left(p_{\mu} p_{\nu} B^{\mu\nu}\right)\nn\\
&+& c \left(p_{\mu}p_{\nu}C^{\mu\nu}\right)+ d \left(p_{\mu} p_{\nu} D^{\mu\nu}\right) = 0.
\label{newimply}
\eea

The latter means that only three of the four constants $(a,b,c,d)$ have to be independent of each other, thus, one of the tensors of Eq.~(\ref{newimply}) is distributed among the rest of the tensor structures. That procedure results in only three tensors being needed to span the whole tensor $\Pi^{\mu\nu}$. A useful orthonormal basis where each element is transverse to satisfy the condition of Eq.~(\ref{WTI}) in a QED-like form ($p_\mu \Pi^{\mu\nu}=0$) is 
\bea
\mathcal{P}_{\p}^{\mu\nu}=\gmn_{\p}-\frac{p_\p^\mu p_\p^\nu}{p_\p^2},
\label{pipara}
\eea
\bea
\mathcal{P}_{\perp}^{\mu\nu}=\gmn_{\perp}+\frac{\pt^\mu \pt^\nu}{\pt^2},
\label{piperp}
\eea
\bea
\mathcal{P}_{0}^{\mu\nu}=\gmn-\frac{p^\mu p^\nu}{p^2}-\mathcal{P}_{\p}^{\mu\nu}-\mathcal{P}_{\perp}^{\mu\nu}.
\label{pizero}
\eea

The selected basis can be regarded as a natural factorization of the whole space into parallel and transverse structures, with respect to the magnetic field four-vector $b^\mu$. To illustrate this, without loss of generality, the magnetic field direction can be chosen as 
\bea
b^{\mu}=(0,\mathbf{b})=(0,0,0,1),
\eea
so that
\bea
\mathbf{b}=\frac{1}{2}\nabla\times(-y,x,0)\equiv\nabla\times\mathbf{a},
\eea
and therefore, by written the vector potential as $\mathbf{A}=B\mathbf{a}$, from the well-known definition $F^{\mu\nu}=\partial^\mu A^\nu -  \partial^\nu A^\mu $, the following relation is found:
\bea
p_{\alpha} p_{\beta} F^{\alpha\mu} F^{\beta\nu} =B^2\begin{pmatrix}
0 & 0 & 0 &0 \\ 
0 & p_2^2 & -p_1 p_2 &0 \\ 
0 &- p_1 p_2 & p_1^2 &0 \\ 
0 & 0 &0  & 0
\end{pmatrix}.
\eea

Also, from Eq.~(\ref{piperp})
\bea
\pt^2\Pt=\begin{pmatrix}
0 & 0 & 0 &0 \\ 
0 & -p_2^2 & p_1 p_2 &0 \\ 
0 &p_1 p_2 &- p_1^2 &0 \\ 
0 & 0 &0  & 0
\end{pmatrix},
\eea
therefore
\bea
B^2\pt^2\Pt=-p_{\alpha} p_{\beta} F^{\alpha\mu} F^{\beta\nu},
\eea
which shows that the choice of $b^\mu$ impacts directly the factorization of the metric into transverse and a parallel structures. Therefore, this basis can be used to  rearrange the tensor structures of Eqs.~(\ref{fs}) (see Appendix~\ref{ApBGluonPolPRD}) as
\bea
i\Pi^{\mu\nu}&=&-\frac{i}{8\pi^2}g^2\int\,d^2x\, f_0(x_1,x_2)\Bigg[\Pi_\parallel\left(x_1,x_2\right)\Pp+\Pi_\perp\left(x_1,x_2\right)\Pt+\Pi_0\left(x_1,x_2\right)\Pcero\Bigg]\nn\\
&-&\frac{i}{8\pi^2}g^2\int\,d^2x\,f_0(x_1,x_2)\Bigg[A_1\left(x_1,x_2\right)\gmn_\parallel+A_2\left(x_1,x_2\right)\gmn_\perp\Bigg],
\label{Pienbaseortonormal}
\eea
where 
\bea
\!\!\!\!\Pi_\parallel=\eB\Bigg[\frac{2x_1x_2\coth\left[\eB(x_1+x_2)\right]}{(x_1+x_2)^3}\pp^2-\frac{x_1\sinh(2\eB x_2)+x_2\sinh(2\eB x_1)}{2(x_1+x_2)^2\sinh^2\left[\eB (x_1+x_2)\right]}\pt^2\Bigg]\nn\\
\label{Pipara}
\eea
\bea
\Pi_\perp&=&\eB\Bigg[\frac{x_1\sinh(2\eB x_2)+x_2\sinh(2\eB x_1)}{2(x_1+x_2)^2\sinh^2\left[\eB (x_1+x_2)\right]}\pp^2\nn\\
&-&\frac{2\sinh(\eB x_1)\sinh(\eB x_2)}{(x_1+x_2)\sinh^3\left[\eB(x_1+x_2)\right]}\pt^2\Bigg]\nn\\
\label{Piperp}
\eea
\bea
\Pi_0=\eB\frac{x_1\sinh(2\eB x_2)+x_2\sinh(2\eB x_1)}{2(x_1+x_2)^2\sinh^2\left[\eB (x_1+x_2)\right]}p^2,
\label{Picero}
\eea
\bea
A_1&=&\eB\Bigg[\frac{x_1\sinh(2\eB x_2)+x_2\sinh(2\eB x_1)}{2(x_1+x_2)^2\sinh^2\left[\eB (x_1+x_2)\right]}\pt^2\nn\\
&+&\frac{\coth\left[\eB(x_1+x_2)\right]}{(x_1+x_2)^3}\left(m_f^2(x_1+x_2)^2-x_1x_2\pp^2\right)\nn\\
&+&\frac{\eB}{(x_1+x_2)\sinh^2\left[\eB(x_1+x_2)\right]}\nn\\
&\times&\left(1-\frac{\tanh(\eB x_1)\tanh(\eB x_2)}{\eB\left[\tanh(\eB x_1)+\tanh(\eB x_2)\right]}\pt^2\right)\Bigg],
\label{coefA}
\eea
and
\bea
A_2&=&\eB\Bigg[\frac{\cosh\left[\eB(x_2-x_1)\right]}{(x_1+x_2)^3\sinh\left[\eB(x_1+x_2)\right]}\left[x_1x_2\pp^2+(x_1+x_2)+m_f^2(x_1+x_2)^2\right]\nn\\
&-&\frac{x_1\sinh(2\eB x_2)+x_2\sinh(2\eB x_1)}{2(x_1+x_2)^2\sinh^2\left[\eB (x_1+x_2)\right]}\pp^2+\frac{\sinh(\eB x_1)\sinh(\eB x_2)}{(x_1+x_2)\sinh^3\left[\eB(x_1+x_2)\right]}\pt^2\Bigg].\nn\\
\label{coefB}
\eea

Equation~(\ref{Pienbaseortonormal})  has a serious issue regarding the appearance of terms that are proportional to $g_\parallel^{\mu\nu}$ and $g_\perp^{\mu\nu}$, which implies a non-transverse structure for the polarization tensor. That situation makes that any calculation based on the projection procedure to compute the coefficients corresponding to each tensor structure is plagued with spurious terms, and therefore, a numerical evaluation of $\Pi^{\mu\nu}$ cannot give the truly physical coefficients~\cite{hattori2013vacuum,hattori2018gluon,Ishikawa:2013fxa}. Explicitly, a projection method yields
\begin{subequations}
\bea
\widetilde{\Pi}_\parallel=\Pi_{\mu\nu}\Pp=\Pi_\parallel+A_1,
\label{PiparaconA1}
\eea
\bea
\widetilde{\Pi}_\perp=\Pi_{\mu\nu}\Pt=\Pi_\perp+A_2,
\label{PiperpconA2}
\eea
and
\bea
\widetilde{\Pi}_0=\Pi_{\mu\nu}\Pcero=\Pi_0-\frac{\pt^2}{p^2}A_1+\frac{\pp^2}{p^2}A_2,
\label{PiceroconA1A2}
\eea
\end{subequations}
where $\Pi_\parallel,\Pi_\perp$ and $\Pi_0$ are given by Eqs.~(\ref{Pipara})-(\ref{Picero}) and $A_1,A_2$ are given by Eqs.~(\ref{coefA})-(\ref{coefB}). Thus, it is necessary to show that $A_1$ and $A_2$ vanishes.

\subsection{Elimination of spurious tensors}
The elimination the so-called spurious terms  is done by following the procedure shown in Ref.~\cite{bjorken1965relativistic}. First, the $x$ parameters are scaled in the form
\bea
x_i\rightarrow\lambda z_i,
\eea
with $(\lambda, \,z_i)\in \mathbb{R}$. Therefore, the integral that involves the coefficient $A_1$ can be wrtitten as
\bea
\mathcal{I}_{A_1}&=&-\lambda\frac{\partial}{\partial\lambda}\int\frac{d^2z}{(z_1+z_2)^2}\coth\left[\lambda\eB(z_1+z_2)\right]\exp\left[\lambda\left(\frac{z_1 z_2}{z_1+z_2}\pp^2-m_f^2(z_1+z_2)\right)\right]\nn\\
&\times&\exp\left[-\frac{\tanh(\lambda\eB z_1)\tanh(\lambda\eB z_2)}{\tanh(\lambda\eB z_1)+\tanh(\lambda\eB z_2)}\frac{\pt^2}{\eB}\right].
\eea

Scaling back $\lambda\,z_1\rightarrow x_i$ gives:
\bea
\mathcal{I}_{A_1}&=&-\lambda\frac{\partial}{\partial\lambda}\int\frac{d^2x}{(x_1+x_2)^2}\coth\left[\eB(x_1+x_2)\right]\exp\left[\frac{x_1 x_2}{x_1+x_2}\pp^2-m_f^2(x_1+x_2)\right]\nn\\
&\times&\exp\left[-\frac{\tanh(\eB x_1)\tanh(\eB x_2)}{\tanh(\eB x_1)+\tanh(\eB x_2)}\frac{\pt^2}{\eB}\right],
\eea
and thus, the derivative is applied to a function independent of $\lambda$. Therefore $\mathcal{I}_{A_1}=0$.

The implementation of the same argument for $\mathcal{I}_{A_2}$ is more involved, given that the function is not a trivial combination of coefficients for $\pp^2$ and $\pt^2$. After the $\lambda$-scaling, the integral is
\bea
\mathcal{I}_{A_2}=\lambda^2\int\frac{d^2z}{\lambda(z_1+z_2)^2}I(\lambda z_1,\lambda_2),
\eea
where
\bea
I(\lambda z_1,\lambda z_2)&=&f_0(\lambda z_1,\lambda z_2) \Bigg{[}\frac{\cosh\left[\lambda\eB(z_2-z_1)\right]}{\sinh\left[\lambda\eB(z_1+z_2)\right]}\left(\frac{z_1z_2}{z_1+z_2}\pp^2+m_f^2(z_1+z_2)+\frac{1}{\lambda}\right)\nn\\
&-&\frac{z_1\sinh(2\lambda\eB z_2)+z_2\sinh(2\lambda\eB z_1)}{2\sinh^2\left[\lambda\eB (z_1+z_2)\right]}\pp^2\nn\\
&+&\frac{(z_1+z_2)\sinh(\lambda\eB z_1)\sinh(\lambda\eB z_2)}{\sinh^3\left[\lambda\eB(z_1+z_2)\right]}\pt^2\Bigg{]},
\eea
from which by expanding in a Taylor series around $\lambda=0$ it is possible to find that
\bea
&&\int I(\lambda z_1,\lambda z_2) d\lambda=-\frac{1}{\eB(z_1+z_2)\lambda}\nn\\
&+&\frac{2\eB^2(z_1+z_2)^2(z_1^2-4z_1z_2+z_2^2)+3\left(z_1z_2p^2-m^2(z_1+z_2)^2\right)^2}{6\eB (z_1+z_2)^3}\lambda\nn\\
&+&\frac{\lambda^2}{6\eB (z_1+z_2)^4}\Bigg{[}\left(3p^4z_1^2z_2^2-3m^2p^2z_1z_2(z_1+z_2)^2+m^4(z_1+z_2)^4\right)(z_1+z_2)^2m^2\nn\\
&-&z_1z_2\left(p^6z_1^2z_2^2-2\eB^2(z_1+z_2)^2\left(p^2(z_1-z_2)^2-\pt^2z_1z_2\right)\right)\Bigg{]}\nn\\
&+&\frac{\lambda^3}{1080\eB(z_1+z_2)^5}\Bigg{[}45\left(p^2z_1z_2-m^2(z_1+z_2)^2\right)^4\nn\\
&+&8\eB^4(z_1+z_2)^4\left(z_1^4+4z_1^3z_2-24z_1^2z_2^2+4z_1z_2^3+z_2^4\right)\nn\\
&-&60\eB^2(z_1+z_2)^2\left(p^2z_1z_2-m^2(z_1+z_2)^2\right)\left(m^2(z_1+z_2)^2(z_1^2-4z_1z_2+z_2^2)\right.\nn\\
&+&\left.z_1z_2\left(p^2(3z_1^2-4z_1z_2+3z_2^2)-6\pt^2z_1z_2\right)\right)\Bigg{]}+\mathcal{O}\left(\lambda^4\right),
\eea
where the desired scaling properties are recovered and hold for all orders in $\lambda$. This means that is possible to write
\bea
\int I(\lambda z_1,\lambda z_2) d\lambda&=&-\frac{1}{\eB(z_1+z_2)\lambda}+h\left(\lambda z_1,\lambda z_2\right),\nn\\
\eea
thus
\bea
I(\lambda z_1,\lambda z_2)&=&\frac{\partial}{\partial\lambda}\left[-\frac{1}{\eB(z_1+z_2)\lambda}+h\left(\lambda z_1,\lambda z_2\right)\right]\nn\\
&=&\frac{\partial}{\partial\lambda}\left[-\frac{1}{\eB(x_1+x_2)}+h\left(x_1, x_2\right)\right]=0,
\eea
and therefore, $\mathcal{I}_{A_2}=0$.

The above argument is valid for all values of $\lambda$. Consequently, the result can be taken as general.

\subsection{Obtaining the Vacuum Polarization Tensor}
It is well established that the polarization tensor in the vacuum contains divergences that have to be isolated from the matter contribution. The simplest way to eliminate such singularities is by identifying the vacuum contribution in the appropriate limits. To perform this, note that at least two vacua can be defined, in terms so f the physical quantities of the problem: 

\begin{itemize}
    \item A vacuum where $p^\mu=0$ and $B=0$, corresponding to a situation where particles and magnetic field appear simultaneously. 
    
    \item A vacuum with $B=0$ and $p^\mu\neq0$, representing a situation where the external field is turned on with pre-existing gluons with four-momentum $p^\mu$.
\end{itemize}

By looking at the computed equations for the polarization tensor, it can be noticed that the magnetic field and the transverse momentum appear within the combination $\pt^2/\eB$ and thus, $\pt^2$ and $B$ cannot be set to zero simultaneously. For this reason, in this work the second vacuum definition is adopted, so that from Eq.~(\ref{Pienbaseortonormal}):
\bea
&&i\Pi^{\mu\nu}(p,\eB\rightarrow0)=-\frac{i}{8\pi^2}g^2\int d^2x\exp\left[\frac{x_1x_2}{x_1+x_2}p^2-m_f^2(x_1+x_2)\right]\nn\\
&\times&\left[\frac{2x_1x_2}{(x_1+x_2)^4}p^2\left(\gmn-\frac{\pmu\pnu}{p^2}\right)+\frac{1}{(x_1+x_2)^3}\left((x_1+x_2)m_f^2-\frac{x_1x_2}{x_1+x_2}p^2+1\right)g^{\mu\nu}\right].\nn\\
\label{vacuum}
\eea

The last equation points out to the fact that the elimination of spurious tensors is of great importance, given that even in the vacuum the tensor is not transverse. The method to eliminate such contributions also applies to Eq.~(\ref{vacuum}), i.e., by scaling the Schwinger parameters, the coefficient of $g^{\mu\nu}$ is
\bea
\mathcal{I}&=&\lambda^2\int\frac{d^2z}{\lambda^2(z_1+z_2)^3}\left(m^2(z_1+z_2)-\frac{z_1z_2}{z_1+z_2}p^2+\frac{1}{\lambda}\right)\,e^{\lambda\left[\frac{z_1z_2}{z_1+z_2}p^2-m_f^2(z_1+z_2)\right]}\nn\\
&=&-\lambda\frac{\partial}{\partial\lambda}\int \frac{d^2x}{(x_1+x_2)^3}e^{\frac{x_1x_2}{x_1+x_2}p^2-m_f^2(x_1+x_2)}\nn\\
&=&0.
\eea

The above leads to identify the vacuum gluon polarization tensor as
\bea
i\Pi^{\mu\nu}_{\text{vac}}(p)&=&-\frac{ig^2p^2}{8\pi^2}\int d^2x \frac{2x_1x_2}{(x_1+x_2)^4}\exp\left[\frac{x_1x_2}{x_1+x_2}p^2-m_f^2(x_1+x_2)\right]\left(\gmn-\frac{\pmu\pnu}{p^2}\right)\nn\\
&\equiv&-\frac{ig^2}{8\pi^2}\int d^2x\,\Pi_{\text{vac}}(x_1,x_2)\left(\Pcero+\Pp+\Pt\right)
\label{vacuumtrue1}
\eea
so that the finite magnetized gluon polarization tensor reads
\bea
i\Pi^{\mu\nu}&=&-\frac{ig^2}{8\pi^2}\int d^2x\,f_0(x_1,x_2)\Bigg\{\left[\Pi_0\left(x_1,x_2\right)-\widetilde{\Pi}(x_1,x_2)\right]\Pcero\nn\\
&+&\left[\Pi_\parallel\left(x_1,x_2\right)-\widetilde{\Pi}(x_1,x_2)\right]\Pp+\left[\Pi_\perp\left(x_1,x_2\right)-\widetilde{\Pi}(x_1,x_2)\right]\Pt\Bigg\},\nn\\
\label{Pifinalwithvacuum}
\eea
where
\bea
\widetilde{\Pi}(x_1,x_2)\equiv\Pi_{\text{vac}}(x_1,x_2)\exp\left[\frac{\tanh(\eB x_1)\tanh(\eB x_2)}{\tanh(\eB x_1)+\tanh(\eB x_2)}\frac{\pt^2}{\eB}-\frac{x_1x_2}{x_1+x_2}\pt^2\right].
\eea

In the following, the analysis is facilitated by performing the change of variables in Eq.~(\ref{Pifinalwithvacuum}) given by
\begin{subequations}
\bea
x_1\equiv \frac{s(1-y)}{m_f^2},\hspace{0.3cm}x_2\equiv \frac{sy}{m_f^2},
\eea
\label{variablesvys}
\bea
\rho^2_{\parallel,\perp}\equiv\frac{p_{\parallel,\perp}^2}{m_f^2},\hspace{0.2cm}\mathcal{B}\equiv\frac{\eB}{m_f^2},
\eea
\end{subequations}
so that the magnetic contribution of  Eq.~(\ref{Pifinalwithvacuum}) becomes
\bea
&&\!\!\!\!\!\!\!\!\!\!i\Pi^{\mu\nu}_{\text{mag}}=-\frac{ig^2m_f^2\B}{8\pi^2}\int_0^1dy\int_0^{\infty}\,ds\,e^{s\left(y(1-y)\rhop^2-1\right)}\exp\left[-\frac{\cosh(\B s)-\cosh\left[\B s (2y-1)\right]}{2\sinh(\B s)}\frac{\rhot^2}{\B}\right]\nn\\
&\times&\Bigg{\{}\frac{(1-y)\B\sinh(2\B sy)}{\B\sinh^2(\B s)}\rho^2\Pcero+\left[2y(1-y)\coth(\B s)\rhop^2-\frac{(1-y)\sinh(2\B sy)}{\sinh^2(\B s)}\rhot^2\right]\Pp\nn\\
&+&\left[\frac{(1-y)\sinh(2\B sy)}{\sinh^2(\B s)}\rhop^2-\frac{\cosh(\B s)-\cosh\left[\B s (2y-1)\right]}{\sinh^3(\B s)}\rhot^2\right]\Pt\Bigg{\}}.
\label{PienSyY}
\eea

To check the validity of the computed expressions, the strong and weak magnetic field limits are studied.

\subsection{Strong Field Limit}

The polarization tensor in an environment with extreme conditions is of special interest both for theoretical and experimental points of view. Particularly, in the QDE sector, the pair production is expected from the Schwinger effect, i.e., the electron-positron creation when an intense electric field is present~\cite{ringwald2001pair,popov2001schwinger}. Moreover, if the electric field varies in time, a strong magnetic field can be generated which interacts with the fermionic vacuum~\cite{PhysRevD.2.1191}. 

In the QCD sector, as it was established in Chapter~\ref{chap:Photons}, strong magnetic fields are created in the early stages of peripheral heavy-ion collisions where the gluon high occupation state, the {\it glasma}, is achieved. Thus, the gluons which form such overpopulated phase moves in a magnetized medium, so that its cloud of virtual quarks can modify the gluon properties. 

In order to get the consequences of a strong magnetic field on the polarization tensor, note that if $\B$ is the dominant energy scale
\bea
\B\coth(\B s)\sim\B,
\nonumber\\
\frac{\B\sinh\left(2\B sy\right)}{2\sinh^2(\B s)}\sim0,
\nonumber\\
\frac{\cosh(\B s)-\cosh\left[\B s (2y-1)\right]}{2\sinh(\B s)}\sim\frac{1}{2\B}
\eea
which hold for all $s$ and $0<y<1$. Therefore
\bea
&&i\Pi^{\mu\nu}=-\frac{ig^2m_f^2\B\rhop^2}{4\pi^2}e^{-\rhot^2/2\B}\int_{0}^{1}dy\;y(1-y)\int_0^{\infty}\,ds\exp\left[s\left(y(1-y)\rhop^2-1\right)\right]\Pp.\nn\\
\label{Pistrong}
\eea

For the kinematical region such that $y(1-y)\rhop^2<1$, the integration over $s$ can be performed, yielding
\bea
i\Pi^{\mu\nu}&=&\frac{ig^2m_f^2\B}{4\pi^2}e^{-\rhot^2/2\B}\int_0^1dy\frac{y(1-y)}{y(1-y)-\rho_\parallel^{-2}}\Pp\nonumber\\
&\equiv&\frac{ig^2m_f^2\B}{4\pi^2}e^{-\rhot^2/2\B}I(\rho_\parallel^2)\Pp.
\label{resultadoFukushima}
\eea

This result is the same that the one obtained in Ref.~\cite{PhysRevD.83.111501} by considering only the Lowest Landau Level (LLL) contribution of Eq.~(\ref{fermionpropdefPi}), namely, $n=0$. A direct calculation starting from the LLL, shows that the polarization tensor is finite and it has not the vacuum piece given by Eq.~(\ref{vacuumtrue1}).  Moreover, the integral over the variable $y$ can be performed analytically, yielding
\bea
I(x)\equiv\int_0^1dy\frac{y(1-y)}{y(1-y)-x}=1-\frac{4x}{\sqrt{4x-1}}\arctan\left(\frac{1}{\sqrt{4x-1}}\right).
\eea

On the other hand, if the spurious terms were not eliminated, the strong magnetic field limit would be
\bea
i\Pi^{\mu\nu}&=&\frac{ig^2m_f^2\B}{4\pi^2}e^{-\rhot^2/2\B}\left[I(\rho_\parallel^2)+I_1(\rho_\parallel^2)\right]\Pp,
\label{StrongFieldwithA1}
\eea
where
\bea
I_1(x)=\int_0^1dy\frac{1}{y(1-y)x-1}=-\frac{4}{\sqrt{x(x-4)}}\arctan\left(\frac{\sqrt{x}}{\sqrt{4-x}}\right).
\eea
\begin{figure}[H]
    \centering
    \includegraphics[scale=0.5]{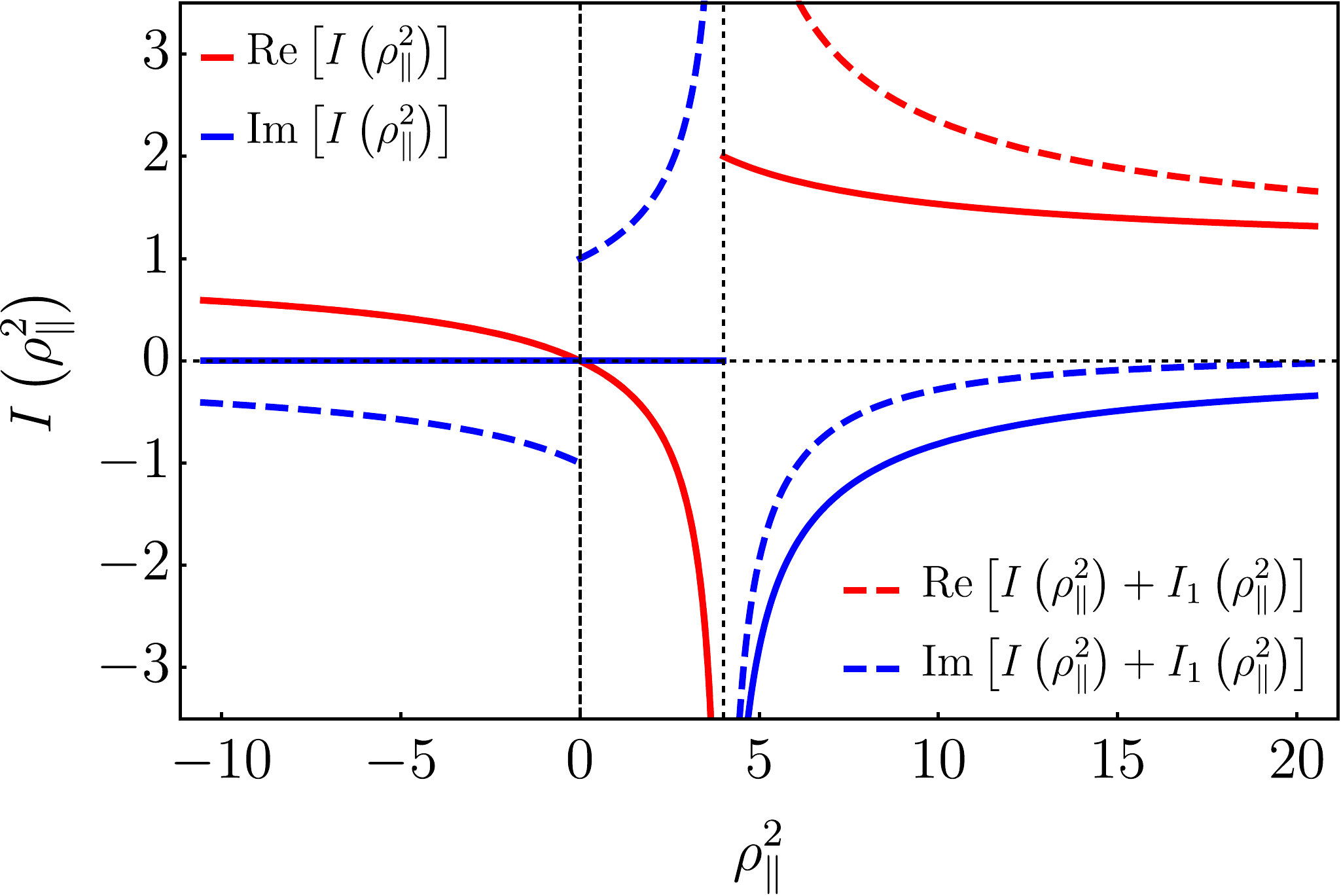}
    \caption{Real and imaginary parts of the function $I(\rho_\parallel^2)$ defined in Eq. (\ref{resultadoFukushima}) compared with the result including the spurious contribution of $I_1(\rho_\parallel^2)$ from Eq.~(\ref{StrongFieldwithA1}).}
    \label{Fig:Fukushima}
\end{figure}

Figure~\ref{Fig:Fukushima} shows the real and imaginary parts of Eq.~(\ref{resultadoFukushima}) compared with the result of Eq.~(\ref{StrongFieldwithA1}) as a function of $\rhop^2$. The former has a discontinuity at the threshold value $\rho_\parallel^2=4$ or equivalently at $p_\parallel^2=4m_f^2$, whereas for the latter this is located at $\rhop^2=0$. Such discontinuities have the interpretation of pair production: the gluon (photon) reaches sufficient energy to decay into a pair quark-antiquark (electron-positron), which in some reference frame has the value $p_0=2m_f$. The main characteristic of this phenomenon is the appearance of an imaginary part in the polarization tensor, related to the decay probability. Thus, it is clear that the threshold of the spurious term located at $\rhop^2=0$ cannot represent a physical situation given the mass-energy conservation.

\subsection{Weak Field Limit}

The hierarchy of energy scales in this limit is taken as $|eB| < m_f^2$, then, a power series of Eq.~(\ref{PienSyY}) around $\B=0$ can be performed, obtaining:
\bea
i\Pi^{\mu\nu}&=&-\frac{ig^2m_f^2\B^2}{8\pi^2}\int_0^1dy\int_0^\infty ds\,\frac{2sy(1-y)}{3}\exp\left[s\left(y(1-y)\rho^2-1\right)\right]\nn\\
&\times&\Bigg{\{}\rho^2\left[sy^2(1-y)^2\rhot^2+2y^2-1\right]\Pcero+\left[sy^2(1-y)^2\rhot^2\rho^2+2(1-y^2)\rhot^2+\rho^2\right]\Pp\nn\\
&+&\left[sy^2(1-y)^2\rhot^2\rho^2+(1+y)\rhot^2-(1-y^2)\rhop^2+y^2\rho^2\right]\Pt\Bigg{\}}\\
&-&\frac{ig^2m_f^2\rho^2}{8\pi^2}\int_0^1dy\int_0^\infty ds\,\frac{2y(1-y)}{s}\exp\left[s\left(y(1-y)\rho^2-1\right)\right]\left(\Pp+\Pt+\Pcero\right),\nn
\label{weakBlimit}
\eea
where the vacuum term is explicitly given by the integral proportional to $\Pp+\Pt+\Pcero$.

The integration over the variables $s$ and $y$ is performed analytically, so that, after subtracting the vacuum contribution, the gluon polarization tensor for a weak magnetic field is given by
\bea
i\Pi^{\mu\nu}_{\text{weak}\;\B}
=-\frac{ig^2m_f^2\B^2}{6\pi^2}\left[\hat{\Pi}_\parallel(\rho^2)\Pp+\hat{\Pi}_\perp(\rho^2)\Pt+\hat{\Pi}_0(\rho^2)\Pcero\right],
\eea
whith
\bea
\hat{\Pi}_\parallel&=&\frac{1}{4-\rho^2}\Bigg[1-\frac{(\rho^2-10)(\rho^2-3)}{4-\rho^2}\frac{\rhot^2}{\rho^4}+\frac{6\rhot^2}{\rho^2}+\Bigg(12\frac{10+(\rho^2-6)\rho^2}{(4-\rho^2)^{3/2}(\rho^2)^{5/2}}\rhot^2\nn\\
&+&2\frac{\rho^2(\rho^2+2)-12}{\sqrt{4-\rho^2}(\rho^2)^{5/2}}\rhot^2+2\frac{(\rho^2-2)\sqrt{(4-\rho^2)\rho^2}}{(4-\rho^2)\rho^2}\Bigg)\arctan\left(\frac{\sqrt{\rho^2}}{\sqrt{4-\rho^2}}\right)\Bigg],\nn\\
\label{Piparaweakfield}
\eea

\bea
\hat{\Pi}_\perp&=&\frac{1}{4-\rho^2}\Bigg[\frac{3\rhot^2}{2\rho^2}-\frac{(\rho^2-10)(\rho^2-3)}{4-\rho^2}\frac{\rhot^2}{\rho^4}-\frac{3\rhop^2}{\rho^4}+\frac{(\rho^2-3)\sqrt{(4-\rho^2)\rho^2}}{\sqrt{4-\rho^2}(\rho^2)^{3/2}}\nn\\
&+&\Bigg(12\frac{10+(\rho^2-6)\rho^2}{(4-\rho^2)^{3/2}(\rho^2)^{5/2}}\rhot^2+3\frac{(\rho^2-2)\sqrt{(4-\rho^2)\rho^2}}{(4-\rho^2)\rho^4}\rhot^2-\frac{\rho^2(\rho^2+2)-12}{\sqrt{4-\rho^2}(\rho^2)^{5/2}}\rhop^2\nn\\
&+&\frac{12+(\rho^2-6)\rho^2}{\sqrt{4-\rho^2}(\rho^2)^{3/2}}\Bigg)\arctan\left(\frac{\sqrt{\rho^2}}{\sqrt{4-\rho^2}}\right)\Bigg],
\label{Piperpweakfield}
\eea

\bea
\hat{\Pi}_0&=&\frac{1}{4-\rho^2}\Bigg[2\frac{(\rho^2-3)\sqrt{(4-\rho^2)\rho^2}}{\sqrt{4-\rho^2}(\rho^2)^{3/2}}-\frac{(\rho^2-10)(\rho^2-3)}{4-\rho^2}\frac{\rhot^2}{\rho^4}-1+\Bigg( 2\frac{12+(\rho^2-6)\rho^2}{\sqrt{4-\rho^2}(\rho^2)^{3/2}}\nn\\
&+&12\frac{10+(\rho^2-6)\rho^2}{(4-\rho^2)^{3/2}(\rho^2)^{5/2}}\rhot^2+-2\frac{(\rho^2-2)\sqrt{(4-\rho^2)\rho^2}}{(4-\rho^2)\rho^2}\Bigg)\arctan\left(\frac{\sqrt{\rho^2}}{\sqrt{4-\rho^2}}\right)\Bigg].
\label{Piceroweakfield}
\eea

\begin{figure}
    \centering
    \includegraphics[scale=0.32]{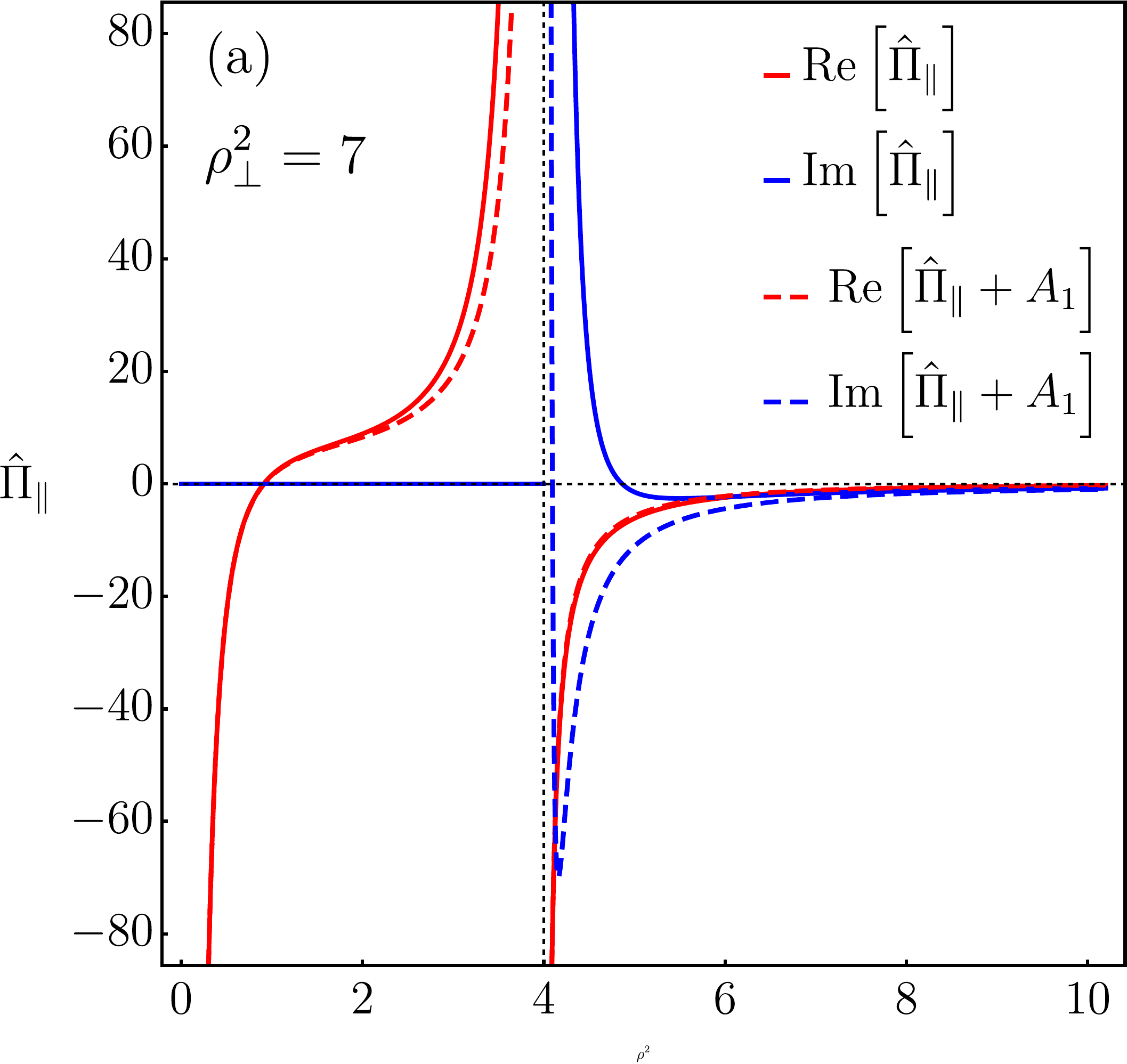}\hspace{0.6cm}\includegraphics[scale=0.32]{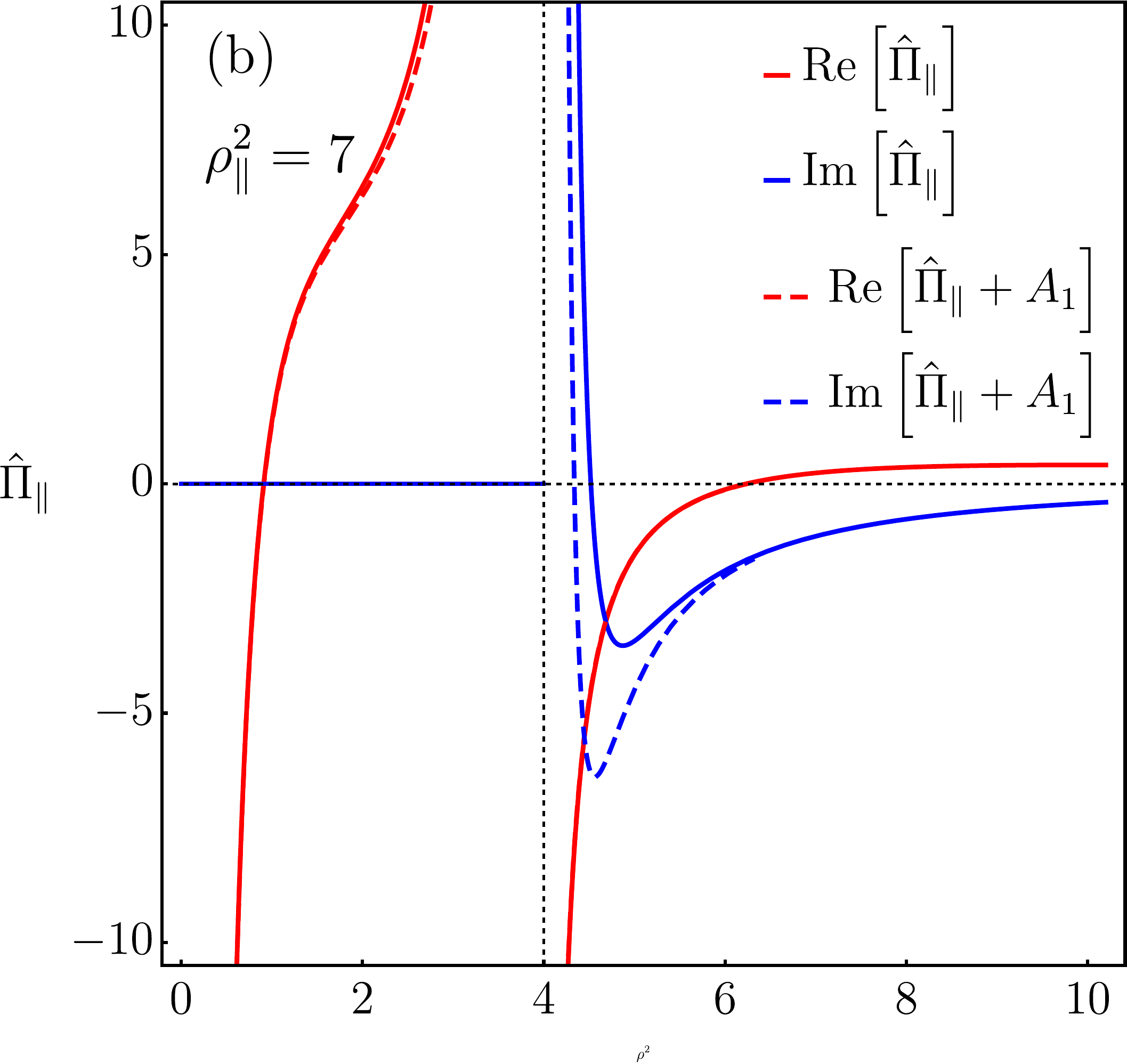}\\
    \vspace{0.3cm}
    \includegraphics[scale=0.32]{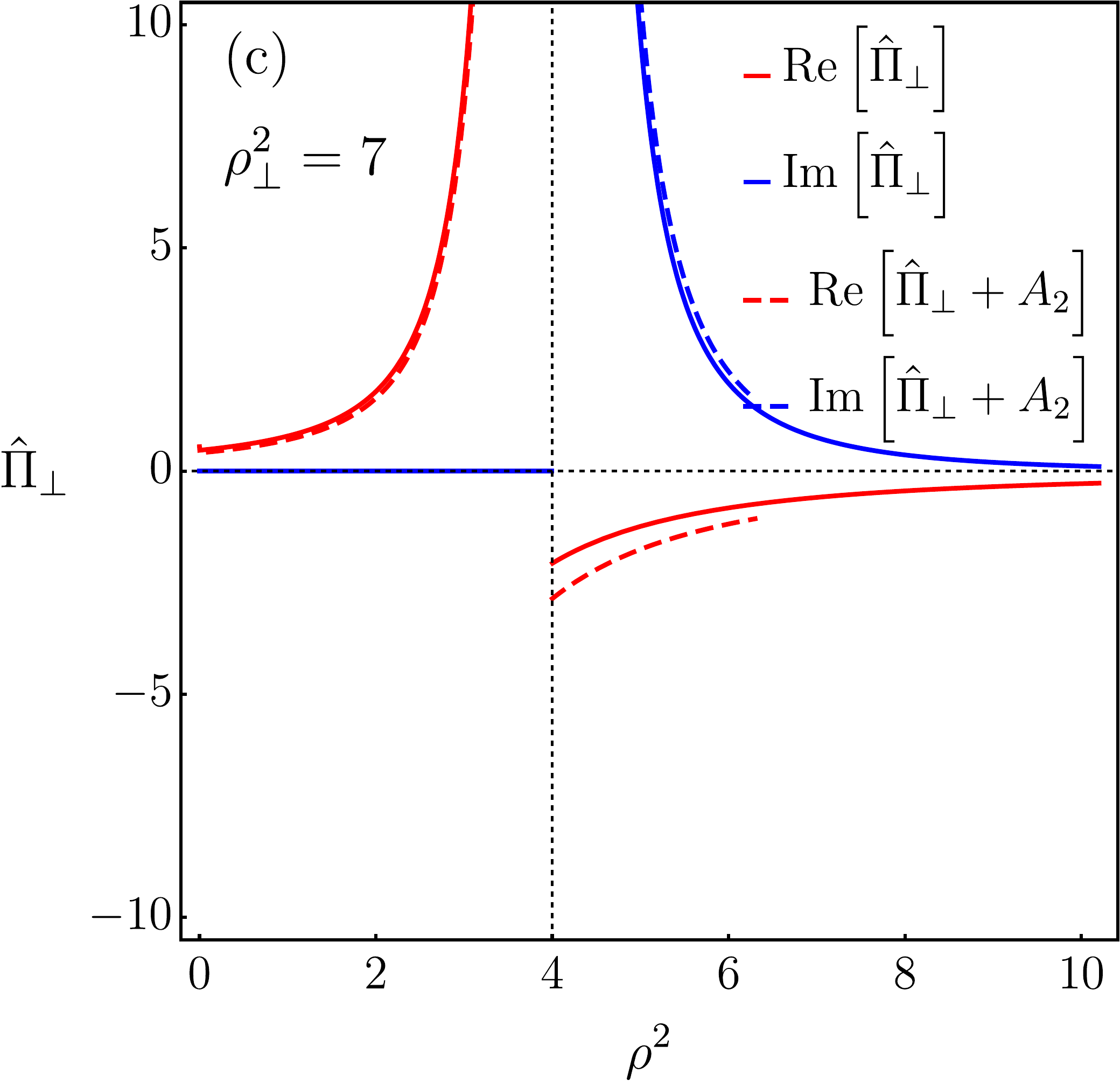}\hspace{0.6cm}\includegraphics[scale=0.32]{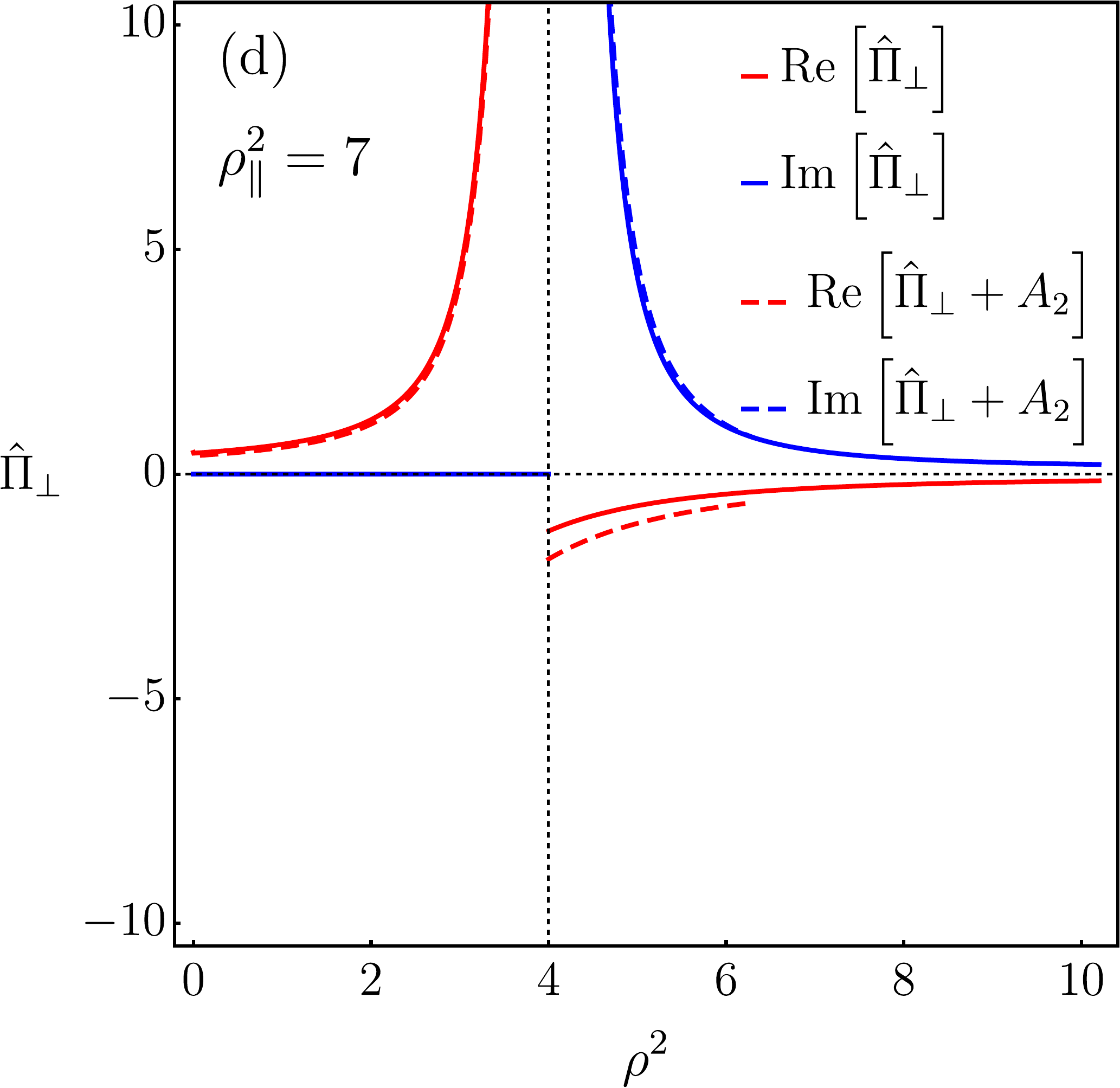}\\
    \vspace{0.3cm}
    \includegraphics[scale=0.32]{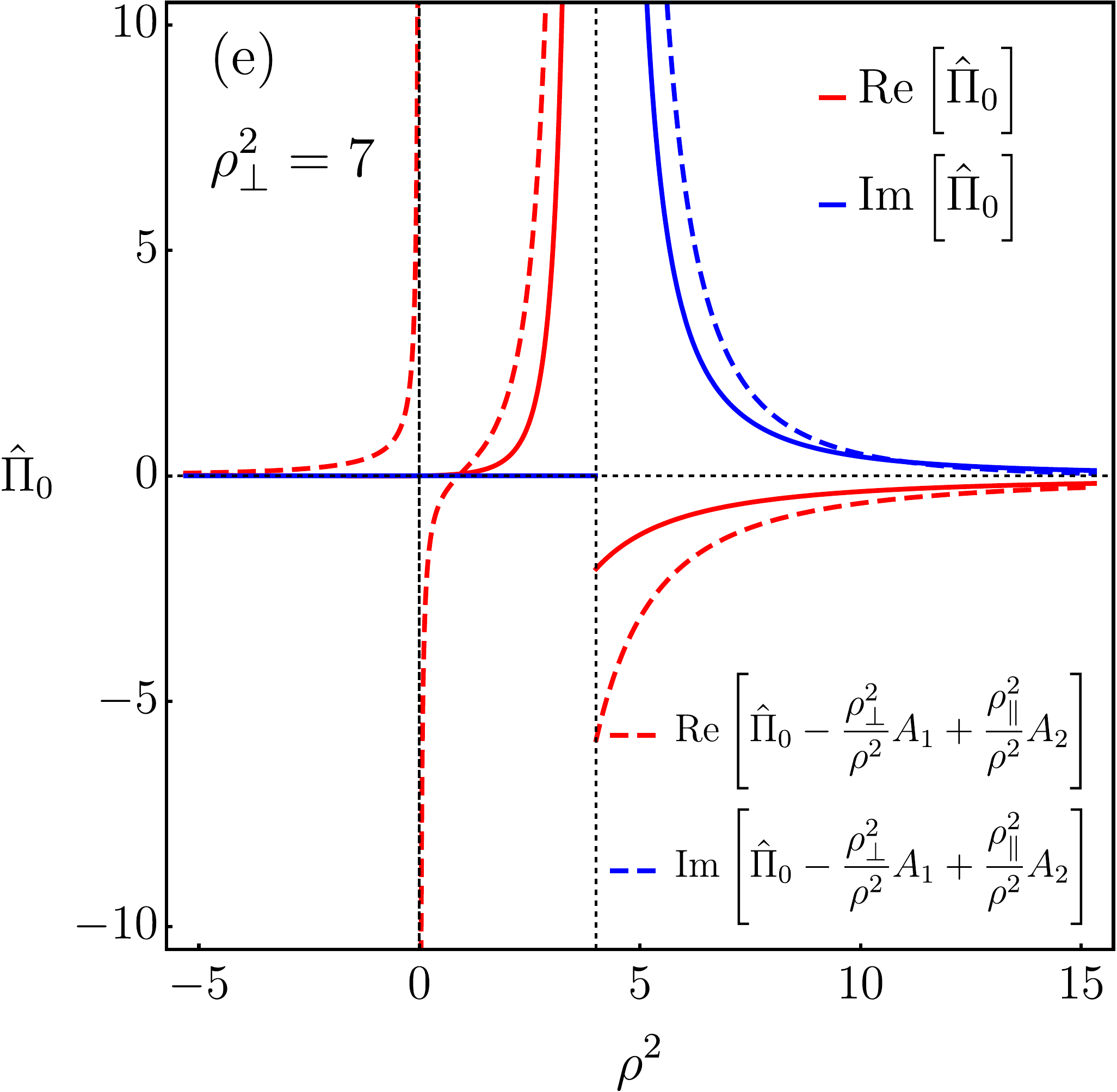}\hspace{0.6cm}\includegraphics[scale=0.32]{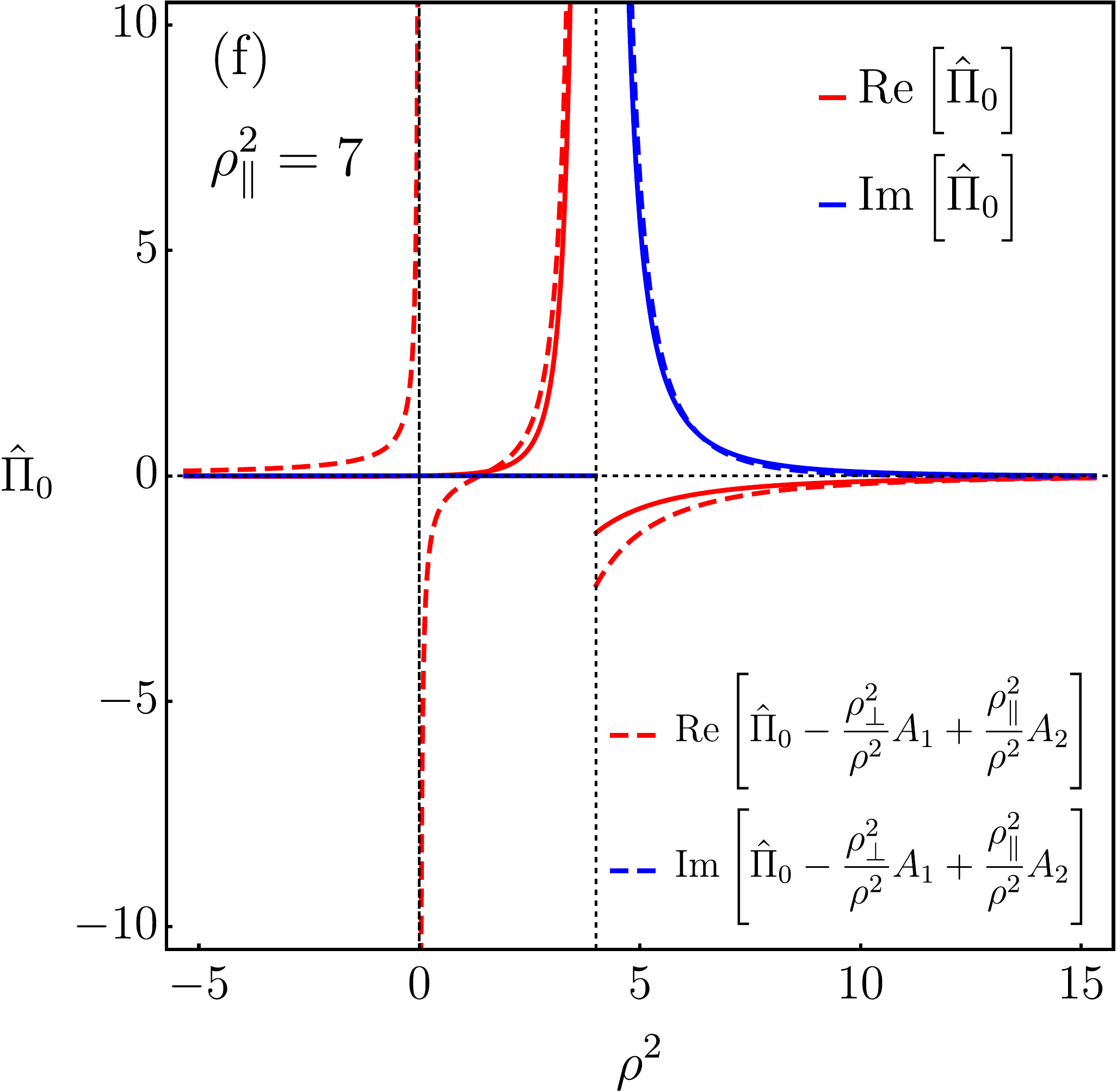}
    \caption{Real and imaginary parts of the coefficients $\hat{\Pi}_\parallel,\,\hat{\Pi}_\perp$ and $\hat{\Pi}_0$ from Eqs.~(\ref{Piparaweakfield})-(\ref{Piceroweakfield}) as functions of $\rho^2$ for fixed values of $\rhop^2$ and $\rhop^2$. For comparison, these coefficients are also plotted including he spurious contributions from $A_1$ and $A_2$, given by Eqs.~(\ref{PiparaconA1})-(\ref{PiceroconA1A2}).}
    \label{Fig:Coefweakfield}
\end{figure}

Figure~\ref{Fig:Coefweakfield} displays the coefficients of Eqs.~(\ref{Piparaweakfield})-(\ref{Piceroweakfield}) as a function of $\rho$ for fixed values of $\rhop^2$ and $\rhot^2$. The spurious coefficients contribution is also plotted for comparison reasons. In the same way that in the strong magnetic field limit, the coefficients $A_1$ and $A_2$ have unphysical thresholds at $\rho^2=0$, producing notable deviations from the correct structures. On the other hand, the weak approximation has a threshold at $\rho^2=4$, in contrast with the strong field limit where the pair production is governed bt $\rhop^2=4$. This feature indicates that all the directions are important in the polarization tensor dynamics, however, as Figs.~\ref{Fig:Coefweakfield}(a)-(b) shows, the tensor structure parallel to the magnetic field dominates over the other directions. 

\subsection{Thresholds for an arbitrary magnetic field}

As was commented on the previous sections, the polarization tensor thresholds are related to the probability of producing a fermion pair from a gluon or photon. In the strong as well as in the weak field limit, there is just one value of $\rho^2$  where such particle production can be reached. The existence of more of these singular points is related to the analytic behavior of the hyperbolic functions in Eq.~(\ref{PienSyY}). For example, its term proportional to $\coth(\B s)$ is
\bea
\mathcal{K}&=&\int_0^1dy\int_0^{\infty}\,ds\,y(1-y)\B\coth(\B s)\exp\left[s\left(y(1-y)\rhop^2-1\right)\right]\nn\\
&\times&\exp\left[-\frac{\cosh(\B s)-\cosh\left[\B s (2y-1)\right]}{2\sinh(\B s)}\frac{\rhot^2}{\B}\right].\nn\\
\label{K},
\eea
such that if $\B\gg 1$
\bea
&&\B\coth(\B s)\exp\left[-\frac{\cosh(\B s)-\cosh\left[\B s (2y-1)\right]}{2\sinh(\B s)}\frac{\rhot^2}{\B}\right]\nn\\
&=&\B\frac{1+e^{-2\B s}}{1-e^{-2\B s}}\exp\Bigg\{\frac{-1-e^{-2\B s}+e^{-2\B s(y-1)}+e^{-2\B sy}}{2\B\left(1-e^{-2\B s}\right)}\rhot^2\Bigg\}\nn\\
&\approx&\B\frac{1+e^{-2\B s}}{1-e^{-2\B s}}+\mathcal{O}(\rhot^2).
\eea

Using that
\bea
\frac{1}{1-e^{-2\B s}}=\sum_{n=0}^{\infty}e^{-2n\B s},
\label{dominant}
\eea
it is easy to prove that
\bea
\frac{1+e^{-2\B s}}{1-e^{-2\B s}}=1+2\sum_{n=1}^{\infty}e^{-2n\B s},
\eea
so that, the dominant term in Eq.~(\ref{K}) is given by
\bea
\mathcal{K}&=&\B\int_0^1dy\int_0^{\infty}\,ds\,y(1-y)\Bigg\{e^{s\left(y(1-y)\rhop^2-1\right)}+2\sum_{n=1}^{\infty}\exp\left[s\left(y(1-y)\rhop^2-2n\B-1\right)\right]\Bigg\}\nn\\
&=&\frac{\B}{\rhop^2}\,I(\rhop^2)+8\B\, J(\rhop^2),
\eea
where $I(x)$ is defined in Eq.~(\ref{resultadoFukushima}) and
\bea
J(x)\equiv-\sum_{n=1}^{\infty}\frac{\arctan\left(\frac{\sqrt{x}}{\sqrt{4(2n\B+1)-x}}\right)}{\sqrt{x\left[4(2n\B+1)-x\right]}}.
\label{J(x)}
\eea
\begin{figure}[H]
    \centering
    \includegraphics[scale=0.5]{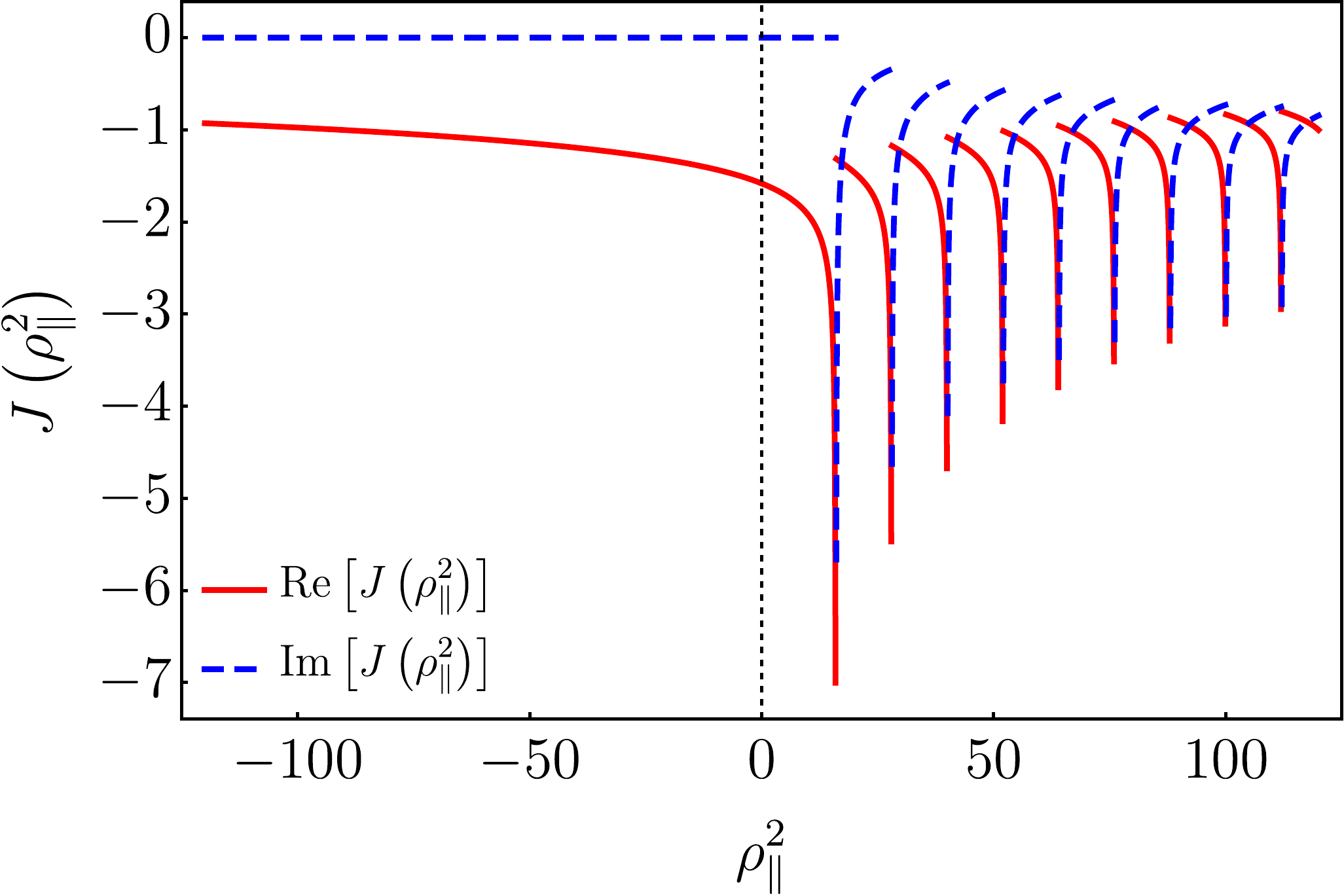}
    \caption{Real and imaginary parts of the function $J(\rhop^2)$ defined in Eq.~(\ref{J(x)}) for $\B=1.5$ and added up to $n_{\text{max}}=100$. Notice the emergence of different resonant thresholds when the quark magnetic mass includes consecutive Landau levels.}
    \label{Fig:UmbralesCoth}
\end{figure}

Figure~\ref{Fig:UmbralesCoth} shows several thresholds of the function $J(\rhop^2)$ in a broad range of $\rhop^2$ for a maximum value of $n$,  $n_{\text{max}}=100$. Those singularities appear from Eq.~(\ref{J(x)}) when $\pp^2=4m_{(Bn)f}^2$; thus, if $n$ labels the n-th Landau level, the pair production is allowed when the gluon momentum not only reaches twice the fermion mass but rather the magnetic mass as well. It can be visualized with a classical picture in which the fermion jumps to the next orbit corresponding to the next Landau Level. The infinite sequence of thresholds is a crucial point to be analyzed. First, the resonant condition with the magnetic field implies that non-linear screening properties become essential in a magnetized medium. Such a regime impacts directly onto the direction of motion for the gauge bosons (photons or gluons); hence, one can expect deviations from the original propagation, which may modify the predictions for the flow coefficients.  Second, at the resonant condition, the polarization tensor becomes divergent, and this implies that the probability of decay in two fermions is also infinite. The latter can be fixed by considering magnetic corrections to the gluon and quark propagators so that the divergence is adequately treated. A calculation that takes into account the named treatment is work in progress.

The next section is dedicated to understanding the effect of the temperature in the gluon polarization tensor. Its computation in a thermomagnetic medium is carried out by considering the strong magnetic field approximation. The thermal corrections to the so-called magnetic Debye mass are discussed. 

\section{Gluon Polarization Tensor in a Hot and Strongly Magnetized Medium}\label{sec:GluonPolTenThermoMag}

It is interesting to explore a situation where the dynamics of quarks and gluons evolves in a magnetized medium with a finite temperature. Recently, ab-initio calculations combined with astrophysical observations suggest hat the core of neutron stars could be composited of free quarks and gluons, which corresponds to a scenario of low temperature and high magnetic field~\cite{annala2020evidence}. To analyze the gluon polarization tensor in a thermo-magnetic medium, it is necessary to take into account the problem's relevant energy scales: the squared fermion-mass $m_f^2$, the squared four-momentum $p^2$, the temperature $T$, and the magnetic field strength $\eB$. Also, the breaking of boost and rotational invariance makes the polarization tensor depend separately on the gluon momentum's longitudinal and perpendicular components. Therefore, in this section, the interplay between the different hierarchies and momentum projections are analyzed.

\subsection{Gluon Polarization Tensor in the High Temperature and Strong Magnetic Field Limits}

 At finite temperature, the loss of boost invariance is achieved in terms of the vector $u^\mu$ that defines the medium's reference frame. Thus, the full-transverse tensor $g^{\mu\nu}-p^\mu p^\nu/p^2$ splits into (three dimensional) transverse $\Pi_{T}^{\mu \nu}$ and longitudinal $\Pi_{L}^{\mu \nu}$ structures, such that
 \begin{equation}
-g^{\mu \nu}+\frac{p^\mu p^\nu}{p^2} = \Pi_L^{\mu \nu}+\Pi_T^{\mu \nu},
\label{split}
\end{equation}
with
\begin{align}
    \Pi_{T}^{\mu\nu}  &= - g^{\mu\nu}+\frac{q^0}{\mathbf{p}^2}(p^\mu u^\nu+u^\mu p^\nu)-\frac{1}{\mathbf{p}^2}(p^\mu p^\nu+p^2u^\mu u^\nu),\nonumber\\
  \Pi_{L}^{\mu\nu}  &=-\frac{p^0}{\mathbf{p}^2}(p^\mu u^\nu+u^\mu p^\nu)
    +\frac{1}{\mathbf{p}^2}\Big[\frac{(p^0)^2}{p^2}p^\mu p^\nu+p^2u^\mu u^\nu\Big], 
\label{structuresLLLHTL}
\end{align}
where, in the medium's reference frame, $u^\mu=(1,0,0,0)$.

In the strong magnetic field limit, only the tensorial structure $\Pp$ survives which is given by Eq.~(\ref{resultadoFukushima}). Such tensorial dominance remains even in the thermalized case when $\eB\gg T^2$. This can be understood from the fact that in a strong magnetized field with finite temperature, the fermion-antifermion pair is created in the same Landau Level (the lowest one) is such a way that if $\eB$ is the dominant scale, neither the virtual quark nor the antiquark has enough thermal energy to transit between Landau levels; thus, no other tensor appear in $\Pi^{\mu\nu}$. 

It is worth of mentioning that the last prescription is only valid for specific vectors $u^\mu$. There exist the possibility of a non-trivial behavior in a given reference frame in such a manner that an electric field appears.  The simultaneous treatment of magnetic and electric field effects on medium's properties has been recently considered in the context of asymmetric collisions or event-by-event fluctuations in heavy-ion reactions~\cite{PhysRevD.101.016017}. Electromagnetic effects have also been studied using chiral kinetic theory with Landau level transitions induced by both boosts and rotations~\cite{gao2020photon}.

Now, by defining the temperature dependent part of the polarization tensor $\widetilde{\Pi}^{\mu\nu}$ as
\begin{equation}
 \widetilde{\Pi}^{\mu \nu}= \widetilde{P}^L \Pi_L^{\mu \nu}+\widetilde{P}^T\Pi_T^{\mu \nu},
 \label{tensorstructureBT}
\end{equation}
the coefficients  $\widetilde{P}^L$ and $\widetilde{P}^T$ It can be computed through the projections:
\begin{eqnarray}
  \Pi_{\mu \nu}^L\widetilde{\Pi}^{\mu \nu}&=& \widetilde{P}^L \nonumber \\
 \Pi_{\mu \nu}^T\widetilde{\Pi}^{\mu \nu}&=&  2\widetilde{P}^T.
 \label{coefficientsT}
\end{eqnarray}

To calculate the above expressions, the quark loop in Fig.~\ref{diagrama} is assumed to be in the LLL, so that the Eq.~(\ref{Pidef}) becomes
\bea
 i\Pi^{\mu\nu}&=&-g^2\sum_f \int \frac{d^2k_\perp}{(2\pi)^2}\exp\left[-\frac{k_\perp^2}{|q_fB|}\right]\exp\left[-\frac{(p_\perp-k_\perp)^2}{|q_fB|}\right] \nonumber \\
 &\times& \int\frac{d^2k_\parallel}{(2\pi)^2}\frac{\text{Tr}[\gamma^\nu(m_f-\slashed{k}_\parallel)\mathcal{O}^+\gamma^\mu(m_f-(\slashed{k}-\slashed{p})_\parallel)\mathcal{O}^+]}{[k_\parallel^2-m_f^2][(k-p)_\parallel^2-m_f^2]},
 \label{twocontributions}
\eea
which, after the perpendicular integration yields
\bea
i\Pi^{\mu\nu}&=&-g^2\sum_f\Big( \frac{\pi |q_fB|}{4\pi^2} \Big)\exp\left(-\frac{p_\perp^2}{2|q_fB|}\right)\nn\\
&\times&\int \frac{d^2k_\parallel}{(2\pi)^2}\frac{[m_f^2-k_\parallel\cdot(k-p)_\parallel]g_\parallel^{\mu\nu}
 +k_\parallel^\mu(k-p)_\parallel^\nu+k_\parallel^\nu(k-p)_\parallel^\mu}{[k_\parallel^2-m_f^2][(k-p)_\parallel^2-m_f^2]}.
\label{PiParallel}
\eea

The temperature effects are included by means the Matsubara imaginary-time formalism: 
\begin{equation}
\int \frac{d^4k}{(2\pi^4)} f(k) \rightarrow i T \sum_{n=-\infty}^{\infty} \int \frac{d^3k}{(2\pi^3)} f(i \tilde{\omega}_n,\vec{k}),
\end{equation}
where the fermion Matsubara frequencies are $\tilde{\omega}_n = (2n +1)\pi T$, $k_0=i \tilde{\omega}_n$ and $q_4=iq_0=\omega$. The thermal vacuum has to be isolated, and as one can expect, it results in the pure magnetic contribution of Eq.~(\ref{resultadoFukushima}). Details are shown in Appendix~\ref{Ap_Vacuum_HTL_and_LLL}. 

It is convenient to introduce the Hard Thermal Loop (HTL) approximation~\cite{LeBellac,rebhan2001hard} which only considers terms where $k~\sim T$. Thus, in the high-temperature limits, the sum can be performed in two cases: 
\begin{itemize}
    \item[(a)] For $p^2\ll m_f^2\ll T^2$:
    
    When the external momentum $p$ is considered as the smallest energy scale, in the HTL approximation, the terms $p^2$ and $k\cdot p$ can be neglected in each of the numerators. The integral over $k_3$ is shown in Appendix~\ref{Ap_Coef_PL_and_PT_1} obtaining that
\begin{align}
	\widetilde{P}^L&=\frac{g^2}{8\pi^2}\sum_f|q_fB| \exp\left(-\frac{p_\perp^2}{2|q_fB|}\right)\Bigg [\frac{p_0^2(p_\perp^2+2p_3^2)}{\mathbf{p}^2 p^2}-1 \Bigg]\left[ \ln\left(\frac{\pi^2 T^2}{m_f^2}\right)-2\gamma_e \right]
    \label{longfinal}
\end{align}
and
\begin{align}	
\widetilde{P}^T&=\frac{g^2}{8\pi^2}\sum_f|q_fB| \exp\left(-\frac{p_\perp^2}{2|q_fB|}\right)\frac{p_{\perp}^2}{\mathbf{p}^2}\left[\ln \left(\frac{\pi^2 T^2}{m_f^2}\right)-2\gamma_e \right].
\label{transfinal}
\end{align}

\item[(b)] For $m_f^2\ll p^2\ll T^2$:

In this limit, the fermion mass can be ignored and maintain the HTL approximation such that:
\begin{align}
    \widetilde{P}^L&=\frac{g^2}{64\pi^2}\sum_f |q_fB|\exp\left(-\frac{p_\perp^2}{2|q_fB|}\right)\left( 2p_0^2+p^2+\frac{p_0^2p_3^2}{p^2}\right)\left( \frac{p_3^2}{\mathbf{p}^2p_\parallel^2}\right)\left(\frac{p_3}{T}\right),
    \label{plhtl211}
\end{align}
and
\begin{equation}
   \widetilde{P}^T=-\frac{g^2}{64\pi^2}\sum_f \eB\exp\left(-\frac{p_\perp^2}{2|q_fB|}\right)\left(\frac{p_\perp^2 p_3^2}{p_\parallel^2 \mathbf{p}^2}\right) \left(\frac{p_3}{T}\right).
    \label{pthtl2}
\end{equation}

The calculation of Eqs.~(\ref{plhtl211}) and (\ref{pthtl2}) can be found in Appendix~\ref{Ap_Coef_PL_and_PT_2}.
\end{itemize}

\subsection{Correction to the gluon Debye mass}

The Debye mass $m_D\equiv p_0$ is defined as the solution of the following equation:
\begin{equation}
  \left[p^2-\widetilde{P}^{L,T} (p_0,p_\perp,p_3)\right]_{\mathbf{p}=0}=0,
    \label{Debye}
\end{equation}
which can be obtained from the coefficients computed in the last section. Notice that in Eqs.~(\ref{longfinal}) and~(\ref{transfinal}) the result may differ depending on whether the parallel or perpendicular momentum component, with respect to the magnetic field, is taken first to zero. This behavior is due to the breaking of the spatial isotropy and is the analog to the purely thermal case, where the limits when either $p_0$ or $|\mathbf{p}|$ goes first to zero, do not commute, due to the loss of Lorentz (boost) invariance.

To implement how the momentum components go to zero, the parameter $a$ is included to ensure that $p_0$ and $p_3$ have different vanishing rates., i.e., $p_3^2=ap_\perp^2$. Thus, the matter contribution in Eqs.~(\ref{longfinal}) and~(\ref{transfinal}) can be written as
\begin{align}
	\widetilde{P}^L=\frac{g^2}{8\pi^2}\sum_f|q_fB| \exp\left(-\frac{p_\perp^2}{2|q_fB|}\right) \left[\ln\left(\frac{\pi^2 T^2}{m_f^2}\right)-2\gamma_e \right]\Bigg [\frac{(1+2a)p_0^2}{(1+a)(p_0^2-(1+a)p_\perp^2)}-1 \Bigg],
    \label{longmatter}
\end{align}
and
\begin{align}	
\widetilde{P}^T= \frac{g^2}{8\pi^2}\sum_f|q_fB| \exp\left(-\frac{p_\perp^2}{2|q_fB|}\right) \left[\ln\left(\frac{\pi^2 T^2}{m_f^2}\right)-2\gamma_e \right] \Bigg [\frac{1}{1+a}\Bigg].
\label{transmatter}
\end{align}

The discussed limits are:
\begin{itemize}
    \item[(i)] $p_\perp\to 0$ and $p_3$ finite or $a\rightarrow0$:
    
In this case, Eqs.~(\ref{longmatter}) and~(\ref{transmatter}) become
\begin{eqnarray}
  \widetilde{P}^L&=&\frac{g^2}{8\pi^2}\sum_f|q_fB|\left(\frac{p_0^2}{p_\parallel^2}\right)\left[ \ln \left( \frac{\pi^2 T^2}{m_f^2} \right)-2\gamma_e \right]\nonumber\\
  \widetilde{P}^T&=&0,
\label{case1}
\end{eqnarray}
which shows that transverse modes are not screened. By taking $p_3\to 0$ in the first of Eqs.~(\ref{case1}), the longitudinal mode develops a Debye mass given by
\begin{eqnarray}
(m_D^2)_L=\frac{g^2}{8\pi^2}\sum_f |q_fB|\left[\ln\left(\frac{\pi^2T^2}{m_f^2} \right)-2\gamma_e \right].
\label{masscase1}
\end{eqnarray}

\item[(ii)] $p_3\to 0$ and $p_\perp$ finite:\\
In this other case, Eqs.~(\ref{longmatter}) and~(\ref{transmatter}) become
\begin{eqnarray}
  \widetilde{P}^L&=&\frac{g^2}{8\pi^2}\sum_f|q_fB|\left[ \ln \left( \frac{\pi^2 T^2}{m_f^2} \right)-2\gamma_e \right]\left(\frac{p_0^2}{p_0^2-p_\perp^2}-1\right) \nonumber \\
  \widetilde{P}^T&=&\frac{g^2}{8\pi^2}\sum_f|q_fB\left[ \ln \left( \frac{\pi^2 T^2}{m_f^2} \right)-2\gamma_e \right],
\label{case2}
\end{eqnarray}
so that by taking $p_\perp \to 0$, $\widetilde{P}^L=0$ and this time it is the transverse modes which develop a Debye mass given by
\begin{eqnarray}
(m_D^2)_T=\frac{g^2}{8\pi^2}\sum_f|q_fB|\left[\ln \left(\frac{\pi^2 T^2}{m_f^2} \right)-2\gamma_E \right],
\label{masscase2}
\end{eqnarray}
whereas the longitudinal one is not screened. Note that for these limits $(m_D^2)_L=(m_D^2)_T$.

\item[(iii)] $p_3$ and$ \ p_\perp \to 0$ at the same rate:

In this last case, Eqs.~(\ref{longmatter}) and~(\ref{transmatter}) become
\begin{eqnarray}
  \widetilde{P}^L=\widetilde{P}^T=\frac{g^2}{16\pi^2}\sum_f|q_fB|\left[\ln \left(\frac{\pi^2 T^2}{m_f^2} \right)-2\gamma_E \right]
\label{case3}
\end{eqnarray}
and both the longitudinal and transverse modes develop a Debye mass given by
\begin{eqnarray}
 (m_{D}^2)_L=(m_{D}^2)_T=\frac{g^2}{16\pi^2}\sum_f\eB\left[\ln \left(\frac{\pi^2 T^2}{m_f^2} \right)-2\gamma_E \right].
 \label{DebyemassLT}
\end{eqnarray}
Notice that when $p_3$ and $p_\perp$ vanish at the same rate, both modes are screened and the Debye mass of the longitudinal mode is equal to that of the transverse mode. Finally, the sum of both transverse and longitudinal masses in the three cases gives:
\begin{eqnarray}
(m_D^2)_L+(m_D^2)_T=\frac{g^2}{8\pi^2}\sum_f |q_fB|\left[\ln \left(\frac{\pi^2 T^2}{m_f^2} \right)-2\gamma_E \right]. \nonumber \\
\label{summasses}
\end{eqnarray}
\end{itemize}
\section{Gluon Polarization Tensor in a Weakly Magnetized Medium}\label{sec:GluonPolTenWeakField}

As discussed in previous sections, the magnetic field in heavy-ion collisions is estimated to have an intensity comparable to the pion-mass squared. Nevertheless, the magnetic field intensity decreases in a short time, and it cannot be considered as the dominant energy scale, which enforces to perform calculations in a weak magnetic field approximation. An example of such a situation can be found in Ref.~\cite{PhysRevD.102.036020}, where heavy quarkonia's properties immersed in a thermal medium of quarks and gluons are computed by assuming a weak magnetic field background. 

Clearly, in a QCD-medium, the magnetic field effects are directly encoded in the quark's (real or virtual) dynamics, which implies that the fermion propagator has to be appropriately written in the desired approximation. The computation of charged fermion and scalar propagators in the weak field limit were computed as an expansion of the full magnetized propagator of Eq.~(\ref{fermionpropdef})~\cite{PhysRevD.62.105014,PhysRevD.71.023004}. Moreover, the general expression shown in Eq.~(\ref{Pifinalwithvacuum}) can be expanded in powers of $\eB$, by considering that momentum and fermion mass are the higher energy scales. Thus, it is natural to assume that the expression of the gluon one-loop self-energy starting from the fermion propagator in a weak magnetic field yields the same results as the ones obtained from the power series expansion of the general polarization tensor result. However, as it will be shown in this section, the former procedure has crucial subtleties regarding gauge invariance encoded in the transversality properties of the polarization tensor. 

\subsection{Weak Field Approximation of the Fermion Propagator Revisited}

An expression for the fermion propagator when the magnetic field is considered as the weakest energy scale is provided in Ref.~\cite{PhysRevD.62.105014}. Nonetheless, although the calculation is correct, the procedure to find it is hard to read, obscuring the origin of the results. Here, a straightforward and more intuitive procedure is presented.

Starting from the Landau Levels representation of the fermion propagator:
\bea
iS(p)=ie^{-\pt^2/\eB}\sum_{n=0}^{\infty}(-1)^n\frac{D_n(q_fB,p)}{p_\p^2-m_f^2-2n\left |  q_fB\right |},
\label{eq:MiranskyPropagator}
\eea
where $D_n(q_fB,p)$ is given by Eq.~(\ref{DnDef}):
\bea
D_n(q_fB,p)=2(\psh_\parallel+m_f)\left[\mathcal{O}^{-}L_n^0\left(\frac{2p_\perp^2}{\left | q_fB \right|}\right)-\mathcal{O}^{+}L_{n-1}^0\left(\frac{2p_\perp^2}{\left | q_fB \right |}\right)\right]+4\psh_\perp L_{n-1}^1\left(\frac{2p_\perp^2}{\left | q_fB \right |}\right),\nn\\\quad
\eea
and the operators $\Op^{\pm}$ are given in Eq.~(\ref{OpDef}), the denominators in Eq.~(\ref{eq:MiranskyPropagator}) can be promoted to a geometric series, by assuming that $\eB$ is the smaller scale, namely,
\bea
\frac{1}{\pp^2-m_f^2-2n\eB}=\frac{1}{\pp^2-m_f^2}\sum_{k=0}^\infty\left(\frac{2n\eB}{\pp^2-m_f^2}\right)^k.
\label{DenominatortoGeomSum}
\eea

From the above, three sums are identified:
\begin{subequations}
\bea
s_1&=&\frac{2ie^{-\alpha}}{\pp^2-m_f^2}\sum_{n=0}^\infty(-1)^n\frac{L_n^0(2\alpha)}{1-\frac{2n\eB}{\pp^2-m_f^2}}\nn\\
&=&\frac{2ie^{-\alpha}}{\pp^2-m_f^2}\sum_{k=0}^\infty\left(\frac{2\eB}{\pp^2-m_f^2}\right)^k\sum_{n=0}^\infty(-1)^n n^k L_n^0(2\alpha),
\label{S1}
\eea
\bea
s_2&=&-\frac{2ie^{-\alpha}}{\pp^2-m_f^2}\sum_{n=1}^\infty(-1)^n\frac{L_{n-1}^0(2\alpha)}{1-\frac{2n\eB}{\pp^2-m_f^2}}\nn\\
&=&-\frac{2ie^{-\alpha}}{\pp^2-m_f^2}\sum_{k=0}^\infty\left(\frac{2\eB}{\pp^2-m_f^2}\right)^k\sum_{n=1}^\infty(-1)^n n^k L_{n-1}^0(2\alpha),
\label{S2}
\eea
\bea
s_3&=&\frac{4ie^{-\alpha}}{\pp^2-m_f^2}\sum_{n=1}^\infty(-1)^n\frac{L_{n-1}^1(2\alpha)}{1-\frac{2n\eB}{\pp^2-m_f^2}}\nn\\
&=&\frac{4ie^{-\alpha}}{\pp^2-m_f^2}\sum_{k=0}^\infty\left(\frac{2\eB}{\pp^2-m_f^2}\right)^k\sum_{n=1}^\infty(-1)^n n^k L_{n-1}^1(2\alpha).
\label{S3}
\eea
\label{sums}
\end{subequations}
with $\alpha=\pt^2/\eB$. The sum in $n$ can easily computed with the help of the following identities:
\begin{subequations}
\bea
e^{-\alpha}\sum_{n=0}^\infty(-1)^n e^{-2inv}L_n^0(2\alpha)=\frac{e^{iv}}{2\cos v}e^{-i\alpha\tan v},
\eea
\bea
e^{-\alpha}\sum_{n=0}^{\infty}(-1)^n e^{-2i(n+1)v}L_{n}^1(2\alpha)=\frac{1}{4\cos^2v}e^{-i\alpha\tan v},
\eea
\label{Identities1def}
\end{subequations}
so that
\begin{subequations}
\bea
e^{-\alpha}\sum_{n=0}^\infty(-1)^n n^k L_n^0(2\alpha)=\frac{1}{(-2i)^k}\lim_{v\rightarrow0}\frac{\partial^k}{\partial v^k}\left(\frac{e^{iv}}{2\cos v}e^{-i\alpha\tan v}\right),
\eea
\bea
e^{-\alpha}\sum_{n=0}^{\infty}(-1)^n (n+1)^ke^{-2i(n+1)v}L_{n}^1(2\alpha)=\frac{1}{(-2i)^k}\lim_{v\rightarrow0}\frac{\partial^k}{\partial v^k}\left(\frac{1}{4\cos^2v}e^{-i\alpha\tan v}\right).
\eea
\label{DerivativeIdentities1}
\end{subequations}

Therefore, 
\begin{subequations}
\bea
s_1=\frac{i}{\pp^2-m_f^2}\lim_{v\rightarrow0}\sum_{k=0}^{\infty}\left(\frac{i\eB}{\pp^2-m_f^2}\right)^k\frac{\partial^k}{\partial v^k}\left(\frac{e^{iv}}{\cos v}e^{-i\alpha\tan v}\right)\equiv\frac{i}{\pp^2-m_f^2}\sigma_1,
\eea
\bea
s_2=\frac{i}{\pp^2-m_f^2}\lim_{v\rightarrow0}\sum_{k=0}^{\infty}\left(\frac{i\eB}{\pp^2-m_f^2}\right)^k\frac{\partial^k}{\partial v^k}\left(\frac{e^{-iv}}{\cos v}e^{-i\alpha\tan v}\right)\equiv\frac{i}{\pp^2-m_f^2}\sigma_2,
\eea
and
\bea
s_3=-\frac{i}{\pp^2-m_f^2}\lim_{v\rightarrow0}\sum_{k=0}^\infty\left(\frac{i\eB}{\pp^2-m_f^2}\right)^k\frac{\partial^k}{\partial v^k}\left(\frac{e^{-i\alpha\tan v}}{\cos^2v}\right)\equiv-\frac{i}{\pp^2-m_f^2}\sigma_3.
\eea
\label{s_and_sigma}
\end{subequations}

The sum in $k$ has not a closed-form, but it can be expanded at the desired order in the parameters $(\eB,\pp^2,\pt^2,m_f^2)$ by simple evaluation. For example, for $k_{\text{max}}=10$ and together with the definitions $x=\pt^2/(\pp^2-m_f^2)$ and $\mathfrak{B}=\eB/(\pp^2-m_f^2)$, one get:
\begin{subequations}
\bea
\sigma_1&=&\left(1+x+x^2+x^3+x^4+x^5+x^6+x^7+x^8+x^9+x^{10}+\cdots\right)\nn\\
&-&\left(1+2 x+3 x^2+4 x^3+5 x^4+6 x^5+7x^6+8x^7+9 x^8+10 x^9+11x^{10}+\cdots\right)\mathfrak{B}\nn\\
&-&2x\left(1+4 x+10 x^2+20 x^3+35 x^4+56 x^5+84 x^6+120 x^7+165 x^8+220 x^9\right.\nn\\
&+&\left.286x^{10}+\cdots\right)\mathfrak{B}^2+2\left(1+8 x+30 x^2+80 x^3+175x^4+336 x^5+588 x^6+960 x^7\right.\nn\\
&+&\left.1485x^8+2200 x^9+3146 x^{10}+\cdots\right)\mathfrak{B}^3+8x\left(2+17 x+77 x^2+252 x^3+672 x^4\right.\nn\\
&+&\left.1554 x^5+3234 x^6+6204x^7+11154 x^8+19019 x^9+31031 x^{10}+\cdots\right)\mathfrak{B}^4\nn\\
&-&8\left(2+34 x+231 x^2+1008 x^3+3360 x^4+9324 x^5+22638 x^6+49632x^7+100386 x^8\right.\nn\\
&+&\left.190190 x^9+341341 x^{10}+\cdots\right)\mathfrak{B}^5-16x\left(17+248 x+1760 x^2+8480 x^3+31790 x^4\right.\nn\\
&+&\left.99704 x^5+273416x^6+674960 x^7+1530815 x^8+3237520 x^9+6456736x^{10}+\cdots\right)\mathfrak{B}^6\nn\\
&+&\Op\left(\mathfrak{B}^7\right),
\eea
\bea
\sigma_2&=&\left(1+x+x^2+x^3+x^4+x^5+x^6+x^7+x^8+x^9+x^{10}+\cdots\right)\nn\\
&+&\left(1+2 x+3 x^2+4 x^3+5 x^4+6 x^5+7x^6+8x^7+9 x^8+10 x^9+11x^{10}+\cdots\right)\mathfrak{B}\nn\\
&-&2x\left(1+4 x+10 x^2+20 x^3+35 x^4+56 x^5+84 x^6+120 x^7+165 x^8+220 x^9\right.\nn\\
&+&\left.286x^{10}+\cdots\right)\mathfrak{B}^2-2\left(1+8 x+30 x^2+80 x^3+175x^4+336 x^5+588 x^6+960 x^7\right.\nn\\
&+&\left.1485x^8+2200 x^9+3146 x^{10}+\cdots\right)\mathfrak{B}^3+8x\left(2+17 x+77 x^2+252 x^3+672 x^4\right.\nn\\
&+&\left.1554 x^5+3234 x^6+6204x^7+11154 x^8+19019 x^9+31031 x^{10}+\cdots\right)\mathfrak{B}^4\nn\\
&-&8\left(2+34 x+231 x^2+1008 x^3+3360 x^4+9324 x^5+22638 x^6+49632x^7+100386 x^8\right.\nn\\
&+&\left.190190 x^9+341341 x^{10}+\cdots\right)\mathfrak{B}^5-16x\left(17+248 x+1760 x^2+8480 x^3+31790 x^4\right.\nn\\
&+&\left.99704 x^5+273416x^6+674960 x^7+1530815 x^8+3237520 x^9+6456736x^{10}+\cdots\right)\mathfrak{B}^6\nn\\
&+&\Op\left(\mathfrak{B}^7\right),
\eea
and
\bea
\sigma_3&=&\left(1+x+x^2+x^3+x^4+x^5+x^6+x^7+x^8+x^9+x^{10}+\cdots\right)\nn\\
&-&2 \left(1+4 x+10 x^2+20 x^3+35 x^4+56 x^5+84 x^6+120 x^7+165 x^8+220x^9\right.\nn\\
&+&\left.286 x^{10}+\cdots\right)\mathfrak{B}^2+8 \left(2+17 x+77 x^2+252 x^3+672 x^4+1554 x^5+3234 x^6\right.\nn\\
&+&\left.6204x^7+11154 x^8+19019 x^9+31031 x^{10}+\cdots\right)\mathfrak{B}^4-16 \left(17+248 x+1760 x^2\right.\nn\\
&+&\left.8480 x^3+31790 x^4+99704 x^5+273416x^6+674960 x^7+1530815 x^8+3237520 x^9\right.\nn\\
&+&\left.6456736 x^{10}\right)\mathfrak{B}^6+\Op\left(\mathfrak{B}^8\right).
\eea
\label{isseries}
\end{subequations}

The power series of $x$ in Eqs.~(\ref{isseries}) can be recognized as polinomial expressions:
\begin{subequations}
\bea
\sigma_1&=&\frac{1}{1-x}-\frac{\mathfrak{B}}{(1-x)^2}-\frac{2x\mathfrak{B}^2}{(1-x)^4}+\frac{2 (3 x+1)\mathfrak{B}^3}{(1-x)^5}
+\frac{8 x (3 x+2)\mathfrak{B}^4}{(1-x)^7}\nn\\
&-&\frac{8 \left(15 x^2+18 x+2\right)\mathfrak{B}^5}{(1-x)^8}-\frac{16 x \left(45 x^2+78 x+17\right)\mathfrak{B}^6}{(1-x)^{10}}+\Op\left(\mathfrak{B}^7\right),
\eea
\bea
\sigma_2&=&\frac{1}{1-x}+\frac{\mathfrak{B}}{(1-x)^2}-\frac{2x\mathfrak{B}^2}{(1-x)^4}-\frac{2 (3 x+1)\mathfrak{B}^3}{(1-x)^5}
+\frac{8 x (3 x+2)\mathfrak{B}^4}{(1-x)^7}\nn\\
&+&\frac{8 \left(15 x^2+18 x+2\right)\mathfrak{B}^5}{(1-x)^8}-\frac{16 x \left(45 x^2+78 x+17\right)\mathfrak{B}^6}{(1-x)^{10}}+\Op\left(\mathfrak{B}^7\right),
\eea
and
\bea
\sigma_3&=&\frac{1}{1-x}-\frac{2\mathfrak{B}^2}{(1-x)^4}+\frac{8 (3 x+2)\mathfrak{B}^4}{(1-x)^7}-\frac{16 \left(45 x^2+78 x+17\right)\mathfrak{B}^6}{(1-x)^{10}}+\Op\left(\mathfrak{B}^8\right).\nn\\
\eea
\label{s_en_x}
\end{subequations}

Thus, the fermion propagator is obtained by replacing Eqs.~(\ref{s_en_x}) into Eqs.~(\ref{s_and_sigma}) and Eqs.~(\ref{sums}), by taking into account the tensor structure given by
\begin{subequations}
\bea
\slashed{p}=\slashed{p}_\parallel-\slashed{p}_\perp,
\eea
\bea
\Op^+ +\Op^-=1,
\eea
and
\bea
\Op^+ -\Op^-=i\,\text{sign}(q_fB)\gamma^1\gamma^2,
\eea
\end{subequations}
so that up to order $\Op\left(\mathfrak{B}^6\right)$ reads
\bea
&&\!\!\!\!\!\!\!\!\!\!iS(p)=\frac{i}{1-x}\,\frac{\ps+m_f}{\pp^2-m_f^2}\nn\\
&-&\left[\frac{\mathfrak{B}}{(1-x)^2}-\frac{2 (3 x+1)\mathfrak{B}^3}{(1-x)^5}+\frac{8 \left(15 x^2+18 x+2\right)\mathfrak{B}^5}{(1-x)^8}\right]\text{sign}(q_f B)\,\gamma^1\gamma^2\,\frac{\ps_\parallel+m_f}{\pp^2-m_f^2}\nn\\
&-&i\left[\frac{2\mathfrak{B}^2}{(1-x)^4}-\frac{8 (3 x+2)\mathfrak{B}^4}{(1-x)^7}-\frac{16 \left(45 x^2+78 x+17\right)\mathfrak{B}^6}{(1-x)^{10}}\right]\frac{x(\ps_\parallel+m_f)-\slashed{p}_\perp}{\pp^2-m_f^2},\nn\\
\eea
which after substituting $x$ and $\mathfrak{B}$ yields
\bea
iS(p)&=&\frac{i\left(\ps+m_f\right)}{p^2-m_f^2}-\mathcal{G}_1(p,B)\,\text{sign}(q_fB)\gamma^1\gamma^2(\ps_\parallel+m_f)\nn\\
&-&2i(\pp^2-m_f^2)\,\mathcal{G}_2(p,B)\left[\frac{\pt^2}{\pp^2-m_f^2}(\ps_\parallel+m_f)-\slashed{p}_\perp\right],
\label{iSfinalfromLandauApp}
\eea
where
\begin{subequations}
\bea
\mathcal{G}_1(p,B)&\equiv&\frac{\eB}{(p^2-m_f^2)^2}-2\frac{3\pt^2+\pp^2-m_f^2}{(p^2-m_f^2)^5}\eB^3\nn\\
&+&8\frac{15\pt^4+18\pt^2(\pp^2-m_f^2)+2(\pp^2-m_f^2)^2}{(p^2-m_f^2)^8}\eB^5,
\eea
and
\bea
\mathcal{G}_2(p,B)&\equiv&\frac{\eB^2}{(p^2-m_f^2)^4}-4\frac{3\pt^2-2(\pp^2-m_f^2)}{(p^2-m_f^2)^7}\eB^4\nn\\
&-&8\frac{45\pt^4+78(\pp^2-m_f^2)+17(\pp^2-m_f^2)^2}{(p^2-m_f^2)^{10}}\eB^6.
\eea
\label{Gdefinitions}
\end{subequations}

Notice that Eq.~(\ref{iSfinalfromLandauApp}) also can be obtained from the propagator in the Schwinger proper-time representation given by
\bea
    iS(p)&=&i\int_0^\infty d\tau\,\exp\left[\tau \left(p_\parallel^2-p_\perp^2 \frac{\tanh(\tau |q_fB|)}{\tau |q_fB|}-m_f^2\right)\right]\nn\\
    &\times&\left\{ [m_f+\slashed{p}_\parallel][1+i\gamma^1 \gamma^2 \tanh(\tau |q_fB|)\text{sign}(q_fB)]  -\frac{\slashed{p}_\perp}{\cosh(\tau |q_fB|)^2}\right\},
    \label{Schwingerprop}
\eea
by performing a Taylor's series expansion for $\tau |q_fB|\sim 0$ and then integrating over the proper time parameter $\tau$. Such a method reveals that the dominant region for the integration over the proper-time parameter is the small $\tau$ region and that, in order to obtain analytical results, this region can be extended to cover the whole original domain $0\leq\tau \leq\infty$.

\subsection{One-Loop Gluon Polarization Tensor}

From the definition of the one-loop gluon polarization tensor:
\bea
    i\Pi^{\mu \nu}_{ab}(p)=-\int \frac{d^4 k}{(2\pi)^4} \text{Tr}\left\{ igt_b\gamma^\nu iS(k)igt_a\gamma^\mu iS(k-p)\right\} + \text{C.C.},
\eea
and from the fermion propagator up to order $\Op\left(\eB^2\right)$:
\bea
iS(k)&=&i \frac{\slashed{k}+m_f}{k^2-m^2}-|q_fB|\frac{\gamma^1 \gamma^2(\slashed{k}_\parallel+m_f)}{(k^2-m_f^2)^2}\text{sign}(q_fB)\nonumber \\ &-&|q_fB|^2\frac{2ik_\perp^2}{(k^2-m_f^2)^4}\Bigg[ m_f+\slashed{k}_\parallel+\slashed{k}_\perp \Bigg(\frac{m_f^2-k_\parallel^2}{k_\perp^2}\Bigg) \Bigg] \nonumber \\ \nonumber \\
&\equiv& iS^{(0)}(k)+iS^{(1)}(k)+iS^{(2)}(k).
\label{propeB2},
\eea
where $iS^{(n)}(k)$ represent the contribution of order $\Op\left(\eB^n\right)$. The magnetized gluon polarization tensor comprises three non-vanishing contributions: one coming from a trace involving the product $S^{(1)}\times S^{(1)}$ (denoted by $i\Pi^{\mu \nu}_{(1,1)}$), other with the product $S^{(2)}\times S^{(0)}$ (denoted by $i\Pi^{\mu \nu}_{(2,0)}$), and its counterpart $S^{(0)}\times S^{(2)}$ (denoted by $i\Pi^{\mu \nu}_{(0,2)}$). The $S^{(0)}\times S^{(0)}$ term originates the vacuum contribution.  

\begin{figure}
    \centering
    \includegraphics[scale=0.8]{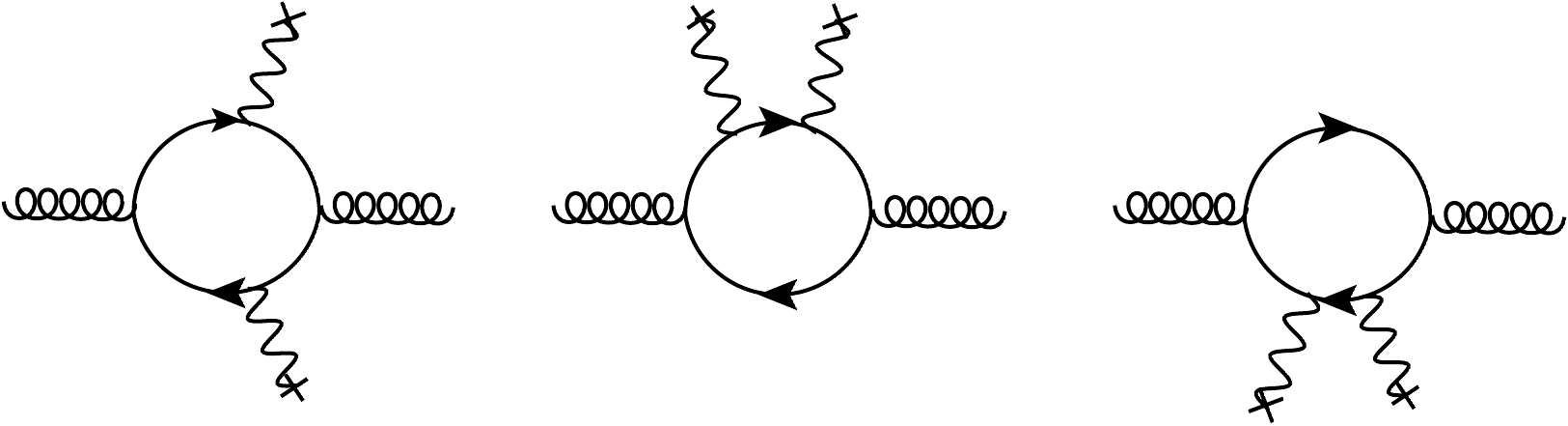}
    \caption{Feynman diagrams contributing to the one-loop gluon polarization tensor in the weak field limit up to $\Op(\eB^2)$. From the left to the right: $i\Pi^{\mu \nu}_{(1,1)}$, $i\Pi^{\mu \nu}_{(2,0)}$ and $i\Pi^{\mu \nu}_{(0,2)}$. }
    \label{Fig:Feyn_Diag_Pol_Tens_Weak_Field}
\end{figure}

The magnetic contributions written in the basis $\left(\mathcal{P}_\parallel^{\mu \nu},\mathcal{P}_\perp^{\mu \nu},\mathcal{P}_0^{\mu \nu}\right)$ defined in Eqs.~(\ref{pipara})-(\ref{pizero}) have the following explicit form (for details see Appendix~\ref{App_Pi11_Pi20_Pi02}):
\begin{eqnarray}
i\Pi^{\mu \nu}_{(1,1)} &=& -\frac{2i\eB^2 g^2}{16 \pi^2}  \int_0^1 dy \int_0^\infty  ds\,s y (1-y) \exp\Bigl[-s(m_f^2 + y(y-1)p^2) \Bigr] \nonumber \\
&\times&\Bigg\{2 y(1-y)  p_\parallel^2  \mathcal{P}_\parallel^{\mu \nu} +g_\parallel^{\mu\nu}\left[m_f^2 -  y(1-y) p_\parallel^2\right] +g_\perp^{\mu\nu}\left[-m_f^2 - \frac{1}{s} - y(1-y) p_\parallel^2\right] \Bigg\},\nn\\
\label{pi11withyands}
\end{eqnarray}
\bea 
i\Pi_{(2,0)}^{\mu\nu}&+&i\Pi_{(0,2)}^{\mu\nu}= -\frac{2}{3}\left(\frac{i}{16\pi^2}\right)g^2\eB^2\int_0^1 dy \int_0^\infty  ds  s  \exp\Bigl[-s(m_f^2 + y(y-1)p^2) \Bigr]\nonumber\\
&\times&\Bigg\{ g_{\parallel}^{\mu\nu} \Bigg[-\frac{1}{ s} + s y^3 (y-1)^3 p_\parallel^2 p_\perp^2  - s y^3 (y-1)^3 p_\perp^4 - y (y-1) (1+ 2y (y-1))p_\perp^2   \nonumber \\
&& +  y (y-1) (1+ 2y (y-1)) p_\parallel^2 + m_f^2 \left( 1- 3y (y-1) +  s y^2 (y-1)^2 p_\perp^2 \right) \Bigg]\nonumber \\
&&+\ g_\perp^{\mu\nu} \Bigg[ \frac{1+ 3y (y-1)}{s }+ s y^3 (y-1)^3  p_\parallel^2 p_\perp^2 - s y^3 (y-1)^3  p_\perp^4 + y (y-1) p_\perp^2
\nonumber\\
&&\ + y (1+ y^2(y-2)) p_\parallel^2 + m_f^2 \left( 1- 3y (y-1) + s y^2 (y-1)^2 p_\perp^2 \right)   \Bigg] \nonumber \\ &&+\mathcal{P}_0^{\mu\nu}\left[ -4 y^2 (y-1)^2 p^2 - 2sy^3 (y-1)^3  p^2 p_\perp^2  \right]  \nonumber \\
&&+\mathcal{P}_{\parallel}^{\mu\nu}\left[ - 2 y(y-1)(1+3y (y-1)) p_\parallel^2 +  4 y^2 (y-1)^2 p_\perp^2 -2 s y^3 (y-1)^3 p^2 p_\perp^2  \right] \nonumber \\
&&+\mathcal{P}_\perp^{\mu\nu}\left[-  4 y^2 (y-1)^2 p_\parallel^2 +  2 y (1+ y^2(y-2))p_\perp^2 -2 s y^3 (y-1)^3 p^2 p_\perp^2 \right] \Bigg\}.
\label{sumPi20and02}
\eea 

Therefore, the gluon polarization tensor in a weak magnetized field is given by
\bea 
i\Pi^{\mu\nu}&=& -\frac{2}{3}\left(\frac{i}{16\pi^2}\right)g^2\eB^2\int_0^1 dy \int_0^\infty  ds  s  \exp\Bigl[-s(m_f^2 + y(y-1)p^2) \Bigr]\nonumber\\
&\times&\Bigg\{ g_\parallel^{\mu\nu} \Biggl[-\frac{1}{ s} + s y^3 (y-1)^3 p_\parallel^2 p_\perp^2  - s y^3 (y-1)^3 p_\perp^4 - y (y-1) (1+ 2y (y-1))p_\perp^2   \nonumber \\
&& +  y (y-1) p_\parallel^2 + m_f^2 \left( 1 +  s y^2 (y-1)^2 p_\perp^2 \right) \Biggl]\nonumber \\
&&+\ g_\perp^{\mu\nu} \Biggl[ \frac{1+6y(y-1) }{s }+ s y^3 (y-1)^3  p_\parallel^2 p_\perp^2 - s y^3 (y-1)^3  p_\perp^4 + y (y-1) p_\perp^2
\nonumber\\
&&\ - y(y-1) (1+ 2y(y-1)) p_\parallel^2 + m_f^2 \left( 1+6y(y-1) + s y^2 (y-1)^2 p_\perp^2 \right)   \Biggr] \nonumber \\ &&+\mathcal{P}_0^{\mu\nu}\left[ -4 y^2 (y-1)^2 p^2 - 2sy^3 (y-1)^3  p^2 p_\perp^2  \right]  \nonumber \\
&&+\mathcal{P}_\parallel^{\mu\nu}\left[ -2y (y-1) p_\parallel^2 +  4 y^2 (y-1)^2 p_\perp^2 -2 s y^3 (y-1)^3 p^2 p_\perp^2  \right] \nonumber \\
&&+\mathcal{P}_\perp^{\mu\nu}\left[-  4 y^2 (y-1)^2 p_\parallel^2 +  2 y (1+ y^2(y-2))p_\perp^2 -2 s y^3 (y-1)^3 p^2 p_\perp^2 \right] \Bigg\}. 
\label{piwithA}
\eea 

Note that the last expression has the form
\begin{equation}
    \Pi^{\mu \nu}= P_\parallel \mathcal{P}_{\parallel}^{\mu \nu}+P_\perp \mathcal{P}_{\perp}^{\mu \nu}+P_0 \mathcal{P}_{0}^{\mu\nu}+A_1 g^{\mu \nu}_\parallel +A_2 g^{\mu \nu}_\perp,
    \label{eq24}
\end{equation}
which as in the case of an arbitrary magnetic field contains terms proportional to $g_\parallel^{\mu\nu}$ and $g_\perp^{\mu\nu}$. That fact breaks the gauge invariance of the tensor, given that the presence of such coefficients implies a non-transverse structure. Nevertheless, coefficients $A_1$ and $A_2$ vanish when the integration over $s$ and $y$ is performed. Hence, the full transverse gluon polarization tensor is
\bea 
i\Pi^{\mu\nu}&=& -\frac{2}{3}\left(\frac{i}{16\pi^2}\right)g^2\eB^2\int_0^1 dy \int_0^\infty  ds\,s  \exp\Bigl[-s(m_f^2 + y(y-1)p^2) \Bigr]\nonumber\\
&\times&\Bigg\{ \mathcal{P}_0^{\mu\nu}\left[ -4 y^2 (y-1)^2 p^2 - 2sy^3 (y-1)^3  p^2 p_\perp^2  \right]  \\
&&+\mathcal{P}_\parallel^{\mu\nu}\left[ -2y (y-1) p_\parallel^2 +  4 y^2 (y-1)^2 p_\perp^2 -2 s y^3 (y-1)^3 p^2 p_\perp^2  \right] \nonumber \\
&&+\mathcal{P}_\perp^{\mu\nu}\left[-  4 y^2 (y-1)^2 p_\parallel^2 +  2 y (1+ y^2(y-2))p_\perp^2 -2 s y^3 (y-1)^3 p^2 p_\perp^2 \right] \Bigg\}.\nn
\label{finalPi}
\eea 

Noticeably, the integrals over $s$ and $y$ can be computed analytically with the constraint $0\leq p^2\leq 4m_f^2$, obtaining the coefficients for each tensor structure:
\begin{subequations}
\bea
&&P_0=\frac{i g^2\eB^2}{6\pi^2}\Bigg \{ \frac{p^2-6 m^2}{p^2 \left(p^2-4 m^2\right)}+\frac{\left(p^2-10 m^2\right) \left(p^2-3 m^2\right)}{p^4 \left(p^2-4 m^2\right)^2}\pt^2\\
&+&\left[\frac{8 m^2 \left(p^2-3 m^2\right) }{(p^2)^{3/2} \left(4 m^2-p^2\right)^{3/2}}-\frac{12 m^2 \left(10 m^4+(p^2-6m^2)p^2\right) }{\left(4m^2-p^2\right)^{5/2} (p^2)^{5/2}}\pt^2\right]\arctan\left(\sqrt{\frac{p^2}{4m_f^2-p^2}}\right) \Bigg \},\nn
\label{coefP0}
\eea
\bea
&&P_\parallel=\frac{i g^2\eB^2}{6\pi^2}\Bigg \{ -\frac{\pp^2}{\left(4 m^2-p^2\right) p^2}-\frac{(p^2-6 m^2)\pt^2}{p^4 \left(p^2-4 m^2\right)}+\frac{\left(p^2-10 m^2\right) \left(p^2-3 m^2\right)\pt^2}{p^4 \left(p^2-4 m^2\right)^2} \nn\\
&+&\left[\frac{-2 \left(p^2-2m^2\right) \pp^2}{\left(4 m^2-p^2\right)^{3/2} (p^2)^{3/2}}-\frac{8 m^2 \left(p^2-3m^2\right)\pt^2}{\left(4 m^2-p^2\right)^{3/2} (p^2)^{5/2}}-\frac{12 m^2 \left(10 m^4+(p^2-6m^2)p^2\right)}{\left(4 m^2-p^2\right)^{5/2} (p^2)^{5/2}}\pt^2\right]\nn\\
&\times&\arctan\left(\sqrt{\frac{p^2}{4m_f^2-p^2}}\right) \Bigg \}, \nn\\
\label{coefPL}
\eea
\bea
&&P_\perp=\frac{i g^2\eB^2}{6\pi^2}\Bigg \{ \frac{\left(p^2-6 m^2\right)\pp^2}{p^4 \left(p^2-4 m^2\right)}-\frac{\left(6 m^2+p^2\right)\pt^2}{2\left(4 m^2-p^2\right) p^4}+\frac{\left(p^2-10 m^2\right) \left(p^2-3 m^2\right)\pt^2}{p^4 \left(p^2-4 m^2\right)^2} \nn\\
&+&\left[\frac{8 m^2 \left(p^2-3m^2\right)\pp^2}{\left(4 m^2-p^2\right)^{3/2} (p^2)^{5/2}}+\frac{2 \left(6 m^4-p^4\right)\pt^2}{\left(4 m^2-p^2\right)^{3/2} (p^2)^{5/2}}-\frac{12 m^2 \left(10 m^4+(p^2-6m^2)p^2\right)}{\left(4 m^2-p^2\right)^{5/2} (p^2)^{5/2}}\pt^2\right]\nn\\
&\times&\arctan\left(\sqrt{\frac{p^2}{4m_f^2-p^2}}\right) \Bigg \}. \nn \\
\label{coefPT}
\eea
\label{coefsweakfieldt}
\end{subequations}
which is the same result of Eqs.~(\ref{Piparaweakfield})-(\ref{Piceroweakfield}).

\subsection{Gluon Dispersion Relation and Polarizations}

\begin{figure}
    \centering
    \includegraphics[scale=0.47]{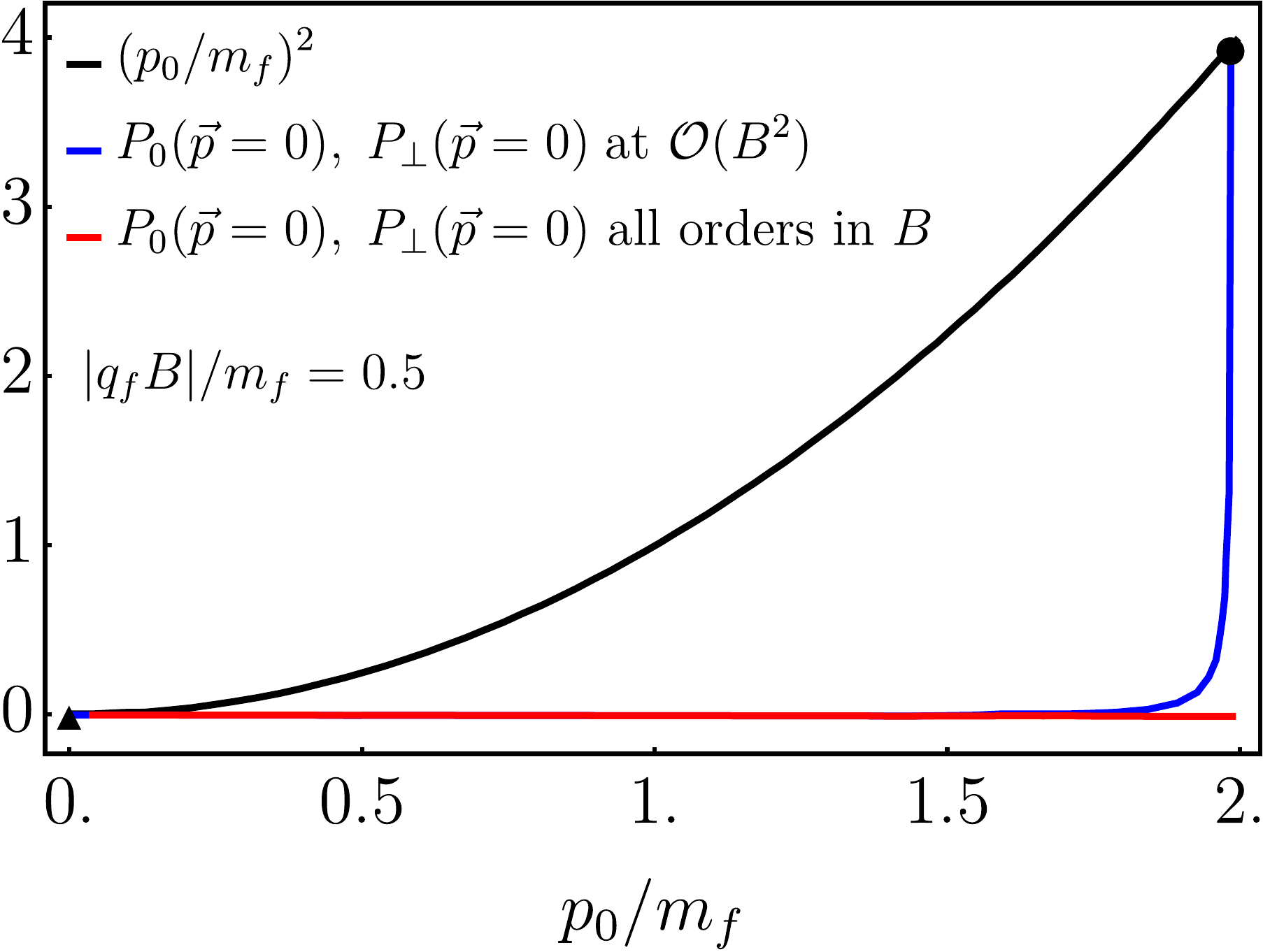}
    \caption{Gluon magnetic mass for the coefficients $P_0$ and $P_\perp$ found by the intersection of the functions at the left and right-side of Eq.~(\ref{solint}) for a fixed value of the magnetic field and for $\alpha_s=g^2/4\pi=0.3$. The triangle represents the physical expected mass whereas the circle is a spurious mass.}
    \label{Fig:Deby_mass_spurios_1}
\end{figure}

The dispersion relation or Debye mass is defined in Eq.~(\ref{Debye}) which is equivalent to
\bea
p_0^2=P_i(p_0)\Big{|}_{p_\perp=0,p_3=0}
\label{solint},
\eea
is computed for the coefficients of Eqs.~(\ref{coefsweakfieldt}). Figure~\ref{Fig:Deby_mass_spurios_1} shows the intersections of the functions that give the coefficients $P_0$ and $P_\perp$ at order $\Op(B^2)$, at all orders (defined by Eq.~(\ref{PienSyY})) and the parabola $(p_0/m_f)^2$. The coefficient $P_\parallel$ is not shown because it has the trivial solution $m_D=0$ in both approximations. As can be noticed, the presented weak field approximation, besides the physical solution $m_D=0$, generates a finite gluon mass for both modes. However, when Eq.~(\ref{solint}) is solved with the expression for all orders in $\eB$, that finite mass disappears, which demonstrates its spurious nature. 
\begin{figure}[H]
    \centering
    \includegraphics[scale=0.47]{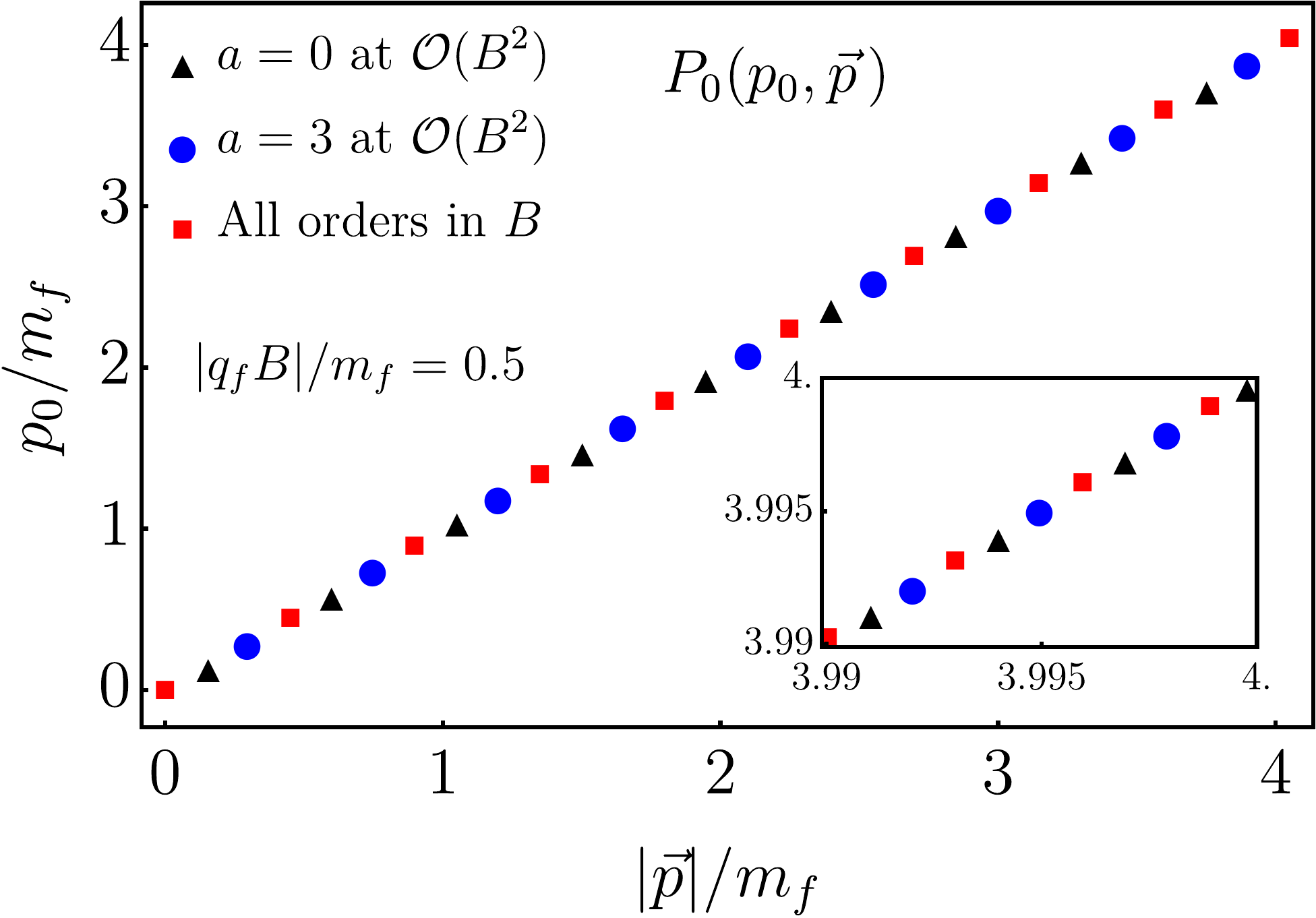}
    \caption{Dispersion relation for the mode $P_0$ obtained with the weak field expansion up to order $\Op(B^2)$ and compared with the general expression at all orders (massless regime). The figure is constructed with the condition $p_3^2=a\pt^2$ for two values of $a$ and a fixed value of $\eB$. The inset shows a complete agreement with the motion along the light cone.}
    \label{Fig:P0conInset}
\end{figure}

The non-physical features are also found in the dispersion relation for the polarization tensor modes. Note that the presence of an external magnetic field breaks Lorentz invariance, which means that the coefficients depend separately on $\pp^2$ and $\pt^2$. Given that the modes do not rely only on $p_0^2$, an analysis has to be made in order to find whether the gluon moves or not along the light-cone. The on-shell condition is studied by the three-momentum parametrization $p_3^2=a\pt^2$, giving the motion along the transverse plane for $a = 0$, and along the magnetic field direction when $a\to\infty$.
\begin{figure}
    \centering
    \includegraphics[scale=0.37]{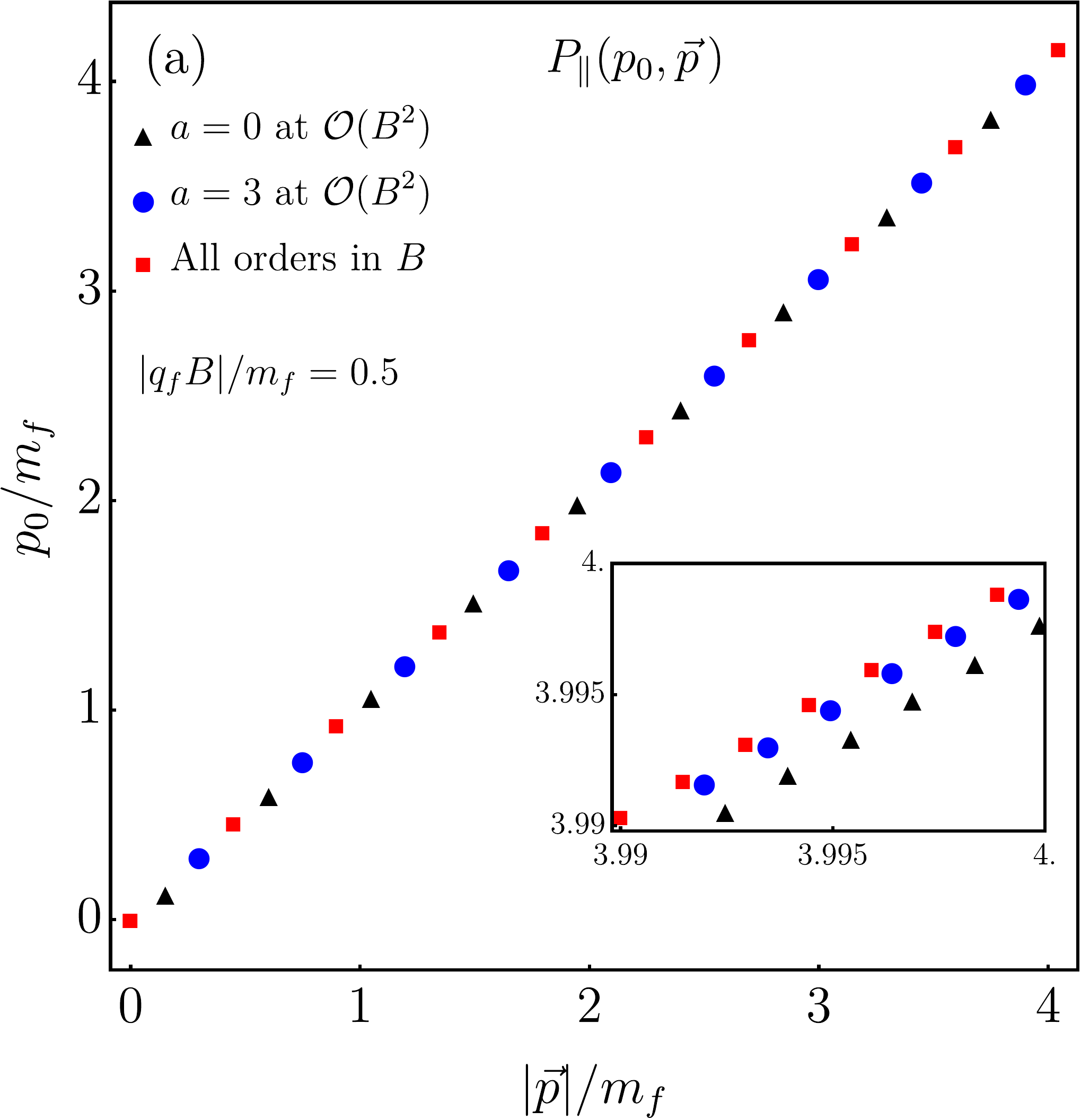}\hspace{0.3cm}\includegraphics[scale=0.37]{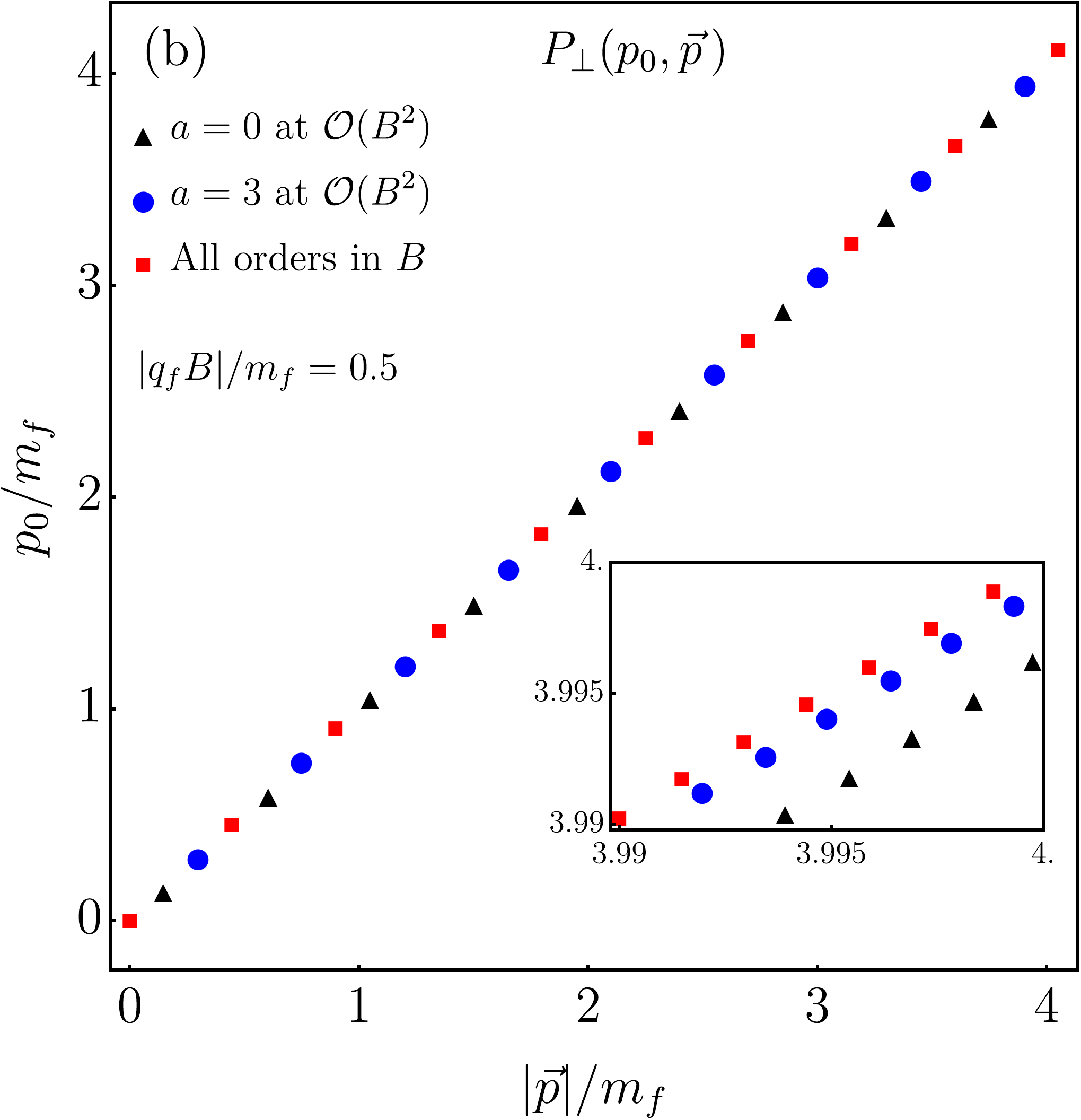}
    \caption{Dispersion relation for the modes $P_\parallel$ and $P_\perp$ obtained with the weak field expansion up to order $\Op(B^2)$ and compared with the general expression at all orders (massless regime). The figure is constructed with the condition $p_3^2=a\pt^2$ for two values of $a$ and a fixed value of $\eB$. The inset shows deviations with respect to the motion along the light cone.}
    \label{Fig:PLandPTconInset}
\end{figure}
\begin{figure}
    \centering
    \includegraphics[scale=0.47]{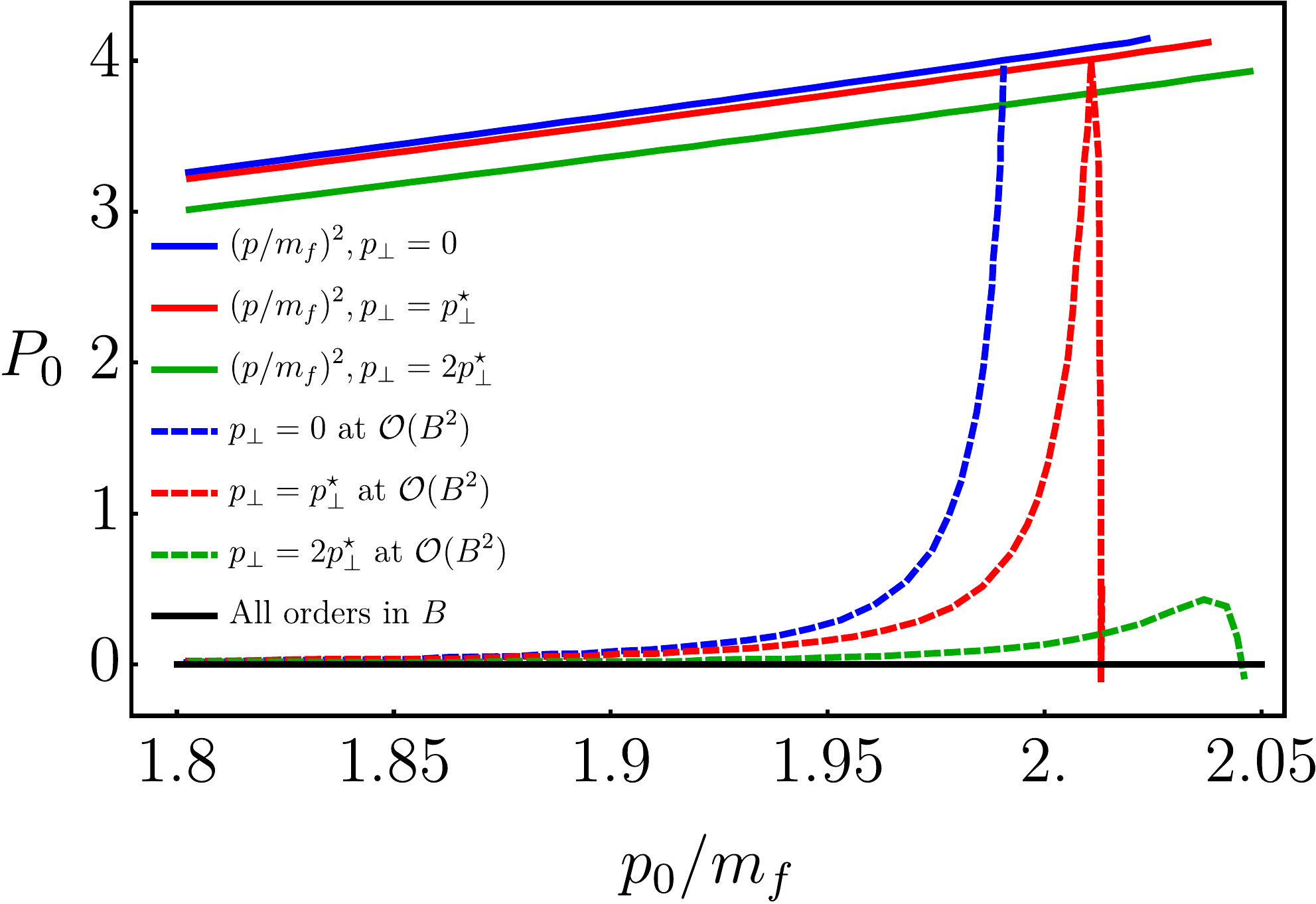}
    \caption{Solutions of Eq.~(\ref{solint}) for the mode $P_0$ as a function of $p_0/m_f$ for several values of $\pt$ with the constriction $p_3^2=a\pt^2$. The plot compares the coefficient obtained from the approximation up to $\Op(B^2)$ with the result obtained to all orders in $B$. As can be noticed the former gives a spurious solution for $\pt\leq\pt^\star\simeq0.1209m_f$, which disappears when the latter is considered. The same behavior is present in the coefficient $P_\perp$.}
    \label{Fig:P0withpstar}
\end{figure}
\begin{figure}
    \centering
    \includegraphics[scale=0.37]{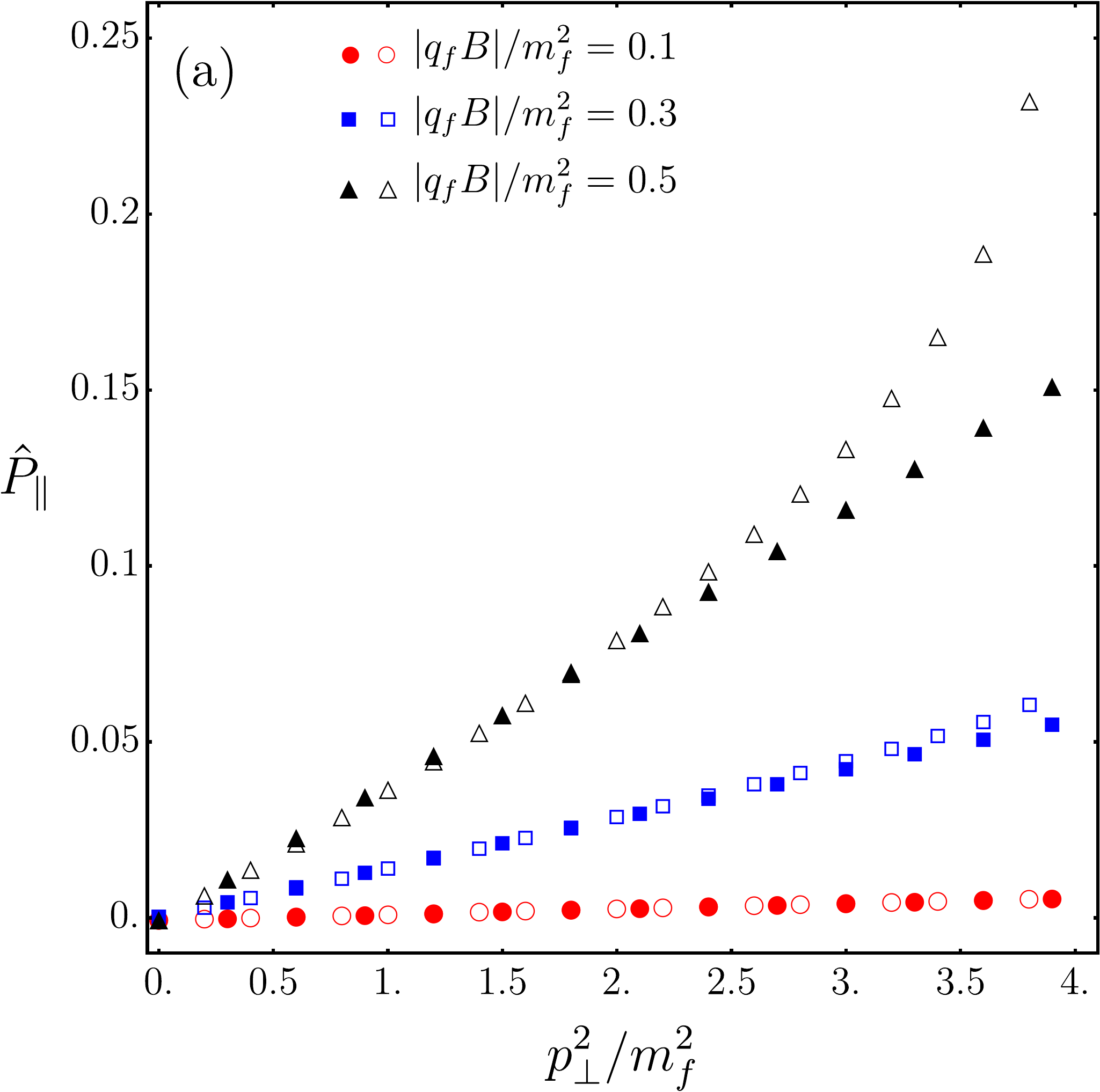}\hspace{0.3cm}\includegraphics[scale=0.37]{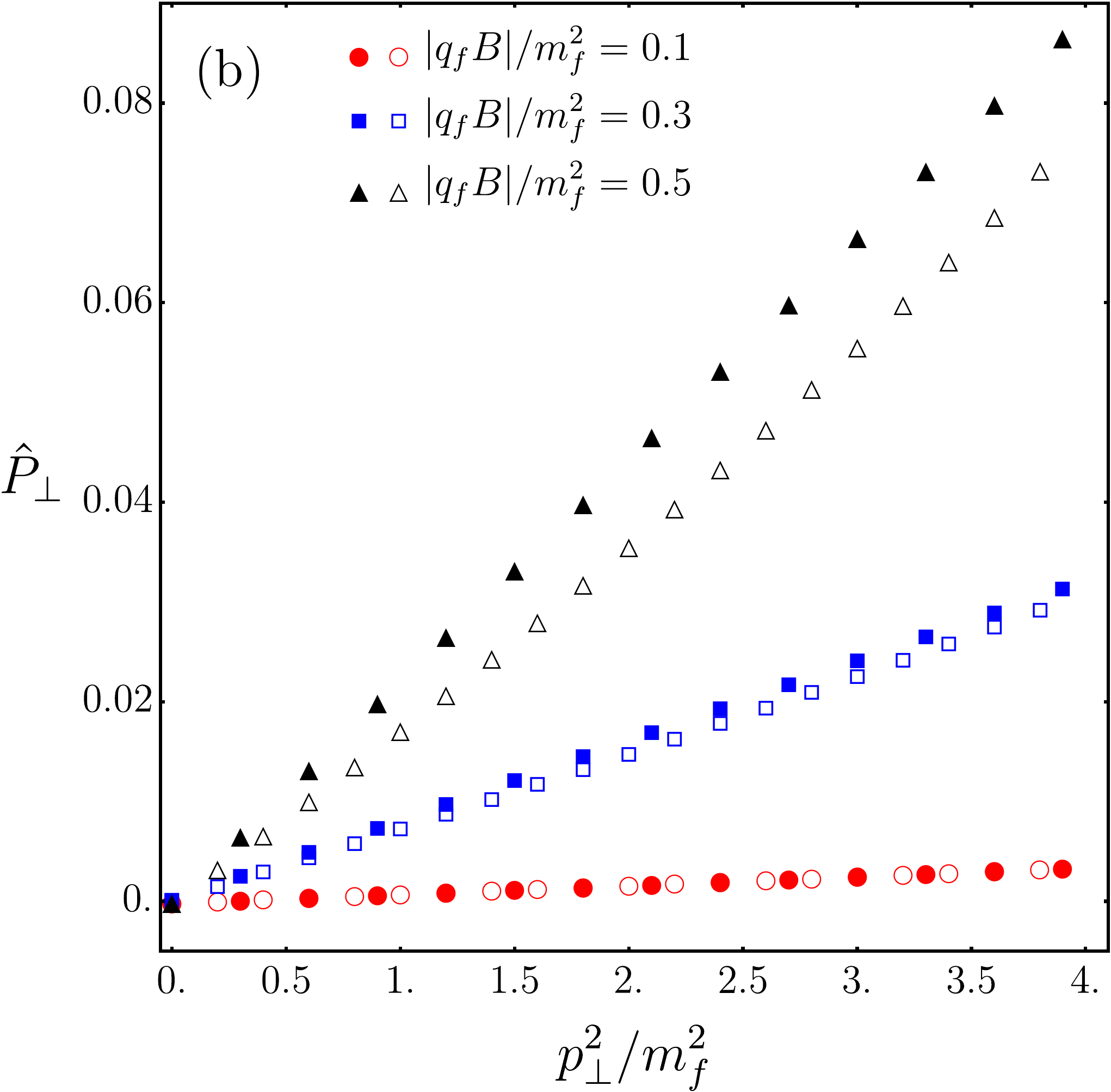}
    \caption{Normalized tensor coefficients $\hat{P}_{\parallel}$ and $\hat{P}_{\perp}$ with the on-shell condition $p^2=0$ as a function of $\pt^2/m_f^2$ computed from the weak field expansion given by Eqs.~(\ref{coefPL})-(\ref{coefPT}) (full symbols) and compared with the general expressions found in Eq.~(\ref{PienSyY}) (open symbols) for three values of $\eB$.}
    \label{Fig:NormalizaedCoefficients}
\end{figure}

Figures~\ref{Fig:P0conInset} and~\ref{Fig:PLandPTconInset} shows the dispersion relation for the mode $P_0$ and $P_\parallel$ and $P_\perp$, respectively, computed from Eqs.~(\ref{coefsweakfieldt}) and compared with the expression to all orders in $\eB$ by considering a motion in the transverse plane ($a=0$) as well as in an arbitrary direction (set by $a=3$). To ensure physically acceptable results, the spurious solution of Eq.~(\ref{solint}) is ignored. The massless regime, i.e., when the gluon moves along the light-cone, is represented by the red squares, and therefore, only the mode $P_0$ reproduce such behavior. As can be noticed, even discarding the finite Debye mass, the approximation up to $\Op(B^2)$  for $P_\parallel$ and $P_\perp$ have slight deviations with respect to the exact result, and the light cone is approached to from below.

As in the case of Debye mass, the solution of Eq.~(\ref{solint}) with the coefficients given by Eqs.~(\ref{coefP0}) and~(\ref{coefPT}) develops spurious results when the transverse momentum is lower than a critical value $\pt=\pt^\star\simeq 0.1209 \ m_f$. As an example, Fig.~\ref{Fig:P0withpstar} shows the dispersion relation for the mode $P_0$ for several values of $\pt$. Such extra solution is not present in the coefficient $P_\parallel$ and is eliminated by considering the magnetic field expansion to all orders. Thus, as far as the gluons' propagation properties are concerned, the approximation for the gluon polarization tensor to the lowest non-trivial order in $ B $ is reliably provided if the second (spurious) solution, corresponding to a finite gluon mass, is discarded.

Figure~\ref{Fig:NormalizaedCoefficients} shows the normalized coefficients $\hat{P}_{\parallel}\equiv\left(8\pi^2/g^2m_f^2\right)P_{\parallel}$ and $\hat{P}_{\perp}\equiv\left(8\pi^2/g^2m_f^2\right)P_{\perp}$ for the on-shell condition $p^2=0$ to analyze the strength of the polarization modes. Such a condition for real gluons implies that the mode $P_0$ vanishes, and the coefficients $P_\parallel$ and $P_\perp$ become only functions of $\pt^2$. The approximation to order $B^2$ and all orders in $B$ are compared for three values of $\eB$. The results in Fig.~\ref{Fig:NormalizaedCoefficients}(a) show that the polarization tensor has a more prominent projection on its parallel component, and the approximation of Eq.~(\ref{coefPL}) is valid for an extended range of $\eB$ and $\pt^2/m_f^2$. In contrast, Fig.~\ref{Fig:NormalizaedCoefficients}(b) implies that Eq.~(\ref{coefPT}) is accurate if small $\pt^2/m_f^2$ or $\eB$ values are considered.

\section{Conclusion}

In this chapter, the gluon-dominated medium's screening effects were computed from the gluon polarization tensor in a magnetized background. Such a calculation was performed in several approximations in order to focus on the study at the relevant energy scales. The calculation presented in Sec.~\ref{sec:Magnetized_Gluon_Polarization_Tensor_from_the_Sum_of_All_Landau_Levels} shows how the resummation of all Landau levels in the fermion propagator splits the polarization tensor into three tensorial structures that encode the translational symmetry breaking. Moreover, the procedure gives the appearance of non-transverse structures proportional to $g_\parallel^{\mu\nu}$ and $g_\perp^{\mu\nu}$. Those non-transverse structures imply that a tensor coefficients calculation starting from projections over the tensor basis may be plagued with non-trivial spurious terms, making it challenging to analyze the gluon properties.  However, a method to systematically eliminate the spurious contribution is provided, and the final expression satisfies the Ward-Takahashi identities. On the other hand, by analyzing the tensor coefficients, several thresholds indicate resonances of the gluon momentum with the external magnetic field Landau levels attributed to pair production. The strong field limit indicates that in a magnetized medium gluons processes the birefringence effect, given the noticeable differences between the tensor coefficients. The birefringence is then associated with the vacuum fluctuations due to the spontaneous emergence of virtual electron-positron (quark-antiquark) pairs that act as dipoles, in analogy with dielectric crystals.

The analysis of both temperature and magnetic field on the gluon screening properties are presented in Sec.~\ref{sec:GluonPolTenThermoMag}. The lowest Landau level approximation, together with the HTL approximations, they were used to find the Debye's mass correction. At the $T=0$ limit, it is shown that the pair-production is oriented along the magnetic field direction. Indeed, since the chromo-electric field thrusts the quark/antiquark motion, only a polarization vector with a component along the external magnetic field can push the virtual pair motion. If temperature $T\ll\eB$ (to ensure thermal fluctuations do not induce transitions between higher Landau levels), the vacuum contribution is isolated, useful for a renormalization group analysis of the gluon polarization tensor. In this hierarchy, temperature and magnetic field effects factorize due to the dimensional reduction in the LLL. The energy scale comparison shows that the final expression is free from infrared divergences, and the Debye mass per mode depends separately on what momentum (longitudinal or transversal) goes first to zero. If the rate is equal, both modes develop the same Debye mass.

Finally, Sec.~\ref{sec:GluonPolTenWeakField} presents the gluon polarization tensor for zero temperature in a weakly magnetized medium. The fermion propagator expansion is revisited to clarify the method presented in previous works. The calculation presents the non-transverse structures again, which after integration over the Schwinger parameters, vanishes. Nevertheless, the weak field limit shows non-physical results in the dispersion relation and the Debye mass by comparison with the general expression. The origin of spurious solutions is identified as a feature of the approximation at order $B^2$, and they are eliminated by resuming all the magnetic contributions. 

\appendix

\chapter{High-Temperature Approximation for the One-Loop Boson and Fermion Effective Potentials}\label{Ap:Veff_bosons_HT}

\section*{High-Temperature Bosonic Potential}

To compute the  high-temperature approximation for the bosonic potential, the starting point is Eq.~(\ref{Vb1int}):
\bea
V_b^{(1)}=\frac{1}{2\pi}\int dk\,k^2\left[\frac{\sqrt{k^2+m_b^2}}{2}+T\ln\left(1-e^{-\sqrt{k^2+m_b^2}/T}\right)\right],
\eea
where the first terms correspond to the vacuum contributions. Sucha a term is computed via dimensional regularization by the following well-known result: 
\bea
\int \frac{d^{d-2\epsilon} k}{(2 \pi)^{d-2\epsilon}} \frac{1}{\left(k^{2}-m_{b}^{2}\right)^{n}} =\mathrm{i}(-1)^{n}\frac{\left(m^{2}\right)^{2-\epsilon-n}}{(4 \pi)^{2-\epsilon}} \frac{\Gamma(n-2+\epsilon)}{\Gamma(n)},
\eea
so that $d=3$ and $n=-1/2$. Thus, by introducing the regulator $\widetilde{\mu}^{3-d}$ the vacuum piece is:
\bea
V_{\mathrm{vac}}^{(1) \mathrm{b}}&=&\frac{\tilde{\mu}^{3-d}}{2} \int \frac{d^{d} k}{(2 \pi)^{d}} \sqrt{k^{2}+m_{b}^{2}}\nn\\
&=&-\frac{m_{b}^{4}}{32 \pi^{2}} \Gamma(\epsilon-2)\left(\frac{4 \pi \tilde{\mu}^{2}}{m_{b}^{2}}\right)^{\epsilon}.
\eea

A power series around $\epsilon=0$ yields:
\bea
V_{\mathrm{vac}}^{(1) \mathrm{b}}=-\frac{m_{b}^{4}}{64 \pi^{2}}\left[\ln \left(\frac{4 \pi \tilde{\mu}^{2}}{m_{b}^{2}}\right)-\gamma_{e}+\frac{3}{2}+\frac{1}{\epsilon}\right],
\eea
where the divergence can be absorbed in the so-called Minimal Subtraction scheme (MS) with the renormalization scale $\tilde{\mu}=ae^{-1/2}$, giving the result:
\bea
V_{\mathrm{vac}}^{(1) \mathrm{b}}=-\frac{m_{b}^{4}}{64 \pi^{2}}\left[\ln \left(\frac{4 \pi a^{2}}{m_{b}^{2}}\right)-\gamma_{e}+1\right].
\eea

The matter contribution to the boson-effective potential is
\bea
V_{\operatorname{matt}}^{(1) \mathrm{b}}=\frac{T}{2 \pi^{2}} \int d k k^{2} \ln \left(1-e^{-\sqrt{k^{2}+m_{b}^{2}} / T}\right),
\eea
which performing the substitution $x=k\beta$, $y=m_b\beta$ with $\beta=1/T$ gives:
\bea
V_{\operatorname{matt}}^{(1) \mathrm{b}}=\frac{1}{2 \pi^{2}\beta^4} \int_{0}^{\infty} d x\,x^{2} \ln \left(1-e^{-\sqrt{x^{2}+y^{2}}}\right).
\eea

To find the integral, the method developed by Dolan and Jackiw is used~\cite{dolan1974symmetry}. That method consists of promoting the integral to a second-order differential equation with the quantities:
\bea
\frac{\partial}{\partial y^2}V_{\operatorname{matt}}^{(1) \mathrm{b}}=\frac{1}{4 \pi^{2} \beta^{4}} \int_{0}^{\infty} d x \frac{x^{2}}{\sqrt{x^{2}+y^{2}}} \frac{1}{\exp \left(\sqrt{x^{2}+y^{2}}\right)-1},
\eea
and
\bea
\frac{\partial^2}{\partial (y^2)^2}V_{\operatorname{matt}}^{(1) \mathrm{b}}=-\frac{1}{8 \pi^{2} \beta^{4}}\int_{0}^{\infty} \frac{d x} {\sqrt{x^{2}+y^{2}}}\frac{1}{\exp\left(\sqrt{x^{2}+y^{2}}\right)-1},
\eea
with the boundary conditions:
\bea
V_{\operatorname{matt}}^{(1) \mathrm{b}}\Big{|}_{y^2=0}=-\frac{\pi}{90\beta^4}
\label{BC1}
\eea
and
\bea
\frac{\partial}{\partial y}V_{\operatorname{matt}}^{(1) \mathrm{b}}\Big{|}_{y^2=0}=\frac{1}{24\beta^4}.
\label{BC2}
\eea

The second-derivative is found by using the following regularized integral:
\bea
I_\epsilon(y)=\int_{0}^{\infty} d x \frac{x^{-\epsilon}}{\left(x^{2}+a^{2}\right)^{1 / 2}}\frac{1}{\exp\left(\sqrt{x^{2}+y^{2}}\right)-1},\; \epsilon<-1,
\eea
which can be written in terms of the series:
\bea
\sum_{n=1}^{\infty} \frac{y}{y^{2}+n^{2}}=-\frac{1}{2 y}+\frac{1}{2} \pi \operatorname{coth} \pi y,
\label{serieforV}
\eea
so that:
\bea
I_{\epsilon}(y)=I_{\epsilon}^{(1)}(y)+I_{\epsilon}^{(2)}(y),
\eea
where
\bea
I_{\epsilon}^{(1)}(y)=\int_{0}^{\infty} d x\,x^{-\epsilon} \sum_{n} \frac{1}{x^{2}+y^{2}+4 \pi^{2} n^{2}},\;\;\; n=0,\pm1,\pm2,\cdots
\label{Ie1}
\eea
and
\bea
I_{\epsilon}^{(2)}(y)=-\frac{1}{2} \int_{0}^{\infty} d x  \frac{x^{-\epsilon}}{\left(x^{2}+y^{2}\right)^{1 / 2}}.
\eea

For $I_{\epsilon}^{(1)}$, the integration over $x$ gives:
\bea
I_{\epsilon}^{(1)}&=&\frac{\pi}{2}\sec\left(\frac{\pi\epsilon}{2}\right)\sum_n\frac{1}{\left(y^2+4\pi^2n^2\right)^{(1+\epsilon)/2}}\nn\\
&=&\frac{\pi}{2}\sec\left(\frac{\pi\epsilon}{2}\right)\left[\frac{1}{y^{1+\epsilon}}+\sum_{n=1}^{\infty} \frac{2}{(2 \pi n)^{1+\epsilon}}
+\sum_{n=1}^{\infty}\frac{2}{(2 \pi n)^{1+\epsilon}}\left(\frac{1}{\left(1+y^{2}/4 \pi^{2} n^{2}\right)^{(1+\epsilon) / 2}}-1\right)\right],\nn\\
\eea
which from the fact that $\sec(\pi\epsilon/2)$ is regular around $\epsilon=0$ can be separated as
\bea
I_{\epsilon}^{(1)}=\frac{\pi}{2y}+2^{-(1+\epsilon)}\pi^{-\epsilon}\zeta(1+\epsilon)+\tilde{I}(y)+\mathcal{O}(\epsilon),
\eea
where $\zeta(x)$ is the Riemann zeta function and
\bea
\tilde{I}(y) =\frac{1}{2} \sum_{n=1}^{\infty} \frac{1}{n}\left[\left(1+\frac{y^{2}}{4 \pi^{2} n^{2}}\right)^{-1 / 2}-1\right]\sim\mathcal{O}\left(y^{2}\right),
\eea
in such a way that by performing a series expansion around $\epsilon=0$ the integral is
\bea
I_{\epsilon}^{(1)}(y)=\frac{1}{2 \epsilon}+\frac{\pi}{2 y}+\frac{1}{2}(\gamma_e-\ln 2 \pi)+\tilde{I}(y)+\mathcal{O}(\epsilon).
\eea

For $I_{\epsilon}^{(2)}(y)$, the integration over $x$ is direct:
\bea
I_{\epsilon}^{(2)}(y)&=&-\frac{1}{2} \int_{0}^{\infty} d x  \frac{x^{-\epsilon}}{\sqrt{x^{2}+y^{2}}}\nn\\
&=&\frac{y^{-\epsilon}}{2\sqrt{\pi}}\,\Gamma\left(\frac{1}{2}-\frac{\epsilon}{2}\right)\Gamma\left(\frac{\epsilon}{2}\right),
\label{Ie2f}
\eea
so that when $\epsilon~\sim0$ behaves as:
\bea
I_{\epsilon}^{(2)}(y)=-\frac{1}{2\epsilon}+\frac{1}{2}\ln\left(\frac{y}{2}\right)+\mathcal{O}(\epsilon).
\eea

In this way, the integral $I_\epsilon(y)$ when $\epsilon\to0$ is
\bea
I(y)=\frac{\pi}{2y}+\frac{1}{2}\ln\left(\frac{y}{4\pi}\right)+\frac{\gamma_e}{2}+\tilde{I}(y),
\eea
and therefore
\bea
\frac{\partial^2}{\partial (y^2)^2}V_{\operatorname{matt}}^{(1) \mathrm{b}}=-\frac{1}{8\pi^2\beta^4}\left[\frac{\pi}{2y}+\frac{1}{2}\ln\left(\frac{y}{4\pi}\right)+\frac{\gamma_e}{2}+\tilde{I}(y)\right].
\label{d2Vmatb}
\eea

Now, for a high-temperature approximation $y\ll1$, so that $\tilde{I}(y)$ can be neglected. The effective potential is computed by integrating Eq.~(\ref{d2Vmatb}) with respect to $y^2$:
\bea
\frac{\partial}{\partial y^2}V_{\operatorname{matt}}^{(1) \mathrm{b}}=-\frac{1}{8\pi^2\beta^4}\left(\pi y+\frac{\gamma_e  }{2}y^2+\frac{1}{4}y^2\left[\ln\left(\frac{y^2}{16 \pi^2}\right)-1\right]\right)+c_1,
\eea
where the constant $c_1$ is given by the boundary condition of Eq.~(\ref{BC2}):
\bea
c_1=\frac{\partial}{\partial y}V_{\operatorname{matt}}^{(1) \mathrm{b}}\Big{|}_{y^2=0}=\frac{1}{24\beta^4}.
\eea

Performing again the integral over $y^2$:
\bea
V_{\operatorname{matt}}^{(1) \mathrm{b}}=\frac{y^2}{24\beta^4}-\frac{1}{8\pi^2\beta^4}\left[\frac{2\pi y^3}{3}+\frac{1}{8}\left(2\gamma_e-\frac{3}{2}+\ln\left[\frac{y^2}{16\pi^2}\right]\right)\right]+c_2,
\eea
with $c_2$ provided by the boundary condition of Eq.~(\ref{BC1}), namely:
\bea
c_2=V_{\operatorname{matt}}^{(1) \mathrm{b}}\Big{|}_{y^2=0}=-\frac{\pi}{90\beta^4}
\eea

Finally, by replacing the value of $y$ and adding all terms, the bosonic high temperature-potential is found. 

\section*{High-Temperature Fermionic Potential}

For the high-temperature fermionic potential the vacuum and matter pieces are separated. Starting from Eq.~(\ref{Vf1int}):
\bea
\quad V_f^{(1)}&=&-\frac{1}{\pi^2}\int dk\,k^2\sqrt{k^2+m_f^2}\nn\\
&+&\frac{T}{\pi^2}\int dk\,k^2\Bigg\{\ln\left[1+e^{-\left(\sqrt{k^2+m_f^2}-\mu\right)/T}\right]+\ln\left[1+e^{-\left(\sqrt{k^2+m_f^2}+\mu\right)/T}\right]\Bigg\},\quad
\eea
it is easy to recognize the vacuum contribution, which has the same functional form as the bosonic vacuum. Therefore, by following the procedure of the last section is straightforward to find that
\bea
V_{\mathrm{vac}}^{(1) \mathrm{f}}=\frac{m_{f}^{4}}{16 \pi^{2}}\left[\ln \left(\frac{4 \pi a^{2}}{m_{f}^{2}}\right)-\gamma_{e}+1\right].
\label{fermionicvacuumap}
\eea

The matter contribution follows the differential equation method discussed above. Making the definitions $x=k\beta$, $y=m_b\beta$ and $z=\mu\beta$ such potential is given by
\bea
V_{\text {matt }}^{(1) \mathbf{f}}=\frac{1}{\pi^2\beta^4}\int_0^\infty dx\,x^2\left[\ln \left(1+e^{-\sqrt{x^{2}+y^{2}}-z}\right)+\ln \left(1+e^{-\sqrt{x^{2}+y^{2}}+z}\right)\right],
\eea
so that
\bea
\frac{\partial V_{\operatorname{matt}}^{(1) \mathrm{f}}}{\partial y^2}=\frac{1}{2\pi^2\beta^4}\int_{0}^{\infty}dx \frac{x^2}{\sqrt{x^{2}+y^{2}}}\sum_{\sigma=\pm} \frac{1}{\exp \left(\sqrt{x^{2}+y^{2}}+\sigma z\right)+1},\nn\\
\eea
\bea
\frac{\partial^2V_{\operatorname{matt}}^{(1) \mathrm{f}}}{\partial (y^2)^2}&=&\frac{1}{4\pi^2\beta^4}\int_{0}^{\infty} \frac{d x} {\sqrt{x^{2}+y^{2}}}\sum_{\sigma=\pm} \frac{1}{\exp \left(\sqrt{x^{2}+y^{2}}+\sigma z\right)+1}.\nn\\
\eea

The last equations provide the boundary conditions:
\bea
V_{\operatorname{matt}}^{(1) \mathrm{f}}\Big{|}_{y^2=0}=-\frac{2}{\pi^2\beta^4}\left[\Li_4\left(-e^{-z}\right)+\Li_4\left(-e^{z}\right)\right]
\label{BC3}
\eea
and
\bea
\frac{\partial V_{\operatorname{matt}}^{(1) \mathrm{f}}}{\partial y^2}\Big{|}_{y^2=0}=-\frac{1}{2\pi^2\beta^4}\left[\Li_2\left(-e^{-z}\right)+\Li_2\left(-e^{z}\right)\right],
\label{BC4}
\eea
where $\Li_s(x)$ is the polylogarithm function of order $s$.

To compute the second derivative it is useful to use the following identity:
\bea
&&\sum_{n=-\infty}^{+\infty} \frac{4 \pi^{2}}{[(2 n+1) \pi+i z]^{2}+x^{2}+y^{2}}\nn\\
&=&\frac{2 \pi^{2}}{\sqrt{x^{2}+y^{2}}}\left[1-\frac{1}{\exp (\sqrt{x^{2}+y^{2}}-z)}-\frac{1}{ \exp (\sqrt{x^{2}+y^{2}}+z)+1}\right],
\eea
so that the desired (regularized) integral is written as
\bea
J_{\epsilon}(y)=J_{\epsilon}^{(1)}(y)+J_{\epsilon}^{(2)}(y),
\eea
with
\bea
J_{\epsilon}^{(1)}(y)=-2 \int_{0}^{\infty} d x \sum_{n=-\infty}^{+\infty} \frac{x^{-\epsilon}}{[(2 n+1) \pi+i z]^{2}+x^{2}+y^{2}},
\eea
and
\bea
J_{\epsilon}^{(2)}(y)=\int_{0}^{\infty} d x \frac{x^{-\epsilon}}{\sqrt{x^{2}+y^{2}}}.
\eea

Note that $J_{\epsilon}^{(2)}(y)$ has the same functional form that Eq.~(\ref{Ie2f}); thus, the calculation of interest is about $J_{\epsilon}^{(1)}(y)$. After integrating over $x$ the function $J_{\epsilon}^{(1)}(y)$ becomes
\bea
J_{\epsilon}^{(1)}(y)=-\pi\sec\left(\frac{\pi \epsilon}{2}\right) \sum_{n=-\infty}^{+\infty}\left\{\frac{1}{[(2 n+1) \pi+i z]^{2}+y^{2}}\right\}^{\frac{1+\epsilon}{2}},
\eea
where the sum can be spitted into positive and negative $n$'s, yielding:
\bea
J_{\epsilon}^{(1)}(y)&=&-\pi \sec\left(\frac{\pi \epsilon}{2}\right)\nn\\
&\times& \sum_{n=0}^{+\infty}\left\{\frac{1}{\left([(2 n+1) \pi+i z]^{2}+y^{2}\right)^{\frac{1+\epsilon}{2}}}+\frac{1}{\left([(2 n+1) \pi-i z]^{2}+y^{2}\right)^{\frac{1+\epsilon}{2}}}\right\}.
\eea

The denominator can be simplified:
\bea
\frac{1}{\left([(2 n+1) \pi \pm i z]^{2}+y^{2}\right)^{\frac{1+\epsilon}{2}}} &=&\frac{1}{(2 \pi)^{1+\epsilon}} \frac{1}{[n+1 / 2 \pm i z / 2 \pi]^{1+\epsilon}} \\
& \times&\left[1+\frac{y^{2} / 4 \pi^{2}}{(n+1 / 2 \pm i z / 2 \pi)^{2}}\right]^{-\frac{1+t}{2}},
\eea
which from the binomial theorem becomes:
\bea
&&\frac{1}{(2 \pi)^{1+\epsilon}} \sum_{l=0}^{+\infty} \frac{(-1)^{l}}{(n+1 / 2 \pm i z / 2 \pi)^{\epsilon+1}}\binom{\frac{\epsilon-1}{2}+l}{\frac{\epsilon-1}{2}}
\frac{(y / 2 \pi)^{2 l}}{(n+1 / 2 \pm i z / 2 \pi)^{2 l}}\nn\\
&=&\frac{1}{(2 \pi)^{1+\epsilon}} \frac{1}{(n+1 / 2 \pm i z / 2 \pi)^{\epsilon+1}}\nn\\
&+&\frac{1}{(2 \pi)^{1+\epsilon}} \sum_{l=1}^{+\infty} (-1)^{l}\binom{
\frac{\epsilon-1}{2}+l}{\frac{\epsilon-1}{2}}\frac{(y / 2 \pi)^{2 l}}{(n+1 / 2 \pm i z / 2 \pi)^{2 l+\epsilon+1}}.
\eea

Preforming the sum over $n$:
\bea
J_{\epsilon}^{(1)}(y)&=&-\pi \frac{\sec\left(\frac{\pi \epsilon}{2}\right)}{(2 \pi)^{1+\epsilon}}\sum_{\sigma=\pm}\left[\zeta\left(1+\epsilon ; \frac{1}{2}+\sigma\frac{i z}{2 \pi}\right)+\tilde{J}_{
\sigma}(y)\right],
\eea
where $\zeta(s,x)$ is the Hurwitz zeta function  of order $s$ and
\bea
\tilde{J}_{\sigma}(y)\equiv\sum_{l=1}^{+\infty}\frac{(-1)^{l}}{(2 \pi)^{2 l}}\binom{\frac{\epsilon-1}{2}+l}{\frac{\epsilon-1}{2}}\zeta\left(2 l+\epsilon+1 ; \frac{1}{2}+\sigma\frac{i z}{2 \pi}\right) y^{2 l}.
\eea

In the same way, as in the boson effective-potential calculation, a high-temperature approximation means $y\ll1$; therefore, the function $\tilde{J}_s(y)$ can be neglected. Moreover, by expanding into power series around $\epsilon=0$:
\bea
J_{\epsilon}^{(1)}(y)\approx-\frac{1}{\epsilon}+\ln(2\pi)+\frac{1}{2}\psi_{0}\left(\frac{1}{2}+\frac{i z}{2 \pi}\right)+\frac{1}{2}\psi_{0}\left(\frac{1}{2}-\frac{i z}{2 \pi}\right)+\Op(\epsilon),
\eea
with $\psi_0(x)$ the zeroth-order polygamma function. 

After joining all the pieces, the second derivative has the form: 
\bea
\frac{\partial^2V_{\operatorname{matt}}^{(1) \mathrm{f}}}{\partial (y^2)^2}=\frac{1}{4\pi^2\beta^4}\left[\ln(2\pi)+\ln\left(\frac{2}{y}\right)+\frac{1}{2}\psi_{0}\left(\frac{1}{2}+\frac{i z}{2 \pi}\right)+\frac{1}{2}\psi_{0}\left(\frac{1}{2}-\frac{i z}{2 \pi}\right)\right],
\eea
and by integration over $y^2$:
\bea
\frac{\partial V_{\operatorname{matt}}^{(1) \mathrm{f}}}{\partial y^2}=\frac{y^2}{4\pi^2\beta^4}\left[\frac{1}{2}+\ln(2\pi)-\frac{1}{2}\ln\left(\frac{y^2}{4}\right)+\frac{1}{2}\psi_{0}\left(\frac{1}{2}+\frac{i z}{2 \pi}\right)+\frac{1}{2}\psi_{0}\left(\frac{1}{2}-\frac{i z}{2 \pi}\right)\right]+c_3,\nn\\
\eea
where $c_3$ is given by the boundary condition of Eq.~(\ref{BC3}). Finally, integrating the last equation:
\bea
V_{\operatorname{matt}}^{(1) \mathrm{f}}&=&\frac{y^2}{4\pi^2\beta^4}\left\{\frac{1}{2}\left[\ln(2\pi)+\frac{1}{2}\psi_{0}\left(\frac{1}{2}+\frac{i z}{2 \pi}\right)+\frac{1}{2}\psi_{0}\left(\frac{1}{2}-\frac{i z}{2 \pi}\right)\right]y^2\right.\nn\\
&+&\left.\frac{y^2}{4}\left[\frac{3}{2}-\ln\left(\frac{y^2}{4}\right)\right]+c_3\right\}+c_4,
\eea
with $c_4$ obtained from Eq.~(\ref{BC4}).

The latter together with Eq.~(\ref{fermionicvacuumap}) defines the high-temperature fermionic potential given in Eq.~(\ref{VfHT}).

\chapter{Schwinger's Phase Factor}\label{ApPhaseFactorCalc}

The Schwinger's phase factor is defined in Eq.~(\ref{PhaseFactorDef}):
\bea
\Phi(x,y)=\exp\left\{i|q_f|\int_{y}^xd\xi^\mu\left[A_\mu+\frac{1}{2}F_{\mu\nu}(\xi-y)^\nu\right]\right\},
\label{IntPhasefac}
\eea
which is is a translational and gauge noninvariant value. Given that $F_{\mu\nu}=\partial_\mu A_\nu-\partial_\nu A_\mu$, the gauge invariance allows to write the 4-potential for a constant magnetic field as:
\bea
A_\mu(x)=-\frac{1}{2}F_{\mu\nu}x^\nu+\partial_\mu\lambda(x),
\label{Achoice}
\eea
where $\lambda(x)$ is an arbitrary well behaved function provided by the choice of gauge. By replacing Eq.~(\ref{Achoice}) in the integral of Eq.~(\ref{IntPhasefac}):
\bea
\mathcal{I}(x,y)&\equiv&\int_{y}^xd\xi^\mu\left[A_\mu+\frac{1}{2}F_{\mu\nu}(\xi-y)^\nu\right]\nn\\
&=&\int_{y}^xd\xi^\mu\left[-\frac{1}{2}F_{\mu\nu}\,y^\nu+\partial_\mu\lambda(\xi)\right]\nn\\
&=&-\frac{1}{2}\left(x-y\right)^\mu F_{\mu\nu}\,y^\nu+\lambda(x)-\lambda(y),
\eea
but $x^\mu F_{\mu\nu}\,y^\nu=-x^\nu F_{\mu\nu}\,y^\mu$ and $y^\mu F_{\mu\nu}\,y^\nu=0$, and therefore:
\bea
\mathcal{I}(x,y)=\frac{1}{2}y^\mu F_{\mu\nu}\,x^{\nu}+\lambda(x)-\lambda(y).
\label{IntPhaseFactFinal}
\eea

From the above to find the term in the exponential of $\Phi(x,y)\Phi(y,z)\Phi(z,x)$ is
\bea
\mathcal{I}(x,y)+\mathcal{I}(y,z)+\mathcal{I}(z,x)&=&\frac{1}{2}y^\mu F_{\mu\nu}\,x^{\nu}+\lambda(x)-\lambda(y)\nn\\
&+&\frac{1}{2}z^\mu F_{\mu\nu}\,y^{\nu}+\lambda(y)-\lambda(z)\nn\\
&+&\frac{1}{2}x^\mu F_{\mu\nu}\,z^{\nu}+\lambda(z)-\lambda(x),
\eea
where the cancellation of the $\lambda$ functions resembles the gauge invariance. To get the desired result, the term $-x^\mu F_{\mu\nu}x^\nu=0$ is added and given that $z^\mu F_{\mu\nu}\,y^\nu=-y^\mu F_{\mu\nu}\,z^\nu$, the exponential factor becomes:
\bea
\mathcal{I}(x,y)+\mathcal{I}(y,z)+\mathcal{I}(z,x)&=&\frac{1}{2}F_{\mu\nu}\left(x^\mu z^\nu-y^\mu z^\nu-x^\mu x^\nu+y^\mu x^\nu\right)\nn\\
&=&\frac{1}{2}(x-y)^\mu F_{\mu\nu}(z-x)^\nu.
\eea

Now, from the choice of $A_\mu$ in Eq.~(\ref{Asymmetric}):
\bea
A^\mu=\frac{B}{2}(0,-y,x,0),
\eea
the only non-vanishing terms of $F_{\mu\nu}$ are $F_{12}=-F_{21}=|B|$ which yields:
\bea
\mathcal{I}(x,y)+\mathcal{I}(y,z)+\mathcal{I}(z,x)=-\frac{|B|}{2}\epsilon_{mj}(z-x)_m(x-y)_j,
\eea
and finally:
\bea
\Phi(x,y)\Phi(y,z)\Phi(z,x)=\exp\left[-i\frac{\eB}{2}\epsilon_{mj}(z-x)_m(x-y)_j\right],
\eea
which is the phase factor of Eq.~(\ref{phasefactor}).

\chapter{Matrix element for the process \texorpdfstring{$gg\rightarrow\gamma$}{} with all the quarks in the LLL}\label{ThreeQuarksinLLL}

\begin{figure}[H]
    \centering
    \includegraphics[scale=0.65]{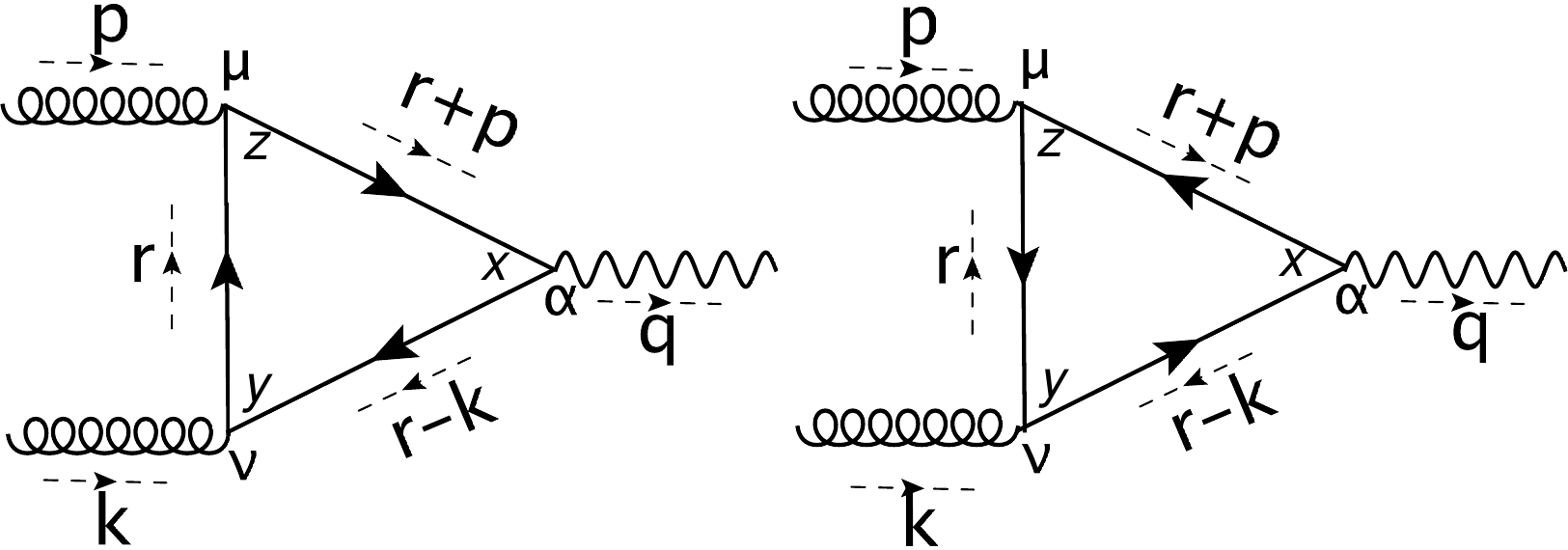}
    \caption{Feynman diagrams contributing to the process $gg\rightarrow\gamma$ with all the fermions in the loop are in the LLL.}
    \label{Fig:todosenLLL}
\end{figure}

In the limit $m_f\to0$ the trace which appears into Eq.~(\ref{amplitude}) related to the left diagram shown in Fig.~\ref{Fig:todosenLLL} is given by
\bea
\text{Tr}\left\{\ga(\rs+\ps)_{\p}\Op^{+}\gm\rs_{\p}\Op^{+}\gn(\rs-\ks)_{\p}\Op^{+}\right\},
\label{diag0a}
\eea
which from the identities
\bea
\ps_{\p}\,\Op^{\pm}&=&\Op^{\pm}\,\ps_{\p}\nn\\
\Op^{\pm}\gm\Op^{\pm}&=&\Op^{\pm}\gm^{\p},
\label{propiedadesOp}
\eea
and given that $\Op^{\pm}$ is idempotent and orthogonal to $\Op^{\mp}$, the operator $\Op^{+}$ can be moved to the left, such that:
\bea
\text{Tr}\left\{\Op^{+}\ga^{\p}(\rs+\ps)_{\p}\gm^{\p}\rs_{\p}\gn^{\p}(\rs-\ks)_{\p}\right\}.
\label{Tr01}
\eea

The charge conjutate of Eq.~(\ref{Tr01}) is related to the right diagram of Fig.~\ref{Fig:todosenLLL}:
\bea
&&\text{Tr}\left\{\ga(-\rs+\ks)_{\p}\Op^{-}\gn\left(-\rs_{\p}\right)\Op^{-}\gm(-\rs-\ps)_{\p}\Op^{-}\right\}\nn\\
&=&-\text{Tr}\left\{\Op^{-}\ga^{\p}(\rs-\ks)_{\p}\gn^{\p}\rs_{\p}\gm^{\p}(\rs+\ps)_{\p}\right\}\nn\\
&=&-\text{Tr}\left\{\Op^{-}\ga^{\p}(\rs+\ps)_{\p}\gm^{\p}\rs_{\p}\gn^{\p}(\rs-\ks)_{\p}\right\},
\label{diag0b}
\eea
such that, the sum of both contributions is:
\bea
\frac{i}{2}\text{Tr}\left\{\gamma_1\gamma_2\ga^{\p}(\rs+\ps)_{\p}\gm^{\p}\rs_{\p}\gn^{\p}(\rs-\ks)_{\p}\right\}.
\eea

The above expression, after the trace is performed is a product combinations of  $g^{1\mu}$, $g^{2\mu}$, $g^{1\nu}$, $g^{2\nu}$, $g^{1\alpha}$, $g^{2\alpha}$, $r_\p^1$, $r_\p^2$, $p_\p^1$, $p_\p^2$ and $k_\p^1,k_\p^2$, which by definition vanishes.

\chapter{Calculation of \texorpdfstring{$\widetilde{M}$}{} for the process \texorpdfstring{$gg\to\gamma$}{}}\label{ApM1Calculation}

For calculation of the amplitude for the process $gg\to\gamma$ depicted in Fig.~\ref{Fig:Diag1} the starting point is Eq.~(\ref{amplitude}):
\bea
   \widetilde{{\mathcal{M}}}&=&-\int \!d^4x\,d^4y\,d^4z\int\!\frac{d^4r}{(2\pi)^4}
   \frac{d^4s}{(2\pi)^4}\frac{d^4t}{(2\pi)^4}e^{-it\cdot (y-x)}e^{-is\cdot (x-z)}e^{-ir\cdot (z-y)}e^{-ip\cdot z}e^{-ik\cdot y}e^{iq\cdot x}\nonumber\\
   &\times&
   \Big\{
   {\mbox{Tr}}\left[ iq_f\gamma_\alpha iS(s) ig\gamma_\mu t^c iS(r) ig\gamma_\nu t^d iS(t) \right]+
   {\mbox{Tr}}\left[ iq_f\gamma_\alpha iS(t) ig\gamma_\nu t^d iS(r) ig\gamma_\mu t^c iS(s) \right]
   \Big\}
   \nonumber\\
   &\times&\Phi(x,y)\Phi(y,z)\Phi(z,x)\epsilon^\mu(\lambda_p)\epsilon^\nu(\lambda_k)\epsilon^{\alpha}(\lambda_q),\quad,
  \label{amplitudeAppendix}
\eea
where the phase factor $\Phi(x,y)\Phi(y,z)\Phi(z,x)$ is given by Eq.~(\ref{phasefactor}), namely:
\bea
\Phi(x,y)\Phi(y,z)\Phi(z,x)=\exp\left[-i\frac{\eB}{2}\epsilon_{mj}(z-x)_m(x-y)_j\right].
\eea

By defining $w=z-x$ y $l=x-y$ the coordinate integral of Eq.~(\ref{amplitudeAppendix}) is
\bea
&&\int d^4x\;d^4w\;d^4l\,\exp\left\{-\frac{i|q_fB|}{2}\epsilon_{mj}w_{m}l_{j}\right\}e^{ix(q-k-p)}e^{-il(r-t-k)}e^{-iw(p+r-s)}\nn\\
&=&\dpi^4\delta^{(4)}(q-k-p)\int d^4w\;d^4l\exp\left\{-\frac{i|q_fB|}{2}\epsilon_{mj}w_{m}l_{j}\right\}e^{-il(r-t-k)}e^{-iw(p+r-s)}.\nn\\
\label{intespacial1}
\eea

The above can be separated into the transverse and longitudinal momentum components because $x\cdot p=x_\p\cdot p_\p-(x_1p_1+x_2p_2)=x_\p\cdot p_\p-x_ip_i$, so that
\bea
&&\dpi^8\delta^{(4)}(q-k-p)\,\delta^{(2)}\left[\left(k+t-r\right)_\p\right]\delta^{(2)}\left[\left(s-p-r\right)_\p\right]\nn\\
&\times&\prod_{m,j=1}^{2}\int d^4w_m\;d^4l_j\exp\left\{-\frac{i|q_fB|}{2}\epsilon_{mj}w_{m}l_{j}\right\}e^{il_j(r-t-k)_j}e^{iw_m(p+r-s)_m}\nn\\
&=&\dpi^{10}\left(\frac{2}{q_fB}\right)^2\delta^{(4)}(q-k-p)\,\delta^{(2)}\left[\left(k+t-r\right)_\p\right]\delta^{(2)}\left[\left(s-p-r\right)_\p\right]\nn\\
&\times&\prod_{j=1}^{2}\int d^4l_j\,\delta^{(2)}\left[-\frac{|q_fB|}{2}\epsilon_{mj}w_{m}l_{j}+(p+s-s)_m\right]e^{il_j(r-t-k)_j}e^{iw_m(p+r-s)_m}\nn\\
&=&\dpi^{10}\left(\frac{2}{q_fB}\right)^2\delta^{(4)}(q-k-p)\,\delta^{(2)}\left[\left(k+t-r\right)_\p\right]\delta^{(2)}\left[\left(s-p-r\right)_\p\right]\nn\\
&\times&\exp\left\{\frac{2i}{\eB}\epsilon_{mj}(p+r-s)_m(r-t-k)_j\right\}.
\eea

\begin{figure}[h]
    \centering
    \includegraphics[scale=0.45]{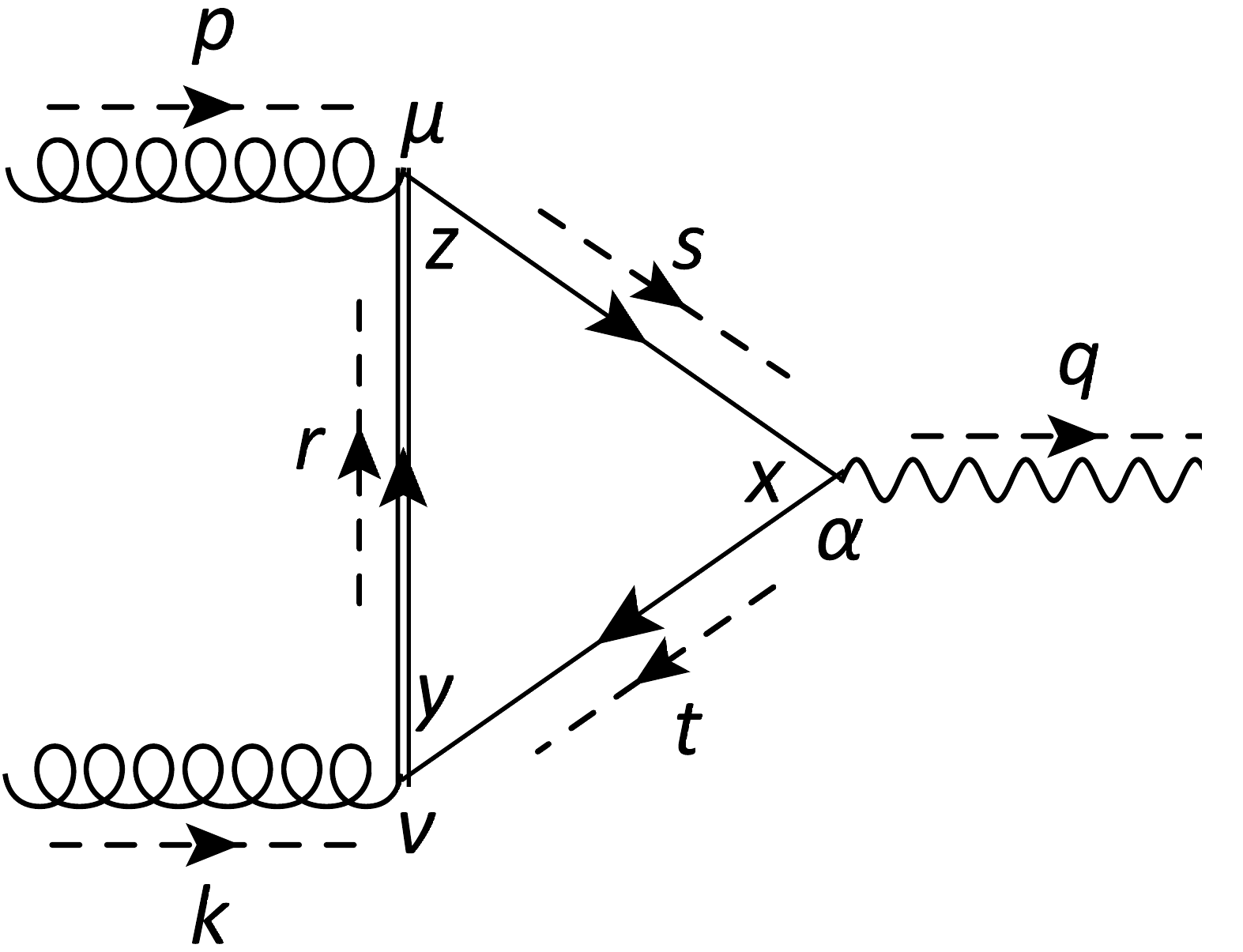}\includegraphics[scale=0.45]{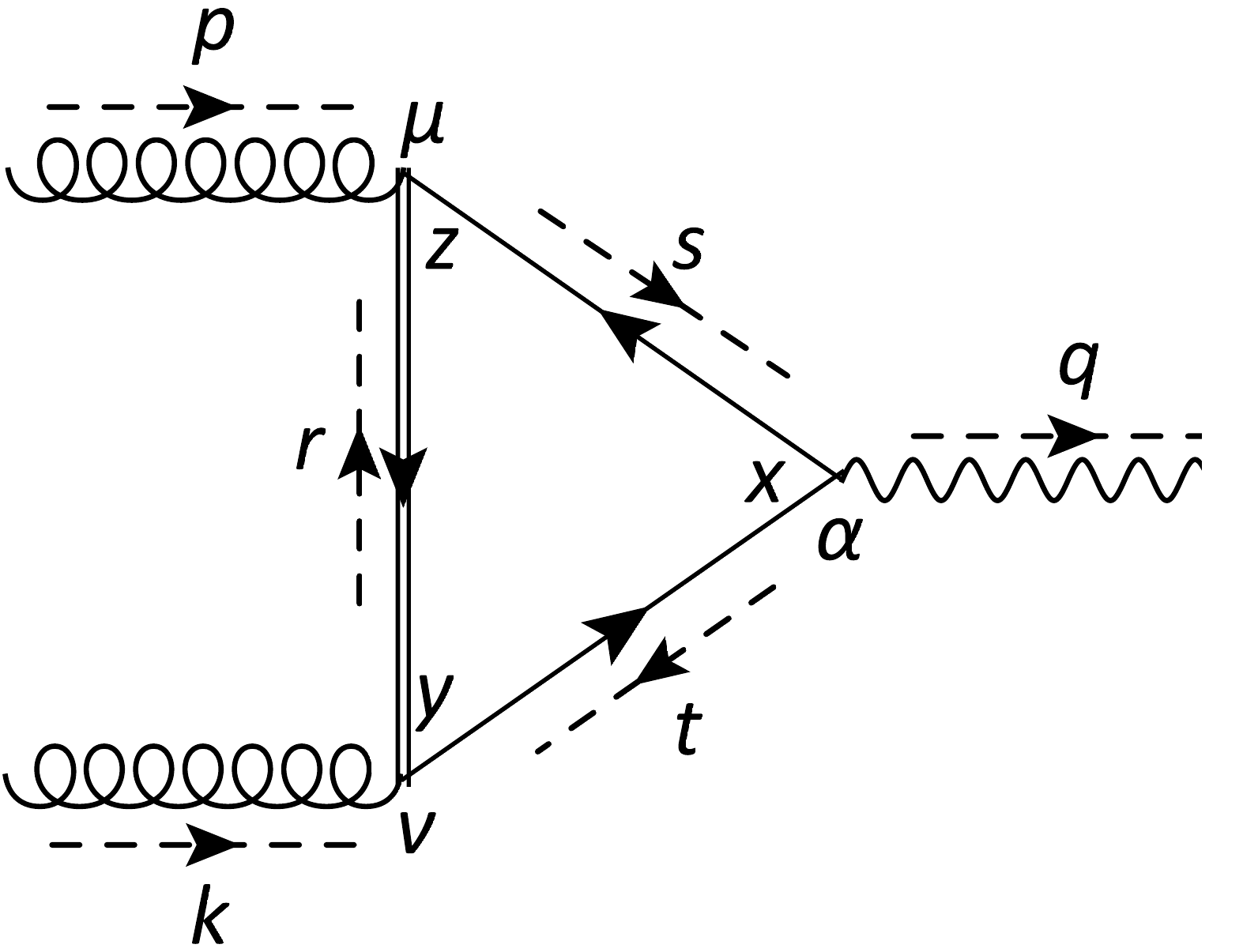}
    \caption{Feynman diagrams contributing to the process when the fermion with momentum $r$ is in the 1LL and the fermions with momentum $s$ and $t$ are in the LLL.}
    \label{Fig:Diag1Diag1a}
\end{figure}

The leading order of the process is given by two quarks in the LLL and one in the 1LL, and for each fermion line in the 1LL there are two contributing diagrams. Thus, by replacing the propagators of Eqs.~(\ref{propLLL1}) and (\ref{propagadores}) the traces of Eq.~(\ref{amplitudeAppendix}) when the fermion with momentum $r$ is in the 1LL and the fermions with momentum $s$ and $t$ (Fig.~\ref{Fig:Diag1Diag1a}) are:
\begin{subequations}
\bea
\text{Tr}\left\{iq_f\ga iS(s)ig\gm t^d iS(r)ig\gn t^ciS(t)\right\}=-16iq_fg^2\delta^{cd}\text{Tr}\left\{\ga\ssl_\p\mathcal{O}^{-}\gm\rs_\perp\gm\ts_\p\mathcal{O}^{-}\right\},
\eea
and
\bea
\text{Tr}\left\{iq_f\ga iS(t)ig\gn t^c iS(r)ig\gm t^d iS(s)\right\}=16iq_fg^2\delta^{cd}\text{Tr}\left\{\ga\ts_\p\mathcal{O}^{+}\gn\rs_\perp\gm\ssl_\p\mathcal{O}^{+}\right\}.
\eea
\end{subequations}

From the properties of Eq.~(\ref{propiedadesOp}) it follows that:
\begin{subequations}
\bea
\text{Tr}\left\{\ga\ssl_\p\mathcal{O}^{-}\gm\rs_\perp\gm\ts_\p\mathcal{O}^{-}\right\}=\text{Tr}\left\{\ga^\p\ssl_\p\gm\rs_\perp\gn\ts_\p\Op^-\right\},
\eea
and
\bea
\text{Tr}\left\{\ga\ts_\p\mathcal{O}^{+}\gn\rs_\perp\gm\ssl_\p\mathcal{O}^{+}\right\}=\text{Tr}\left\{\ga^\p\ssl_\p\gm\rs_\perp\gn\ts_\p\Op^+\right\},
\eea
\end{subequations}
thus, the trace in Eq.~(\ref{amplitudeAppendix}) is
\bea
i\text{Tr}\left\{\gamma^1\gamma^2\ts_\p\gn\rs_\perp\gm\ssl_\p\ga^\p\right\}.
\eea

The same procedure can be applied for the other diagrams so that
\bea
 \widetilde{{\mathcal{M}}}&=&16i\dpi^{10}q_fg^2\left(\frac{2}{q_fB}\right)^2\delta^{cd}\nn\\
&\times&\int\frac{d^4r}{\dpi^4}\frac{d^4s}{\dpi^4}\frac{d^4t }{\dpi^4}\,\delta^{(4)}(q-k-p)\,\delta^{(2)}\left[\left(k+t-r\right)_\p\right]\delta^{(2)}\left[\left(s-p-r\right)_\p\right]\nn\\
&\times&\exp\left\{\frac{2i}{\eB}\epsilon_{mj}(p+r-s)_m(r-t-k)_j\right\}\exp\left\{-\frac{r_\perp^2+s_\perp^2+t_\perp^2}{\eB}\right\}\nn\\
&\times&\text{Tr}\left\{\frac{\gamma^1\gamma^2\ts_\p\gn\rs_\perp\gm\ssl_\p\ga^\p}{t_\p^2s_\p^2\left(r_\p^2-2\eB\right)}+\frac{\gamma^1\gamma^2\rs_\p\gm\ssl_\perp\ga\ts_\p\gn^\p}{r_\p^2t_\p^2\left(s_\p^2-2\eB\right)}+\frac{\gamma^1\gamma^2\ssl_\p\ga\ts_\perp\gn\rs_\p\gm^\p}{r_\p^2s_\p^2\left(t_\p^2-2\eB\right)}\right\}\nn\\
&\times&\epsilon^{\mu}(\lambda_p)\epsilon^{\nu}(\lambda_k)\epsilon^{\alpha}(\lambda_q).
\label{MsinInt}
\eea

\section*{Integration over $t_\perp$}

The terms in the exponentials of Eq.~(\ref{MsinInt}) can be grouped as 
\bea
-\frac{2i}{\eB}\epsilon_{mj}(p+r-s)_mt_j-\frac{t_j^2}{\eB}=-\frac{1}{\eB}\left[(\T{t}_j+i\epsilon_{jm}(p+r-s)_m)^2+(p+r-s)_m^2\right],\nn\\
\eea
so that by defining
\bea
\T{t}_j=t_j+i\epsilon_{jm}(p+r-s)_m,
\eea
the integration over $t$ is given by
\bea
\int\frac{d^2t_\p}{\dpi^2}\frac{d^2\T{t}_j}{\dpi^2}\exp\left\{-\frac{\T{t}_j^2}{\eB}\right\}\text{Tr}\left\{\frac{\gamma^1\gamma^2\ssl_\p\ga\gamma^l\left[\T{t}_l-i\epsilon_{lm}(p+r-s)_m\right]\gn\rs_\p\gm^\p}{r_\p^2s_\p^2\left(t_\p^2-2\eB\right)}+\cdots\right\}.\nn\\
\label{C11}
\eea

The odd powers do not contribute to the perpendicular integration, thus the first term of Eq.~(\ref{C11}) becomes
\bea
\!\!-\frac{\pi\eB}{\dpi^2}\exp\left\{-\frac{(p+r-s)_m^2}{\eB}\right\}\int\frac{d^2t_\p}{\dpi^2}\text{Tr}\left\{\frac{\gamma^1\gamma^2\ssl_\p\ga\gamma^l\left[i\epsilon_{lm}(p+r-s)_m\right]\gn\rs_\p\gm^\p}{r_\p^2s_\p^2\left(t_\p^2-2\eB\right)}\right\}.\nn\\
\quad
\label{C12}
\eea

From the above:
\bea
\widetilde{\mathcal{M}}&=&16i\dpi^{8}(\pi\eB)\left(\frac{2}{q_fB}\right)^2q_fg^2\delta^{cd}\nn\\
&\times&\int\frac{d^4r}{\dpi^4}\frac{d^4s}{\dpi^4}\frac{d^2t_\p}{\dpi^2}\;\delta^{(4)}(q- k-p)\delta^{(2)}\left[\left(t+ k-r\right)_\p\right]\delta^{(2)}\left[\left(s-p-r\right)_\p\right]\nn\\
&\times&\exp\left\{-\frac{(p+r-s)_m^2}{\eB}\right\}\exp\left\{-\frac{r_\perp^2+s_\perp^2}{\eB}\right\}\exp\left\{\frac{2i}{\eB}\epsilon_{mj}(p+r-s)_m(r- k)_j\right\}\nn\\
&\times&\text{Tr}\left\{\frac{\gamma^1\gamma^2\ts_\p\gn\rs_\perp\gm\ssl_\p\ga^\p}{t_\p^2s_\p^2\left(r_\p^2-2\eB\right)}+\frac{\gamma^1\gamma^2\rs_\p\gm\ssl_\perp\ga\ts_\p\gn^\p}{r_\p^2t_\p^2\left(s_\p^2-2\eB\right)}\right.\nn\\
&-&\left.\frac{\gamma^1\gamma^2\ssl_\p\ga\gamma^l\left[i\epsilon_{lm}(p+r-s)_m\right]\gn\rs_\p\gm^\p}{r_\p^2s_\p^2\left(t_\p^2-2\eB\right)}\right\}\epsilon^{\mu}(\lambda_p)\epsilon^{\nu}(\lambda_k)\epsilon^{\alpha}(\lambda_q).
\label{MCsinIntenS}
\eea

\section*{Integration over $s_\perp$}

The terms in the exponential of Eq.~(\ref{MCsinIntenS}) are
\bea
&&-\frac{1}{\eB}\left[s_m^2-2(p+r)_m s_m+(p+r)_m^2+s_m^2+2i\epsilon_{mj}(r- k)_js_m\right]\nn\\
&=&-\frac{2}{\eB}\left[\left(s_m+\frac{1}{2}\left[i\epsilon_{mj}(r- k)_j-(p+r)_m\right]\right)^2+\frac{1}{2}(p+r)_m^2\right.\nn\\
&-&\left.\frac{1}{4}\left[i\epsilon_{mj}(r- k)_j-(p+r)_m\right]^2\right],
\eea
so that by defining $\T{s}_m=s_m+\frac{1}{2}\left[i\epsilon_{mj}(r- k)_j-(p+r)_m\right]$ the integration over $s_\perp$ is easily performed in the same way as in Eq.~(\ref{C12}), i.e., all the odd powers of $s_\perp$ are ignored giving the following expression:
\bea
&&\frac{1}{\dpi^2}\frac{\pi\eB}{2}\int\frac{d^2s_\p}{\dpi^2}\,e^{-r_\perp^2/\eB}\exp\left\{\frac{2i}{\eB}\epsilon_{mj}(p+r)_m(r- k)_j\right\}\nn\\
&\times&\exp\left\{\frac{1}{2\eB}\left[i\epsilon_{mj}(r- k)_j-(p+r)_m\right]^2-\frac{1}{\eB}(p+r)_m^2\right\}\nn\\
&\times&\text{Tr}\left\{-\frac{\gamma^1\gamma^2\ssl_\p\ga\gamma^li\epsilon_{lm}\left[(p+r)_m+i\epsilon_{mj}(r- k)_j\right]\gn\rs_\p\gm^\p}{2r_\p^2s_\p^2\left(t_\p^2-2\eB\right)}\right.\nn\\
&+&\left.\frac{\gamma^1\gamma^2\ts_\p\gn\rs_\perp\gm\ssl_\p\ga^\p}{t_\p^2s_\p^2\left(r_\p^2-2\eB\right)}-\frac{\gamma^1\gamma^2\rs_\p\gm\gamma^l\left[i\epsilon_{lj}(r- k)_j-(p+r)_l\right]\ga\ts_\p\gn^\p}{2r_\p^2t_\p^2\left(s_\p^2-2\eB\right)}\right\}.
\eea

By collecting results:
\bea
\widetilde{\mathcal{M}}&=&8i\dpi^{6}(\pi\eB)^2\left(\frac{2}{q_fB}\right)^2q_fg^2\delta^{cd}\nn\\
&\times&\int\frac{d^4r}{\dpi^4}\frac{d^2s_\p}{\dpi^2}\frac{d^2t_\p}{\dpi^2}\;\delta^{(4)}(q- k-p)\delta^{(2)}\left[\left(t+ k-r\right)_\p\right]\delta^{(2)}\left[\left(s-p-r\right)_\p\right]\nn\\
&\times&\exp\left\{-\frac{r_\perp^2}{\eB}\right\}\exp\left\{\frac{2i}{\eB}\epsilon_{mj}(p+r)_m(r- k)_j\right\}\nn\\
&\times&\exp\left\{\frac{1}{2\eB}\left[i\epsilon_{mj}(r- k)_j-(p+r)_m\right]^2-\frac{1}{\eB}(p+r)_m^2\right\}\nn\\
&\times&\text{Tr}\left\{-\frac{\gamma^1\gamma^2\ssl_\p\ga\gamma^li\epsilon_{lm}\left[(p+r)_m+i\epsilon_{mj}(r- k)_j\right]\gn\rs_\p\gm^\p}{2r_\p^2s_\p^2\left(t_\p^2-2\eB\right)}\right.\nn\\
&+&\left.\frac{\gamma^1\gamma^2\ts_\p\gn\rs_\perp\gm\ssl_\p\ga^\p}{t_\p^2s_\p^2\left(r_\p^2-2\eB\right)}-\frac{\gamma^1\gamma^2\rs_\p\gm\gamma^l\left[i\epsilon_{lj}(r- k)_j-(p+r)_l\right]\ga\ts_\p\gn^\p}{2r_\p^2t_\p^2\left(s_\p^2-2\eB\right)}\right\}\nn\\
&\times&\epsilon^{\mu}(\lambda_p)\epsilon^{\nu}(\lambda_k)\epsilon^{\alpha}(\lambda_q).
\eea

\section*{Integration over $r_\perp$}

Similarly, at the last steps, the exponential factor is
\bea
&&-\frac{1}{\eB}\left(r_m^2-2i\epsilon_{mj}(p+r)_m(r- k)_j-\frac{1}{2}\left[i\epsilon_{mj}(r- k)_j-(p+r)_m\right]^2+(p+r)_m^2\right)\nn\\
&=&-\frac{2}{\eB}\left[r_m^2+\frac{1}{2}c_m r_m\right]-\frac{1}{\eB}\left[\frac{1}{2}\left(p_m^2+k_m^2\right)+ i\epsilon_{mj}p_mk_j\right]\nn\\
&=&-\frac{2}{\eB}\left[r_m+\frac{1}{4}c_m\right]^2+f(p_\perp,k_\perp),
\eea
where
\bea
c_m&=&p_m- k_m +i\epsilon_{mj}(p_j+ k_j)\nn\\
f(p_\perp,k_\perp)&=&\frac{1}{8\eB}\left[a_m\right]^2-\frac{1}{2\eB}\left(p_m^2+k_m^2+ i\epsilon_{mj}p_mk_j\right).
\eea

Defining $\T{r}_m=r_m+\frac{1}{4}c_m$ and ignoring the linear terms of $r_\perp$ the desired integral gives:
\bea
&&\frac{1}{\dpi^2}\frac{\pi\eB}{2}\frac{1}{4}\,e^{f(p_\perp,k_\perp)}\nn\\
&\times&\int\frac{d^2r_\p}{\dpi^2}\text{Tr}\left\{\frac{\gamma^1\gamma^2\ssl_\p\ga\slashed{a}\gn\rs_\p\gm^\p}{r_\p^2s_\p^2\left(t_\p^2-2\eB\right)}+\frac{\gamma^1\gamma^2\rs_\p\gm\slashed{b}\ga\ts_\p\gn^\p}{r_\p^2t_\p^2\left(s_\p^2-2\eB\right)}-\frac{\gamma^1\gamma^2\ts_\p\gn\slashed{c}\gm\ssl_\p\ga^\p}{t_\p^2s_\p^2\left(r_\p^2-2\eB\right)}\right\},\nn\\
\eea
where $a,b,c,$ and $f(\pt,\kt)$ are given in Eq.~(\ref{habc_def}).

To obtain Eq.~(\ref{matrint}) the approximation $2|q_fB|\gg t_\p^2,\ s_\p^2,\ r_\p^2$ is performed, so that it allows to compute the parallel momentum integrals in terms of a Feynman parametrization:
\bea
\frac{1}{AB}=\int_0^1\frac{dx}{\left[xA+(1-x)B\right]},
\eea
in such a way that the only integrals are reduced to the following identities~\cite{peskin1995quantum}:
\bea
\int\frac{d^dl}{\dpi^d}\frac{1}{(l^2-\Delta)^n}&=&\frac{(-1)^n}{(4\pi)^{d/2}}i\frac{\Gamma(n-d/2)}{\Gamma(n)}\left(\frac{1}{\Delta}\right)^{n-d/2}\nn\\
\int\frac{d^dl}{\dpi^d}\frac{l^2}{(l^2-\Delta)^n}&=&\frac{(-1)^{n-1}}{(4\pi)^{d/2}}i\frac{d}{2}\frac{\Gamma(n-d/2-1)}{\Gamma(n)}\left(\frac{1}{\Delta}\right)^{n-d/2-1}\nn\\
\int\frac{d^dl}{\dpi^d}\frac{l^\mu l^\nu}{(l^2-\Delta)^n}&=&\frac{(-1)^{n-1}}{(4\pi)^{d/2}}i\frac{d}{2}g^{\mu\nu}\frac{\Gamma(n-d/2-1)}{\Gamma(n)}\left(\frac{1}{\Delta}\right)^{n-d/2-1}.
\label{PeskinIntegrals}
\eea

\chapter{Invariant Momentum Distribution and Flow Coefficient Calculation for the Process \texorpdfstring{$gg\to\gamma$}{}}\label{Ap_M_squared}

To compute the invarian momentum distribution and the flow coefficient $v_2$, the starting point is the energy-momentum conservation given by:
\bea
\wq&=&\wwp+\wk\nn\\
\n{q}&=&\n{p}+\n{k},
\label{conservacion}
\eea
so that, if the particles are in the mass-shell, i.e., $|\mathbf{p}|=\wwp$ it is easy to find that
\bea
\wq^2=(\wwp+\wk)^2=\wwp^2+\wk^2+2\wwp\wk\cos\delta,
\eea
where $\delta$ is the angle between the gluon's three-momentum. The above equation is satisfied only if $\delta=0$, and therefore, the spatial momentum of the gluons are parallel.

\section*{Squared Matrix Element}\label{Ap_M2_calculation_exp}

From the tensorial structure of Eq.~(\ref{matrint}):
\bea
\Gamma^{\mu\nu\alpha}&=&\left(\gma_\p-\frac{p^\mu_\p p^\alpha_\p}{p_\p^2}\right)h^\nu(a)-\left(\gmn_\p-\frac{p^\mu_\p p^\nu_\p}{p_\p^2}\right)h^\alpha(a)+\left(\gmn_\p-\frac{k^\mu_\p k^\nu_\p}{k_\p^2}\right)h^\alpha(b)\nn\\
&-&\left(\gan_\p-\frac{k^\alpha_\p k^\nu_\p}{k_\p^2}\right)h^\mu(b)
+\left(\gan_\p-\frac{q^\alpha_\p q^\nu_\p}{q_\p^2}\right)h^\mu(c)-\left(\gma_\p-\frac{q^\mu_\p q^\alpha_\p}{q_\p^2}\right)h^\nu(c),
\label{Gamma}
\eea
where
\bea
T_\p^{\mu\nu}&=&\gmn_\p-\frac{q^\mu_\p q^\nu_\p}{q_\p^2}\nn\\
S_{1\perp}^\nu&=&h^\nu(a)-h^\nu(c)=\frac{i}{\pi}\epsilon_m^\nu\left(2p^m+k^m-i\epsilon^{mn}k^n\right)\nn\\
S_{2\perp}^\alpha&=&h^\alpha(b)-h^\alpha(a)=\frac{i}{\pi}\epsilon_m^\alpha\left(p^m-k^m-i\epsilon^{mn}(p^n+k^n)\right)\nn\\
S_{3\perp}^\mu&=&h^\mu(c)-h^\mu(b)=\frac{i}{\pi}\epsilon_m^\mu\left(-3p^m+i\epsilon^{mn}(p^n+2k^n)\right),
\eea
the squared matrix element is
\bea
\sum_{\text{pol}}\left |  \widetilde{\mathcal{M}}\right |^2&=&
\left |-i\pi^3q_fg^2\delta^{(4)}(q-p-k)e^{f(p,k)} \right |^2\sum_{\text{pol}}\Gamma^{\mu\nu\alpha}\Gamma^{*\mu'\nu'\alpha'}\epsilon_\mu\epsilon_\nu\epsilon_\alpha\epsilon_{\mu'}^*\epsilon_{\nu'}^*\epsilon_{\alpha'}^*\nn\\
&=&-\left |-i\pi^3q_fg^2\delta^{(4)}(q-p-k)e^{f(p,k)} \right |^2g_{\mu\mu'}g_{\nu\nu'}g_{\alpha\alpha'}\Gamma^{\mu\nu\alpha}\Gamma^{*\mu'\nu'\alpha'}\nn\\
&=&-\dpi^4\delta^{(4)}(q-p-k)\mathcal{VT} \left |\pi^3q_fg^2e^{f(p,k)} \right |^2\sum_i\left |S_{i\perp}\right |^2
\label{M2expandedindices}
\eea
where where it has been used the fact that in an unpolarized cross section the Ward identity implies that~\cite{peskin1995quantum}:
\bea
\sum_{\text{pol}}\epsilon_\mu^*\epsilon_\nu\to-g_{\mu\nu}.
\eea
together with the identity:
\bea
\left|\delta^{(4)}(q-p-k) \right |^2=\dpi^4\delta^{(4)}(q-p-k)\mathcal{VT}.
\eea

Note that after sum over flavours, Eq.~(\ref{M2expandedindices}) corresponds to Eq.~(\ref{Msquareddiractdeltafac}).

On the other hand:
\bea
\left |S_{1\perp}^\nu\right|^2&=&S_{1\perp}^\nu S^{1\perp\,*}_\nu\nn\\
&=&\frac{1}{\pi^2}\epsilon_m^\nu\epsilon^l_\nu\left(2p^m+k^m-i\epsilon^{mn}k^n\right)\left(2p_l+k_l+i\epsilon_{lj}k_j\right)\nn\\
&=&\frac{1}{\pi^2}\delta_m^l\left(2p^m+k^m-i\epsilon^{mn}k^n\right)\left(2p_l+k_l+i\epsilon_{lj}k_j\right)\nn\\
&=&-\frac{4\pt^2+2\kt^2+4\pt\cdot\kt}{\pi^2},
\label{S1squared}
\eea
where the following conventions were used:
\begin{subequations}
\bea
p^m p_m=-p_1^2-p_2^2=-\pt^2
\eea
and
\bea
p^m k_m=-p_1k_1-p_2k_2=-\pt\cdot\kt.
\eea
\end{subequations}

Similarly:
\bea
\left |S_{2\perp}^\alpha\right|^2=-\frac{2(\pt^2+\kt^2)}{\pi^2},
\label{S2squared}
\eea
and
\bea
\left |S_{3\perp}^\mu\right|^2=-\frac{10\pt^2+4\kt^2+4\pt\cdot\kt}{\pi^2}.
\label{S3squared}
\eea

By using Eq.~(\ref{conservacion}) and the fact that the three-momentum for the involved particles are collinear it is straightforward to obtain the rule of Eq.~(\ref{energymomentumconservation}):
\bea
   p^\mu&=&\omega_p(1,\hat{p})
   =\left(\omega_p/\omega_q\right)q^\mu,\nonumber\\
   k^\mu&=&\omega_k(1,\hat{k})
   =\left(\omega_k/\omega_q\right)q^\mu,
\eea
in such a way that the quantities shown in Eqs.~(\ref{S1squared})-~(\ref{S3squared}) can be written in terms of $\qt^2$, namely:
\bea
\pt^2&=&\left(\frac{\wwp}{\wq}\right)^2\qt^2\nn\\
\kt^2&=&\left(\frac{\wk}{\wq}\right)^2\qt^2,
\eea
so that after sum over flavors:
\bea
\sum_{\text{pol},f}\left |  \widetilde{\mathcal{M}}\right |^2&=&\dpi^4\delta^{(4)}(q-p-k)\mathcal{VT}\frac{2\alpha_{\text{em}}\alpha_{\text{s}}^2q_\perp^2}{\pi\omega_q^2}\left(2\omega_p^2+\omega_k^2+\omega_p\omega_k\right)\nn\\
&\times&\sum_{f}q_f^2\exp\left\{-\frac{q_\perp^2}{\eB\omega_q^2}\left(\omega_p^2+\omega_k^2+\omega_p\omega_k\right)\right\},
\eea
which is the result of Eq.~(\ref{Msquared}). 

\section*{Invariant Momentum Distribution}\label{Nmag_calc_exp_1}
From the invariant momentum distribution definition of Eq.~(\ref{invdist}) and from the squared  matrix element of Eq.~(\ref{Msquared}) it follows that:
\bea
\wq\frac{d^3N}{d^3q}&=&\frac{1}{2\dpi^3}\int\frac{d^3p}{2\dpi^3\wwp}\int\frac{d^3k}{2\dpi^3\wk}n(\wwp)n(\wk)\sum_{\text{pol},f}\left |  \widetilde{\mathcal{M}}\right |^2\nn\\
&=&\dpi^4\frac{\mathcal{VT}}{2\dpi^3}\;\frac{2q_f^2\alpha_{\text{em}}\alpha_{\text{s}}^2}{\pi\omega_q^2}q_\perp^2\int\frac{d^3p}{2\dpi^3\wwp}\int\frac{d^3k}{2\dpi^3\wk}n(\wwp)n(\wk)\nn\\
&\times&\delta\left(\wq-\wwp-\wk\right)\delta^{(3)}\left(\n{q}-\n{p}-\n{k}\right)\left(2\omega_p^2+\omega_k^2+\omega_p\omega_k\right)\nn\\
&\times&\exp\left\{-\frac{q_\perp^2}{\eB\omega_q^2}\left(\omega_p^2+\omega_k^2+\omega_p\omega_k\right)\right\},
\eea
where in the last line the sum over $f$ is implicit and it was used the fact that
\bea
\delta^{(4)}(P)=\delta^{(0)}(P^0)\,\delta^{(3)}(\mathbf{P}).
\eea

By using the mass-shell condition for gluons and photon the integral of interest is given by:
\bea
I&=&\int\frac{d^3p}{2\dpi^3\np}\int\frac{d^3k}{2\dpi^3\nk}\,\delta\left(\wq-\wwp-\wk\right)\delta^{(3)}\left(\n{q}-\n{p}-\n{k}\right)\nn\\
&\times&n(\np)\,n(\nk)\left(2\np^2+\nk^2+\n{p}\cdot\n{k}\right)\exp\left\{-\frac{q_\perp^2}{\eB\omega_q^2}\left(\np^2+\nk^2+\n{p}\cdot\n{k}\right)\right\}\nn\\
&=&\frac{1}{2\dpi^3}\int\frac{d^3p}{2\dpi^3\np\,|\n{q}-\n{p}|}\,\delta\left(\wq-\wwp-\sqrt{\n{p}^2+\n{q}^2-2\np\nq\cos\alpha}\right)\nn\\
&\times&n(\np)\,n\left(|\n{q}-\n{p}|\right)\left(2\np^2+|\n{q}-\n{p}|^2+\n{p}\cdot(\n{q}-\n{p})\right)\nn\\
&\times&\exp\left\{-\frac{q_\perp^2}{\eB\omega_q^2}\left(\np^2+|\n{q}-\n{p}|^2+\n{p}\cdot(\n{q}-\n{p})\right)\right\}.
\label{ItocomparewithJ}
\eea

The integration over $p$ is performed in a spherical coordinate system and given the momentum's collinearity condition:
\bea
\delta\left(\wq-\wwp-\sqrt{\n{p}^2+\n{q}^2-2\np\nq\cos\alpha}\right)=\frac{|\n{q}-\n{p}|}{\wwp\wq}\delta(\cos\alpha-1),
\eea
it follows that:
\bea
\wq\frac{d^3N}{d^3q}&=&\frac{q_f^2\alpha_{\text{em}}\alpha_{\text{s}}^2\mathcal{VT}}{2\dpi^5}\frac{q_\perp^2}{\wq^3}\int d\wwp\left(2\wwp^2+\wq^2-\wwp\wq\right)n(\wwp)\,n(|\wq-\wwp|)\nn\\
&\times&\exp\left\{-\frac{q_\perp^2}{\eB\wq^2}\left(\wwp^2+\wq^2-\wwp\wq\right)\right\}.
\label{wqdndq1}
\eea

Now, given that:
\bea
\frac{\wq}{d^3q}=\frac{1}{d^2q_\perp dy}=\frac{1}{\wq d\wq d\theta dy},
\eea
with $q_\perp=\wq\sin\theta$, and $\theta$ the angle between $\mathbf{q}$ and the magnetic field, the Eq.~(\ref{wqdndq1}) becomes:
\bea
\frac{dN}{\wq d\wq d\theta dy}=\frac{q_f^2\alpha_{\text{em}}\alpha_{\text{s}}^2\mathcal{VT}}{2\dpi^5}\frac{\sin^2\theta}{\wq}\,\mathcal{I}(\wq,\theta),
\label{dNenI}
\eea
where
\bea
\mathcal{I}(\wq,\theta)&=&\int d\wwp\left(2\wwp^2+\wq^2-\wwp\wq\right)n(\wwp)\,n(|\wq-\wwp|)\nn\\
&\times&\exp\left\{-\frac{\sin^2\theta}{\eB}\left(\wwp^2+\wq^2-\wwp\wq\right)\right\}.
\eea

From Eq.~(\ref{dNenI}) the number of photons for central rapidity $\Delta y\simeq 1$ can be obtained as follows:
\bea
\frac{dN}{\wq d\wq}&=&\frac{q_f^2\alpha_{\text{em}}\alpha_{\text{s}}^2\mathcal{VT}}{2\dpi^5\wq}\int_0^{2\pi}d\theta\sin^2\theta\,I(\wq,\theta).
\eea

Finally, by using:
\bea
\int_0^{2\pi}d\theta\sin^2\theta\,e^{-f(x)\sin^2\theta}=\pi e^{-f(x)/2\theta}\left[I_0\left(\frac{f(x)}{2}\right)-I_1\left(\frac{f(x)}{2}\right)\right],
\eea
with $I_k(x)$ the Bessel functions of order $k$, and sum over $f$ the invariant momentum distribution of Eq.~(\ref{yieldexpl}) is obtained.

\section*{Flow Coefficient}\label{v2_calc_exp_1}
The photon angular distribution is obtained from Eq.~(\ref{dNenI}):
\bea
\frac{dN}{d\theta}=\frac{q_f^2\alpha_{\text{em}}\alpha_{\text{s}}^2\mathcal{VT}}{2\dpi^5}\sin^2\theta\int_0^\infty d\wq I(\wq,\theta).
\eea

The $v_2$ coefficient usually is measured from the collision plane (see Fig.~\ref{Fig:Photon_production_geometry}), but in the present calculation the angle $\theta$ is refereed to the magnetic field. Thus, in order to compare with the experimental data it is necessary to perform the shift $\theta=\pi/2-\phi$  so that:
\bea
\frac{dN}{d\phi}&=&\frac{q_f^2\alpha_{\text{em}}\alpha_{\text{s}}^2\mathcal{VT}}{2\dpi^5}\cos^2\phi\int_0^\infty d\wq I(\wq,\pi/2-\phi)\nn\\
&=&\frac{N}{2\pi}\left[1+\sum_n v_n(\wq)\cos(n\phi)\right].
\eea

From the identity:
\bea
\int_0^{2\pi}d\phi\cos^2\phi\cos(2\phi) e^{-f(x)\cos^2\phi}=\pi e^{-f(x)/2}\left[I_0\left(\frac{f(x)}{2}\right)-\frac{2+f(x)}{f(x)}I_1\left(\frac{f(x)}{2}\right)\right],
\eea
the flow coefficient of Eq.~(\ref{v2expl}) is achieved.

\chapter{Proof of Eq.~(\ref{yieldexpl2})}\label{Mk=M-k}

From thre cross symmetry of Eq.~(\ref{Mgg_gamma=Mg_ggamma}):
\bea
\mathcal{M}_{g\rightarrow g\gamma}(p,k,q)=\mathcal{M}_{gg\rightarrow\gamma}(p,-k,q),
\eea
and taking into account the result of Eq.~(\ref{M2expandedindices}) the problem is resumed to evaluate the expression given by
\bea
\left|e^{f(p,-k)}\right|^2\sum_i\left |S_{i\perp}(-k)\right |^2,
\eea
so that the sign change is reflected in quantities:
\begin{subequations}
\bea
\kt^2=\left(\frac{\wk}{\wq}\right)^2\qt^2
\eea
and
\bea
\kt\cdot\pt=-\frac{\wwp\wk}{\wq^2}\qt^2,
\eea
\end{subequations}
then,
\bea
\sum_{\text{pol},f}\left |  \widetilde{\mathcal{M}}_{g\to g\gamma}\right |^2&=&\dpi^4\delta^{(4)}(q-p+k)\mathcal{VT}\frac{2\alpha_{\text{em}}\alpha_{\text{s}}^2q_\perp^2}{\pi\omega_q^2}\left(2\omega_p^2+\omega_k^2-\omega_p\omega_k\right)\nn\\
&\times&\sum_{f}q_f^2\exp\left\{-\frac{q_\perp^2}{\eB\omega_q^2}\left(\omega_p^2+\omega_k^2-\omega_p\omega_k\right)\right\}.
\eea

From the above, the integral in the invariant momentum distribution is:
\bea
J&=&\int\frac{d^3p}{2\dpi^3\np}\int\frac{d^3k}{2\dpi^3\nk}\,\delta\left(\wq-\wwp+\wk\right)\delta^{(3)}\left(\n{q}-\n{p}+\n{k}\right)\nn\\
&\times&n(\np)\left[1+n(\nk)\right]\left(2\np^2+\nk^2-\n{p}\cdot\n{k}\right)\exp\left\{-\frac{q_\perp^2}{\eB\omega_q^2}\left(\np^2+\nk^2-\n{p}\cdot\n{k}\right).\right\}\nn\\
\eea

The integration over $d^3k$ implies $\n{k}=\n{p}-\n{q}$, therefore:
\bea
J&=&\frac{1}{2\dpi^3}\int\frac{d^3p}{2\dpi^3\np\,|\n{q}-\n{p}|}\,\delta\left(\wq-\wwp-\sqrt{\n{p}^2+\n{q}^2-2\np\nq\cos\alpha}\right)\nn\\
&\times&n(\np)\left[1+n\left(|\n{q}-\n{p}|\right)\right]\left(2\np^2+|\n{q}-\n{p}|^2+\n{p}\cdot(\n{q}-\n{p})\right)\nn\\
&\times&\exp\left\{-\frac{q_\perp^2}{\eB\omega_q^2}\left(\np^2+|\n{q}-\n{p}|^2+\n{p}\cdot(\n{q}-\n{p})\right)\right\},
\eea
which apart from the form of the gluon distributions has the same form than Eq.~(\ref{ItocomparewithJ}), so that the steps to compute such integral yields Eq.~(\ref{yieldexpl2}).

\chapter{Derivation of tensor structures of Eqs.~(\ref{fs})}\label{ApAGluonPolTensorPRD}

Recalling the general expression of the gluon polarization tensor of Eq.~(\ref{Pidef}):
\bea
i\Pi^{\mu\nu}_{(ab)}&=&-\frac{1}{2}\int\frac{d^4k}{\dpi^4}\text{Tr}\left\{igt_b\gamma^\nu iS^{(n)}(k)igt_a\gamma^\mu iS^{(m)}(q)\right\}+ {\mbox{C.C.}}.
\eea
The trace in the above expression involves two fermion propagator factors, each given by Eqs.~(\ref{fermionpropdef})-(\ref{DnPi}), so that their product produces nine terms. By defining the function
\bea
\widetilde{E}\left(\pt,\kt\eB\right)\equiv\exp\left[-\frac{\kt^2+(k-p)_{\perp}^2}{\eB}\right],
\eea
such terms are explicitly given by
\bea
t_1^{\mu\nu}&=&-g^2\int\frac{d^4k}{\dpi^4}\widetilde{E}\left(\pt,\kt\eB\right)\sum_{n,m=0}^{\infty}(-1)^{n+m}L_n^0\left(\frac{2\kt^2}{\eB}\right)L_{m}^0\left[\frac{2(k-p)_{\perp}^2}{\eB}\right]\nn\\
&\times&\frac{\text{Tr}\left\{ \gamma^\nu (\ks_\p+m_f)\Op^{-} \gamma^\mu (\ks_\p-\ps_\p+m_f)\Op^{-}\right\}}{\left[k_\p^2-m_f^2-2n\left |  q_fB\right |\right]\left[(k-p)_{\parallel}^2-m_f^2-2m\left |  q_fB\right |\right]}+\text{C.C.}
\label{t1}
\eea

\bea
t_2^{\mu\nu}&=&g^2\int\frac{d^4k}{\dpi^4}\widetilde{E}\left(\pt,\kt\eB\right)\sum_{n=0,m=1}^{\infty}(-1)^{n+m}L_n^0\left(\frac{2\kt^2}{\eB}\right)L_{m-1}^0\left[\frac{2(k-p)_{\perp}^2}{\eB}\right]\nn\\
&\times&\frac{\text{Tr}\left\{ \gamma^\nu (\ks_\p+m_f)\Op^{-} \gamma^\mu (\ks_\p-\ps_\p+m_f)\Op^{+}\right\}}{\left[k_\p^2-m_f^2-2n\left |  q_fB\right |\right]\left[(k-p)_{\parallel}^2-m_f^2-2m\left |  q_fB\right |\right]}+\text{C.C.}
\label{t2}
\eea

\bea
t_3^{\mu\nu}&=&-2g^2\int\frac{d^4k}{\dpi^4}\widetilde{E}\left(\pt,\kt\eB\right)\sum_{n=0,m=1}^{\infty}(-1)^{n+m}L_n^0\left(\frac{2\kt^2}{\eB}\right)L_{m-1}^1\left[\frac{2(k-p)_{\perp}^2}{\eB}\right]\nn\\
&\times&\frac{\text{Tr}\left\{ \gamma^\nu (\ks_\p+m_f)\Op^{-} \gamma^\mu (\ks_\perp-\ps_\perp)\right\}}{\left[k_\p^2-m_f^2-2n\left |  q_fB\right |\right]\left[(k-p)_{\parallel}^2-m_f^2-2m\left |  q_fB\right |\right]}+\text{C.C.}
\label{t3}
\eea

\bea
t_4^{\mu\nu}&=&g^2\sum_{n=1,m=0}^{\infty}\int\frac{d^4k}{\dpi^4}\widetilde{E}\left(\pt,\kt\eB\right)(-1)^{n+m}L_{n-1}^0\left(\frac{2\kt^2}{\eB}\right)L_{m}^0\left[\frac{2(k-p)_{\perp}^2}{\eB}\right]\nn\\
&\times&\frac{\text{Tr}\left\{ \gamma^\nu (\ks_\p+m_f)\Op^{+} \gamma^\mu (\ks_\p-\ps_\p+m_f)\Op^{-}\right\}}{\left[k_\p^2-m_f^2-2n\left |  q_fB\right |\right]\left[(k-p)_{\parallel}^2-m_f^2-2m\left |  q_fB\right |\right]}+\text{C.C.}
\label{t4}
\eea

\bea
t_5^{\mu\nu}&=&-g^2\int\frac{d^4k}{\dpi^4}\widetilde{E}\left(\pt,\kt\eB\right)\sum_{n=1,m=1}^{\infty}(-1)^{n+m}L_{n-1}^0\left(\frac{2\kt^2}{\eB}\right)L_{m-1}^0\left[\frac{2(k-p)_{\perp}^2}{\eB}\right]\nn\\
&\times&\frac{\text{Tr}\left\{ \gamma^\nu (\ks_\p+m_f)\Op^{+} \gamma^\mu (\ks_\p-\ps_\p+m_f)\Op^{+}\right\}}{\left[k_\p^2-m_f^2-2n\left |  q_fB\right |\right]\left[(k-p)_{\parallel}^2-m_f^2-2m\left |  q_fB\right |\right]}+\text{C.C.}
\label{t5}
\eea

\bea
t_6^{\mu\nu}&=&2g^2\int\frac{d^4k}{\dpi^4}\widetilde{E}\left(\pt,\kt\eB\right)\sum_{n=1,m=1}^{\infty}(-1)^{n+m}L_{n-1}^0\left(\frac{2\kt^2}{\eB}\right)L_{m-1}^1\left[\frac{2(k-p)_{\perp}^2}{\eB}\right]\nn\\
&\times&\frac{\text{Tr}\left\{ \gamma^\nu (\ks_\p+m_f)\Op^{+} \gamma^\mu (\ks_\perp-\ps_\perp)\right\}}{\left[k_\p^2-m_f^2-2n\left |  q_fB\right |\right]\left[(k-p)_{\parallel}^2-m_f^2-2m\left |  q_fB\right |\right]}+\text{C.C.}
\label{t6}
\eea

\bea
t_7^{\mu\nu}&=&-2g^2\int\frac{d^4k}{\dpi^4}\widetilde{E}\left(\pt,\kt\eB\right)\sum_{n=1,m=0}^{\infty}(-1)^{n+m}L_{n-1}^1\left(\frac{2\kt^2}{\eB}\right)L_{m}^0\left[\frac{2(k-p)_{\perp}^2}{\eB}\right]\nn\\
&\times&\frac{\text{Tr}\left\{ \gamma^\nu \ks_\perp \gamma^\mu (\ks_\p-\ps_\p+m_f)\Op^{-}\right\}}{\left[k_\p^2-m_f^2-2n\left |  q_fB\right |\right]\left[(k-p)_{\parallel}^2-m_f^2-2m\left |  q_fB\right |\right]}+\text{C.C.}
\label{t7}
\eea

\bea
t_8^{\mu\nu}&=&2g^2\int\frac{d^4k}{\dpi^4}\widetilde{E}\left(\pt,\kt\eB\right)\sum_{n=1,m=1}^{\infty}(-1)^{n+m}L_{n-1}^1\left(\frac{2\kt^2}{\eB}\right)L_{m-1}^0\left[\frac{2(k-p)_{\perp}^2}{\eB}\right]\nn\\
&\times&\frac{\text{Tr}\left\{ \gamma^\nu \ks_\perp \gamma^\mu (\ks_\p-\ps_\p+m_f)\Op^{+}\right\}}{\left[k_\p^2-m_f^2-2n\left |  q_fB\right |\right]\left[(k-p)_{\parallel}^2-m_f^2-2m\left |  q_fB\right |\right]}+\text{C.C.}
\label{t8}
\eea

\bea
t_9^{\mu\nu}&=&4g^2\int\frac{d^4k}{\dpi^4}\widetilde{E}\left(\pt,\kt\eB\right)\sum_{n=1,m=1}^{\infty}(-1)^{n+m}L_{n-1}^1\left(\frac{2\kt^2}{\eB}\right)L_{m-1}^1\left[\frac{2(k-p)_{\perp}^2}{\eB}\right]\nn\\
&\times&\frac{\text{Tr}\left\{ \gamma^\nu \ks_\perp \gamma^\mu (\ks_\perp-\ps_\perp)\right\}}{\left[k_\p^2-m_f^2-2n\left |  q_fB\right |\right]\left[(k-p)_{\parallel}^2-m_f^2-2m\left |  q_fB\right |\right]}+\text{C.C.}
\label{t9}
\eea

In order to perform the sum over Landau levels, a Schwinger parametrization is performed to the denominators:
\bea
\frac{1}{y}=\int_0^\infty e^{-y x} dx.
\eea

For the Eq.~(\ref{t1})

\bea
t_1^{\mu\nu}&=&-g^2\int\frac{d^4k}{\dpi^4}\widetilde{E}\left(\pt,\kt\eB\right)\text{Tr}\left\{ \gamma^\nu (\ks_\p+m_f)\Op^{-} \gamma^\mu (\ks_\p-\ps_\p+m_f)\Op^{-}\right\}\nn\\
&\times&\sum_{n,m=0}^{\infty}\frac{(-1)^{n+m}L_n^0\left(\frac{2\kt^2}{\eB}\right)L_{m}^0\left[\frac{2(k-p)_{\perp}^2}{\eB}\right]}{\left[k_\p^2-m_f^2-2n\left |  q_fB\right |\right]\left[(k-p)_{\parallel}^2-m_f^2-2m\left |  q_fB\right |\right]}\nn\\
&=&-g^2\int d^2x\int\frac{d^4k}{\dpi^4}\widetilde{E}\left(\pt,\kt\eB\right)\text{Tr}\left\{ \gamma^\nu (\ks_\p+m_f)\Op^{-} \gamma^\mu (\ks_\p-\ps_\p+m_f)\Op^{-}\right\}\nn\\
&\times&e^{\alpha(\kp)x_1+\beta(\kp)x_2}\sum_{n,m=0}^{\infty}r_1^nL_n^0(s_1)r_2^mL_m^0(s_2)+\text{C.C.}
\eea
where
\begin{subequations}
\bea
\alpha(\kp)&=&\kp^2-m_f^2,
\label{alphadef}
\eea
\bea
\beta(\kp)&=&(\kp-p_\p)^2-m_f^2,
\label{betadef}
\eea
\bea
r_i&=&-e^{-2\eB x_i},\; i=1,2,
\label{ri}
\eea
\bea
s_1=\frac{2\kt^2}{\eB},
\label{s1}
\eea
and
\bea
s_2=\frac{2(k-p)_{\perp}^2}{\eB}.
\label{s2}
\eea
\end{subequations}

By using the generating function of the Laguerre Polynomials, given by
\bea
\sum_{n=0}^\infty r^n L_n^{b}(s)=\frac{1}{\left(1-r\right)^{b+1}}\exp\left(-\frac{r}{1-r}s\right),
\label{GeneratingFunction}
\eea
it can be found that
\bea
t_1^{\mu\nu}&=&-g^2\int \frac{d^2x}{\left(1+e^{-2\eB x_1}\right)\left(1+e^{-2\eB x_2}\right)}\int\frac{d^4k}{\dpi^4}\widetilde{E}\left(\pt,\kt\eB\right)e^{\alpha(\kp)x_1+\beta(\kp)x_2}\nn\\
&\times&\exp\left[ n(x_1)\frac{2\kt^2}{\eB}\right]\exp\left[ n(x_2)\frac{2(k-p)_{\perp}^2}{\eB}\right]\nn\\
&\times&\text{Tr}\left\{ \gamma^\nu (\ks_\p+m_f)\Op^{-} \gamma^\mu (\ks_\p-\ps_\p+m_f)\Op^{-}\right\}+\text{C.C.}
\eea
where it has been defined
\bea
n(x)\equiv\frac{1}{e^{2\eB x}+1}.
\label{ndef}
\eea

Now, for the trace computation, note that
\begin{subequations}
\bea
\left[ \gamma^\mu _\p,\Op^{(\pm)}\right]=0,
\label{Ogammaconmmutator}
\eea
\bea
\Op^{(\pm)} \gamma^\mu \Op^{(\pm)}=\Op^{(\pm)} \gamma^\mu _\p,
\label{Ogammapara}
\eea
\end{subequations}
and therefore
\bea
&&4\text{Tr}\left\{ \gamma^\nu (\ks_\p+m_f) \gamma^\mu _\p(\ks_\p-\ps_\p+m_f)\Op^{-}\right\}+\text{C.C.}\nn\\
&=&4\text{Tr}\left\{ \gamma^\nu (\ks_\p+m_f) \gamma^\mu _\p(\ks_\p-\ps_\p+m_f)\right\}\nn\\
&=&16\left[\left(\kp\cdot\pp+m_f^2-\kp^2\right) g^{\mu\nu} _\p+2\km_\p\kn_\p-\km_\p p^\nu_\p-\kn_\p p^\mu_\p\right].
\eea

Putting all together
\bea
t_1^{\mu\nu}&=&-4g^2\int \frac{\mathcal{I}_1(x_1,x_2)\mathcal{J}_1^{\mu\nu}(x_1,x_2)}{\left(1+e^{-2\eB x_1}\right)\left(1+e^{-2\eB x_2}\right)}d^2x,\nn\\
\eea
with
\bea
\mathcal{I}_1&=&\int\frac{d^2k_\perp}{\dpi^2}\exp\left[-\frac{\kt^2+(k-p)_{\perp}^2}{\eB}\right]\exp\left[ n(x_1)\frac{2\kt^2}{\eB}\right]\exp\left[ n(x_2)\frac{2(k-p)_{\perp}^2}{\eB}\right]\nn\\
\label{I1}
\eea
and
\bea
&&\mathcal{J}_1^{\mu\nu}=\int\frac{d^2\kp}{\dpi^2}e^{\alpha(\kp)x_1}e^{\beta(\kp)x_2}\left[\left(\kp\cdot\pp+m_f^2-\kp^2\right) g^{\mu\nu} _\p+2\km_\p\kn_\p-\km_\p p^\nu_\p-\kn_\p p^\mu_\p\right].
\label{J1}
\eea

The transverse integral $\mathcal{I}_1$ is performed by making the shift
\bea
\kt=\qt+\frac{1-2n(x_2)}{2\left[1-n(x_1)-n(x_2)\right]}\pt,
\label{qtshift}
\eea
which turns the integral into a simple Gaussian form. It is straightforward to prove that
\bea
\mathcal{I}_1&=&\frac{\pi}{\dpi^2}\frac{\eB}{\tanh\left(\eB x_1\right)+\tanh\left(\eB x_2\right)}\nn\\
&\times&\exp\left[-\frac{\tanh(\eB x_1)\tanh(\eB x_2)}{\tanh(\eB x_1)+\tanh(\eB x_2)}\frac{\pt^2}{\eB}\right].\nn\\
\label{I1result}
\eea

For the parallel integral $\mathcal{J}_1^{\mu\nu}$, the appropriate shift is
\bea
l=\kp-\frac{x_2}{x_1+x_2}\pp
\label{ldef},
\eea
and by performing a rotation to Euclidean space, the integral becomes of a Gaussian form in the variable $l^2_E=l_4^2+l_3^2$, and thus
\bea
\mathcal{J}_1^{\mu\nu}&=&\frac{i\pi}{\dpi^2}\exp\left[\frac{x_1x_2}{x_1+x_2}\pp^2-m_f^2(x_1+x_2)\right]\nn\\
&\times&\left[\left(\frac{x_1x_2}{(x_1+x_2)^3}\pp^2+\frac{m_f^2}{x_1+x_2} g^{\mu\nu} p\right)-\frac{2x_1x_2}{(x_1+x_2)^3}\pp^\mu\pp^\nu\right].
\eea

Collecting terms
\bea
t_1^{\mu\nu}&=&-\frac{i\eB}{16\pi^2}g^2\int d^2x\frac{e^{\eB(x_1+x_2)}}{\sinh\left[\eB(x_1+x_2)\right]}\exp\left[\frac{x_1x_2}{x_1+x_2}\pp^2-m_f^2(x_1+x_2)\right]\nn\\
&\times&\exp\left[-\frac{\tanh(\eB x_1)\tanh(\eB x_2)}{\tanh(\eB x_1)+\tanh(\eB x_2)}\frac{\pt^2}{\eB}\right]\nn\\
&\times&\left[\left(\frac{x_1x_2}{(x_1+x_2)^3}\pp^2+\frac{m_f^2}{x_1+x_2}\right) g^{\mu\nu} p-\frac{2x_1x_2}{(x_1+x_2)^3}\pp^\mu\pp^\nu\right].
\eea

Note that the term $t_5^{\mu\nu}$ of Eq. (\ref{t5}) has the same tensor structure as $t_1^{\mu\nu}$. By means of the variable shifts $m'=m-1$ and $n'=n-1$, which produce a factor $e^{-2\eB (x_1+x_2)}$, this gives rise at the same set of transverse and parallel integrals than for the case of $t_1^{\mu\nu}$, i.e.,
\bea
t_5^{\mu\nu}&=&-\frac{i\eB}{16\pi^2}g^2\int d^2x\frac{e^{-\eB(x_1+x_2)}}{\sinh\left[\eB(x_1+x_2)\right]}\exp\left[\frac{x_1x_2}{x_1+x_2}\pp^2-m_f^2(x_1+x_2)\right]\nn\\
&\times&\exp\left[-\frac{\tanh(\eB x_1)\tanh(\eB x_2)}{\tanh(\eB x_1)+\tanh(\eB x_2)}\frac{\pt^2}{\eB}\right]\nn\\
&\times&\left[\left(\frac{x_1x_2}{(x_1+x_2)^3}\pp^2+\frac{m_f^2}{x_1+x_2}\right) g^{\mu\nu} p-\frac{2x_1x_2}{(x_1+x_2)^3}\pp^\mu\pp^\nu\right].
\eea

Adding up these two terms:
\bea
t_1^{\mu\nu}+t_5^{\mu\nu}&=&-\frac{i\eB}{8\pi^2}g^2\int d^2x\coth\left[\eB(x_1+x_2)\right]\exp\left[\frac{x_1x_2}{x_1+x_2}\pp^2-m_f^2(x_1+x_2)\right]\nn\\
&\times&\exp\left[-\frac{\tanh(\eB x_1)\tanh(\eB x_2)}{\tanh(\eB x_1)+\tanh(\eB x_2)}\frac{\pt^2}{\eB}\right]\nn\\
&\times&\left[\left(\frac{x_1x_2}{(x_1+x_2)^3}\pp^2+\frac{m_f^2}{x_1+x_2}\right) g^{\mu\nu} p-\frac{2x_1x_2}{(x_1+x_2)^3}\pp^\mu\pp^\nu\right].\nn\\
&\equiv&\factorglobal f_0(x_1,x_2)f_1^{\mu\nu}(x_1,x_2).
\eea

For the term $t_2^{\mu\nu}$ of Eq.~(\ref{t2}) the trace involved is computed by using Eq.~(\ref{Ogammaconmmutator}) and the relation
\bea
\Op^{(\pm)} \gamma^\mu \Op^{(\mp)}=\Op^{(\pm)} \gamma^\mu _\perp,
\label{Ogammaperp}
\eea
so that
\bea
\text{Tr}\left\{ \gamma^\nu _\perp(\ks_\p+m_f) \gamma^\mu (\ks_\p-\ps_\p+m_f)\right\}+\text{C.C.}=4\left(\kp\cdot\pp-\kp^2+m_f^2\right) g^{\mu\nu} _\perp.
\eea

This results implies that after introducing the Schwinger parametrization, the integration over the transverse momentum gives the same results as those in Eq.~(\ref{I1result}). Moreover, in order to apply  Eq.~(\ref{GeneratingFunction}) it is necessary to perform the shift $m'=m-1$. That shift extracts a factor $-e^{-2\eB x_2}$ from the sum , thus
\bea
t_2^{\mu\nu}&=&-\frac{4\pi\eB}{\dpi^4}g^2\int \,d^2x\,\exp\left[-\frac{\tanh(\eB x_1)\tanh(\eB x_2)}{\tanh(\eB x_1)+\tanh(\eB x_2)}\frac{\pt^2}{\eB}\right]\nn\\
&\times&\frac{e^{-2\eB x_2}}{\left(1+e^{-2\eB x_1}\right)\left(1+e^{-2\eB x_2}\right)}\frac{\eB}{\tanh\left(\eB x_1\right)+\tanh\left(\eB x_2\right)}\nn\\
&\times&\int d^2\kp\left(\kp\cdot\pp-\kp^2+m_f^2\right)e^{\alpha(\kp)x_1}e^{\beta(\kp)x_2}g^{\mu\nu}_\perp.
\eea

The parallel integration is carried out with the help of the momentum shift of Eq.~(\ref{ldef}) which in Euclidean space gives
\bea
t_2^{\mu\nu}&=&-\frac{4i\pi^2\eB}{\dpi^2}g^2\int \,d^2x\,\exp\left[\frac{x_1x_2}{x_1+x_2}\pp^2-m_f^2(x_1+x_2)\right]\nn\\
&\times&\exp\left[-\frac{\tanh(\eB x_1)\tanh(\eB x_2)}{\tanh(\eB x_1)+\tanh(\eB x_2)}\frac{\pt^2}{\eB}\right]\nn\\
&\times&\frac{e^{-2\eB x_2}}{\left(1+e^{-2\eB x_1}\right)\left(1+e^{-2\eB x_2}\right)}\frac{\eB}{\tanh\left(\eB x_1\right)+\tanh\left(\eB x_2\right)}\nn\\
&\times&\nn\\
&\times&\left[\frac{x_1x_2}{(x_1+x_2)^3}\pp^2+\frac{m_f^2}{x_1+x_2}+\frac{1}{(x_1+x_2)^2}\right] g^{\mu\nu} _\perp.
\eea

From the fact that the term $t_4^{\mu\nu}$ has the same tensor structure of $t_2^{\mu\nu}$, it is easy to show that both expressions are related to each other after the exchange $x_1\leftrightarrow x_2$, so that
\bea
t_4^{\mu\nu}&=&-\frac{4i\pi^2\eB}{\dpi^2}g^2\int \,d^2x \frac{e^{-2\eB x_1}}{\left(1+e^{-2\eB x_1}\right)\left(1+e^{-2\eB x_2}\right)}\nn\\
&\times&\exp\left[\frac{x_1x_2}{x_1+x_2}\pp^2-m_f^2(x_1+x_2)\right]\exp\left[-\frac{\tanh(\eB x_1)\tanh(\eB x_2)}{\tanh(\eB x_1)+\tanh(\eB x_2)}\frac{\pt^2}{\eB}\right]\nn\\
&\times&\left[\frac{x_1x_2}{(x_1+x_2)^3}\pp^2+\frac{m_f^2}{x_1+x_2}+\frac{1}{(x_1+x_2)^2}\right]g^{\mu\nu} _\perp,
\eea
and therefore, after manipulating the exponential, it reads
\bea
t_2^{\mu\nu}+t_4^{\mu\nu}&=&-\frac{i\eB}{8\pi^2}g^2\int d^2x\,\frac{\cosh\left[\eB(x_2-x_1)\right]}{\sinh\left[\eB(x_1+x_2)\right]}\exp\left[\frac{x_1x_2}{x_1+x_2}\pp^2-m_f^2(x_1+x_2)\right]\nn\\
&\times&\exp\left[-\frac{\tanh(\eB x_1)\tanh(\eB x_2)}{\tanh(\eB x_1)+\tanh(\eB x_2)}\frac{\pt^2}{\eB}\right]\nn\\
&\times&\left[\frac{x_1x_2}{(x_1+x_2)^3}\pp^2+\frac{m_f^2}{x_1+x_2}+\frac{1}{(x_1+x_2)^2}\right] g^{\mu\nu} _\perp\nn\\
&\equiv&\factorglobal f_0(x_1,x_2)f_2^{\mu\nu}(x_1,x_2).
\eea

For the term $t_3^{\mu\nu}$, the trace is computed with the help of Eqs.~(\ref{Ogammaconmmutator}),(\ref{Ogammapara}) and (\ref{Ogammaperp}):
\bea
&&\text{Tr}\left\{ \gamma^\nu (\ks_\p+m_f)\Op^{-} \gamma^\mu (\ks_\perp-\ps_\perp)\right\}+\text{C.C.}=4\left[\kp^\mu\left(\kt^\nu-\pt^\nu\right)+\kp^\nu\left(\kt^\mu-\pt^\mu\right)\right].
\eea

After introducing Schwinger's parametrization and using the generating function for the Laguerre polynomials (with the shift $m'=m-1$), it yields
\bea
t_3^{\mu\nu}&=&-\frac{8}{\dpi^4}g^2\int d^2x\int d^4k\frac{e^{-2\eB x_2}e^{\alpha(\kp)x_1}e^{\beta(\kp)x_2}}{\left(1+2e^{-2\eB x_1}\right)\left(1+2e^{-2\eB x_1}\right)^2}\nn\\
&\times&\exp\left[-\frac{\kt^2+(k-p)_{\perp}^2}{\eB}\right]\exp\left[ n(x_1)\frac{2\kt^2}{\eB}\right]\exp\left[ n(x_2)\frac{2(k-p)_{\perp}^2}{\eB}\right]\nn\\
&\times&\left[\kp^\mu\left(\kt^\nu-\pt^\nu\right)+\kp^\nu\left(\kt^\mu-\pt^\mu\right)\right].
\eea

The change of variable in Eq.~(\ref{qtshift}) leads to the result
\bea
t_3^{\mu\nu}&=&-\frac{8}{\dpi^4}g^2\int d^2x\int d^4k\frac{e^{-2\eB x_2}\left[\mathcal{I}_2^{\mu\nu}(x_1,x_2)+\mathcal{I}_2^{\nu\mu}(x_1,x_2)\right]}{\left(1+2e^{-2\eB x_1}\right)\left(1+2e^{-2\eB x_1}\right)^2}\nn\\
&\times&\exp\left[-\frac{\tanh(\eB x_1)\tanh(\eB x_2)}{\tanh(\eB x_1)+\tanh(\eB x_2)}\frac{\pt^2}{\eB}\right],
\eea
where
\bea
\mathcal{I}_2^{\mu\nu}&=&\int d^2\kp e^{\alpha(\kp)x_1}e^{\beta(\kp)x_2}\int d^2q_\perp e^{-\eta\,\qt^2}\kp^\mu\left[\qt^\nu+(\sigma-1)\pt^\nu\right],
\eea
with
\begin{subequations}
\bea
\eta\equiv\frac{\tanh(\eB x_1)+\tanh(\eB x_2)}{\eB},
\eea
and
\bea
\sigma\equiv\frac{\tanh(\eB x_2)}{\tanh(\eB x_1)+\tanh(\eB x_2)}.
\eea
\label{etaandsigma}
\end{subequations}

The perpendicular integration has a simple Gaussian form for which the linear term in $\qt$ integrates to zero, yielding
\bea
\mathcal{I}_2^{\mu\nu}&=&\frac{\pi(\sigma-1)}{\eta}\pt^\nu\int d^2\kp e^{\alpha(\kp)x_1}e^{\beta(\kp)x_2}\kp^\mu.\nn\\
\eea

The shift of variable in Eq.~(\ref{ldef}) also implies a Gaussian integration (in Euclidean space), where the linear terms in $l$ vanish after integration. In this way
\bea
\mathcal{I}_2^{\mu\nu}&=&\frac{\pi^2(\sigma-1)}{\eta}\frac{x_2}{(x_1+x_2)^2}\pp^\mu\pt^\nu\exp\left[\frac{x_1x_2}{x_1+x_2}\pp^2-m_f^2(x_1+x_2)\right]\nn\\
&=&-\frac{i\pi^2\eB x_2}{(x_1+x_2)^2}\frac{\tanh(\eB x_1)\pp^\mu\pt^\nu}{\left[\tanh(\eB x_1)+\tanh(\eB x_2)\right]^2}.
\eea

Putting together these results
\bea
t_3^{\mu\nu}&=&-\frac{i\eB\pi^2}{\dpi^4}g^2\int d^2x\frac{x_2 e^{\eB x_1}\sinh(\eB x_1)}{(x_1+x_2)^2\sinh^2\left[\eB (x_1+x_2)\right]}\nn\\
&\times&\exp\left[\frac{x_1x_2}{x_1+x_2}\pp^2-m_f^2(x_1+x_2)\right]\exp\left[-\frac{\tanh(\eB x_1)\tanh(\eB x_2)}{\tanh(\eB x_1)+\tanh(\eB x_2)}\frac{\pt^2}{\eB}\right]\nn\\
&\times&\left(\pp^\mu\pt^\nu+\pp^\nu\pt^\mu\right).
\eea

The structure $t_6^{\mu\nu}$ is obtained from $t_3^{\mu\nu}$ after the shift $n'=n-1$ wichs means introducing a factor $-e^{-2\eB x_1}$, thus
\bea
t_6^{\mu\nu}&=&-\frac{i\eB\pi^2}{\dpi^4}g^2\int d^2x\frac{x_2 e^{-\eB x_1}\sinh(\eB x_1)}{(x_1+x_2)^2\sinh^2\left[\eB (x_1+x_2)\right]}\nn\\
&\times&\exp\left[\frac{x_1x_2}{x_1+x_2}\pp^2-m_f^2(x_1+x_2)\right]\exp\left[-\frac{\tanh(\eB x_1)\tanh(\eB x_2)}{\tanh(\eB x_1)+\tanh(\eB x_2)}\frac{\pt^2}{\eB}\right]\nn\\
&\times&\left(\pp^\mu\pt^\nu+\pp^\nu\pt^\mu\right),
\eea
and therefore
\bea
&&t_3^{\mu\nu}+t_6^{\mu\nu}=-\frac{i\eB}{8\pi^2}g^2\int d^2x\frac{x_2 \cosh(\eB x_1)\sinh(\eB x_1)}{(x_1+x_2)^2\sinh^2\left[\eB (x_1+x_2)\right]}\nn\\
&\times&\exp\left[\frac{x_1x_2}{x_1+x_2}\pp^2-m_f^2(x_1+x_2)\right]\exp\left[-\frac{\tanh(\eB x_1)\tanh(\eB x_2)}{\tanh(\eB x_1)+\tanh(\eB x_2)}\frac{\pt^2}{\eB}\right]\nn\\
&\times&\left(\pp^\mu\pt^\nu+\pp^\nu\pt^\mu\right).
\eea

Coming now to the terms $t_7^{\mu\nu}$ and $t_8^{\mu\nu}$, it can be notice that they share a common tensor form. Starting from $t_3^{\mu\nu}$, the expression for $t_7^{\mu\nu}$ is obtained by replacing $x_1\rightarrow x_2$ and $p\rightarrow-p$. Moreover, $t_8^{\mu\nu}$ is obtained from $t_7^{\mu\nu}$ by performing the shift $m'=m-1$ which amounts to introducing a factor $-e^{-2\eB x_2}$. Implementing these observations, the following results is obtained:
\bea
t_7^{\mu\nu}+t_8^{\mu\nu}&=&-\frac{i\eB}{8\pi^2}g^2\int d^2x\frac{x_1 \cosh(\eB x_2)\sinh(\eB x_2)}{(x_1+x_2)^2\sinh^2\left[\eB (x_1+x_2)\right]}\nn\\
&\times&\exp\left[\frac{x_1x_2}{x_1+x_2}\pp^2-m_f^2(x_1+x_2)\right]\exp\left[-\frac{\tanh(\eB x_1)\tanh(\eB x_2)}{\tanh(\eB x_1)+\tanh(\eB x_2)}\frac{\pt^2}{\eB}\right]\nn\\
&\times&\left(\pp^\mu\pt^\nu+\pp^\nu\pt^\mu\right),
\eea
then
\bea
&&t_3^{\mu\nu}+t_6^{\mu\nu}+t_7^{\mu\nu}+t_8^{\mu\nu}=-\frac{i\eB}{8\pi^2}g^2\int \frac{d^2x}{2(x_1+x_2)^2\sinh^2\left[\eB (x_1+x_2)\right]}\nn\\
&\times&\exp\left[\frac{x_1x_2}{x_1+x_2}\pp^2-m_f^2(x_1+x_2)\right]\exp\left[-\frac{\tanh(\eB x_1)\tanh(\eB x_2)}{\tanh(\eB x_1)+\tanh(\eB x_2)}\frac{\pt^2}{\eB}\right]\nn\\
&\times&\Big[x_1\sinh(2\eB x_2)+x_2\sinh(2\eB x_1)\Big]\left(\pp^\mu\pt^\nu+\pp^\nu\pt^\mu\right)\nn\\
&\equiv&\factorglobal f_0(x_1,x_2)f_3^{\mu\nu}(x_1,x_2).
\eea

Finally, the trace in the term $t_9^{\mu\nu}$ is given by
\bea
\text{Tr}\left\{ \gamma^\nu \ks_\perp \gamma^\mu (\ks_\perp-\ps_\perp)\right\}=4\left[\left(\kt\cdot\pt+\kt^2\right) g^{\mu\nu} +2\kt^\mu\kt^\nu-\left(\pt^\mu\kt^\nu+\pt^\nu\kt^\mu\right)\right].
\eea

After introducing the Schwinger parametrization and performing the sum together with the shift in Eq.~(\ref{qtshift}) is obtained that:
\bea
t_9^{\mu\nu}&=&-\frac{2}{\dpi^4}g^2\int\frac{d^2x}{\cosh^2(\eB x_1)\cosh^2(\eB x_2)}\nn\\
&\times&\exp\left[-\frac{\tanh(\eB x_1)\tanh(\eB x_2)}{\tanh(\eB x_1)+\tanh(\eB x_2)}\frac{\pt^2}{\eB}\right]\int d^2\kp e^{\alpha(\kp)x_1}e^{\beta(\kp)x_2}\nn\\
&\times&\int d^2\qt e^{-\eta\qt^2}\Big[\left(\qt^2+\sigma(\sigma-1)\pt^2\right)g^{\mu\nu}+2\qt^\mu\qt^\nu+2\sigma(\sigma-1)\pt^\nu\pt^\mu\Big],
\eea
where the linear terms in $\qt$ were ignored, and the variables $\eta$ and $\sigma$ are defined in Eqs.~(\ref{etaandsigma}). In Euclidean space, by means of the change of variable given in Eq.~(\ref{ldef}), the parallel integral is easily performed, yielding
\bea
t_9^{\mu\nu}&=&-\frac{2i\pi}{\dpi^4}g^2
\int\frac{d^2x}{(x_1+x_2)\cosh^2(\eB x_1)\cosh^2(\eB x_2)}\mathcal{J}_2^{\mu\nu}(x_1,x_2)\nn\\
&\times&\exp\left[\frac{x_1x_2}{x_1+x_2}\pp^2-m_f^2(x_1+x_2)\right]\exp\left[-\frac{\tanh(\eB x_1)\tanh(\eB x_2)}{\tanh(\eB x_1)+\tanh(\eB x_2)}\frac{\pt^2}{\eB}\right],\nn\\
\eea
where
\bea
\mathcal{J}_2^{\mu\nu}&=&\int d^2\qt e^{-\eta\qt^2}\Big[\left(\qt^2+\sigma(\sigma-1)\pt^2\right)g^{\mu\nu}+2\qt^\mu\qt^\nu+2\sigma(\sigma-1)\pt^\nu\pt^\mu\Big].
\eea

The last integral has a simple Gaussian form and it is straightforward to compute it, yielding
\bea
\mathcal{J}_2^{\mu\nu}&=&\frac{\pi\eB^2}{\left[\tanh(\eB x_1)+\tanh(\eB x_2)\right]^2}\nn\\
&\times&\left[\left(1-\frac{\tanh(\eB x_1)\tanh(\eB x_2)\pt^2}{\eB\left[\tanh(\eB x_1)+\tanh(\eB x_2)\right]}\right) g^{\mu\nu}- g^{\mu\nu} _\perp\right.\nn\\
&-&\left.\frac{2\tanh(\eB x_1)\tanh(\eB x_2)}{\eB\left[\tanh(\eB x_1)+\tanh(\eB x_2)\right]}\pt^\nu\pt^\mu\right].
\eea

Putting all of this together, we get
\bea
t_9^{\mu\nu}&=&-\frac{i\eB^2}{8\pi^2}\int\frac{d^2x}{(x_1+x_2)\sinh^2\left[\eB(x_1+x_2)\right]}\nn\\
&\times&\exp\left[\frac{x_1x_2}{x_1+x_2}\pp^2-m_f^2(x_1+x_2)\right]\exp\left[-\frac{\tanh(\eB x_1)\tanh(\eB x_2)}{\tanh(\eB x_1)+\tanh(\eB x_2)}\frac{\pt^2}{\eB}\right]\nn\\
&\times&\left[\left(1-\frac{\tanh(\eB x_1)\tanh(\eB x_2)\pt^2}{\eB\left[\tanh(\eB x_1)+\tanh(\eB x_2)\right]}\right) g^{\mu\nu}-g^{\mu\nu} _\perp\right.\nn\\
&-&\left.\frac{2\tanh(\eB x_1)\tanh(\eB x_2)}{\eB\left[\tanh(\eB x_1)+\tanh(\eB x_2)\right]}\pt^\nu\pt^\mu\right]\nn\\
&\equiv&\factorglobal f_0(x_1,x_2)f_4^{\mu\nu}(x_1,x_2).
\eea

\chapter{Tensor manipulation of Eqs.~(\ref{fs})}\label{ApBGluonPolPRD}

In order to obtain the tensor structure of Eq.~(\ref{Pienbaseortonormal}), the terms $f_1^{\mu\nu}(x_1,x_2)$, $f_3^{\mu\nu}(x_1,x_2)$ and $f_4^{\mu\nu}(x_1,x_2)$ in Eqs.~(\ref{fs}) have been factorized in a convenient way, so as to avoid the projection procedure which can lead non-physical contributions. The tensor $f_2^{\mu\nu}(x_1,x_2)$ remains unchanged and the  manipulation is made by direct inspection.

For $f_1^{\mu\nu}(x_1,x_2)$:
\bea
f_1^{\mu\nu}(x_1,x_2)&=&\eB\coth\left[\eB(x_1+x_2)\right]\left[\left(\frac{x_1x_2}{(x_1+x_2)^3}\pp^2+\frac{m_f^2}{x_1+x_2}\right)\gmn_\parallel-\frac{2x_1x_2}{(x_1+x_2)^3}\pmu_\parallel\pnu_\parallel\right]\nn\\
&=&\eB\coth\left[\eB(x_1+x_2)\right]\left[\left(\frac{x_1x_2}{(x_1+x_2)^3}\pp^2+\frac{m_f^2}{x_1+x_2}\right)\gmn_\parallel\right.\nn\\
&+&\left.\frac{2x_1x_2}{(x_1+x_2)^3}\left(\pp^2\gmn_\parallel-\pp^2\gmn_\parallel-\pmu_\parallel\pnu_\parallel\right)\right]\nn\\
&=&\eB\frac{\coth\left[\eB(x_1+x_2)\right]}{(x_1+x_2)^3}\left[2x_1x_2\pp^2\,\Pp+\left(m_f^2(x_1+x_2)^2-x_1x_2\pp^2\right)\gmn_\parallel\right].\nn\\
\eea

For $f_3^{\mu\nu}(x_1,x_2)$:
\bea
f_3^{\mu\nu}(x_1,x_2)&=&\frac{\eB}{2(x_1+x_2)^2\sinh^2\left[\eB (x_1+x_2)\right]}\nn\\
&\times&\Big[x_1\sinh(2\eB x_2)+x_2\sinh(2\eB x_1)\Big]\left(\pmu_\parallel\pnu_\perp+\pnu_\parallel\pmu_\perp\right),
\eea
Notice that
\bea
\pmu\pnu=\left(\pp^\mu-\pt^\mu\right)\left(\pp^\nu-\pt^\nu\right)=\pp^\mu\pp^\nu+\pt^\mu\pt^\nu-\left(\pp^\mu\pt^\nu+\pp^\nu\pt^\mu\right),
\eea
therefore,
\bea
\left(\pp^\mu\pt^\nu+\pp^\nu\pt^\mu\right)&=&\pp^\mu\pp^\nu+\pt^\mu\pt^\nu-\pmu\pnu\nn\\
&=&\pp^\mu\pp^\nu+\pt^\mu\pt^\nu-\pmu\pnu+p^2\gmn-p^2\gmn\nn\\
&=&p^2\left(\gmn-\frac{\pmu\pnu}{p^2}\right)+\pp^\mu\pp^\nu+\pt^\mu\pt^\nu-(\pp^2-\pt^2)\left(\gmn_\parallel+\gmn_\perp\right)\nn\\
&=&p^2\left(\gmn-\frac{\pmu\pnu}{p^2}\right)-\pp^2\Pp+\pt^2\Pt-\pp^2\gmn_\perp+\pt^2\gmn_\parallel\nn\\
&=&p^2\left(\gmn-\frac{\pmu\pnu}{p^2}-\Pp-\Pt\right)+p^2\Pp+p^2\Pt\nn\\
&-&\pp\Pp+\pt^2\Pt-\pp^2\gmn_\perp+\pt^2\gmn_\parallel\nn\\
&=&p^2\Pcero-\pt^2\Pp+\pp^2\Pt-\pp^2\gmn_\perp+\pt^2\gmn_\parallel.
\eea
Thus,
\bea
f_3^{\mu\nu}(x_1,x_2)&=&\frac{\eB\Big[x_1\sinh(2\eB x_2)+x_2\sinh(2\eB x_1)\Big]}{2(x_1+x_2)^2\sinh^2\left[\eB (x_1+x_2)\right]}\nn\\
&\times&\left(p^2\Pcero-\pt^2\Pp+\pp^2\Pt-\pp^2\gmn_\perp+\pt^2\gmn_\parallel\right).
\eea

Finally, for $f_4^{\mu\nu}(x_1,x_2)$, from
\bea
\pmu_\perp\pnu_\perp=\pmu_\perp\pnu_\perp+\pt^2\gmn_\perp-\pt^2\gmn_\perp=\pt^2\Pt-\pt^2\gmn_\perp,
\eea
it can be obtained that
\bea
f_4^{\mu\nu}(x_1,x_2)&=&\frac{\eB^2}{(x_1+x_2)\sinh^2\left[\eB(x_1+x_2)\right]}\nn\\
&\times&\Bigg[
\left(1-\frac{\tanh(\eB x_1)\tanh(\eB x_2)}{\eB\left[\tanh(\eB x_1)+\tanh(\eB x_2)\right]}\pt^2\right)\gmn-\gmn_\perp\nn\\
&-&\frac{2\tanh(\eB x_1)\tanh(\eB x_2)}{\eB\left[\tanh(\eB x_1)+\tanh(\eB x_2)\right]}\Pt\nn\\
&+&\frac{2\tanh(\eB x_1)\tanh(\eB x_2)}{\eB\left[\tanh(\eB x_1)+\tanh(\eB x_2)\right]}\pt^2\gmn_\perp\Bigg].\nn\\
\eea

By collecting the common terms of the structures $\Pp, \Pt$ and $\Pcero$, the coefficients of Eqs.~(\ref{Pipara})-(\ref{coefB}) can be found.

\chapter{Vacuum contribution in the HTL and LLL approximations}\label{Ap_Vacuum_HTL_and_LLL}

Here the thermal vacuum contribution of the gluon polarization tensor of Sec.~\ref{sec:GluonPolTenThermoMag} is computed in the LLL and HTL approximations. The result coincides with the obtained after performing the integration over $k_0$ in Eq.~(\ref{PiParallel}), which means that the pure magnetic contribution given in Eq.~(\ref{resultadoFukushima}) remains at $T=0$. Recalling Eq.~(\ref{PiParallel}):
\bea
i\Pi^{\mu\nu}&=&-g^2\sum_f\Big( \frac{\pi |q_fB|}{4\pi^2} \Big)\exp\left(-\frac{p_\perp^2}{2|q_fB|}\right)\nn\\
&\times&\int \frac{d^2k_\parallel}{(2\pi)^2}\frac{[m_f^2-k_\parallel\cdot(k-p)_\parallel]g_\parallel^{\mu\nu}
 +k_\parallel^\mu(k-p)_\parallel^\nu+k_\parallel^\nu(k-p)_\parallel^\mu}{[k_\parallel^2-m_f^2][(k-p)_\parallel^2-m_f^2]},
\eea
the component $k_0$ is isolated to apply the Matsubara formalism:
\bea
i\Pi^{\mu\nu} &=&g^2\sum_f\Big( \frac{\pi |q_fB|}{4\pi^2} \Big)\exp\left(-\frac{p_\perp^2}{2|q_fB|}\right)\int \frac{d^2k_\parallel}{(2\pi)^2}
\frac{1}{[k_\parallel^2-m_f^2][(k-p)_\parallel^2-m_f^2]}\nonumber \\
 &\times& \Big[(m_f^2+k_3^2-k_3p_3)g_\parallel^{\mu\nu}+2(k_3^2-k_3(p_3+p_0))\nn\\
 &+&k_0(p_0 g_\parallel^{\mu\nu}+2(k_3-p_3-p_0)) +k_0^2(-g_\parallel^{\mu\nu}+2) \Big],
\eea
and by the introduction of the imaginary-time method with $p_0=i\omega$ and $k_0=i\widetilde{\omega}_n$ the above equation becomes:
\begin{align}
i\Pi^{\mu\nu}&=i g^2\sum_f\Big(\frac{\pi |q_fB|}{4\pi^2}  \Big)\exp\left(-\frac{p_\perp^2}{2|q_fB|}\right)\nn\\
&\times T \sum_{n=-\infty}^{\infty}\int \frac{dk_3}{(2\pi)^2}\Bigg[
\frac{(m_f^2+k_3^2-k_3p_3)g_\parallel^{\mu\nu}+2(k_3^2-k_3(p_3+i \omega))}{[\tilde{\omega}_n^2+k_3^2+m_f^2][(\tilde{\omega}_n-\omega)^2+(k_3-p_3)^2+m_f^2]} \nonumber \\
&-\frac{(i\widetilde{\omega}_n)(i\omega g_\parallel^{\mu\nu}+2(k_3-p_3-i\omega))}{[\tilde{\omega}_n^2+k_3^2+m_f^2][(\tilde{\omega}_n-\omega)^2+(k_3-p_3)^2+m_f^2]}\nn\\ &+\frac{(i\widetilde{\omega}_n)^2(-g_\parallel^{\mu\nu}+2)}{[\tilde{\omega}_n^2+k_3^2+m_f^2][(\tilde{\omega}_n-\omega)^2+(k_3-p_3)^2+m_f^2]}\Bigg].
\label{tensorsum}
\end{align}

The Matsubara frequencies sums can be performed by the following identities~\cite{LeBellac}:
\begin{align}
I_0 &= T \sum_{n=-\infty}^{\infty}\frac{1}{[\tilde{\omega}_n^2+k_3^2+m_f^2][(\tilde{\omega}_n-\omega)^2+(k_3-p_3)^2+m_f^2]}\nonumber\\ &=- \sum_{s_1,s_2 = \pm1} \frac{s_1s_2}{4E_1 E_2} \Bigr[\frac{1-\tilde{f}(s_1 E_1)-\tilde{f}(s_2 E_2)}{i \omega -s_1 E_1-s_2 E_2} \Bigl] \nonumber \\
I_1 &= T \sum_{n=-\infty}^{\infty}\frac{i \widetilde{\omega}_n}{[\tilde{\omega}_n^2+k_3^2+m_f^2][(\tilde{\omega}_n-\omega)^2+(k_3-p_3)^2+m_f^2]}\nonumber\\ &= -\sum_{s_1,s_2 = \pm1} \frac{s_1s_2 (s_1 E_1)}{4E_1 E_2} \Bigr[\frac{1-\tilde{f}(s_1 E_1)-\tilde{f}(s_2 E_2)}{i \omega -s_1 E_1-s_2 E_2} \Bigl] \nonumber \\
I_3 &= T \sum_{n=-\infty}^{\infty}\frac{(i \widetilde{\omega}_n)^2}{[\tilde{\omega}_n^2+k_3^2+m_f^2][(\tilde{\omega}_n-\omega)^2+(k_3-p_3)^2+m_f^2]}\nonumber\\ &=-\sum_{s_1,s_2 = \pm1} \frac{s_1s_2 (s1 E_1)(s_2E_2-i \omega)}{4E_1 E_2} \Bigr[\frac{1-\tilde{f}(s_1 E_1)-\tilde{f}(s_2 E_2)}{i \omega -s_1 E_1-s_2 E_2} \Bigl].
\label{mastersums}
\end{align}
where $E_1^2=k_3^2+m_f^2$, $E_2^2=(k_3-p_3)^2+m_f^2$ and
\begin{eqnarray}
\tilde{f}(x)=\frac{1}{(e^{x/T}+1)}.
\end{eqnarray}

The vacuum pieces, i.e., the terms independent of the Fermi-Dirac distribution function $\tilde{f}$ are isolated and their contributions are given by:
\begin{align}
i\Pi^{\mu\nu}_{\text{vac}}&=-i g^2\sum_f\Big(\frac{\pi |p_fB|}{4\pi^2}  \Big)\exp\left(-\frac{p_\perp^2}{2|q_fB|}\right) \nonumber \\
&\times\int \frac{dk_3}{(2\pi)^2}\Big(\frac{1}{i \omega-E_1-E_2}- \frac{1}{i \omega+E_1+E_2}\Big)\Bigg[\frac{2-g_\parallel^{\mu\nu}}{4}\left(\frac{i\omega}{E_2}-1\right)
\nonumber \\
&+
\frac{(m_f^2+k_3^2-k_3p_3)g_\parallel^{\mu\nu}+2(k_3^2-k_3(p_3+i \omega))}{4E_1E_2}- \frac{i\omega g_\parallel^{\mu\nu}+2(k_3-p_3-i\omega) }{4E_2}\Bigg].\nonumber\\
\label{finalsumando}
\end{align}

The above result is the same obtained by performing the integration over $k_0$ in Eq.~(\ref{PiParallel}) and performing the analytic continuation $p_0\to i\omega$.

\chapter{Calculation of \texorpdfstring{$\widetilde{P}^L$}{} and \texorpdfstring{$\widetilde{P}^T$}{} when \texorpdfstring{$q^2\ll m_f^2\ll T^2$}{}}\label{Ap_Coef_PL_and_PT_1}

Here the explicit steps that lead to the results for the matter contributions in Eqs.~(\ref{longfinal}) and~(\ref{transfinal}) are shown. First, the temperature effects are introduced into the corresponding projections onto $\widetilde{\Pi}^{\mu \nu}$, Eq.~(\ref{coefficientsT}), using the Matsubara formalism of thermal field theory, obtaining:
\begin{align}
 \widetilde{P}^L&=-g^2T\sum_{n=-\infty}^{\infty}\sum_f\Big( \frac{\pi \eB}{4\pi^2} \Big)\exp\left(-\frac{p_\perp^2}{2\eB}\right) \nonumber \\
 &\times \int \frac{dk_3}{(2\pi)}  
\frac{1}{[\tilde{\omega}_n^2+k_3^2+m_f^2][(\tilde{\omega}_n-\omega)^2+(k_3-p_3)^2+m_f^2]} \nonumber  \\
 &\times \left\{  \Big[4  \tilde{\omega}_n( -\tilde{\omega}_n \omega -k_3 p_3 ) +2\tilde{\omega}_n (\omega_n^2+p_3^2) + 2 \omega (\tilde{\omega}_n^2+k_3^2+m_f^2) \Big] \frac{\omega}{\mathbf{p}^2}
 \nonumber \right.\\ 
 &+ \left. \Big[2(\tilde{\omega}_n^2 \omega^2+k_3^2 p_3^2+2\tilde{\omega}_n \omega k_3 p_3) - (\omega_n^2+p_3^2) (\tilde{\omega}_n \omega +k_3 p_3+m_f^2) \Big] \frac{\omega^2}{\mathbf{p}^2 p^2}
 \nonumber \right. \\
 &+ \left. \Big[\tilde{\omega}_n^2+2\tilde{\omega}_n \omega -k_3^2 +\tilde{\omega}_n \omega +k_3p_3-m_f^2 \Big] \frac{p^2}{\mathbf{p}^2}\right\}\label{paralelterm2}
\end{align}
and
\bea
\widetilde{P}^T&=&-g^2T\sum_{n=-\infty}^{\infty}\sum_f\Big( \frac{\pi\eB}{4\pi^2}\Big)\exp\left(-\frac{p_\perp^2}{2\eB}\right)\nn\\
&\times&\int \frac{dk_3}{(2\pi)}\frac{-\tilde{\omega}_n^2+k_3^2-m_f^2-\tilde{\omega_n}\omega-k_3p_3}{[\tilde{\omega}_n^2+k_3^2+m_f^2][(\tilde{\omega}_n-\omega)^2+(k_3-p_3)^2+m_f^2]}.
\label{paralelterm1}
\eea

To compute the matter contribution the following integral:
\begin{equation}
\int_{-\infty}^\infty dk_3 \frac{\tilde{f(E_1)}}{E_1} = \int_{-\infty}^\infty dk_3\frac{1}{(k_3^2+m_f^2)^{1/2}} \frac{1}{e^{\sqrt{(k_3^2+m_f^2)}/T}+1},
\label{mattercontribution}
\end{equation}
is performed by using the well-known result~\cite{kapusta2006finite}:
\begin{equation}
f_n(\tilde{y})=\frac{1}{\Gamma(n)} \int_0^{\infty}\frac{dx x^{n-1}}{\sqrt{x^2+\tilde{y}^2}} \frac{1}{e^{\sqrt{x^2+\tilde{y}^2}}+1},
\label{Kapustafn}
\end{equation}
so that Eq.~(\ref{mattercontribution}) correspondings to the case with $n=1$ and $\tilde{y}=m_f/T$. In the limit where  $\tilde{y}$ is small, Eq.~(\ref{Kapustafn}) becomes
\begin{equation}
f_1(m_f/T)=-\frac{1}{2} \ln\left(\frac{m_f}{\pi T}\right) - \frac{1}{2} \gamma_e + \ldots,
\end{equation}
therefore
\begin{eqnarray}
\int_{-\infty}^\infty dk_3 \frac{\tilde{f(E_1)}}{E_1} = -\ln\left(\frac{m_f}{\pi T}\right) -\gamma_e. \nonumber \\
\end{eqnarray}

Using the above expression it is straightforward to get
\bea
	\widetilde{P}^L=-\frac{g^2}{8\pi^2}\sum_f\eB \exp\left(-\frac{p_\perp^2}{2\eB}\right)\Bigg [\frac{-\omega^2(p_\perp^2+2p_3^2)}{\mathbf{p}^2 p^2}-1 \Bigg]\left[\ln \Big( \frac{m_f^2}{\pi^2 T^2} \Big)+2\gamma_e \right]
    \label{longfinalC}
\eea
and
\begin{align}	
\widetilde{P}^T&=-\frac{g^2}{8\pi^2}\sum_f\eB \exp\left(-\frac{p_\perp^2}{2\eB}\right) \left(\frac{p_{\perp}^2}{\mathbf{p}^2}\right)\left[ \ln \Big( \frac{m_f^2}{\pi^2 T^2} \Big)+2\gamma_e \right].
\label{transfinalC}
\end{align}
Finally, to obtain Eqs.~(\ref{longfinal}) and~(\ref{transfinal}) an analytical continuation $i\omega\to q_0$ is performed back to Minkowski space.

\chapter{Calculation of \texorpdfstring{$\widetilde{P}^L$}{} and \texorpdfstring{$\widetilde{P}^T$}{} when \texorpdfstring{$m_f^2\ll p^2\ll T^2$}{}}\label{Ap_Coef_PL_and_PT_2}

The change of hierarchy implies that the momentum $k$ is the largest energy scale, and the quark mass $m_f$ is the smallest.  The matter contribution is computed in the same fashion as Appendix~\ref{Ap_Coef_PL_and_PT_1}. 

The temperature effects into the corresponding projections onto $\widetilde{\Pi}^{\mu \nu}$, Eq. (\ref{coefficientsT}), are introduced by using the Matsubara formalism of thermal field theory, obtaining:
\begin{align}
 \widetilde{P}^L&=-g^2T\sum_{n=-\infty}^{\infty}\sum_f\Big( \frac{\pi \eB}{4\pi^2} \Big)\exp\left(-\frac{p_\perp^2}{2\eB}\right)\nn\\
 &\times\int \frac{dk_3}{(2\pi)}
\frac{1}{[\tilde{\omega}_n^2+k_3^2+m_f^2][(\tilde{\omega}_n-\omega)^2+(k_3-p_3)^2+m_f^2]}\nonumber  \\
 &\times \left\{  \Big[4  \tilde{\omega}_n( -\tilde{\omega}_n \omega -k_3 p_3 ) +2\tilde{\omega}_n (\omega_n^2+p_3^2) + 2 \omega (\tilde{\omega}_n^2+k_3^2+m_f^2) \Big] \frac{\omega}{\mathbf{p}^2}
 \nonumber \right.\\ 
 &+ \left. \Big[2(\tilde{\omega}_n^2 \omega^2+k_3^2 p_3^2+2\tilde{\omega}_n \omega k_3 p_3)- (\omega_n^2+p_3^2) (\tilde{\omega}_n \omega +k_3 p_3+m_f^2) \Big] \frac{\omega^2}{\mathbf{p}^2 p^2}
 \nonumber \right. \\
 &+ \left. \Big[\tilde{\omega}_n^2+2\tilde{\omega}_n \omega -k_3^2 +\tilde{\omega}_n \omega +k_3p_3-m_f^2 \Big] \frac{p^2}{\mathbf{p}^2}\right\}\label{paraleltermd}
\end{align}
and  
\bea   
 \widetilde{P}^T&=&-g^2T\sum_{n=-\infty}^{\infty}\sum_f\Big( \frac{\pi \eB}{4\pi^2} \Big)\exp\left(-\frac{p_\perp^2}{2\eB}\right)\nn\\
 &\times&\frac{p_\perp^2}{\mathbf{p}^2} \int \frac{dk_3}{(2\pi)}
\frac{(-\tilde{\omega}_n^2+k_3^2-m_f^2-\tilde{\omega_n}\omega -k_3p_3)}{[\tilde{\omega}_n^2+k_3^2+m_f^2][(\tilde{\omega}_n-\omega)^2+(k_3-p_3)^2+m_f^2]}. \label{paralelterm1d}
\eea

Note that since in the HTL approximation terms proportional to $m_f^2$, $\tilde{\omega}_n$, and $k_3$ in the numerators do not contribute, the calculation of Eqs.~(\ref{paraleltermd}) and~(\ref{paralelterm1d}) involves only two kinds of sums over the Matsubara frequencies given in Eqs.~(\ref{mastersums}), namely:
\begin{equation}
\chi_0\equiv\sum_{s_1,s_2 = \pm1} \frac{-s_1s_2}{4E_1 E_2} \Bigr[\frac{1-\tilde{f}(s_1 E_1)-\tilde{f}(s_2 E_2)}{i \omega -s_1 E_1-s_2 E_2} \Bigl]
\label{suma2ddd},
\end{equation}
and
\begin{equation}
\chi_1\equiv \sum_{s_1,s_2 = \pm1} \frac{s_1s_2 E_1}{4E_2} \Bigr[\frac{1-\tilde{f}(s_1 E_1)-\tilde{f}(s_2 E_2)}{i \omega -s_1 E_1-s_2 E_2} \Bigl] \label{sumad},
\end{equation}
so that $\chi_1=-E_1^2\chi_0$; thus, one should only find an expression for the sum in Eq.~(\ref{sumad}).
In this approximation ($k>q>m_f$) Eq.~(\ref{sumad}) is given by
\begin{equation}
\chi_1= \frac{-p_3}{2 p^2_{\parallel}} \Bigl[\frac{p_3 e^{k_3/T}}{T(e^{k_3/T}+1)^2}+\frac{1}{2} \frac{p_3^2 e^{k_3/T}(e^{k_3/T}-1)}{T^2 (e^{k_3/T}+1)^3}+...\Bigr] \label{sumad2},
\end{equation}
from which Eqs.~(\ref{paraleltermd}) and~(\ref{paralelterm1d}) take the following form
\begin{align}
 \widetilde{P}^L&=-g^2T\sum_{n=-\infty}^{\infty}\sum_f\Big( \frac{\pi \eB}{4\pi^2} \Big)\exp\left(-\frac{p_\perp^2}{2\eB}\right)\nn\\
 &\times\int \frac{dk_3}{(2\pi)}  
\frac{1}{[\tilde{\omega}_n^2+k_3^2+m_f^2][(\tilde{\omega}_n-\omega)^2+(k_3-p_3)^2+m_f^2]}  \nonumber  \\
 &\times \left\{ \Bigl[-4\tilde{\omega}_n^2 \omega + 2 \omega (\tilde{\omega}_n^2+k_3^2)\Bigr]\frac{\omega}{\mathbf{p}^2}
 + \Bigl[2(\tilde{\omega}_n^2 \omega^2+k_3^2p_3^2)\Bigr] \frac{\omega^2}{\mathbf{p}^2 p^2}
 + \Bigl[\tilde{\omega}_n^2-k_3^2\Bigr] \frac{p^2}{\mathbf{p}^2}\right\}
\label{paraleltermdd}
\end{align}
and
\bea  
 \widetilde{P}^T&=&-g^2T\sum_{n=-\infty}^{\infty}\sum_f\Big( \frac{\pi \eB}{4\pi^2}\Big)\exp\left(-\frac{p_\perp^2}{2\eB}\right)\nn\\\nn\\
 &\times&\frac{p_\perp^2}{\mathbf{p}^2} \int \frac{dk_3}{(2\pi)} \left\{ 
\frac{-\tilde{\omega}_n^2+k_3^2}{[\tilde{\omega}_n^2+k_3^2+m_f^2][(\tilde{\omega}_n-\omega)^2+(k_3-p_3)^2+m_f^2]} \right\}. \label{paralelterm1dd}
\eea

Finally, by using Eqs.~(\ref{paraleltermdd}) and~(\ref{paralelterm1dd}) the desired result is achieved. 

\chapter{Calculation of \texorpdfstring{$i\Pi^{\mu \nu}_{(1,1)}$}{}, \texorpdfstring{$i\Pi^{\mu \nu}_{(2,0)}$}{} and \texorpdfstring{$i\Pi^{\mu \nu}_{(0,2)}$}{}}\label{App_Pi11_Pi20_Pi02}

\subsection*{Calculation of $i\Pi^{\mu\nu}_{(1,1)}$}

The contribution from $\Pi^{\mu \nu}_{(1,1)}$ is given by
\bea
    i\Pi^{\mu \nu}_{(1,1)}=-\int \frac{d^4k}{(2 \pi)^4}\text{Tr}\left\{igt_b \gamma^\nu iS^{(1)}(k-p)igt_a \gamma^\mu iS^{(1)}(k)\right\},
    \label{pi11initial}
\eea
which is computed by introducing two Schwinger parameters $x_1$ and $x_2$ as follows:
\begin{equation}
   \frac{1}{k^2-m^2}=-\int_0^\infty dx \ e^{x(k^2-m^2)}.
\label{idenSchwinger},
\end{equation}
so that, after the trace computation yields
\bea
i\Pi^{\mu \nu}_{(1,1)}&=&2\eB^2 g^2\int \frac{d^4k}{(2 \pi)^4}\frac{1}{(k^2-m_f^2)^2\left[(k-p)^2-m_f^2\right]^2}\nn\\
&\times&\left[k_\parallel^\nu(k_\parallel^\mu-p_\parallel^\mu)+k_\parallel^\mu(k_\parallel^\nu-p_\parallel^\nu)-(g_\parallel^{\mu\nu}-g_\perp^{\mu\nu})\left(k_\parallel \cdot\left(k_\parallel-p_\parallel\right) -m_f^2 \right)\right].
\label{ApP11afterSchwinger}
\eea

By applying the Schwinger parametrization, the last equation is
\bea
i\Pi^{\mu \nu}_{(1,1)}&=&2\eB^2 g^2\frac{\partial}{\partial m_{1}^2} \frac{\partial}{\partial m_{2}^2}\int d^2x \int \frac{d^4k}{(2 \pi)^4}\,\mathbb{E}\left(\mathbf{x},\mathbf{m},p,k\right) \nn\\
&\times& \left[ k_\parallel^\nu(k_\parallel^\mu-p_\parallel^\mu)+k_\parallel^\mu(k_\parallel^\nu-p_\parallel^\nu)
-(g_\parallel^{\mu\nu}-g_\perp^{\mu\nu}) \left[ k_\parallel \cdot(k_\parallel-p_\parallel) -m_f^2 \right]\right],
\eea
where
\bea
\mathbb{E}\left(\mathbf{x},\mathbf{m},p,k\right)\equiv\exp\left[(x_1+x_2)\left(k-\frac{x_1 }{x_1+x_2} p\right)^2 + \frac{x_1 x_2}{x_1+x_2}p^2-(m_{1}^2 x_1 + m_{2}^2 x_2)\right].
\eea

The integral over $k$ is performed after the change of variable
\begin{equation}
    l = \left(k-\frac{x_1 }{x_1+x_2} p\right), 
    \label{cambio}
\end{equation}
so that by performing the derivatives with respect $m_1$ and $m_2$, and by replacing $m_1=m_2=m_f$ it follows that
\begin{eqnarray}
i\Pi^{\mu \nu}_{(1,1)}&=& 2\eB^2 g^2 \int\frac{d^4l}{(2 \pi)^4}\int d^2x\, x_1 x_2 \exp\Bigl[(x_1+x_2) (l^2-m_f^2) + \frac{x_1 x_2}{x_1+x_2}p^2 \Bigr] \nonumber \\
&\times& \Bigg\{ 2 l_\parallel^\mu l_\parallel^\nu - 2\frac{x_1 x_2}{(x_1+x_2)^2} p_\parallel^\mu p_\parallel^\nu 
-(g_\parallel^{\mu\nu}-g_\perp^{\mu\nu})\left[l_\parallel^2 -\frac{x_1 x_2}{(x_1+x_2)^2}p_\parallel^2 -m_f^2\right]\Bigg\}.
\label{L7}
\end{eqnarray}

Now, from the fact that
\begin{equation}
p_\parallel^\mu p_\parallel^\nu = - p_\parallel^2 \mathcal{P}_\parallel^{\mu \nu} + p_\parallel^2 g_\parallel^{\mu \nu},
\end{equation}
equation~(\ref{L7}) is
\begin{eqnarray}
i\Pi^{\mu \nu}_{(1,1)} &=& 2\eB^2 g^2\int \frac{d^4l}{(2 \pi)^4}\int d^2x\, x_1 x_2 \exp\left[(x_1+x_2) (l^2-m_f^2) + \frac{x_1 x_2}{x_1+x_2}p^2 \right] \nonumber \\
&\times&\left\{- 2 l_\parallel^\mu l_\parallel^\nu - 2\frac{x_1 x_2}{(x_1+x_2)^2} p_\parallel^2  \mathcal{P}_\parallel^{\mu \nu}+g_\parallel^{\mu\nu}\left(m_f^2+l_\parallel^2 +\frac{x_1 x_2}{(x_1+x_2)^2}p_\parallel^2\right)\right.\nn\\
&-&\left.g_\perp^{\mu\nu}\left(m_f^2+l_\parallel^2 -\frac{x_1 x_2}{(x_1+x_2)^2}p_\parallel^2\right) \right\}.
\end{eqnarray}

To carry out the integration on $l_\parallel$ it is useful to make the replacement:
\begin{equation}
l_\parallel^\mu  l_\parallel^\nu\to \frac{1}{2} g^{\mu \nu}_\parallel l_\parallel^2,
\end{equation}
so that the integral over $l$ factorizes as $d^4l=d^2l_\parallel d^2l_\perp$. By following the same procedure presented in Appendix~\ref{ApAGluonPolTensorPRD}, i.e., a Wick's rotation allows compute the integral as a Gaussian one, to that:
\begin{eqnarray}
i\Pi^{\mu \nu}_{(1,1)} &=& -\frac{2i\eB^2 g^2}{16 \pi^2}  \int d^2x \frac{x_1 x_2}{(x_1+x_2)^4} \exp\left[-(x_1+x_2)m_f^2 + \frac{x_1 x_2}{x_1+x_2}p^2 \right] \nonumber \\
&\times& \left\{2 x_1 x_2  p_\parallel^2  \mathcal{P}_\parallel^{\mu \nu}+g_\parallel^{\mu\nu}\left(m_f^2 (x_1+x_2)^2- x_1 x_2 p_\parallel^2\right)\right.\nn\\
&+&\left.g_\perp^{\mu\nu}\left(-m_f^2 (x_1+x_2)^2- (x_1+x_2) - x_1 x_2 p_\parallel^2\right) \right\}.
\end{eqnarray}

Finally, the variable changes $x_1=s(1-y)$ and $x_2=sy$ gives the result shown in Eq.~(\ref{pi11withyands}).

\subsection*{Calculation of $i\Pi^{\mu\nu}_{(2,0)}$ and $i\Pi^{\mu\nu}_{(0,2)}$}

The second non-vanishing magnetic contribution is given by
\bea
i\Pi_{(2,0)}^{\mu\nu}=-\int \frac{d^4k}{(2\pi)^4}\text{Tr}\left\{ igt_b\gamma^\nu iS^{(0)}(q)igt_a\gamma^\mu iS^{(2)}(k)\right\},
\label{Pi_20_def}
\eea
which after a Schwinger parametrization and the change of variable of Eq.~(\ref{cambio}) results in
\bea
i\Pi_{(2,0)}^{\mu\nu} &=&\eB^2 g^2   \int \frac{d^4l}{(2 \pi)^4} \int d^2x\exp\Bigl[(x_1+x_2) (l^2-m_f^2) + \frac{x_1 x_2}{x_1+x_2}p^2 \Bigr] \nonumber \\
 &\times&\Biggl[\frac{x_2^2}{2!} \text{Tr}\Bigg\{
\gamma^\nu\gamma^\alpha\gamma^\mu\gamma^\beta \left(l_\perp-\frac{x_2 p_\perp}{x_1+x_2}\right)_\alpha\left(l_\perp+\frac{x_1 p_\perp}{x_1+x_2}\right)_\beta\Bigg\} \nonumber \\
&-&\frac{x_2^3}{3!}  \text{Tr} \Bigg\{
\gamma^\nu\gamma^\alpha\gamma^\mu\gamma^\beta \left(l-\frac{x_2 p}{x_1+x_2}\right)_\alpha \left(l+\frac{x_1 p}{x_1+x_2}\right)_\beta \left(l_\perp+\frac{x_1 p_\perp}{x_1+x_2}\right)^2\nn\\
&+&\gamma^\nu\gamma^\mu \left(l_\perp+\frac{x_1 p_\perp}{x_1+x_2}\right)^2m_f^2 \Bigg\} \Biggr]\nn\\
&\equiv&I_1^{\mu\nu}+I_2^{\mu\nu}+I_3^{\mu\nu},
\label{pi20l}
\eea
where
\begin{subequations}
\bea
 I_1^{\mu\nu} &=&\eB^2 g^2  \frac{1}{2!} \int \frac{d^4l}{(2 \pi)^4} \int d^2x\, x_2^2\exp\left[(x_1+x_2) (l^2-m_f^2) + \frac{x_1 x_2}{x_1+x_2}p^2 \right]\nn\\
 &\times&\text{Tr}\Bigg\{
\gamma^\nu\gamma^\alpha\gamma^\mu\gamma^\beta \left(l_\perp-\frac{x_2 p_\perp}{x_1+x_2}\right)_\alpha \left(l_\perp+\frac{x_1 p_\perp}{x_1+x_2}\right)_\beta\Bigg\},
\eea
\bea
I_2^{\mu\nu} &=&-\eB^2 g^2 \frac{1}{3!} \int \frac{d^4l}{(2 \pi)^4} \int d^2x\, x_2^3\exp\left[(x_1+x_2) (l^2-m_f^2) + \frac{x_1 x_2}{x_1+x_2}p^2 \right]\nn\\
&\times&\text{Tr}\Bigg\{
\gamma^\nu\gamma^\alpha\gamma^\mu\gamma^\beta \left(l-\frac{x_2 p}{x_1+x_2}\right)_\alpha \left(l+\frac{x_1 p}{x_1+x_2}\right)_\beta\left(l_\perp+\frac{x_1 p_\perp}{x_1+x_2}\right)^2\Bigg\},
\eea
and
\bea
I_3^{\mu\nu} &=& -\eB^2 g^2 \frac{m_f^2}{3!} \int \frac{d^4l}{(2 \pi)^4} \int d^2x\, x_2^3 \exp\left[(x_1+x_2) (l^2-m_f^2) + \frac{x_1 x_2}{x_1+x_2}p^2 \right]\nn\\
&\times&\text{Tr}\left\{\gamma^\nu\gamma^\mu \left(l_\perp+\frac{x_1 p_\perp}{x_1+x_2}\right)^2\right\}.
\eea
\end{subequations}

After computing the trace, $I_1^{\mu\nu}$ is
\bea
I_1^{\mu\nu} &=&2 \eB^2 g^2 \int \frac{d^4l}{(2 \pi)^4} \int d^2x\, x_2^2 \exp\left[(x_1+x_2) (l^2-m_f^2) + \frac{x_1 x_2}{x_1+x_2}p^2 \right] \nonumber \\
 &\times& \left[ g_\parallel^{\mu \nu } l_\perp^2 -\frac{x_1 x_2}{(x_1+x_2)^2} \left(p^\nu p_\perp^\mu+p^\mu p_\perp^\nu - p_\perp^2g^{\mu \nu}\right)  \right],
\eea
and given that
\bea
p^\nu p_\perp^\mu+p^\mu p_\perp^\nu &=& p_\perp^2 \mathcal{P}_\parallel^{\mu \nu} -(p_\parallel^2-2p_\perp^2)\mathcal{P}_\perp^{\mu \nu} -p^2\mathcal{P}_0^{\mu \nu}- p_\perp^2 g_\parallel^{\mu \nu} + (p_\parallel^2-2p_\perp^2)g_\perp^{\mu \nu}, \label{tensor}
\eea
the integral becomes:
\bea
I_1^{\mu\nu} &=&2 \eB^2 g^2 \int \frac{d^4l}{(2 \pi)^4} \int d^2x\, x_2^2 \exp\left[(x_1+x_2) (l^2-m_f^2) + \frac{x_1 x_2}{x_1+x_2}p^2 \right] \nonumber \\
 &\times& \Bigg[ g_\parallel^{\mu \nu } l_\perp^2 +\frac{x_1 x_2}{(x_1+x_2)^2} \left(p^2\mathcal{P}_0^{\mu \nu} - p_\perp \mathcal{P}_\parallel^{\mu \nu}+(p_\parallel^2-2p_\perp^2)\mathcal{P}_\perp^{\mu\nu}- g_\perp^{\mu \nu}p^2\right)  \Biggr],
\eea
which can be promoted to Gaussian integrals over $l$, so that
\bea
I_1^{\mu \nu} &=&2 \left(\frac{i}{16\pi^2}\right) \eB^2 g^2  \int d^2x\, x_2^2 \exp\left[-(x_1+x_2)m_f^2 + \frac{x_1 x_2}{x_1+x_2}p^2 \right] \nonumber \\
 &\times& \Bigg[ \frac{g_\parallel^{\mu \nu }}{(x_1+x_2)^3}  +\frac{x_1 x_2}{(x_1+x_2)^4} \left(p^2\mathcal{P}_0^{\mu \nu} - p_\perp \mathcal{P}_\parallel^{\mu \nu} +(p_\parallel^2-2p_\perp^2)\mathcal{P}_\perp^{\mu \nu} - g_\perp^{\mu \nu}p^2\right)  \Biggr].\nn\\
\eea

The integral $I_2^{\mu\nu}$, after trace operation is
\bea
I_2^{\mu \nu}&=&-\eB^2 g^2 \frac{2}{3} \int \frac{d^4l}{(2 \pi)^4} \int d^2x\, x_2^3 \exp\left[(x_1+x_2) (l^2-m_f^2) + \frac{x_1 x_2}{x_1+x_2}p^2 \right]\left(l_\perp+\frac{x_1 p_\perp}{x_1+x_2}\right)^2 \nonumber \\
&\times& \Bigg\{ -g^{\mu \nu} \left[l^2 - \frac{x_1 x_2}{(x_1+x_2)^2}p^2 + \frac{x_1-x_2}{x_1+x_2}(l \cdot p)  \right]+ 2l^{\mu}l^{\nu}- \frac{2x_1 x_2}{(x_1+x_2)^2}p^{\mu}p^{\nu}\nn\\
&+&\frac{x_1-x_2}{x_1+x_2} \left(l^{\mu} p^{\nu}+l^{\nu}p^{\mu}\right) \Bigg\}.
\eea

By using Eq.~(\ref{tensor}) and discarding odd powers in $l$ looks like
\bea
I_2^{\mu \nu} &=&-\eB^2 g^2 \frac{2}{3} \int \frac{d^4l}{(2 \pi)^4} \int d^2x\, x_2^3\exp\left[(x_1+x_2) (l^2-m_f^2) + \frac{x_1 x_2}{x_1+x_2}p^2 \right] \nonumber \\
&\times& \Bigg\{\left(l_\perp^2+\frac{x_1^2 p_\perp^2}{(x_1+x_2)^2}\right)\Bigg[ g_\parallel^{\mu \nu} l_\perp^2 -  g_\perp^{\mu \nu} l_\parallel^2 -\frac{x_1 x_2}{(x_1+x_2)^2} g^{\mu \nu} p^2\nn\\
&+&\frac{2 x_1 x_2}{(x_1+x_2)^2} p^2 \left( \mathcal{P}_0^{\mu \nu}+ \mathcal{P}_\parallel^{\mu \nu}+\mathcal{P}_\perp^{\mu \nu}  \right) \Bigg] \nonumber \\
&+& \frac{x_1(x_1-x_2)}{(x_1+x_2)^2}l_\perp^2 \left[ g_\perp^{\mu \nu}p^2  -p^2 \mathcal{P}_0^{\mu \nu}+p_\perp^2 \mathcal{P}_\parallel^{\mu \nu} - (p_\parallel^2-2p_\perp^2)\mathcal{P}_\perp^{\mu \nu}\right] \Bigg\}, 
\eea
and the remaining integral over $l$ yields
\bea
I_2^{\mu \nu} &=&-2\left(\frac{i}{16\pi^2}\right)\eB^2 g^2 \frac{1}{3} \int d^2x\, x_2^3 \exp\left[-(x_1+x_2) m_f^2 + \frac{x_1 x_2}{x_1+x_2}p^2 \right] \nonumber \\
&\times& \Biggl\{  g_\parallel^{\mu \nu} \left[\frac{2}{(x_1+x_2)^4}+\frac{x_1^2}{(x_1+x_2)^5}p_\perp^2\right] -  g_\perp^{\mu \nu} \left[-\frac{1}{(x_1+x_2)^4}-\frac{x_1^2}{(x_1+x_2)^5}p_\perp^2\right] \nonumber \\
&-&\frac{x_1 x_2}{(x_1+x_2)^2}\left[\frac{1}{(x_1+x_2)^3}+\frac{x_1^2}{(x_1+x_2)^4}p_\perp^2\right]\left[ g^{\mu \nu} p^2-2p^2 \left( \mathcal{P}_0^{\mu \nu}+ \mathcal{P}_\parallel^{\mu \nu}+\mathcal{P}_\perp^{\mu \nu}  \right) \right]\nonumber \\
&+& \frac{x_1(x_1-x_2)}{(x_1+x_2)^5} \left[ g_\perp^{\mu \nu}p^2  -p^2 \mathcal{P}_0^{\mu \nu}+p_\perp^2 \mathcal{P}_\parallel^{\mu \nu} - (p_\parallel^2-2p_\perp^2)\mathcal{P}_\perp^{\mu \nu}\right] \Biggr\}. 
\eea

For the integral $I_3^{\mu\nu}$ follows the same procedure: change the momentum integration by an integral over $l$, discard odd powers of $l$ and promote the remaining integration as a Gaussian integral. That computation yields:
\bea
I_3^{\mu \nu} &=&  -\eB^2 g^2 \frac{2}{3} \int \frac{d^4l}{(2 \pi)^4} \int d^2x\, x_2^3 \exp\left[(x_1+x_2) (l^2-m_f^2) + \frac{x_1 x_2}{x_1+x_2}p^2 \right]\nn\\
&\times& \left(l_\perp+\frac{x_1 p_\perp}{x_1+x_2}\right)^2m_f^2\,g^{\mu\nu}\nn\\
&=&-2\left(\frac{i}{16\pi^2}\right)\eB^2 g^2 \frac{1}{3} g^{\mu\nu} m_f^2\nn\\
&\times&\int d^2x\, x_2^3 \exp\left[-(x_1+x_2)m_f^2 + \frac{x_1 x_2}{x_1+x_2}p^2 \right]\left[\frac{1}{(x_1+x_2)^3}+ \frac{x_1^2}{(x_1+x_2)^4}p_\perp^2\right]\nn\\
\eea

Adding $I_1^{\mu\nu}$, $I_2^{\mu\nu}$ and $I_3^{\mu\nu}$ results in an expression for $i\Pi_{(2,0)}^{\mu\nu}$ in terms of $(x_1,x_2)$, which from $ i\Pi_{(0,2)}^{\mu\nu}$ is obtained by the exchange $x_1\leftrightarrow x_2$. Finally, the change of variable $x_1=s(1-y)$ and $x_2=sy$ reproduces the result of Eq.~(\ref{sumPi20and02}).

\printbibliography
\end{document}